\documentclass[11pt,a4paper]{article}
\usepackage{amsmath}
\usepackage{amsfonts}
\usepackage{mathrsfs}
\usepackage{latexsym}
\usepackage{amssymb}
\usepackage{graphicx} 
\usepackage{sidecap}
\usepackage{empheq}
\usepackage[scr=euler]{mathalfa}
\usepackage{xcolor}
\usepackage{hyperref}
\usepackage{caption}
\usepackage{datetime}
\usepackage{eso-pic}% http://ctan.org/pkg/eso-pic
\usepackage{multirow,cite}
\hypersetup{
    colorlinks,
    linkcolor={blue!20!black},
    citecolor={blue!50!black},
    urlcolor={blue!80!black}
}

\usepackage{titlesec}
\usepackage{tocloft}

\usepackage{etoc}

\usepackage{enumitem}

\setlist[description]{font=\normalfont}

\numberwithin{equation}{section}
\makeatletter
\renewcommand{\@seccntformat}[1]{%
  \csname the#1\endcsname.\ }
\makeatother

\counterwithin*{footnote}{section}

\def\a{\alpha}
\def\b{\beta}
\def\g{\gamma}
\def\e{\epsilon}
\def\l{\lambda}

\def\s{\sigma}
\def\d{\delta}
\def\om{\omega}
\def\D{\Delta}
\def\Dmu{\mathcal{D}_\mu}

\def\O{\mathcal{O}}
\def\L{\mathcal{L}}
\def\P{\mathcal{P}}
\def\J{\mathcal{J}}
\def\H{\mathcal{H}}
\def\X{\mathcal{X}}
\def\N{\mathcal{N}}
\def\r{\rightarrow}
\def\p{\partial}

\newcommand\bi{\begin{itemize}}
\newcommand\ei{\end{itemize}}
\newcommand\be{\begin{equation}}
\newcommand\ee{\end{equation}}
\newcommand\bea{\begin{eqnarray}}
\newcommand\eea{\end{eqnarray}}

\title{From black holes to solvable irrelevant deformations and back
\vspace{6mm}}

\date{}

\author{ 
Monica Guica$^{\S,\dag,\sharp }$\vspace{1mm}\\
\\\vspace{1mm}
${}^\S$\emph{\small Universit\'e Paris-Saclay, CNRS, CEA,} 
\emph{\small Institut de Physique Th\'eorique, 91191 Gif-sur-Yvette, France} \\ %\vspace{1mm}
 \vspace{1mm}
${}^\dag$\emph{\small Institute of Physics, Ecole Polytechnique Fed\'erale de Lausanne, CH-1015 Lausanne, Switzerland} \\ 
${}^\sharp$\emph{\small Theoretical Physics Department, CERN, CH-1211 Geneva 23, Switzerland }}

\begin{document}

\maketitle

%\abstract{Can't have this, fucks up the spaces}
%
%\AddToShipoutPictureBG*{
%  \AtPageUpperLeft{
%    \hspace{\paperwidth}
%    \raisebox{-\baselineskip}{
%      \makebox[0pt][r]{\today ~~ \currenttime ~~~~~}
%}}}

\vskip0.5cm

\begin{center}
\textbf{Abstract}
\end{center}

\vskip10mm

\noindent This is a combined review on the Kerr/CFT correspondence on the one hand and solvable irrelevant deformations  of two-dimensional QFTs - specifically, the $T\bar T$ and $J\bar T$ deformations - on the other. These  subjects are interconnected, since the microscopic description of general black holes can be linked to 
very special irrelevant deformations of two-dimensional CFTs; conversely, one may draw interesting
 insights  into  black hole microscopics   from  the study of the exactly solvable $T\bar T$ and $J\bar T$ deformations. 

\vskip2mm

\noindent The Kerr/CFT part emphasizes the conceptual challenges faced by this proposed holographic description of general extremal black holes, especially in light of  recent advances indicating that the classical geometry of extremal  black holes is unreliable.  The review of the $T\bar T$ and $J\bar T$ deformations is self-contained and presented from a purely field-theoretical perspective.  It covers core topics such as the finite-size spectrum, thermodynamics, scattering, non-perturbative definition and the holographic dictionary for these deformations. Particular emphasis is placed upon the extended symmetries of  $T\bar T$ and $J\bar T$ - deformed CFTs, including the perfect match  between the symmetries  derived via field-theoretic methods and the asymptotic symmetries of the dual spacetimes. These symmetries are also central to understanding the precise relationship between single-trace $T\bar T$ and $J\bar T$ - deformed CFTs and three-dimensional asymptotically linear dilaton and, respectively,  warped AdS backgrounds.

\newpage

\section*{Author's note}

The initial incentive to write this review came from an administrative requirement to prepare a thesis encompassing the author's post-PhD research.  I used this opportunity to write two reviews, one for each  of the two main subjects I worked on: the Kerr/CFT correspondence on the one hand, and solvable irrelevant deformations of two-dimensional QFTs on the other. These subjects are interrelated, so another goal of this work is to lay out the arguments connecting them.  

A soul-searching review of the Kerr/CFT correspondence seems particularly timely. There are currently two main review articles\footnote{See also the relevant sections of \cite{Simon:2011zza}.} on Kerr/CFT: one short and more philosophical \cite{Bredberg:2011hp} and one extensive and more complete  \cite{Compere:2012jk}, which summarize much of the pre-2016 status of the field. However, 2016 marked the beginning of a mini-revolution in AdS$_2$ holography  \cite{Maldacena:2016upp}, which eventually led to the conclusion that the classical geometry of extremal non-supersymmetric black holes -  which plays a central role in the Kerr/CFT correspondence - suffers from large quantum fluctuations and is thus not trustworthy \cite{Iliesiu:2020qvm}.  It would therefore seem that the entire Kerr/CFT reasoning is invalidated by this result. 

Of course, long before these results appeared, Kerr/CFT was known to suffer from many conceptual problems that I will summarize, some of them related to the so-called `AdS$_2$ non-dynamics problem' - a harbinger of the issues uncovered in \cite{Iliesiu:2020qvm}. To avoid this problem, and also to understand the microscopic origin of black hole entropy, part of the Kerr/CFT subcommunity turned to studying `warped AdS$_3$ holography', which can be viewed as a toy model for the Kerr/CFT correspondence. In this review,  we conceptually separate the Kerr/CFT problem (which does suffer from the issue of  large quantum fluctuations) from the warped AdS holography  problem (which does not) and suggest ways in which the former correspondence may survive once a small physical temperature is introduced. 

Warped AdS$_3$ holography has been studied using both top-down and bottom-up methods. While  the bottom-up approach is invaluable in making progress in non-AdS holography, the warped AdS example provides an instructive opportunity to expose its many potential pitfalls. This underscores the importance of using complementary approaches, such as a string-theoretical embedding or explicit, independently-defined field-theoretical examples, to ensure one remains on the right track. 

%The top-down approach indicates that the holographic dual to warped AdS$_3$ spacetimes is a UV-complete two-dimensional QFT that is local and conformal on the left, non-local on the right, and can be viewed as a finely-tuned irrelevant deformation of a $2d$ CFT by operators starting with  dimension $(1,2)$ one. I will denote such a theory as a \emph{dipole CFT}.  Studies of non-extremal black holes links them to $2d$ CFTs deformed  also by $(2,2)$ operators. 

$T\bar T$ and $J\bar T$ - deformed CFTs are explicit examples of  non-standard QFTs (UV- complete, yet non-local) that are useful for understanding  non-AdS holography. In particular, 
$J\bar T$ - deformed CFTs possess the same basic `\emph{dipole CFT}' structure  as top-down holographic duals to warped AdS spacetimes. While  several reviews of the $T\bar T$ deformation exist already: \cite{cernlect}, upon which I expand, and \cite{Jiang:2019epa,He:2025ppz},  I tried to complement these works by  including a comprehensive review of also $J\bar T$ deformations and a detailed discussion of the extended symmetries of these theories, which are   interesting in their own right. I also included a detailed account of the relation between single-trace $T\bar T$/$J\bar T$ deformations and asymptotically linear dilaton/warped AdS spacetimes, a topic the literature often presents in a confusing manner.

% often presented confusingly in the literature. 

Readers only interested in the field-theoretic aspects of the $T\bar T$ and $J\bar T$ deformations can focus on  sections \ref{solvirrelsec}, \ref{infsymmsec} and \ref{corrfsec}. 
Those interested  in Kerr/CFT and non-extremal black holes can  read sections \ref{bhstoirrelsec} and \ref{newperspold}; parts of section \ref{holostr}  are also relevant.  For precision holography for $T\bar T$ and $J\bar T$ - deformed CFTs, it is sufficient to read  parts of section \ref{solvirrelsec} and \ref{holointdtr}-\ref{infsymmsec}. Section \ref{holostr}, as well as parts of section \ref{bhstoirrelsec} and \ref{afvsald} deal with non-AdS holography. Throughout this review, I have tried to focus more on conceptual aspects than on reproducing the calculations in the literature. I hope the reader will find these notes useful. 

Given the vast literature on both  $T\bar T$  and Kerr/CFT, I have  adopted a selective approach to citations. References are limited to works essential for understanding the text, excluding papers I have not studied in sufficient depth,  or whose results or interpretations I found unconvincing or redundant. This choice is intended to aid readability and keep the bibliography manageable. I apologize to authors whose work is not cited for these reasons;   citations to the broader literature can be found in the references.

 %I thus mostly referred to those works that are necessary to understand the main text,  and   refrained from citing articles that I have not read or understood to a reasonable degree,  as well as articles with whose results or interpretation I disagree or find superfluous.    This is both in the interest of the readers, and to bring references to  a manageable size.   Citations to all the relevant literature can be found in the references. I appologize in advance to authors whose work I have not cited for these reasons. 

%\newpage

\tableofcontents

\newpage

\section{Introduction}

Black holes are some of the most fascinating objects in physics, from both an observational and a theoretical point of view. Their study -   in particular, the study of their thermodynamic properties - has been key to
 progress in quantum gravity,  culminating in the discovery of the holographic principle \cite{tHooft:1993dmi,Susskind:1994vu},  believed to be a fundamental property of quantum gravity. 
 
 The aforementioned  thermodynamic considerations \cite{Bekenstein:1972tm,Bekenstein:1973ur,Bardeen:1973gs} strongly suggest that black holes be assigned an entropy proportional to their horizon area
\be
S_{BH} = \frac{\mathcal{A}_H}{4 G } \label{bhentintro}
\ee
This formula,
known as the Bekenstein-Hawking entropy, is  completely \emph{universal}, as it  applies to all black hole solutions of Einstein gravity coupled to matter fields. %It suggests that the degrees of freedom describing the black hole - which are invisible from a gravitational point of view - are somehow distributed on the surface of its horizon.
Understanding the microscopic origin of this entropy as the log of the number of microstates of an underlying quantum system is often an essential step in  understanding  quantum gravity in the given black hole background. 

 One of the great successes of string theory has been to provide the first microscopic accounting of the Bekenstein-Hawking entropy of a black hole \cite{Strominger:1996sh}. In this original work, the black hole was charged, supersymmetric and five-dimensional, and was
  realised as  a stack of string-theoretical extended objects knowns as D1 and D5 - branes.  Their (supersymmetry-preserving) excitations could be easily counted at weak coupling, whereas  the black hole description became more appropriate   at strong coupling. The logarithm of the leading number of microstates of the D1-D5 system at large charge precisely matched the  Bekenstein-Hawking entropy of the corresponding  black hole, thus providing a statistical interpretation for the latter.  
 
Of course, this explanation was for a very specific black hole within the string theory context. As pointed out later in \cite{Strominger:1997eq}, the microscopic accounting of \cite{Strominger:1996sh}  needed not invoke the details of the string theory construction, and it only depended on universal properties of the black holes in question: namely,  the fact that the full description of its near-horizon region contained a factor of a three-dimensional anti-de Sitter (AdS$_3$) spacetime. The microscopics of such black holes can  always be accounted for using universal features of the AdS$_3$/CFT$_2$ holographic correspondence.
 
One is nonetheless still far from  understanding, in microscopic terms, the Bekenstein-Hawking entropy of  \emph{general, asymptotically flat} black holes, which constitutes  the main motivation behind the  work reviewed in this article. In the remainder of the introduction, I will give a selective overview\footnote{
This overview will be expanded upon in section \ref{bhstoirrelsec}, where more details and  references will also be given. } of the status of our microscopic understanding of such black holes, which will in turn help motivate the study of tractable irrelevant deformations of $2d$ CFTs, to which the bulk of this article is dedicated. %[Below, a logical thread and list of questions, see section \ref{} for a more detailed explanation].
% 

%
% \etocsetnexttocdepth{5}
%    \etocsettocstyle{\subsubsection*{Contents of this section: }}{}
%    \cftsubsubsecindent 34pt
%    \localtableofcontents
% 
\subsection*{Brief overview  of black hole microscopics}

The epitome of a successful, \emph{universal} microscopic explanation of a black hole's entropy is without doubt the ability to reproduce the Bekenstein-Hawking entropy of (asymptotically flat) black holes with an AdS$_3$ factor in their near-horizon region \cite{Strominger:1997eq}. A near-extremal  such black hole can be mapped, via a low-energy decoupling limit,   to a  \emph{non-extremal} BTZ black hole in the  AdS$_3$ near-horizon region.  Above the Hawking-Page temperature, this  black hole - which is thermodynamically stable -  dominates the thermal ensemble, and its entropy can be shown to precisely equal   the \emph{universal} Cardy formula for the entropy of a two-dimensional CFT at high temperatures. %As beautifully shown in \cite{} for CFTs with a large central charge and a sparse light spectrum, the entropy is universal for energies above $c/12$, and also for temperatures above HP (also below, but there it's thermal AdS). 
 Thanks to modular invariance, the derivation of this formula is is analytic and elegant, and also valid at strong coupling. %The evaluation of the entropy is analytic and elegant, and 
 There is no need to invoke string theory, as long as the low-energy gravitational theory in the AdS$_3$ region 
 is known - or assumed - to admit a consistent UV completion. %\footnote{Bla-bla AdS$_3$/CFT$_2$ outside strings; universality vs. control.}. %This type of calculation models the entropy match for all (AF, etc.) black holes in higher-dimensions with an AdS$_3$ factor in their near-horizon (decoupling limit?).  For too low temperatures, Cardy doesn't apply and the BTZ geometry becomes unreliable. 

One should be wary of applying these arguments to black holes that are too close to extremality, namely that reduce to a (near)-extremal BTZ black hole in the near-horizon region.  For such black holes, one may consider an additional low-energy `very near-horizon' limit, whereby one obtains a near-AdS$_2$ spacetime, the dynamics of which is well-modeled by JT gravity. A beautiful set of works \cite{Iliesiu:2020qvm,Heydeman:2020hhw} on the JT gravitational path integral in such backgrounds  showed 
quantum corrections to the geometry  become  important around the energy scale where the thermodynamic description breaks down \cite{Preskill:1991tb}, even to the point of washing out the entropy of extremal non-supersymmetric black holes, together with the classical extremal black hole geometry. %Relatively recently it has been understood  that it is best to not take this limiy completely, and focus instead on near-AdS$_2$ dynamics, which is well-modeled by JT gravity.  
On the CFT side, the applicability of Cardy's formula at such low temperatures depends quite sensitively on how the limit is taken \cite{Pal:2025yvz,Pal:2023cgk}.

It is important to point out %, especially in view of our subsequent discussion,
 that JT gravity - and the Schwarzian mode  it reduces to - only captures the dynamics and entropy contribution  of the near-extremal fluctutations, but not the leading piece of the full black hole entropy, which is an input from the JT  viewpoint. This is related to the fact that the very-near-horizon limit focuses the attention on a \emph{subset} of states in the original CFT,  being similar to a near-lightcone `DLCQ' limit. 
 Since it relates high and low temperatures,
%
%can only model departures from the leading entropy at extremality, not the extremal entropy itself. 
the modular invariance that is essential in deriving Cardy's formula is no longer visible in this limit;  %similar remarks apply to the symmetries of the UV CFT$_2$. %, which become difficult to see from the very near horizon limit.
% Thus,  the JT approach does not address the question we are trying to answer: the microscopic origin of the near-extremal BTZ entropy. For that, % it can only be answered in the ``UV CFT'' dual to 
%For that, 
to recover it, 
one instead needs the original AdS$_3$ decoupling limit, which  yields a fully dynamical and UV-complete \emph{theory}\footnote{We are using the term `theory' to mean a  UV-complete quantum theory, in particular one to which arguments based on modular invariance can be applied. By contrast, a `subset of states' is a truncation of a theory to a particular subsector, which may allow for a nice  effective description, but which is not complete, e.g. is not closed under modular transformations. }. %, wherein arguments based upon modular invariance can be used to derive the entropy. % that describes the low-energy dynamics of the system. 
%[\emph{Later:} The near-extremal limit is delicate also because of  issues  of dominance, as it is no longer clear the black hole alone dominates the gravity partition function, while on the CFT side, it is not clear the entropy is   any longer universal (thus, even if $S_{BH} = Cardy$, it is hard to argue this must be so). This issue is particularly visible in the (rich) supersymmetric setting, where the black hole only exists for a restricted set of charges (Cardy may though still work). \emph{Understand!}] 
%
%The key to this success is the existence of a near-horizon  \emph{decoupling limit}, which yields a fully dynamical and UV-complete \emph{theory}  that describes the very low-energy dynamics of the system.  Different near-extremal black holes correspond to different thermal states in the CFT. The fact one can recover the full spectrum justifies one to use the Cardy formula (a UV-IR relation) to reproduce the entropy. 
%
%Studies in higher dimensions suggest that the extremal black holes does not dominate the microcanonical ensemble (gas does) and that in the susy case, dressed black hole configurations are the gravity dual of supersymmetric CFT states that don't satisfy the charge constraint.
% While this is very rich and interesting, we prefer to stick to universal explanations, and thus higher $T$.  Moral: if we'd like a simple and universal explanation of bh entropy, should stick to classes of slightly near-extremal black holes.

So far for  black holes with an AdS$_3$ factor in their near-horizon region. The next level of difficulty are \emph{general near-extremal} black holes, which include the  near-extremal rotating Kerr black holes that may in principle occur in our own universe. For such black holes,  a  near-horizon  limit exists, in which the geometry simplifies dramatically and universally contains a  so-called \emph{warped} AdS$_3$ factor (a spacetime that is a deformed - squashed or stretched - version of AdS$_3$ and possesses $SL(2,\mathbb{R}) \times U(1)$ isometry) together with some compact directions. The  Kerr/CFT correspondence \cite{Guica:2008mu} draws insight from %exploits 
 the asymptotic symmetries of this near-horizon spacetime to propose that the extremal Kerr black hole is holographically described by a chiral half of a two-dimensional CFT. As a check, the Cardy entropy of this putative dual CFT is shown to match the Bekenstein-Hawking entropy \eqref{bhentintro} of  extreme Kerr. % black hole. 
 The same set of ideas can then be used  to reproduce the entropy of general extremal black holes, in various dimensions and carrying various charges \cite{Compere:2012jk}. 

A word of caution is in place concerning the older literature on the Kerr/CFT correspondence,  produced before the recent advances in JT gravity. In these older works, one oftentimes considers the exactly extremal black hole geometry.  The near-horizon limit for (near)-extremal black holes mentioned above is very similar to the very-near-horizon limit we discussed for (near)-extremal BTZ, and leads again to a (near)-AdS$_2$ factor that  can be treated within the JT gravity approximation. As before, quantum effects become large %\emph{Check!} 
for energies below that where the thermodynamic approximation breaks down so, strictly speaking, the exactly extremal black hole classical geometry does not  exist. One can nonetheless imagine extrapolating %most results of 
the correspondence to slightly non-zero temperature, for energies above extremality that are above the breakdown scale, but still well below the mass of the black hole.  It is in this slightly generalised sense that we will be referring to the Kerr/CFT correspondence throughout. % these notes. 

%
%. %\emph{True that taking near-extremal near horizon can solve quantum corrections problem, but not frozen temperature one?} 
%The reason to consider near-extremal black holes, rather than the original extremal ones is that, as before, the gravitational approximation breaks down in the strict extremal limit (also EFT).  This limit, which most resembles the very near horizon limit of the previous paragraph, is not without its conceptual problems/difficulties, 
%related to the appearance of a (non-dynamical) $AdS_2$ factor, to a frozen subset of states  and to the fact that the resulting geometry may not be stable (the states one focuses upon being metastable)/ entropically dominant (even in AdS is it not clear the  Higgs branch configuration dominates, rather than simply being the chosen one).   However, it has the merit of singling out a universal type of spacetime, \emph{warped AdS$_3$}, whose (presumably also universal) holographic dual should provide a universal microscopic description of extremal black holes (minus the decoupling limit caveats above, which have to do with embedding the decoupled subsector into a UV-complete theory). 

As before, while  JT gravity and the corresponding Schwarzian term can well capture  certain perturbations of near-extremal black holes, it can only account for certain contributions to the near-extreme black hole entropy \cite{Kapec:2023ruw,Rakic:2023vhv}, but not the full answer. The essential difference between the Kerr/CFT correspondence and the JT analysis of the near-extreme Kerr throat is that the former aims to be a full holographic  description of the black hole degrees of freedom, whereas the latter is just an effective description of the near-extremal fluctuations, in which the bulk of the entropy is an input.

 A known problem in realising the Kerr/CFT goal is the absence, for most near-extremal asymptotically flat black holes, of  an intermediate decoupling limit yielding  a warped AdS$_3$ spacetime dual to an entire \emph{theory}, where modular invariance arguments can be used, rather than just a subset of states. For this reason,
one works instead with toy models that allow for similar warped AdS$_3$  solutions, now assumed to be dual to full theories. Within these toy models, one tries %has a potentially well-defined framework for addressing the question: decoupled, dominant corresponding to certain non-AdS decoupling limits \emph{Careful!} in string theory. The leftover task is 
to identify the universal properties of the holographic duals to warped AdS$_3$ spacetimes using bottom-up holographic methods, such as black hole thermodynamics, asymptotic symmetries, etc. Interestingly, one finds two \emph{distinct} universality classes of toy models for warped AdS$_3$: one that suggests  a so-called \emph{warped CFT} structure for the dual theory, and one that we will refer to as a \emph{dipole CFT} structure. Since all known toy models that can be embedded in a consistent theory of quantum gravity (i.e., string theory) are of the latter type, we will advocate that the holographic description of near-extremal black holes should be in terms of a dipole CFT. As we explain in the following subsection, these should correspond to UV-complete, non-local QFTs obtained via a certain,  special type of finely-tuned irrelevant deformation of  a two-dimensional CFT.

%How these theories are related to the original black hole is not as straightforward,  having to do with the decoupling limit: extreme Kerr may not be the dominant, perhaps just metastable, not clear how to embed the near-horizon wAdS into one with full dynamics. Maybe this makes sense as long as energies are small compaerd to irrelevant coupling. [The main subject of this thesis is to study in detail QFT toy models that share some essential universal features to the theories of interest.] Whether the NHEK decoupling limit actually yields a theory rather than a collection of states  should be better studied. \emph{In the stringy black hole example true it's a theory? What is $\l$ and what is the temperature (after proper parametrisation)? }

The last general  class of asymptotically flat black holes whose entropy we would like to understand are  non-extremal ones, which are clearly of the most phenomenological interest, but also conceptually the most difficult. A notorious problem is the fact that their specific heat is negative, which renders modeling their thermodynamics difficult. Another problem is the absence  of a decoupling limit  of their near-horizon region from the asymptotically flat one.  Despite this, there exist various hints of a CFT structure, such as the presence of an  approximate  conformal symmetry  at low energies and a suggestive rewriting of the non-extremal Kerr entropy as a Cardy formula \cite{Castro:2010fd}.  Attempts to make sense of this picture \cite{Cvetic:2011hp,Cvetic:2011dn} led to the idea that  non-extremal black holes are states in theories obtained by several irrelevant deformations of a CFT,  possibly with the special property that the entropy of  thermal states is unaffected by the deformations \cite{Baggio:2012db}.  This indicates  that non-extremal black holes, too,  may
%fall under  the framework put forth in these notes: namely, that  the black holes of interest are
be  described in the framework of special
irrelevant deformations of $2d$ CFTs.

The  link between the microscopic description of realistic black holes and finely-tuned
irrelevant deformations of two-dimensional CFTs %that may perhaps be tractable, 
constitutes the main motivation - at least from the point of view of these notes -  for studying explicit, tractable examples of   such irrelevant deformations in detail. %We spell out our conclusions and lessons learned in section \ref{}.  
But first, let us explain how this link is motivated. 

%. The scale of the  irrelevant deformation is the same as that of the temperature, so the former can certainly not be ignored. 

\subsection*{Tackling warped AdS$_3$ holography} 
 
As explained above, the question of the microscopic description of near-extremal black holes can be reduced - at least as far as capturing its gist  is concerned -   to a question about warped AdS$_3$, for which the universal behaviour of near-extremal black holes suggests a universal holographic description. More precisely, one would like to know: what type of field theory is dual to a warped AdS$_3$ spacetime, assuming the existence of a consistent UV completion?

Before delving into the details, let us make a few general comments about how one may approach such a question in non-AdS holography, emphasizing the  subtleties one may encounter.  % than is usually discussed. 

%\bigskip

\subsubsection*{General comments on non-AdS holography}

%\medskip 
 
\noindent Generally speaking,  given a holographic duality that one would like to understand - by which we mean finding the microscopic  description of a given class of asymptotic backgrounds\footnote{We will concentrate on set-ups where the supergravity description is reliable asymptotically, thus excluding the well-known $D_{p<3}$ - brane backgrounds, where stringy corrections become important in the asymptotic region. } -  one may consider a top-down or a bottom-up approach to
uncovering it.  Here, by top-down we mean that there exists an explicit decoupling limit of string theory that yields the spacetime of interest; when this is the case, one has a reasonable guarantee of the existence of the holographic dual, and that checks of the proposed duality should go through, the correspondence being embedded inside a theory of quantum gravity (i.e., string theory) that is widely believed to be consistent. These are the main advantages of the top-down approach.  On the down side, this approach works on a case-by-case basis, producing specific examples of holographically dual pairs, and it is not always clear which of their properties are universal, or generalisable, or necessary for the correspondence to work. 

By contrast,
%
% The top-down is  via explicit decoupling examples , having the advantage of control and existence; 
 %
 the bottom-up approach consists of studying various properties of a spacetime with given asymptotics and matching them to  corresponding properties of a putative dual theory. These properties can be either universal - such as symmetries,  the  density of states at high energy - or not, such as the low-lying energy spectrum.  A common assumption is that the spacetime asymptotics encode the `type' of dual boundary theory, while different representative boundary theories can be obtained by changing the low-energy gravitational action that admits this spacetime  as a solution, assuming it allows for a UV-completion.  Conversely, a given  `type' of boundary theory (defined  via  a set of axioms, as one has  e.g., for  CFTs) will be holographically dual to a particular kind of spacetime, assumed to be emergent from it.

  In the case of AdS/CFT, the top-down approach and the associated precision tests of the duality were essential in establishing, to a high degree of confidence, the existence of a non-trivial equivalence between a theory of gravity in AdS and a lower-dimensional CFT; the top-down approach provides a \emph{derivation} - albeit limited -  of the AdS/CFT correspondence. The bottom-up approach is nonetheless key to understanding universal aspects of the correspondence, such as the emergence of AdS space-time from the dual CFT \cite{El-Showk:2011yvt}. %, etc., 
Another advantage of this approach is   
  that, in principle, it goes beyond the string-theoretical context, its conclusions also holding for any other consistent theory of quantum gravity that is yet to be found. %  (though no examples so far - how about Vasiliev?).
   The strength of the bottom-up approach in the AdS/CFT context  relies on the fact  that the boundary side of the correspondence  involves an extremely well-studied set of field theories, namely CFTs, for which an axiomatic universal definition at strong coupling is available and possible to work with. In particular, the idea that many different bulk actions are possible
 %  
%   the gravitational action doesn't matter
    is rooted in the fact that, at strong coupling, large $N$ bootstrap   does not constrain the spectrum or the interactions of the AdS theory, besides requiring locality \cite{Heemskerk:2009pn}, though  more refined constraints do exist.  %\emph{True? Status matter vs gravity?}

The reasons  holography beyond AdS is difficult are manifold. First, string theory does not provide examples of decoupling limits to many spacetimes of interest: for example, there is no known decoupling limit that yields flat space, and the string-theoretical construction of de Sitter backgrounds is a notoriously hard and subtle problem.
Second, even when string theory arguments do signal the existence of a duality, the boundary side of it may prove to be intrinsically intractable%[in its own terms]
, with the only insights into its properties coming from
% weren't it for 
 the dual gravitational description. %[In many  of the string-theoretical examples of non-AdS decoupling limits (where the supergravity approximation is asymptotically valid), the dual theories are non-local and their full structure is not well understood (note need to go to strong coupling).] %, and holography may be in fact the best way to investigate its properties. 
 One  thus  has the choice of either studying holography for a handful of non-AdS spacetimes, whose interest is not immediately obvious, and for which string theory provides at best some helpful hints, or attempt a bottom-up approach for the spacetimes of interest, such as asymptotically flat, de Sitter, etc.
 
 In the bottom-up approach to non-AdS holography,  one seeks to infer from the gravitational side of the duality the universal properties of the boundary theory.  In the following, I will assume this is a quantum theory, so one has a notion of a Hilbert space and of operators acting on it. 
 
\begin{figure}[h]
\centering
\includegraphics[height=4.5cm]{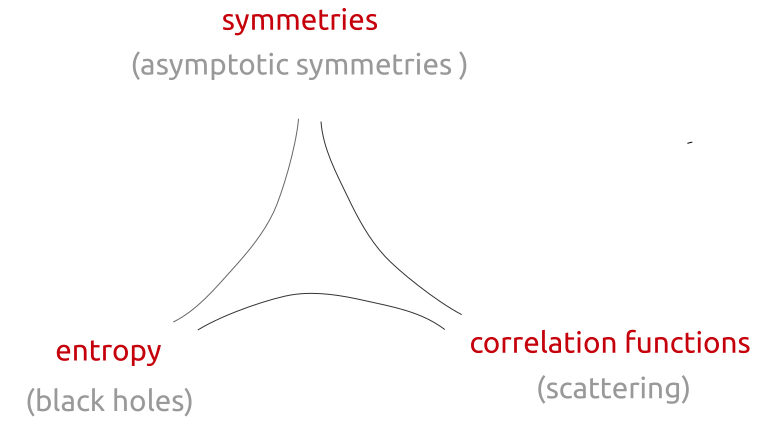} 
\caption{\small{The triad of properties that we will be using to characterise  theories.} %\textcolor{red}{\emph{Rewrite!}}
}
\label{generaltriad}
 \end{figure}

 Throughout these notes, I will focus on three representative properties, depicted in figure \ref{generaltriad},  that one can easily access in holography: the symmetries of the dual theory -  encoded in the asymptotic symmetries of the bulk; the entropy and other thermodynamic quantities - which can be read off from black hole  thermodynamics in the bulk -  and correlation functions - which can be inferred from bulk scattering experiments.   Of course, there exist many other observables of interest, such as entanglement entropies, Wilson loops  etc, but we will mainly concentrate on the above triad. The  main challenge of bottom-up holography is then to put together these pieces of data into a consistent whole,  namely a logically self-contained and mathematically consistent theory that would present all these different, yet interrelated properties.

\medskip

\vskip-2mm 
 
\noindent The reason this task is less straightforward than may na\"{i}vely appear is two-fold: i) there are ambiguities in extracting the boundary observables from the gravitational data and ii) the possible consistent trios of `quantum theory properties', namely the  pairing between symmetries, thermodynamics and correlation functions is not \emph{a priori} known for the boundary dual of some given spacetime.  In absence of a concrete realisation of the boundary theory via e.g. an explicit example,  one does not know what to expect, nor can one fully check whether a given trio  originates from a consistent theory, or how  should it be interpreted for this to be the case.

Concerning the first point, an important source of ambiguities in holography %duality 
stems from a lack of knowledge of \emph{the} correct boundary conditions for the fluctuations of the bulk fields. As will be thoroughly exemplified in these notes, changing the boundary conditions for the bulk fields can completely change the nature of the holographic dual, so they are a key part of the definition of the bulk theory. Nonetheless, in bottom-up holography the boundary conditions are usually the result of an educated guess, while even in top-down holography it is not entirely clear what they should be, though additional clues are usually present. The asymptotic symmetries are highly sensitive to changes in the boundary conditions\footnote{A situation that is not uncommon in low-dimensional holography is one where the boundary conditions one imposes select a subsector of the theory, with restricted dynamics. The symmetries that act on this subsector may be enlarged, or simply easy to be given a different interpretation than if one had access to general states in the theory. For this reason, choosing boundary conditions that select a complete, dynamical phase space is extremely important.  }, while the  thermodynamics and correlators can be affected, too. 

The second problem has to do with the fact that, given a quantum theory and a number of `basic properties' for it, such as symmetries, locality properties, etc., one expects the various observables, and in particular the triad we discussed, to be constrained by them in interrelated ways. For example, in $2d$ CFTs one starts with the assumption of a local $2d$ QFT possessing  global conformal symmetry, and derives the fact that the symmetry extends to the full Virasoro algebra, that modular invariance plus scaling symmetry implies Cardy's formula for the entropy, while conformal covariance and locality imply that the form of two - and three-point functions of local primary operators is entirely fixed. In the bottom-up approach to  non-AdS holography, the `basic properties' are sometimes not obvious, or subject to assumptions\footnote{For example, in the celestial holography approach to flat space holography, 
 that the dual theory lives on the celestial sphere  is an assumption.  }, which can translate into further ambiguities in the derived properties, such as those in figure \ref{generaltriad}. Even in top-down examples,  one may not be able to uncover all the hidden symmetries or structures of the boundary theory that constrain its observables, due to lack of analytical control. Finally, even when the properties are spelled out, the full list of constraints and interrelations between observables that they imply could take much additional work to be properly understood. 

\subsubsection*{Warped AdS$_3$ holography} 
 
% \medskip
 
\noindent Compared with the general status of non-AdS holography we overviewed, warped AdS$_3$ holography is in relatively good standing, in that both  top-down and bottom-up approaches are available. 

The top-down construction involves a particular warped AdS$_3$ spacetime that is obtained from a  decoupling limit of the D1-D5 system in presence of a non-trivial background $B$-field. On the gravitational side, one obtains a background that interpolates between an AdS$_3$ spacetime in the interior and a decoupled warped AdS$_3$ background asymptotically, which can be viewed as a non-normalisable deformation of AdS$_3$. The field theory side is intractable in concrete terms; however, one may use the holographic dictionary to infer  the properties of the dual theory from the structure of the bulk solution. 

\medskip

\begin{figure}[h]
\centering
\includegraphics[height=5.5cm]{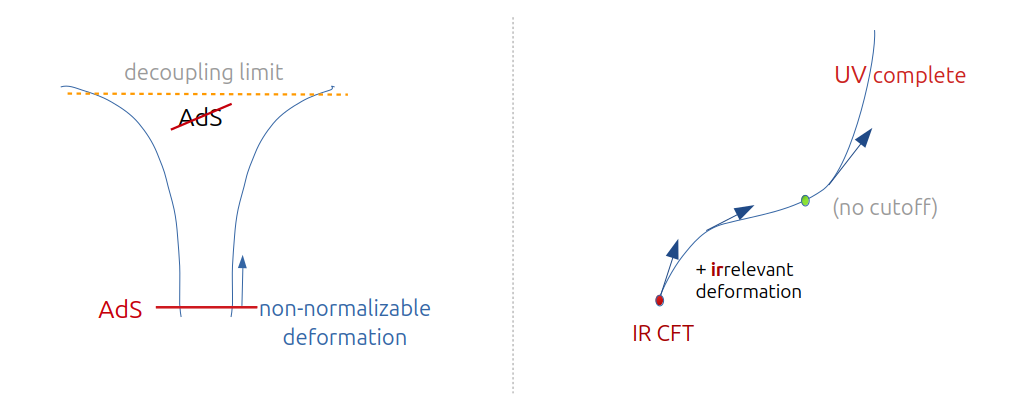}
\caption{\small{The typical holographic situation we will consider. The non-AdS spacetime of interest is obtained via a decoupling limit and corresponds to a non-normalisable deformation of an   AdS spacetime in the interior. The holographically dual picture is that of a UV-complete but non-standard theory that can be obtained via a finely-tuned irrelevant deformation of a CFT.  } }
\end{figure}

\vskip-1mm

\noindent First, the fact that the warped AdS$_3$ spacetime is obtained  via a decoupling limit implies that the dual  theory is very  likely UV-complete. The fact that the geometry becomes AdS$_3$ in the interior implies that the boundary theory flows to the CFT$_2$ dual to it in the IR. The particular form of the non-normalisable deformation of AdS$_3$ that seeds the flow to warped AdS$_3$  maps, using the standard AdS/CFT dictionary, to a source for   an irrelevant  operator of dimension $(1,2)$, whose addition to the CFT action preserves locality and conformal invariance on the left, but not on the right. This, together with the UV-completeness of the full theory, indicates the latter will be non-local on the right, at the scale set by the irrelevant deformation parameter.  Additional irrelevant operators that respect the   $SL(2,\mathbb{R}) \times U(1)$ symmetries of the problem may need to be added, with finely-tuned coefficients, at higher orders in the irrelevant flow,  in order to achieve the expected UV-completeness of the full theory; it is well-known that generic irrelevant deformations typically lead to theories with a cutoff.

%implies that it is given by an irrelevant deformation of the dual CFT$_2$  by a leading operator of dimension $(1,2)$, plus possible additional operators that respect the same symmetries, namely . 

%Since the deformation is exactly marginal on the left, but irrelevant on the right, it will preserve locality and conformal invariance on the left, and makes the theory non-local on the right . The existence of a decoupling limit indicates that the resulting boundary theory is UV-complete, non-local and  decoupled from  gravity -  an argument we will often use.

 While this argument that predicts such a special structure of the boundary theory
   may not appear convincing, it is significantly strengthened by the fact that the same type of decoupling limit applied to D3 - branes in a $B$ field
   yields a tractable boundary theory \cite{Bergman:2000cw}, which can be written as a deformation of $\mathcal{N} = 4$ SYM by a certain     infinite set of irrelevant operators  with finely-tuned coefficients,    in such a way that planar diagrams are unaffected. % [dipole deformation corresponds to adding an infinite number of irrelevant operators to $\N=4$ SYM, with finely tuned coefficients. Hidden property that large N diagram same, leading to the same thermodynamics.]
%
%The situation for warped AdS holography is markedly better. First,  there exist examples of decoupling limits within string theory that yield a warped AdS$_3$ spacetime. This is important, because it means one has the right to expect/hope for a dual description that is decoupled from gravity. The string theory constructions also  indicate the QFT structure of these dual theories: they are irrelevant deformations of $2d$ CFTs that lead to UV complete yet non-local theories with invariance. 
We  refer to UV-complete theories that are obtained via such finely-tuned irrelevant  deformations, which render the theory non-local along one null direction, while preserving locality and conformality along the others, as  \emph{dipole CFTs}.  % \textcolor{red}{\emph{Emphasize more definition?}}
%
 
% definition  xxx.  Thus, look for such a theory in 2d. Looks fully intractable. 
 
It should be rather clear that constructing a UV-complete theory via a finely-tuned irrelevant flow is a virtually  impossible task,  due to the  ambiguities  that proliferate at higher perturbative orders in  the irrelevant  parameter; an alternate definition of the theory, as is the case in \cite{Bergman:2000cw}, is needed. 
 Unfortunately, string theory does not provide any tractable example of a dipole CFT in $2d$, so one must appeal to bottom-up holography methods to infer their properties. Concentrating on the triad of properties we have advocated for, the result is  
 
\begin{figure}[h]
\centering
\includegraphics[height=5.5cm]{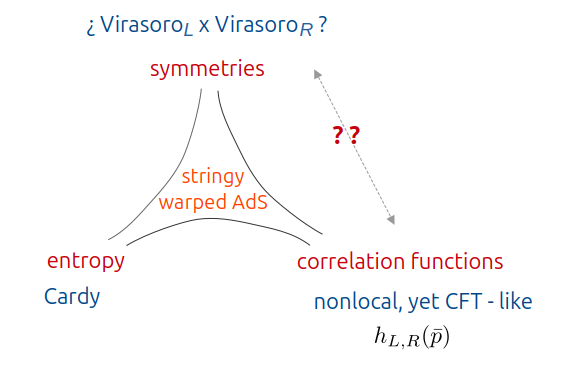}
\caption{\small{Properties of the holographic duals to warped AdS$_3$ spacetimes embedded in string theory, inferred using bottom-up methods.  If it is true that the symmetries contain a Virasoro$_R$ factor, there is a clear tension with  the non-localities exhibited by correlation functions on this side.}}
\label{triadstringsold}
\end{figure}

\noindent As it turns out, since the boundary conditions for warped AdS$_3$  spacetimes are not known, the results of the asymptotic symmetry group (ASG) analyses -  which should indicate the symmetries of the dual theory -   
   are extremely inconclusive:  depending on the chosen boundary conditions and counterterms used,   one can  find an ASG consisting of a single right-moving copy of the Virasoro algebra (same as in Kerr/CFT) \cite{ElShowk:2011cm},  a single left-moving Virasoro \cite{Song:2011sr},  both a left-moving and a right-moving   Virasoro \cite{Compere:2014bia}, or a  set of left Virasoro-Kac-Moody symmetries known as  warped conformal symmetries \cite{Detournay:2012pc}, etc. Black hole entropy follows Cardy's formula.  As for scattering observables, %as in Kerr/CFT \cite{Bredberg:2009pv},
    the momentum-space two-point  functions resemble CFT ones, but with the would-be conformal dimension replaced by a function of the right-moving momentum, in agreement with the expected non-locality of the dual theory on the right. %A drawing of the `trio' is, with our favourite ASG result 

 Very similar trios of properties are obtained for other warped AdS$_3$  backgrounds in string theory -  most of them \emph{not} derived from decoupling limits -  and for near-extremal  black holes, including Kerr\footnote{These models derived from string theory were not the only toy  models proposed to study warped AdS holography - in fact, most of the activity concentrated  on simple, phenomenological models that allow for warped AdS$_3$ solutions, such as topologically massive gravity (TMG), new massive gravity (NMG), or gravity coupled to topological electromagnetism. Besides the fact that these theories likely do not admit a consistent UV completion (TMG and NMG  contain ghosts), it turns out that the triad of properties one extracts from the study of warped AdS solutions in them is quite different from the one extracted from the string-theory toy models (figure \ref{triadstringsold}). 
 We are thus in a case where the models that descend from UV-complete theories are qualitatively different, already at the EFT level, from those that are phenomenological and involve minimalistic couplings. Therefore, we concentrate on the models descending from string theory only, which are dual to dipole CFTs. 
}.  If it is true that the correct asymptotic symmetries contain a right-moving 
 Virasoro factor,  there is a clear tension between them and  the non-local structure displayed by the correlation functions on the same side. 
 In order to determine whether the Virasoro symmetry can be  compatible with the non-locality, it is necessary to  have a concrete handle over the boundary theory, i.e. an explicit QFT construction\footnote{It is also possible (but less interesting) that the correct boundary conditions do not lead to  Virasoro$_R$ symmetries.}.

The main advance in the Kerr/CFT problem that is recounted in these notes is the elaboration of an explicit quantum-field-theoretical toy model for the  correspondence. It is a model in the sense that it has precisely the
 structure  of a dipole CFT -  which string theory suggested is the correct characterisation of the holographic duals to warped AdS$_3$ -  and it is toy because it is extremely simple and solvable, but also because its holographic dual is not a warped AdS$_3$ spacetime, but rather just AdS$_3$ with non-standard boundary conditions. Nonetheless, this toy model will allow us to show explicitly %, as a  proof of concept, 
 that the Virasoro symmetries and non-local correlation function can, in fact, be compatible with each other in  a dipole theory. Let us now say a few words about it. 
 
%  Incidentally, a set of closely-related theories turn out to be related to a different non-AdS holography problem, which is equally interesting, and may shed light on the microscopic description of a different  type of near-extremal black holes. Thus the subject becomes much richer,a nd we recount it below. 

% \emph{Explain here bottom-up approach, structure vs theory, and the need of complementary approaches. ASG ambiguous, Cardy-like formulae produced by different theories. Theory: existence argument via decoupling, existence proof of concept via explicit example.}

\subsection*{Exactly solvable irrelevant deformations of two-dimensional QFTs}

The examples \cite{Smirnov:2016lqw,Cavaglia:2016oda} of UV-complete, non-local two-dimensional QFTs obtained from a finely-tuned sequence of irrelevant deformations  of  local ones came from an entirely different corner of the theoretical physics community: the integrability one, though much of their physics had already been understood from (QCD-motivated) attempts  to view the fundamental string from an EFT perspective \cite{Dubovsky:2012wk,Dubovsky:2013ira}. %, motivated by understanding the QCD string. 

% with possibly some inspiration from the phenomenology subfield.  %groups working to understand the QCD string.

The question leading to the construction of these theories may be posed as a two-dimensional S-matrix bootstrap problem: given an integrable S-matrix $\mathcal{S}_0$ (associated with some local, integrable QFT) that satisfies analyticity, crossing and unitarity, it is well known that these axioms continue to be satisfied if the solution is multiplied by a so-called CDD factor, which corresponds to an infinite set of momentum-dependent phases, with arbitrary coefficients.
One may wonder whether there exist QFTs whose S-matrix is given by $\mathcal{S}_0$ multiplied by such a CDD factor. Smirnov and Zamolodchikov \cite{Smirnov:2016lqw} proposed a series of irrelevant deformations of the original QFT Lagrangian, constructed as bilinears of certain higher spin conserved currents, that would produce precisely this modification of the S-matrix.  

The \emph{universal} $T\bar T$ deformation, which is a bilinear constructed from the stress tensor components, corresponds to one such factor, which coincides with the  S-matrix for scattering  worldsheet modes on an infinitely long Nambu-Goto string in static gauge. 
 Much of the interest in the  $T\bar T$ deformation  comes from its solvability - indeed, the finite size spectrum can be solved for exactly in terms of the one of the undeformed QFT. The same holds for the S-matrix, as we just explained. Having, in principle, its exact expression up to arbitrarily high energies indicates the resulting theory is UV-complete \cite{Dubovsky:2013ira}. On the other hand, the theory is clearly non-local at the scale set by the irrelevant deformation, as can be seen in many obseravbles, including the ground state energy in finite volume, the  behaviour of the S-matrix at large momentum, the Hagedorn growth of the density of states, a.s.o. %[Recently, progress on correlation functions. ] 

 What is absolutely remarkable about the $T\bar T$ deformation is that it has found applications in many different subfields of theoretical physics, including integrability, the QCD string, the hierarchy problem,  holography with effective Dirichlet boundary conditions at finite radius and non-AdS holography. In this review, we will touch upon most of these aspects, to a  greater or lesser extent. Our main motivation will be to use these exactly solvable irrelevant deformations to make progress in non-AdS holography. 
 
 Soon after the introduction of the solvable $T\bar T$ deformation,  another universal, solvable  Smirnov - Zamolodchikov current bilinear deformation %  the study of  SZ deformation,
   - termed $J\bar T$ - was proposed  \cite{Guica:2017lia}. As its name indicates, it is constructed from a $U(1)$ current  and the generator of right-moving translations.  This deformation is Lorentz-breaking  but, if it acts on a seed CFT$_2$, it leads to a theory that is local and conformal on the left, while being non-local on the right, with a global spacetime symmetry group $SL(2,\mathbb{R})_L \times U(1)_R$. Moreover,  this deformation can be argued to lead to a UV-complete theory. We thus see that $J\bar T$ - deformed CFTs fulfill all requirements to correspond to  dipole CFTs, and  one can therefore use them as toy models  for the  Kerr/CFT correspondence. Their solvability allows one to study  in detail the triad  of  properties  in figure \ref{generaltriad} and how they relate to each other. 
%\textcolor{blue}{\emph{$J\bar T$ toy model that yields Virasoro and CFT-like correlators in a non-local theory. While details momentum-dependence don't match, important as proof of principle. }}  

After reviewing the definition and some well-known, basic properties of    the  $T\bar T$ and $J\bar T$ deformations, our emphasis will shift to holography and how the triad of observables is matched across the duality. We pay special attention to the extended symmetries of    $T\bar T$ - and $J\bar T$ - deformed CFTs, which turn out to be extremely rich  and to precisely coincide with the symmetries of the undeformed CFT (two copies of the Virasoro or Virasoro-Kac-Moody algebra) in an appropriate basis. Thus, these theories provide a proof of concept that Virasoro symmetries can co-exist with non-locality; nonetheless, differences with the standard Virasoro action are clearly visible. 
  % 
%  extended symmetries, which will turn out to be modified version of Virasoro symmetries, that have adapted to the non-locality of the theory.
   We will find a beautiful precision match for these symmetries between the boundary and the bulk. %Given the importance of extended symmetries to three-dimensional holographic dualities, I will put special emphasis on them. We will find a beautiful precision match between the bulk and the boundary, resulting in fully consistent precision holography for AdS$_3$ with non-standard boundary conditions. 
We will also show, for the case of $J\bar T$ - deformed CFTs, that the Virasoro symmetries are compatible with the non-local correlation    functions in the theory, and  can be used to compute them.

\subsection*{Back to black holes}

 From the point of view of the non-AdS holographic applications that motivate this review, it turns out that the bulk duals of the standard $T\bar T$ and $J\bar T$ deformations of a holographic CFT are a tad too trivial: they simply correspond to changing the boundary conditions for the metric and possibly  gauge fields in the AdS$_3$ spacetime dual to the undeformed CFT, but do not induce any local change in the geometry. For this reason, one is led to studying single-trace versions of  these deformations, which should correspond to proper non-AdS holographic scenarios: asymptotically linear dilaton backgrounds in the case of $T\bar T$ and warped AdS$_3$ backgrounds in that of $J\bar T$.  Nonetheless, the corresponding 
 deformations again  become intractable from a field-theoretical point of view in the strongly-coupled regime dual to supergravity in the bulk, leading one to rely on holographic inference. The result is rather interesting, especially for the linear dilaton backgrounds,  as it indicates the existence of  classes of theories with the same non-trivial extended symmetries and entropy as single-trace $T\bar T$ - deformed CFTs%\footnote{While one would expect that the universal properties of a class of warped AdS$_3$ spacetimes should be similaarly captured by single-trace  $J\bar T$, more work is needed to ascertain this picture.}
 , but with a different low-lying spectrum. We will be calling these theories \emph{little string CFTs}, for reasons that will be obvious once the details are given.
 
The analogous studies that are expected to link a class  of string-theoretical  warped AdS$_3$ backgrounds to single-trace $J\bar T$-deformed CFTs - which are of obviously high relevance to understanding the Kerr/CFT correspondence - are currently still inconclusive; the  main reason is that the corresponding backgrounds are not obtained via decoupling limits of string theory, which makes their parametrisation uncertain. Nonetheless, the discovery of $J\bar T$ - deformed CFTs has clearly sharpened (and slightly shifted) the questions that surround the warped AdS$_3$ duality, as well as the tools used to answer them, and one is clearly much closer to understanding the two-dimensional dipole CFTs that appear on the boundary side of the correspondence.

%This study does, nonetheless, lead to an interesting, somewhat modified perspective on Kerr/CFT. While the Kerr/CFT question is far from settled, but sharpened and tools are provided to study it in a precise setting.  As it turns out, also single-trace $T\bar T$ theories appear to play a role in understanding the microscopics of certain AF black holes. Comparison to flat space analyses are also interesting.  

 In addition, one may argue that the near-horizon dynamics of non-extremal black holes that are not too far from extremality is described by QFTs that loosely resemble 
 generalisations of $T\bar T$ and $J\bar T$ - deformed CFTs.  This connection needs to be made much more precise, but it already suggests: i) the existence of universality classes of theories that share some of the properties of $T\bar T$ and $J\bar T$ - deformed CFTs, which would be very interesting to understand better, and ii) 
that these theories may be relevant in describing general physical phenomena, such as certain  contributions to the entropy of black holes. Since the latter are nothing but a holographic description of  thermal, strongly-coupled quantum systems, it would be very interesting to understand what general type of quantum theories or phenomena can be universally modelled using properties similar to those of $T\bar T$ and $J\bar T$.

% could be  describing ; if this is the case, then it suggests that appropriately-defined universality classes of such theories 

 % then the uses of these irrelevant deformations in holography will far surpass the initial purpose. 

% [Also mention ALD. The $T\bar T$ subject has turned out to be extremely rich and beautiful, which is  encouraging:  after all, CFTs describe anything from phase transitions to RG fixed points, the string worldsheet and quantum gravity in AdS, so it is a good sign if $T\bar T$ - deformed CFTs also appear all over the place, as it may mean there is a universal set of physical phenomena that may fit under its description. $ J\bar T$ is a simple generalisation in the same vein, more adapted to extremal black holes. ]

\subsection*{Organisation of this review }

Section \ref{bhstoirrelsec} expands on the black hole overview given in the introduction, providing additional conceptual and technical details. This section presents the status of black hole microphysics prior to the advent of 
$J\bar T$ and $T\bar T$ deformations; conclusions drawn from these developments are deferred to section 
\ref{backtobh}. Section \ref{solvirrelsec} reviews standard results on the $T\bar T$ and $J\bar T$ deformations  and their single-trace analogues, focusing on well-studied topics such as the spectrum and the S-matrix.
%
%In section 2, we expand upon the black hole review presented in this introduction, explaining more of the conceptual and  technical  details.  We  only  present the pre- $J\bar T/T\bar T$ status of black hole microscopics  in this section,   postponing to section 8 to draw conclusions of what has been learned. In section 3, we present standard results in $T\bar T$, $J\bar T$ and their single-trace analogues, concentrating on the well-roded subjects of the spectrum and S-matrix. 
%
In Section \ref{holointdtr}, we discuss the holographic interpretation of the double-trace $T\bar T$ and $J\bar T$  deformations, demonstrate the exact matching of the thermodynamics and the spectra with the field theory results, and get a first glimpse of the extended symmetries of $T\bar T$ and $J\bar T$ - deformed CFTs from the asymptotic symmetries of the dual spacetimes. Section \ref{infsymmsec} analyzes these symmetries from a field-theoretic perspective, again finding perfect agreement. Section \ref{corrfsec} is devoted to discussing correlation functions and explains how, in the case of $J\bar T$-deformed CFTs, a natural basis of operators can be  fixed using the above symmetry structure.
%
%In section 4 we discuss the holographic interpretation of the double trace deformations, show the spectrum is perfectly matched, and having a first glimpse at the extended symmetries from asymptotic symmetries of the dual spacetimes. In section 5, we study symmetries from field theory point of view, showing perfect match. In section 6, we discuss correlation functions and how, in the case of $J\bar T$, the definition of the operators is fixed by the symmetries we discussed.
%
In section \ref{holostr} we discuss non-AdS holographic correspondences that have been linked to single-trace $T\bar T$ and $J\bar T$ -deformed CFTs. Finally,  section \ref{backtobh} revisits the issues raised in section \ref{bhstoirrelsec} in light of the insights gained from studying these exactly solvable irrelevant deformations,  discusses some new directions they point to, including  suggestions for asymptotically flat holography.

% we look back upon what has been learned about the problems raised in section 2 from the study of these exactly solvable irrelevant deformation, and all the interesting new windows/roads that they may be opening, including a perspective on AF holography. 

 %Finally, Section 8 
 
% Some of the above sections are quite lengthy, and are organised in sub- and subsubsections. The table of contents for each section including subsubsections is displayed at the beginning of each section. 

Some sections are necessarily lengthy and are therefore organized into subsections and subsubsections. For ease of navigation, the full table of contents for each section—including all subsubsections—is provided at the beginning of that section.

\section{From black holes to irrelevant deformations of two-dimensional CFTs\label{bhstoirrelsec}}

In this section, we review the status of the universal, near-horizon approaches to understanding black hole entropy, where the universality is related to the type of geometry of the near-horizon region.  Our main goal is to show that the microscopic description of general asymptotically flat black holes is  linked to two-dimensional CFTs deformed by certain irrelevant operators. 

As in the introduction, we start with near-extremal black holes with an AdS$_3$ factor in their near-horizon region, which are ubiquitous in string theory  - 
 the original Strominger-Vafa black hole is a textbook example.  Their  microscopic description is extremely well understood,  as  it can often be reduced to that of  a non-extremal BTZ black hole in the near-horizon region. The latter corresponds to a thermal state in a dual CFT$_2$, fact that can  be established to a reasonable level of confidence  using just bottom-up holographic techniques. 
%
% [We will then attempt to understand more general black holes as variations/complications of the BTZ model.] 

 Moving on to \emph{general} near - extremal black holes, whose  near-horizon  region universally contains a warped AdS$_3$ factor, we present the Kerr/CFT conjecture, which is an attempt to explain the entropy of near-extremal general black holes as a variation upon the BTZ result. Upon discussing the various pieces of evidence and puzzles that arise, we argue that the  dual theory should be related to  \emph{dipole} CFTs. In their purest form, these are UV-complete  two-dimensional QFTs that preserve $SL(2,\mathbb{R}) \times U(1)$ spacetime symmetry and are obtained via  certain  $SL(2,\mathbb{R})$ - preserving irrelevant deformations of a CFT$_2$. Finally, we describe a number of analyses of \emph{non-extremal} black holes that similarly indicate the presence of a CFT$_2$ deformed by irrelevant operators, a picture that can be made relatively precise for black holes embedded in supergravity.
 
%  \emph{The gravity side looks much more universal than the microscopic description, to which there are many limitations.}

 \etocsetnexttocdepth{5}
    \etocsettocstyle{\subsubsection*{Contents of this section: }}{}
    \cftsubsubsecindent 34pt
    \localtableofcontents

\subsection{Microscopic understanding of black  holes with an AdS$_3$ near horizon\label{microads3}}

As we aready mentioned, this can be reduced to an exercise in the AdS$_3$/CFT$_2$ correspondence, by mapping the near-extremal asymptotically flat black hole to a non-extremal BTZ black hole in the near-horizon region. Below, we exemplify how this map works in the concrete case of the near-extremal D1-D5-P black hole. 

\subsubsection{From the asymptotically flat black hole to BTZ}

 It is useful to take a step back and start with the non-extremal D1-D5-P black hole. This is a solution of type IIB supergravity on  $M_4= T^4$ or K3 with just RR three - form flux turned on. The  ten-dimensional string-frame metric is given by \cite{Maldacena:1998bw}
\be
ds^2 = \frac{1}{\sqrt{f_1 f_5}} \left(-  dt^2 + d\s^2 + \frac{r_0^2}{r^2} (\cosh \d_n dt +\sinh \d_n d\s)^2 \right) + \sqrt{f_1 f_5}\, \left(\frac{dr^2}{f} + r^2 d\Omega_3^2\right) + \sqrt{\frac{f_1}{f_5}}\, ds^2_{M_4}
\ee
where the so-called blackening factor $f$  and the $f_i$ are harmonic functions that parametrize the solution
\be
f = 1 - \frac{r_0^2}{r^2} \;, \;\;\;\;\; f_i = 1+ \frac{r_0^2 \sinh^2 \d_i}{r^2} 
\ee
The `boost parameters' $\d_i$ are related to the quantized D1, D5 and momentum charges as 

\be
r_0^2 \sinh 2 \d_1 =\frac{ 2\a' g_s}{v} p \;, \;\;\;\;\;\; r_0^2 \sinh 2\d_5 = 2\a' g_s k  \;, \;\;\;\;\; r_0^2 \sinh 2\d_n= \frac{2\a'^2 g_s^2 n}{R^2 v} \label{bhcharges}
\ee
where  $g_s$ is  the $10d$ string coupling,  $R$ is the radius of the $\s$ circle and  $(2\pi)^4 \, v$ is the volume  of the internal manifold in string units. 
The dilaton is $e^{2\Phi} = f_1/f_5$.
The solution also carries $k$  units of magnetic  three-form flux, and $p$ units of electric flux.

This metric corresponds to  an asymptotically flat black string, namely asymptotically $R^{4,1} \times S^1 \times M_4$.  One can reduce on $S^1 \times M_4$ to obtain an electrically charged  $5d$  black hole, see e.g. \cite{David:2002wn} for details. 
The metric one obtains is  %\emph{Re-check!}
\be
ds^2  = - (f_1 f_5 f_n)^{-2/3}  f dt^2 + (f_1 f_5 f_n)^{1/3} \left(\frac{dr^2}{f} + r^2 d \Omega_3^2 \right)
\ee
This metric can be easily put into the standard $5d$ Reissner-Nordstr\"{o}m form when the three functions are equal, via the redefinition $\tilde r = r \sqrt{f_1} = \sqrt{r^2+r_1^2}$. 

The Hawking temperature and angular potential of this black string  are %\emph{Do the left/right temperatures also have a nice form before decoupling? See 9705192}
%\emph{Factors 2?}

\be
T_H = \frac{1}{2\pi r_0 \cosh \d_1 \cosh \d_5 \cosh \d_n} \;, \;\;\;\;\;\; \frac{ \Omega_H}{T_H}=  2 \pi r_0 \sinh \d_n \cosh \d_1 \cosh \d_5
\ee
and its mass is given by %\emph{Factors v?}
\be
M = \frac{r_0^2 v R }{2 g_s^2 \a'^2} (\cosh 2 \d_1 +\cosh 2  \d_5+\cosh 2\d_n) 
\ee
It is interesting to split this into the mass of the extremal D1-D5 configuration ($n= r_0=0$)

\be
M_{extr} = \frac{r_0^2  v R}{2 g_s^2 \a'^2} (\sinh 2 \d_1 +\sinh 2  \d_5) = \frac{R}{g_s \a'} (n_1 + v\,  n_5)
\ee
with its usual $g_s^{-1}$ scaling, and the energy above extremality  

\be
E = M-M_{extr} = \frac{r_0^2 v R }{2 g_s^2 \a'^2} (e^{- 2 \d_1} + e^{- 2  \d_5}+\cosh 2\d_n) 
\ee
The entropy is given by 
\be
S= \frac{2\pi R v r_0^3}{g_s^2 \a'^2} \cosh  \d_1 \cosh \d_5 \cosh \d_n
\ee
The limit in which one recovers an AdS$_3$ factor is the near-extremal near-horizon \emph{decoupling} limit, relevant for deriving the AdS$_3$/CFT$_2$ correspondence \cite{Maldacena:1997re}. It is obtained by taking $\a'\r 0$ with

\be
\frac{r}{\a'} \, \;\;\;\; \frac{r_0}{\a'} \;,\;\;\;g_s, v ... \label{declimd1d5}
\ee
held fixed. Note that in this limit, $\d_{1,5} \r \infty$ as $e^{\d_{1,5}} \propto 1/\sqrt{\a'}$, while $\d_n$ is finite. Such a hierarchy of the boost parameters is sometimes referred to as the `dilute gas' approximation \cite{Maldacena:1996ix}.  This limit produces an entire spectrum of (black hole) states in AdS$_3$, with a wide range of energies and momenta, corresponding to  a full theory.

 Note $T_H$ and the energy above extremality remain finite in this limit. However,  the original asymptotically flat  black hole does become near-extremal, since $M_{extr} \r \infty$, and thus $E/M \r 0$. This limit effectively amounts to dropping the ones in the $f_{1,5}$ harmonic functions, but not in $f_n$. The $3d$ part of the metric becomes \cite{Maldacena:1998bw} that of the non-extremal BTZ black hole \cite{Banados:1992wn} solution to pure $3d$  gravity with a negative cosmological constant, $\Lambda = - 2/\ell^2$, with $\ell^2 = g_6 \sqrt{pk}, \; g_6=g_s/\sqrt{v}$

\be
ds^2 = \a' \left[ \frac{u^2}{\ell^2} (-dt^2+d\s^2) + \frac{u_0^2}{\ell^2} (\cosh \d_n dt+\sinh \d_n d\s)^2 + \frac{\ell^2}{u^2-u_0^2}  du^2\right]  \label{btzinnh}
\ee
Above, $u = r/\a'$, and similarly for $u_0$. Introducing a new radial coordinate
$r = u \sqrt{f_n}/\ell$ (different from the previous one!) and letting , %$r_+ r_- = u_0^2 \sinh \d_0 \cosh \d_0/\ell^2$ and $r_+^2-r_-^2 = u_0^2/\ell^2$
 $r_\pm = \frac{u_0}{2\ell} (e^{\d_n} \pm e^{-\d_n})$, \eqref{btzinnh} can be put in the  more standard form 

\be
ds^2  = - \frac{(r^2-r_+^2)(r^2-r_-^2)}{r^2} dt^2 + \frac{\ell^2 r^2 dr^2}{(r^2-r_+^2)(r^2-r_-^2)} +  r^2 \left(d\sigma + \frac{r_+r_-}{r^2} dt\right)^2 \label{btzmet}
\ee
This solution corresponds to a quotient of global AdS$_3$ \cite{Banados:1992gq}, as follows from the fact that all solutions to pure $3d$ gravity with $\Lambda <0$ are locally AdS$_3$. The quotient  is related to  the left/right temperatures of the black hole \cite{Maldacena:1998bw} that couple, by definition,  to the left/right-moving energies $E_{L,R} \equiv \frac{1}{2}(E\pm P)$

\be \label{TLRbtz}
\frac{1}{ T_{L,R}}= \frac{1\pm \Omega_H}{T_H} = \frac{2\pi\ell^2 e^{\pm \d_n}}{u_0} = \frac{2\pi \ell}{r_+\mp r_-}% \;, \;\;\;\; r_+ \pm r_- = 2 \pi \ell R T_{L,R}
\ee
This concludes our review of how the non-extremal BTZ black hole arises in the near-horizon decoupling limit of the D1-D5-P system. More
 generally, one would like to forgo the use of string theory, and devise this limit on a family of black holes without invoking $\a'$. Since in a theory of gravity coupled to electromagnetic fields (such as  supergravity), the charges of the black hole  are given by  just the left-hand sides of \eqref{bhcharges}, the prescription is to take two of these charges to be large and fixed, while the third (assumed to be a Kaluza-Klein charge corresponding to momentum along  a sixth direction) is finite and allowed to vary. %This is sometimes called the `dilute gas' approximation. \emph{Correct?}
    Note this is different from the strictly extremal limit, which is $r_0\r 0$, $\d_i \r \infty$ with  the charges held fixed, whereby one drops the constant term in all the $f_i$ - this yields an extremal BTZ black hole in the near-horizon. 
    
%     Unless we tweak the overall constants, $M_{extr}$ is fixed (it may make sense to add in the momentum contribution, $n/R$, and subtract it from $E$), while the corresponding energy above extremality vanishes. All charges are treated on the same footing.  Then, the excitations carrying the third charge cannot be treated as small perturbations on top of the background created by the other two. We will thus refer to the previous decoupling. 

Thus,  the microscopic interpretation of a black hole with an AdS$_3$ factor in their near-horizon region  can be reduced to an exercise in the AdS$_3$/CFT$_2$ correspondence, namely explaining the entropy of a non-extremal BTZ black hole from the CFT$_2$ viewpoint.

\subsubsection{Bottom-up holography for AdS$_3$}

\begin{figure}[h]
\centering
\includegraphics[height=5 cm]{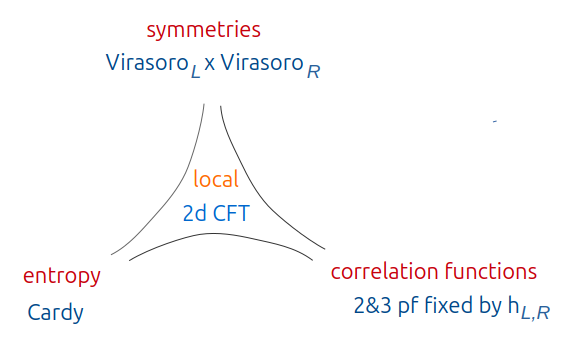}
\caption{\small{The property triad for $2d$ unitary CFTs. They all follow from global conformal invariance and locality.}}
\label{cfttriad}
\end{figure}

Gravitational theories in an AdS$_3$ spacetime with Dirichlet boundary conditions at infinity are holographically described by  two-dimensional conformal field theories (CFT$_2$). While there is plenty of precision evidence for the AdS$_3$/$CFT_2$ holographic correspondence, we would like to present this duality from a bottom-up perspective, which would be the one that most easily generalises to the attempts to understand the non-AdS case. 
In this bottom-up approach, one uses  the  (classical) gravitational action in AdS$_3$   to uncover the basic properties of the dual theory, such as its symmetries and their associated conserved charges, its thermodynamics, its correlation functions, and so on. For a CFT$_2$, these properties are summarized in figure \ref{cfttriad}. 

\noindent In these notes, we will mostly use the classical gravity approximation, which is usually valid when the radius of curvature of the geometry is much larger than the Planck or the string length\footnote{As we discuss shortly, the situation is slightly more subtle for near-AdS$_2$ spacetimes.}. We will consider that, in principle, matter fields are also present, as is the case in the instances of  AdS$_3$ holography derived from  string theory. 

 Let us now describe in turn how one may uncover the above features of the dual theory using  classical $3d$ gravity, possibly coupled to  matter fields. 

\vskip3mm

\noindent \emph{The symmetries}

\vskip2mm

\noindent In holography,  conserved currents in the boundary theory correspond to gauge fields in the bulk; their associated global symmetries correspond to asymptotic symmetries of the bulk gauge theory, as we now discuss.  For convenience, we focus on  
 diffeomorphisms; gauge transformations are treated in an entirely analogous manner.

In a theory of gravity,  the only non-trivial contributions to the  conserved charges associated to diffeomorphisms come from boundary terms, since the bulk contribution  vanishes on-shell, as can be seen in either the Hamiltonian or the Lagrangian formalism. (We assume that the spacetime of interest has a timelike or null boundary, and that boundary conditions for all the fields are specified).  Consequently, only `large' diffeomorphisms, which do not vanish at infinity, contribute to the charges. Diffeomorphisms that, due to their rapid fall-off, yield vanishing charges for generic allowed field configurations are called `trivial' and  carry no physical meaning, simply corresponding to redundancies of description. Diffeomorphisms that do correspond to physical symmetries of the  gravitational theory - the so-called \emph{asymptotic symmetries} -  are those diffeomorphisms that are allowed by the boundary conditions % one needs to impose on the asymptotic metric
and  whose associated charges  are finite, conserved, (integrable) and  generically non-zero on the allowed field configurations.  By definition, they depend on the  boundary conditions, $\mathcal{C}$,  on the fields.
Since  asymptotic symmetries correspond to symmetries of the \emph{theory},  they are present for arbitrary allowed field configurations; they should not be confused with the isometries of a particular geometry, which correspond to symmetries of a particular \emph{state} in that theory.

 There are many different formalisms one may use to compute the conserved charges; for the purposes of this article we will mostly refer to the covariant phase space one \cite{Lee:1990nz,Iyer:1994ys} (see \cite{Compere:2018aar} for a review), which is consistently very powerful, irrespective of the spacetime asymptotics. In this formalism,  the difference in the conserved charge $Q_\xi$ associated with some diffeomorphism $\xi$ between two nearby backgrounds, denoted $\Phi$ and $\Phi + \d \Phi$, where $\Phi$ stands for  all the fields in the theory (metric, gauge fields, scalars, etc), is given by 
 \be
 \slash\!\!\! \d Q_\xi = \int_{\p \Sigma} \boldsymbol   k_\xi^{(D-2)} (\Phi, \d \Phi) \label{covphspch}
 \ee
 $\boldsymbol{k}_\xi$ is a $D-2$ - form that is algorithmically constructed from the Lagrangian for the theory (see e.g. \cite{Compere:2018aar}), and the integral is performed over the $D-2$ dimensional boundary of a spatial slice, $\Sigma$. The result needs to be integrated in field space from the initial ($\Phi_i$) to the final ($\Phi_f$) configuration, and for this the charge difference defined above needs to be integrable.

What is nice about this formalism is that it is very general, and the only necessary input are  an action and a set of boundary conditions, $\mathcal{C}$,   that the fields and their variations obey. Once this input is given, the formula \eqref{covphspch} algorithmically produces the conserved charges and their algebra, defined as 

\be
\{ Q_\chi, Q_\xi \} \equiv \d_\xi Q_\chi = \int_{\p \Sigma}\boldsymbol  k_\chi (\Phi, \d_\xi \Phi ) \label{chalgcovphs}
\ee 
On the other hand, how to determine the appropriate, consistent boundary conditions for a gravitational theory in a given spacetime  is in general a difficult question.  For asymptotically AdS$_3$ spacetimes, a natural set of boundary conditions \cite{Brown:1986nw} follows by inspecting the falloffs of conical defect geometries, or BTZ black hole solutions. These Dirichlet\footnote{More precisely, one should fix the conformal class of the metric on the boundary - see e.g. \cite{Papadimitriou:2005ii}. } - or Brown-Henneaux -  conditions  turn out to be the appropriate boundary conditions for the standard AdS$_3$/CFT$_2$ correspondence. As we will discuss later in this review, there do exist other consistent boundary conditions for the metric of asymptotically locally AdS$_3$ spacetimes, which correspond to non-trivial modifications of the boundary theory.

In the following, we assume the  Brown-Henneaux boundary conditions on the AdS$_3$ metric are given. The asymptotic form of the metric then reads, in radial gauge %\textcolor{red}{\emph{Defn $\rho$!}}

\be
ds^2 = \frac{dU d V}{\rho}+ \ell^2 \frac{d\rho^2}{4\rho^2}  + \L(U)\, dU^2 + \bar{\L} (V) \, dV^2 + \ldots \label{asybanad}
\ee
where $U,V \equiv \s \pm t$ are null coordinates on the boundary at $\rho \r 0$, and the $\ldots$ stand for subleading terms in the $\rho$ expansion. The gravitational phase space (seen as the space of solutions in this gauge) is parametrised by two  arbitrary functions, $\L$ and $\bar{\L}$, of the respective coordinate, as follows from the asymptotic equations of motion. It is assumed that matter fields, if present, fall off sufficiently rapidly to not change, via backreaction,  the above leading asymptotic behaviour. 

The asymptotic symmetries correspond to diffeomorphisms of the form  (see e.g.  \cite{Compere:2018aar} for details) 

\be
\xi^m_L = e^{i m U} (\p_U + i m \rho\p_\rho )\;\;\;\; \mbox{and} \;\;\;\;\; \xi^m_R = - e^{-i m V} (\p_V - i m  \rho\p_\rho ) \label{asgads3}
\ee
and their associated conserved charges on the background \eqref{asybanad} equal the Fourier modes of $\L (U)$ and, respectively, $\bar{\L}(V)$, which we denote as $L_m, \bar L_m$. The algebra of these charges 
%It was shown a long time ago \cite{} that asymptotically AdS$_3$ spacetimes with Dirichlet  boundary conditions so that the associated  asymptotic symmetries 
organises into two commuting copies of the  Virasoro algebra \cite{Brown:1986nw}, given by\footnote{In order to obtain the standard central extension $\frac{c}{12} m (m^2-1)$, we need to shift $L_0$ by the ground state energy $-\frac{c}{24}$.}  %\emph{include reference for detailed computation. Should we be on the cylinder?}

\be
[L_m, L_n] = (m-n) L_{m+n} + \frac{c}{12} m^3 \d_{m+n}  \;, \;\;\;\;\; c = \frac{3\ell}{2G_3}
\ee
The above is precisely the extended symmetry algebra of a two-dimensional CFT, where $c$ is the central charge of the latter, associated with the conformal anomaly. While in the CFT the central charge is due to a quantum effect (see e.g. \cite{DiFrancesco:1997nk}), in gravity it is captured  by an entirely classical  computation, whose result
%
%, whereas in the dual CFT, it corresponds to a quantum effect, not entirely trivial to compute \cite{}. Moreover, the classical gravity calculation is also able to capture the central extension of the Virasoro, which is given by the ratio of the AdS length to the $3d$ Planck length times a numerical coefficient.
% This central charge 
 precisely matches the central charge of the dual CFT whenever it can be independently computed. %, as for example in the string-theoretical setup mentioned above. 
 We will always be interested in the regime where the classical gravitational approximation is reliable ($\ell_{AdS} \gg \ell_{Planck} \propto G_3$) which, using the above formula, requires that the CFT have a parametrically large central charge.
 
Of course, the simple fact of  having uncovered a Virasoro$_L \times$ Virasoro$_R$ asymptotic symmetry algebra does not automatically mean that we have proven that the dual theory is a $2d$ CFT - the very theories we study in this review are  ones that possess Virasoro symmetries, but are not standard CFTs. More data is required to establish this (such as information about locality), and preferably a decoupling argument. Nor does this calculation imply that one has identified a single, or consistent dual theory: the ASG analysis applies to all theories whose low-energy effective description is given by $3d$  Einstein gravity (possibly coupled to matter fields), and the issue of whether this low-energy effective theory admits a consistent UV completion must be settled independently. Attempts to understand the holographic description of pure $3d$ gravity with a negative cosmological constant illustrate the difficulty of addressing these types of problems from a bottom-up perspective (see e.g. \cite{DiUbaldo:2023hkc} and references therein).

The Virasoro symmetries of $2d$ CFTs can be viewed as an enhancement of the global conformal group  in two dimensions, $SO(2,2) \simeq SL(2,\mathbb{R})_L \times SL(2,\mathbb{R})_R  $%\cite{Polchinski:1987dy}
. The latter are usually also the symmetries of the vacuum state, and are realised holographically as isometries of the vacuum AdS$_3$ solution. Then,  from a bulk perspective, one sometimes writes that the isometries of the vacuum solution are enhanced to the Virasoro $\times$ Virasoro asymptotic symmetries
 %
% Thus, even though the isometries of the AdS$_3$ vacuum are just $SL(2,\mathbb{R}) \times SL(2,\mathbb{R}) $, i.e. the global part of the two-dimensional conformal group
% , the ASG calculation being able to capture its full infinite-dimensional extension. 

\medskip
\begin{figure}[!h]
\centering
\includegraphics[height=1.6cm]{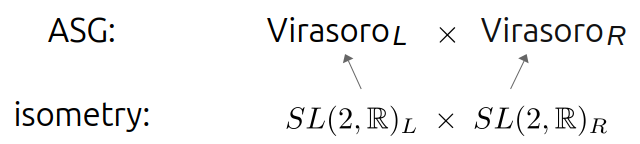}
\end{figure}
%
%\bea
%&& ASG: Virasoro_L \times Virasoro_R \\
%&& isometries: SL(2,\mathbb{R})_L \times SL(2,\mathbb{R})_R 
%\eea

\noindent The interpretation of such an enhancement is slightly different for spacetimes containing an AdS$_2$ factor, as we review below. Also note the Virasoro $\times$ Virasoro ASG can be reproduced using many different formalisms: Hamiltonian \cite{Brown:1986nw}, covariant phase space \cite{Compere:2018aar}, Brown-York \cite{Balasubramanian:1999re}, and so on.

%Note that in the case of AdS$_3$ there is no ambiguity as to what the boundary conditions should be: fix conformal class of the metric on the boundary. Once this is established, then any formalism (holographic renormalisation, covariant phase space, ADM, etc) will reproduce the expected ASG. 

%\emph{Some comment on pure 3d gravity vs a theory with a proper holographic dual?} Note existence is a different issue; the interpretation of the ASG   calculation being that it correspond to the symmetries of the dual theory if the latter exists. Similar comments would hold for the black hole entropy calculations, except that the holographic  interpretation of the horizon area is unclear in theories of pure gravity. 

\vskip4mm

\noindent \emph{The entropy}
\vskip2mm

\noindent We expect that  BTZ black holes, with metric \eqref{btzinnh} or \eqref{btzmet}, holographically represent thermal states in the dual CFT. %They are solutions of $3d$ pure Einstein gravity with a negative cosmological constant and, as such, they are locally AdS$_3$, nonetheless, they differ from vacuum AdS$_3$ at the global level, which is what makes them non-trivial. 
The energy and temperature of the system are  encoded in the geometry of the corresponding black hole, the first as the conserved charge associated with the asymptotic time translation symmetry, and the second in the identification of the Euclidean time.  
Similarly,  angular momentum is associated with $\s$ translations and the angular velocity to a twist in the euclidean identifications. As we already mentioned, it is very convenient to work in terms of the left/right-moving temperatures $T_{L,R}$ defined in \eqref{TLRbtz}, which are conjugate to the left/right-moving energies $E_{L/R}$. In the BTZ background, the relation between them is $E_{L/R} = \frac{\pi^2 c}{6} T_{L/R}^2 R$, where $R$ is the radius of the $\s$ circle.  %\emph{Define R or set it to one!} 

BTZ black holes are thermodynamically stable, as their specific heat is positive; they also dominate the asymptotically AdS$_3$ gravitational path integral  at large enough temperatures, more precisely when  $T_L T_R  >1/(2\pi R)^2 $. % This holds of course in the classical gravity approximation, where $c\sim \ell/G_3>>1$. 
We thus expect that  their Bekenstein-Hawking entropy agrees with the leading CFT entropy at these temperatures. The latter is  %The entropy is given by 
%
%\be
%S_{BTZ} = \frac{\mathcal{A_H}}{4 G_3}  = 2\pi \sqrt{\frac{c E_L R}{6}} +  2\pi \sqrt{\frac{c E_R R}{6}} = \frac{\pi^2 c}{3} (T_L+T_R) R
%\ee
% where we used \eqref{} to rewrite the formula in terms of energies/temperatures. 
%
%On the CFT side, the entropy is 
given by Cardy's formula \cite{Cardy:1986ie}

 \be
 S_{Cardy} = 2\pi \sqrt{\frac{c E_L R}{6}} +  2\pi \sqrt{\frac{c E_R R}{6}} = \frac{\pi^2 c}{3} (T_L+T_R) R \label{cardyfintro}
 \ee
which is a universal formula for the degeneracy of states in a two-dimensional CFT, derived using modular invariance of the CFT partition function, which relates $Z(\b_L, \b_R) = Z(4\pi^2R^2/\b_L, 4\pi^2R^2/\b_R)$. This formula is a priori valid for $E\gg c$, but the regime can be extended to $T_L T_R >1/(2\pi)^2$ or $E_L E_R > (c/24)^2$ % \emph{Check! Also explain zero point energy} 
for CFTs with a large central charge and a sparse light spectrum \cite{Hartman:2014oaa}, i.e. precisely those CFTs that can be dual to a spacetime where the classical gravity approximation holds.

 Thus, the gravitational side knows about the basic ingredients appearing in Cardy's formula: it knows about the central charge through the central extension of the asymptotic symmetry group, and about the energies/temperatures via the conserved charges/identifications of the euclidean solution.  
 Working out the details, one can show \eqref{cardyfintro} always reproduces the Bekenstein-Hawking entropy of the BTZ black hole \cite{Strominger:1997eq}. Of course, this match is ultimately a consequence of the  AdS$_3$/CFT$_2$ holographic duality; the point here is that one can show it  using only universal ingredients that are easily available on the gravitational side. As before, one needs to assume that the entropy of BTZ black holes has a state-counting interpretation; the example of $3d$ pure gravity is revealing in this sense  \cite{Harlow:2018tqv}. %\emph{Write better!} 
Moreover, the modular invariance used to derive Cardy's formula can be seen directly on the gravitational side; for this,  it is important that both high-temperature and low-temperature states are in principle present. 
 
%  From the gravitational point of view, the modular group is simply a subset of very large diffeomorphisms. Important that both high and low temperatures allowed. 

Note that  Cardy's formula captures only the leading asymptotic behaviour of the entropy. Corrections to it can in principle  be computed and sometimes matched to  corresponding gravitational corrections. 
%
 %a lot more can be understood about various corrections to its entropy, especially for small black holes. Also, for general black holes one can understand log corrections. \emph{Where do Iliesiu's results fit?} 
 However, for the purposes of this review, we are mostly interested in accounting for the leading  black hole entropy, but for black holes that are as general  as possible.

\vskip3mm

\noindent \emph{Correlation functions}

\vskip2mm

\noindent  Another important set of observables that can be matched are correlation functions, which can be inferred from emission or absorption rates for various propagating matter fields that are assumed to be present in the black hole background. These rates can be related to finite-temperature two-point functions of the dual theory \cite{Maldacena:1997ih}, evaluated in momentum space. The universal and excellent fit of this type of scattering amplitudes off near-BPS black holes has been one of the early hints for the AdS/CFT correspondence \cite{Maldacena:1997ih, Maldacena:1996ix, Das:1996wn%Das:1996jy
}.

For a black hole with an AdS$_3$ factor in their near-horizon region, these calculations match to a finite-temperature momentum-space correlator in a CFT$_2$ on a line

\be
\mathcal{G}_{\mbox{\tiny{$T_{L,R}$}}}(p,\bar p) \sim \frac{T_L^{2h_L-1} }{\Gamma(2h_L)}  e^{- \frac{p}{2T_L}} \left|\Gamma\left(h_L + \frac{i p}{2\pi T_L}\right)\right|^2 \; \times \;  \frac{T_R^{2h_R -1}}{\Gamma(2h_R)}  e^{- \frac{\bar p}{2T_R}} \left|\Gamma\left(h_R + \frac{i \bar p}{2\pi T_R}\right)\right|^2 \label{bh2pf} 
\ee
%
%
%\be
%\mathcal{G}_{\mbox{\tiny{$T_{L,R}$}}}(p,\bar p) \sim T_L^{2h(\bar p)-1}  e^{- \frac{p}{2T_L}} \left|\Gamma\left(h(\bar p) + \frac{i p}{2\pi T_L}\right)\right|^2 \; \times \;  T_R^{2h (\bar p)-1}  e^{- \frac{\bar p}{2T_R}} \left|\Gamma\left(h(\bar p) + \frac{i \bar p}{2\pi T_R}\right)\right|^2 \label{bh2pf} 
%\ee
up to numerical prefactors%\emph{Should we care about their $h$ - dependence?}
, where $T_{L/R}$ are the left/right-moving temperatures in the CFT, and $p,\bar p$ are the left/right-moving momenta. The dimension of the CFT operator can be read off from the large $r$ behaviour of the corresponding bulk field (here taken to be a scalar) and is related to its mass  as 
\be
 h_L =  h_R  = \frac{1+ \sqrt{1+ m^2 \ell^2}}{2}
\ee
 Of course, one could also be matching correlation functions in vacuum AdS, which are significantly simpler. However, given that we will be working with black hole backgrounds, for which the vacuum state may not be easily accessible, we prefer to stick to these finite-temperature observables. %[in view of future comparisons we will concentrate on two-point functions (\emph{and three?}) on a black hole background, ]
 % Higher-point functions can also be matched. % \emph{Refs?} 

%So, how about black holes whose near-horizon does not contain an AdS$_3$ factor? %Some important progress has been achieved for supersymmetric black holes (in AdS?) by using $F$ - maximization. \emph{True?} However, here supersymmetry is key, and thus does not address the question of more realistic black holes. 

\bigskip

\noindent To recapitulate, the fact that black holes with an AdS$_3$ factor in their near-horizon region are microscopically described by a two-dimensional CFT can be readily inferred from the bottom-up approach that we have described: first, the ASG analysis yields an extended symmetry algebra that  is precisely that of a $2d$ CFT with a specific central charge; second, the gravitational Bekenstein-Hawking entropy precisely matches the CFT prediction. Further checks of the CFT picture can be performed using scattering, and they yield a perfect match to correlation functions of a local $2d$ CFT.  

\subsubsection{Near-extremal BTZ black holes}

In order for the entropy match discussed above to hold, the black hole should be  held sufficiently far from extremality.  Close to  it, there are at least two issues that  one should take into account: whether the black hole still dominates the thermal ensemble, and whether its thermodynamic description still makes sense.

The condition that the black hole dominate the thermal ensemble is 
\be
T_L T_R > \frac{1}{4\pi^2 }
\ee
where now $T_{L,R}$ are the dimensionless temperatures, satisfying $2/T_H = 1/T_L + 1/T_R$.  One can make the black hole near-extremal, $T_H \ll 1$ by  e.g. taking $T_L$ to be small\footnote{Note that this is the opposite left/right convention with respect to much of the literature, but we chose it to harmonise with the rest of the article.}. However, if one would like the black hole to still dominate the thermal ensemble, $T_R$ should be made  correspondingly large. If one considers instead the microcanonical ensemble, one finds that - still for CFTs with a large central charge and a sparse light spectrum -  the density of states is universal and agrees with the black hole answer for $E_L E_R > (c/24)^2$ \cite{Hartman:2014oaa}. Below this energy, the entropy is not universal, and `enigmatic' configurations may dominate. Of course, this does not prevent configurations with a Cardy growth of states from existing, they  may just no longer represent the leading contribution to the CFT entropy.% \emph{Think better what you want to say.}

The second issue is that at very small temperatures, one expects the  thermodynamic description  of the black hole to break down, as  first pointed out in \cite{Preskill:1991tb}.   This happens when the energy of a single  Hawking quantum is comparable with the   available energy above extremality of the black hole  

\be
E - E_{extr} \sim  c\,  T^2 \sim T \; \;\;\;\;\; \Rightarrow\;\; \;\;\; T_{\mbox{\footnotesize{break}} }  \sim 1/c
\ee
and thus its emission can significantly change the system temperature. What happens to the black hole at  around this temperature was understood only recently, thanks to progress in evaluating the gravitational path integral on the black hole geometry. 

%
%Around and below this temperature, the classical (saddle point) description of the black hole is no longer reliable, due to quantum fluctutations becoming large/strongly coupled.  Fortunately, these fluctuations are nonetheless under control, as they can be captured by a solvable JT gravity path integral. 
%
To set  up this problem, we note that close to extremality (namely $r_+-r_- \sim \epsilon$ with $\e$ small), the black hole develops a long (near) - AdS$_2$ throat, as can be seen by expanding $r^2 = r_+^2 + \e \, \rho$ in \eqref{btzmet}. As long as the second term is much smaller than the first ($\e \,\rho \ll r_+^2$, or $r - r_+ \ll r_+$), the AdS$_2 \times U(1)$ approximation to the geometry is valid; the throat region is long if $\epsilon  \ll r_+$. The size of the $U(1)$  fibre parametrised by $\s$ is approximately constant but - importantly - not exactly so, lest all matter configurations  have zero energy\cite{Maldacena:1998uz} (see also \cite{Balasubramanian:2009bg}). This is the (in)famous `AdS$_2$ non-dynamics' problem that has pervaded much of the older literature on AdS$_2$ holography and extremal black holes. As has been since understood  \cite{Almheiri:2014cka,Maldacena:2016upp},  in order to have a dynamical gravity problem one should consider instead  near-AdS$_2$ configurations, by allowing the linearly growing mode in the size ($g_{\s\s}$) of the $U(1)$ fibre, at the expense of slightly breaking the AdS$_2$ conformal symmetry.

% One problem that was much-discussed in the older literature was the , namely the fact that the $2d$ dilaton (corresponding to the $g_{\s\s}$ component of the $3d$ metric) blows up at the AdS$_2$ boundary/cannot be bounded, if any finite amount of (matter) energy is placed inside AdS$_2$. It has since been understood that this is simply a feature, and one should thus just consider `near-AdS$_2$' holography. \emph{Expand a bit. }

The dynamics of gravity inside this near-AdS$_2$ throat  
 can be well-modeled using JT (super)gravity coupled to matter fields obtained via KK reduction along the $U(1)$ fibre.  A proper treatment \cite{Iliesiu:2020qvm} of the JT path integral finds that quantum fluctuations of the JT mode become large  for temperatures $T_L \lesssim 1/c$, leading to important corrections to the classical gravity picture. This is exactly the temperature at which the statistical description of the black hole breaks down, as described above.    At even smaller temperatures, non-perturbative corrections to $JT$ gravity are to be expected, which are needed to reproduce a discrete dual spectrum.

One can use the JT path integral  to estimate the density of states of the system.    The end result of this computation is qualitatively  different  for supersymmetric \cite{Heydeman:2020hhw} and non-supersymmetric \cite{Iliesiu:2020qvm} black holes, see   \cite{Turiaci:2023wrh} for a summary.
In the latter case,   the density of states %\emph{smoothly (?)}
 goes to zero at extremality, %\emph{How is this justified from scaling with $c$? Does the dominance assumption have to break down?} 
 showing that the system does not, after all, possess a large ground state degeneracy, as the extremal black hole solution was na\"{i}vely  indicating. On the other hand, for supersymmetric black holes the entropy at extremality is non-zero, and there is a $\O(1/c)$ gap to the first excited state, reinforcing the picture that had been long held in the literature. 
 % \emph{Up?} 

Since the   description in terms of JT gravity coupled to matter is only valid inside the near-AdS$_2$ throat, it only captures a \emph{subset} of the original degrees of freedom of the dual CFT$_2$. More precisely,   the limit one takes in the canonical ensemble corresponds to 

\be
T_L \sim 1/c \;, \;\;\;\;\; T_R \;\;\mbox{fixed} \;\; ( \, \mbox{and}\; \gtrsim \O(c)) \label{vnhfocus}
\ee
implying that  the operators it focusses on have left/right dimension  $h_L = \frac{c-1}{24} + \O(1/c) \,, \; h_R \sim c \, T_R^2 $. 
In fact, it is also possible \cite{Ghosh:2019rcj} to derive the Schwarzian density of states and correlators that JT gravity predicts directly  from the point of view of the UV  CFT$_2$,  by simply expanding its partition  and correlation functions  in the limit above   which - under reasonable assumptions -  is dominated by the dual identity block. %This  makes precise  how to compute observables in this limit. The chirality comes from the fact that the identity block dominates (in some channel) and not that the subset of operators is chiral in any (other) sense. Note that a large range of $T_R$ is available in the focussing limit. 
 Changing to an ensemble where the temperature and spin, $J$,  are fixed, one finds  precisely the Schwarzian answer

\be
Z_J^{CFT}(\b) \approx e^{S_0 - \b E_0} Z_{Schw} \left(\frac{12 \,\b}{c} \right) = e^{S_0 - \b E_0} \, \left(\frac{\pi c}{ 12 \b} \right)^{3/2} e^{\frac{\pi^2 c}{12\b}}  \;, \;\;\;\;\;\; E_0 = J-\frac{1}{12}
\ee
where the parameter  $S_0$  takes the value $S_0 = 2\pi \sqrt{\frac{c}{6} J} $ that corresponds to the na\"{i}ve extremal  entropy. Note this parameter is an \emph{input} from the point of view of JT gravity\footnote{Obviously, it can be computed from the gravity description, but here the question is whether its value can be explained within the JT/Schwarzian framework.}  but can be determined if one can access the UV CFT$_2$ description, where it is fixed using  modular invariance. Naturally,  modular invariance - which relates low to high temperatures - is no longer visible upon focussing on the subset of low-temperature states  \eqref{vnhfocus}.  

%It is possible \cite{} to derive the Schwarzian density of states \emph{Any need for $T_L$ to scale as $1/c$?} and  behaviour of the matter field correlators by taking an appropriate limit on the UV CFT$_2$ density of states and correlators, which are dominated by a particular block in the limit. The states that end up contributing are dual to operators with 

Another interesting question - which is relevant to our upcoming discussion of general extremal black holes - is whether the extended Virasoro $\times$ Virasoro symmetries of the UV CFT are visible in the above near-horizon limit.  A priori, this is expected to be so, since in empty AdS$_3$ or on a fixed BTZ background,  the  conserved charges - appropriately defined - do not depend on the radial surface on which  they are evaluated \cite{Compere:2015knw}. Note, however, that from the purely $2d$ perspective of JT gravity coupled to a gauge field and KK matter, both $3d$ Virasoro generators would be acting non-trivially on the KK modes. This is in accord with the analysis of \cite{Cvetic:2016eiv} who, upon restricting to the KK  
zero  modes,  only found an ASG consisting of two   $U(1)$ charges for  the relevant  `running dilaton' subsector. % (the only ones that can have non-trivial conserved charges). }. %Note that for the so-called constant dilaton solutions, there is an infinite set of allowed 2d diffeos accompanied by gauge transformations, whose Lie bracket is Witt $\times U(1)$. However, their associated conserved charges are zero, as expected from general arguments when the dilaton is constant. \emph{Maybe quickly review argument at beginning of the section.}
% \emph{Any relation between this Virasoro$_L$ and that in the parent theory?}  The reason that one does not expect the left Virasoro to be visible from the very near horizon analysis is that it changes the left energy by a finite (quantised) amount, so its action does not fit within the vnh focussing limit.
Note that the action of the left Virasoro (which increases $h_L$ by one) is not compatible with the strict focussing limit \eqref{vnhfocus}, whereas the right-moving Virasoro is in principle  visible.

Consequently, while JT gravity coupled to the infinite tower of KK matter modes should in principle see the extended symmetries of the UV CFT, the way they may be uncovered in practice looks quite complicated and it is preferrable - at least in what concerns the specific question of the extended symmetries - to be working directly with the three-dimensional geometry.

Before moving on, it is worth  mentioning also the older picture of extremal black holes. %, which should still hold for the supersymmetric case.
 We
 concentrate  again on the simple BTZ example. The limit focussing on the AdS$_2$ near-horizon region is known as the `very near horizon' limit \cite{Strominger:1998yg}. On the $T_H=0$ black hole geometry, it was argued to correspond to a 
 a DLCQ (discrete light-cone quantisation) limit of the original CFT \cite{Strominger:1998yg,Balasubramanian:2009bg}. This is most easily seen by  rewriting \eqref{btzmet} with $r_-=r_+$ and the replacement  $r^2 = r_+^2 +\e \, \tilde r$ as
 
\be
ds^2 =  \frac{\ell^2}{4} \frac{ d\tilde r^2}{\tilde r^2} + \e \, \tilde r \, (d\s^2-dt^2) + r_+^2 (d \s+dt)^2 \;, \;\;\;\;\;\;\;\; \s \sim  \s  + 2\pi  R
\ee 
Introducing the lightcone coordinates $U,V \equiv \s \pm t$, they are simultaneously identified mod $2\pi R$.

One way to take the near-horizon limit  is to keep $\tilde r$ fixed, and send $\e \r 0$. The resulting geometry will be the same as the one with $\e=1$, provided we introduce the rescaled coordinate $U'=\e\, U$

\be
ds^2 = \frac{\ell^2}{4} \frac{ d\tilde r^2}{\tilde r^2} +  \tilde r \,dU' dV + r_+^2 dV^2\;, \;\;\;\;\;\;\;\; U'\sim U' + 2\pi \e\, R \;, \;\;\; V \sim V + 2\pi R \label{dlcqads3}
\ee
Such identifications can be obtained by compactifying the CFT on a space of radius $R_s= R \sqrt{\e}$, followed by a boost with parameter $e^\g = 1/\sqrt{\e}$. 
In the $\e \r 0$ limit, the boost becomes infinite and the identification only acts on the null $V$ coordinate, precisely realising 
 Seiberg's definition of DLCQ \cite{Seiberg:1997ad}.   
 The CFT energies are affected by the boost as 
%
%In fact, a better way to think about DLCQ is as the infinite boost limit of a spatial compactification of radius $R_s \r 0$. After a boost with strength $\g$,   Taking $R_s \r 0, \g \r \infty$ with $R_s e^\g = R = $ fixed, we obtain precisely the identification above. The boost acts on the left/right-moving energies in the CFT as 
$E_{L,R} \r e^{\pm\g}  E_{L,R} = \e^{\mp \frac{1}{2}}  E_{L,R}$.  In order for $E_L$ to be finite after the large boost, the initial $E_L \approx 0$, or $h_L \approx c/24$. Also, any potential small  gap for the left-movers in the original CFT would become infinite after the boost, resulting into a  `frozen'  left-moving sector, where only the ground states survive. 
%
%Thus, the initial $E_L$ needs to be very tiny in order for the energy after the boost to be finite-  the DLCQ limit effectively freezes the left sector (by freezing $h_L=c/24$). It is also said to generate an infinite gap in this sector, since $P_L = (h-c/24) e^\g/R$, in the sense that any 
 %
%  - however, this is not so, because the expected gaps are exponential, while the boost is not truly infinite. 
 % 
  On the other hand, $E_R \r E_R e^{-\g} = (E_L + n/R_s) \sqrt{\e}$ is finite and quantized in lightcone units.  From the point of view of the original black hole, the DLCQ limit is also fixing the third charge, denoted as $J$ above, 
 though one can consider situations in which it varies.

In the strict $\e \r 0$ limit,  the spacetime \eqref{dlcqads3} can be written  as a $U(1)$  fibre over AdS$_2$, parametrized by $V$. This spacetime is known as   self-dual AdS$_3$ and has  $SL(2,\mathbb{R})_L \times U(1)_R$ isometry. It can also be obtained as a quotient of AdS$_3$, the parameter of the quotient being related to the identification of the $V'\equiv r_+ V$ coordinate, which also encodes the right-moving temperature \cite{Strominger:1998yg}.  An ASG analysis of this spacetime reveals the following enhancement of the isometries

\begin{figure}[!h]
\centering
\includegraphics[height=2.2cm]{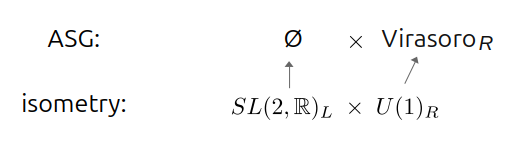}
\end{figure}

\vskip-3mm

\noindent in agreement with expectations: since this is effectively an AdS$_2$ setup, the charges associated with the $SL(2,\mathbb{R})_L$ generators are identically zero (and thus, $SL(2,\mathbb{R})_L$ is not a global symmetry of the theory, unlike in vacuum AdS$_3$). The right-moving Virasoro is an IR vestige of the corresponding symmetry of the UV CFT$_2$ - the only one that survives the very-near-horizon limit, since it has finite energy from the very-near-horizon point of view \cite{Balasubramanian:2009bg}.

 In light of the recent developments \cite{Ghosh:2019rcj}, the picture above needs to be updated at least for the non-supersymmetric case, as the classical extremal geometry is not at all reliable.  One can easily adapt the focussing argument to 
  near-extremal  black holes, %\footnote{ Writing \eqref{btzmet} as
%\be
%ds^2 = \frac{\ell^2}{4} \frac{\e \, d\rho^2}{\rho ( \e \, \rho + r_+^2-r_-^2 )} + \left(\e\, \rho + \frac{r_+^2-r_-^2}{2}\right) (d\s^2-dt^2) + \frac{(d\s+dt)^2}{4} (r_+-r_-)^2 + \frac{(d\s-dt)^2}{4} (r_++r_-)^2 
%\ee
%one notes 
%the geometry is the same as $\e =1$ geometry if one fixes $r_++r_-$ and $(r_+-r_-)/\e$, and the identifications are as above. Geometries with the same $r_++r_-, \s-t$ and $r_+-r_-, \s+t$ related by rescaling are the same.}
 with the difference that now the boost parameter is not to be  taken infinite, but only up to a maximum value set by the temperature of the system.  It is an interesting question whether the ASG analysis can be redone on this small, but finite $T$ geometry, and in particular whether it is sufficient to only keep the linear in $T$ departure of the geometry  from the very-near-horizon limit in order to see the expected ASG results.  If so, then it is interesting to ask whether the above-mentioned ASG analysis  of the self-dual geometry  can be viewed as a limit of a meaningful near-extremal computation. % - the CFT expectation would be that the answer is yes. 

 To sum up, while the conclusions of the bottom-up analyses  - namely,  the ASG computation 
and the Cardy entropy match -  would na\"{i}vely appear to also apply to the near-horizon limit of the (near)-extremal BTZ black hole, one should be careful whether to trust them. Besides issues of dominance one should keep under check, there are important quantum corrections to the classical gravity picture. Fortunately, these can be brought under control using a dimensionally-reduced description in terms of JT gravity coupled to KK matter\footnote{These corrections can also be  computed using the full asymptotically flat/AdS geometry  \cite{Kapec:2024zdj,Kolanowski:2024zrq,Castro:2025itb}.}. Nonetheless, the JT description - or any other description of the (very)-near-horizon limit only - cannot explain the entropy near extremality, due to their focussing on a subset of states, which is not closed under modular transformations. 
 % 
% In particular, they ,which comes from using modular invariance in the parent UV CFT, as modular transformations take one out of the subset of states considered in the limit; 
 Cardy's formula is nonetheless still obeyed, thanks to the embedding in the UV CFT. 
%invariance  relates small and large $T_L$. \emph{Correct? Recheck!}
In addition, the JT description  makes it difficult to see the symmetries of the parent UV theory from a near-horizon perspective, even if they are expected to be present. The full $3d$ description of the  near-horizon  geometry is able to capture at least the right-moving Virasoro  of the UV CFT, but the calculation needs to be updated to take into account the proper, quantum-corrected geometry.   It appears likely that the symmetries - which live at the boundary of the region - will not be significantly  affected by these corrections.

\subsection{General near-extremal black holes and the Kerr/CFT correspondence}

Unfortunately,  black holes with an AdS$_3$ factor in their near-horizon region are mostly confined to the realm of string theory and toy models. The black holes  observed in our universe are of  the Kerr type and are parametrized by their mass $M$ and angular momentum $J$, with $G M^2 \geq |J| $;  if $M$ is close to this lower bound, the black hole is said to be near-extremal.

The Kerr/CFT correspondence \cite{Guica:2008mu} (see also the reviews \cite{Bredberg:2011hp,Compere:2012jk}) is an attempt to provide a microscopic explanation for the entropy of the extreme %\footnote{While the initial proposal concentrated on the exactly extremal Kerr black hole, which was not known at the time to be problematic, most of the important results in the correspondence  extend to the near-extremal case. In view of this revised understanding, we will talk about the Kerr/CFT correspondence as applied to near-extremal Kerr, at the expense of being slightly anachronistic.} 
Kerr black hole. %, which  was not known at the time to be problematic. 
In the following, we review the basic statement and evidence for the correspondence as it was originally proposed;
 then, we discuss two classes of  issues that arise, and also comment upon the extent to which the conclusions of the extremal analysis are to be affected by the recent developments in evaluating the gravitational path integral on the (near)-extremal black hole geometry \cite{Rakic:2023vhv,Kapec:2023ruw}.

\subsubsection{Statement of the correspondence}

The Kerr/CFT correspondence again makes use of the existence of a near-horizon  limit and a special, more symmetric geometry that appears in the near-horizon  region of the extreme Kerr black hole \cite{Bardeen:1999px}. The latter is obtained by  expanding the Boyer-Lindquist radial coordinate in the extremal Kerr geometry as $r_+ + \e\,r$  and scaling $\e \r 0$, upon 
 rescaling $t \r t/\e$ and appropriately shifting the azimuthal angle $\phi$. The near-horizon extremal Kerr (NHEK)  spacetime one obtains in this limit  takes the schematic form 
\be
ds^2 = 2 J \Omega^2  \left[\,  - r^2 dt^2 + \frac{dr^2}{r^2} + a^2   (d\phi + r dt)^2  + \; \ldots \;\right]% \;, \;\;\;\;\;\; \phi \sim \phi + 2 \pi 
\label{wadsm}
\ee
where $\Omega(\theta)$ and $a(\theta)$
%  $\Omega^2= \frac{1}{2}(1+\cos^2\theta)$ and $a = \sin \theta/\Omega^2$
   are specific functions\footnote{Concretely, for Kerr $\Omega(\theta) = \frac{1}{2} (1+\cos^2 \theta)$ and  $a(\theta) =  \frac{\sin \theta}{\Omega (\theta)}$.%, in addition to an overall factor of the angular momentum, which is the only dimensionful parameter in the geometry.
   } 
of the polar angle $\theta$,  and the $\ldots$ simply stand for  $d\theta^2$. The
three-dimensional part of the geometry that we exhibited in parantheses is known as `space-like self-dual \emph{warped AdS$_3$}' and corresponds to   a $U(1)$ Hopf fibre, parametrized by $\phi \sim \phi + 2 \pi $, over AdS$_2$, whose coordinates are $r,t$.
The isometries of this spacetime are thus  $SL(2,\mathbb{R})_L \times U(1)_R$.  The qualificative `warped'  refers to the fact that, for  $a\neq 1$, the Hopf fibre is either squashed or stretched, and the spacetime differs from AdS$_3$ by a non-normalizable term; `self-dual' simply refers to the $\phi$ quotient.  When $a=1$, the spacetime becomes a quotient of AdS$_3$, which is nothing but the self-dual AdS$_3$  discussed in the previous subsection. Note that  $t,\phi$ are now \emph{null} coordinates on the boundary, while $r$ is the square of the usual holographic coordinate.  The appearance of an AdS$_2$ factor in \eqref{wadsm} indicates that one has all reasons to worry about the presence of large quantum fluctuations in the geometry; we leave this discussion to the following subsection.

  %\emph{Careful NHEK throat unstable to superradiance!}

 Note the near-horizon limit yielding NHEK is directly analogous to the very-near-horizon limit \eqref{dlcqads3}, rather than to the standard decoupling limit \eqref{declimd1d5}. %In particular, in view of the recent developments that we have just reviewed,
%and one has thus all rights to worry about the reliability of the classical extremal geometry one is focussing on. % This will be discussed in the next subsection, so far we concentrate upon the original proposal/ This section is written following the older literature, which referred to the exactly extremal geometry; in view of the more recent developments is should instead be read with the near-extremal geometry in mind.   
While the modes inside the throat na\"{i}vely decouple from those outside, the  decoupling from the far region is rendered subtle by the presence of superradiant modes in  NHEK  \cite{Bardeen:1999px}, with their associated instabilities, and of IR divergences in the NHEK partition function \cite{Rakic:2023vhv,Kapec:2023ruw}, which ultimately signal its failure. % We will be glossing over these issues until the next subsection. % of  decoupling from the far region.

 The Kerr/CFT reasoning closely parallels that for  black holes with an AdS$_3$ factor in their near-horizon region:  %, at least at a superficial level. %, leaving aside for a moment the subtle issues mentioned above. 

\bigskip
\noindent \emph{The asymptotic symmetry group computation}
\medskip

\noindent First, one shows that there exist boundary conditions on the fluctuations of the near-horizon metric that lead to non-trivial asymptotic symmetries, whose algebra consists of one centrally-extended copy of the Virasoro algebra, with central charge $12 J$.

These boundary conditions - which are posited, rather than  argued for or derived  from some fundamental principle - are quite peculiar, in that the $tt$ component of the near-horizon metric is allowed to have fluctuations that diverge as 
$\O(r^2)$. Notwithstanding, these boundary conditions are consistent, in the sense that the asymptotic charges \eqref{covphspch}  associated with the allowed diffeomorphisms    (which asymptotically take the form $\xi_n \sim e^{i n \phi} (\p_\phi + i n r \p_r) +\ldots, \, \forall n \in \mathbb{Z}$) are all  finite, integrable and conserved\footnote{To obtain  this  result, the use of the  covariant phase space formalism \cite{Lee:1990nz,Iyer:1994ys,Barnich:2001jy,Barnich:2007bf}  was crucial, as no other currently available formalism seems powerful enough to reproduce it. }.  The charges associated with the $SL(2,\mathbb{R})$ isometries of the spacetime vanish, as expected from the fact that the reduction of the geometry along the squashed sphere yields AdS$_2$ with a constant dilaton, to which the general arguments of \cite{Balasubramanian:2009bg} apply. The enhancement of the background isometries to asymptotic symmetries is 

\vskip-2mm
%
%\be
%SL(2,\mathbb{R})_L \times U(1)_R \;\;\; \rightarrow \;\;\; Virasoro_R
%\ee
\begin{figure}[!h]
\label{kcftasg}
\centering
\includegraphics[height=2.2cm]{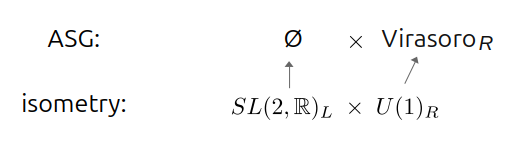}
\end{figure}

\vskip-3mm
\noindent Interestingly, the Virasoro enhances the $U(1)$ factor of the isometry group, despite the fact that the  asymptotics of the spacetime differ from AdS$_3$ by a non-normalisable term.

Based on these ASG results,  \cite{Guica:2008mu}  conjectured that the near-horizon dynamics  inside the NHEK throat is captured by a `chiral half' of a $2d$ CFT, where the  chirality  is related to the exact extremality of the background; one does expect to see conformal left-moving excitations and symmetries in the near-extremal case.  
% \emph{Is this also true for the toy mackgnds where the warping is det by the coupling?} 

\bigskip
\noindent \emph{Entropy match}
\medskip

\noindent The closest analogue of the Hartle-Hawking state for the Kerr black hole is the so-called Frolov-Thorne  vacuum \cite{Frolov:1989jh}, which looks thermal for observers that co-rotate with the horizon (see \cite{Duffy:2005mz,Ottewill:2000qh} for further discussions of various subtleties).  The state of the quantum fields in extreme Kerr is obtained by taking the extremal limit of the Frolov-Thorne result, upon changing to the $t,\phi$ coordinates that are natural in the very-near-horizon limit and to their associated momenta. The  left/right temperatures that couple to the momenta along $t, \phi$ are
\be
T_L = 0 \;, \;\;\;\;\;\;\; T_R = \frac{1}{2\pi}
\ee
If one now plugs these temperatures into Cardy's formula \eqref{cardyfintro}, using the central charge $c=12 J$ that results from the ASG analysis, one precisely reproduces the Bekenstein-Hawking entropy,  $S_{BH}=2\pi J$, of the extreme Kerr black hole. % The match can be easily extended to near-extremal black holes \emph{What is $S(T,J)$ for Kerr?}, where one simply assumes the same central charge for the left-movers as for the right (without any justification except the absence of a gravitational anomaly).
 This perfect entropy match supports the conjecture suggested by the ASG anaysis.

Remarkably, the above reasoning and the matching of the macroscopic and microscopic entropy formulae can be extended to almost all extremal black holes - be they higher-dimensional, charged, with higher derivative corrections, a.s.o. - which turn out to generically exhibit \cite{Kunduri:2007vf} a warped AdS$_3$ factor in their near horizon\footnote{In higher dimensions there can be various $U(1)$ symmetry directions, and thus various different ways of singling out a warped AdS$_3$ factor. }, as long as they possess some  rotation. % \emph{Does susy (how much) require AdS$_3$? True that only spacelike (and possibly null) warped AdS appear in the nh of extremal black holes?} % Moreover, it can be easily generalized to all sorts of extremal black holes ). All these black holes share a special structure of their near-horizon  decoupling region, since all extremal rotating black holes have a  in their near horizon region. \emph{Careful RN!} 
%
%This match has been quickly generalised to other extremal rotating black holes. As it turns out, the near-horizon geometry of any extremal black hole with non-vaninshing area always contain a wAdS$_3$ factor. 
 The ASG analysis easily generalises and the results for the central charge and temperature are always such that the extremal Cardy formula reproduces the Bekenstein-Hawking entropy, even for black holes where the ASG analysis can be associated with several possible $U(1)$ factors \cite{Lu:2008jk}. % \emph{Is there any ASG analysis near extremality?}

\bigskip
\noindent \emph{Match of correlation functions}
\medskip

\noindent The correspondence has  been extended  beyond the matching of the symmetries and  the entropy, to the matching of the scattering amplitudes in the extreme Kerr background to thermal CFT$_2$ two-point \cite{Bredberg:2009pv} and extremal three-point functions \cite{Becker:2010jj}. To obtain a non-zero temperature for the left-movers, one  simultaneously takes the near-horizon and near-extremal limit. This results in a finite left-moving temperature from the very-near-horizon point of view, while the asymptotic temperature vanishes.

On the gravity side, the scattering amplitude is obtained by focussing on modes
of a probe field (e.g., a massless scalar) near the superradiant bound $\om = \Omega_H (m + \e \, n_L)$, where $\om$ is the frequency of the mode from the viewpoint of asymptotically flat space, $m$ is the angular momentum quantum number, $\Omega_H$ is the angular potential of the black hole, and $n_L$ is the left-moving frequency from the throat point of view.  Upon stripping off factors associated with scattering faraway from the black hole and passing to the decay rate, \cite{Bredberg:2009pv} found a perfect-looking match to the CFT scattering amplitudes \eqref{bh2pf}.  The CFT  null directions  were identified with the NHEK $t,\phi$ coordinates, the CFT null momenta $p, \bar p$ -- with the momenta along these directions,  and the conformal dimensions of the would-be dual operators  were given in terms of the angular quantum numbers as %usual read off from the large $r$ behaviour of the scalar perturbation, and read 
%\emph{Correct for any mass?} % Interestingly, now the dimensions do not just depend on the mass of the field, but also on the angular quantum number

\be
h_L = h_R = \frac{1}{2} +\sqrt{\frac{1}{4} + \bar K_\ell - 2 m^2} \label{kcftdims}
\ee
where $\bar K_\ell (\ell, m)$ is the eigenvalue of the corresponding prolate spheroidal harmonic. This match is quite remarkable, given that only an $SL(2,\mathbb{R})_L$ symmetry is present in the geometry. We will comment on the $m$ - dependence of this answer shortly\footnote{In \cite{Bredberg:2009pv}, a match to a charged CFT two-point function is claimed. However, introducing the charge potential does not seem to be necessary - especially since no gauge field is present  in the Kerr background. %There also seems to be a type-o in eqn $(5.12)$.
 }
. The  dimensions \eqref{kcftdims} can become imaginary for $m \lesssim \ell$, a phenomenon related to near-horizon superradiance. 

One can similarly match finite-temperature extremal three-point scalar correlators in the NHEK geometry, where the extremal choice is made to simplify the integrals involved \cite{Becker:2010jj}. Given the  dependence of the would-be dimensions on the angular quantum number, $m$, one may need to fine-tune the mass(es) of the scattered fields in an $m$ - dependent fashion to satisfy the extremal correlator condition. % on the dimensions. %\emph{Correct?}

\subsubsection{The fine print}

At a first glance, the computations described above appeared to give a robust and consistent picture of the microscopic description of extremal Kerr black holes. However, upon closer inspection this correspondence exhibited a number of puzzling features, already at the time %even before taking into account the issue of quantum corrections

\bi
\item[i)]  the near-horizon geometry was \emph{not dynamical}, i.e. the boundary conditions we discussed only allow for  perturbations  of the near-horizon spacetime that are diffeomorphic to the unperturbed NHEK background \cite{Amsel:2009ev,Dias:2009ex}%; in addition, the Kerr/CFT boundary conditions did not respect the natural falloffs of metric perturbations in the respective background [\emph{Comment on flexibility in ASG.} ] 

\item[ii)] the left temperature vanished and  the right-moving temperature could not be changed without introducing conical deficits on the sphere, implying that Cardy's formula could not be derived 

\item[iii)] since  the would-be CFT$_2$ conformal dimensions depend on the
 angular quantum number, $m$, and the latter actually corresponds to the  momentum along the $\phi$ direction, $\bar p$, the scattering amplitude cannot correspond to the momentum-space two-point function in a standard, \emph{local} $2d$ CFT.
\ei
%Another problem that was also in principle known at the time, but perhaps less appreciated (likely due to the commonly-accepted assumption of a black hole mass gap) was that of the breakdown of the thermodynamic description for the near-extreme Kerr black hole, which should occor at $T_{bkdn} \sim 1/{\sqrt{G_N J^3}}$. In addition, breakdown of EFT for near-extremal black holes was observed recently. Both point to the fact that the exactly extremal geometry is unreliable. 

\noindent The first two  problems stem from the fact that the NHEK limit is a very-near-horizon limit, which is dominated by the presence of the AdS$_2$ factor.
 How to deal with the non-dynamics problem has since been understood, and  the basic physics is very similar to the near-extremal BTZ case we already explained: to recover some dynamics one should consider perturbations that break the exact AdS$_2$ asymptotics by letting the size of the $U(1)$ fibre diverge; the corresponding Schwarzian mode was identified\footnote{ One should note that, contrary to near-extremal BTZ and spherically symmetric black holes, for near-extreme Kerr the reduction to JT gravity is non-trivial due to the $\theta$ - dependent squashing of the geometry.} in \cite{Castro:2019crn}.  The quantum fluctuations of this mode become large at temperatures below the thermodynamic breakdown scale for Kerr,  $T_{break} \sim  1/{\sqrt{G_N J^3}}$; in particular,  the  partition function on  the exact NHEK background is IR  divergent. These divergences can be regulated by considering the linear - in - temperature correction to the black hole geometry, and the final answer for the gravitational partition function reproduces the universal Schwarzian behaviour \cite{Rakic:2023vhv,Kapec:2023ruw}.

 Thus, the apparent lack of dynamics associated with exact extremality (a problem that has since been superseded)  and the breakdown of the classical description of the black hole 
  can be avoided by turning on a small temperature\footnote{There is no astrophysical reason to consider smaller temperatures than $T_{break}$, but in principle one does have the formalism to deal  with quantum corrections at this
  and smaller temperatures. Then, one should also be wary of  possibly large tidal forces in the near-horizon region, due to eventual   higher-derivative corrections to the gravitational action \cite{Horowitz:2023xyl,Horowitz:2024dch}, and also take into account the possibility of superradiant decay \cite{Rakic:2023vhv}. }  $T_H > T_{break}$. This allows one to recover  some  dynamics and also reproduces the correct near-extremal physics%[and captures the quantum corrections to the classical geometry]
  , at the expense of violating the NHEK boundary conditions.  %The immediate question is then  whether the interesting  Virasoro ASG (\ref{kcftasg}) is still present in some form. We  will return to it  shortly.
  %\emph{Also mention Horowitz somewhere.}
%
%\bi
%\item Mukund comment on timescale of Schwarzian correction versus superradiant instability
%\ei

Let us now address the second point. As in the near-extremal BTZ case, the focussing involved in the very-near-horizon limit concentrates on a subset of states in a theory, rather than a full theory containing both high and low-energy states, and so one cannot derive Cardy's formula from the bulk information provided \cite{Balasubramanian:2009bg}. Moreover, while in BTZ states with arbitrary $T_R$ could be considered, in NHEK only $T_R =1/(2\pi)$ corresponds to a gemetry that is smooth at the poles of the deformed $S^2$.  The latter problem may be remedied by considering more general black holes, for example ones that carry both charge and angular momentum (see e.g. \cite{Compere:2012jk}); in that case, the temperature can be smoothly varied without introducing conical deficits.

% \emph{True? Does it also work for Kerr-Newman, uplifted?} In these cases (of charged black holes) it seems more natural to consider the KK charge direction as the one where the would-be theory lives, the ASG analysis working for any choice. Of course, this doesn't mean any such choice is correct.  Very unclear if several possible descriptions. (The question is of what class of black holes to consider as part of the same family/theory).  

As in near-extremal BTZ,  in order to justify the applicability of Cardy's formula  - assuming some variant of modular invariance is at play - it is not sufficient  to  turn on a little non-extremality; one   needs instead a proper decoupling limit, which isolates a full \emph{theory}. In practice, one would need to find an intermediate decoupling limit of the asymptotically flat near-extremal black holes - analogous to \eqref{declimd1d5} - that would yield a UV-complete such theory, whose properties could then be systematically studied. 
% Unlike BTZ, for general near-extremal black holes it is generally unclear how to take an intermediate decoupling limit that yields a wAdS spacetime, whose asymptotic symmetries could be rigorously studied. 
%
  However, even upon embedding Kerr-like black holes inside string theory \cite{Guica:2010ej}, it is not clear how to find such an intermediate decoupling limit. % for general near-extremal black holes. 
  
 %[ There do,  nonetheless, 
%
%disentangle a decoupling limit that yields a full theory (with a complete set of states) from one that yields a frozen subsector, though it can be done in a few special cases. It is quite clear that the physics of interest of the Kerr/CFT correspondence lies within this UV theory that is hard to isolate. [Part of the reason is that the Kerr black hole does not dominate the microcanonical ensemble (nor the thermal one, when it can be defined). \emph{Read about metastable states?} Q: Is there a clear range of energies where ``UV" warped AdS physics should dominate (i.e., flat space physics hasn't yet taken over)? Maybe wave eqn can give a clue.] However,
  % exist examples in the string theory setting    of %black holes allowing for
   % decoupling limits that yield a dynamical warped AdS$_3$ spacetime,  expected to be dual to a UV-complete theory, where the very-near-horizon limit of near-extremal black holes is a self-dual warped AdS$_3$ spacetime, as expected on general grounds. ]

%Given this state of affairs, 
In view of these points, 
the interpretation of the Virasoro symmetries found by \cite{Guica:2008mu} - assuming the conclusions of the ASG analysis on NHEK can be trusted - 
is not as the potential symmetries of a theory, but rather as the symmetries of a certain low-energy subsector thereof.  In order for this subsector to be non-empty, one would need the symmetries to survive beyond the strictly extremal limit.  The most reasonable optimistic interpretation one could have of them  is as an IR vestige of a similar extended symmetry that would exist in the putative  decoupled UV  description of the Kerr black hole that we previously invoked. %, where they can be used to constrain it and derive interesting properties. 
%
%which can be transported to any radius, as we saw can be done for BTZ\footnote{It would be nice to check this statement in the string theory settings where both an UV and an IR warped AdS$_3$ exist. [ In the BTZ geometry, this was trivial to see, as the fully decoupled geometry was available and very simple.] }. 
If this picture is correct, then one may use it to argue for the survival of the Virasoro symmetries in the near-NHEK geometry: first,  one would expect that the UV symmetries of the theory should also make sense at a small non-zero  temperature and, if it is true that they can be transported at any radius in the interior of the spacetime%\footnote{It would be nice to check this statement in the string theory settings where both an UV and an IR warped AdS$_3$ exist.}
, as  in  BTZ \cite{Compere:2015knw}, they should also be reproducible from an analysis of the near-NHEK near-horizon geometry that is classically more   reliable. While for the Kerr black hole this picture is hard to verify, it can  in principle be checked in string-theoretical examples of geometries that interpolate between a decoupled warped AdS$_3$ background in the UV and a space-like self-dual background in the IR \cite{Song:2011sr}. If correct, then one may be able to reproduce the right-moving Virasoro by working at  linear order in  the temperature around the NHEK background. On the linearly corrected geometry, one could also ask about the existence of 
  a potential left-moving Virasoro%\footnote{In the Kerr/CFT literature, there exist boundary conditions that single out a left Virasoro (at the expense of excluding the right one), but the charges are only nonzero if one integrates over time.}
 ;  the details of this exercise are  currently unclear even for near-extremal BTZ.

Let us now move on to the third  problem: that, despite the fact that the scattering amplitudes in NHEK take precisely the form of a finite-temperature CFT$_2$ correlation function, the would-be  CFT  conformal dimensions \eqref{kcftdims} depend on the angular quantum number $m$, which corresponds to the right-moving momentum in the dual theory\footnote{This momentum is quantized due to the near-DLCQ limit  NHEK corresponds to, but can  in principle be continuous.}.  This is a clear indication of non-locality of the dual theory along the $\phi$ direction;  however, the  structure of this non-locality is highly constrained, lest the amplitude would not take the CFT-like form \eqref{bh2pf}. 
%
% are momentum-dependent, which indicates non-locality. \cite{} considered a model where the momentum equalled some $U(1)$ charge, but does not appear to resolve it. Thus, whatever the UV theory whose near-DLCQ sector describes NHEK, it needs to be a non-local theory, but with a highly constrained non-local structure. 
This non-locality is clearly in tension  with the right-moving Virasoro structure uncovered by the ASG analysis and implies that, whatever the theory that describes the near-extremal Kerr black hole is, it cannot be a standard, \emph{local} $2d$ CFT. For this reason, we will henceforth  put  quotation marks around the `CFT' in Kerr/`CFT'.
 
\medskip

Let us  summarize our conclusions so far. While the Kerr/`CFT' reasoning ostensibly follows that for BTZ black holes, the correspondence suffers from multiple problems that stem from the fact that the NHEK limit is essentially a very-near-horizon limit. While some of them (such as the `old' non-dynamics problem and the updated  large quantum fluctuations one) can be brought under control by working with a near-NHEK geometry, the problem of  deriving Cardy's formula, which is important for explaining the near-extremal Kerr entropy, necessitates the existence of a proper decoupling limit that isolates a (UV-complete) \emph{theory}, and not just a subset of states.  

\begin{figure}[!h]
\centering
\includegraphics[height=5.5cm]{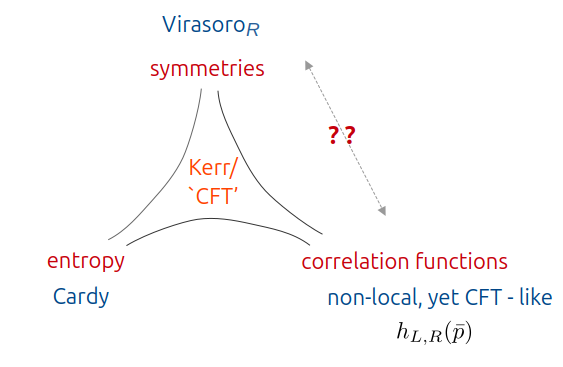}
\caption{\small{The triad of properties of the  dual to near-extreme Kerr, as inferred via bottom-up methods. }}
\label{kcft_triangle}
\end{figure}

If the results of   the bottom-up analyses \cite{Guica:2008mu,Bredberg:2009pv}  of the NHEK spacetime  are IR vestiges of the properties of such a decoupled  UV  theory, one infers the latter has the  properties listed in figure \ref{kcft_triangle}
  that one may hope to hold beyond the extremal limit in which the supporting calculations were performed.  The characteristics summarised  above   were, at least at the time, unlike the properties of any known two-dimensional QFT. Of course, this picture is just a suggestion,  given all the ambiguities  affecting bottom-up holographic analyses  that we discussed in the introduction.

%\vskip-4mm

The Kerr/`CFT' problem' may be split into  several logically distinct pieces, which are related to each other,  but that can be  independently formulated:

\begin{itemize}
\item[i)] what are the QFTs  dual to (decoupled) warped AdS$_3$ spacetimes, and what are their properties?
\item[ii)]  what is the precise relation between near-extremal black holes and such decoupled spacetimes?
\item[iii)] what is the near-horizon IR dynamics of near-extremal Kerr black holes?
\end{itemize}
\noindent   The first  question is the one that drives the research reviewed in this article. From the Kerr/`CFT' point of view, it is a question about the putative UV theory that explains the Kerr microscopics;  in practice, it has however only been studied for three-dimensional toy models of the correspondence.  

The main focus of this question is to understand whether QFTs with the properties summarised in figure \ref{kcft_triangle}  do exist, and in what sense these properties (or, possibly corrected versions thereof) are compatible with each other. One  also hopes to derive  a universal entropy formula by studying  the interplay   of the symmetries  with the entropy,  as well as understand the very special form of the non-localities revealed by the correlation functions by studying their action on operators. In the next subsection, we review the earlier work that indicated what type of QFT one should look for; the bulk of this review is dedicated to understanding the properties of such  QFTs, using a concrete example.

 %understanding whether it is possible to have QFTs that exhibit the set of incongruous-looking properties summarised in  figure \ref{}. %Since one expects to deal with a UV-complete theory, an important goal of this research strand is to For a discussion of the current status of this effort, see section \ref{}. \emph{Emphsize further Virasoro vs non-loc?}

There is, of course, much more to the Kerr/`CFT' correspondence than understanding the holographic duals to warped AdS$_3$ spacetimes. While the first question directly addresses the microscopic derivation of the entropy formula, as a matter of principle, the second aims to  apply this microscopic understanding to general near-extremal black holes, by understanding how to exhibit at least a partially decoupled warped AdS$_3$ spacetime in their geometry.  To help disentangle the presumed QFT directions from the non-QFT ones, the study of  higher-dimensional black holes may help, as would  embedding  the very-near-horizon geometry into an asymptotically Taub-NUT spacetime \cite{Song:2011ii,Bena:2012wc}, instead of   flat space.

Finally, another very  interesting  set of questions are  related to the IR near-horizon dynamics of near-extremal Kerr black holes and their generalisations.  It is here that the recent developments involving the Schwarzian play a crucial role in showing that the entropy near extremality tends to zero  \cite{Rakic:2023vhv,Kapec:2023ruw}, despite the classical  intuition,  and that at very low energies the  Hawking emission rate is significantly modified from  the semiclassical prediction \cite{Brown:2024ajk}. 
%
%The above properties were derived from analyses on the near-extremal very-near-horizon geometry, which suffers from various problems (non-dynamics, superradiance), most notably large quantum fluctuations. Some of these problems can be brought under control by considering the near-extremal black hole geometry  for $T > T_{bkdn}$. For example, one can compute the universal Schwarzian correction to the entropy that shows the degeneracy at extremality is not large, after all. 
%
As we briefly discussed, it is interesting to ask whether  other observables that are relevant for Kerr/`CFT', such as the ASG,  need to be corrected in view of these developments. A full picture of the IR dynamics would also  need to
 take into account  potential superradiant instabilities and effects of higher-derivative corrections. 
 The aim of this class of questions would be to obtain a \emph{universal}  low-energy effective description of a wide class of black holes, to the extent this is possible \cite{Castro:2021wzn}. This description would ideally capture the extended symmetries uncovered by the Kerr/`CFT' ASG analysis; however, it would not be a UV-complete  description, and thus would  have no bearing on the microscopic explanation of the full black hole entropy. 
 
 Each class of questions listed above is tuned to a particular range of energies: UV, intermediate and,  respectively, IR;  this review is exclusively focussed on the UV question. If one is able to answer them,    putting the responses together would presumably give a satisfactory picture of the near-extreme Kerr microscopics, which can then be used to give a dual description to many gravitational processes that occur close to the horizon of such a black hole, including those that may be observed in current and future experiments. It should however be clear that one still has a long way to go to arrive at  a satifactory microscopic picture of rotating near-extreme black holes. On the positive side  this is, among the holographic correspondences with roots in the real world, the one   closest to being understood,   at least in our opinion. 
In the following subsection, we attempt to motivate why.

\subsection{Warped AdS$_3$ toy models for  Kerr/`CFT' \label{wads3toym}}

There has been a significant amount of work dedicated to better understanding of the Kerr/`CFT' correspondence.  Much of it focussed precisely on the first question mentioned above, namely: what type of QFTs are  involved in the Kerr/`CFT' correspondence, and how do they  explain the Kerr entropy?

Since 
%
% which is the first proposed microscopic description for an extremal black hole, while being reasonably  precisely formulated. As explained above, this correspondence comports a number of subtleties: some of them relate to the non-dynamical nature of the decoupled region, but these are by now well understood features of physics in  AdS$_2$ \cite{}; others refer directly to the warped AdS$_3$ structure. Since this structure is necessary to understand the leading extremal entropy \cite{}, correlation functions, etc, (the triangle diagram) we will therefore concentrate on understanding the triangle.   
%
the direct study of the  near-horizon extreme Kerr geometry is difficult, both technically (due to the polar angle dependence of the warp factors) and conceptually (due to the very-near-horizon nature of the NHEK limit), the vast majority of  works circumvented 
%
%second because $\phi$, being  the azimuthal angle on the sphere, is required to have a fixed identification, which maps to a fixed right-moving temperature $T_R=1/(2\pi)$.
 these problems by studying  genuinely $3d$ toy models for the warped AdS$_3$ factor, believed to be  playing  the essential role in the duality.  In these toy models,  the warp factor ($a$ in \eqref{wadsm}) is a constant and the  problems associated to AdS$_2$ are avoided either by considering a decoupled theory, or by decompactifying  
 the analogue of the $\phi$ direction. % could be decompactified, thus 
 %
% : i) the temperature can be \emph{varied} \emph{Is this a problem in general, or just for pure Kerr?} and ii) the warp factor is a constant.
  Such geometries with constant warping can be obtained from the near-horizon limit of the extreme five-dimensional  Myers-Perry black hole (the direct higher-dimensional analogue of $4d$ Kerr) and its charged generalizations \cite{Cvetic:1996xz}, and are thus directly related to Kerr/`CFT'. %Most are \emph{not} decoupling limits.

From a three-dimensional perspective,  warped AdS$_3$ solutions need to be supported by an extra propagating degree of freedom  in addition to the non-dynamical graviton of $3d$ Einstein gravity with a negative cosmological constant -  usually either a massive graviton, or a massive vector field. One can distinguish two main  classes of models: 

\begin{enumerate}
\item[i)] \emph{bottom-up models}. These include topologically massive gravity  (TMG) \cite{Deser:1981wh, Deser:1982vy}, new massive gravity (NMG) \cite{Bergshoeff:2009hq} and simple  massive or topologically massive vector models coupled to Einstein gravity.  These theories contain warped AdS solutions where the warping factor is determined by a \emph{coupling} in the Lagrangian. Sometimes, they also  contain black holes, which are quotients of  global warped AdS \cite{Anninos:2008fx}. The existence of  a UV completion for these theories is unclear, in particular for TMG/NMG, whose propagating mode  is a ghost.

\item[ii)]  \emph{top-down models.} These are consistent trucations of string theory that admit warped AdS$_3$ solutions. Among their advantages is the fact that they are related to actual  black hole solutions in string theory, which also guarantees the existence of a UV completion. Their general features are different from those above: in particular, the warp factor is \emph{temperature-dependent}. These $3d$
 truncations resemble massive vector models coupled to $3d$ gravity and scalars, but with extremely finely-tuned scalar potentials. 
\end{enumerate}

\noindent In the following, we start by discussing observables that can be computed in these backgrounds using bottom-up holography: black hole thermodynamics, ASG analyses and  scattering experiments. As we emphasize, 
the properties of the putative dual theories that emerge from these analyses are rather different for the two classes of models above, indicating they must correspond to two different universality classes for warped AdS$_3$ holography, whose properties we discuss and contrast.

We subsequently concentrate on the top-down models  and discuss the holographic interpretation of the backgrounds  from the point of view of AdS/CFT, as well as explicit string-theoretical  realisations, which result in a certain  characterisation of the dual QFTs. % that could amount to a formal definition. 

\subsubsection{Holographic properties of the warped AdS$_3$ backgrounds }

As we already mentioned, the two different types of toy models we introduced above turn out to behave rather differently from a thermodynamic and an asymptotic symmetry perspective.  We start with the bottom-up models, whose simplicity made them into a prime candidate for modelling Kerr/`CFT'.

\bigskip
\noindent 
 \emph{Topologically massive  gravity}

\medskip

\noindent   These are $3d$ gravitational theories with a negative cosmological constant where the Einstein term is supplemented by a gravitational Chern-Simons term \cite{Deser:1981wh}. 
 The theory contains one propagating degree of freedom - the massive graviton - which is generally a ghost. It is sometimes excluded by imposing restrictive boundary conditions, though questions arise as to the consistency of such procedures \cite{Skenderis:2009kd}. 

 TMG possesses various warped AdS$_3$ solutions \cite{Anninos:2008fx}, namely geometries with (local) $SL(2,\mathbb{R}) \times U(1)$  isometries that can be written as a spacelike/timelike/null $U(1)$ Hopf fibre over Lorentzian or Euclidean AdS$_2$. The warping factor - the coefficient of the Hopf fibre, analogous to  $a$ in \eqref{wadsm} - is determined by the coupling of the topologically massive term in the Lagrangian\footnote{More precisely, letting $\nu = \frac{\mu \ell}{3}$ with $\mu$ the graviton mass (or inverse TMG coupling) and $-\frac{2}{\ell^2}$ the cosmological constant, one has timelike/spacelike stretched solutions for $\nu>1$,  squashed ones for $\nu <1$, and null warped AdS ones only for $\nu=1$. }. Timelike stretched AdS$_3$ contains closed timelike curves (CTCs). Black holes are free of CTCs in the spacelike stretched case, and correspond to  quotients of global spacelike stretched AdS. Near infinity, the $U(1)$ Hopf fibre maps to a linear combination of the black hole Schwarzschild time ($t$) and the spatial coordinate ($\theta$), while global time in the AdS$_2$ base is related to $\theta$ only.

% The black holes - which are free of CTCs if spacelike stretched -  are quotiends of global warped AdS$_3$ . % \emph{True for both?} In these models, one can also obtain qualitatively different warped AdS solutions, where the $U(1)$ fibre is  timelike (timelike stretched has CTCs) or null (at $\nu=1$ only). 

The entropy of  the spacelike stretched black holes in TMG takes the form \cite{Detournay:2012pc} %\emph{Redo factors!}
 
\be
S_{wBH} = - \frac{2 \pi \ell^2}{3k G} \,  M + 2\pi \sqrt{ \frac{c_L}{6}\left( - \frac{ M^2 \ell^2}{k} -J\right)} \label{tmgentropy}
\ee 
where  $M = Q_{\p_t}$ is the mass, $J = Q_{\p_\theta}$ the angular momentum, while
$c_L$ and $k \, (<0)$ are given in \eqref{cexttmg}. 
Note this is rather far from looking like the Cardy formula \eqref{cardyfintro}.

% Notwithstanding, in the early/original work \cite{} this entropy of  spacelike stretched black holes was suggestively written as a Cardy formula  in terms of some `temperatures' (not related to the Hawking temperature and angular potential in the standard way, but instead rel to the quotients of global wAdS that yield the bhs) and some $\nu$ - dependent central charges, one of which is \eqref{} below, and the other differing from it by the TMG anomaly (they also agree with the TMG central charges when $\nu=1$). Note this is mostly/just numerology, as subsequent analyses - reviewed below - point instead towards a ``warped CFT'' description, from which this Cardy-like formula can also be derived, without the theory being a CFT.

  In TMG, there exists a natural set of boundary conditions that preserve the asymptotic form of the metric and allow for all the black hole solutions \cite{Compere:2008cv}.    The associated asymptotic symmetry algebra is %\textcolor{red}{\emph{Careful meaning global symm!} }
\begin{figure}[h]
\centering
\includegraphics[height=2.2cm]{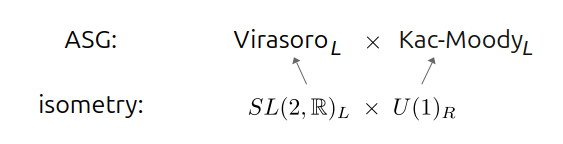}
  \caption{The symmetry enhancement pattern that defines a warped CFT.}
  \label{wcftenh}
\end{figure}

\vskip-1mm
\noindent where, in  our conventions, the left-moving sector is the one associated with the $SL(2,\mathbb{R})$ isometry. Somewhat peculiarly, the spatial translations generator - whose spectrum is unbounded -  belongs to this sector, whereas time traslations are associated with the $U(1)_R$, which is decompactified. This ASG has  central extensions
\be
c_L = \frac{(5 \nu^2+3) \ell}{\nu (\nu^2+3) G} \;, \;\;\;\;\; k=-\frac{(\nu^2+3) \ell}{6\nu G} \label{cexttmg}
\ee
The symmetry enhancement pattern \ref{wcftenh} is known as a \emph{\textbf{warped CFT}} structure.  Warped CFTs are \emph{putative} two-dimensional QFTs with $SL(2,\mathbb{R})_L \times U(1)_R$ global spacetime symmetry (left conformal symmetry and right translations) where, as expected, the $SL(2,\mathbb{R})_L$ symmetry is enhanced to Virasoro$_L$ but, interestingly, the right-moving translations are enhanced to a \emph{left-moving}  $U(1)$ Kac-Moody algebra. Note these theories are not Lorentz-invariant. This interesting symmetry enhancement possibility  was uncovered in \cite{Hofman:2011zj}, via  an abstract analysis that, in particular, used locality\footnote{Strictly speaking, \cite{Hofman:2011zj} only assumed left translation and scaling symmetry;   special conformal  follows from enhancement.  }. So far, the only known concrete example of warped CFTs are free \cite{Hofman:2014loa,Castro:2015uaa}%(and are dual to smth known as lower spin gravity)
, and  it is not known whether interacting QFTs with such enhanced symmetries exist.

The ASG analysis quoted above suggests that TMG with appropriate boundary conditions is holographically dual to a non-trivial warped CFT. Nonetheless, behind such a statement there lies a very important assumption, namely that TMG makes sense as a quantum theory, beyond the semiclassical limit. The likelyhood that this assumption is correct is rather small, given that even perturbatively, TMG has ghost-like excitations. Consequently, one should not immediately take such a `duality' as evidence for the existence of non-trivial warped CFTs. Note that, if TMG  were indeed dual to a warped CFT, the latter would  need to be both non-unitary - as indicated by the negative level above -  and non-local on the right (if  dynamical at all), as indicated by the fact that the $SL(2,\mathbb{R})_L$ dimensions of any propagating mode\footnote{The analysis of \cite{Anninos:2009zi} found that the TMG propagating mode in spacelike stretched AdS is excluded by the Comp\`ere-Detournay boundary conditions, suggesting this could be a theory with no propagating degrees of freedom. The general philosophy emphasized in \cite{Skenderis:2009kd}, according to which half of the modes of the fields should be kept, suggests instead these boundary conditions are overly restrictive. If so, then the ASG  may be even larger for the `correct' boundary conditions.   } depend on the right-moving momentum, as they  also did  in  NHEK. % background.% \emph{Are the propagating ghost modes allowed or not by the Compere-Detournay bnd cond?}

A very nice result that can be derived for abstractly-defined warped CFTs
%
%The main piece of evidence that TMG with these boundary conditions is described by a wCFT 
is that  theories with such symmetries have a universal high-energy density of states, given by  \cite{Detournay:2012pc}
\be
S= - \frac{4\pi i P_0 P_0^{vac}}{k} + 4\pi \sqrt{- \left(L_0^{vac} -\frac{P_{0,vac}^2}{k}\right) \left(L_0 - \frac{P_0^2}{k} \right)} \label{wcftent}
\ee
where $L_0, P_0$ are the zero modes of the $SL(2,\mathbb{R})_L$ and, respectively, $U(1)_R$ charges, and  $L_0^{vac}, P_0^{vac}$ are the corresponding charges of the vacuum state. This formula is derived using a warped analogue of modular invariance.

As it turns out, the entropy \eqref{tmgentropy} of black holes in TMG precisely matches this formula, upon identifying $L_0$ with (minus) the angular momentum and $P_0$ with the energy, while the vacuum state in which  $L_0^{vac}, P_0^{vac}$ are computed is identified as   global timelike warped AdS - a CTC-containing spacetime with an imaginary value of $P_0$. See \cite{Detournay:2012pc} for the detailed argument leading to this choice.

%  Plugging in $L_0^{vac} = (P_0^{vac}0^2 = - c/24$, one  can precisely reproduce the wBH entropy formula \eqref{}, so everything looks consistent. [Moreover, it was shown in \cite{} using a variant of modularity arguments that their entropy is also universal. \emph{Notation!}]

In order to not have to deal with imaginary conserved charges, \cite{Detournay:2012pc}  introduce an alternate basis for the symmetries of warped CFTs, which are related to the original Virasoro-Kac-Moody generators $L_n,P_n$ via
\be
\tilde L_n = L_n - \frac{2}{k} P_0 P_n +\frac{1}{k} P_0^2 \d_{n,0} \;, \;\;\;\;\; \tilde P_n = \frac{2}{k} P_0 P_n - \frac{1}{k} P_0^2 \d_{n,0} \label{Lttmg}
\ee
The relation between the two sets of generators formally corresponds to a spectral flow by $\frac{2}{k} P_0$, where $P_0$ is  the generator of right-moving translations\footnote{We will  encounter precisely this  spectral flow by the right-moving Hamiltonian in our study of $J\bar T$ - deformed CFTs. }, accompanied by  a rescaling of the $U(1)$ current modes by the same operator-valued factor. %This corresponds to a non-local reparametrisation of the theory. 
The algebra of the redefined generators follows from that of  $L_m, P_m$ and  corresponds again to a Virasoro-KM algebra, except for the presence of 
 certain state-dependent central terms proportional to $\tilde P_0$.  In particular, the Kac-Moody level is proportional  to $\tilde P_0$.

The  main advantage of using the generators \eqref{Lttmg} is that now all backgrounds, including the timelike warped vacuum, have real charges. Note the rescaling of the $P_n$ by $P_0$ effectively amounts to rescaling time by an energy-dependent factor. %, the generators \eqref{Lttmg} are associated with boundary conditions  where  the leading metric at infinity can fluctuate. \emph{Is this actually true?} 
Given these new generators, it is natural to define an ensemble - denoted as the `quadratic' ensemble  in \cite{Detournay:2012pc} -  where one traces over the Hilbert space in presence of  chemical potentials for $\tilde L_0, \tilde P_0$. In this ensemble, the asymptotic density of states is governed by a Cardy-like formula 
\be
S = 4\pi \sqrt{-\tilde P_0^{vac} \tilde P_0} + 4 \pi \sqrt{-\tilde L_0^{vac} \tilde L_0}
\ee
which only involves real conserved charges. This formula explains, in particular, why the early work \cite{Anninos:2008fx} could suggestively write the TMG black hole entropy in the Cardy form $S= \frac{\pi^2 \ell}{3} (c_L T_L  + c_R  T_R )$, where $T_{L,R}$ are not the potentials coupling to the left/right-moving energies, but rather the parameters of the global warped AdS quotient that yields the black holes: these $T_{L,R}$ correspond to the chemical potentials for the charges in the quadratic ensemble, in which the entropy takes a Cardy form. Of course, by no means does this `suggestive rewriting' imply that the dual theory is a CFT: if it exists, then the evidence we have reviewed indicates it should be a warped CFT.

Note that the perfect match between the universal warped CFT entropy \eqref{wcftent} and that of black holes in TMG \eqref{tmgentropy} does not require the existence of a consistent holographic dual to TMG: the warped modular invariance argument can likely be run at the level of large diffeomorphisms acting on  the (Euclideanised) warped AdS$_3$ boundary, where the gravitational partition function - whether it exists or not - is approximated by the classical TMG action. Therefore, a microscopic interpretation of the entropy of  black holes in TMG in terms of state counting  need not exist. If it does, then most likely it is the entropy of a thermal state in a `holographic warped CFT'.

%Some comments about scattering: clearly momentum-dependent conformal dimensions (typical wAdS), however, propagating modes are often excluded because they are ghosts. \emph{Comment implications for dual theory and how proper holography is done.}

 The overall conclusion of this analysis is that, \emph{if} there exists a holographic dual QFT to TMG with Comp\`ere-Detournay boundary conditions, then the bottom-up holographic analyses  we reviewed strongly suggest a `warped CFT' structure: a theory with extended left-moving Virasoro-KM symmetry, where the KM factor enhances right translations. The entropy in such a theory takes the universal warped CFT form \eqref{wcftent}, as implied by the warped variant of modular invariance. If the theory has dynamical degrees of freedom, represented via propagating fields in the bulk, then their dynamics is clearly non-local%\footnote{This non-locality should also be present at the level of the  boundary graviton correlation functions, but is less obvious. }
, as indicated by the dependence of the $SL(2,\mathbb{R})$ conformal dimensions on the right-moving momentum, which is a universal signature of the warped AdS geometry.   
 
 \begin{figure}[!h]
\centering
\includegraphics[height=5.5cm]{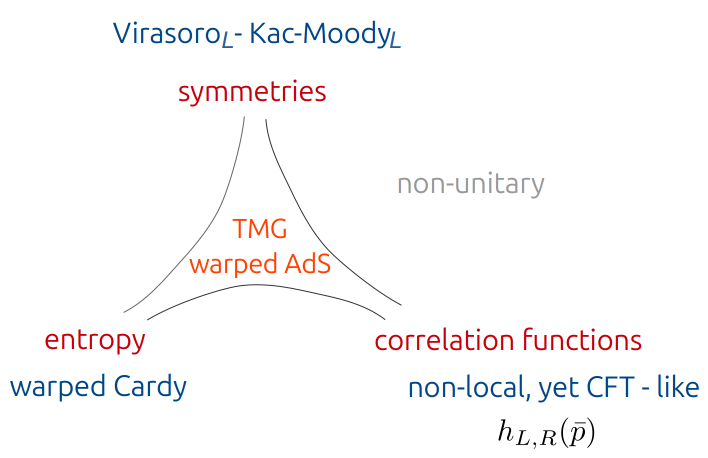}
\caption{\small{Property triad of the putative holographic dual to waped AdS solutions of TMG.}}
\label{tmg_triad}
\end{figure}

\vskip-1mm
 
\noindent The putative warped CFT dual to TMG would be non-unitary, as indicated by the negative level \eqref{cexttmg} and the presence of ghosts in the bulk. While the latter can be banished (at linear level) via restrictive boundary conditions,  it is absolutely unclear whether the resulting theory does not suffer from other, less obvious inconsistencies. Such restrictive boundary conditions are also not in the general spirit  of holography \cite{Skenderis:2009kd}. To the extent that boundary conditions in holography should fix just half of the data that characterises the asymptotic behaviour of the fields, it appears likely the correct boundary conditions for  TMG in an asymptotically warped AdS spacetime are more lax than those of \cite{Compere:2008cv}. Even if the `correct' ASG is larger\footnote{One reason to be expecting this is that, if the correspondence is dynamical (i.e., not describing just boundary gravitons that can be restricted at will) then one expects fields with non-trivial dependence on the right-moving coordinate. This dependence will be non-local, as indicated by the asymptotic falloffs of propagating modes. If the boundary theory is two-dimensional and non-local on the right, it seems  peculiar that all components of the stress tensor current are local and only depend on the left-moving coordinate.  Note also that the conclusions of \cite{Hofman:2011zj} only apply to local theories.   } than \ref{wcftenh}, the derivation \cite{Detournay:2012pc} of the black hole entropy from warped modular invariance is still expected to hold.

% \emph{It is somewhat strange that stress tensor modes only depend on $x^+$. The Strom-Hofman arguments don't apply because of non-locality. Even if restricted this ASA sufficient for studying just the bh subsector.}

%  there is evidence of wCFT ASG and thermodynamics in these theories, but there are ghosts everywhere. Also not clear ASG is the full one \emph{Check!}, and whether unstable solutions/ghosts must be included in the boundary conditions.  Even if excluded, they strike back. Also, not often mentioned:

\bigskip
\noindent 
 \emph{Other bottom-up models}

\medskip

\noindent Other bottom-up models that admit warped AdS solutions have holographic features rather similar to TMG. One such model is new massive gravity \cite{Bergshoeff:2009hq}, which consists of $3d$ Einstein gravity %(with a wrong-sign kinetic term)
 with a particular higher-curvature coupling, which 
propagates two massive helicity-two modes.  This model also contains spacelike stretched black hole solutions (with the same metric as those in TMG), whose warping is determined by the parameters in the Lagrangian. A very similar ASG analysis \cite{Donnay:2015iia} to the TMG one finds again a centrally-extendend left-moving Virasoro-Kac-Moody algebra, where $c_L$ and $k$ have opposite signs.

% of chosen mass (two massive spin two fields, plus a massive ghost corresp to the trace of the metric). \emph{Understand wrong sign of EH term necessary in these theories.}  For NMG ASG see 1302.6643 and Donnay-Giribet, which seems to be a crappy paper, where the find wCFT ASG with a negative Virasoro cc.  For general black hole sols in TMG + NMG see Tonni.

%\emph{[Need to figure out whether these also fall under the wCFT framework, or were simply not studied enough.]} 

Another class of bottom-up three-dimensional theories one can study are massive or topologically  massive vector models coupled to Einstein gravity with a negative cosmological constant. Unlike in the gravitational toy models, here the issue of ghosts does not arise. These theories admit warped AdS$_3$ solutions, with the warping determined by the massive vector coupling in the Lagrangian \cite{Banados:2005da};  reality of the vector field restricts the deformation to be squashed only.  As a consequence, the warped black hole solutions in these theories - which, again, are locally equivalent to the global spacetime -  exhibit CTCs asymptotically. They are sometimes referred to as G\"{o}del black holes.

The asymptotic symmetries of the warped AdS spacetimes in Einstein gravity coupled to a topologically massive gauge field were studied in \cite{Compere:2007in}, using boundary conditions directly analogous to those in TMG.  The resulting asymptotic symmetry algebra consisted of a left-moving Virasoro - (Kac-Moody)$^2$ algebra, where the second copy of the affine $U(1)$ generators is associated with large gauge transformations of the vector field. The Virasoro-Kac-Moody enhancement of the spacetime symmetries takes the standard warped CFT form and,  as before, the central extension of the affine $u(1)$ algebra  has an opposite sign from the Virasoro central extension, indicating again a conflict with unitarity.  The entropy of the G\"{o}del black holes again takes the form \eqref{tmgentropy} characteristic of warped CFTs (up to signs), and thus it can likely  be derived from a similar argument based on warped modular invariance.

Motivated by the ASG results found in these toy models, \cite{Compere:2013bya}  (henceforth CSS) 
investigated the question whether  asymptotically AdS$_3$ spacetimes - viewed as a limiting case of warped AdS$_3$ -  could also  admit boundary conditions that yield an ASG of the form \ref{wcftenh}. %\emph{Is the dynamics restricted in warped AdS/ is there a smooth limit to AdS?} 
To achieve this, a particular component of the AdS boundary metric was allowed to fluctuate, while a particular subleading component had to be restricted\footnote{Note that these boundary conditions are somewhat less natural than their warped AdS counterparts , as they \emph{a priori} restrict the right-moving energy of the black holes,  whereas in warped AdS  the black hole parameters were free to fluctuate.}, lest charge integrability was violated. While this set of boundary conditions did yield the desired Virasoro-KM  asymptotic symmetries characteristic of warped CFTs, their holographic meaning % - which goes beyond just checking the finiteness and integrability of the conserved charges -
 is far from clear: given their almost complete freezing of the right-moving sector, do these  boundary conditions isolate a \emph{theory}, or just a truncated subsector thereof? % that happens to admit some extra symmetries? 
 Should one  interpret the CSS boundary conditions in terms of alternate quantization for half the metric components\footnote{In the sense that the usual non-normalisable mode fluctuates, while the  normalisable one  is fixed, except its zero mode.} and,  if so, what is the holographic interpretation of simultaneously considering backgrounds with different `fixed' modes? At a more technical level, it is not entirely clear whether these boundary conditions are compatible with the inclusion of generic matter field profiles which, even if they fall off fast near infinity, will generically backreact in a coordinate-dependent fasion on the metric component that \cite{Compere:2013bya} would like to hold constant.  %\emph{Correct?}
 
  As we will discuss later in these notes, alternate boundary conditions for AdS$_3$ that resemble the CSS ones,
but allow for full dynamics    do exist, and they are dual to a boundary QFT that can be independently formulated; however, the interpretation of the left-moving Virasoro-Kac-Moody piece of their associated ASG is not in terms of a warped CFT.

To summarize, bottom-up toy models for the Kerr/`CFT' correspondence  are simple, $3d$ theories of gravity coupled to  some other degree of freedom that admit warped  AdS solutions. In these models, the warping is determined by a coupling in the Lagrangian, and   natural boundary conditions  on the asymptotic metric suggest a warped CFT structure. This structure can be used to explain the entropy of warped AdS black holes. There are many questions regarding the consistency of these theories that we have briefly touched upon. Also note that no alternate boundary conditions on the NHEK geometry have  been found, that yield a left-moving Virasoro-Kac-Moody asymptotic symmetry algebra.

\bigskip

\noindent \emph{Top-down models}
\medskip

\noindent In this approach, one studies solutions of string theory (usually type II$B$) that contain a warped AdS$_3$ factor. Sometimes, these solutions can also be described within a consistent truncation of the $10d$ supergravity action to $3d$, which generally results into a  massive vector model coupled to $3d$ gravity and various scalars, with non-trivial potentials \cite{Detournay:2012dz}. % that can be obtained as , usually type IIB. These truncations are designed to capture warped AdS$_3$ solutions of string theory.  One can also work directly with the $10d$ action and its solutions.

 One large class of warped AdS$_3$ backgrounds in string theory is obtained by  starting with AdS$_3 \times S^3 \times M_4$  solutions of type IIB supergravity,   supported by either NS-NS or RR  flux, and applying  a
solution generating technique known as TsT (T-duality, shift, T-duality) along an asymptotically  null isometry direction in AdS and  one inside the compact space (along the $S^3$ or $T^4$), possibly combined with other S and T dualities \cite{Bena:2012wc,ElShowk:2011cm}. If one starts from Poincar\'e AdS, then one only generates null warped AdS backgrounds in this way; spacelike warped  solutions can be obtained by starting instead with a BTZ black hole as seed geometry\footnote{Of course, one can also find the solutions by directly solving the $10d$ supergravity equations of motion, inputting the isometries of warped AdS$_3$ and certain isometries of the $S^3$ \cite{Bobev:2011qx,ElShowk:2011cm}. If the deformation of AdS$_3$ is null, then the sphere is usually undeformed, whereas if it is spacelike warped, then the sphere needs to be deformed, too. The vacuum solutions can preserve $4$ or $8$ supersymmetries, depending on the  spherical symmetry breaking pattern.}. 
 The warping factor is \emph{not} related to the couplings in the Lagrangian, but rather depends on the product of the TsT parameter and the right-moving temperature of the black hole; since the former is continuous, one can  in principle understand these backgrounds as \emph{continuous deformations} of the CFT dual to AdS, which can at least be studied in perturbation theory. Note this important handle on the theory  is not available  for most  bottom-up models, as those  warped AdS solutions are not continuously connected to AdS.

 The above discussion only refers to how to construct warped AdS solutions in string theory. An important question is whether the backgrounds generated in this fashion can also be obtained via a decoupling limit in string theory: on the one hand, this  
 would strongly suggest they allow for a UV-complete dual description; on the other, such a limit would clarify which are the  fixed parameters and coordinates in the background,  %, which characterise the theory, 
 and which  the fluctuating ones,
 % which characterise the states in it
 an information upon  which  the thermodynamics and asymptotic symmetries of  the system depend rather sensitively.

  So far, only those backgrounds obtained by applying a TsT transformation to 
AdS$_3$ supported by purely RR three-form flux are known to correspond to a  decoupling  limit (of the backgrounds generated by the TsT acting on the full, asymptotically flat D1-D5 solution). This decoupling limit: $\a' \r 0$, with an appropriate scaling of the $B$ - field \cite{Bergman:2000cw,Alishahiha:2003ru},  is known to yield the dipole deformation of the gauge theory living on a D-brane's worldvolume \cite{Bergman:2000cw}. Consequently, we will refer to the associated decoupled warped AdS$3$ background as the D1-D5 `dipole' background \cite{ElShowk:2011cm,Song:2011sr}. While  the decoupling limit does give some clues into the nature of the holographically dual QFT - to be discussed shortly - it does not provide tractability in this case.

The remaining backgrounds 
%
%the action of the TsT on this underlying brane system is partially understood.
% this is, however, not the case for most backgrounds 
 studied in \cite{Bena:2012wc} are known to correspond  at best to very-near-horizon limits of various black string/black hole  solutions \cite{Guica:2010ej} - including the uplift of the five-dimensional Myers-Perry black hole with two equal angular momenta -  but not  to proper decoupling limits. To avoid the problems associated  to the AdS$_2$ that appears in the very near horizon limit, one simply decompactifies the warped AdS$_3$ Hopf fibre direction, which is unproblematic in this case. %- at least for $5d$ black holes with two equal angular momenta that lead to an enhanced $SU(2)$ symmetry - is always possible.
  One is also free to recompactify the spatial boundary direction % \emph{Can this be obtained from a decoupling limit?}
   which, in the decoupled warped AdS$_3$  spacetime dual to the dipole deformation of the D1-D5 CFT, corresponds to placing the boundary theory on a circle; in this case, the spacetime will present CTCs near infinity (as expected from a vector model, which only admits squashed AdS solutions), but not near the horizon \cite{Song:2011sr}. %, , where only squashed AdS solutions are possible. 

Thus, despite  these backgrounds being embedded in string theory, one does not have  independent, field-theoretical control over the dual theories, and  the only available way to access their properties  is via bottom-up holographic methods. The results of the bottom-up holographic analyses of such backgrounds depend on the assumptions one makes concerning the parametrisation and the boundary conditions. For example,  in \cite{Detournay:2012dz}, a two-parameter family of  warped BTZ black string solutions supported by purely NS-NS flux were studied.  Their entropy per unit length (measured along a direction that originated from the AdS$_3$ spatial coordinate, $\s$) came out to be given precisely by Cardy's formula when the two parameters were equal in absolute value;  the result does not appear to take on a nice form otherwise\footnote{One of these backgrounds has been intensively studied more recently in the $J\bar T$ context, where a differet choice of parametrisation seemes more natural, leading to a different answer for the entropy. We discuss this in section \ref{wads3stjtbsec}.}. The entropy per unit length is also given by the Cardy formula for the backgrounds - supported by both NS-NS and RR flux - obtained via the dipole  decoupling limit, whose parametrisation should  be trustworthy. %By redefining   what the fixed coordinates should be one  can also find agreement \cite{Detournay:2012pc} with the entropy formula in a warped CFT in the quadratic ensemble; this only reinforces our point regarding the importance of knowing what the decoupled coordinates are.  
Of course, one could aso have turned on more charges on the original BTZ solution; whether these impacted the Cardy form of the entropy was not studied in these older works.

 As for the ASG analyses, there is no consensus in the literature as to what the  boundary conditions that determine the ASG generators should be. For example, \cite{ElShowk:2011cm} employed boundary conditions that were directly analogous to the NHEK ones and found a right-moving Virasoro algebra upon integration on the null boundary direction of the  warped AdS$_3$ spacetime (the charges integrated along the spatial direction $\s$ were divergent). For the case of the dipole backgrounds, \cite{Song:2011sr} proposed boundary conditions that yielded either a left-moving or a right-moving Virasoro algebra%(with charges again integrated along the null directions)
, but not both.   On the other hand, \cite{Detournay:2012pc} proposed  boundary conditions for the same backgrounds that produced a warped CFT algebra, Virasoro$_L \times$ KM$_L$, with the charges  more satsifactorily integrated over the  spatial coordinate. The Kac-Moody level was again  found to be negative (and temperature-dependent), despite the fact that  the underlying theory is now expected to be unitary. Finally, a slight modification of the definition of the charges allowed \cite{Compere:2014bia} to simultaneuously find a left and a right-moving Virasoro algebra (with charges integrated along space) for several examples of warped AdS background in string theory\footnote{More precisely, this was achieved by adding a specific counterterm to the symplectic form, which cancelled some unwanted divergences. One should generally be careful  when adding  such counterterms, as they may afect the positive-definiteness of the symplectic norm \cite{Andrade:2011dg}.}.    Of course,  one would expect that a given boundary theory (e.g., the dipole-deformed D1-D5 CFT) is dual to a bulk theory with \emph{one} specific set of boundary conditions. The challenge is how to determine them, as there is no general prescription for how to find the boundary conditions in non-AdS holography\footnote{For example, \cite{vanRees:2012cw,Guica:2011ia} proposed two  different holographic dictionaries for warped AdS$_3$ spacetimes, by making different choices for the `sources', i.e., boundary conditions: one followed the natural structure of conformal perturbation theory, whereas the other was based on the general radial symplectic form proposal of \cite{Papadimitriou:2010as}. }.

\begin{figure}[t]
\centering
\includegraphics[height=5.5cm]{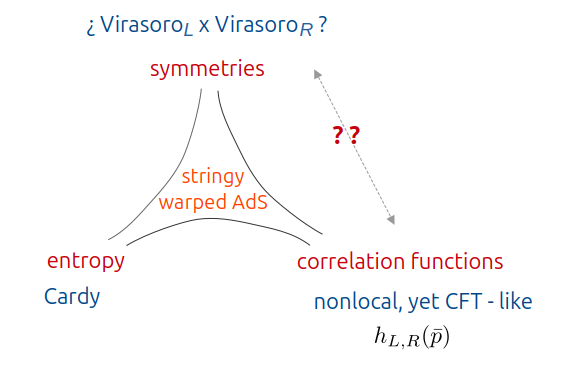}
\caption{\small{Triad of properties for the holographic duals to warped AdS backgrounds in string theory, inferred via bottom-up methods. There is no consensus in the literature on what the ASG should be. }}
\label{topdnwadstriad}
\end{figure}

To summarize, the properties of the holographic duals to warped AdS$_3$ in top-down constructions, as inferred from bottom-up holographic analyses, are depicted in figure \ref{topdnwadstriad}. 
 As we have tried to emphasize, there are large  potential errorbars to this picture; above, we have simply picked the answers that subjectively looked the most reasonable. As we saw, the ASG analyses are entirely inconclusive: there are several, different boundary conditions  that yield charges which are
 integrable and conserved, indicating these  requirements 
  are too weak  to select a preferred set among them.%\footnote{Or, perhaps the ASG analyses must be performed with significantly more care.}
   In particular, unlike in the bottom-up models, the warped CFT structure does not appear to be preferred; on the contrary, its results appear at odds with the expected unitarity of the dual theory.   %Choice of theory directions also important. 
   This example  thus provides a good illustration of the limitations of the bottom-up approach to holography.

\vskip-4mm

%[In the following, we will concentrate on the analysis of massive vector models that arise from consistent truncations of string theories. The reason is that\footnote{In addition, TMG has ghosts and its holography appears to work in a significantly different manner; the symmetry enhancement of warped CFTs was never shown to occur in extreme black holes. \emph{True??} } some of them describe the actual near-horizon region of proper extremal black holes in string theory (extreme Myers-Perry, or the direct $5d$ analogue of Kerr).  

%[ Most pressing is to find boundary conditions, and we do not, in fact, have a general rule in non-AdS holography\footnote{E.g. Balt vs me, vs all Schroed holography stuff (disregard dipole structure?).}.]

\subsubsection{The structure of the dual theory in top-down models}

%\subsubsection*{}

The main advantage of having a top-down model (via a string-theoretical brane construction) is the possibility to obtain an independent handle on the dual theory by taking the decoupling limit in the open string description of the branes. 

As we discussed, only a particular subclass of warped AdS$_3$ backgrounds in string theory are obtained via a known decoupling limit. Within this subclass, the dual field theory is known only for a higher-dimensional analogue of the system of interest. Even so, having a concrete glimpse into the structure of the QFT dual to this type of spacetime is invaluable. In the following, we  review this concrete example of non-AdS holography and then  argue, using the fact that the warped AdS backgrounds in top-down models are continuously connected to AdS, that their QFT duals should fit within the same framework.

\bigskip

\noindent \emph{Dipole-deformed field theories}% and Schr\"{o}dinger holography}
\medskip 
 
\noindent  The `dipole' D1-D5  warped AdS$_3$ background is generated by applying a TsT transformation  to the D1-D5 black string solution, with one  of the TsT directions   along the D1-D5 worldvolume and the other is perpendicular to the branes, and then taking the decoupling limit. More generally, one may consider the action of such a  TsT on $Dp$-branes.
At the level of supergravity, this corresponds to the standard solution-generating technique.  
 % To understand the dual theory, should understand action of TsT on the branes' worldvolume.  
As it turns out, the action of such a TsT transformation at the level of the gauge theory living on a D-brane's worldvolume is also understood \cite{Bergman:2000cw} and   corresponds to the so-called \emph{dipole deformation} of the gauge theory.  Many of its properties are a direct counterpart of the more involved non-commutative deformation \cite{Seiberg:1999vs}. 

The  TsT transformation generates a B-field, whose presence affects the boundary conditions for open strings ending on the brane. 
%
% As is well known, TsT transformations along a stack of D-branes (consider $D3$) lead to a non-commutative deformation of the corresponding gauge theory \cite{Seiberg:1999vs}. The field-theory interpretation of TsT transformations with one leg along a stack of D-branes and one leg perpendicular to them has been studied in \cite{Bergman:2001rw}.
 %
  The net effect at the level of the gauge theory living on the branes 
   is that the product of fields appearing in the Lagrangian  is replaced by a non-local star product
\be
(\phi_1 \star \phi_2) (x^\mu) = \phi_1 (x^\mu- L_2^\mu) \phi_2 (x^\mu+L_1^\mu)
\label{dipprod}
\ee
where $L_i^\mu$ are the `dipole lengths' associated with the fields $\phi_i$. They equal the product of the R-charges, $q_i$, of the fields under  the $U(1)$ R-symmetry direction involved in the TsT and a fixed vector along the TsT direction (denoted $y$) that points along the branes: $L_i^\mu =\frac{1}{2} q_i L^\mu$, $L^\mu = \l \delta^\mu_y$, where
 $\l$ is the TsT shift parameter. 
% $x^-$ is the other direction involved in the TsT. % \emph{Comment on why null is special.} 
  Associated  with this deformed gauge theory is a modified notion of gauge transformations, which act on the  fields  $\phi_i$ as
  
\be
\phi_i(x) \r U(x) \star \phi_i(x) \star U^{-1}(x) = U(x-L_i) \phi_i(x) U^{-1}(x+L_i)
\ee  
This is precisely the transformation of  a dipole extending from $x^\mu-L_i^\mu$ to $x^\mu+L_i^\mu$, hence the name.   Gauge-invariant operators built from $\phi_i$ are  non-local, and one needs to add a Wilson line of the appropiate length in order for the trace to be invariant \cite{Gross:2000ba,Guica:2017mtd}. A field redefinition known as the Seiberg-Witten map \cite{Seiberg:1999vs,Bergman:2000cw} allows one to rewrite dipole field in terms of one  transforming  locally at $x^\mu$, attached to a Wilson line.

Despite their non-locality, the theories obtained by  dipole deformations of renormalisable QFTs have been argued to be themselves renormalisable in a specific sense \cite{Dasgupta:2001zu}; in particular, no new information is required to absorb the divergences that arise in various diagrams.  At the level of the momentum-space Feynman rules, the non-local star product \eqref{dipprod} simply adds momentum-dependent phases to the vertices. Using momentum and R-charge conservation, these phases can be shown to almost completely cancel in planar diagrams: the latter are identical to their undeformed counterparts, up to an overall phase factor that only depends on the external charges and momenta. In particular, the planar free energy is the same, with important consequences for the thermodynamics of the system.

\bigskip

\noindent \emph{Dipole-deformed $\mathcal{N} =4$ super Yang-Mills}
\medskip 
 
\noindent  Let us now concentrate on dipole-deformed D3-branes (in the decoupling limit), which have the advantage that the undeformed gauge theory is a UV-complete theory and enjoys conformal symmetry.  % In the case of D3 (\emph{only?}) 
In general, by singling out the vector $L^\mu$, the dipole deformation breaks rotations and conformal transformations along the corresponding directions, as well as the $SU(4)$  R-symmetry group. However, when the vector $L^\mu$ is null, $L^\mu\propto \d^\mu_-$ where $x^-$ is a null direction, the theory preserves a non-relativistic conformal (a.k.a  Schr\"{o}dinger) subgroup of the original conformal group -
%
%the theory preserves non-relativistic (Schr\"{o}dinger) conformal symmetry,
 the higher-dimensional analogue of $SL(2,\mathbb{R}) \times U(1)$.
 The  Schr\"{o}dinger group consists of translations, spatial rotations and Galilean boosts acting on the coordinates $(x^+, x^i)$ - where the remaining null coordinate, $x^-$, plays the role of non-relativistic time - together with  dilatations

\be
 x^+ \r \eta^2 x^+\;, \;\;\;\;\;\; x^i \r \eta x^i \;, \;\;\;\;\; x^- \r x^- \label{nrscal}
 \ee 
 % - with respect to which the deformation is exactly marginal. 
and one special conformal transformation, all non-relativistic. %The momentum along $x^-$ plays the role of mass.  
 We will henceforth concentrate on the  null dipole deformation only.

As is clear from \eqref{dipprod}, the deformed theory is \emph{non-local} along $x^-$ and  \emph{local} along the remaining null direction $x^+ $ and spatial ones,  $ x^i$. The Seiberg-Witten map mentioned above allows one to rewrite dipole-deformed $\N=4$ SYM as the standard $\N=4$ SYM deformed by an infinite number of  irrelevant operators of ever increasing dimension, with finely-tuned coefficients\footnote{The equivalence of the two descriptions can only be established perturbatively in $\l$, up to arbitrary  finite order \cite{Seiberg:1999vs}. }. If the deformation is null, then all these deforming operators preserve Schr\"{o}dinger symmetry -  being exactly marginal with respect to the Schr\"{o}dinger group - providing another way to ascertain that the theory is local along $x^+, x^i$ and non-local along $x^-$. 
 % Despite the presence of these irrelevant deformations, dipole-deformed $\N=4$ is UV complete and,
While the arguments of \cite{Dasgupta:2001zu}, which rely on the dipole star product structure,  indicate that the theory is UV-complete,
%
 %which is hard to understand from  this infinite fine-tuning picture; in fact, the dipole-deformed theory is secretely renormalisable \cite{}, which is easier to see . 
 this fact is entirely obscured in the irrelevant deformation picture.

\begin{figure}[h]
\hspace{8mm}
\begin{minipage}{0.45\linewidth}
\flushleft
%\vspace{6mm}
\includegraphics[height=3.5cm]{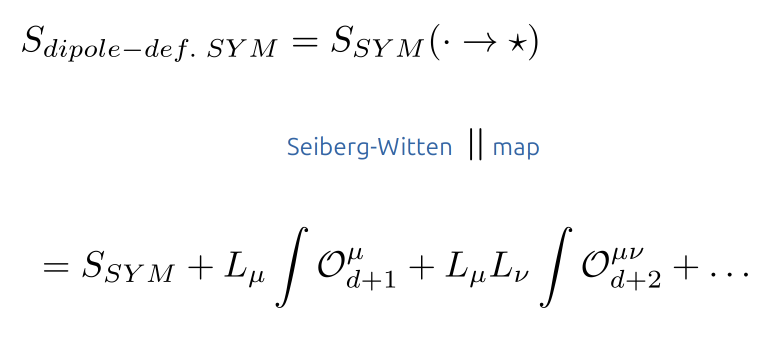}
%{\color{blue} This definition specifies the operator that is turned on at each point along the flow. \emph{Correct? Renorm. operator?} }
% \vspace{2mm}
\end{minipage}
\hspace{1.5cm}
\begin{minipage}{0.45\linewidth}
%\flushright
\includegraphics[height=5cm]{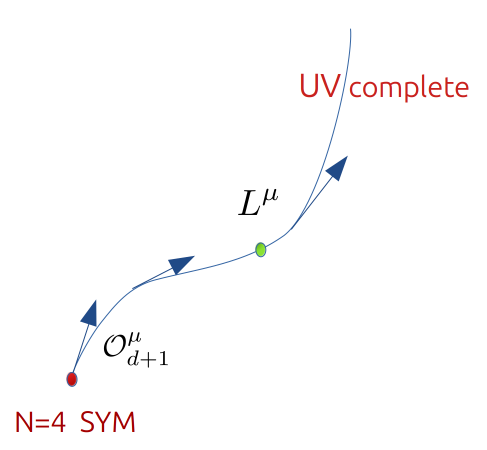}
\end{minipage}
\caption{\small{The star product presentation of dipole-deformed $\N=4$ SYM makes it clear the theory is UV-complete and non-local. Using the Seiberg-Witten map, the dipole theory can be rewritten in terms of fields with standard transformation properties under $SU(N)$; in this presentation, it corresponds to $\N=4$ SYM deformed by an infinite number of higher-spin, higher-dimension operators, with finely-tuned coefficients. }}
\end{figure} 

\vskip-3mm
 
%\begin{figure}[h]
%\centering
%\includegraphics[height=5cm]{ttbarflow}
%\caption{\textcolor{red}{\emph{Picture! Irrelevant flow, and smth about the star product.}  }
%}
%\end{figure}

\noindent The dipole star product perspective provides 
 an important computational handle over  observables in the theory.  For example, it implies that % planar diagrams are identical to their undeformed counterparts, up to an overall phase factor; in particular,
  the planar free energy is the same as in the undeformed theory, %, and the energy also turns out to be the same, then so is the thermodynamics. \emph{Can the identification of temperature fail?}
  a property that would be very hard to see in the irrelevant deformation picture. 
Another important observable are the scaling  dimensions of operators under the unbroken non-relativistic conformal group.  General arguments based on conformal perturbation theory show they will generically become momentum-dependent; more precisely, they will be functions of the Lorentz-invariant combination $ L_\mu \, p^\mu = \l \bar p$ \cite{Guica:2010sw}. While one can use conformal perturbation theory in the irrelevant deformation
 to compute the leading anomalous dimension, it is the dipole picture  of $\N=4$ SYM  and the fact that the  dipole deformation preserves integrability that allow one to compute  the one-loop anomalous dimensions of  certain classes of operators up to arbitrary order in the $\l \bar p$ expansion  %, including the large-momentum asymptotics 
 \cite{Guica:2017mtd}.

%
%\bi
%\item the left conformal dimensions are still well-defined, but receive corrections that are functions of $\l \k$, where $\k$ is the $U(1)_R$ momentum. This matches well with the space-time picture.
%\item they are non-universal (appear to simplify at strong coupling)
%\ei

\bigskip

\noindent \emph{Schr\"{o}dinger spacetimes}
\medskip 
 
\noindent The background that is holographically dual  to  null dipole-deformed $\N=4$ SYM  is a so-called Schr\"{o}dinger spacetime (the higher-dimensional analogue of warped AdS$_3$) - more precisely, $Schr_5 \times S^5$ - and
 is obtained by applying the TsT  transformation along $x^-$ and a direction  on the sphere to AdS$_5 \times S^5$ \cite{Alishahiha:2003ru}. The metric of a $Schr_{d+1}$ spacetime reads
  
\be 
ds^2 = - \frac{\tilde \l^2 (dx^+)^2}{z^4} + \frac{2 dx^+ dx^- + dx^i dx_i + dz^2}{z^2} \;, \;\;\;\;\;\;\; i \in \{1, \ldots, d-2\}
\ee
and one can easily check that its isometries realise the non-relativistic conformal group. In the above example of $Schr_5 \times S^5$ obtained via TsT, it is important to note that the parameter $\tilde \l^2$ that appears in the supergravity solution differs by a factor of the 't Hooft coupling, $g_{YM}^2 N$, from the corresponding squared field theory shift parameter.  The $Schr_3$ spacetime is the same as null warped AdS$_3$. From the $d+1$ - dimensional perspective, the solution is usually supported by a massive vector field. Note the dual dipole-deformed $\mathcal{N}=4$ theory does not live on the conformal boundary of the Schr\"{o}dinger spacetime (which is one-dimensional), but rather on the flat space spanned by $x^\pm, x^i$. 

The finite temperature version of this spacetime can be obtained by applying the TsT transformation to a planar AdS black hole \cite{Maldacena:2008wh}. The thermodynamics is identical with that of the undeformed AdS$_5$ background, in agreement with the large $N$ field theory argument.  The non-relativistic conformal dimensions of the fields can be simply computed from the small $z$ falloffs of the solutions to the wave equation in the Schr\"{o}dinger  background, and for a probe scalar field they take the form

\be
\Delta = \frac{d}{2} + \sqrt{\frac{d^2}{4} + m^2 \ell^2 + \tilde \l^2 \bar p^2}
\ee
A very similar formula holds for spinning BMN strings, for which $\Delta = (J^2 + \tilde \l^2 \bar p^2)^{1/2}$, where remember the  $\tilde \l^2 \bar p^2$ term is proportional to the 't Hooft coupling. This strong-coupling expression is thus much simpler and more universal than the weakly coupled one; the two match at leading order in  large $J$ \cite{Guica:2017mtd}. 

% More concretely, the spacetime \eqref{} is  the lowest-dimensional example of a Schr\"{o}dinger spacetime $Schr_{d+1}$, where in the general case  the $2d$ boundary metric should be replaced by a $d$-dimensional one $2 dx^+ dx^- \r 2 dx^+ dx^- + \sum_i dx_i^2$. In a string-theoretical context, such spacetimes can be obtained by applying a certain solution generating technique known as  TsT (T-duality, shift, T-duality) to AdS$_{d+1} \times S^p $. To produce a Schr\"{o}dinger spacetime, one of the TsT directions lies along AdS ($x^-$) and the other one along the compact space. 
 
While the duality between dipole-deformed $\N=4$ and $Schr_5 \times S^5$ was understood when the former theories were constructed,    Schr\"{o}dinger spacetimes have  received subsequent interest  in the context of the so-called `non-relativistic holography',  aimed at modeling features of strongly-coupled non-relativistic CFTs using  holographic setups, usually bottom-up ones \cite{Son:2008ye}. Since, if one takes the non-relativistic symmetry seriously, $p^-$ plays the role of mass, one generally performs a null compactification of the Schr\"{o}dinger spacetime to discretize the momentum, and then works in a fixed momentum sector. Upon this, the spacetime becomes `non-distinguishing' \cite{Hubeny:2005qu}, while in the dual theory one is implementing a DLCQ procedure along the null direction \cite{Maldacena:2008wh}. %\emph{Stringy corrections?} 

The result is  a  codimension-two holographic correspondence between Schr\"{o}dinger spacetimes and non-relativistic CFTs.  While this is fine as far as modeling strongly-coupled NR CFTs using holography goes, if one were interested in the inverse problem - namely, of finding the holographic dual of the Schr\"{o}dinger spacetime - then the manifest NR conformal symmetry would only provide a tiny glimpse into the properties of the system. Indeed, it is 
the special  structure associated to the dipole star product that determines many of the properties of the system, such as the fact that the thermodynamics is $\l$ - independent and the highly restricted form  of the non-locality  along $x^-$. Unfortunately, it is completely unclear how to uncover this special structure from the bulk point of view (except via its consequences), which is the problem that plagues most of the holographic correspondences discussed in these notes. It should nonetheless be clear that working in a top-down holographic set-up, with its finely-tuned interactions among bulk fields, is very important  for being able to  reproduce this special structure from the bulk perspective.

% \emph{More comments bottom-up vs top-down?}

%. While this is not incorrect, one does ignore (by fixing the null momentum) all the structure lying along the null non-local direction. Another comment is that the dual theory (precisely because of this) does not live on the conformal boundary of the dual spacetime. From a bottom-up perspective, the dimensions are also fixed, b/c they are determined from the $\N=4$ spectrum by the Drinfeld twist. 

\bigskip

\noindent \emph{Summary}

\medskip

\noindent There exist explicit string-theoretical examples, such as null dipole-deformed $\mathcal{N}=4$ SYM, of theories that are holographically dual to higher-dimensional analogues of warped AdS$_3$. These theories are given by finely-tuned \emph{irrelevant} deformations of a CFT, are non-local along a null direction, yet they are UV complete thanks to a structure (the dipole star product) that is largely hidden from the irrelevant deformation perspective. They  preserve non-relativistic conformal symmetry along the local directions. The dipole structure implies their thermodynamic properties at large $N$ are identical to those of the original CFT.  Integrability can be used to understand the momentum-dependent conformal dimensions.

 Performing TsT on other D-brane solutions yields similar results, except that now one needs to deal with the fact that the theories are non-conformal, and their gauge  theory description may be valid only in certain energy ranges. For the case of the D1-D5 system,  one  considers the dipole deformation of parallel D1 and D5 branes, and then flows to the IR, which corresponds to the decoupling limit.  Even in the absence of the deformation, this flow is extremely hard to track down. % \emph{Can one see 2d conformal symmetry emerging from the D1-D5 gauge theory description?} 
%
% and even more so in its presence.
  However, it appears that the end result will again be a very special irrelevant deformation of the D1-D5 CFT, though the field theory side is no longer tractable. To show this, the simplest is  to analyse the spacetime description of the deformed theory and interpret it holographically.

% Note decoupling commutes with TsT. Thus, the dipole background in type IIB is conceptually nice because we in principle know what the dual theory is, and it is given by a decoupling limit. Other backgrounds are less nice, both because there's no proposed theory, and no known decoupling limit (these are independent issues).     

\bigskip

\noindent \emph{AdS/CFT interpretation of (null) warped AdS$_3$ spacetimes}

\medskip

\noindent  Let us turn back  to  generic null warped AdS$_3$ spacetimes in string theory (not necessarily obtained via the TsT of a purely RR AdS$_3$ background). Even though we concentrate on three dimensions, the discussion below trivially generalises  to $Schr_{d+1}$ solutions of string theory, $\forall \, d$.

  The most salient common features of the top-down constructions of warped AdS$_3$ spacetimes are the existence of (at least) one \emph{continuous} parameter that connects them to AdS, and the fact that the nature of the warping (spacelike versus null) depends on the state considered; in particular, null warped AdS corresponds to deforming AdS in Poincar\'e coordinates. %\emph{Shorten?} 
  In the following, we  concentrate on vacuum solutions only, whose symmetries will provide useful guidance to the nature of the holographic duals.

% Above, we discussed a particular instance of Schr\"{o}dinger holography, where the decoupling argument can be made, and one ha some idea about the observables. However, there exist many more examples of Schr\"{o}dinger backgrounds, obtained via TsT + other dualities (\emph{any other methods?}).  We will call all these examples dipole CFTs, even if we are agnostic about the decoupling limit. One can see some features  of dipole-like thories from the dual spacetimes, viewed as  \emph{continuous} one-parameter (can be more) irrelevant deformations of an AdS/CFT setup. Note TMG toy model is not of this kind. Advantage: much more universal. 

%As remarked above,  $3d$ Einstein gravity coupled to massive vectors of the appropriate mass  admits warped AdS$_3$ solutions. (This argument also works for TMG and is still bottom-up)
 The vacua correspond to \emph{null warped} AdS$_3$ spacetimes, with metric and vector field

\be
ds^2 = - \a \l^2 r^2 (d x^+)^2 + 2 r dx^+ dx^- + \frac{dr^2}{r^2}  \;, \;\;\;\;\;\;\;\; A = \l  r dx^+  \label{schr}
\ee
where $\l$ is an arbitrary (vector) parameter, $\a$ is a numerical factor determined by the particular couplings in the Lagrangian, and we have traded $t,\phi$ for the more standard parametrization  $x^\pm $ of the null boundary coordinates.  The form of the solutions is determined by the $SL(2,\mathbb{R}) \times U(1)_{null}$ isometries of the problem. In principle, there can be several vectors, but we will concentrate on just one, for simplicity.  All scalar fields vanish in the vacuum.  If $x^\pm$ are non-compact, then $\l$ can be rescaled away, but the fact it is a physical parameter can be easily seen by introducing a scale (radius or temperature). %Spacelike warped AdS solutions are obtained when considering finite energy. To ensure consistency/existence of a dual, we consider warped AdS as string theoretical solutions, case in which they are obtained via decoupling limits of the form $\a' \r 0, B\r \infty$, with $\a' B$ fixed. 

  The holographic interpretation of the vacuum  follows from the fact that $\l$ is a tunable parameter:

\bi
\item for $\l=0$, the spacetime is AdS$_3$, which is dual to a CFT$_2$
\item for $\l$ infinitesimal, we can use the usual AdS$_3$/CFT$_2$ holographic dictionary to interpret the background value of $A$  as turning on a \emph{source} for an \emph{irrelevant} $(1,2)$ operator

\be
S_{\l}= S_{CFT_2} + \l \int \mathcal{O}_{(1,2)} + \ldots \label{defdip}
\ee
 Adding the $(1,2)$ operator to the action breaks the $SL(2,\mathbb{R})_R$ symmetries of the CFT, but preserves $SL(2,\mathbb{R})_L \times U(1)_R$, in agreement with the isometries of the background \eqref{schr}. 
 Thus, the deformed theory is still at a conformal fixed point with respect to the left conformal symmetries; turning on $\l$ corresponds to an exactly marginal deformation with respect to $SL(2,\mathbb{R})_L$ \cite{Guica:2010sw}.

\item at higher orders in $\l$,  the %backreaction of $A$ on the spacetime metric terminates at $\mathcal{O} (\l^2)$ \cite{Kraus:2011pf}.   The 
exactness of the  background \eqref{schr} at $\mathcal{O} (\l^2)$  \cite{Kraus:2011pf} suggests that all the higher-dimension operators that may appear in $\ldots$ will continue to preserve these symmetries. %, which %. $SL(2,\mathbb{R})_L$ invariance 
%implies 
Their dimensions are thus of the form $(1,n)$. %, whose spins are $n-1$.
 Note that most such operators would be invisible in the weakly-coupled gravity approximation, where all single-trace operators  have spin at most $2$.

\item  $\l$ finite:  \emph{if} the  background \eqref{schr} is obtained from a decoupling limit in string theory,  it appears reasonable to assume that the finite $\l$ theory resulting from the irrelevant deformation  that starts as \eqref{defdip} is UV complete  
 \ei
% \emph{Clear it won't work for e.g. spacelike warped? Any ideas? Comment! }
%%
%Thus, the theory dual to the spacetime  \eqref{schr} which,   for reasons that will soon become clear,  we call a \emph{dipole CFT},  is given by  
%%
%\be
%S_{dipole \, CFT}= S_{CFT_2} + \l \int \mathcal{O}_{(1,2)} + \ldots \label{defdip}
%\ee
%where the $\ldots$ stand for higher-dimension operators. 

\begin{figure}[h]
\centering
\includegraphics[height=5cm]{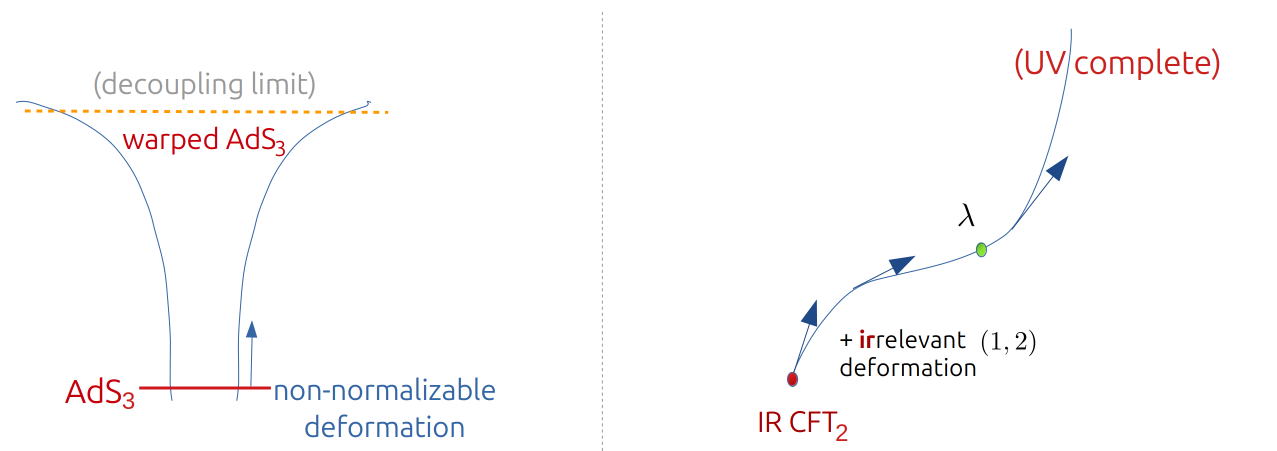}
\caption{\small{Spacetime argument for the structure of the dual dipole CFT$_2$. If the warped AdS$_3$ spacetime is obtained from a decoupling limit, then one can argue the dual theory is UV-complete.}}
\end{figure}
  
\vskip-4mm

\noindent  Let us make a few comments.  First,  simple arguments based on conformal perturbation theory  show that the requirement of preserving the $SL(2,\mathbb{R})_L$ symmetry at higher orders in $\l$ imposes restrictions on which operators may appear in the OPEs of the deforming operator(s) \cite{Kraus:2011pf}. These can be mapped between the two sides of the correspondence.   
 
Second, there is no reason, in general, to expect that the $\ldots$ in \eqref{defdip} will be zero. When the CFT is strongly coupled,  all the low-lying single-trace operators (which belong to short supersymmetry multiplets) have spin at most two, 
 %
%  One can wonder whether there are any corrections to the $(1,2)$ operator above which, as explained above, must have dimension $(1,n)$ and thus spin $n-1$. Since for a weakly coupled gravity dual the spin of low-lying single-trace operators has to be $\leq 2$,
 % 
and so   the only  additional single-trace operators % from a short multiplet
    that could appear have dimension $(1,3)$; however, one can explicitly check  that in e.g. the D1-D5 CFT, there are no such  operators with all the required quantum numbers \cite{ElShowk:2011cm}. This argument does not exclude possible  contributions from multitrace operators, or from  operators belonging to long multiplets; note, however, that in order for them to be exactly marginal with respect to  $SL(2,\mathbb{R})_L$, their standard dimension must be an integer - a non-trivial requirement for operators whose dimension is not, in general,  protected by supersymmetry\footnote{In the context of maximal supersymmetry-preserving irrelevant deformations of $\N=4$ SYM - which are  similar to the ones discussed here -  it has been proposed that the leading deforming operator  is exact, in the sense that the deformation is generated by a  supersymmetric descendant of a particular chiral primary operator \cite{Intriligator:1999ai}, with the higher corrections coming from corrections to the supersymmetry generators; however, the status of this proposal is currently unclear \cite{Caetano:2020ofu}. }.

 %   \emph{Look up Kraus-Perlmutter. How about the fact that their (anomalous) dimension can depend continuously on the coupling?  E.g., what happens perturbatively around the free orbifold pt?}% This issue is discuused in xxx. 

  Third, the requirement that a theory obtained via an irrelevant deformation of a CFT be UV-complete is extremely non-trivial as, more often than not, irrelevant deformations produce theories with a cutoff. One may imagine that, by finely-tuning the coefficients in the $\ldots$ one may achieve UV-completeness, but how to do this in practice in the irrelevant deformation picture is entirely non-obvious. In particular, it is not known which special properties of the undeformed CFT and of the deforming operator(s) are needed for the existence of the UV completion,  whether there exists a principle that allows for completing the theory in the desired way, nor whether the perturbative series in $\l$ would converge to yield a useful definition of the theory as an irrelevant deformation of a CFT. Consequently, the UV-completeness  will be assumed, or argued for, separately from the irrelevant deformation.

 %  The     UV-completeness requirement implies  that the higher-dimension operators  entering the $\ldots$ have finely-tuned coefficients, whose dependence on $\l$  such  that they conspire to produce it. The only reason to believe that such a scenario is possible is that there exist concrete string-theoretical realisations of holography for spacetimes that are higher-dimensional analogues of warped AdS where the dual theory is known, and has precisely this structure.

\begin{figure}[t]
\hspace{8mm}
\begin{minipage}{0.45\linewidth}
\flushleft
\begin{description}
\item[\emph{Dipole CFT$_2$: }] UV-complete $2d$ QFT with \\
~~~~~~~~~~~~ $SL(2,\mathbb{R})_L \times U(1)_R$ invariance \\
~~~~~~~~~~~~ that can be written as
\be
    S_{dipole  \;CFT}= S_{CFT_2} + \l \int \O_{(1,2)} + \ldots
    \label{defndip}
    \ee
 \end{description} 
\end{minipage}
\hspace{1.5cm}
\begin{minipage}{0.45\linewidth}
%\flushright
\includegraphics[height=5cm]{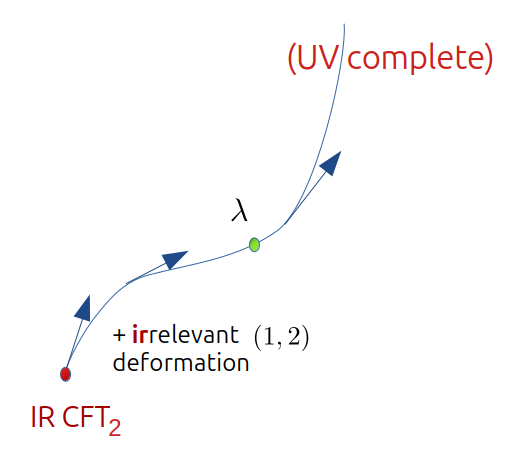}
\end{minipage}
\caption{\small{Definition of two-dimensional  dipole CFTs as UV-complete $2d$ QFTs that can be obtained via a finely-tuned irrelevant deformation of an IR CFT.  }}
\label{defndipcft}
\end{figure}

It thus appears that the most reliable definition of the theories dual to such null warped AdS backgrounds is via the decoupling limit, when such a construction is known. Existence of the decoupling limit, even when 
 the field theory side  is intractable in  practical terms, should be sufficient for defining the theory. %;  the latter can then be modeleled, using the fact that it flows to a $2d$ CFT in the IR, by the above series of irrelevant deformations around the IR fixed point. 
 %
% the three-dimensional case. Nonetheless, the spacetime analysis just discussed reveals that this theory flows to a $2d$ CFT in the IR, and allows one to easily identify the leading irrelevant deformation that goes back up the flow. \emph{Careful flow preserves integrability and number of degrees of freedom! Also for generalisations of TsT}
 %
 We will denote a theory  that is holographically dual to   a warped AdS$_3$ spacetime obtained from a decoupling limit inside string theory as a  \emph{dipole CFT}, by analogy with the previous stringy example. 
 In two dimensions, it is a UV-complete theory
 %  
%In any case, \emph{if} there exists an independent reason to believe that the theory obtained by the irrelevant deformation starting as \eqref{defdip} exists and is UV-complete, as would e.g. be suggested by a bulk decoupling argument, then we would like to interpret        \eqref{defdip}, with an appropriate, finely-tuned completion for the $\ldots$, as \emph{defining} a theory
 with $SL(2,\mathbb{R})_L \times U(1)_R$ invariance that is local on the left, but non-local on the right. The latter fact follows the presumed UV-completeness of the theory, together with the existence of a scale for the right-movers, namely the null vector $\l$.  This non-locality is in full agreement with the fact that the left conformal dimensions acquire momentum - dependent corrections.     
%
 %The RHS of \eqref{defdip} is thus correct up to possible multitrace corrections, and operators dual to  heavy string modes . 
 %   We denote such a putative theory as a \emph{dipole CFT}, by analogy with the previous stringy example. 
%
 This theory flows to a CFT$_2$ in the IR; to return to the UV one needs to turn on a series of irrelevant operators that are exactly marginal with respect to  $SL(2,\mathbb{R})_L$, starting with  one of dimension $(1,2)$, as depicted in figure \ref{defndipcft}. 
 
%The fact that the deformation is irrelevant on the right, together with the assumed UV completeness of the theory, imply that the latter is non-local on the right. 

      Let us emphasize that the existence of a decoupling limit, which is expected to ensure UV-completeness of the dual theory and is thus an essential assumption in our definition of dipole CFTs, is by no means guaranteed. As reviewed, only a particular class of warped AdS backgrounds are known to be  obtainable from a decoupling limit; we return to this point in sections \ref{gencommentsint} and \ref{newperspold}. On the other hand, for all of the top-down backgrounds, the low-energy expansion looks as above. It would herefore be very interesting if one could define the theory via the irrelevant deformation, even though currently 
      it is not clear whether \eqref{defndip} can provide  a useful definition of the theory; certainly, it can  model it for many different purposes. In these notes, we  may sometimes, via abuse of language, talk about the dipole CFT as being defined via the irrelevant deformation, but this is not known how to do in general.   
      
    %
%\emph{Talk about ill-definiteness of irrelevant defs. in general. } The latter statement, which is rather non-trivial, is that there exists a set of operators $\ldots$ so that the resullting theory is UV-complete. While the picture above is appealing, the best way to  believe it is  true is  thanks to an analogy to a five-dimensional setup where this phenomenon is known to happen. The reason behind this definition is the decoupling limit.  However, for most backgrounds discussed there the decoupling limit is not known. Thus, status not great, but can be hopeful these additional dec lims can be found. 
%

The above definition/modeling of dipole CFTs is perfectly consistent with the fact that the theory dual to warped AdS  lives along the $x^\pm$ directions, as before the deformation, and not on the conformal boundary of the deformed  spacetime.  This picture is perfectly consistent with and predicts the fact that the anomalous dimensions of fields become momentum-dependent. On the other hand, it offers no insight into any properties of the holographic dual that do not follow straightforwardly from the manifest symmetries of the problem, such as the special structure of the non-locality, the Cardy form of the entropy, or the 
appearance of a Virasoro symmetry on the right-moving, non-local side, which is nonetheless predicted by various ASG analyses. 
If anything, by confirming the non-local structure of the dual theory, all these properties become all the more  mysterious!

% that a non-local theory should have Virasoro symmetry on the non-local side.

% One of the main achievements of the work presented is to exhibit a theory with this non-local structure, where one also finds a right-moving Virasoro symmetry. 

%[ In any case, these symmetries were associated with specific boundary conditions on the wAdS asymptotic region. In absence of a good understanding/definition of the theory, one can also in principle consider other sets of boundary conditions, as long as they are consistent. \emph{Later?}]

 % \textcolor{red}{\emph{Comment somewhere dual doesn't live on conformal boundary!}}

\subsubsection{Summary}

In this subsection,  we have provided an overview of  the various toy models that have been used to try to understand the Kerr/`CFT' correspondence. These were organised into two main classes: bottom-up models, which used minimalistic theories of gravity that allowed for warped AdS solutions, and top-down ones, which descended from string theory. Somewhat surprisingly, their properties turned out to be qualitatively rather different, as summarized in the table below%\footnote{The asterixes indicate reservations about the results, see the main text for the explanation. } 

\medskip

\begin{center}
\begin{tabular}{|c|c|c|}\hline
& TMG $\&$ other bottom-up & top-down constructions \\[2pt] \hline \hline
warping (spacelike/timelike/null) & coupling & state\\[1pt] \hline
black holes & quotients global $w$AdS & not quotients\\[1pt] \hline
relation to AdS & discontinuous & continuous deformations AdS\\ [1pt] \hline
entropy & warped Cardy & Cardy \\ [1pt] \hline
ASG & Viraroso$_L$-KM$_L$* & ¿ Vir$_L$ and/or Vir$_R$ or KM$_L$ ? \\ [1pt] \hline
proposed dual & \emph{warped} CFT & \emph{dipole} CFT \\ \hline
\end{tabular}
\end{center}

\medskip

\noindent  Therefore, we seem to find two \emph{different} universality classes of holography for warped AdS spacetimes, depending on the type of action that admits them as a solution: 
 the first corresponds to simple (and possibly generic) bottom-up actions that allow for warped AdS solutions, whereas  the actions in the second class %are highly fine-tuned from the three-dimensional point of view. %   (in those cases in which the $10d$ action allows for a consistent truncation to $3d$ that includes the solution of interest). 
 have, from the three-dimensional point of view,  extremely finely-tuned couplings between the vectors and the scalars. We are thus in a case where the models that descend from UV-complete theories are qualitatively different, already at the EFT level, from those that are phenomenological and involve minimalistic couplings. 
This is rather different from AdS, where the coarse dual features one extracts from holography are the same, whether one  uses a 
bottom-up or a top-down action. The difference appears related to the fact that in AdS,  the only  background field is the metric, whereas in warped AdS, the background is supported by both the metric and matter fields.  More precisely,
the different properties of the holographic duals one reads off from the last three rows of the table above are directly related to the rather different   structure (indicated on the first three rows) of the warped AdS solutions in the two types of theories, which is directly influenced by the  matter couplings in the Lagrangian.

%[ Later: The ASG proposals  for the top-down constructions have been Virsoro$_R$ and/or Virasoro$_L$, as well as left-moving Virasoro-KM. For Virasoro$_R$ ASG, the boundary conditions look extremely different from the warped CFT ones, but similar to the orginal Kerr/CFT boundary conditions. \emph{Put up!} ]

The holographic duals of the theories in each universality class are denoted as warped CFTs and, respectively, dipole CFTs. While the literature not always very clear  about their respective definitions, we have attempted to settle for  ones that are likely to capture the consensus:
\bi
\item \emph{warped} CFTs are  two-dimensional  QFTs with global $SL(2,\mathbb{R})_L \times U(1)_R$   spacetime symmetries that are enhanced as in 
\eqref{wcftenh}. The only known examples are free, and it is not known whether interacting such theories can be built.  Warped CFTs that appear in the holographic context (put forth by the ASG calculation) are only \emph{putative}, since their existence depends on that of a  consistent UV completion of the  gravitational theory in question, as well as of the correctness of the boundary conditions. %, to be able to define them in this way.
 The former seems unlikely in the toy models where they appear, which often exhibit pathologies such as ghosts. % \emph{Look up Ale-Stephane}
\item \emph{dipole} CFTs are  UV-complete two-dimensional QFTs that preserve $SL(2,\mathbb{R})_L \times U(1)_R$  symmetry and flow to a CFT in the IR.  Currently, the best  way to ensure their  UV-completeness is to define them 
via a decoupling limit of string theory. % that, in principle, guarantees the existence and  of the theory.
 The latter is unfortunately intractable on the field theory side; however,  these theories can be also understood as  finely-tuned irrelevant deformations of a CFT$_2$ that are exactly marginal with respect to $SL(2,\mathbb{R})_L \times U(1)_R$, as corroborated by
  the dual spacetime picture, which corresponds to a continuous non-normalisable deformation of AdS$_3$. Dipole CFTs are thus non-local on the right. %Their extended symmetries are not understood, but certain ASG computations suggest they could be Virasoro$_L \times$ Virasoro$_R$
The only current method for investigating their properties, such as extended symmetries, thermodynamics and correlation functions, is via holographic bottom-up methods, which unfortunately suffer from various ambiguities. 
\ei

\noindent Of the  two different universality classes above,  the one that models most closely the behaviour of (near)-extremal black holes is the dipole one - indeed, some of the top-down backgrounds correspond  to nothing but the near-horizon regions  of extremal charged rotating   black holes, embedded in string theory. These similarities persist at the level of the ASG analyses. 
Since the properties of the bottom-up models are qualitatively very different from those of the top-down ones, which: i) are better at modeling extremal black holes and ii) are likely the only ones to have  a consistent UV completion, it is reasonable to propose that the  models whose predictions we should trust as good toy models for the Kerr/`CFT' correspondence are the 
 top-down ones. 

%This is good, as these likely are the only models that admit a consistent UV completion.  
%
%[Why stringy truncations: in all other models, warping is determined by the theory, not the state; TMG has ghosts and massive vector models weren't studied because for real gauge kinetic terms one only gets spacelike squashed black holes which have CTCs (should we be bothered, though?); the ASG one naturally gets in TMG (and NMG) is the wCFT one. By contrast, the stringy truncations give state-dependent warping, and a more likely Virasoro algebra (the CTCs are still there, see Wei \& Strom).  Note that when a 3d trucation is possible, the top-down actions look very peculiar (finely-tuned) from a $3d$ point of view, as compared with the bottom-up ones. However, there is no reason to a priori expect universality. ]
%
Thus, our best current proposal for the `UV theory' that captures the microscopics of near-extremal rotating black holes is a dipole CFT. Note we do not currently have a definition of dipole CFTs that is independent of their string theory embedding.  While this dipole CFT is expected to explain the entropy of the black hole, note that the geometry need not contain a fully decoupled warped AdS spacetime factor, but rather just its IR limit above near-extremal states. How to understand this embedding is a separate problem that we do not address in these notes.

Having settled on dipole CFTs, the next task is to understand their properties. Unfortunately, the decoupling limit - when it exists - is untractable and the irrelevant deformation  picture
%
%The dipole CFT/irrelevant deformation/ by deforming a $2d$ CFT by an \emph{irrelevant} $(1,2)$ operator, and flowing \emph{up the RG}  picture of warped AdS 
%
%The holographic analysis of warped AdS toy models strongly suggests the holographic dual is a dipole CFT, i.e. a theory obtained by deforning a CFT$_2$ by an irrelevant $(1,2)$ operator. This
 can only explain the non-locality, including why the $SL(2,\mathbb{R})$ dimensions become momentum-dependent. Bottom-up holography provides the most powerful methods for inferring the properties of the dual theory (figure \ref{topdnwadstriad}), which are extremely interesting: the possible presence of a  Virasoro symmetry on the non-local side, the Cardy form of the entropy,  the CFT-like right-moving piece of the correlation functions, but entirely mysterious from the point of view of 
what is known about dipole CFTs (i.e.,   the irrelevant deformation).

However, bottom-up methods are known to be plagued with ambiguities, as can be seen from the many different proposals for the ASG of warped AdS in the top-down models\footnote{ In particular, if the correct ASG turns out to be \ref{wcftenh}, then  dipole CFTs could be warped CFT, via the above definition. }, so they aren't reliable methods to study the properties of the dual theory.  Instead, what is needed is an \emph{explicit field-theoretical toy model}, where all the fascinating properties that are hinted at by the holographic analyses can be tested and properly understood. From the discussion above, it is clear that the minimal properties that one would like to require of this toy model are: a finely-tuned irrelevant deformation of a $2d$ CFT that preserves $SL(2,\mathbb{R})_L \times U(1)_R$ and leads to a UV-complete theory. In these notes, we will present and study precisely such a theory.

%The upshot/global picture: the holographic analysis of the massive vector toy models of the Kerr/CFT correspondence that emerge from string thoery suggest the following global picture:

%Picture: dipole CFT = decoupling limits, UV-complete irrelevant deformations, Cardy?, Virasoro? Scattering: special non-local. 

\medskip

In conclusion, the holographic duals to extremal black holes may perhaps be understood as irrelevant deformations of a $2d$ CFT that start with a $(1,2)$ deformation. It is interesting to ask whether  non-extremal black holes can be understood within a similar framework. As we review below, in this case  the $(1,2)$ deformations will be supplemented by $(2,2)$ ones, also at leading order. Before we  explain this, we start with some general comments and brief review of attempts to understand the microscopic description of non-extremal black holes.

%Some of these conclusions may also be extended to non-extremal black holes, which to leading order may be given as irrelevant deformations by $(2,2)$ operators. where $(2,2)$ also appear.

%In the following, we turn to exactly solvable irrelevant deformations of 2d CFTs by $(1,2)$ and $(2,2)$ operators, and hope to address these puzzles. 

\subsection{General non-extremal black holes \label{nonextrsec}}

%Start perhaps with some serious issues that need to be taken into account when discussing thermodyn: instability, ensemble, dominance. Then, note near-extremality may be ok. Then, numerology. 

The entropy one would ideally like to explain is that of the Schwarzschild black hole; 
%[One well-known issue is that of thermodynamic instability, as black holes sufficiently off extremality have $C<0$.  Thus, thermodynamically unstable and cannot couple them to infinite (?) heat bath.]  
 this seems, however, to be the hardest case, partly due to its negative specific heat, and partly to the fact that it is furthest from the `ground state' of the system. On the other hand, near-extremal black holes, whose temperature is very small, are  at least partially understood from a microscopic perspective, as we have reviewed. It thus makes sense to try to understand non-extremal black holes as a departure from the extremal limit. This leads us to disccuss generalisations of the Schwarzschild black hole, such as non-extremal Kerr or Reissner-Nordstrom, or  Kerr-Newman.
 %We will stick to asymptotically flat  black holes.
% 
   We start with  some general thermodynamic considerations that are relevant to  understanding of the entropy of generic non-extremal black holes. 
 
% ,  which should be part of the 
 %   dicussion of reproducing microscopically the entropy of non-extremal black holes.% \emph{True?}

%\bigskip

\subsubsection{General thermodynamic considerations}

%\medskip

\noindent  To talk about thermodynamics, one should first define an ensemble.  Consider a Kerr-Newman black hole held in equilibrium at some temperature $T$ in a heat bath\cite{Davies_royal}. %, with which it can reversibly exchange energy. 
Neglecting the effects of superradiance, the angular momentum $J$ and charge $Q$ of the black hole are constant.  
The entropy as a function of the black hole mass $M$ and $ Q,J$ is  ($G=1$)

\be
S = \pi (r_+^2 + a^2) \;, \;\;\;\;\; r_\pm = M\pm \sqrt{M^2-a^2 -Q^2} \;, \;\;\;\;\; a \equiv \frac{J}{M}
\ee
The Hawking temperature is given by 

\be
T = \left(\frac{\p S}{\p M} \right)^{-1}_{J,Q} = \frac{r_+-r_-}{4\pi (r_+^2+a^2)}
\ee
If we plot $T(M)$ at fixed $J, Q$, we obtain 

\begin{figure}[h]
\centering
\includegraphics[height=4cm]{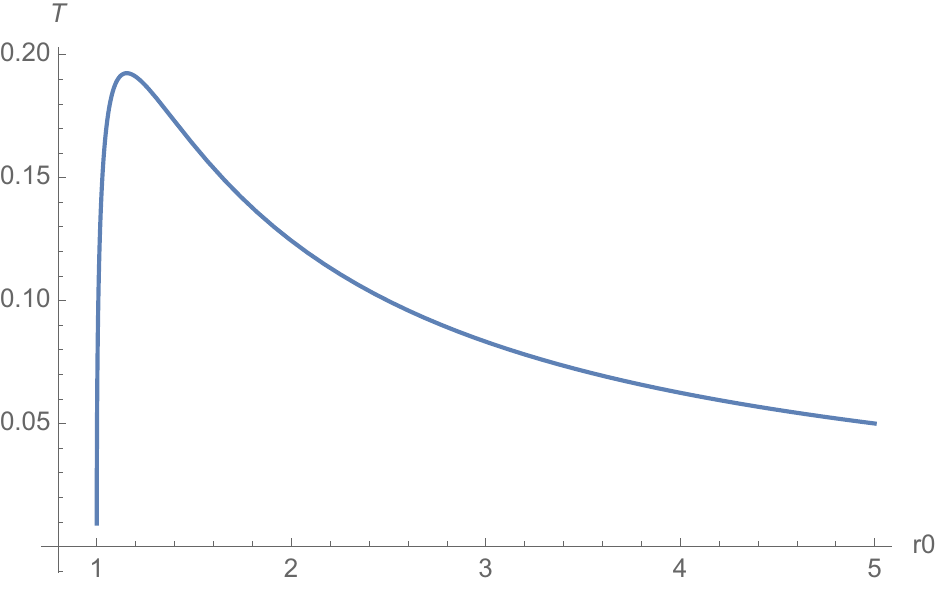}
\caption{\small{The Hawking temperature of a charged (rotating) black hole with fixed $Q$ (respectively, $J$) as a function of the horizon radius ($\propto$ energy) measured in  units of the  radius at extremality. }}
\label{tempvsengnonextr}
\end{figure}

\noindent namely there is a maximum temperature $T_{max}$ for each fixed $J,Q$. Letting $J^2=\a M^4$ and $Q^2=\b M^2$ with $\a+\b \leq 1$, the position of the maximum is at $\a^2+6\a+4 \b =3$, which corresponds to $J \sim 0.68 M^2$  for Kerr and $Q \sim 0.86 M$ for Reissner-Nordstr\"{o}m. The value of the black hole mass corresponding to the maximum temperature will be denoted as $E_{max}$.

For each $T < T_{max}$ at fixed $J,Q$, there are two possible black hole solutions: one small and one large. The 
%
%To study stability, one may compute  the relevant specific heat is  
%
%\be
%C_{J,Q} = T \left(\frac{\p S}{\p T}\right)_{J,Q} = \frac{8 M S^3 T}{J^2 + Q^4/4 - 8 T^2 S^3}
%\ee
%\emph{Explicit formula above not needed, just explain $S(T,J,Q)$.}
%
%Fixing $J$ and/or $Q$, the
 specific heat is positive on the small black hole branch ($M < E_{max}$) and negative  for the large black holes ($M > E_{max}$), with a discontinuity in-between.  
%has a discontinuity where it goes from positive infinite to negative infinite. 
%\emph{Check discontinuity at $T_{max}$}. %, and then it's negative on the large black hole branch. 
 The small black hole is thus stable, minimizing the free energy, %thus dominating the thermal ensemble,
   but also  has smaller entropy.   The large black hole is unstable, and thus it does not make sense in the canonical ensemble\footnote{Indeed,  a fluctuation up in energy lowers its temperature, which means it will absorb even more from the thermal gas around it, which must be in contact with a heat reservoir in order to be able to keep $T$ constant. Conversely, a downwards fluctuation in energy will trigger more emmisions from the black hole, until it disappears. }.  Thus, thermal equilibrium with a bath at $T < T_{max}$ %(which requires minimising $F$ for the equilibrium to be stable), and neglecting superradiance effects, implies that the black hole needs to be the small one, which has positive specific heat.]
  appears to select the small black hole as the configuration minimising the free energy. 

% Gibbs free energy and its first derivatives continuous, so second order phase transition.  \emph{Interpretation? Does it also appear for RN-AdS? Footnote Davies comment on calling $S$ extensive. Note also $C_{\Omega, \Phi}$ are conthermaltinuous $\r$ what does this mean?}

% \emph{Correct? What to do with gas of gravitons in infinite space? If regulated, instabilities? }  

This conclusion must, however, be modulated in view of the fact that the 
%
%even for the small black holes, the thermal ensemble  does not truly make sense, as the
 black holes would  be in equilibrium with a thermal gas of gravitons in asymptotically flat space, which is known to suffer  from   Jeans and black hole nucleation instabilities \cite{Gross:1982cv}, see also discussion in \cite{Atick:1988si}. %, in addition to carrying infinite energy. \emph{Which one, exactly and, if unstable, does it make sense to talk about the infinite energy?}
  One could still imagine ignoring the effects of the gas by working in the strict  $G \r 0$ limit, using the fact that the black hole entropy scales with $G^{-1}$, while that of the gas does not.  %\emph{What happens to the gas instabilities as $G \r 0$? The Jeans one happes for $k< k_J \r 0$, while the decay rate goes to zero as $e^{-1/(\ell_P T)^2}$. }

To study the system including the gas, it is  better  to use the microcanonical ensemble, which is anyhow a must in the case of the large black hole.  As a warm-up, let us briefly review the well-known case of the Schwarzschild black hole \cite{Hawking:1976de}. Consider such a black hole in equilibrium with thermal radiation inside a box of volume $V$, with a fixed total energy $E$. The radiation hosts a fraction $x$ of the total energy,  $M_r = x E$, while the black hole has mass $M=(1-x) E$.  The total entropy is 

\be
S = 4\pi G M^2 + \frac{4}{3} (a V M_r^3)^{1/4}
\ee
where $a$ is the usual radiation constant. 
Extremizing $S$ with respect to $x$, one finds $x (1-x)^4 = \frac{ a V}{ (8\pi G)^4 E^5}$. This equation only has a solution if  $V < V_{max} = (32\pi G)^4 (E/5)^5/ a$. For this solution to correspond to a maximum of the entropy, one needs $x<1/5$, i.e. the radiation only carries a small fraction of the total energy.   %So, the Schwarzschild bh mass must be at least $4/5 E$/ the radiation must be at most $1/5 E$, which implies there is a maximum volume that the box can have so that the black hole be in (stable) thermal equilibrium with the radiation. 
As explained in \cite{Hawking:1976de}, the reason the equilibrium is stable is that, when the black hole absorbs  energy from the gas, its temperature drops, but the gas's temperature drops even more - given  its finite volume - for the corresponding energy loss. If $V > V_{max}$, then the stable configuration is all gas.

%[It is in thermal equil with gas, which has infinite volume, so this has to be regulated. AS log as the gas volume  is smaller than the light crossing time \emph{Correct?} it's eaten up by the black hole. If we try to take $V\r \infty$ with constant density, then either (infinite) black hole or all gas. (This also holds in ALD if one considers gas).]

A similar analysis can be performed upon turning on angular momentum and/or  charge, see e.g. \cite{Schumacher:1991ts} for a detailed study of a Kerr black hole immersed in a rotating heat bath. Starting with the setup of the previous paragraph, namely with a Schwarzschild black hole in equilibrium with thermal radiation inside a box, 
%
%The entire system is placed inside a finite-size box, fixing the total energy and angular momentum; for an appropriate range of these external parameters, the black hole and radiation  equilibrate at a common  temperature and rotational frequency.  To understand what happens,  it is simplest 
%
 % While the radiation-only state is a local maximum of the entropy, the nucleation of small black holes being enthropically suppressed, what the global maximum is  depends on the parameters.
%
one
 raises the total angular momentum, holding  the total energy and the box size fixed. As the angular momentum  is increased, most of it initially  goes into the black hole;  however, at some point it becomes entropically  favourable for the gas to carry it, while  the central black holes spins slower.  If one  increases the total angular momentum further, the  equilibrium state contains rotating radiation only. This situation is rather similar to that of rotating configurations in AdS \cite{Kim:2023sig}. The central black hole is never on the small/fast-spinning branch, though the assumption of  \cite{Schumacher:1991ts}  that its horizon radius be much smaller than the radius of the box may be effectively putting an \emph{a priori} upper limit on its angular velocity, as the speed of the gas at the rim of the box should not exceed that  of light.

In conclusion, it seems hard to make sense of  thermodynamic equilibrium for  non-extremal black holes in asymptotically flat space, both due to the negative specific heat of black holes on the `large' branch and  to the instabilities of thermal gas in flat space. One can isolate the black hole by placing it inside a box, but note that any holographic interpretation of the system would also need to take into account the box, which is not provided `for free', as in AdS.  If one can neglect the gas, then  black holes that are sufficiently close to extremality - namely, on the `small' branch, so that their specific heat is positive - do appear to make sense in the thermal ensemble. If not,  one can simply treat  black holes  as long-lived metastable %\emph{Correct?}
  states,  interpretation that should be reflected in the dual description\footnote{Note our discussion has completely ignored potential superradiant instabilities. }.

\medskip

So far for specifying the ensemble. The next obvious question is: ensemble of states in \emph{which} theory?   As is well known, there is no proper decoupling for non-extremal black holes, and  
 flat space holography is still a faraway dream in what concerns black hole microscopics. One may hope, nonetheless, that in certain limits there may be a theory that captures the  `intrinsic' black hole degrees of freedom, through a perhaps non-standard isolating procedure such as, for example, modifying the asymptotics of the spacetime (i.e., the  boundary conditions) in a controlled fashion.

%It is desireable to have some sort of   [decoupling limit] from flat space/that isolates the near-horizon region, which would correspond to fitting/describing the degrees of freedom of the black hole  within some simpler theory.  \emph{(For Kerr, would that correspond to considering the same state in a different UV completion, whatever that means?)} I guess some way to isolate the degrees of freedom pertaining to the black hole is of interest, but not clear if it makes sense.  Below we review some work that suggests a universal interpretation of general BH dof. \emph{Why universality?}]
 
 Given this state of affairs, it should not come as a surprise that very little is known concretely about the microscopic description of non-extremal asymptotically flat black holes\footnote{ There is a suggestion \cite{Horowitz:1996ay}, based on a particular way to express  the entropy of the non-extremal D1-D5-P black hole,  that it may be interpreted in terms of brane and anti-brane contributions. 
%\textcolor{red}{\emph{Does this observation have any reasonable interpretation, say in the LST decoupling limit?}}
 }. In the following we review an attempt, inspired by the Kerr/`CFT' correspondence, to give them an interpretation in terms of a $2d$ CFT, as well as two further lines of work attempting to put this proposal on a firmer footing. 

%\bigskip

%[Concerning the microscopic description of non-extremal asymptotically flat black holes, relatively little concrete is known.
%
%the entropy can be written as a product of brane and antibrane contributions. 
% [Of course, non-extremal black holes in AdS, if they are large, corresponds to the thermal entropy of the CFT, but this was not the   question (as they map to near-extremal in flat space).] 

\subsubsection{Hidden conformal symmetries}

%\medskip

\noindent As discussed earlier in this section, the Kerr/`CFT' correspondence, and in particular its associated conformal symmetries,  relies on the existence of a decoupling limit (albeit imperfect) of the near-horizon region of the extreme Kerr black hole from the asymptotically flat geometry.  Such a decoupling of the near-horizon physics is not available for non-extremal black holes.

 In their study of the non-extremal Kerr black hole, \cite{Castro:2010fd}  propose instead to look at the symmetries of the wave equation - for, say, a scalar field - in the low-frequency ($\om M \ll 1$), `near' ($r\ll 1/\om$) limit,\footnote{% In the low frequency limit, the eigenvalues of the deformed sphere Laplacian, which depend on $\omega a$, become $\ell$.
  Note that, for a near-extreme black hole, this is different from the low-energy limit used in Kerr/`CFT' where one takes instead $\omega \r m \Omega $, which is finite and momentum-dependent. This difference translates into a difference in the $SL(2,\mathbb{R})$ conformal weights of perturbations, which are momentum-dependent in Kerr/`CFT', but simply constants in \cite{Castro:2010fd}.  } where $M$ is the mass of the black hole. They argue that a certain truncation of this  equation motivated by the above limit possesses $SL(2,\mathbb{R})_L\times SL(2,\mathbb{R})_R $ conformal symmetry,  termed `hidden conformal symmetry'.  The authors also identify a set of vector fields associated to these symmetries, which are entirely analogous to the local isometries of a BTZ background,  as well as a set of  `temperatures' that are related, just as in BTZ,  to the  lack of global definiteness of the symmetries. %\emph{Check if this said in BTZ section!} 
  Plugging  these `temperatures' (which are not related to the Hawking temperature of the Kerr black hole in any immediate way) into Cardy's formula \eqref{cardyfintro}, using the central charge input $c=12 J$  derived from the Kerr/CFT correspondence (which nonetheless, only applies to the near - extremal case), \cite{Castro:2010fd} are able to  precisely reproduce the Bekenstein-Hawking entropy of the non-extreme Kerr black hole.

The vector fields one infers from the truncated wave equation are not isometries of the Kerr spacetime, hence the `hidden' qualifier. Nor are they  asymptotic symmetries %(i.e., true symmetries of the theory) 
in any obvious way. In particular, some of them take  one out of the low-frequency approximation that was necessary for their existence in the first place; an additional issue is that exact\footnote{Here, by `exact' we mean locally exact, namely the symmetries are allowed to be broken by global effects. In \cite{Charalambous:2021kcz}, a different set of $SL(2,\mathbb{R})$ symmetries -  this time globally exact - of a somewhat differently truncated wave equation  were used to explain the vanishing of the  static Love numbers for the Kerr black hole.  It is currently unclear whether these `Love symmetries' are also related to the black hole's thermodynamics; it  would be interesting to further explore this point. } conformal symmetry of the wave equation is present  only
if certain $\O(\om^2)$  terms are removed from it, but not others.

Given this state of affairs, one may ask:  even if the conformal symmetry is broken at finite energies,  can one nonetheless identify a CFT description at very low energies, and in particular 
 read off the parameters (energies and temperatures) in this `CFT' from the low-energy limit of the Kerr data?
Concretely, if one tries to match the wave equation/scattering amplitudes in the Kerr geometry  to leading order in $\om$ (with the angular quantum number $m$ fixed) to the wave equation in BTZ/a CFT scattering amplitude \eqref{bh2pf} for some CFT temperatures  $T_{L,R} $ and momenta $  \om_{L,R} $,  one can  readily identify
\be
\frac{\omega_L}{2\pi T_L} = 2 M \omega \;, \;\;\;\;\; \frac{\omega_R}{2\pi T_R} = \frac{4 M^2 \omega}{r_+-r_-} - \frac{2 a}{r_+-r_-} m \;, \;\;\;\;\; \;\;  a = \frac{J}{M}%T_H = \frac{1}{8\pi} \frac{r_+-r_-}{M r_+} \;, \;\;\; \Omega = \frac{a}{2 M r_+}
\ee 
This match is obtained without  invoking any of the $\O(\om^2)$ terms. While it is clear that the above equations can only determine 
$T_{L,R}$ and $ \om_{L,R}$ up to an overall multiplicative factor, if one additionally requires that the would-be `CFT' spatial momentum, $\om_L -\om_R$, be quantized (and thus equal to $m$), then one immediately reproduces %\footnote{Note energies in 6.9 of CMS not integrable, unless $J$ fixed. } 
the results of \cite{Castro:2010fd}
\be
\om_L =\frac{2 M^2}{a} \om \;, \;\;\;\;\;  \om_R = - m +  \frac{2 M^2}{a}  \om \;, \;\;\;\;\; T_{L,R} = \frac{r_+\pm r_-}{4\pi a} \label{hiddentempmom}
\ee 
% The coefficient is assumed to be the same, because one still wants the difference in left and right energies to be $m$. Also not that in NHEK, $\om \approx m$, because the need to go to a corotating frame to penetrate the near horizon region - a rather different limit.  In any case, one can easily see that $T_R$ is fixed by the matching of the CFT and Kerr amplitudes at $\om=0$. The match to first order in $\om$ should be then able to fix $\a$ and $T_L$ via 
%
As already explained, plugging these `temperatures' into Cardy's formula using $c=12 J$ precisely reproduces the Bekenstein-Hawking entropy of the non-extremal Kerr black hole. The use of Cardy's formula is, however, not justified, since 
%
%Note that the `hidden CFT' temperatures have nothing, a priori, to do with Kerr thermodynamics. Also, 
the CFT structure is only present at leading order in $\om$, being explicitly broken by $\O(\om^2)$ terms. %Nonetheless, a piece of black magic shows 
%
%Then $T_R$ is fixed to equal $T_H/\Omega$, while $T_L/T_R = 2 M/(r_+-r_-)$.  As emphasized, the  additional constraint CMS impose that actually fixes $T_{L,R}$ is that $\omega_L-\omega_R =m$. \emph{} The black magic then notices 
%that the Kerr entropy precisely coincides with the Cardy formula with these values of $T_{L,R}$ and $c=12J$.  
The would-be `CFT' energy, obtained by integrating \eqref{hiddentempmom} at fixed\footnote{Fixing $J$ is necessary for integrability. } $J$, is related to the spacetime one as   $E_L \sim M^4/J$,  explaining why 
the specific heat is positive with respect to this new `energy'.  % To discuss the interpretation of this numerological observation, we review some subsequent work. 
%
%\emph{Note energies in 6.9 of CMS not integrable, unless $J$ fixed.}

%[, the Cardy formula is shown to match the Bekenstein-Hawking entropy of the black hole. There are many extensions of this work to other back holes. \emph{Status? Form of correlation functions?}  However, due to the absence of a proper decoupling limit, no ASG analysis can be properly performed to study the symmetries of the system, and thus no natural set of diffeos (hidden symmetries) can be singled out. \emph{True?}

%More precisely, in CMS they  and truncate the wave equation. The effective mass is angular-omentum dependent. The lack of global definiteness of the vectors is identified with temperatures. 
%The Cardy formula with $c_L=c_R=12J$ mysteriously matches the BH entropy.
 %] %(\emph{ In fact, it is quite clear the vectors mean nothing (except pretending to give a conformal structure), one just has the numerological observation that the temperatures into $c$ give Cardy's formula and that low-energy scattering is CFT-like.}).]
%
%
%\bi
%\item note however that $S(T_H, J)$ is a rather involved function
%\ei 

In conclusion, the hidden conformal symmetries do, indeed, appear to be associated to CFT$_2$ symmetries that are spontaneously broken by certain temperatures, but only in the deep IR limit.  Their significance beyond this limit is unclear. %, as that of the associated spacetime  vectors which, given the lack of any decoupling limit to the near region, are not singled out as 
 The  relation between the properties of the radial wave equation in a black hole background,  which determines scattering data, and the thermodynamic ones were further explored in \cite{Castro:2013kea}, focussing on the monodromies of the solutions  around the singular points, which provide a connection between the two. For the simple BTZ case, these monodromies were given an attempted CFT interpretation in \cite{CarneirodaCunha:2016zmi}.  %Hidden conformal symmetries in a Kerr background  (globally well-defined this time) can also be used to explain the vanishing of Love numbers for the Kerr black hole, but seem completely unrelated to the previous ones. 
%\emph{Anything about Love?} 
 Thus, it seems  fair to classify the observation of \cite{Castro:2010fd} as a very interesting clue into the microscopic interpretation of non-extremal black holes that, however, 
still waits to be put on a firmer footing.

%Before we go in this direction, let us mention a few ramifications related to hidden conformal symmetries: Dubovsky for Love numbers and Alejandra for further study of wave equation via monodromy. A possible CFT interpretation was given in \cite{}. Note their choice of terms to keep in the near region wave equation differ from Dubovsky, who found instead a globally well defined set of $SL(2,\mathbb{R})$ vectors that could be used to relate solutions to the wave equation. It is not clear how many such vectors (non-isometries) can be found. 
% UV/IR mixing by the generators - what does it mean? Review also near-horizon symmetries, at least to trash them.  Also monodromies and relation to Virasoro blocks. 

%\bigskip 

\subsubsection{Black hole in a `conformal box'}

%\medskip

\noindent One such attempt \cite{Cvetic:2011hp,Cvetic:2011dn} was to modify the black hole geometry in a  way that rendered the   wave equation   exactly conformal, while preserving the `intrinsic' thermodynamic properties of the black hole.   

Concretely, \cite{Cvetic:2011hp,Cvetic:2011dn} considered general non-extremal black hole solutions of  certain four and five-dimensional $\N=2$ supergravity theories, which can be written as a timelike fibration over a fixed base.  By modifying just the warp factor of the fibration in an appropriate way, the wave equation can be made to locally exhibit exact  $SL(2,\mathbb{R})_L\times SL(2,\mathbb{R})_R$ symmetry. %\emph{Careful: squared.} 
This modification does not affect the thermodynamic potentials and the entropy   of the black hole, while it does significantly alter   the asymptotics of the solution.  
%
%performing a particular modification to the black hole geometry, namely in a particular warp factor. It is important these were families of STU black holes in $4d$, or rotating D1-D5-p in $5d$, though uncharged black holes are among them. 

The physical significance attached to this procedure was a `subtraction' of the ambient Minkowski space, which does not affect the `intrinsic'  thermodynamic properties of the black hole, while it does change the asymptotic notion of energy and, with it, the specific heat, which becomes positive.  
It  was also interpreted as isolating the black hole\footnote{Thus, one expects to be dealing with the `same system', but within   different surroundings. For this, one would generally expect the charges to be the same, but it is not clear the subtraction procedure always leaves them unchanged. % Same question about moduli. The fact that some system can be mapped to a CFT of the same temperature, while not keeping the parameters constant is by itself not very impressive. \emph{ How about subtraction for Schwarzschild? Doesn't it actually change the charge? }
 } from its surrounding space by `placing it in a conformal box'. % \emph{True that the specific heat for black hole in a standard box becomes positive for small enough box? Nonetheless, that of the black hole itself is still negative. } The intuition is that, while the "intrinsic" thermodynamic potentials did not change, the (asymptotic) notion of energy does, leading to a flip in the sign of the specific heat.
%
%[aimed to isolate the black hole from its surrounding space by placing it in a ``conformal box''.% This was achieved by modifying the warp factors of the solution (of the STU model), which
One of the advantages of this construction   is that it provides a transparent geometric explanation for the conformal symmetry of the wave equation,  as the subtracted geometry simply corresponds to a dimensional reduction of AdS$_3 \times S^p$, $p=2,3$.  This, together with the fact that the thermodynamic potentials $T_H, \Omega$ are held fixed in the subtraction procedure, implies  the black hole entropy can  be written in Cardy form. % \emph{Write better!}

%  An advantage of this method is that it relates the modification of the wave equation to a modification of the metric in a transparent way.] 
  
 % In $5d$, the modification of the warp factor consists of simply dropping certain ones in the harmonic functions (whose argument is nonetheless affected by the rotation, but only that). The quantities $\Pi_{c,s}$ are held fixed (or, rescaled), which ensures the Hawking temperature is unchanged; also, the subtracted warp factor only depends on them. By contrast,  in stringy realisations it is  the number of branes that is supposed to be held fixed. In $4d$, the dropping of ones is less clear; however, these a clearly Harrisson transformations. The reason the entropy is Cardy is simply that
  
%  , so obviously the entropy can be written as Cardy if it could so be written in the extremal limit. It is also noted in \cite{withMirjam} \emph{Also earlier?} that the central charge entering the subtracted Cardy formula is different from the hidden (or Kerr)/CFT central charge $c=12 J$; in fact, in the scaling limit the subtracted central charge $\# r_0^3$ $\r 0$, while $\Pi_{s,c}$ diverge. Thus, subtracted $\neq$ hidden. \emph{Bla-bla different frames? Would these frames be so different for Kerr vs KN in $4d$? Does $Q$ fixed allow for integrable variations of the momenta?}

Subsequent work \cite{Cvetic:2012tr} showed  that the subtraction procedure can be understood as performing a scaling limit of the original black hole, which resembles  the dilute gas approximation, albeit symmetrized with respect to the charges. This implies that the CFT dual to the dimensionally uplifted subtracted geometry  can be understood as a low-energy limit of the theory dual to the uplift of the original black hole. In turn, this leads  to the idea  that the  holographic dual to the original geometry can be understood as a CFT deformed by certain irrelevant operators.

% \bigskip
 \subsubsection{Interpolating geometries}
 %\medskip
 
\noindent Further light on this problem was shed by \cite{Baggio:2012db}, who  constructed explicit interpolating solutions between a certain class of non-extremal black holes and their subtracted geometries. The  $4d$  black holes they discussed were static and carried one electric and three magnetic charges, the latter of which were fixed in the interpolating flow (lest it would be hard to think of the subtracted geometry as representing the `same' system).  The interpolating solutions were fully specified by four harmonic functions, and all that  changed in the flow were the constant terms therein:
  the subtracted geometry mapped to the case where the constants in the magnetic functions were dropped, while they were finite  for the original black hole, more precisely $(\sinh \d_i)^{-1}$, where the $\d_i$ are defined similarly to \eqref{bhcharges}. 
 
 Considering the interpolating flow infinitesimally around the subtracted solutions and uplifting to five dimensions allowed \cite{Baggio:2012db} to identify its holographic interpretation, which simply corresponds to adding sources for certain $(2,2)$ irrelevant operators in the CFT$_2$ dual to the uplifted subtracted geometry, i.e. AdS$_3 \times S^2$. 
%
%The subtraction procedure was then interpreted as generated by solution generating transformations in this model (less nice than interpolating, as it screws up identifications and possibly charges), which generates irrelevant deformations of a CFT.  In the static case first studied in \cite{deboer} these irrelevant deformations were by $(2,2)$ operators. 
%
As the authors point out, the irrelevant deformations associated to the endpoint of this flow are only small when the original black hole is  near-extremal, which corresponds to $\d_i \gg 1$. In this case, there is a large range of energies (below  the scale set by the irrelevant deformation) where the black hole's behaviour is captured by a CFT. Nonetheless, for a general black hole, for which $\d_i \sim \O(1)$, the scale associated to the irrelevant deformations is the same as that set by the temperature, so there is no energy regime  where the CFT description can usefully predict anything. The fact that the entropy of the original black hole is the same as that in the  CFT dual to the subtracted geometry is attributed to a particular choice of `scheme' for dealing with the irrelevant deformations  from an IR perspective. 

 In the more general case  of rotating black holes, $(1,2)$  irrelevant deformations are also present.
%
%It is also comented in \cite{deboer} that the source for an irrelevant operator is, in general, scheme dependent, and that the irrelevant deformations associated with the subtraction procedure are, likely, in a scheme where the entropy is unmodified
 %\emph{I don't understand this, if entropy counts the number of states, then how can one decide if it's fixed? Maybe this only makes sense before any new degrees of freedom are introduced. Indeed, having the freedom to keep entropy fixed looks like a big assumption - doesn't it effectively encode all possible deviation of $S(T)$ from the CFT result into a temperature dependence of the irrelevant coupling (or, more possibly, radius, since the irrelevant coupling is absent in the deep IR)? Note these expectations do not appear borne out in LST (the scale of the deformation is a dimensionful constant, which predicts CFT entropy for scales much smaller, but the entropy is modified away from low energies). }
% 
 % We further discuss this issue in section \ref{}. [As we will see, in a case where the parameter in the interpolating geometry does have an independent UV interpretation, one can meaningfully measure changes in $S(T)$ away from the CFT Cardy's formula. ]
 %
 As it turns out, one can also generate the interpolation via solution-generating techniques \cite{Virmani:2012kw,Cvetic:2013cja,Sahay:2013xda}. %, though one should be generally  careful about their action on the charges.
  By analogy with the case of dipole-deformed $\N=4$ SYM, this suggests there might exist a `hidden' explanation for why the entropy stays the same.

While the general picture put forth in \cite{Baggio:2012db} may well be correct, there are still many questions that remain to be answered: i) even if the number of states stays the same due to the adiabaticity of the deformation, these are states in a completely different theory (given by an irrelevant deformation of the  CFT dual to the subtracted geometry) whose microscopic origin in the new theory needs to be explained; ii) it is not  completely clear whether  the number of states remains the same, as the argument regarding the freedom to choose a scheme  applies from an IR point of view only; iii) the subtraction procedure makes the most sense for black holes embedded in supergravity, with a particular matter content and a geometric extra dimension, and one needs to ascertain the same conclusions will apply universally to all black holes in any theory of gravity, as suggested by \cite{Cvetic:2011hp}. 

%at least in the string theory context, it is possible to show that  some of these irrelevant deformations lead to theories decoupled from gravity, but the final one that leads to flat  space (and also seems connected to the negative specific heat)  no decoupling limit is known, bringing up the question whether one should expect a decoupled holographic description at all; iv) finally, 
%
%[while for non-extremal ones one should work in the microcanonical ensemble. It in fact makes sense to put the system in a box, and look at the configuration of black hole + gas that maximizes the entropy. As in the AdS box case, one finds that the more entropically favourable configuration has the gas carry the $J$, and thus the central black hole is never spinning very fast. \emph{Does it get to have $C>0$? Is there a bomb made? How was this avoided in AdS?} Nonetheless, no one prevents us from considering a long-lived metastable black hole very close to extremality, as there is sufficient experimental evidence such black holes exist in the sky, despite the difficulty with defining an ensemble in which such states would dominate.  \emph{Does a metastable ensemble make sense? Read Biroli-Kurchan. } Thus, possibly canonical near extremality, and microcanonical otherwise.  ]
%
%The important thing about the interpolation seems to be the ability to provide a small parameter in the expansion around the near-horizon, even for black holes far from extremality. 
%

\medskip

%\noindent 
To conclude, the absence of decoupling of the near-horizon region  of non-extremal black holes from the asymptotically flat one, as well as the negative specific heat of most such black holes, pose important challenges  to even  developing a framework in which their microscopic entropy  should be understood.  In this context, it
is encouraging  that
 an analysis of the near-horizon physics - through the lens of the wave equation and the associated scattering amplitudes - 
%
% expansion of the geometry (even in absence of decoupling) 
 %
  suggests that, similarly to the  Kerr/`CFT' correspondence, the holographic dual can be modelled as a $2d$ CFT in the deep IR, which is then deformed by 
   a series of irrelevant  operators of dimensions  $(1,2)$ and/or $(2,2)$, which mysteriously appear to leave the entropy invariant. Thus,  irrelevant deformations  of $2d$ CFTs with special,  hidden  properties make an appearance also in the problem of understanding  non-extremal black hole microscopics, reinforcing our motivation to  study such theories in detail.

\section{Solvable irrelevant deformations of two-dimensional QFTs \label{solvirrelsec}}

As discussed in the introduction, the solvable irrelevant deformations put forth by Smirnov and Zamolodchikov \cite{Smirnov:2016lqw} have given renewed hope of understanding the kind of non-local  field theories dual to warped AdS$_3$ spacetimes that are relevant to finding the microscopic description of (near) - extremal black holes.

Generally speaking,  the study of irrelevant deformations of a QFT is challenging, for at least two reasons: first, it is increasingly difficult to specify the deforming operator at higher and higher orders in perturbation theory, due to the proliferation of counterterms; second, when adding an irrelevant operator to a QFT, the deformed theory does not usually make sense in the UV on its own - rather, it has a cutoff.  What makes the so-called Smirnov-Zamolodchikov  deformations special is that the deformed QFT appears to be \emph{well-defined} up to arbitrary scales, is \emph{solvable} in a certain sense, and the effect of the deformation on various physical observables can be computed without much effort for \emph{finite} values of the deformation parameter%\footnote{By contrast, the theories we discussed in the previous section can sometimes be argued to be UV-complete,  and can also be thought of as being given by a finite irrelevant deformation of a CFT, but  the precise deformation is not known all along the flow (at arbitrary coupling) and is generally intractable. }
.  In the following, we review the definition and basic properties of these deformations, and give a few examples.

 \etocsetnexttocdepth{5}
    \etocsettocstyle{\subsubsection*{Contents of this section: }}{}
    \cftsubsubsecindent 34pt
    \localtableofcontents

\subsection{Smirnov-Zamolodchikov deformations\label{szgensec}}

The Smirnov-Zamolodchikov deformations \cite{Smirnov:2016lqw} are  deformations of a two-dimensional QFT by an operator constructed as a bilinear in two conserved currents, $J^A, J^B$. In the most interesting cases, this bilinear operator is \emph{irrelevant} and leads to novel UV asymptotics for the deformed QFT. 

\subsubsection{Definition}

 To define the Smirnov-Zamolodchikov deformations (henceforth  `SZ'), one proceeds in two steps. The first step is to  construct the Smirnov-Zamolodchikov operator, denoted $\O_{J^A \wedge J^B}$. This construction relies upon the observation that 
 the derivative  with  respect to either $x^\rho$ or $y^\rho$ 
   of the  antisymmetric combination  $\e^{\a\b} J_\a^A (x) J_\b^B (y)$ of the  components of two conserved currents $J^A, J^B$ is a total $(x+y)$ - derivative\footnote{This can be easily shown by using the two-dimensional identity $\e_{\a\b} \p_\g + \e_{\b\g} \p_\a + \e_{\g \a} \p_\b =0$.} %\emph{Check!}
\be
\p_{x^\rho} [ \e^{\a\b} J^A_\a (x) J^B_\b (y) ]  =  \e_{\a\rho} (\p_{x^\b} + \p_{y^\b}) \, J^\a{}_A(x) J^\b{}_B (y) \label{szident}
\ee
upon explicitly using the current conservation equation.  Taking $x\r y$ and  assuming that the above product of currents has a standard OPE expansion%\textcolor{red}{\emph{Issues defining OPE?}}
, one immediately concludes that either the coefficient functions in the OPE  of $\e^{\a\b} J^A_\a (x) J^B_\b(y)$ - which, by translation invariance, are only functions of $x-y$ - are constant, or they multiply an operator that is a total derivative. Given this, one can \emph{define} the Smirnov-Zamolodchikov operator as the term in the above OPE that appears with a constant coefficient

\be
\lim_{x\r y} \e^{\a\b} J^A_\a (x) J^B_\b (y) \sim \O_{J^A \wedge J^B} (y) + \mbox{\emph{total derivatives}} \label{defszop}
\ee
 up to total derivative terms. Thanks to the property \eqref{szident}, it  follows that the SZ operator  has the nice property  of factorization in translationally-invariant states, such as energy eigenstates\footnote{\label{factE}Indeed, taking the expectation value of \eqref{szident} in an energy eigenstate, one deduces that $\langle E |  \e^{\a\b} J^A_\a (x) J^B_\b (y) |E\rangle $ is independent of both $x$ and $y$. By taking the insertions far apart and using cluster decomposition, this correlator is expected to factorize. By taking the coincidence limit, the correlator reduces to the expectation value of the SZ operator in the corresponding eigenstate. }.

The second step in the definition of Smirnov-Zamolodchikov deformations
 consists of  incrementally adding  the SZ operator to the action, as %\textcolor{red}{\emph{Factors 2 measure!}}

\begin{figure}[h]
\hspace{8mm}
\begin{minipage}{0.36\linewidth}
%\flushright
\vspace{8mm}
\be
\frac{\p S}{\p \mu} =  \int d^2 z \, \O_{J^A\wedge  J^B}^{[\mu]} 
 \label{ttbdefintro}
\ee
\vskip 4mm
\noindent where $\O_{J_A \wedge J_B}^{[\mu]}$ is the SZ operator in the theory already deformed by an amount $\mu$. Note that the conserved currents that enter its definition do, in general, depend non-trivially on $\mu$. 
%
%{\color{blue} This definition specifies the operator that is turned on at each point along the flow. \emph{Correct? Renorm. operator?} }
 \vspace{5mm}
\end{minipage}
\hspace{5mm}
\begin{minipage}{0.45\linewidth}
\flushright
\includegraphics[height=5cm]{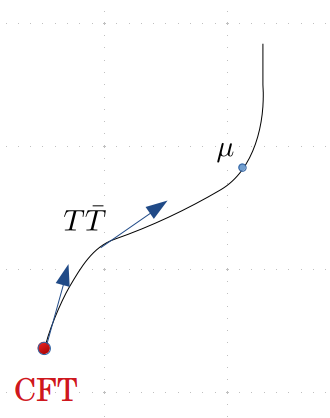}
\end{minipage}
\caption{\small{Definition of the $T\bar T$ deformation. The undeformed theory can be either a $2d$ CFT or  QFT. }}
\end{figure}
\noindent This definition makes sense in a perturbative expansion around the undeformed QFT, where the deformed theory can still be considered %`quasi-local', by which we mean l
local on scales much larger than the one set by $\mu$. Below this scale,
%
% below the scale associated with the non-locality of the deformed theory, as
  it is not clear how to associate a conserved current to the given symmetry.  It is important to note that SZ deformations \emph{do not} correspond to an RG flow \cite{zoharlectures} - rather, they simply correspond to deforming the QFT action  as in \eqref{ttbdefintro}. In particular,  one does not integrate in or out of degrees of freedom.  
  
  % \textcolor{red}{Discuss the fact this is not an RG flow upwards (cite Zohar). }

%One can of course wonder whether this construction will break down once the non-locality scale is reached, since 

\subsubsection{Examples}

\noindent There are several well-studied examples of Smirnov-Zamolodchikov  deformations 
\begin{enumerate}
\item[i)] \emph{The $T\bar T$ deformation} \cite{Smirnov:2016lqw,Cavaglia:2016oda}, obtained by taking $J^A_\a = T_\a{}^A$, where $T_{\a A}$ is the stress tensor, i.e. the generator of translations in the direction $\hat x^A$. Defining $J^B_\b = T_{\b}{}^B $  and 
then anti-symmetrizing with respect to the indices $A,B$, one obtains the Lorentz-invariant `$T\bar T$' operator\footnote{It can also be written as $\O_{T\bar T} = \frac{1}{2} ( T_{\a\b} T^{\a\b} - (T^\a_\a)^2)$. Expanding around a CFT, the trace of the stress tensor vanishes, while the first term is proportional to $T_{zz}  T_{\bar z \bar z}$, hence the name. Note that the trace term does not vanish in general. The operator is of course defined via the limiting procedure \eqref{defszop}, but we will often omit writing it, for simplicity.}

\be
\O_{T\bar T} \equiv   \frac{1}{2} \e^{\a\b} \e^{A B} T_{\a A} T_{\b B} 
\ee

   This is the  prototypical Smirnov-Zamolodchikov 
 deformation. It is a \emph{universal} deformation of two-dimensional  QFTs,   only requiring the existence of a stress tensor.  Since the deforming operator has dimension  $(2,2)$  in two dimensions, the corresponding coupling, $\mu$, has dimensions of $(length)^2$. 
 \item[ii)]\emph{The $J\bar T$ deformation} \cite{Guica:2017lia}. One can also construct a SZ deformation from the components of a $U(1)$ current and those of the right-moving stress tensor. This is a Lorentz-breaking operator of dimension $(1,2)$, whose coupling is a vector $\l^a = \l \d^a_{\bar z}$, where $\l$ has dimensions of length. The fact that the deforming operator is exactly marginal on the left implies that, if 
the deformation is applied to a CFT, the resulting theory will be local and conformal on the left. This not only makes $J\bar T$ - deformed CFTs much easier to study than $T\bar T$ deformed ones, but also shows they have 
the same symmetry  and locality structure as dipole CFTs.
  %
  %Also, please note the similarities with the structure of the irrelevant deformation 
%  appearing in the context of extremal black holes. 
  
 \item[iii)]  One may also consider the  $JT^a$ deformation \cite{LeFloch:2019rut,Frolov:2019xzi,Anous:2019osb}, built from the components of a $U(1)$ current and the generator, $T_{\a a}$, of translations in some fixed direction, $\hat x^a$. This deformation is very similar to $J\bar T$, except that it does not generally preserve any conformal symmetry.

 \item[iv)]It is also possible to consider arbitrary combinations of $T \bar T$, $J T^a$ and their generalizations. If one considers several deformations simultaneously, then one must take into account the fact that new current-current deformations can be generated from the OPE of the original deformations; this problem has been discussed in \cite{LeFloch:2019rut,Frolov:2019xzi}.

  \item[v)]\emph{Generalised $(T\bar T)_s$} \cite{Smirnov:2016lqw}. In integrable QFTs, one can consider SZ deformations constructed from the higher-spin conserved currents associated to integrability. 
\end{enumerate}

\noindent All these deformations are \emph{universal}, in the sense that they require a rather minimal or generic structure %\footnote{As opposed to a theory-specific structure.} 
in order to be defined: the first four only require the existence of  a stress tensor and of certain conserved $U(1)$ currents; the latter requires the existence of an integrable structure. Since most results in the literature are for the $T\bar T$ and $J \bar T$ deformations, including from the point of view of holography, we will henceforth mostly concentrate on  these deformations. % only.  Also sufficient for our ultimate holographic goal. 

\subsubsection{Some universal properties}

Smirnov-Zamolodchikov deformations have a number of remarkable properties. Many can be  seen by  studying the deformed theory on a cylinder (of circumference\footnote{While $R$ has so far denoted the radius of the cylinder where the theory lives, from now on we will use the standard $T\bar T$ notation, where $R$ is the circumference.} $R$) under the assumption that it can then be treated as a simple quantum-mechanical system with discrete  energy levels. %\textcolor{red}{\emph{Careful!}}
 One focuses on the flow of energy, momentum and  charge eigenstates, which we will generically denote as  $|n\rangle$

%One of the major reasons these theories are interesting is their solvability. Before delving into the details of each  deformation in part, we would like to discuss an observable that can be universally computed, which is the finite-size spectrum of the deformed QFT.% A main observable is the finite-size  spectrum, whose derivation we will now briefly  review. 

%For this, one places the two-dimensional QFT on an euclidean cylinder of circumference $R$, with $z = \s - i \tau$, $\s \sim \s +R$. Consider an eigenstate $|n\rangle$ of the energy and momentum and charges

\be
H |n\rangle = E_n |n\rangle \;, \;\;\;\;\;\; 
P |n\rangle = P_n |n\rangle \;, \;\;\;\;\;\;  Q|n\rangle= Q_n |n\rangle \label{eigenstdef}
\ee 
The properties  common to all SZ deformations are: 
\bi
\item %factorisability in energy-momentum eigenstates/However, it was shown that
 the expectation value of the SZ operator in energy eigenstates is independent of position and  factorizes (see footnote \ref{factE})  %\emph{Include proof from lecture notes.}
\item as $\mu$ is \emph{infinitesimally} changed, the definition \eqref{ttbdefintro} of the deformed QFT %by the addition of the instantaneous SZ operator
 implies that the change in the energy of the $n^{th}$ energy eigenstate is given by  
\be
\p_\mu E_n^{[\mu]} (R) =   \langle n | \int_0^R\!\!  d\s \,\O_{J^A \wedge J^B} ^{[\mu]} |n\rangle 
\label{genfloweng}
\ee
as follows from usual quantum-mechanical first order perturbation theory.  %\textcolor{red}{\emph{Discuss Hamiltonian vs Lagrangian}}
Thus, the energy eigenvalues obey the flow equation%\footnote{See \cite{zoharlectures} for a more rigorous derivation.} 

\be
\p_\mu E_n = \e^{\a\b} \langle n | J^A_\a |n \rangle \langle n| J^B_\b | n \rangle  \label{engflow}
\ee 
where only the zero modes of the various current components on the spatial cylinder contribute and we omitted writing the superscript $[\mu]$.  The expectation value of the time component of the currents  simply equals the corresponding conserved charges \eqref{eigenstdef} in the given eigenstate, while the expectation value of the integrated spatial components can be computed by coupling to a constant background field. %: e.g. for the $\s\s$ component of the  stress tensor, this is just the radius.
 This method has lead to a complete solution for the finite-size spectrum of $T\bar T$, $ J \bar T$ - deformed QFTs and combinations thereof, %; for  $(T\bar T)_s$ this was not possible because. \emph{Why?}
which only depends on the spectrum of the undeformed QFT in presence of background fields.

\item  SZ deformations preserve  integrability, if  present in the undeformed QFT \cite{Smirnov:2016lqw}
 %//[In integrable theories, the deformed spectrum determines the deformation of the S-matrix via the TBA equation. This leads precisely to the expression \eqref{}. \emph{Do we need explicit spectrum solution - see generalized $T\bar T$?} For the case of the ``universal'' deformations, the deformation of any S-matrix element can be argued for independently of integrability.  \emph{Spell out argument!}]

\item all SZ deformations have been argued to lead to a simple deformation of the  S-matrix, by just a phase factor that depends on the external momenta, and sometimes also the charges. Concentrating on  $2 \r 2$ scattering and 
Lorentz-invariant SZ deformations,  this phase corresponds precisely to a so-called CDD factor \cite{Castillejo:1955ed}, of the general form
\be
\mathcal{S}^{[\mu]} (\theta) = e^{i \sum_k \a_k^{[\mu]} \sinh k\theta} \mathcal{S}^{[0]} (\theta)
\ee
where $\mathcal{S}^{[0]} (\theta)$   is the S-matrix of the undeformed QFT and $\theta$ is the relative rapidity.   Specific deformations select specific values of $s$; for example, the $T\bar T$ deformation corresponds to the case where all $\a_s$ but $\a_1$ vanish. 

\bi
\item[$\star$] in  integrable theories, the
 $2 \r 2$ S-matrix is sufficient to determine also higher-point scattering processes.   In addition, one may use the TBA equations \cite{Zamolodchikov:1991vx} to relate changes in this S-matrix  to the change in the spectrum \eqref{engflow} for general $(T\bar T)_s$ deformations \cite{Hernandez-Chifflet:2019sua}.  %\emph{Does one know the S-matrix deformation in generalised $T\bar T$? If so, can one get/solve the spectrum via TBA? Relation b/w the S-matrix asymptotics and Hagedorn spectrum? Maybe later.} Also in this case, the $2 \r 2$ S-matrix 

\item[$\star$] in the case of $T\bar T$ and $JT^a$ deformations and combinations thereof, one can  use the path integral definition of the deformations \cite{Dubovsky:2017cnj,Anous:2019osb} to directly show the dressing of an arbitrary S-matrix element by the corresponding phase factor, with no need to  resort to integrability.
\ei
\item since the above  S-matrix is  well-defined up to \emph{arbitrarily high energies}, this is a strong indication that the deformed theory is \emph{UV complete}. 
\item  the UV behaviour of the deformed theories (assuming UV completeness) is not captured by a local UV fixed point\footnote{This new type of UV behaviour has been termed \emph{asymptotic fragility} \cite{Dubovsky:2012wk,Dubovsky:2013ira}. It refers to a QFT that looks na\"{i}vely non-renormalisable from the IR point of view, but is in fact UV-complete, albeit in a non-standard, non-local fashion.   %\textcolor{red}{\emph{Correct?}}
 }. %Non-locality may have to do with UV asymptotics of the S-matrix.
Instead, the theory is  intrinsically non-local in the UV, where the non-locality   scale is set by the dimensionful irrelevant coupling ($\sqrt{\mu}$, $\l$, etc.)

%\item at least/all  the deformations constructed from the stress tensor and spin one currents  are solvable \emph{Updates generalised $T\bar T$?}

\item  one can also study the flow of the energy eigenstates  under the SZ deformations, which is adiabatic.  If the seed theory is a  CFT,  then the operator flowing the states can be used to also flow the Virasoro generators, which remain conserved. At least in the case of $T\bar T$ and $J\bar T$, these generators can be easily related to appropriate integrals of the quasilocal stress tensor components. This suggests the deformed theories still  possess infinite Virasoro(-Kac-Moody) symmetries.  %\emph{True also for gen $T\bar T$?}
\ei 
 
\noindent  The above is an inexhaustive list of properties of SZ deformations that are common to most examples of such theories that have been studied.  Of course, there are also many deformation-specific properties, such as the preservation - or not - of Lorentz invariance, and deformation-specific applications (e.g., $T\bar T$-deformed QFTs are  closely related to the worldsheet theory of strings, $J\bar T$-deformed CFTs are relevant  to  understanding holography for extremal black holes, etc.), which makes them interesting to study in their own right.

In the following, we will concentrate exclusively on  $T\bar T$, $J\bar T$, $JT^a$ deformations and combinations thereof, which 
seem more directly relevant to the question advertised in the introduction, namely that of understanding non-AdS holography. The general statements  we will  make will thus only be guaranteed to apply to these deformations; nonetheless, some of them are likely true  more generally,  including for 
 the generalised $(T\bar T)_s$ deformations and their Lorentz-breaking counterparts. These  have many interesting properties in their own right; however, in the interest of keeping these notes self-contained, the readers will be asked to check themselves the relevant literature.

\subsubsection{Hamiltonian formulation}

The  definition \eqref{ttbdefintro} of the SZ deformations was as a change in  the action of the system. There is also much insight to be gained from a Hamiltonian formulation of the deformation.  We thus consider the QFT on the cylinder and define the deformation as an infinitesimal change in the Hamiltonian 

\be
\p_\mu H^{[\mu]} = - \int d\s\,  \O_{J^A\wedge J^B}^{[\mu]}
\ee
Classically, one can show \cite{Kruthoff:2020hsi} this definition is equivalent  to \eqref{ttbdefintro}. Quantum-mechanically, the relationship between the two definitions is less clear; however, the same authors showed, using  current  conservation %\textcolor{red}{\emph{Check!}}
 $\p_t J_t + i [H, J_t] = \p_\s J_\s$, that this operator can always be written as (dropping ${}^{[\mu]}$)

\be
\p_\mu H = \frac{1}{R} \e^{\a\b} \int d\s J^A_\a(\s) \int d\s J^B_\b (\s) +  \left[H,\int d\s d\tilde \s G(\s-\tilde \s) J^A_t (\s) J^B_t(\tilde \s)\right]
\ee
%
%\bea
%\p_\mu H & = & - \int d \s d\tilde \s \d(\s-\tilde \s) ( J^A_t (\s) J^B_\s (\tilde \s) -J^A_\s (\tilde \s) J^B_t ( \s)) = \int d \s \d \tilde\s ( \p_{\tilde \s} G(\s-\tilde \s) - \frac{1}{R} ) \ldots \nonumber\\
%&=& \frac{1}{R} \e^{\a\b} \int d\s J^A_\a(\s) \int d\s J^B_\b (\s) +  \p_t \int d\s d\tilde s  G(\s-\tilde s) J_t^A (\s) J_t^B(\tilde \s)
%\eea
where $G(\s-\tilde \s)$ is a Green's function on the cylinder, given in \eqref{greensfcyl},  and appropriate symmetrization is required. This expression automatically reproduces the flow \eqref{genfloweng} of the energy eigenvalues, as the second term does not contribute to the expectation value. One can also use this to derive a flow equation for the energy eigenstates themselves, see section \ref{clslimflow} for the $T\bar T$ and $J\bar T $ explicit expressions. Note that any derivative ambiguities in the SZ operator either drop out from $\p_\mu H$ (if they can be written as total space derivatives) or can be absorbed into the last term, and thus do not affect the spectrum. % \textcolor{red}{\emph{Check with Ruben.}}

%Note that when promoting these to operators, there are ambiguities, which are either total space derivatives (don't matter in $H$) or time ones (which will only correct the commutator). Note that in the quantum theory, the spectrum comes out as expected, because the second term does not contribute to the diagonal matrix elements. Moreover, the term in the commutator yields the flow of the energy eigenstates.  The remaining term contributes to the flow of the eigenstates. 

In the classical limit, one may directly write a flow equation for the Hamiltonian density 

\be
\p_\mu \H = - \e^{\a\b} J^A_\a J^B_\b
\ee
where the currents are built from  the canonical coordinates and their conjugate momenta.  This equation can be solved explicitly for general $T\bar T/ J\bar T/JT^a$ - deformed CFTs, and shows the Hamiltonian density is entirely determined by the undeformed Hamiltonian and currents. This fact will turn out to be extremely useful for performing exact computations in $\mu$ in the classical theory.

\subsubsection{Holographic interpretation}

 The  holographic interpretation of SZ deformations has, again, certain universal and certain deformation-specific features. Note that, in holographic parlance, the SZ deformations are double-trace, as they involve a product of two single-trace current operators.

All SZ deformations are \emph{tractable}  irrelevant deformations of $AdS_3/CFT_2$,
since two-dimensional conserved currents are dual to non-dynamical fields in the bulk, and the double-trace deformations correspond to turning on mixed boundary conditions for these fields. The resulting holographic dictionary can be studied at precision level; in fact, it is the first instance of holography for AdS$_3$ with mixed asymptotic boundary conditions for the metric, where one has \emph{independent} control over both the bulk and the boundary side of the duality

\bi
\item  $T\bar T$ - deformed CFTs   are holographically dual to AdS$_3$ gravity with mixed boundary conditions for the  metric.  For negative $T\bar T$ coupling, the $T\bar T$-deformed observables coincide, in typical states, with those measured by bulk observers at a fixed  radial position in AdS$_3$ (BTZ) 
 \item  $J\bar T$ - deformed CFTs  are holographically dual to AdS$_3$ gravity with mixed boundary conditions for the  graviton and a Chern-Simons gauge field. In the metric sector, they are very similar to the `CSS' boundary conditions \cite{Compere:2013bya} and shed a new light on the interpretation of the latter (see \ref{asysymmdtr})  
\ei
While it is clearly valuable to have a precise holographic dictionary, which   can be independently tested, for AdS$_3$ with specific mixed boundary conditions for the fields, the gravitational backgrounds that are dual to the double-trace SZ deformations are 
 somewhat uninteresting from a holographic perspective.  Rather, one would like  to consider deformations of AdS$_3$ triggered by non-normalisable perturbations; these are dual to sources for \emph{single-trace} irrelevant operators.  

\subsubsection{Single-trace Smirnov-Zamolodchikov deformations}

This motivates  considering  a simple generalization of SZ deformations,
 which go under the name of single-trace  $T\bar T$ or $J\bar T$  - deformed QFTs, to quote the most common types. To define them, one  considers  a symmetric product orbifold of two-dimensional QFTs and deforms it by an operator that is a sum over the corresponding SZ operators, one for each copy of the QFT %\textcolor{red}{\emph{Notation!}}
\be
\p_\mu S^{[\mu]} = \sum_i \int d^2 \s \, \O_{J^{A}_i \wedge  J^{B}_i}^{[\mu]}
\ee
\vskip-2mm
\noindent Note this is different from the double-trace deformation, which in this notation would be $\sum_{i,j} J^A_i \wedge J^B_j  $.

 The theory that results from this deformation is a symmetric product orbifold of $T\bar T$ or $J\bar T$ - deformed CFTs.  The observables in these theories are  slightly different from those of their double-trace counterparts, though equally universal and determined by them. Single-trace SZ deformations thus
 provide another interesting example of UV-complete theories with unusual UV behaviour. % from the double-trace ones. % belonging to the ``space of integrable QFTs'', as well as of sovable non-conformal symmetric orbifolds. 

As we already mentioned, these  single-trace deformations are especially interesting from the point of view of holography. As expected, the exact symmetric product orbifold theories defined above are  holographically dual to highly stringy bulk spacetimes, where the notion of geometry may not be very clearly defined. %\textcolor{red}{\emph{Look up str $T\bar T$ wsheet papers!}} 
 However, there also exist `links' - whose precise form is spelled out in  section \ref{holostr} -    between certain types of genuinely non-AdS spacetimes where the supergravity approximation is valid and single-trace SZ deformations. More specifically:

\bi
\item the \emph{single-trace}  $T\bar T$ deformation has been linked \cite{Giveon:2017nie} to holography in an asymptotically linear dilaton background of string theory,  thus
 showing interesting connections to little string theory

\item the \emph{single-trace}  $J\bar T$ deformation has been linked \cite{Apolo:2018qpq,Chakraborty:2018vja} to holography for certain warped AdS$_3$ spacetimes, and  are thus relevant to understanding the Kerr/`CFT' correspondence
\ei
%The precise form of these `links' is discussed in section \ref{holostr}. \noindent {\color{blue}These examples also suggest that there exist generalizations of $T\bar T$ and $J\bar T$ - deformed QFTs that are less universal, yet share the same type of UV behaviour.} We will discuss this in the concluding section.   

%\medskip

\noindent This concludes our brief, yet general overview of SZ deformations. In the rest of this section,  we will  discuss the properties of  each deformation of interest ($T\bar T$, $J\bar T$, $JT^a$) in part, from a field theoretical point of view. The holographic interpretation of these deformations is discussed in section \ref{holointdtr}.

\subsection{The $T\bar T$ deformation \label{ttbsubsec}}

\begin{figure}[h]
\centering
\includegraphics[height=5.6cm]{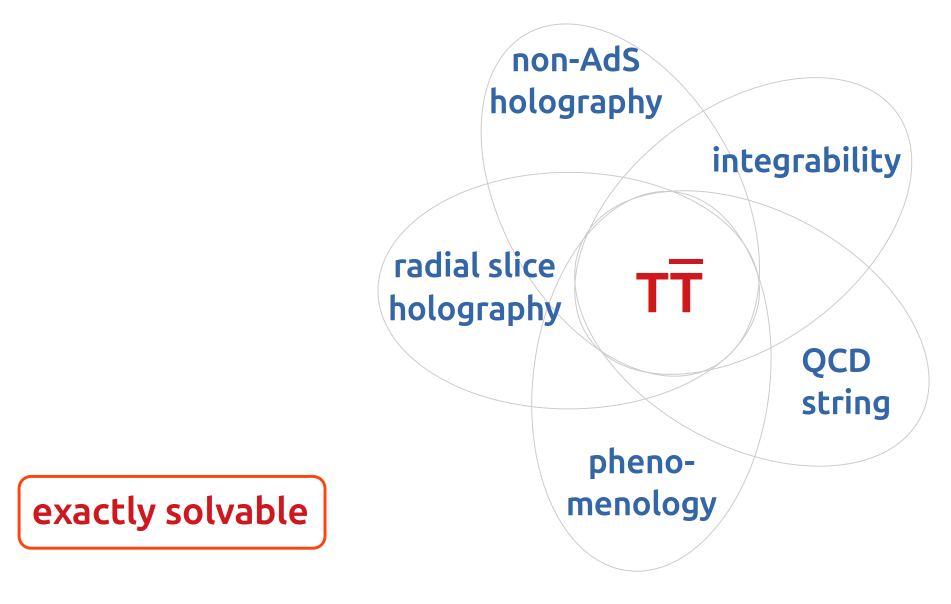}
\caption{\small{The many applications of the $T\bar T$ deformation to different subfields. Much of the interest in this deformation is related to its exact solvability. }}
\label{ttbapps}
\vskip-1mm
\end{figure}

The $T\bar T$ deformation is the best-studied and, of the Smirnov-Zamolodchikov deformations introduced so far, aguably the one with the richest physics. Besides its solvability and other universal features it shares with the other SZ deformations, the applications of the $T\bar T$ deformation to widely different subfields of theoretical physics, listed in figure \ref{ttbapps},   make it particularly attractive to study.

\noindent Some of the field-theoretical  applications  of the $T\bar T$ deformation are directly related to the special form of the S-matrix in these theories, which is given by a universal momentum-dependent factor times the original S-matrix \eqref{genttbsmat}.  For example, $T\bar T$ is the answer to the following integrable S-matrix bootstrap question: given an integrable QFT whose $2\r 2$ S-matrix, $\mathcal{S}_0$, that satisfies analyticity, crossing and unitarity, what is the theory whose S-matrix is given by multiplying $\mathcal{S}_0$ by the CDD factor in \eqref{genttbsmat}? Of course, since the asymptotic behaviour of the S-matrix is modified, one does not expect the resulting theory to be local, and indeed, it is not. The applications to phenomenology \cite{Dubovsky:2013ira} consist of analysing the hierarchy problem in theories defined via an S-matrix of the form \eqref{genttbsmat}.

As it turns out, there exists a close connection between  strings and  $T\bar T$: indeed, the $T\bar T$ deformation of  $D-2$ free bosons yields 
the static-gauge Nambu-Goto action for a string propagating in $D$ flat target space dimensions; in the critical dimension, this relation is  true at full quantum level. While this setup is integrable, it  provides the departure point for connections between $T\bar T$ and the QCD string - where integrability is broken - and a  computationally effective approach for comparison with lattice QCD data. 

In accord with this connection, the S-matrix of $T\bar T$ - deformed QFTs  exhibits certain features expected of theories of $2d$ quantum gravity (represented by the theory on the string worldsheet) such as  a time delay in scattering  proportional to the energy, as well as the presence of a minimum length \cite{Dubovsky:2012wk}. However,  the lack of a sum over topologies in defining these theories keeps them closer in spirit to QFTs, % \textcolor{red}{\emph{Agree?}}
 as do the various existing proposals for defining off-shell observables (i.e., correlation functions) in $T\bar T$ - deformed CFTs. 
%
%Among these, perhaps the most interesting is the connection between $T\bar T$ and strings and  theories of $2d$ gravity, as reflected in the following features \emph{Maybe rethink}
%\bi
%\item the \textcolor{blue}{However, it is not yet clear whether  $T\bar T$-deformed QFTs entirely lack off-shell observables, as expected in a theory of gravity }  \textcolor{red}{Ask Victor}
%\item  by extension, they are related to the  effective string theory approach to studying the QCD string and provide a  computationally effective approach for comparison with lattice QCD data \emph{Check!}
%\ei
%
In the context of non-AdS holography, $T\bar T$ is again connected to strings, but without gravity, more precisely in modeling the behaviour of compactified  `little string' theory\footnote{One may naturally wonder, given the multitude of applications of $T\bar T$ to very different theories containing strings, whether the latter  may display a universal UV  behaviour, that $T\bar T$ is able to capture. }. It is probably fair to say that $T\bar T$ - deformed theories are in-between theories of $2d$ gravity and non-local QFTs and, depending on the context, they can be interpreted   more as  one, or more as  the other. See \cite{dubcosmo} for interesting recent work on further gravitational features of $T\bar T$.

As for the $T\bar T$ - deformed observables, they turn out to be given by extremely simple, yet non-local deformations of the original QFT or CFT ones, as can be observed from the table  below %\ref{ttbproptable}. 

\medskip

\begin{table}[h]
\begin{center}
\begin{tabular}{|c|c|}\hline
deformed finite-size spectrum & $E(R) = E^{[0]} (R+\mu E)$ \\[2pt]  \hline
entropy & $S(E) = S^{[0]} (E^{[0]})$\\[2pt] \hline
~~~~~~~~~$\sim ~~~$ (CFT) & Cardy $\r$ Hagedorn\\[2pt] \hline
S-matrix & $\mathcal{S}^{[\mu]} (p_i) = e^{i \mu \sum_{i<j} \e^{\a\b} p^i_\a p^j_\b} \mathcal{S}^{[0]} (p_i)$\\[2pt] \hline
correlation functions (CFT) &  integral transform CFT correlator \\ [2pt] \hline
extended symmetries (CFT) & (non-linear modification of) Virasoro $\times$ Virasoro\\ \hline
\end{tabular}
\end{center}
\caption{\small{The observables in a $T\bar T$ - deformed QFT/CFT are related in a simple, yet non-local fashion to those of the undeformed theory (marked in paranthesis if it is a CFT). }}
\label{ttbproptable}
\end{table}
%
%\begin{figure}[h]
%\centering
%\includegraphics[height=6cm]{ttbprop_table}
%\caption{\textcolor{red}{\emph{Turn this into a proper table, and write Cardy $\r$ Hagedorn.}}}
%\end{figure}
%
%

\noindent How the upper three universal properties    of the $T\bar T$  deformation are derived  will be  explained now; the extended symmetries and correlation functions are discussed in sections \ref{infsymmsec} and, respectively, \ref{corrfsec}. 
The calculations involved are very similar across deformations.  We divide the discussion into three main themes: spectrum and thermodynamics,  relation to strings, and relation to (topological) gravity. 
\bigskip

\noindent\textbf{\large \emph{I. Spectrum and thermodynamics}}

\medskip

\noindent In this first part, we discuss the most well-studied observable in $T\bar T$ - deformed theories: their finite-size spectrum. The thermodynamic properties follow, as does a  generalised notion of modular invariance. 

\subsubsection{Spectrum of $T\bar T$-deformed QFTs on the cylinder}

 As discussed in the introduction, using first order quantum-mechanical perturbation theory and factorisation of
%
 % (see also \cite{Cavaglia:2016oda}), using the  result \eqref{ttbarcyl} for 
 the expectation value %$\langle n | \O_{T\bar T}|n\rangle$
  of SZ operators in an energy eigenstate on the cylinder, one can easily obtain an equation, \eqref{engflow}, for how the deformed energy spectrum $E_n^{[\mu]}(R)$ varies as $\mu$
 is {infinitesimally} changed. In the specific case of $T\bar T$,
  the %definition of the deformed QFT by the addition of the instantaneous $T\bar T$ operator implies that the 
change in the energy of the $n^{th}$ energy eigenstate is given by  %usual quantum-mechanical first order perturbation theory%\footnote{The reason that it is sufficient to use first order perturbation theory is that the deformation is infinitesimally defined. \emph{Rewrite!}} formula
%\textcolor{red}{\emph{Factors!}}

\be
\frac{\p E_n^{[\mu]} (R)}{\p \mu} %= - 4  \langle n | \int_0^R\!\!  d\s \,(\O_{T\bar T})_\mu |n\rangle  = - 4 R \langle n | (\O_{T\bar T})_\mu |n\rangle 
= R \left(\langle n| T_{\tau\tau} |n\rangle \langle n| T_{\s\s}|n \rangle - \langle n| T_{\tau\s} |n\rangle^2\right) 
\ee
%where the additional factor of $2$ with respect to \eqref{defseucl} comes from the change of measure, $d^2 z = 2 d\tau d\s$.
%
%where we used that the expectation value of the $T\bar T$ operator in an energy eigenstate on the cylinder factorizes in terms of the one-point functions of the stress tensor components. %This relation can be alternatively written as 
%
%\be
%\langle n| \O_{T\bar T} |n\rangle = \frac{1}{8}  \left(\langle n| T^{\a\b} |n\rangle \langle n| T_{\a\b}|n \rangle - \langle n| T^\a{}_\a |n\rangle^2\right)= - \frac{1}{4}  \left(\langle n| T_{\tau\tau} |n\rangle \langle n| T_{\s\s}|n \rangle - \langle n| T_{\tau\s} |n\rangle^2\right) \label{cylfact}
%\ee
where we are working in Euclidean signature. 
In turn, the one-point functions $\langle n| T_{\a\b} |n\rangle$ of the stress tensor components are related to the global conserved charges of the state as \cite{Zamolodchikov:2004ce} %\textcolor{red}{\emph{Signs! Looks correct...}}
\be
\langle n| T_{\tau\tau} |n\rangle = - \frac{E_n(R)}{R} \;, \;\;\;\;\;\; 
\langle n| T_{\tau\s} |n\rangle =  - \frac{i P_n(R)}{R}\;, \;\;\;\;\;\; 
\langle n| T_{\s\s} |n\rangle = - \frac{\p E_n(R)}{\p R} \label{onepfep}
\ee
%which simply follow from the definition of the energy and momentum conserved charges and translational invariance of the one-point functions. 
%
%
% \bigskip
%
%\noindent \begin{minipage}[t]{0.05\textwidth}  \emph{Exercise:} \end{minipage} \hspace{7mm} \begin{minipage}[t]{0.9\textwidth} Show that $\langle n| T_{\s\s} |n\rangle = - \p_R E_n(R)$. (\emph{Hint:} consider the QFT on the euclidean torus and interchange the labeling of space and time.)  
%\end{minipage}
%\bigskip
%
%\noindent To summarize, what we have shown in this subsection is that the expectation value of the $T\bar T$ operator in an energy eigenstate on the cylinder factorizes and is determined solely by the global conserved charges of the state, as
%%
%\be
%\langle n| \O_{T\bar T} |n\rangle = - \frac{1}{4R} \left( E_n \frac{\p E_n}{\p R} + \frac{P_n^2}{R}\right) \label{ttbarcyl}
%\ee
%a relation we obtained by combining \eqref{cylfact} and \eqref{onepfep}. This relation is \emph{universally} valid. 
%
Plugging  this into the above, we find\footnote{While the  derivation we presented is very simple and intuitive, rigorous-minded readers might complain that  in its first step, \eqref{genfloweng}, we passed from the Lagrangian to the Hamiltonian definition of the deformation in a somewhat cavalier fashion, as the deforming `potential' contains an infinite number of derivatives. A more rigorous derivation can be found in e.g. \cite{zoharlectures},  who carefully evaluate the torus partition function of the deformed QFT to extract the flow of the energies. }

\be
\frac{\p E_n^{[\mu]}(R)}{\p \mu} = E_n^{[\mu]}(R) \frac{\p E_n^{[\mu]}(R)}{\p R} + \frac{P_n^2(R)}{R} \label{burger}
\ee
\vskip2mm
\noindent The above can be recognised as the inviscid Burger's equation, with the momentum squared playing the role of a forcing term. It is an  entirely \emph{universal} equation governing the flow of energy levels in a $T\bar T$ - deformed QFT, as it does not depend at all on the details of the QFT that is being deformed - the latter  only enter as initial conditions for the solution to this equation. Note, in particular, that each energy level is evolving independently.  

To solve \eqref{burger}, one notes that the dependence of the momentum on $\mu$ and $R$ is entirely fixed by the fact that $\s \sim \s+R$, so it must obey the quantization condition 

\be
P_n = \frac{2\pi k_n}{R} \;, \;\;\;\;\; k_n \in \mathbb{Z}
\ee
In particular, the momentum cannot depend on the continuous parameter $\mu$. 

If the finite-size spectrum of the seed QFT at $\mu =0$ is known (as a function of $R$), then one can integrate the Burger's equation  to find the spectrum at arbitrary finite $\mu$. This is particularly easy to see for those states that have $P_n=0$, %for which the following holds: 
%
%
%
% \bigskip
%
%\noindent \begin{minipage}[t]{0.05\textwidth}  \emph{Exercise:} \end{minipage} \hspace{7mm} \begin{minipage}[t]{0.9\textwidth} Show that for states with $P_n=0$,  the function
where  % \textcolor{red}{\emph{Notation!}}

\be
E_n^{[\mu]}(R) = E_n^{[0]} (R+\mu E_n^{[\mu]}(R)) \label{relengsttb}
\ee
solves  Burger's equation \eqref{burger} with the correct initial condition.
% \end{minipage}
%\bigskip
%
 For general $P_n$, the solution can be obtained via similar manipulations \cite{Cavaglia:2016oda}. In the following, we discuss the explicit solution for the case of $T\bar T$-deformed \emph{conformal} field theories, where the $R$ - dependence   of the undeformed energy spectrum is particularly simple, being fixed by conformal invariance. 

\subsubsection{The spectrum of $T\bar T$-deformed \emph{C}FTs}

In a CFT, the state-operator correspondence maps the energy and momentum of a  state on the cylinder to the conformal dimension $\D= h_L+h_R$ and spin $s= h_L-h_R$ of the corresponding CFT operator on the plane, as %\textcolor{red}{\emph{Notation!}}
\be
E^{[0]} (R) = 2 \pi \, \frac{h_L+h_R - \frac{c}{12}}{R} \;, \;\;\;\;\;\; P (R)  =   2 \pi\, \frac{ h_L-h_R}{R} \label{engcylcft}
\ee
where the label $n$ has now been replaced by ${\D,s}$ or $(h_L,h_R)$. 
 The shift by $-c/12$ is the usual Casimir energy on the cylinder, where $c$ is the CFT's central charge. For states with zero momentum \eqref{relengsttb}, together with \eqref{engcylcft} immediately imply that $E^{[0]}(R) \, R = E^{[\mu]}(R)\, (R+\mu E^{[\mu]}(R))$; as it turns out,  non-zero momentum states satisfy a similar equation for the deformed energy which, using the constancy of the momentum along the flow,  can be written in the following suggestive way 
 \be
 E_L^{[0]} R = E_L^{[\mu]} \left(R + 2 \mu E_R^{[\mu]}\right) \;,\;\; \;\;\;\;\; E_R^{[0]} R = E_R^{[\mu]} \left(R + 2 \mu E_L^{[\mu]}\right) \;, \;\;\;\;\;\;\; E_{L,R} \equiv \frac{E\pm P}{2}  \label{defengeqttb}
 \ee
 where we have dropped the labels of the state, which will be self-understood from now on. The solution  for the deformed energy can be written as%, subject to the above initial condition, is \textcolor{red}{\emph{Maybe write it as $E^{[0]} R = E^{[\mu]} R_u$.}}
%
%
%Let us first concentrate on zero-momentum states, $s=0$. Then, the general result proven in the last exercise implies that 
%%
%\be
%E_{\D}(\mu,R) =2 \pi \, \frac{\D - \frac{c}{12}}{R+\mu \, E_{\D}(\mu,R) }
%\ee
%which yields a quadratic equation for the energy, with solution 
%
%\be
%E_{\D}(\mu,R) = - \frac{R}{2\mu} \pm \sqrt{ \left(\frac{R}{2\mu}\right)^2+ \frac{2\pi(\D - \frac{c}{12})}{\mu}}
%\ee
%In order to have a smooth limit to the CFT spectrum as $\mu \r 0$, one needs to select the upper sign if $\mu >0$, and the lower sign if $\mu <0$. 
%
%For states with non-zero momentum, the solution for the deformed energy is

%
\be
E^{[\mu]}(R) = \frac{R}{2\mu} \left(-1+ \sqrt{1 + \frac{4 \mu E^{[0]}(R)}{R} + \frac{4 \mu^2 P^2}{R^2}}\right) \label{defengcft}
\ee
where we have chosen the branch with a smooth $\mu \r 0$ limit.
%
%\be
%\boxed{E_{\D,s}(\mu,R) =  \frac{R}{2\mu} \left( -1 + \sqrt{ 1+ \frac{8\pi \mu(\D - \frac{c}{12})}{R^2}+ \left(\frac{4\pi \mu  s}{R^2} \right)^2}\, \right)} \label{defengcft}
%\ee
%
%\be
%E_{h,\bar h}(\mu,R) = - \frac{R}{2\mu} \pm \sqrt{ \left(\frac{R}{2\mu}\right)^2+ \frac{2\pi(h+\bar h - \frac{c}{12})}{\mu}+ \left(\frac{2\pi (h-\bar h)}{R} \right)^2}
%\ee
%
This expresses the energy of each eigenstate  in the deformed theory in terms of the energy and momentum  of the corresponding undeformed eigenstate. Note that the deformed energy levels do not cross. 
The  full
 $R$ - dependence of the deformed energies is obtained by pluging in the formula \eqref{engcylcft} for $E^{[0]}$ and $P$, with the explicit result
 
\be
E_{\D,s}^{[\mu]}(R) =  \frac{R}{2\mu} \left( -1 + \sqrt{ 1+ \frac{8\pi \mu(\D - \frac{c}{12})}{R^2}+ \left(\frac{4\pi \mu  s}{R^2} \right)^2}\, \right) \label{defengcftDs}
\ee 
   One immediately notices that the form of this $R$ - dependence, 
  in particular  that for the ground state energy, is not compatible with a CFT fixed point in the UV.  Indeed, if that were the case, then in the $R \r 0$ limit the ground state energy should have behaved as $-c_{UV}/(12 R)$, which is not the case for either sign of $\mu$. Thus, $T\bar T$ - deformed CFTs have a rather unusual behaviour in the UV; in particular, they are not governed by a local CFT fixed point. 

Let us now briefly discuss the salient properties of this solution, focussing for simplicity on the case $s=0$. If $\mu>0$, then for  those states  which have $\D < c/12$ (such as the ground state, which has $\D=0$) the deformed energy becomes complex if  $\mu/R^2> 1/(8\pi |\D-c/12|)$. For a seed CFT with a discrete spectrum, this represents a finite number of states. In particular, the ground state energy becomes complex for $R \leq R_c$, with
\be
 R_c \equiv \sqrt{\frac{2 \pi \mu c}{3}} \label{rcrit}
\ee 
Since the energy of excited states with $\D < c/12$ would acquire an imaginary part at radii smaller than $R_c$, we conclude that for positive $\mu$, $T\bar T$ - deformed CFTs can be consistent when placed on a circle, provided its circumference is larger than $R_c$. %Notice that for CFTs with a large central charge, $R_c$ is much larger than the non-locality scale $\sqrt{\mu}$ set by the $T\bar T$ coupling. \textcolor{red}{ \emph{Relevant?}}

For $\mu<0$  on the other hand, the formula for the deformed energy  implies that for fixed $|\mu|$ and $R$, an \emph{infinite} number of energy eigenstates, namely all levels with 

\be
\D  > \D_{max}  \equiv \frac{c}{12}+  \frac{R^2}{8\pi |\mu|}
\ee
acquire  an imaginary energy.  This is a far more worrisome behaviour, as it is present for any finite $R$, no matter how large it is\footnote{For $\mu<0$, the branch of solutions where the sign of the square root is flipped may also become relevant. Extremal states on this branch are connected to their standard branch counterparts, but not general states. See e.g.  \cite{Guica:2019nzm} for details. }. 

\bigskip

\begin{figure}[h]
\begin{minipage}{0.45 \linewidth}
\flushleft
\includegraphics[height=4cm]{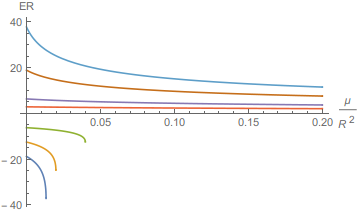}
\caption{The dimensionless energy $E R$ as a function of the dimensionless coupling $\mu/R^2$ for $\mu>0$. }
\end{minipage}
\hspace{1cm}
\begin{minipage}{0.45 \linewidth}
\flushright
\includegraphics[height=4cm]{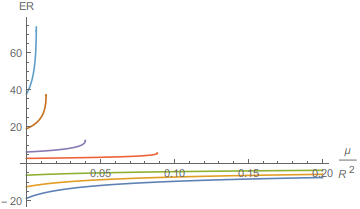}
\caption{The dimensionless energy $E R$ as a function of the dimensionless coupling $\mu/R^2$ for $\mu<0$. }
\end{minipage}
\end{figure}

\noindent The above maximum value of $\D$ corresponds to an upper bound on the energy 

\be
E_{max} = \frac{R}{2|\mu|} \label{emax}
\ee
attained just before the complex energy states set in - see figure above. This maximum energy is 
 sometimes referred to as the `UV cutoff' of  $T\bar T$ - deformed CFTs with $\mu<0$, since the maximum allowed energy density, $E_{max}/R$, is precisely the one set by the irrelevant coupling parameter, $1/|\mu|$. 
% States of the finite volume system whose energies are below $E_{max}$ are  to be kept, while states that acquire complex energies are to be discarded.
 
There has been much discussion in the literature concerning the proper treatment and interpretation of the above imaginary energy modes. Clearly, it would be most satisfactory if one could point out to some instability or other physical consequence associated with them, which may invalidate the derivation of the formula \eqref{defengcftDs} in this regime. A proposal, mainly motivated from holography, is that these modes should be simply truncated away, keeping only the states with real energies predicted by \eqref{defengcftDs}. However, it is not at all clear whether this procedure would yield a consistent quantum-mechanical theory; if the imaginary energy modes are to signal an inconsistency of the theory, simply removing this particular problem does not necessarily prevent  other possible inconsistencies from surviving. % \textcolor{red}{\emph{Understand!} modular covariance may be affected, etc. }

A physical explanation of the appearance of the complex energies for $\mu <0$ was given in \cite{Cooper:2013ffa}, who pointed out that the $\mu<0$ theory exhibits superluminal propagation around positive energy backgrounds (see discussion around eqn. \eqref{clstimedelay}). While in two dimensions, superluminality does not immediately imply the existence of closed timelike curves (CTCs), it does lead to them if the theory is placed in finite volume. In particular, $\D_{max}$ corresponds to the value of the energy for which the time advance gained in propagating once around the circle is comparable to the circumference of the circle. Since, for any given $\mu$ and $R$, there are always states in the original CFT with $\D > \D_{max}$, CTCs will always form if $R$ is finite. Thus, it is inconsistent to 
 place the $\mu <0$ $T\bar T$ - deformed CFT in finite volume; the complex energy states are simply a manifestation of this inconsistency. %It is not clear whether manually removing the states around which CTCs appear makes the truncated theory well-defined.  \textcolor{red}{\emph{Redundant!}}
Even if  $\mu <0$ $T\bar T$ - deformed CFTs may not make sense in finite volume,  they may still be well defined in infinite volume\footnote{This is precisely what happens for $J\bar T$ - deformed CFTs, for which the spectrum on the cylinder also contains an infinite number of complex energy states, but the spectrum of conformal dimensions on the plane, while non-trivially deformed, does not suffer from any obvious inconsistency \cite{
Guica:2019vnb}. }.   Nonetheless, as discussed in \cite{Cooper:2013ffa}, such theories with superluminal propagation  are rather peculiar.

In the above, we only discussed the spectrum for $P=0$. The analysis can be straightforwardly extended to general states, see \cite{Cooper:2013ffa} or \cite{Guica:2019nzm} for a discussion.

%imaginary energies also can be found for general states. For extremal states, $\bar h =0$, the energy spectrum appears to be real; however, this is slightly misleading. The plot is; we do get real but funny energies for $ h > 1/\mu$. \emph{Exercise?}

\subsubsection{Thermodynamics}

%When the parameter $\mu$ of the irrelevant deformation is small, $T\bar T$-deformed CFTs should have a behaviour close to that of the original CFT. It is though interesting  to ask how the entropy $S(E)$ behaves for large energy and finite $\mu$. 

We will concentrate on $T\bar T$ - deformed CFTs. According to the exact formula \eqref{defengcft},  each energy level of the CFT on the cylinder -  labeled by $(E^{[0]}, P)$ or, equivalently, $(\D,s)$ -  is   smoothly deformed under $T\bar T$ and does not cross other levels as $\mu$ is varied.  Consequently, the number of states within a fixed interval $\d \D, \d s$ centered around a given $(\D,s)$ does not change with $\mu$, since these are the invariant labels of the state.

In the undeformed CFT, the degeneracy of states at large enough energies/conformal dimensions is given by Cardy's formula \eqref{cardyfintro} with $R \r R/(2\pi)$.
%
%
%Given that what appears under the square roots in Cardy's formula are precisely the combinations $E^{[0]}_{L,R} R$
%%
%\be
%S(h,\bar h)= 2 \pi \sqrt{\frac{c}{6} \left(h-\frac{c}{24}\right)}+2 \pi \sqrt{\frac{c}{6} \left( \bar h-\frac{c}{24} \right)} \;, \;\;\;\;\;\;\; h, \, \bar h = \frac{\D \pm s}{2} \label{cardyfhhb}
%\ee
%which we are now writing in terms of conformal dimensions, rather than energies \eqref{cardyfintro}, using the CFT relation \eqref{engcylcft} between the energies of states on the cylinder and operator conformal dimensions. Setting  the momentum to zero, for simplicity, the `Cardy' behaviour at high energies is given by  $S(E) \propto \sqrt{c E R}$. 
%
 To compute the entropy $S(E,P)$ in the $T\bar T$ - deformed CFT, i.e. the log of the degeneracy of levels per unit energy and momentum interval $\d E, \d P $, we simply need to replace $E_{L,R} \leftrightarrow  E_{L,R}^{[0]}$  in \eqref{cardyfintro} by their expressions \eqref{defengeqttb}  in terms of the deformed  left- and right-moving energies,  with the result\footnote{The fact that we now measure the number of states in an interval $\d E$, as opposed to the number of states in an interval $\d \D$ does not affect the exponential factor that yields the leading contribution to the entropy.  }

 \be \label{entropyttb}
S_{T\bar T} (E_{L,R}) %=S_{Cardy} (E^{[0]}(E)) = % 2\pi \sqrt{\frac{c\, E_L^{(0)}R}{12 \pi}} + 2\pi \sqrt{\frac{c \, E_R^{(0)}R}{12\pi}} 
%\ee
%but we need to express it in terms of the physical energies in the deformed theory. Using \eqref{}, we obtain% The relation is given by:
%\be
%E_L^{(0)}  = E_L \left( 1 +   \frac{2\mu  E_R}{ R}\right) \;, \;\;\;\;\;\; E_R^{(0)}  = E_R \left(1 +  \frac{2\mu E_L}{ R}\right) 
%\ee
%Plugging this into the equation above, we obtain
%\be
%S_{T\bar T} = 
 = 2\pi \sqrt{\frac{c  E_L (R +2 \mu E_R) }{12\pi}} + 2\pi \sqrt{\frac{c  E_R(R+ 2\mu E_L)}{12\pi}}
\ee  
where  $E_{L,R}$ now stand for the deformed energies, previously denoted as $E_{L,R}^{[\mu]}$. To gain some physical intuition into this formula, it is useful to concentrate on states with zero momentum. The above formula simplifies to 
\be
S(E) %= 2 \pi \sqrt{\frac{c}{3} \left(\D(E) - \frac{c}{12} \right)} 
= 2 \pi \sqrt{\frac{c}{6\pi} (E R + \mu E^2) } \label{defent}
\ee
For the case $\mu>0$, for which the finite-size spectrum is well-defined at large energies, this formula interpolates between Cardy behaviour for $c \ll E \ll R/\mu$,  
and Hagedorn behaviour, $S(E)= \b_H E$ at very high energies. 
The Hagedorn temperature is given by
\be
\b_H = T_H^{-1} = \sqrt{\frac{2 \pi \mu c}{3}}
\ee
An entropy proportional to the energy means that in the canonical ensemble, the system cannot be in equlibrium at temperatures larger than $T_H$. Notice that $\b_H$ precisely equals the critical radius $R_c$  \eqref{rcrit} below which the ground state energy becomes imaginary: thus, the fact that $T\bar T$ - deformed CFTs do not make sense on circles whose size is smaller than $R_c$ maps (via an appropriately-defined modular transformation) to  the fact that  the temperature
cannot be made higher than $T_H$.

 The Hagedorn growth of states is of course familiar from the behaviour of the partition function of a free string at high temperatures. In fact, as we will soon see, there is a close connection between $T\bar T$ - deformed CFTs and the worldsheet theory on a free string. Note, however, that while in string theory, the Hagedorn growth  is superseded by non-perturbative  effects and there is a 
phase transition before $T_H$ is reached, it is not clear this would be the case in $T\bar T$. The behaviour of the 
 heat capacity close to the transition can give us more insights into its behaviour.  A simple computation \cite{Dubovsky:2012wk} yields
\be
 C_v = \frac{\pi c}{3} \frac{T R}{\left( 1 - \frac{T^2}{T_H^2}\right)^{3/2}}
 \ee
 showing the heat capacity, as well as its integral with respect to $dT$, diverge as $T \r T_H$.  This indicates $T_H$ is a maximum temperature, impossible to reach by adding a finite amount of energy to the system. 
 One may also consider corrections to $T\bar T$ thermodynamics that originate from universal logarhitmic corrections \cite{Carlip:2000nv} to the Cardy formula in the seed CFT. Interestingly, \cite{Barbon:2020amo} found that the 
  heat capacity in the vicinity of $T_H$ becomes negative as a result of including this correction\footnote{Heuristically, the shape of the $T(E)$ curve one obtains is similar to \ref{tempvsengnonextr}, except that it asymptotes to a constant, $T_H$. }. It would be interesting to better understand the physical implications of this result for the $T\bar T$ - deformed CFT viewed as a self-standing, UV-complete theory.

%For future reference %\emph{\textcolor{blue}{Compare subtracted!}}
 One may also compute the entropy as a function of temperature
  \be
  S(T) = \frac{\pi c}{3} \frac{T R}{\sqrt{1-T^2/T_H^2}} \label{ttbenttemp}
  \ee 
which clearly shows departures from the CFT formula \eqref{cardyfintro} when $T$ becomes a sizeable fraction of $T_H$. 
  %

 %Calculate the speed of sound and show it vanishes as $T\r T_H$.}
%%
%\bigskip
%
%\noindent \begin{minipage}[t]{0.05\textwidth}  \emph{Exercise:} \end{minipage} \hspace{7mm} \begin{minipage}[t]{0.9\textwidth} Compute the heat capacity and show that it is positive and divergent as $T\r T_H$. What happens if one includes the first logarithmic correction to the Cardy entropy formula? 
% \end{minipage}
%%\bigskip
%
%\noindent
%
%\be
%C_V = T \left(\frac{\p S}{\p T}\right)_R = \frac{R T T_H}{2 \mu (T_H^2 -T^2)^{3/2}}
%\ee
%which diverges at $T_H$. \emph{Does this signal a phase transition?} One can also study the case $\mu < 0$, case in which one has $E_{max} = R/|\mu|$ before reaching the imaginary energies. 
%
For $\mu <0$, it does not appear to be physically sensible to consider the theory on a cylinder. Nonetheless, one may still consider the thermodynamic properties of those states (a finite number of them) with energies below $E_{max}$, given in \eqref{emax}, with the parameters chosen so that   $E_{max} \gg c$.  Most thermodynamic quantities are obtained by simply flipping the sign of $\mu$ in the previous expressions. The temperature formally diverges at $E_{max}$, and the specific heat (which is generally positive) vanishes. These properties have been given  a gravitational interpretation in \cite{McGough:2016lol}. %\textcolor{blue}{\emph{Revisit.}}

\subsubsection{Generalised modular invariance}

Modular invariance of the torus partition function is a very important and powerful property of CFTs. It allows, in particular, to \emph{derive} the universal behaviour for the high-energy density of states - given by Cardy's formula \cite{Cardy:1986ie}, as we discussed - and other interesting constraints, for example on certain averaged OPE coefficients \cite{Kraus:2016nwo}.  It is thus interesting to inquire what are the modular properties of $T\bar T$ - deformed CFTs, i.e. the properties of their torus partition function  under large diffeomorphisms. 

The torus partition function of the $T\bar T$ - deformed CFT is defined as usual via the Hilbert space trace
\be
Z^{[\mu]} (\tau, \bar \tau, R) = \sum_n e^{-\tau_2 R E_n^{[\mu]} + i \tau_1 R P_n} \label{defttbpf}
\ee
where $\tau=\tau_1 + i \tau_2 $ is the complex structure modular parameter and $R$ is the length of the $a$-cycle
of the torus, here designated as the spatial one.  The metric on the torus can be written as
\begin{align}
ds^2&= R^2 |dx+\tau dy|^2=R^2 dz d\bar{z}
\end{align}
where  $x,y$ are real coordinates of unit periodicity  and  the complex coordinates $z,\bar z$ are defined as $z=x+\tau y,\,\bar{z}=x+\bar{\tau}y$.  This metric is invariant under  large   $PSL(2,\mathbb{Z})$ diffeomorphisms of the torus 
\be
\begin{pmatrix}x \\y \end{pmatrix}\mapsto\begin{pmatrix}\; a & - b \\- c & \;d \end{pmatrix}\begin{pmatrix}x \\y \end{pmatrix} \;, \;\;\;\;\;\;\; 
\tau\mapsto\frac{a\tau+b}{c\tau+d}\;, \;\;\;\;\;\; ad-bc=1 \label{psl2z}
\ee
%
%
%\be
%\begin{pmatrix}x \\y \end{pmatrix}=\begin{pmatrix}d & b \\ c & a \end{pmatrix}\begin{pmatrix}x' \\y' \end{pmatrix} \;, \;\;\;\;\;\;\; 
%\tau\rightarrow\frac{a\tau+b}{c\tau+d}
%\ee
which leave the coordinate periodicities  intact, provided we also transform

\be
%\tau\rightarrow\frac{a\tau+b}{c\tau+d} \;, \;\;\;\;\;\; 
R\r |c\tau+d|R
\ee
Note this ensures that the area of the torus, $R^2 \tau_2$, is invariant. %Under \eqref{psl2z}, the complex coordinates change as
The holomorphic coordinate transforms as 
$
z\mapsto \frac{z}{c\tau+d}$, and $\bar z$ as the complex conjugate of this. 
% \;, \;\;\;\;\;\; \bar z\mapsto \frac{\bar{z}}{c \bar \tau+d} $
%together with an $SL(2,\mathbb{Z})$ transformation of the real coordinates $(x,y)$:
%\begin{align}
%x+\frac{a\tau+b}{c\tau+d}y=\frac{(xd+by)+\tau(ay+xc)}{c\tau+d}=\frac{x' + \tau y'}{c\tau+d}, \hspace{0.4cm} \begin{pmatrix}x' \\y' \end{pmatrix}=\begin{pmatrix}d & b \\ c & a \end{pmatrix}\begin{pmatrix}x \\y \end{pmatrix}
%\end{align}
%Since the area of the torus remains the same, it follows that at a modular transformation, the characteristic length of the torus changes:
%\begin{align}
%R\mapsto |c\tau+d|R
%\end{align}

Assuming the partition function \eqref{defttbpf} can also be computed via an Euclidean path integral over the torus, which is naturally invariant under the diffeomorphisms  discussed above, we conclude that

\be \label{modularttbar}
Z^{[\mu]} \bigg(\frac{a\tau + b}{c\tau+d},\frac{a\bar{\tau}+b}{c\bar{\tau}+d},R |c\tau+d|\bigg) = Z^{[\mu]}(\tau,\bar{\tau},R)
\ee 
While we wrote this relation with $T\bar T$ - deformed CFTs in mind, it applies to \emph{any} UV-complete theory with a single scalar dimensionful parameter\footnote{For example, one can explicitly check that the partition function $Z^{[m]}$ of a $2d$ massive scalar   \cite{Aggarwal:2024axv} satisfies this relation. } \cite{Aharony:2018bad}, which is to be held fixed as the torus coordinates are relabeled. In a CFT, the radius dependence drops out by scale invariance, resulting in the usual modular invariance requirement;  \eqref{modularttbar} may then be referred to as `generalised modular invariance'.

 If the dimensionful parameter has dimensions of $(length)^2$, as in $T\bar T$, then $Z^{[\mu]}$ only depends on $R$
via the dimensionless combination $ \mu/R^2$, so \eqref{modularttbar} can be rewritten as

%
%whose partition function depends on a single dimensionful parameter $\mu$, it 
% should hold in any UV-complete two-dimensional QFT with dimensionful scalar couplings - collectively denoted as `$[\mu]$' - whose partition function [is well defined and] can be computed via a path integral over the euclidean torus. %{\color{ForestGreen}(mention here that this is the "generalised modular invariance" we refer to later)}%, and it simply corresponds to the freedom to assign different Hilbert space interpretations to it/diff invariance. 
% 
%  It simply states the invariance of the partition function under a relabeling of the torus coordinates, and as such it is natural that these  transformations relate theories defined on tori  with different sizes of the $a$-cycle, where the scalar couplings (which may be dimensionful) are held fixed. 

%For a $T\bar T$ - deformed CFT, , since $\mu$ is the only dimensionful parameter in the theory\emph{Shorten:} Let us now discuss the modular transformation properties of this partition function.
%
%where $a,b,c,d\in \mathbb{Z}$ with $ad-bc=1$, 

 % In a $T\bar T$ - deformed CFT, the partition function $Z^{[\mu]} (\tau,\bar \tau, R) =  Z_{T\bar{T}}(\tau,\bar \tau, \mu/R^2)$, and so the above relation reads %\emph{Notation!}

\be \label{TTbarmodinv}
Z \bigg(\frac{a\tau + b}{c\tau+d},\frac{a\bar{\tau}+b}{c\bar{\tau}+d}, \frac{\mu}{R^2 |c\tau+d|^2}\bigg) = Z \bigg(\tau,\bar{\tau},\frac{\mu}{R^2}\bigg)
\ee
One may then reinterpret \eqref{TTbarmodinv} as relating theories on a circle of the  same radius, but with different dimensionless couplings. The above relation was checked explicitly in \cite{Datta:2018thy} for $T\bar T$ - deformed CFTs, but please note it should hold in any \emph{UV-complete} theory with a single irrelevant coupling of this dimension. 

%[This was derived in \cite{} by expanding the $T\bar T$ partition function \eqref{} in $\mu$, using the explicit solution \eqref{} for the spectrum, and showing that each term in the expansion is a modular form of weight equal to twice the power of $\mu$. Now, we see this had to work this way by the mere existence of a path integral formulation of the torus partition function. ]

%As emphasized, this result holds in \emph{any UV-complete} QFT with a single dimensionful scale, $\mu$. The non-trivial requirement/assumption is that of UV-completeness.

One may therefore ask what makes  $T\bar T$ - deformed CFTs special. In \cite{Aharony:2018bad}, it was shown that if one makes the additional assumption that the energy of a state at finite $\mu$  only depends on $\mu$ and the energy of the undeformed state - which is assumed to be generic - then  one can derive a flow equation for the partition function with respect to the dimensionless coupling $\mu/R^2$ and $\tau, \bar \tau$, whose 
unique\footnote{Single-trace $T\bar T$-deformed CFTs appear to provide a counterexample to this statement, likely because their spectrum is not generic. We thank S. Georgescu for this observation.} solution  for $\mu>0$ is\footnote{For $\mu<0$, \cite{Aharony:2018bad} also discussed certain non-perturbative ambiguities in the partition function. It  also pointed out that the  partition function where the spectrum is  truncated to real energies only does not obey generalised modular invariance. }  the $T\bar T$ - deformed spectrum \eqref{defengcft}.  Of course, such a requirement is extremely constraining; this argument shows, therefore, that the only $(2,2)$ deformation where the energies evolve independently of each other and agree with the CFT spectrum at $\mu=0$ is $T\bar T$; other irrelevant deformations will not lead to this very special property.

%however, that the deformed energies in the generalised $(T\bar T)_s$ deformations or in LST must take significantly more complicated forms in terms of the undeformed spectrum\footnote{In the latter case, the undeformed spectrum is of course not generic.} than they do in $T\bar T$. \textcolor{blue}{\emph{Correct? How about single-trace $T\bar T$?}}

Another  interesting question is  whether the high-energy density of states in  a theory  satisfying \eqref{TTbarmodinv} takes on a universal Cardy $\r$ Hagedorn form \eqref{entropyttb}.  In \cite{Apolo:2023aho}, a variant of this question was studied, and it was shown that the generalised $S$ modular transform can indeed predict (under certain assumptions) this asymptotic degeneracy of states if one inputs the  $T\bar T$-deformed vacuum energy in the crossed channel. % Of course, there is no known argument to establish that generic theories with an irrelevant parameter $\mu$ of length dimension two should have precisely this form of the vacuum energy, but it would be interesting to find one\footnote{In this context, it is interesting to note that the vacuum energy of the asymptotically linear dilaton background \eqref{} does take this form \cite{Apolo:2019zai}. }. 

% The answer is affirmative. \emph{Is there a conclusion also without assuming sparseness?}  

%In \cite{} the modular property plus a certain sparseness condition of the light spectrum of the theory was used to argue for vacuum dominance in the cross channel, which in turn yielded a universal entropy of $T\bar T$ - deformed CFTs.

%In fact, this should be true of any theory with one dimensionful parameter. \emph{Correct? Comparison single and double-trace?}

%Another interesting question concerns the uniqueness of the $T\bar T$ deformation. It is also possible to show \cite{aharony} that the $T\bar T$ - deformed energy spectrum is the only generalised modular invariant solution with a single dimensionful parameter, under the assumption that the deformed energies on depend on the initial ones and $\mu$. This has been established in \cite{} by expanding the modular in/covariance relation \eqref{} order by order in $\mu$, and equating teh various terms in the expansion. 
%I think this argument does (implicitly) use genericity of the initial spectrum (or lack of correlation between the energies of the different states). \emph{Correct?} \emph{What does this result imply about other deformations with a single parameter?} Note that assuming the deformation exists for an arbitrary initial CFT  spectrum is a very contraining assumption. 

Finally, remember from our previous discussion that $T\bar T$ - deformed CFTs with  $\mu>0$ placed on a cylinder only make sense if the circumference of the latter is larger than a minimum one \eqref{rcrit}, below which the ground state energy becomes imaginary. Since the radius changes under generalised modular transformations, one might rightfully worry whether, having started with a theory where $R> R_{c}$, this condition is preserved by modular transformations.  
Note that under an $S$ transform this condition maps to the Hagedorn condition $R|\tau| > R_c = \b_H$, which can be obtained more generally by requiring the partition sum be well-defined

\be
\lim_{E_{L,R} \r \infty} Z_{T\bar T} \approx e^{- \b_L E_L - \b_R E_R} e^{2\sqrt{\frac{2 \pi \mu c}{3} E_L E_R}} \leq e^{2 \sqrt{E_L E_R} (\b_H - \sqrt{\b_L \b_R})}
\ee
Above, $\b_{L,R}$ are the left/right-moving temperatures, which satisfy $\b_L \b_R = R^2 |\tau|^2$.  Thus, the partition function will be  well-defined provided  also $R |\tau| > R_c$.  One may check - by appropriately choosing the integer part of $\tau_1$ - that 
% 
%$\b > R_{min}$. \emph{What is the general condition for a tilted torus?}  If all is to make sense, 
modular transformations do not take us out of this regime.

%have this property. %e. \textcolor{blue}{\emph{Flow eqn wrt $\mu$?}}

%
%\bi
%%\item can the $T\bar T$ entropy be interpreted as the entropy of stringy dof?
%\item how about well-definiteness of the partition function for $\mu <0$?
%\ei

\bigskip
\noindent\textbf{\emph{II. Relation to strings: classical analysis}}
\medskip

\noindent  As prefigured by the asymptotic Hagedorn behaviour of the entropy, there exists 
 a close relationship between the $T\bar T$  deformation and strings. In this subsection, we explore the  relation, at a classical level,  between $T\bar T$ - deformed free bosons and the Nambu-Goto string and use it to extract certain characteristic properties of $T\bar T$. % from it.\textcolor{blue}{\emph{Uniform lightcone gauge? How about QCD?} }

\subsubsection{$T\bar T$-deformed free boson(s) and the Nambu-Goto string}

Let us now discuss the simplest example of a $T\bar T$-deformed CFT: the $T\bar T$-deformed free boson(s) \cite{Cavaglia:2016oda}. Despite its simplicity, this model has surprisingly rich physics, and captures all of the representative physical properties of $T\bar T$. %: the preservation of integrability, the universal modification of the S-matrix by  an energy-dependent phase, the field-dependent coordinate transformation that relates the deformed and undeformed theories \cite{Dubovsky:2012wk}.
 Moreover, it displays a rather interesting connection to the worldsheet theory of an infinitely long free  bosonic string. 

%\textcolor{blue}{Despite its simplicity, this example  illustrates/captures all of the representative/quintessential physical properties of $T\bar T$-deformed QFTs: the preservation of integrability, a universal modification of the S-matrix (by multiplication by an energy-dependent phase), the relation between the deformed and undeformed theories via a field-dependent coordinate transformation, which in the string picture has a particularly nice interpretation as a change of gauge in the Nambu-Goto action.}

%\textcolor{blue}{In the following, we explicitly derive the \emph{classical} Lagrangian of a $T\bar T$-deformed free boson (the generalization to several  bosons being straightforward) and relate it to the Nambu-Goto action for a string in a particular gauge. Using the Nambu-Goto perspective, we show that the deformed and the undeformed theories are related via a field-dependent coordinate transformation. In \ref{smatsec}, we turn to the quantum case and describe the calculation of the deformed S-matrix, which is just a  phase, using integrability techniques.  Next, in \ref{physmanif} we discuss   the basic physical manifestations of the scattering phase: a time delay in scattering proportional with the energy and  the existence of  a minimum length, both of which point towards a (quantum) - gravitational  interpretation of $T\bar T$ - deformed QFTs. A large part  of this section is based on \cite{Dubovsky:2012wk}.  In \ref{genqft}, we very briefly  comment on  deforming more general QFTs. }

The analysis in this section is purely classical. We start from a free scalar in $2d$, with Lagrangian 

\be
S_L^{[0]} = \frac{1}{2} \int dt d\s (\dot \phi^2-\phi'^2)
\ee 
where $\;\dot{} = \frac{d}{dt}$ and ${}'=\frac{d}{d\s}$, and solve the flow equation \eqref{ttbdefintro}  for the deformed Lagrangian , where the stress tensor  computed via the Noether procedure

\be
\p_\mu L = \frac{1}{2} (T^{\a\b} T_{\a\b} - T^2) \;, \;\;\;\;\; T^\a{}_\b = \frac{\p L}{\p(\p_\a \phi)} \p_\b \phi - L \, \d^\a_\b
\ee
Since, by Lorentz invariance and dimensional analysis, the Lagrangian must take the form $\p_\a \phi \p^\a \phi f(\mu  \p_\a \phi \p^\a )$, one finds that the solution for the  action is 

\be
S_L^{[\mu]} = \frac{1}{2\mu} \int dt d\s \left(1 - \sqrt{1+2 \mu\, (\phi'^2-\dot \phi^2)}\right)  \label{sttbfb}
\ee
When expanding this action in $\mu$, it contains an infinite number of `higher derivative' terms. 
% so the $T\bar T$ - deformed free boson naively appears to be a  non-renormalizable theory, with cutoff  set by $\mu$.
  However, these  terms turn out to have a highly constrained structure since, up to a constant shift,  \eqref{sttbfb} precisely coincides with the Nambu-Goto action 

\be
S_{NG} = - \frac{1}{\ell_s^2} \int d^2 \s \sqrt{-\det \g_{\a\b}} \;, \;\;\;\;\;\; \g_{\a\b} = \eta_{\mu\nu} \frac{\p X^\mu}{\p \s^\a} \frac{\p X^\nu}{\p \s^\b} \;, \;\;\;\;\;\;   \ell_s^2 = 2\pi \a'
\ee
in three Minkowski target space dimensions in \emph{static gauge}, i.e.

\be
 X^0 = t \;, \;\;\;\; X^1 = \s\;, \;\;\;\; X^2 = \sqrt{2\mu}\, \phi \label{stgauge}
\ee
provided we identify $\ell_s^2 = 2 \mu$. Indeed, it can be easily checked that in this gauge, the induced metric on the string worldsheet is
\be
\g_{\a\b} %= \frac{\p X^\mu }{\p \s^\a} \frac{\p X_\mu}{\p \s^\b}
 = \left(\begin{array}{cc} - 1+ 2\mu\, \dot{\phi}^2 & 2\mu\, \dot \phi\, \phi' \\ 2\mu \, \dot \phi\, \phi'  & 1 + 2\mu\, \phi'^2  \end{array}\right)
\ee
%and the square root of the metric determinant  agrees  with \eqref{sttbfb} (up to a constant shift),  provided we identify $\mu=\ell_s^2/2$. 
%
Thus, at the classical level, the action of one $T\bar T$-deformed free boson is nothing but the Nambu-Goto action for an infinitely long string embedded in three-dimensional Minkowski space, in static gauge. It is not hard to show that the connection between the $T\bar T$ deformation  and the Nambu-Goto action persists if we deform several free bosons \cite{Cavaglia:2016oda}. More precisely, the $T\bar T$ deformation of $D-2$ free bosons yields the Nambu-Goto action for a string propagating in $D$ target space dimensions, in static gauge.

 Below, we discuss some physics that can be extracted from the relationship between $T\bar T$ and strings, much of it developed in the prescient article \cite{Dubovsky:2012wk}.

\subsubsection{The field-dependent coordinates}

As is well known, 
 in conformal  gauge ($\g_{\a\b} \propto \eta_{\a\b}$),  the  dynamics of the bosonic string  reduces to that of $D-2$ free bosons, which parametrize fluctuations in the  directions  transverse to the string. Thus, from the point of view of the $T\bar T$ deformation, the undeformed CFT (describing  $D-2$ free  bosons) corresponds to the Nambu-Goto action in conformal gauge, while the deformed CFT corresponds to the Nambu-Goto action in static gauge. We conclude that there should be a change  of coordinates on the string worldsheet that takes the deformed CFT to the undeformed one. At the purely classical level, this statement holds for any target space dimension, $D$, and 
we will exemplify it below for a string in three target space dimensions (single boson). % At the quantum level, one needs to work in the critical dimension.  

 Let
 \be
  U = \s + t = X^1+X^0\;, \;\;\;\;\; V=\s-t= X^1-X^0
  \ee be the  worldsheet coordinates in static gauge (which are identified with two null coordinates in the target space), and $u,v$ be the corresponding null worldsheet coordinates in conformal gauge. In the static gauge \eqref{stgauge},  the  induced worldsheet line element takes the form

\be
ds^2 = dU dV + 2 \mu (\p_U \phi\, dU + \p_V \phi \, dV)^2
\ee  
Let us now change the coordinates on the worldsheet to $u,v$
\bea
ds^2  &=& \left(\p_u U\, du + \p_v U \, dv \right)\left(\p_u V\, du + \p_v V \, dv \right) + 2 \mu (\p_u \phi\, du + \p_v \phi \, dv)^2 \\
&& \hspace{-1.1cm} = \; \left(\p_u U\p_u V +  2 \mu (\p_u \phi)^2 \right) du^2 +  \left(\p_v U\p_v V +  2 \mu (\p_v \phi)^2 \right) dv^2 + \left(\p_u U\p_v V + \p_v U\p_u V + 4 \mu \p_u \phi \p_v \phi \right) du dv \nonumber
\eea
and set $\g_{uu} = \g_{vv} =0$, as required by the conformal gauge condition. These equations are not sufficient to completely fix the map between $U,V$ and $u,v$, so we additionally require that $\p_u U = \p_v V =1$, a condition that is compatible with both the conformal gauge equations of motion for the target space coordinates $U,V$  and the requirement that as $\mu \r 0$, we have $U=u$ and $V=v$.  We thus find, on-shell % (note this does not hold off-shell) 
\be
U = u - 2 \mu \int^v T_{vv} dv \;, \;\;\;\;\;\;V = v - 2 \mu \int^u T_{uu} du \label{fdepcootrfb}
\ee
where $T_{uu} = (\p_u \phi)^2$ and $T_{vv}=(\p_v \phi)^2$ are the non-zero components of the free boson stress tensor. In this map, we have assumed the theory is on $\mathbb{R}^{1,1}$, ignoring the possible identification of the spatial coordinate\footnote{On a compact space, there exists a subtle zero-mode contribution to these relations, discussed  in sections \ref{asysymmdtr}, \ref{infsymmsec}.}. The coordinates $u,v$ that solve the above equation are sometimes called the `field-dependent' coordinates, as they depend on the configuration of the system. 
  One can easily check that the equation of motion for the field $\phi(U,V)$ in static gauge simply maps to the free wave equation  $\p_u \p_v \phi =0$ in terms of the field-dependent coordinates, and so do its solutions.

Written in the form above, the transformation \eqref{fdepcootrfb}  generalises to any $T\bar T$ - deformed CFT.  Thus, the full dynamics of the deformed theory can be reduced to that of the original CFT, written  in terms of the field-dependent coordinates\footnote{See e.g. \cite{Conti:2018tca} for a detailed map of the classical solutions.}. Conversely, from the point of view of the original CFT (which is analogous to the CFT found in conformal gauge, with coordinates $u,v$), the target space coordinates $U,V$ in  which the $T\bar T$ - deformed theory lives are sometimes called `dynamical coordinates' \cite{Dubovsky:2017cnj}. One can thus state that the $T\bar T$ dynamics simply corresponds to viewing the dynamics of the original theory through the prism of these dynamical coordinates \cite{dubovskytalk}. This viewpoint provides a simple physical intuition into why every single observable that has been computed in a $T\bar T$ - deformed theory is related in a simple -  yet non-local - way to the corresponding observable in the undeformed QFT. % \textcolor{blue}{\emph{Any mention of emergence from $T\bar T$ flow here? Note dynamical coordinates particularly clear from path integral pdv.}}

%These coordinates can also be defined in Hamiltonian formalism, where the  relation \eqref{} translates into simply $\p_\s u = 1 + 2 \mu \H_R$ and $\dot u - \{ H,u\} =??? $. However, there  are very subtle zero mode contributions when integrating these expression, with important effects on the charge algebra. 

\subsubsection{Time delay and signal propagation speed}

 As we will see, much of the physics of the $T\bar T$ deformation is encoded in the field-dependent coordinate transformation \eqref{fdepcootrfb}. A signature effect that is visible already at the classical level is   a universal  time delay  proportional to the energy \cite{Dubovsky:2012wk}.
% in scattering processes proportional to the energy. At classical level, this can be seen as a time delay when passing through a non-trivial classical background. 

  To see this, consider for simplicity  a purely left-moving classical background $\phi(U)=\phi(u)$, which  can be easily seen to solve the $T\bar T$-deformed equations of motion.   The relation between the $T\bar T$ coordinates $U,V$ and the conformal gauge ones is, in this background
\be
U = u \;, \;\;\;\;\; V = v - 2 \mu \int T_{uu} \, du
\ee
Consider now the propagation of left/right-moving waves on the worldsheet immersed in  this background, which by definition propagate on lines of $u=const$, $v \in (\infty, -\infty)$ and respectively $v=const$, $u \in (-\infty,\infty)$. 
It is clear that the presence of the background does not at all affect the left-moving excitations. As for the right-moving ones, they acquire a shift
\be
\Delta V %= \D \s - \D t
= - 2 \mu \int_{-\infty}^\infty T_{uu} du = - 2 \mu \Delta E_L  
\ee
where $\Delta E_L$ is the total left-moving energy of the background. 
If we wait for the wave to arrive at the same position  $\s$ near infinity, $\D V =-\D t$, where

\be
\Delta t = 2 \mu \D E_L \label{clstimedelay}
\ee
is the \emph{time delay} (for $\mu >0$) for the wave to arrive, as compared to if no background had been present. A time delay proportional to the energy is typical of gravitational scattering, reason for which it has been proposed that the $T\bar T$-deformation turns the original QFT into a  gravitational theory.  The time delay is present if $\mu >0$; for $\mu <0$ one  obtains instead a time advance; this is related to the fact, discussed below equation \ref{emax}, that $T\bar T$-deformed CFTs with $\mu<0$ exhibit superluminal propagation around positive energy backgrounds \cite{Cooper:2013ffa}. While this behaviour is unusual, it is  not immediately inconsistent, provided one stays in infinite volume.  %\emph{Write explicitly the signal prop speed and refer back to spectrum discussion! Time advance necessary for CTCs seems to precisely coincide with imaginary energies onset - True?}

\subsubsection{$T\bar T$ and  strings in uniform lightcone gauge}

As it turns out, there exist also other ways of relating a $T\bar T$ - deformed sigma model to strings. As nicely explained in \cite{Frolov:2019nrr}, it is possible to realise the $T\bar T$-deformed solution for the energies  $E^{[\mu]}( R) = E^{[0]}(R+\mu E)$   by studying strings in the so-called uniform lightcone gauge  

\be
X^+ = \tau + a \s \;, \;\;\;\;\;  p_+ =1
\ee
where $a$ is a constant,  $X^\pm$ are  certain $a$ - dependent linear combinations of the target   time and space (both assumed to be isometry directions in the target space) and $p_+$ is the lightcone momentum density \cite{Arutyunov:2009ga}.  Requiring that the target space energy  not depend on the choice of gauge parameter $a$ then implies the above relation between the energies.  The radius of the $\s$ coordinate in this gauge does, nonetheless, depend on the uniform lightcone gauge parameter, which is to be identified with the $T\bar T$ coupling up to a numerical shift. 

The above method for reproducing the spectrum of $T\bar T$ - deformed CFTs is quite general, as it works for  $\s$ models involving scalars and fermions, with arbitrary potential.  One interesting output of this way of understanding the $T\bar T$ deformation is that it allowed writing a general formula for the deformed Hamiltonian density in terms of the undeformed one.  The same formula has been obtained later in \cite{Jorjadze:2020ili} by simply solving the $T\bar T$ flow equation for the classical Hamiltonian density, as we now explain.

%
%\bi
%\item can one perhaps use the `string' formulation to derive the same for general initial $\H$ (Frolov/Theisen/Sfo)?
%\item in particular, this allows us to write the deformed Hamiltonian in terms of the undeformed one \emph{Check if true!}
%\item then, make transition to theisen and understand lefloch (they simply noted that replacing $E_0$ by $\mathcal{H}_0$ solved the equation)
%\ei

%One may also start with a string propagating in a more general target space metric (not necessarily flat, but still with certain constraints, i.e. isometry directions) and derive the deformed Hamiltonian by changing the uniform lightcone gauge. \emph{Does this work in arbitrary spacetimes? What happens to the boundary conditions?}

\subsubsection{The general classical deformed Hamiltonian}

 The setup of \cite{Jorjadze:2020ili} consisted of a classical two-dimensional field theory with classical fields $\phi_k$ and conjugate momenta $\pi_k$, treated in the Hamiltonian formalism. They considered the $T\bar T$ flow of the classical Hamiltonian density 
 \be
 \p_\mu \H = - (T_{tt} T_{\s\s} - T_{t\s}^2)% = - \H (\pi_k \p_{\pi_k} \H + \phi'_k \p_{\phi'_k} \H - \H) + \P^2 
 \label{hamflowttb}
 \ee  
where the components of the stress tensor are obtained from the Hamiltonian density via
\be
T_{tt} = \H \;, \;\;\;\;\; T_{t\s } = \P = \pi^k \phi'_k \;, \;\;\;\;\; T_{\s t} = \p_{\pi^k} \H \, \p_{\phi'_k} \H \;, \;\;\;\;\; T_{\s\s} = \pi^k \p_{\pi^k} \H + \phi'_k \p_{\phi'_k} \H - \H \label{stresstcompham}
\ee 
with repeated indices  summed over. If the trace of the stress tensor in the undeformed theory vanishes identically, then $ \H^{[0]}$ is a homogenous function of degree two in $\pi^k, \phi'_k$  and the flow equation implies that the deformed Hamiltonian density is just a function of the undeformed one, $\H^{[0]}$ and the momentum density, $\P$. Once this is established, the solution to \eqref{hamflowttb} is trivial to find  in closed form 
\be
\H^{[\mu]} (\pi^k, \phi_k) = \frac{1}{2\mu} \left( \sqrt{1+4 \mu \H^{[0]} (\pi, \phi) + 4 \mu^2 \mathcal{P}^2}-1 \right) \label{solnhamttb}
\ee
Interestingly, the functional form of the relation is identical to the relation \eqref{defengcft} between deformed and undeformed energies on the cylinder, though it does not follow from it in any simple way\footnote{It was already observed in \cite{LeFloch:2019rut} that the relation \eqref{solnhamttb}, guessed from the relation between the energies, satisfies the flow equation \eqref{hamflowttb}. }.  It is also trivial to check that if one inputs the Hamiltonian for a free boson, $\H^{[0]} = \frac{1}{2} (\pi^2 + \phi'^2)$, into this formula, one obtains the one that follows from the Lagrangian \eqref{sttbfb}.

%As expained in the introductory section, at the classical level the Hamiltonian density flow equation is an equivalent definition of the $T\bar T$ - deformed theory. The solution for the deformed Hamiltonian density turns out to be identical to the energy formula

%Note though one formula doesn't follow from the other. So far, this has been achieved at the classical level, and yield a complementary yet interesting perspective. 
%
%\bi
%\item restriction on initial CFT to be able to do this?
%\ei

Given the fixed relation \eqref{solnhamttb} between the deformed and undeformed Hamiltonian, it follows that the Poisson brackets of $\H$ and $\P$ are \emph{universal} - even though not particularly simple -  since they follow from those of the undeformed CFT via this relation.  This fact will be extremely important in section \ref{infsymmsec}, where we work out the classical symmetries of $T\bar T$ - deformed CFTs. %See \cite{} for the explicit expressions. 

Given the explicit form for $\H^{[\mu]}$ in terms of  $\H^{[0]}$, one may compute the $T_{\s\s}$ component of the deformed stress tensor explicitly and use the result to show that classical $T\bar T$ - deformed CFTs satisfy a trace relation 
\be
tr \, T= - 2 \mu \, \O_{T\bar T} = - \mu (T^{\a\b} T_{\a\b} - T^2) \label{ttbtrrel}
\ee
Thus, despite the fact that the $T\bar T$ deformation breaks scale invariance and generates a trace for the stress tensor, the latter still only has two independent components off-shell, similarly to a $2d$ CFT \cite{Guica:2019nzm}. 

Last but not least, let us point out that the  general form of the solution \eqref{solnhamttb} can also be obtained by assuming the deformed Hamiltonian is a function of just the undeformed one and the momentum, and requiring consistency of the equal-time Poisson brackets \cite{Guica:2022gts}.  This entirely fixes the form of \eqref{solnhamttb}, up to a constant shift, and the $T\bar T$ flow equation is obeyed automatically, upon correctly identifying the non-trivial arbitrary constant in the solution with $1/\mu$.  %\textcolor{blue}{ \emph{Can one relax this requirement for more general QFTs, e.g. by making dependence on also $T_{\s\s}^0$?}} 
It would be interesting if  this analysis could be extended to seed field theories where the trace of the stress tensor does not vanish. 

%More precisely, one can show that in any classical $T\bar T$ - deformed CFT (\emph{Does one need the Hamiltonian formulation to show this?}) there are only two independent components of the stress tensor off-shell.

\bigskip

\noindent \textbf{\emph{III. Scattering and gravitational dressing}}

\medskip

\noindent  Let us now turn back to the quantum theory. In integrable QFTs, $2 \r 2$ scattering is characterised by a phase, which is related to the ground state energy in finite size  via the Thermodynamic Bethe Ansatz (TBA) equations \cite{Zamolodchikov:1989cf}. Since the $T\bar T$ deformation induces a universal change in the energy levels of the finite-size system and preserves integrability if initially present, one would expect  the universal modification of the spectrum translates to a universal modification of the scattering phase. This is indeed the case, and turns out to be true beyond the realm of integrable QFTs.

To have a well-defined S-matrix in the first place, one should consider either a QFT containing massive particles, or the scattering of non-interacting massless ones \cite{Fendley:1993jh}. Much of the early work on this subject  \cite{Dubovsky:2012sh,Dubovsky:2012wk} was performed for the latter case, where the problem of scattering  $T\bar T$ - deformed free bosons can be mapped to the scattering of worldsheet excitations
around a long string background. Despite the simplicity of the system, this work has gone a long way to clarify the physical meaning of the  $T\bar T$ scattering phase and how it reflects the non-locality of the underlying theory: its basic physical manifestations are  a time delay in scattering proportional with the energy and  the existence of  a minimum length, both of which point towards a (quantum)-gravitational  interpretation of $T\bar T$ - deformed QFTs. The same works investigated whether the relation between $T\bar T$ and strings survives at the quantum level, %the extent to which the SZ prescription \eqref{ttbdefintro} provides a definition of the \emph{quantum} $T\bar T$ - deformed theory, etc,
 and    paved the way  for developing i) a non-perturbative definition of general $T\bar T$ - deformed QFTs in terms of coupling the original QFT to topological gravity and ii)  the relation between  $T\bar T$ and the effective  theory of  e.g. the QCD string. The latter research strand has not only proven to be a successful approach for matching lattice QCD data, but also shows that the study of $T\bar T$ - like theories is  relevant not just for abstract theoretical considerations, but also for the study of Nature.

 In this subsection, we address these issues roughly in the order mentioned above.

\subsubsection{Extracting the scattering phase from the $T\bar T$ - deformed free boson spectrum}

The free boson theory is a trivial example of an integrable theory - in particular, it has an infinite number of conserved charges, roughly given by $\int d\s (\p \phi)^n $. Since, as we have seen above, the $T\bar T$-deformed free boson is related to the undeformed free boson by a simple change of gauge, the deformed theory is also integrable; else, one can use the general arguments of \cite{Smirnov:2016lqw} to show integrability.  %\emph{Can we show this explicitly?} 
%\footnote{It is in fact possible to show, on very general grounds, that the $T\bar T$ deformation always preserves integrability \cite{smzam}.}. 
This can be also seen explicitly from the exact finite-size spectrum  \eqref{defengcftDs} of $D-2$ $T\bar T$ - deformed free bosons %\textcolor{red}{\emph{Check!}}
\be
E (N, \tilde N,R)=  \sqrt{ \left( \frac{R}{2\mu}\right)^2 + \frac{2\pi}{\mu} \left(N+\tilde N - \frac{D-2}{12}\right) + \left(\frac{2\pi (N-\tilde N)}{R}\right)^2 } - \frac{R}{2\mu} \label{entnr}
\ee
where $N, \tilde N$ are the levels for the left/right-movers\footnote{The same formula is obtained by quantizing a bosonic string  in $D$ target space dimensions in lightcone gauge, in the winding one sector \cite{Dubovsky:2012wk}; the exact relation between the two systems will be discussed in the next subsubsection.}. Since  states with a definite particle number are exact energy eigenstates, one concludes there should be no particle production in scattering. %\textcolor{red}{\emph{What if one considered strings with higher winding?}} % \textcolor{blue}{\emph{Level?}}

Consider now %, following  \cite{Dubovsky:2012wk}, 
the $2\r 2$ scattering of these bosons, taken to be identical for simplicity. Before the $T\bar T$ deformation, the S-matrix is simply one, and can be defined despite the absence of a mass gap because the particles do not interact. Since the deformed CFT is integrable, the $2\r 2$  S-matrix will be given by a phase, $S= e^{i \d(p_i)}$, where $p_i$ are the particles' momenta, and this phase  can be related to the ground state energy in finite-size, $E_0(R)$, via the TBA equations mentioned above. We will not use  the full TBA equations herein but, following \cite{Dubovsky:2012wk},  we will use a simpler argument to find the effect of the deformation on the S-matrix. 

%Remember that  the energy $E (N,\tilde N,R)$ of $N$ left-moving and $\tilde N$ right-moving  $T\bar T$-deformed free bosons  on a cylinder of circumference $R$ is given by  \eqref{entnr}. 
Consider a two-particle eigenstate of the Hamiltonian on the cylinder (e.g., $a^\dag_{-\mbox{\tiny{$N$}}} \tilde a^\dag_{-\mbox{\tiny{$ N$}}} |0\rangle$) with zero total momentum, so $\tilde N = N$. In the Schr\"{o}dinger picture, the time evolution of this state is
\be
|N,N,t \rangle = e^{-i E(N,N,R)\, t} |N,N,0\rangle
\ee 
The same state can be thought of as describing two massless particles, one left-moving and one right-moving, which circle around the cylinder in opposite directions and interact every $\Delta t = R/2$. At each interaction,  the two-particle wavefunction picks up a phase shift    $e^{i \d (p_i)}$, which can be taken to be the flat space scattering phase if $R$ is very large. It will be convenient  to work in terms of the Mandelstam variable  
\be
s = (p_1 +p_2)^2 = E_{cm}^2 = \frac{16\pi^2 N^2}{R^2}
\ee
The scattering phase only depends on the external momenta  $p_i$ though $s$ due to the special $2d$ kinematics. From the point of view of the two particles, the state at time $t\gg R$ can be written as
 
\be
|N,N,t \rangle = e^{-i (\Delta E(N,0,R)+\Delta E(0,N,R)+E(0,0,R)) t +  i \d(s) 2 t/R} |N,N,0\rangle
\ee 
where $\D E (N,\tilde N, R)$ is the energy of the respective state with respect to the vacuum, whose energy is $E(0,0,R)$. Equating the two exponents, we find that 

\be
 \d(s) = \lim_{R\r \infty} \frac{R}{2} (\Delta E(N,0,R)+\Delta E(0,N,R)+E(0,0,R)-E(N,N,R))
\ee
i.e., the scattering phase is proportional to the binding energy of the two-particle state in the large radius limit. As we send $R \r \infty$, we would like to keep the particles' momenta, $2\pi N/R$ fixed, which yields in the following result for $\d(s)$

\be
 \d(s) = \frac{8 \pi^2 N^2 \mu}{R^2} = \frac{\mu s}{2}
\ee
Thus,  the exact  S-matrix  takes the extremely simple form 

\be
S = e^{i \mu s/2} \label{fbsmat}
\ee
This expression can also be derived using the TBA equations \cite{Dubovsky:2012wk}. %\textcolor{red}{\emph{Talk about asymptotic fragility}}

Despite its unusual form, the above S-matrix is perfectly consistent with the S-matrix axioms: unitarity, analyticity, crossing, as described in e.g. \cite{Zamolodchikov:1991vx}. More precisely, it corresponds to a so-called CDD factor  \cite{Castillejo:1955ed}, %  which in the massless case is simply a phase factor $e^{i P(s)}$, \textcolor{red}{\emph{Not correct!}} with $P(s)$   an odd polynomial. More 
which generally takes the form 

\be
\Phi_{\{\mu\}} (\Delta \b) = \exp \left( - i \sum_{ k \in 2\mathbb{Z} +1} \mu_k \sinh (k\, \Delta \b)\right) \label{cdd}
\ee
where $\Delta \b$ is the relative rapidity of the two particles being scattered; see \cite{Zamolodchikov:1991vx} for the corresponding massless expression, which involves a phase that is an odd polynomial in $s$.  CDD factors represent ambiguities that arise in  the $S$ - matrix bootstrap approach, since they can be added (in a multiplicative fashion) to a consistent $2 \r 2$ scattering amplitude without violating the analyticity, unitarity and crossing symmetry conditions.  In general, such terms are set to zero in order to avoid that  the S-matrix  behave  non-polynomially   as $s \r \infty$, which is associated with non-locality of the corresponding QFT. 

In the $T\bar T$ case, one is  interested in precisely  the non-local behaviour associated with the CDD phase factor, given by \eqref{fbsmat} for free bosons.  In the sequel we show that, indeed, a theory with such an S-matrix  
%
% Its peculiar, non-polynomial behaviour as $s \r \infty$ indicates that the S-matrix does not correspond  to the S-matrix of a local QFT. This will be clearly seen in the next section, where we show the theory
 %
  exhibits a minimum length that can be probed in scattering.

\subsubsection{Physical manifestations of the scattering phase \label{physmanif}}

%We now go back to the physics of the deformation. We discuss for simplicity the free boson case, but it  is of course generalized.  \emph{Refer also to time delay discussion from previous subsection.}

In this subsection, which closely follows \cite{Dubovsky:2012wk}, we discuss two important physical effects of the scattering phase, still in the context of the $T\bar T$-deformed free bosons:  a universal time delay proportional to the energy - already prefigured by the classical analysis leading to \eqref{clstimedelay} -  and the presence of a minimum length. %The time delay due to scattering in a classical background was already discussed in section \ref{clasttbfb}. 

To see them, we consider the scattering of two Gaussian wavepackets, one left-moving and one right-moving, with profile functions $f_L(p_L)$ and $f_R(p_R)$ given by\footnote{For  $\bar p \gg \D p$, we can replace the integration range to be $(-\infty,\infty)$, instead of $(0,\infty)$; an overbar denotes an average in this subsection.}

\be
f_L (p_L) = \frac{1}{\sqrt{\Delta p_L \sqrt{\pi}}} \exp \left( - \frac{(p_L - \bar p_L)^2}{2 (\Delta p_L)^2} \right)
\ee
and similarly for $f_R$. The \emph{in} state, prepared at $t \r - \infty$, takes the form 

\be
|in \rangle = \int_0^\infty dp_L dp_R f_L(p_L) f_R (p_R) \, a^\dag(p_L) \tilde a^\dag (p_R) | 0\rangle
\ee
where $a^\dag$ and $\tilde a^\dag$ are the creation operators for left/right-moving modes.  As $t \r \infty$, we have

\be
|out \rangle = \int_0^\infty dp_L dp_R f_L(p_L) f_R (p_R)\, e^{2 i \mu \, p_L p_R} \, a^\dag(p_L) \tilde a^\dag (p_R) | 0\rangle
\ee
since $s = 2 \, \eta^{\a\b} p_{L,\a} p_{R,\b}  = 4 \, p_L p_R$. The momentum-space reduced density matrix for the left-movers, obtained after tracing out the right-moving modes,  is given by 
\bea
\rho(p_L, p_L')& = & Tr_{RM} |out\rangle \langle out | = f(p_L) f_L^\star(p_L') \int_0^\infty dp_R |f_R(p_R)|^2 e^{2 i \mu (p_L - p_L') p_R} \nonumber \\
&= & f(p_L) f_L^\star(p_L')  \, e^{2i \mu (p_L-p_L') \bar p_R - \mu^2 (p_L-p_L')^2 \Delta p_R^2} \label{pspacedm}
 \eea
The probability density to find the left-moving wavepacket %(which, in the particle limit, travels at $x +t = const$)
 at some $x_L$ (large and negative) for $t$ large%(i.e., after the scattering)
, is given by the Fourier transform of the above expression %\emph{Re-check!}
\be
\rho(x_L,x_L,t) = \int d p_L e^{-i p_L (t+ x_L)} \int d p'_L e^{-i p'_L (t+ x_L)} \rho(p_L, p_L')  \approx e^{- \frac{(t+x_L - 2 \mu \bar p_R)^2}{(\Delta x_L)^2}}
\ee
where $\Delta x_L$ is given by 
\be
(\Delta x_L)^2 = \mu^2 (\Delta p_R)^2 + \frac{1}{4 (\Delta p_L)^2}  \label{widthscatttb}
\ee
This answer  has several rather interesting features. The first feature is the time delay: if in absence of the deformation, the probability would have peaked at $t \approx - x_L$, in its presence  it peaks \emph{later} by an amount $2 \mu \bar p_R$, proportional to the energy of the right-moving wavepacket.  This is very  reminiscent of gravitational scattering, which %typically/universally
 also produces time delays proportional to the energy. It is the quantum version of the effect \eqref{clstimedelay} we have seen earlier at classical level, which was induced by the field-dependent coordinate transformation. The reason  this coordinate transformation can affect the S-matrix is that it is a `large' coordinate transformation, which acts non-trivially  on the asymptotic states of the theory. 

   %\textcolor{blue}%{\emph{The non-trivial S-matrix is due to non-trivial action of the field-dependent coordinate transformation  on asymptotic states/observers}}. 

The second very interesting effect is the existence of a minimum length. As we see from \eqref{widthscatttb}, the width of the wavepacket after the scattering is given by $\Delta x_L$, and there will be a similar expression for the post-scattering width
$\Delta x_R$ of the right-moving wavepacket. The two satisfy the inequality

\be
\Delta x_L \Delta x_R \geq \mu
\ee
which was nicknamed a `stringy uncertainty principle' in \cite{Dubovsky:2012wk}, and indicates that it is  impossible to resolve lengths smaller than $\sqrt{\mu}$. %\emph{Which experiment?}
 A third interesting effect one can already notice  from the momentum-space expression \eqref{pspacedm} is that the off-diagonal matrix elements of the density matrix are suppressed, implying that the outgoing wavepackets are  highly entangled with each other after the scattering. In \cite{Dubovsky:2012wk}, this has been likened with a very toy version of black hole evaporation. 

Thus,  even though the S-matrix of $T\bar T$ - deformed free bosons is an extremely simple phase, it encodes some surprisingly  rich physics, which shows that the $T\bar T$ deformation produces a non-local  theory with certain gravitational properties. Whether this should be considered as a theory of quantum gravity in $2d$ is unclear, not least because gravity in $2d$ has no propagating degrees of freedom.  %\textcolor{blue}{\emph{Could the lack of sum over geometries be responsible? How to show it?}}
 Before moving on to more general $T\bar T$ - deformed QFTs and exploring their link to $2d$ gravitational theories, we would like to briefly review what happens to the link between  $T\bar T$ - deformed free bosons and the worldsheet theory of the bosonic string at the quantum level. 

\subsubsection{ $T\bar T$ - deformed free bosons vs. strings in flat space: quantum analysis}

As we saw in the previous subsection, $D-2$  $T\bar T$-deformed   free bosons are classically equivalent to  the  Nambu-Goto action in $D$ target space dimensions in  static gauge. In the quantum theory, a few interesting and interrelated questions arise:

\bi
\item[i)] the above identification implies that the $T\bar T$ - deformed free boson theory  possesses  non-linearly realised target space Poincar\'e symmetry, even though only an $ISO(1,1) \times SO(D-2)$   subgroup is manifest. What happens to this symmetry upon quantization?
\item[ii)] the $T\bar T$ - deformed free boson theory is UV-complete, as indicated by the exact S-matrix \eqref{fbsmat}. This fact is not obvious from the string perspective, as  the static-gauge Nambu-Goto action  na\"{i}vely appears to correspond to a non-renormalizable theory, due to the presence of  an infinite number of higher derivative terms. Of course, at least in the critical dimension $D=26$, this action describes the bosonic string and is thus  UV-complete, but how about other dimensions?
%\item[iii)] in the EFT approach to  e.g. the QCD string,  Nambo-Goto  is the first term in the effective action, followed by an infinite series of higher-derivative interactions, all respecting the target space Poincar\'e symmetry. In the planar limit of the underlying gauge theory, this EFT applies in a large range of energies \cite{Dubovsky:2015zey}, and one may wonder whether its leading UV behaviour is controlled by an integrable, asymptotically fragile theory such as  $T\bar T$ \textcolor{red}{\emph{Correct? }} 
\ei
The question  whether the %symmetry survives in the full quantum theory. Since strings in flat space must realise the target space symmetries, this question is equivalent to asking whether the
identity between $T\bar T$ - deformed free bosons and strings in flat space survives at the quantum level  was effectively addressed in   \cite{Dubovsky:2012sh}.   From the $T\bar T$ perspective, a failure of this relation would be seen as a breaking of the non-linearly realised $ISO(1,D-1)$,  while integrability and  the scattering amplitude \eqref{fbsmat} are preserved; from the  perspective of the string - which must realise the target space isometries - it would be seen as a departure from \eqref{fbsmat}, i.e.  a  breaking of integrability.

The authors studied the scattering of worldsheet excitations on an infinitely long string in $D$ target space dimensions perturbatively in $\a'=\ell_s^2$, taking an effective field theory approach.  While this brute-force calculation  becomes
quickly   tedious\footnote{It has been carried up to four loops in \cite{Conkey:2016qju}. }, and one can only reproduce a few terms in the expansion of the scattering phase \eqref{fbsmat}, the usefulness of this exercise is that it renders more transparent the
various assumptions at play, the structure of divergences one encounters and the possible choices of
counterterms needed to subtract them, the role played by integrability in the definition of $T\bar T$, and what is special about the
critical dimension.

At low  energies, the relevant degrees of freedom correspond to the  $D-2$  Goldstone bosons, $X^i$, associated to the translational symmetries broken by the presence of the string.  The first  term in the effective action is the Nambu-Goto area term; in static gauge about the long string background, one is thus studying the scattering of $D-2$ massless bosons,  with manifest $SO(D-2)$ `flavour' symmetry,  in this na\"{i}vely non-renormalizable theory.
%
 %found that the two theories were equivalent only in the critical dimension, $D=26$. 
%
% Conversely, the $T\bar T$ deformation has been shown \cite{} to preserve integrability, so a failure of string scattering - as computed from the  perturbative expansion Nambu-Goto action plus counterterms, which can be dealt with in gory detail - to satisfy integrability (in particular, the absence of particle production) signifies a breakdown of the equality between $T\bar T$ and the NG string.
%It is the latter route that we will be taking here.  We will take $D$ to be arbitrary. The static gauge Nambu-Goto action has
   The amplitude for scattering particles $i,j$ into $k,l$  takes the general form 
\be
\mathcal{M}_{ij \r kl} = A \d_{ij} \d_{kl} + B \d_{ik} \d_{jl} + C \d_{il} \d_{jk}\;, \;\;\;\;\;\;\;i,j,k,l\in \{1,\ldots ,D-2\} \label{mijkl}
\ee
where $A$ is the amplitude for annihilation, $B$ for transmission, and $C$ for reflection. They are functions of the Mandelstam variables $s,t,u$ and the string length $\ell_s$, and are related by crossing. Due to a peculiarity of $2d$ kinematics, one has either $t=0$ with $u=-s$, or $u=0$ with $t=-s$.

%The answer is generally no, except if $D=3, 26$.

Taking the `$T\bar T$ perspective', we note that 
%
% The rough argument is that 
 %
  the formula \eqref{entnr}, which represents the exact spectrum of $D-2$ $T\bar T$-deformed free bosons, indicates that states with the same level, but different $SO(D-2)$ quantum numbers, are exactly degenerate. The interaction Hamiltonian is thus proportional to the identity on the degenerate subspace, from which one concludes that the annihilation part, $A$, of the amplitude  should be zero. %\textcolor{red}{\emph{Correct?}} 
  In \cite{Dubovsky:2012sh}, the coefficients $A,B,C$ above were computed up to one loop, using the derivative expansion of the Nambu-Goto action only % (up to quartic order). 
%
% S-matrix for $2\r 2$ scattering in the Nambu-Goto theory has been computed by brute force up to one loop. One starts by expanding the Nambu-Goto action in powers of the dimensionful coupling $\mu$ up to the desired order
%
%\be
%L_{NG} = -\frac{1}{2l_s^2}\int d^2\s \left[ (\p_\a X^i)^2 + \frac{1}{4} (\p_\a X^i \p^\a X^i)^2 - \frac{1}{2}  \p_\a X^i \p_\b X^i \p^\a X^j \p^\b X^j + \ldots \right]
%\ee
and keeping track of divergences via dimensional regularization. %The only divergence that appeared could be absorbed by an evanescent counterterm $\int R\sqrt{h}$, i.e. one that vanished in $d=2$, but is non-zero in $2-\e$ dimensions. 
It was found that  %\emph{Careful!}
\be
A_{tree} =0 \;, \;\;\;\;\; B_{tree} = \frac{l_s^2}{2} s^2\;,\;\;\;\;\;\; A_{1-loop} = - \frac{D-26}{192\pi} l_s^4 s^3 \;, \;\;\;\;\;\; B_{1-loop} = i \frac{l_s^4}{16} s^3 
\ee 
while $C$ can be determined by crossing. 
Precisely in the critical dimension, $D=26$, the annihilation part of the amplitude vanishes and the Nambu-Goto action plus counterterms can %(and does)
 equal the $T\bar T$ - deformed bosons. The case $D=3$ is special, as  the index $i$ only takes one value, so it only makes sense to talk about the analogous contribution to the full amplitude $A+B+C$, which does vanish. One can easily check that the transmission  part of the amplitude, $B$, precisely equals the expansion of \eqref{fbsmat} with $\mu = \ell_s^2/2$ up to this order\footnote{Taking into account normalisations, the relationship between the two is $e^{2 i \d(s)} = 1 + i B/(2s)$.}. 

In all other dimensions, in order to have $A=0$, one needs to explicitly add to the action a counterterm that cancels the contribution of the so-called Polchinski-Strominger (PS) term \cite{Polchinski:1991ax}. To the given order in static gauge, this counterterm takes the form %\textcolor{red}{\emph{What does it correspond to in the covariant expansion? Or, does it come from the measure? Where is the $K_{\a\b} K^{\a\b}$ term?}}
\be
S_{PS} = - \frac{D-26}{192\pi} \int d^2\s \, \p_\a \p_\b X^i \p^\a\p^\b X^i \,\p_\g X^j \p^\g X^j \label{psctterm}
\ee
which explicitly breaks the non-linearly realised Lorentz symmetry. Thus, we conclude that the non-linearly realised Poincar\'e symmetry of the classical $T\bar T$ - deformed action is \emph{not} preserved at the quantum level, except for $D=3,26$.  In all other dimensions,
the $T\bar T$ deformation of $D-2$ free bosons  is just an integrable theory, whose exact S-matrix is given by \eqref{fbsmat} and is associated, via integrability, to the exact spectrum \eqref{entnr}. The action that perturbatively reproduces this amplitude is 
  $S_{NG} + S_{ct} + S_{PS}$, which also contains the higher order analogues of the Lorentz-breaking PS term required to cancel the Nambu-Goto contribution to the annihilation amplitude up to the desired order. % The finite part of the counterterms is fixed by invoking integrability, i.e. match to \eqref{fbsmat}.  
 As for the counterterms,  one needs ever new ones as the loop level is increased (an infinite number of them), and the only rule that appears to fix their finite part is to require integrability, i.e. that the scattering amplitude they predict  match the result \eqref{fbsmat} predicted from integrability.  %\textcolor{red}{\emph{Correct? Doesn't $T\bar T$ definition help?}} 
 In principle, one could ask whether the SZ definition  \eqref{ttbdefintro} can be used to perturbatively build the action using the renormalised stress tensor (in some chosen scheme) at the previous order; to our knowledge, this question has not been addressed in the literature. % \textcolor{red}{\emph{Correct?}}
 
%\textcolor{red}{ Note that even in $D=3,26$ one needs integrability to fix counterterms. Thus, theory defined by dressing, rather than by action.} Note that even in the critical dimension, one needs counterterms 

Thus, one sees that integrability %\textcolor{blue}{(or, possibly, just the $T\bar T$ definition, if properly understood)}
 provides a guide %principle %/\textcolor{blue}{an algorithm}
  for how to incorporate ever higher-derivative terms to the $D-2$ free boson action in order to produce a UV-complete theory, albeit one with a non-standard UV behaviour indicative of short-range non-locality. %, termed asymptotic fragility. 
 While integrability %\textcolor{blue}{(or $T\bar T$)} 
 appears essential for being able to define the theory, it is an interesting question whether this UV behaviour can be maintained while slightly breaking the integrable structure. %, especially in view of  the following point. 

Requiring integrability for any $D$ at the expense of the target space Lorentz symmetry is natural from the $T\bar T$ point of view. However,  one is sometimes interested in an effective string description of, e.g.,  the QCD string, for which the Nambu-Goto action provides the departure point. In that case, one would instead like to preserve the non-linearly realized Lorentz symmetry, %(which fixes counterterms to be built from curvature invariants) 
%at the expense of integrability, 
and therefore does not add the $PS$ counterterm  \eqref{psctterm}. % to the action. 
Assuming approximate integrability, the effective action can  be inferred, to low orders, by comparison to lattice QCD data on the finite-size spectrum %, and  is still quite constrained by approximate integrability 
 \cite{Dubovsky:2013gi,Dubovsky:2014fma,Dubovsky:2016cog}; interestingly, evidence is found for an  additional light, but massive excitation on the string worldsheet. 
  %See \cite{} for developments and related works for developments along this line. 
% , as e.g. non-polynomial contributions to the amplitude only start appearing at three loops \cite{Conkey:2016qju}. % This paper also contains an attempt to discuss asymptotic fragility outside the integrable context. % (start from free-PS, and then dress??). 
%\emph{Shorten! Discuss applications to effective/QCD string and Poincar\'e vs. integrability.  Also write down explicitly scattering phase for the free boson. }  
%
An interesting %and phenomenologically relevant 
question is whether such a theory, which is not exactly integrable, may nonetheless exhibit asymptotically fragile behaviour at high energies, as suggested by the very nature of the underlying string. This question is most meaningfully formulated in the planar limit of the %underlying
 gauge theory, where the string worldsheet description applies in a large range of energies \cite{Dubovsky:2015zey}. %, and one may wonder whether its leading UV behaviour is controlled by an integrable, asymptotically fragile theory such as  $T\bar T$ \textcolor{red}{\emph{Correct? }} 

In the following, we return to  $T\bar T$  and  address the obvious question whether more general $T\bar T$ - deformed QFTs  also exhibit a universal asymptotic behaviour for their S-matrix. If the original QFT is integrable, then this question can be approached using TBA. In the general case, one needs to resort to an alternate, non-perturbative definition of   $T\bar T$ - deformed QFTs  put forth in \cite{Dubovsky:2017cnj,Dubovsky:2018bmo}.

\subsubsection*{Universal $T\bar T $ deformation of an integrable S-matrix via TBA}

%As we saw in the first section, the $T\bar T$ deformation induces a \emph{universal} change in the finite-size spectrum of the deformed QFT, captured by the Burger's equation. 

The Thermodynamic Bethe Ansatz (TBA)  reasoning that allowed us to derive the scattering phase for the $T\bar T$ - deformed free bosons using their known spectrum can be easily generalised to arbitrary $T\bar T$ - deformed integrable QFTs. To have a general, well-defined S-matrix, we concentrate on massive QFTs. % and start with the brief review of the  TBA equations.
%

%For simplicity, we take the mass of all scattered particles to be the same $m$. 

%In integrable QFTs, the finite-size spectrum is related to the scattering phase via the Thermodynamic Bethe Ansatz (TBA) equations. Since, as shown in \cite{smzam},  the $T\bar T$ deformation preserves integrability if initially present, we can hope to relate the universal modification of the spectrum to a universal modification of the scattering phase. Below, we briefly sketch the argument. 

Consider the scattering of massive particles with mass $m$  (taken to be the same, for simplicity) in an integrable QFT. It is convenient to write the particles' momenta in terms of the rapidities, $\b$ %\emph{Position index?}
\be
p^0_i = m \cosh \b_i \;, \;\;\;\;\;\; p^1_i = m \sinh \b_i
\ee
In the integrable case, the $2\r 2 $ S-matrix is a pure phase, $S= e^{i\d(\b_i)}$.   As beautifully explained in \cite{Zamolodchikov:1989cf}, 
this scattering phase is related to the finite-size spectrum of the theory via the so-called TBA equations, which we now briefly review.

% Let us now very briefly review the TBA approach. As beautifully explained in \cite{Zamolodchikov:1989cf}, the derivation of these equations proceeds in two steps. 
 
As a first step, one 
considers the partition function of the $2d$ QFT on a torus of size $(L,R)$, where $L\gg R$. Depending on which torus direction is interpreted as 
euclidean time, this partition function can be approximated as either the ground state energy in finite volume, or as the finite-temperature free energy in infinite volume %\textcolor{red}{\emph{Notation!}}
\be
Z(L,R) \approx e^{-L E_0(R)} \approx e^{-R f(R) L}
\ee   
which leads to an equality between the two quantities in the exponent. 

%Since the mirror theory lives in approximately infinite volume, one has a well-defined notion of asymptotic states. The second step of the derivation consists in doing statistics over a large number of  (mirror) particles that scatter,  and find the minimum of their free energy, subject to the constraint that the momentum of each particle, $i$, satisfies a quantization condition of theform

In the second step, 
the free energy  is estimated by noting  that in approximately infinite volume, one has a well-defined notion of asymptotic states, and doing statistics over a large number of particles that scatter. The scattering phase $\d(p_i,p_j)$ enters in the quantization condition for their momenta as 

\be
m  L \sinh \b_i + \sum_{j \neq i}  \d(\b_i,\b_j) = 2 \pi n_i \;, \;\;\;\;\; n_i \in \mathbb{Z} \label{betheansatz}
\ee
 In the limit of a large number of particles, it is convenient to introduce the particle and level densities, which are constrained by the above  equation. The free energy is computed using usual thermodynamic considerations and then minimized. Specializing to the `bosonic' case, at the minimum one finds

\be
 E_0 (R) = Rf(R) = \frac{m}{2\pi} \int d\b \cosh \b \ln (1-e^{-\varepsilon(\b)}) \label{engtba}
\ee
where the `pseudoenergy' $\varepsilon(\b)$ is constrained to obey %\textcolor{red}{\emph{Check factor $R$! Derivation?}}

\be
\varepsilon(\b) = m R \cosh \b + \frac{1}{2\pi} \int d\beta' \frac{\p \d (\b',\b)}{\p \b'} \ln (1- e^{-\varepsilon(\b')}) \label{pseudoeng}
\ee
This provides the advertised TBA  relation between the finite-size ground state energy and the scattering phase. This relation can also be generalised to excited states \cite{Dorey:1996re}.

Let us now turn to the $T\bar T$ deformation of an integrable QFT.  The scattering phase $\d^{[0]} (\b',\b)$  of the undeformed QFT satisfies the TBA equations for some pseudoenergy $\varepsilon^{[0]}$. Let us    assume, for simplicity,  that the momentum $P = m \int d\b \sinh \b \ln (1-e^{-\varepsilon(\b)}) =0 $. %\textcolor{red}{\emph{Can we relax?}}
After the deformation, the pseudoenergy takes on a new value, determined by  the change in the scattering phase, which in turn determines the modification of the ground state energy. Since the latter satisfies $E_0^{[\mu]}(R)= E_0^{[0]} (R+\mu E_0^{[\mu]})$, we see that the deformed pseudoenergy should satisfy \eqref{pseudoeng} with the same $\d^{[0]}(\b',\b)$, but with the radius shifted as 
 $R \r R + \mu E_0^{[\mu]}$.  $E_0^{[\mu]}$ is given by \eqref{engtba}, computed  with the deformed pseudoenergy.   From the explicit expression \eqref{pseudoeng}, it is easy to see that the shift in the radius can be traded for a shift in the scattering phase that satisfies  $\p_{\beta'} \Delta \d(\b',\b) = \mu m^2 \cosh \beta' \cosh \b$, up to terms proportional to $\sinh \b'$ that integrate to zero, by assumption. This implies that the scattering phase in the deformed theory takes the form

\be
\d^{[\mu]}(\b_i)= \d^{[0]}(\b_i) + \Delta \d(\b_i)\;, \;\;\;\;\; \Delta \d (\b',\b) = \mu m^2 \sinh (\b'-\b)  = \mu \, \e_{ab} \, p^a p'^b \label{scattph}
\ee
with $\e_{01} =1$. 
Thus, the $T\bar T$ deformation leads to a universal shift in the scattering phase, given by \eqref{scattph}, which is linked to the universal $T\bar T$ deformation of the spectrum as we just explained. 
%
%where $\d_{QFT}$ is the scattering phase of the undeformed QFT and $\d_\mu$ is the additional shift due to the $T\bar T$ deformation. 
%
%Let us now turn to the $T\bar T$ deformation. Using the expression  \eqref{scattph} for the  scattering phase, it is easy to  see that the solution for $\varepsilon(\b)$ in the deformed QFT is related to $\varepsilon(\b)$ in the undeformed one by the formal replacement $R \r R + \mu E$. Remember from section \ref{defqftttb}~ that the finite-size energies of the deformed theory (with momentum set to zero, for simplicity), could be obtained from the finite-size energies of the undeformed QFT by replacing $R \r R + \mu E (\mu,R)$. %We immediately notice that for $\d (\b',\b) = \mu\,  m^2 \sinh (\b-\b')$ \emph{Is this the same as $\mu \, s$?}, the equation  for $\varepsilon(\b)$ is related to the one in the undeformed theory by a shift of the radius $R \r R- \mu E$.
%This establishes the link between the scattering phase \eqref{scattph} and the deformed spectrum. 
 A generalization of the TBA argument to non-zero momentum and to excited states is also possible \cite{Cavaglia:2016oda}, with the same result for $\Delta \d$.% the scattering phase.  %\emph{Write explicitly the result for the S-matrix.}
%
%\textcolor{blue}{\emph{Are the above valid just for $T\bar T$, or all (scalar) SZ deformations?} }

%
%\bi
%\item is it true that this is the unique answer that follows from integrability + $2\r 2$ answer?
%\ei

One recognises that the scattering phase \eqref{scattph} corresponds to the CDD factor \eqref{cdd} with only the $k=1$ term non-zero. Thus, also in the massive case, we find that the S-matrix is multiplied by a factor that is fully consistent with the S-matrix bootstrap equations. Assuming the original QFT was UV-complete  (with the UV behaviour governed by a CFT), the exact expression for the $T\bar T$ - deformed S-matrix  indicates the deformed theory will also be UV-complete,  albeit with a non-local   behaviour in the UV.  For an arbitrary factorisable scattering process, the phases add up, and the deformed S-matrix reads  %\textcolor{red}{\emph{Check!}}
%
 %The change in the S-matrix can also be written as  \textcolor{blue}{\emph{Shall I write this for arbirary \# of particles?}}
%
\be
\mathcal{S}^{[\mu]}(p_i) = e^{i \mu \sum_{i<j} \e_{ab} p_{i}^a p_{j}^b} \, \mathcal{S}^{[0]} (p_i) \label{genttbsmat}
\ee
where we have taken all particles to be incoming, and ordered according to their rapidities. 
%making the analogy with  the phase in \eqref{} manifest.  The effect on an arbitrary S-matrix follows. 

Therefore, we see that, in integrable theories, the effect of the $T\bar T$ deformation on the S-matrix is to simply multiply it by the above universal phase factor.
 It is  interesting to ask what is the effect of the $T\bar T$ deformation on the S-matrix of an arbitrary  $2d$ QFT, which need not be integrable. Of course, one can simply define \cite{Dubovsky:2013ira} the deformed S-matrix via the dressing procedure above but, in absence of integrability, its relation to the $T\bar T$ - deformed QFT is unclear. In the following, we describe an alternate definition of the  $T\bar T$ deformation that  is able to reproduce both the universally-deformed spectrum and the universal modification of the S-matrix, and applies to general QFTs.

%Note that in the regime where the scattering phase becomes important, the regime of validity of the SZ argument appears to fail. 

\subsubsection{Non-perturbative definition of the $T\bar T$deformation}

The definition \eqref{ttbdefintro} of the $T\bar T$ deformation discussed at the beginning of this section has the advantage of being based on an action, and allows one to compute the exact spectrum of the deformed QFT; however, this definition only makes sense at distances much longer than the non-locality scale $\sqrt{\mu}$, where the QFT can be treated quasi-locally and one is able to associate a  stress tensor  to the translational symmetries. 

The fact that the deformed QFT could actually be UV complete was instead seen, at least for the case of integrable QFTs, from the exact expression \eqref{genttbsmat} for the S-matrix (\eqref{fbsmat} in the massless case), which is well-defined up to arbitrarily high energies. Given this, one may be lead \cite{Dubovsky:2013ira} to provide an alternative definition of the $T\bar T$ deformation, based on its effect on the S-matrix  \cite{Dubovsky:2013ira}. 
%
%\be
%S_\mu(p_i) = e^{i \mu \sum_{i<j} \e^{ab} p_{i a} p_{j b}} S_{QFT} (p_i) \label{genttbsmat}
%\ee
%where  the particles are naturally ordered according to their rapidities. This expression is the natural generalization of the scattering phase to massive particles. 
This S-matrix-based definition makes the UV completeness manifest; it nonetheless has the drawback of  not being based on an action principle.

In the following, we briefly sketch how these two approaches can be unified via a general, path-integral definition of the $T\bar T$ deformation of \emph{arbitrary} QFTs on flat space, which is able to reproduce both the deformed spectrum and the deformed S-matrix. 
%
%It is interesting to ask whether the action-based  definition of $T\bar T$ valid at large scales and the S-matrix-based non-perturbative definition \eqref{genttbsmat} can be unified  also for non-integrable QFTs.  The $T\bar T$ deformation can be
 %non-perturbatively defined by a
 This was achieved via a particular coupling of the original quantum
field theory to a special theory of topological gravity \cite{Dubovsky:2018bmo}
%
%Since the SZ definition of the $T\bar T$ deformation holds for arbitrary QFTs, we would like to understand whether the change \eqref{chsmat} in the S-matrix is universally valid, too, i.e. also for non-integrable QFTs. So far, we have only shown this for integrable QFTs, where the two are related via the TBA equations. One could of course \emph{define} the $T\bar T$ deformation at the level of the S-matrix via \eqref{chsmat}, as was done in \cite{Dubovsky:2013ira}, but it is not clear whether this definition corresponds to the same theory, and it is also unsatisfactory because it is not a definition coming from an action.
%
% This  was achieved in \cite{Dubovsky:2018bmo}, who proposed a path integral definition of the $T\bar T$ deformation %was given\footnote{This definition strictly \emph{True?} holds for $T\bar T$-deformed theories  on flat manifolds. It is not clear whether the $T\bar T$ operator can be unambiguously defined on curved manifolds, though a similar definition does work in the classical large $N$ limit. }
%
\be
Z_{T\bar T}[\mu] = \int \mathcal{D} X^a \mathcal{D} e^a_\a \exp \left[- \frac{1}{2\mu} \int d^2 x \, e \, \e_{ab} \, \e^{\a\b} (\p_\a X^a -  e^a_\a) (\p_\b X^b - e^b_\b) + S_{QFT} (e^a_\a) \right] \label{ttbpathintdef}
\ee
where $e^a_\a$  is a nontrivial background vielbein and the $X^a$ are a pair of auxiliary fields whose equations of
motion set $\p_{[\a} e^a_{\b]}=0$, implying that the curvature is exactly zero. This is important for the quantum
theory and is ultimately responsible for the solvability of this deformation, as the semiclassical sum over
topologies only receives a contribution from the torus (in the compact case), with the higher-genus Riemann surfaces being
precluded by the Gauss-Bonnet theorem. % The above  can be thought of as coupling the original quantum field theory to a  topological theory of  gravity. 
% \textcolor{red}{
 The absence  of a sum over topologies in the above definition   suggests that $T\bar T$-deformed QFTs are more like non-local $2d$ QFTs than  theories of two-dimensional quantum gravity, despite the presence of many gravitational features.

The above definition can be justified by treating the $T\bar T$ deformation  using the  Hubbard-Stratonovich method \cite{Cardy:2018sdv}, which   in this case amounts to coupling the original QFT to a dynamical metric\footnote{This formulation is also the departure point for understanding the holographic interpretation of the $T\bar T$ deformation.}. However, as shown in that work, the conservation of the stress tensor implies that the path integral reduces to one only over flat metrics, at least infinitesimally. When passing from metrics to vielbeine, a simple way to enforce this constraint is to introduce the auxiliary fields $X^a$ as above, whose equations of motion impose the flatness condition $\p_{[\a} e^a_{\b]} =0$.
%
%splitting the $T\bar T$ deformation via the Hubbard-Stratonovich trick, along the lines of what was done in \cite{Cardy:2018sdv} in the metric formalism, which introduces a coupling to a dynamical vielbein, and then showing that only non-dynamical vielbeine contribute. In the above formulation, this is ensured by the $X^a$ equations, which impose $\p_\a e^a_\b =0$, i.e. the spin connection must vanish. Thus, the integral is reduced to over only flat vielbeine.   
%
The vielbein equations of motion impose

\be
\p_\a X^a = e^a_\a + \mu \, \e^{ab} \e_{\a\b} T^\b_b
\ee
which is nothing but a generalization of the  field-dependent coordinate transformation from static $(\p_\a X^a=\d^a_\a)$ to conformal gauge ($e^a_\a = \d^a_\a$) that we have seen before in the Nambu-Goto action. 
%Taking $e^a_\a = \d^a_\a$ corresponds to conformal gauge in the string picture, whereas choosing $\p_\a X^a = \d^a_\a$ is the analogue of static gauge. The $X^a$ are dynamical coordinates. \emph{In which sense?} 
%
This equation points towards an interpretation of the $T\bar T$ deformation as providing a set of dynamical coordinates, $X^a$, through which the underlying QFT dynamics is seen \cite{dubovskytalk}.

By computing the torus partition function using the above definition, \cite{Dubovsky:2018bmo} were able to derive a flow equation for the partition function\footnote{The flow equation can also be found  using the metric formulation  \cite{Cardy:2018sdv}.}, which precisely reproduces  the flow equation \eqref{burger} on the energy levels. Concommitently, in \cite{Dubovsky:2017cnj} it was shown that the effect of coupling to the non-dynamical vielbein precisely reproduces the dressing \eqref{genttbsmat} on the S-matrix. Thus, the path integral above is indeed able to  unify the  two previous definitions of the $T\bar T$ deformation.
%definition (valid only for small $\mu$, (compared to energy scale) where the QFT can still be considered quasilocal) to the large $\mu$ S-matrix  definition, while providing an action principle. 

%It is an interesting question to what extent this definition (and the time delay, etc), makes $T\bar T$-deformed QFTs into a theory of $2d$ quantum gravity. Perhaps a test would be the existence of well-defined off-shell observables such as correlation functions, see \cite{} for recent progress in this direction. 

%
%
%\bi
%\item comment connection to massive gravity
%\item more on the S-matrix?
%\ei
%
%
%
%\subsubsection*{Does  the $T\bar T$ deformation yield a theory of $2d$ quantum gravity? }
%
%
%
%
%\bi
%\item S-matrix: gravitational time delay, min length
%\item however, one-to-one map between CFT and $T\bar T$ observables, in part well-def corr f
%\item off-shell obs: corr f $\r$ good degree of control (Cardy, Aharony, Rosenhaus?)
%\item symmetries? 
%\item is $S(R)$ extensive? $S(R,T) = \frac{c T R}{\sqrt{1-\mu T^2}}$, so ext in $R$ \emph{Is this the correct notion of extensiveness?}
%\ei
%
%
%Is there UV/IR mixing in $T\bar T$/LST? 
%
%
%

\subsection{The $J\bar T$ and $JT^a$ deformations}

The $J\bar T$ deformation  \cite{Guica:2017lia} is another very well-studied Smirnov-Zamolodchikov deformation. As its name indicates, the deforming operator is constructed from the components of a $U(1)$ current, $J$,  and those of the generator of right-moving translations, $T_{\a \bar z}$

\be
\O_{J\bar T} = \lim_{x' \r x} \left( J_z (x') T_{\bar z \bar z} (x) - J_{\bar z} (x') T_{z \bar z}  (x) \right) + ...
\ee
which only has a $\bar T \equiv T_{\bar z \bar z}$ component when considered in a CFT. 

Since the $J\bar T$ operator has spin one, the corresponding deformation breaks Lorentz invariance, which in turn implies that the deformed stress tensor is no longer symmetric.  As for all SZ deformations, the change in the action is given by 

\be
\frac{\p S}{\p \l} =  \int d^2 x \, \O_{J\bar T}(\l)  \label{jtbfldef}
\ee
We have denoted the $J\bar T$ coupling constant by $\l$, to differentiate it from the $T\bar T$ coupling. It has dimensions of length and, due to the non-zero spin of the $J\bar T$ operator, should best be thought as a vector $\l^a = \l \d^a_{\bar z}$ that points in the right-moving direction (which is null in Lorentzian signature).  One can similarly define the $JT^a$ deformation, constructed from a $U(1)$ current and the components of the translation generator in the direction $\hat x^a$, $T_{\a a}$ -  whose coupling constant is a fixed, but arbitrary vector $\l^a$  - and has very similar properties to $J\bar T$.  As in the $T\bar T$  case, these theories are expected to be UV - complete.

One special feature of the $J\bar T$ deformation that distinguishes it from the generic $JT^a$ ones is that it preserves an $SL(2,\mathbb{R})$ subroup of the original conformal group if the seed theory that is being deformed is a CFT. This follows from the fact that the $J\bar T$ operator has dimension $(1,2)$, i.e., it is exactly marginal on the left,  whereas a general $JT^a$ one contains both $(1,2)$ and $(2,1)$ contributions.  Consequently, a $J\bar T$ - deformed CFT will be \emph{local} and \emph{conformal} on the left-moving side, and all its non-locality will be concentrated along the null right-moving direction.  Note these are precisely the properties expected of a  two-dimensional \emph{dipole CFT}, introduced in section \ref{bhstoirrelsec}, which is a UV-complete  QFT that can be obtained via a series of $SL(2,\mathbb{R})_L$- invariant irrelevant deformations of a $2d$ CFT, starting with a   $(1,2)$ operator.   Our main motivation for studying $J\bar T$ - deformed CFTs is thus that

\vskip7mm

\noindent $\star$ \emph{$J\bar T$-deformed  CFTs represent the first concrete, tractable example of a two-dimensional dipole CFT} $\star$
\vskip5mm
 
\noindent As per our discussion in section \ref{bhstoirrelsec}, they consequently  represent the first solvable toy model for  the Kerr/`CFT' correspondence. Of course, this is not to mean that extremal black holes are actually described by $J\bar T$ - deformed CFTs; the claim is simply that $J\bar T$ - deformed CFTs are the first concrete example of QFTs whose structure (i.e.,  the not-yet-formalized axioms that dipole CFTs obey) may be the same as that of theories  holographically describing near-extremal black holes.  As such,   one may hope they will shed light on  some of the puzzles raised in the Kerr/`CFT' or, rather, warped AdS$_3$   context, such as the presence of Virasoro symmetry and CFT-like correlation functions in what appears to be a non-local theory.  We will show that $J\bar T$ - deformed CFTs can, indeed, offer an interesting perspective on these issues; see  section \ref{infsymmsec} and, respectively, \ref{corrfsec}. As we will see, the `half-local' and conformal structure  of $J\bar T$ - deformed CFTs    make the study of  precisely these type of  observables   much simpler than in $T\bar T$, which is a fully non-local theory.

 %This will explain how Virasoro coexists with non-locality, as well as special form of correlation functions. Nonetheless, not Cardy.   Consequently,    our   attention will mainly be concentrated on this deformation. \emph{Correct QFT type, for holography one needs to work harder.}  

In the current section, we will only  concentrate on very basic properties of $J\bar T$ and $JT^a$-deformed QFTs that  follow from their special status as Smirnov-Zamolodchikov deformations and are directly analogous to the $T\bar T$ properties we previously studied:  connection to integrability, universal deformation of the spectrum and of the S-matrix,  non-perturbative definition by coupling to topological background fields.
%
%are in one-to-one correspondence to the properties of  we have  reviewed: finite-size spectrum, thermodynamics,  integrability, S-matrix, non-perturbative definition. 
%These results are obtained through  manipulations that are directly analogous to those we used in the previous subsection for $T\bar T$ - an unsurprising fact, given that both are universal SZ deformations. 
 While in the $T\bar T$ case, there was a strong physical motivation -  the connection to (lattice QCD data  on) the confining QCD string -  to be studying the finite-size spectrum and its relation to the S-matrix via TBA,  for $J\bar T$ the most interesting physics application is to holography where, as explained, the most relevant observables are the symmetries and the form of correlation functions\footnote{The reason is that these observables  are expected to be (more) universal, whereas the finite-size spectrum is not.  }.

%Besides this connection that will be of central interest to us, $J\bar T$ and $JT^a$ - deformed QFTs have many nice properties that

 %These observables can be computed, as they were for $T\bar T$, but they are of less intrinsic interest.  

 Before we enter the details, let us make a technical remark regarding the computation of the finite-size spectrum of $J\bar T$ - deformed CFTs. 
 In the original formulation \cite{Guica:2017lia} of the $J\bar T$ deformation, the $U(1)$ current was restricted to be chiral, which was believed to lead to important simplifications. 
 However, these apparent simplifications led to a number of subtleties in the quantum analysis, due to anomalies. As it turns out, the theories defined with a chiral versus non-chiral  $J$ are in many ways equivalent, and for many purposes it is conceptually clearer to   simply take $J$  non-chiral.  In what concerns the spectrum computation, the correct general way to approach it is by studying the deformed theory in presence of constant background fields \cite{LeFloch:2019rut}, case in which there is no particular simplification to considering $J\bar T$ vs. $JT^a$, or CFT vs. QFT; we will present this computation towards the end of this subsection. For the special case of $J\bar T$ - deformed CFTs with $J$ chiral, there turns out to exist a much simpler, indirect argument for obtaining the correct spectrum \cite{Chakraborty:2018vja}, which we now present.  %\textcolor{blue}{\emph{(Can one have QFT with chiral J?)}}

\bigskip 

\noindent \textbf{\emph{I. Spectrum - take one - and thermodynamics}}

\medskip
\noindent In this subsection, we present a simple - if slightly indirect - derivation of the  spectrum of a $J\bar T$ - deformed CFT on the cylinder. The current $J$ is assumed to be chiral and the steps in the argument closely follow 
those used for deriving the  $T\bar T$ flow equation for the spectrum. A more complete and general analysis can be found at the end of this subsection. 

\subsubsection{The spectrum of $J\bar T$ - deformed CFTs on a cylinder}

As explained in section \ref{szgensec}, under a SZ deformation the energy levels $E_n(\l,R)$ of the system put on a cylinder  will be changing as %\textcolor{red}{\emph{Extra overall factor of 2!}}

\be
\frac{\p E_n^{[\l]} (R)}{\p \l} = 2 R \,\left(  \langle n | J_z | n \rangle \langle n | T_{\bar z \bar z} | n \rangle - \langle n | J_{\bar z} | n \rangle \langle n | T_{z \bar z} | n \rangle \right)\label{model}
\ee
where we simply plugged in the specific $J\bar T$ deformation into  \eqref{engflow} and replaced $\mu$ by $\l$.  As in \cite{Smirnov:2016lqw,Cavaglia:2016oda}, this will be the essential equation allowing us to compute the exact finite-size spectrum of the deformed theory, provided we can relate the expectation values of the current components in the energy eigenstate to conserved charges. 

For the stress tensor, this is a rather simple task. The expectation values of $T_{\tau\tau}, T_{\tau \s}$ and $T_{\s\s}$   in the translationally-invariant state $| n \rangle$ are, as before, related to the corresponding conserved charges as  in  \eqref{onepfep}. 
%
%\be
%\langle n | T_{\tau\tau} | n \rangle = -\frac{E_n}{R} \;, \;\;\;\;\;\; \langle n | T_{\s\s} | n \rangle = -\frac{\p E_n}{\p R} \;,\;\;\;\;\;\; \langle n | T_{\tau \s} | n \rangle = {\color{red} -}\frac{i P_n}{R} %\;, \;\;\;\;\;\; \langle n | J_{\tau} | n \rangle = \frac{ i Q_n}{ R}
%\ee
A priori, this leaves undetermined the $\s \tau$ component of the stress tensor which, due to the broken Lorentz invariance, no longer equals the $\tau \s$ one. However, this problem can be circumvented in the case of $J\bar T$ - deformed \emph{CFTs}, if one assumes that the $SL(2,\mathbb{R})_L$ symmetry is preserved at the full quantum level, which requires that %\textcolor{red}{\emph{Check!}}

\be
T_{\bar z z} = \frac{1}{4} [T_{\s\s} + T_{\tau\tau} +i (T_{\tau\s} - T_{\s \tau})] =0
\ee 
thus determining the $T_{\s \tau}$ component in terms of the rest. 
To solve the flow equation, one also needs the current components. As usual, the time component of the current is related to the conserved charge as  %\textcolor{red}{\emph{Check!}}
\be
\langle n | J_{\tau} | n \rangle = \frac{ i Q_n}{ R} \label{expjt}
\ee
but  we still need  an expression for $\langle n | J_{\varphi} | n \rangle$. There are two ways of proceeding:  one is to make the additional simplifying assumption that the current is purely holomorphic, as in \cite{Guica:2017lia,Chakraborty:2018vja}; the other, following \cite{LeFloch:2019rut,Frolov:2019xzi} is to compute $\langle n | J_{\varphi} | n \rangle$ by coupling to background gauge potential.  In this subsection, we will follow the first path, which only applies to seed CFTs, but is also much  simpler. 

  Assuming that the current is purely holomorphic, which implies $J_\varphi = - i J_\tau$ in Euclidean signature, and plugging into the flow equation for the energies, we find \textcolor{red}{\emph{Factors of 2!}}
%
%\be
%\langle n | \O_{J\bar T} | n \rangle =  - \frac{Q_n}{2 R} \left( \frac{\p E_n}{\p R} + \frac{ P_n}{R} \right)
%\ee
%Thus, when changing $\mu$ infinitesimally, the energy levels change as

\be
\frac{\p E_n^{[\l]} ( R)}{\p \l} =  - Q_{L,n} \left( \frac{\p E_n}{\p R} + \frac{ P_n}{R} \right)% = - Q_n \p_R (E -P)
\label{jtbfle}
\ee
%where we used the fact that $P$ is inversely proportional to $R$, being quantized in units of the radius.  This 
which is the equation that we need to solve. The subscript `$L$' that we placed on $Q$ indicates it is the charge associated with the chiral left-moving current.  
Since %$P_n$ is quantized, 
$P_n R \in \mathbb{Z}$,  $P_n$ cannot vary with $\l$, while its $R$ - dependence is fixed. It is useful to introduce the left/right-moving energies $E_{L,R}$ via

\be
E  = E_L + E_R     \;, \;\;\;\;\;\; P = E_L - E_R 
\ee
where we dropped the subscript $n$ on the eigenstates. In terms of these, the flow equation simplifies to  $
\p_\l  E_R =- Q_L  \p_R  E_R. $ To solve it, we need to know how the left-moving charge depends on $\l$. As it turns out, if we insist upon chirality of $J$ and that the level of the $U(1)$ is non-zero (as we generically expect it to be in a non-trivial theory) then, due to the chiral anomaly, the charge starts flowing with $\l$. 

To  properly understand this flow, one would need a better grasp on how the current behaves in conformal perturbation theory. %, and in particular of our assumption that $J$ equals the chiral part of $\hat J$ at any point along the flow. 
There is, however, a  rigorous \emph{indirect} argument for obtaining the modified spectrum, given in  \cite{Chakraborty:2018vja}.  The argument  is based upon splitting the left-moving stress tensor into a  contribution from the chiral current, which takes the Sugawara form, and an independent `coset' contribution, whose OPE with the current vanishes along the flow. \cite{Chakraborty:2018vja} then argue that the coset part is unaffected by the deformation, which implies a spectral-flow type equation for the left-moving energy\footnote{Our notation is as follows: $E_{L/R}$, which should be written as  $E_{L/R}^{[\l]}$, for consistency, are the left/right-moving energies in the deformed theory; we dropped  the superscript to  unclutter the notation. Same comment applies to $Q_L$. $J_0$, which should have been written as $Q_L^{[0]}$, is the charge of the left chiral current in the undeformed CFT; when a right-moving current will be considered, $\bar J_0$ will be the corresponding right chiral current, and the winding will be defined as $w = J_0-\bar J_0$.   }

\be
E_L R - \frac{2\pi Q_L^2}{k} = const = E_L^{[0]} R -  \frac{2\pi J_0^2}{k}
%2\pi \left(h_L - \frac{c}{24} - \right) 
\equiv \hat{\mathcal{E}}_L\label{model}
\ee 
where $E_L^{[0]} R$ is related to the left conformal dimension as in \eqref{engcylcft} and $J_0$ is the undeformed chiral charge, which we assume to be quantized. Using constancy of $P = E_L - E_R$ along the flow, one finds that also $E_R R -2\pi Q_L^2/k = const$. Taking a $\l$ derivative and using the fact that  the only dimensionless parameter in the problem is $\l/ R$, which implies that $E_R = R^{-1} f(\l/ R)$, one finds the following equation for the flow of $Q_L$
\be
 \p_\l Q_L = -  \frac{k}{4\pi} R \, \p_R E_R=
\frac{k}{4\pi} (E_R + \l \, \p_\l E_R) =
\frac{k}{4\pi} \p_\l (\l E_R)
\ee
% Combining this with \eqref{vare}, it is possible to show that that $Q$ precisely satisfies  the equation \eqref{flowq} above. Trading the $R$ derivative for a $\mu$ derivative,
 %
Consequently,   one finds that the charge varies along the flow as %\textcolor{red}{\emph{Notation!!!}}

\be
Q_L= J_0 + \frac{\l k}{4\pi} E_R \label{newq}
\ee
Plugging in the expression for $Q_L$, the  %first equation becomes 
%
%\be
%%R E_R - \frac{2\pi Q^2}{k } = 
% E_R - \frac{2\pi}{k R} \left( Q_0 + \frac{\l k}{4\pi} E_R\right)^2 = \frac{2\pi}{R} \left(  h_R - \frac{c}{24}- \frac{Q_0^2}{k} \right)
%\ee
%and thus the
 solution for the right-moving energy in terms of the original CFT data is %\textcolor{blue}{\emph{Term in last () is $R /2\pi$}}
\be
E_R = \frac{4\pi}{\l^2 k} \left( R - \l  J_0 - \sqrt{\left(R-\l J_0 \right)^2 - \frac{\l^2 k R}{2\pi} E_R^{[0]} }\right)  \label{moder}
\ee
which can also be written in terms of $h_R$ using \eqref{engcylcft}. 
The expression for $E_L$  follows from momentum conservation
\be
 E_L = E_R + \frac{2\pi (h_L - h_R)}{R} \label{elinter}
\ee
Plugging \eqref{moder} into \eqref{newq}, the expression for the chiral charge $Q_L$ is given by

\be
Q_L = \frac{1}{\l} \left(R-\sqrt{ (R-\l J_0)^2 - \l^2 k \left(h_R- \frac{c}{24}\right)}\right) \label{chiralchjtb}
\ee
Let us now comment on the properties of this spectrum.  One immediately notes that when $h_R$/$E_R^{[0]}$ becomes large at fixed $J_0$, the deformed energies become imaginary,   a feature reminiscent of  the behaviour of the finite-size spectrum of    $T\bar T$ - deformed CFTs with $\mu <0$. Note that quantum effects, captured by the anomaly coefficient $k$, are very important for reaching this conclusion. 

%
%\footnote{In the $k\r 0$ limit, the energy formula reduces to  
%%
%\be
%E_R =2 \pi  \frac{ \left( h_R - \frac{c}{24} \right) }{R-\l Q_0}
%\ee
%representing  the solution to the flow equation \eqref{jtbfle} obtained 
%  by ignoring the flow of the chiral charge $Q_L$ with $\l$, as originally derived in \cite{Guica:2017lia}. This energy is mostly problematic at $R \approx \l Q_0$.   % \emph{Is there an upper bound on $E_R$?}
%  }.

%We can again relate the imaginary energies to the appearance of CTCs in this background.

 The interpretation of the imaginary energy states is the same as that put forth in \cite{Cooper:2013ffa} for $\mu<0$ $T\bar T$: the propagation speed in backgrounds with   large $h_R$ exceeds the speed of light\footnote{Note that superluminal propagation does not in itself pose a problem, as the deformed theory is not Lorentz invariant.},  leading to the formation of CTCs in finite size; the imaginary energies are a direct manifestation of this inconsistency. To show the CTCs occur at precisely the expected energy, one needs to use the fact (discussed later in this section), that the $J\bar T$ dynamics can also be reduced to that of the original CFT in terms of the field-dependent coordinates\footnote{Note the transformation $V \r V - \l \phi(U)$ would be a symmetry of warped CFTs, though it is not a symmetry here. }
\be
u =U \;, \;\;\;\;\; v = V - \l \phi
\ee
where $\phi$ is the bosonisation of the $U(1)$ current. On a background characterised by a large chiral charge $Q_L$, we have\footnote{Even in the non-chiral case, we have $\phi = Q_L U/R-Q_R v/R_v$, which leads to the same $\Delta V$.} $\phi = Q_L U/R$. Assuming propagation is on  the CFT $v = V -\l \phi = const$ lines, the accumulated time advance on such a background is

\be
\Delta t =  \frac{1}{2}(\Delta U - \Delta V) = \frac{\Delta U}{2 R} (R-\l Q_L)
\ee
which becomes negative if $ Q_L> R/\l$ (note  $Q_L$ increases as $E_R^{[0]}$ is raised, at fixed $J_0$).  Using \eqref{chiralchjtb}, we see the CTC limit is reached precisely when the square root vanishes, i.e. at the onset of the imaginary energy states. %(\emph{I imagine we are fixing $Q_0$ and raising $E_R^{(0)}$ to reach this bound; $Q_L$ is expected to increase.})

\begin{figure}[h]
    \centering
    \includegraphics[width=7.5cm]{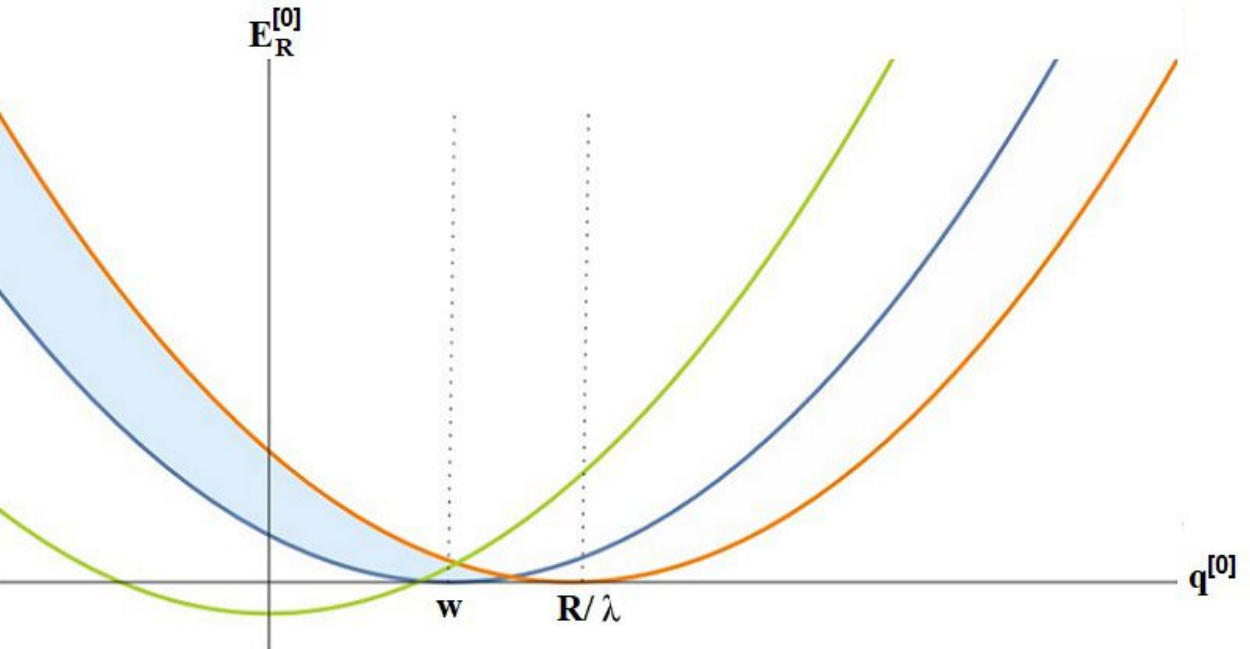}
    \caption{\small{Range of  undeformed right-moving energies (shaded region) that lead to real energies in  $J\bar T$ - deformed CFTs and are allowed by  the  cosmic censorship bound on the left/right-moving energy (green/blue parabola). Note this range  extends to infinite energy. The region becomes larger if we use instead the BPS bounds.}}
     \label{figallowed}
\end{figure}

 Given  that $Q_L(\l)$ is bounded above by $R/\l$, it may be more natural to keep this deformed charge fixed instead as $h_R$ is taken to be large.
%   The relationship between the deformed and undeformed right-moving energy at fixed $Q_L$ is given by 
%\be
% E_R = \frac{4 \pi}{\l^2 k} \left(\sqrt{(R-\l Q_L)^2 + \frac{}}\right)
% \ee 
Reality of $Q_L$ requires that $\hat h_R$ lie below the parabola $(R/\l - J_0)^2$. As can be seen from figure \ref{figallowed}, in which $q^{[0]}=J_0$, this still allows access to infinite energies, even if one imposes a BPS or a cosmic censorship bound (blue parabola) on the right-moving energy.  Thus, the situation is better than in $\mu<0$ $T\bar T$, where only states very close to being extremal can be followed to high energies, and there is little sense to be made of an asymptotic formula for the entropy. %\textcolor{red}{\emph{Discuss in $T\bar T$ section?}}  

As will turn out, even in the case when $J$ is not chiral from the outset, it will be possible to construct a chiral current out of it and the stress tensor. Moreover, the deformation of the theory  appears to be the same whether it is driven by the original, non-chiral current, or this chiral derivate. As a result, the flow equation for the spectrum will always take the form \eqref{jtbfle}, and there will be again a spectral flow on the left energies. Allowing for potentially different initial left/right-moving charges, it follows that the right-moving energies also satisfy a spectral flow-like equation, of the form 

\be
E_R \big(R-\l(J_0-\bar J_0)\big) - \frac{2\pi}{k } \left(  \bar J_0 + \frac{k \l}{4\pi}E_R\right)^2 = R E_R^{[0]} - \frac{2\pi}{k }  \bar J_0^2 \equiv \hat{\mathcal{E}}_R \label{jtbspflr}
\ee
This simply follows by adding $P$ to the flow equation \eqref{model} for the left energy, and requiring that the `initial' value for the right-moving charge be in principle different from that for the left-movers.   Note, however, that the solution for $E_R$ is still given by \eqref{moder}, which does not depend on $\bar J_0$. This result will be derived  in full generality at the end of this section.  It will  be convenient to introduce the notation 

\be
R_v \equiv R-\l (J_0-\bar J_0) = R-\l w
\ee
which corresponds to the radius of the field-dependent coordinate $v$ above, where $w$ is the winding of the scalar field around the compact direction.

\medskip

\subsubsection{Thermodynamics} 

As in the $T\bar T$ case, the entropy  of $J\bar T$ - deformed CFTs follows from the fact that
%
%Even though we have - strictly speaking - only derived the deformed spectrum for the case of a purely holomorphic current, let us assume (as we will prove in section \ref{} ) that both left and currents are present. The defomed spectrum is then given by the formula \eqref{}. The purely chiral case is obtained for $\bar J_0=0$. 
%
%Let us start with the case of a purely holomorphic current, where the energy levels are non-trivially displaced, as in \eqref{newlev}. Since 
the spectrum is smoothly deformed, so we  expect that the degeneracy  still equals given by the  degeneracy in the undeformed CFT\footnote{This is because the levels of fixed $ h_R$ and $J_0$ do not cross as we vary $\l$. %\textcolor{red}{\emph{Notation!}} %Note that if $Q$ were not fixed, then levels with  $h'>h$ may cross if $Q'<Q$.
 }  written 
in terms of the deformed energies \eqref{moder} - \eqref{elinter}.

Taking into account the fact that now there is charge, the undeformed CFT degeneracy is given in terms of the charged Cardy formula 

\be
S_{ch.\, Cardy} = \sqrt{\frac{\pi c}{3}}\bigg( \sqrt{R E_L^{[0]} - \frac{2\pi}{k}J_0^2}+\sqrt{R E_R^{[0]} -\frac{2\pi}{k} \bar J_0^2} \bigg) \label{schcardy}
\ee
where $E_{L,R}^{[0]}$ are related to the CFT conformal dimensions via the usual relation \eqref{engcylcft}. %\textcolor{blue}{ This holds for $h, \bar h >> c$. Here $k$ is the level of the left Kac-Moody chiral algebra.} 
We immediately note that the quantities under the two square roots are nothing but the spectral flow invariants $\hat{\mathcal{E}}_{L,R}$ \eqref{model} and \eqref{jtbspflr},  which do not change under the $J\bar T$ flow. Consequently, the entropy can be written in terms of the $J\bar T$ - deformed energies and charges as %\textcolor{blue}{ $\hat{\mathcal{E}}_L \equiv R E_L - \frac{2\pi}{k} Q_L^2$ and $\hat{\mathcal{E}}_R \equiv R_v E_R - \frac{2\pi}{k} Q_R^2 $ are invariant under the $J\bar T$ flow\footnote{The relation between the boundary and the bulk notation is $\frac{12\pi}{c}\hat{\mathcal{E}}_L=\left(\frac{R}{\ell}\right)^2 \! \L$ and $\frac{12\pi}{c}\hat{\mathcal{E}}_R=\left(\frac{R_v}{\ell}\right)^2\bar{\L}$. }.}

\be \label{entropyJTbarFT}
S  = \sqrt{\frac{\pi c}{3}}\bigg( \sqrt{R E_L - \frac{2\pi}{k}Q_L^2}+\sqrt{R_v E_R -\frac{2\pi}{k} Q_R^2} \bigg)
\ee
where $Q_{L,R}$ are related to the charges $J_{0}, \bar J_0$ in the undeformed CFT -  which are assumed to be integer-quantized - via %\textcolor{red}{\emph{Definition $Q_R$?}}
\begin{align}\label{flowofcharges}
Q_L&=J_0+\frac{\lambda k}{4\pi} E_R\hspace{1cm}Q_R=\bar{J}_0+\frac{\lambda k}{4\pi} E_R
\end{align}
For now, the second equation can be considered a definition of $Q_R$. We will give a more appropriate one in the sequel.
We remark that the difference between the $J\bar T$ - deformed charged Cardy formula and its CFT counterpart is rather minimal, if we write the former in terms of the deformed charges $Q_{L,R}$; indeed, the only difference is the factor of $R_v = R-\l Q_L+\l Q_R$. Things stand rather differently, of course, if we have access to the charge quantization condition, and interpret \eqref{entropyJTbarFT} in terms of $J_0, \bar J_0$, case in which the difference from the CFT formula is significant.  Of course, in this case one also needs to take into account the discussion of the previous section, regarding the fact that at fixed $J_0$, $E_R$ will eventually become imaginary, so the regime of large $E_R$ is not reached. 

The relation between the left/right-moving energies and their associated temperatures  $T_{L,R}$ depends on which set of charges one holds fixed in the deformed theory. Assuming $J_0, \bar J_0$ are fixed, % \textcolor{blue}{which is rather natural from a holographic perspective \emph{Why??}}, 
the  first law of thermodynamics 
\begin{align}
\delta S&= \frac{1}{T_L}\delta E_L +\frac{1}{T_R}\delta E_R + \mu_L \delta J_0 + \mu_R \delta \bar{J}_0
\end{align}
allows us to compute 
the temperatures and chemical potentials as  

\begin{align}
\frac{1}{T_{L/R}}&=\frac{\partial S}{\partial E_{L/R}}\bigg|_{J_0,\bar{J}_0,E_{R/L}=fixed}%\hspace{0.5cm}\frac{1}{T_R}=\frac{\partial S}{\partial E_R}\bigg|_{J_0,\bar{J}_0,E_L=fixed}, 
\hspace{0.5cm}\mu_L=\frac{\partial S}{\partial J_0}\bigg|_{E_{L,R},\bar{J}_0=fixed}, \hspace{0.5cm}\mu_R=\frac{\partial S}{\partial \bar{J}_0}\bigg|_{E_{L,R},J_0=fixed}
\end{align}
The undeformed thermodynamic potentials are related to the  $J\bar T$ flow invariants $\hat{\mathcal{E}}_{L/R}$ %introduced above 
and the charges as 
\vskip-7mm
\begin{align}
\frac{1}{T_{L/R}^{[0]}}=\sqrt{\frac{\pi c}{12}}\frac{R}{\sqrt{\hat{\mathcal{E}}_{L/R}}}\;, %\hspace{0.7cm}\frac{1}{T_R^{[0]}}=\sqrt{\frac{\pi c}{12}}\frac{R}{\sqrt{\hat{\mathcal{E}}_R}}
\hspace{0.7cm}\mu_L^{[0]}=-\sqrt{\frac{\pi c}{12}}\frac{4\pi}{k}\frac{J_0}{\sqrt{\hat{\mathcal{E}}_L}}\;, \hspace{0.7cm}\mu_R^{[0]}=-\sqrt{\frac{\pi c}{12}}\frac{4\pi}{k}\frac{\bar{J}_0}{ \sqrt{\hat{\mathcal{E}}_R}} \label{undeftpot}
\end{align}
Using the expressions \eqref{flowofcharges} for the $U(1)$ charges
and the $J\bar T$ flow invariants $\hat{\mathcal{E}}_{L,R}$, the deformed thermodynamic potentials  evaluate to  \cite{Georgescu:2024ppd}
%\textcolor{ForestGreen}{(rescale everything by $\sqrt{\frac{\pi c}{12}}$? leave only second expr for $\mu_R$?)}
\be\label{fieldtheorytempapp}
\frac{1}{T_L}=\sqrt{\frac{\pi c}{12}}\frac{R}{\sqrt{\hat{\mathcal{E}}_L}} = \frac{1}{T_L^{[0]}}\;, \hspace{0.71cm}\frac{1}{T_R}=\sqrt{\frac{\pi c}{12}}\bigg(\frac{R-\lambda Q_L}{\sqrt{\hat{\mathcal{E}}_R}}-\frac{\lambda Q_L}{\sqrt{\hat{\mathcal{E}}_L}}\bigg) = \frac{R-\l Q_L}{R T_R^{[0]}} - \frac{\l Q_L}{R T_L^{[0]}}
\ee
and 
\bea
\mu_L&=&\sqrt{\frac{\pi c}{12}}\bigg(-\frac{4\pi Q_L/k}{\sqrt{\hat{\mathcal{E}}_L}}-\frac{\lambda E_R}{\sqrt{\hat{\mathcal{E}}_R}}\bigg) =  \mu_L^{[0]}- \frac{\lambda E_R}{R} \bigg(\frac{1}{T_L^{[0]}}+\frac{1}{T_R^{[0]}}\bigg) \nonumber\\[4pt]
\mu_R&=&\sqrt{\frac{\pi c}{12}}\bigg(\frac{\lambda E_R - \frac{4\pi}{k}Q_R}{\sqrt{\hat{\mathcal{E}}_R}}\bigg)=-\sqrt{\frac{\pi c}{12}}\frac{4\pi}{k} \frac{\bar{J}_0}{\sqrt{\hat{\mathcal{E}}_R}} =\mu_R^{[0]}\label{fieldtheorypotentialapp}
\eea
where for the second set of equalities we used \eqref{undeftpot}.  

To obtain the corresponding thermodynamic quantities for the chiral case with fixed $J_0$, we simply set $\bar J_0 =0$ in the formulae above.  Again, one should be careful about the fact that the deformed energies become imaginary at fixed $J_0$. One can alternatively consider the ensemble with fixed $Q_L$ and $\bar J_0$ (with $\bar J_0=0$ in the chiral case).  The entropy \eqref{entropyJTbarFT} then reads

\be
S  = \sqrt{\frac{\pi c}{3}}\bigg( \sqrt{R E_L - \frac{2\pi}{k}Q_L^2}+\sqrt{(R-\l Q_L) E_R + \frac{\l^2 k}{8\pi} E_R^2 -\frac{2\pi}{k} \bar J_0^2} \bigg)
\ee
which displays an interesting Hagedorn behaviour (for the right-movers only) in the regime of large $E_R$ that, as discussed, is accessible if $Q_L$ is fixed.

Finally, one can consider the thermodynamic potentials when $Q_{L,R}$ are held fixed. The temperatures one obtains then are %\textcolor{red}{\emph{Different notation?}}

\be
\frac{1}{T_L'}=\sqrt{\frac{\pi c}{12}}\frac{R}{\sqrt{\hat{\mathcal{E}}_L}} = \frac{1}{T_L^{[0]}}\;, \hspace{0.71cm}\frac{1}{T_R'}=\sqrt{\frac{\pi c}{12}}\bigg(\frac{R-\lambda Q_L +\l Q_R}{\sqrt{\hat{\mathcal{E}}_R}}\bigg) = \frac{R_v}{R T_R^{[0]}} 
\ee
In terms of these deformed temperatures, the entropy takes the form

\be
S= \frac{\pi c}{6} (R T'_L + R_v T'_R)
\ee
again, a minimal departure from Cardy's formula, despite the fact that the theory is rather different. In particular, this departure would be invisible if one concentrates on states that had $J_0=\bar J_0$ in the undeformed CFT.

\subsubsection{Modified modular invariance}

Consider the partition function of a $J \bar T$ - deformed CFT  on a torus with modular parameter $\tau$ and length of the $a$-cycle $R$ %\textcolor{red}{\emph{Notation!}}
\be
Z_{J\bar T} \left(\tau, \bar \tau, \nu , \frac{\l}{R}\right) = \sum e^{-\tau_2 R E_n^{[\l]} (R) + i \tau_1 R P_n + 2 \pi i \nu q_n^{[\l]}}
\ee
where $\nu$ is a chemical potential that couples to the chiral $U(1)$ current. We explicitly indicated the fact that the partition function depends on the dimensionless ratio $\l/R$
%
% We wrote the coupling $\l$
as an argument - rather than a label - because it changes under diffeomorphisms, due to its vectorial
nature. 
Due to the presence of imaginary energy modes, it is currently not
well understood to what extent this partition function is well defined; however, the fact that the
theory admits a non-perturbative definition \cite{Anous:2019osb} yields hope that its study is meaningful.

The modular transformation properties of this partition function were first discussed in \cite{Aharony:2018ics}. To derive them, it is useful to introduce the complex coordinates on the torus, $z = x + \tau y, \, \bar z = x + \bar \tau y$, in terms of which the metric takes the form $ds^2 = R^2 dz d\bar z$. Under the modular transformation \eqref{psl2z}, $z,\bar z$ transform as 
\be
z \r \frac{z}{c\tau + d} \;, \;\;\;\;\;\; \bar z \r \frac{\bar z}{c\bar \tau +d}
\ee
 Since  the $J\bar T$ coupling $\l^a$ transforms as a vector, $\l^{\bar z}$ , under modular transformations, which has the same transformation
properties as $R\bar z$, it follows that the dimensionless combination $\l/R$ transforms exactly as $\bar z$.  A similar argument
can be used to derive the well-known transformation properties\footnote{The chemical potential is related to a background gauge field, $a^z$ , that couples to the left current as $\nu=\b a^z = \tau_2 R a^z$.
Invariance of the action under diffeomorphisms implies that $a^z$ transforms in the opposite way from $R\bar z$, which leads to $R a^z \r (c\bar \tau+d) R a^z$. Taking into account the transformation of $\tau_2$, we find the quoted result. } of the chemical potential $\nu$. With this in mind, the partition function has the standard anomalous transformation under diffeomorphisms of the torus 
\be
Z \left(\frac{a\tau+b}{c\tau +d}, \frac{a\bar \tau + b}{c\bar \tau +d}, \frac{\nu}{c\tau+d}, \frac{\l}{R (c\bar \tau +d)}\right) = e^{\frac{2i\pi k c \nu^2}{c\tau+d}} Z \left(\tau, \bar \tau, \nu, \frac{\l}{R}\right)
\ee  
Note the above transformation properties hold for \emph{any} UV-complete two-dimensional QFT with a single (null) vector coupling with length dimension one, namely for any two-dimensional dipole CFT, as per our definition. %\textcolor{blue}{It would be interesting to understand whether such theories all share the same asymptotic density of states}. 

If, in addition to modular covariance - which is simply a statement about the well-definiteness of the torus partition function  - %\textcolor{blue}{\emph{Move up?}}
 one makes the additional, rather strong assumption that the energy spectrum of the deformed theory only depends on $\l$ and the undeformed value of the respective level (taken to be generic), then one can again show \cite{Aharony:2018ics} that the unique solution satisfying these assumptions (and the initial condition) is that of $J\bar T$ - deformed CFTs. This partition function turns out to obey a flow equation with respect to $\l$, which can be in principle be reproduced using the non-perturbative definition of $J\bar T$ - deformed CFTs. %\textcolor{blue}{\emph{Later?} This flow equation is rather ugly, presumably due to tracking the chiral charge.}
These considerations can be easily generalised to UV - complete QFTs whose  dimensionful coupling is  a vector that is not necessarily  null. %\textcolor{blue}{\emph{More?}}

\subsubsection{The spectrum of conformal dimensions on the plane }

As we have discussed, the spectrum of $J\bar T$ - deformed CFTs on a cylinder exhibits imaginary energy states, which we have linked to the appearance of closed timelike curves (CTCs) in the theory in any finite size. Since these CTCs are absent in the planar limit, there is  a logical possibility that  $J\bar T$ - deformed CFTs on the plane are inconsistency-free, following the suggestion of \cite{Cooper:2013ffa}.

In this subsubsection, we bring evidence that this is indeed the case, by computing the spectrum of $SL(2,\mathbb{R})_L$ conformal dimensions  of the theory on the plane  - a natural observable, since $J\bar T$ - deformed CFTs have left-moving conformal invariance.  Given the  structure of the $J\bar T$ coupling, the general expectation \cite{Guica:2010sw} is that these dimensions depend on the right-moving momentum as  $h(\l \bar p)$.

 These conformal dimensions can be  obtained from known spectrum of left-moving energies on a cylinder via an infinite boost. The latter is necessary because the cylinder identification, unless it is null, breaks the left conformal symmetry. Let $x^\pm = \s \pm t$ be the coordinates on the cylinder, originally identified as $x^\pm \sim x^\pm + R$. Under a boost 
 \be
 x^\pm \r \tilde x^\pm = e^{\mp \g} x^\pm
 \ee
the identification radii of the tilded coordinates will be $\tilde R_\pm = R e^{\mp \g}$,  while the 
 %
%  We would like to derive the spectrum on a boosted cylinder, with the identifications
%
%\be
%\tilde x^+ \sim \tilde x^+ + \tilde R_+ \;, \;\;\;\;\;\; \tilde x^- \sim \tilde x^- + \tilde R_-
%\ee
%where $\tilde R_+$ is finite and $\tilde R_- \r \infty$. The spectrum of $J\bar T$ deformed CFTs on a space with these identifications can be related to the spectrum on the usual cylinder via a boost
%
%\be
%\tilde x^\pm \r x^\pm = e^{\pm \g}  \tilde x^\pm
%\ee
%with the boost parameter chosen such that %the coordinate identifications are
%%
%\be
%%x^\pm \sim x^\pm + R \;, \;\;\;\;\;\;\;
%  \tilde R_+ \, e^\g = \tilde R_-\, e^{-\g} =R \label{boostedrad}
%\ee
%To obtain the spectrum on an infinitey boosted cylinder,  we will be interested in the limit $\g \r \infty$ with $\tilde R_+$ fixed.
relation between energies and $J\bar T$ coupling are
%
% The various quantities before and after the boost are related as
%
\be
\l = \tilde \l \, e^{-\g} \;, \;\;\;\;\; E_L = e^{-\g} \tilde E_L \;, \;\;\;\;\;\; E_R = e^{\g} \tilde E_R \label{repl}
\ee
as follows from their Lorentz transformation properties. 
%
%where the transformation law for $\mu$ follows from the fact that it is a constant null vector with a lower `$+$' component.
 $E_{L,R}$ are the energies on the unboosted cylinder, given by 
\eqref{moder} - \eqref{elinter}. We will be interested in the limit $\g \r \infty$ in which $\tilde R_-$ decompactifies, thus restoring the $SL(2,\mathbb{R})_L$ symmetry. We would like  the parameters  $\tilde \l$, $\tilde R_+$ and the right-moving energy  $\tilde E_R$ in the boosted theory to be fixed in this limit. Plugging  \eqref{repl} into the expression for $ E_L$ we find 

\be
\tilde E_L = e^\g \left(E_R + \frac{2\pi ( h_L -  h_R)}{R} \right) =\frac{2\pi \left( h_L -c/24\right) }{\tilde R_+} + e^{2\g} \left(\tilde E_R- \frac{2 \pi ( h_R -c/24)\, e^{-2\g}}{\tilde R_+}\right) \label{expEL}
\ee
The only way that $\tilde E_L$ could be finite is if $\tilde E_R$ cancels against $2\pi ( h_R -c/24)\, e^{-2\g}/\tilde R_+$ with precision $e^{-2\g}$. Since $\tilde E_R \equiv \bar p$ itself is finite, we conclude that we should scale %\textcolor{red}{\emph{Careful order!}}
\be
 h_R - \frac{c}{24} = \frac{\bar p \tilde R_+}{2\pi} \, e^{2\g} +  a + \O(e^{-2\g}) \label{ho}
\ee
where $a$ is some $\O(1)$ dimensionless quantity.
 Plugging this expansion of $ h_R$ into the expression
 \eqref{moder} for $E_R$ and expanding in the $\g \r \infty$ limit, we find 
\bea
\tilde E_R & = &  \frac{4\pi e^{-\g}}{\tilde \l^2 k e^{-2\g}} \left(\tilde R_+ e^\g - \tilde \l \,  q^{[0]} e^{-\g} - \sqrt{\left(\tilde R_+ e^\g - \tilde \l \, q^{[0]} e^{-\g}\right)^2 -  \frac{\tilde \l^2 k}{2\pi} \bar p   \tilde R_+ - \tilde \l^2 k a e^{-2\g} +\ldots}\right) \nonumber \\
& \approx & \bar p +  \frac{1}{\tilde R_+} \left(2 \pi a + \tilde \l \bar p q^{[0]}+ \frac{k \tilde \l^2 \bar p^2}{8\pi } \right) \, e^{-2\g} +\O(e^{-4\g})
\eea
where we have relabeled $J_0 \r q^{[0]}$. 
Plugging this expansion back into \eqref{expEL} and taking $\g \r \infty$, we arrive at
\be
\tilde E_L =  \frac{2\pi}{\tilde R_+} \left( h_L - \frac{c}{24} +  \frac{\tilde \l}{2\pi} \bar p q^{[0]}+ \frac{k \tilde \l^2 \bar p^2}{16\pi^2 } \right)
\ee
the factors of $a$ cancelling out. 
Using the usual map from the cylinder  to the  plane, which maps energies to conformal dimensions, we find that the deformed $SL(2, \mathbb{R})_L $ dimensions are %\textcolor{red}{ \emph{Redundant!}}

\be
h^{[\l]} = h^{[0]} + \frac{\l}{2\pi} q^{[0]} \bar p + \frac{\l^2 k}{16 \pi^2} \bar p^2 \label{jtbdims}
\ee
where we have dropped the tilde from $\l$ and rewrote the undeformed left dimension as $h_L \r h^{[0]}$.  The expression for the charge in the deformed theory is given by \eqref{newq}, which in our new notation  reads

\be
q^{[\l]} = q^{[0]} + \frac{\l k}{4 \pi} \bar p \label{chshift}
\ee
The above expressions correspond to an exact formula for the conformal dimensions and charges in $J\bar T$-deformed CFTs as a function of the deformation parameter. As expected, these data only depend  on the combination $\l \bar p$ and, unlike the spectrum on the cylinder, show no obvious pathology as $\l$ becomes large.    Remarkably, these exact expressions terminate at $\O(\l^2)$. The combination $\hat h \equiv h^{[\l]} - q_{[\l]}^2/k$
%
%Let us make a few remarks about this formula. First, note it depends only on the combination $\mu \bar p$, as expected on general grounds. It is quite remarkable that it terminates  at order $\mu^2$, and it would be very nice to be able to reproduce this from conformal perturbation theory and see that/why the corrections terminate. Second, the quantity
%
%\be
%\hat h \equiv h(\mu) - \frac{q(\mu)^2}{k} =  h - \frac{q^2}{k}
%\ee
is independent of $\l$, in perfect agreement with the argument presented in \cite{Chakraborty:2018vja}, 
and hints towards a possible interpretation of the $J\bar T$ deformation as  spectral flow.% transformation, with a momemtum-dependent spectral flow parameter $\mu \bar p$. %This can be understood from the argument of \cite{kutasov}, that the conformal dimension associated with $\hat T$ is unaffected by the deformation. 
%Note that in order to have an exact interpretation as spectral flow, the OPE coefficients should stay unchanged. % Next, note that unlike the expression \eqref{} for the spectrum of energies on the cylinder, which become imaginary at large enough $\mu$, there is no apparent problem with the spectrum of conformal dimensions on the plane.  

A simple consequence of the above formulae is that
purely holomorphic quantities do not acquire an anomalous dimension, as they have $\bar p =0$; however, initially purely antiholomorphic quantities, such as the right-moving stress tensor,  receives a non-trivial left-moving anomalous dimension and charge

\be
h_{\bar T}^{[\l]}= \frac{\l^2 k \bar p^2}{16\pi^2}\;, \;\;\;\;\; q_{\bar T}^{[\l]} = \frac{\l k \bar p}{4\pi} \label{hqtbar}
\ee
These formulae can be checked  to leading order(s) using  conformal perturbation theory, finding perfect agreement \cite{Guica:2019vnb}. The precise expression for the corresponding operator in terms of the fields in the theory is however not entirely clear.

%\textcolor{blue}{\emph{This is just the CPT result - clear to which (quasi-local) operator it pertains?}}

\bigskip

\noindent \textbf{\emph{II. Classical $J\bar T$ - deformed CFTs: Lagrangians and Hamiltonians}}

\medskip

\noindent 
So far, we have discussed quantum properties of $J\bar T$ - deformed CFTs, such as their spectrum. In this subsection, we turn to studying in detail the classical action and Hamiltonian of certain $J\bar T$ - deformed CFTs, paying special attention to the conserved currents that exist in the theory. These will be essential for understanding the holographic dictionary in section \ref{holodictjtbdtr} and the extended symmetries in section \ref{infsymmsec}. We start with the $J\bar T$ - deformed free boson, and then discuss general  $J\bar T$ - deformed CFTs in the Hamiltonian formalism.

\subsubsection{The $J\bar T$ - deformed free boson}

The $J\bar T$ - deformed free boson was first studied in \cite{Guica:2017lia}. The $SL(2,\mathbb{R})_L \times U(1)_R$ symmetry of the problem require the action to take the form 
\be
S_{J\bar T}= - \int dU dV \p_U \phi \p_V \phi\, \mathcal{F}(\l \p_V \phi) \label{jtbfb}
\ee
for some function $\mathcal{F}$ satisfying $\mathcal{F}(0)=1$.   The $J\bar T$ flow equation \eqref{jtbfldef}, where the $U(1)$ current is e.g. taken to be the shift symmetry current\footnote{It turns out one obtains the same result for the deformed Lagrangian if one uses instead the topological current \eqref{jtildef}, or the chiral one \eqref{eq:JTbarKUKV}. }, determines it to be 
\be
\mathcal{F}(x) = \frac{1}{1-x} \label{exprF}
\ee
The equations of motion are simply
\be
\p_V \left( \frac{\p_U \phi}{1-\l \p_V \phi} \right) =0 \label{eomjtb}
\ee
%with general solution $\phi (U,V) = f(U) + g(V-\l f(U))$, for two arbitrary functions $f,g$. 
%
As in the $T\bar T$ case, there exists a field-dependent coordinate transformation\footnote{In this section, as well as in the next one, we will be abusively  using the same notation $u,v$ for the field-dependent coordinates, even though their expression in terms of the fields basic changes from \eqref{fdepcootrfb} in $T\bar T$, to \eqref{uvcoojtb} in $J\bar T$. %, to \eqref{uvjta} in $JT_a$ and \eqref{fdepcoojtta} in $\tilde J T_a$. 
We hope that this choice, made for ease of notation, will not cause confusion to the reader.} 
%\textcolor{red}{\emph{Comment choice constant?}}

\be \label{uvcoojtb}
u=U\;, \;\;\;\;\;\;v= V - \l \phi (U,V) +const.  %\;\;\;\; \Leftrightarrow\;\;\;\; U=u\;, \;\;\;\;\; V= v + \l \phi(u,v)
\ee
that maps the system to a free boson, this time already at the level of the action. 
Indeed, using the relation between the first derivatives of $\phi$ 
\be
\p_U \phi = \frac{\p_u \phi}{1+\l \p_v \phi} \;, \;\;\;\;\;\; \p_V \phi = \frac{\p_v \phi}{1+\l \p_v \phi}
\ee
one can easily check that the action \eqref{jtbfb} becomes $\int du dv \, \p_u \phi \p_v \phi$.
%
%As in the case of the $T\bar T$ deformation, the field-dependent coordinates in terms of which the dynamics reduces to that of the undeformed CFT are also the ones appearing in the infinite-dimensional pseudo-conformal symmetries. 

Let us now discuss the conserved currents in this theory. Using the Noether procedure, one can  compute the  components of the stress tensor, finding  
\be
T_{UU} = \frac{(\p_U \phi)^2}{(1-\l \p_V \phi)^2} \;, \;\;\;\;\; T_{VU} =0\;, \;\;\;\;\; T_{UV} = \frac{\l \p_U \phi (\p_V \phi)^2}{(1-\l\p_V \phi)^2} \;, \;\;\;\;\;T_{VV} = \frac{(\p_V \phi)^2}{1-\l \p_V \phi} \label{jtbstresst}
\ee
The stress tensor is not symmetric because  the $J\bar T$ deformation breaks Lorentz invariance. The $T_{VU}$ component vanishes as a consequence of the $SL(2,\mathbb{R})$ symmetry preserved by the deformation. 
%
%
%Given the equations of motion, one notes that $f(U) T_{UU}$ is conserved for any $f$, as is $\bar f(v) T^\a{}_V$, upon noting that 
%% 
%\be
%\frac{\p_U v}{\p_V v} = - \frac{\l \p_U \phi}{1-\l \p_V \phi} = - \frac{T_{UV}}{T_{VV}}
%\ee
% The first transformation is simply the infinite enhancement of the left-moving conformal symmetries. The second transformation corresponds to a field-dependent, right-moving pseudo-conformal transformation. We postpone the construction of conserved charges, which turns out to be somewhat subtle. 
%

The $U(1)$  Noether current associated to the shift symmetry of $\phi$ is
 
\be
J_{\mbox{\tiny{$U$}}}^{sh} = \frac{\p_U \phi}{(1-\l \p_V \phi)^2} \;, \;\;\;\;\; J_{\mbox{\tiny{$V$}}}^{sh} = \frac{\p_V \phi}{1-\l \p_V \phi} 
\ee
There is, additionally, a topologically conserved current, $J^{top} = \star d\phi$, with components  %\textcolor{red}{\emph{Notation?}}
 
\be
 J_{{\mbox{\tiny{$U$}}}}^{top} = \p_U\phi\;, \;\;\;\;\;  J_{{\mbox{\tiny{$V$}}}}^{top}=-\p_V\phi \label{jtildef}
\ee
The conserved charges associated to the two currents will be denoted as 

\be
Q_0 = \int d\s (J^{sh}_{\mbox{\tiny{$U$}}}-J_{\mbox{\tiny{$V$}}}^{sh}) \equiv J_0 +\bar J_0 \;, \;\;\;\;\; w = \int d\s (J^{top}_{\mbox{\tiny{$U$}}}-J_{\mbox{\tiny{$V$}}}^{top}) \equiv J_0 -\bar J_0
\ee
Both will be assumed to be quantized.
Note that in a CFT, the combinations $J^\pm = \frac{1}{2}(J^{sh}\pm  J^{top})$ are chiral and, respectively, anti-chiral, with charges denoted as $J_0, \bar J_0$, as in the previous subsection.

 Due to the $SL(2,\mathbb{R})_L$ symmetry of the  deformed theory, we would still expect \cite{Hofman:2011zj} to be able to construct a purely left-moving conserved current. Indeed, it is not hard to notice that the combination 
\be
K_\a \equiv\frac{1}{2}( J^{sh}_\a + J^{top}_{\a} - \l T_{\a V}) \label{chircurjtb}
\ee
has a vanishing right-moving component. Concretely, 
\be
 K_{U} = \frac{\p_U \phi}{1-\l \p_V \phi} \;, \;\;\;\;\;  K_V =0
\label{eq:JTbarKUKV}
\ee 
The left-moving component is holomorphically conserved, as  seen by using the equations of motion \eqref{eomjtb}. We also  note that  $T_{\mbox{\tiny{$UU$}}} = K_{\mbox{\tiny{$U$}}}^2$, so the combination $T_{\mbox{\tiny{$UU$}}}-K_{\mbox{\tiny{$U$}}}^2$ is trivially independent of $\l$. As we will see, the $\l$ - independence of this combination generalizes and is related to the fact that the $J\bar T$ deformation induces an operator-valued spectral flow transformation on the left-movers \cite{Chakraborty:2018vja,Guica:2019vnb}. 

The conserved charge associated to the chiral current $K_U$ will be denoted\footnote{In our conventions, $k=2\pi$ for a free boson, explaining the difference with \eqref{newq}. We will sometimes use $\hat k = k/(2\pi)$ to ease the notation.} %\textcolor{red}{\emph{Factors $2\pi$?}}

\be
Q_L = \int d\s K_U = J_0 + \frac{\l}{2} E_R
\ee
One can also construct an analogous `right-moving' current\footnote{This terminology should not be taken literally, as the on-shell  solution for $\bar K_\a$ is is not just a function of $V$; in the case of the  stress tensor, 
by `right-moving' we simply mean the generator of translations in the right-moving null direction V. } 

\be
\bar K_\a = \frac{1}{2}( J^{sh}_\a - J^{top}_{\a} - \l T_{\a V}) \;, \;\;\; \mbox{or} \;\;\;\; \;\;\bar K_U = \frac{\l \p_U\phi \p_V\phi}{1-\l \p_V \phi} \;, \;\;\;\;\; \bar K_V = \p_V \phi \label{kbarfb}
\ee
The associated conserved charge will be denoted as %\textcolor{red}{\emph{Sign? Check!}}

\be
Q_R =  \int d\s (\bar K_U - \bar K_V) = \bar J_0 + \frac{\l}{2} E_R
\ee
We note that the difference $Q_L -Q_R = J_0 - \bar J_0 = w$ is $\l$ - independent. 

Since the currents $T_{UU}$ and $K_U$ are both chiral and conserved, it follows that they will still be so if we multiply them by arbitrary functions of $U$, thus obtaining an infinite set of conserved currents, one for each Fourier mode of these functions. As far as the right-moving currents are concerned, it is not hard to note that
%
% Since the $V$ component of the current $K_\a$ vanishes, the current $\chi(U)  K_\a$ is conserved for any function $\chi(U)$. Thus, we find an infinite set of conserved charges associated with this chiral $U(1)$ symmetry, given by 
%% 
%\be
%P_\chi = \int d\s \chi(U) K_U  \label{defPjtb}
%\ee
%It is interesting to ask whether there also  exists an infinite-dimensional enhancement of the right-moving $U(1)$ symmetry. A natural candidate is the current $J^- $, which is purely right-moving in the undeformed CFT. In presence of the deformation, its components are 
% 
%\be
%%\bar{K}_\a  \equiv \frac{1}{2}( J_\a -\tilde J_{\a})\; \;\; \Rightarrow\;\;\;\; 
%J^-_U = \frac{\l \p_U \phi \p_V \phi}{(1-\l \p_V \phi)^2} \left(1-\frac{\l}{2} \p_V \phi\right) \;, \;\;\;\;\;  J^-_V =\frac{ \p_V \phi}{1-\l \p_V \phi} \left(1-\frac{\l}{2} \p_V \phi\right) 
%\ee
%These have the important property that
%% 
\be
\frac{T_{UV}}{T_{VV}} = \frac{\bar K_U}{\bar K_V} = \frac{\l \p_U \phi}{1-\l \p_V \phi}
 = - \frac{\p_U v}{\p_V v} \label{specialKb}
\ee
where $v$ is the field-dependent coordinate introduced in \eqref{uvcoojtb}.  This implies that the currents $\bar f(v) T_{\a V} $ and 
 $\bar \chi(v) \bar K_\a$ are  conserved for any functions $\bar f(v), \bar \chi (v)$, as is any linear combination of $J^-_\a$ and $T_{\a V}$, when multiplied by an arbitrary function of $v$. This observation will be very important in building an infinite set of conserved charges in this theory.   
%
% current of the form $\bar \chi(v) \bar K^{(\xi)}$ with
%
%\be
%\bar K_\a^{(\xi)} = J^-_\a - \xi \, T_{\a V}
%\ee
%%will be conserved, for any constant $\xi$. 
% The conserved charges associated with the $\bar \chi(v) \, J^-_\a$ currents are 
% 
%\be
% \bar P_{\bar \chi} = \int d\s \bar \chi(v) (J^-_U-J^-_V) \label{defPbjtb}
%\ee

\subsubsection{Classical Hamiltonian of general $J\bar T$ - deformed CFTs}

We can easily compute the Hamiltonian of the $J\bar T$ - deformed free boson. However, it is possible to have the answer for a general seed Hamiltonian, under certain assumptions about the initial CFT.

We consider a two-dimensional classical field theory with fields $\phi_k$, with $k = 1, \ldots, n$. We assume that the field theory has a  $U(1)$ symmetry, which for simplicity will be taken to be a shift symmetry of the first scalar field, $\phi_1$. The Hamiltonian $\H (\pi_k, \phi_k)$  thus only depends on $\phi_1$ through its spatial derivatives. %\textbf{\emph{How about higher derivatives?}}\RM{Higher spatial derivatives requires also higher time-derivatives due to rotational/boost invariance. AFAIK the Hamiltonian formalism then requires the Ostrogradsky formalism or an approach with Lagrange multipliers and constraints.} 
 The $U(1)$ shift current has components
\be
J^{sh}_t = \pi_1 \;, \;\;\;\;\;\; J^{sh}_\s = \p_{\phi_1'} \H \label{J1ham}
\ee 
The  components of the topologically conserved current \eqref{jtildef} are
\be
 J^{top}_t= \phi_1' \;, \;\;\;\;\;\; J^{top}_\s = \p_{\pi_1}\H
\ee
 It is convenient to introduce the following combinations
of the time components of these two currents

\be
\mathcal{J}_\pm = \frac{\pi_1 \pm \phi_1'}{2}
\ee
Of course, this only leads to a $U(1)$ current algebra with (rescaled) level $\hat k =1$; nonetheless,
the level can be easily set to any desired value by rescaling $\mathcal{J}_\pm$ by $\hat k^{1/2}$, which we will view
as a corresponding rescaling of $\phi_1, \pi_1$.  
We  will furthermore assume that in the undeformed CFT, the currents $J^\pm \equiv \frac{1}{2}(J^{sh}\pm J^{top})$ are chiral and, respectively, anti-chiral, which will impose certain constraints on the undeformed Hamiltonian density $\H^{[0]}$. More precisely, the vanishing of their (anti)chiral components   requires that $\H_R^{[0]} = \frac{1}{2} (\H^{[0]}-\P)$ satisfy

\be
J^{+}_{{{\mbox{\tiny{$V$}}}}}=\frac{1}{2}(\p_{\pi_1} + \p_{\phi'_1}) \H_R^{[0]} =0 \label{chircJ1p}
\ee
%
%\be
%(J^1_t+ \tilde J^1_t)-(J^1_\s+ \tilde J^1_\s)= \pi_1 +\phi'_1 -\p_{\pi_1} \H^{(0)} -\p_{\phi'_1} \H^{(0)} = -2 (\p_{\pi_1} + \p_{\phi'_1}) \H_R^{(0)} =0
%\ee
%whereas $J^1- \tilde J^1$ is anti-chiral, and so
and 
\be
%(J^1_t- \tilde J^1_t)+(J^1_\s- \tilde J^1_\s)
J^{-}_{{\mbox{\tiny{$U$}}}}= \frac{1}{2}(\pi_1 -\phi'_1) -\frac{1}{2} (\p_{\pi_1} - \p_{\phi'_1}) \H_R^{[0]}=0 \label{achircJ1m}
\ee
which ultimately imply that $\H^{[0]}$ has  a decoupled free boson subsector, $\H^{[0]} = \frac{1}{2} (\pi_1^2 + \phi_1'^2) + \H^{[0]} (other)$. One also  assumes  that the undeformed Hamiltonian has dimension $2$. %\emph{Check need!}}

We first consider deforming the Hamiltonian by the $J^{sh} \bar T$ operator % $%J_1\bar T = J^1_\s T_{t V}- J_t^1 T_{\s V}$
% = \frac{\pi_1}2 (\H - \pi_k \p_{\pi_k} \H - \phi'^{ k} \p_{\phi'^{ k}} \H + \p_{\pi_k} \H \p_{\phi'^k} \H)- \frac{\p_{\phi_1'} \H}2 (\H - \P) 
constructed from the above current and the components of the stress tensor which,  in terms of the deformed Hamiltonian density, are given  by \eqref{stresstcompham}. The flow equation for the Hamiltonian density is %\textcolor{red}{\emph{Minus..}}
 
\be
\p_\l \H = - \O_{
J^{sh}\bar T } = J^{sh}_t T_{\s V} - J^{sh}_\s T_{t V}\label{plhjtb}
\ee
In terms of the right-moving energy density $\H_R = \frac{1}{2} (\H - \P)$, the expression for $J^{sh} \bar T$ simplifies to

\be
\O_{J\bar T} = 2(\pi_1 \p_{\pi_k} \H_R \p_{\phi'_k} \H_R -  \H_R\, \p_{\phi'_1} \H_R  )
\ee
resulting in a flow equation for $\H_R$ only.
One can then use this explicit expression to show that, at least in the single-field case,   the $VU$ component of the stress tensor  will stay zero along the flow, if it starts out being zero. This implies that the $J\bar T$ flow preserves the initial $SL(2,\mathbb{R})_L$ symmetry, as expected \cite{Guica:2020uhm}.  
%
%As explained in the previous subsection, the $J\bar T$ deformation preserves left conformal invariance, so we expect 
%% 
%\be
%T_{VU}= -\H_R  + \pi_k \p_{\pi_k} \H_R + \phi'^k \p_{\phi'^k} \H_R + \p_{\pi_k} \H_R \p_{\phi'^k} \H_R \label{tvuconstr}
%\ee
%to be zero along the flow.  At least in the single-field (i.e., $J\bar T$ - deformed free boson) case, this is indeed the case, as  $T_{VU}$ can be shown to obey 
%% 
%\be
%2\p_\l T_{VU} + \pi \p_\pi \H \, \p_{\phi'}  T_{VU}+\pi \p_{\phi'} \H \,\p_\pi  T_{VU} -\pi^2 \p_\pi  T_{VU} -\H \, \p_{\phi'}  T_{VU} +\pi \,  T_{VU} -\p_{\phi'} \H \, T_{VU}=0
%\ee
%which shows that if $T_{VU}$ 
%\textcolor{blue}{We expect  an appropriate generalization of this equation to multiple fields to hold as well. \emph{Was proof for just a single field?}} 

The authors of \cite{Guica:2020uhm} also showed that, if one assumes that $\H_R$ in the deformed theory is a function of only $\l^2 \H_R^{[0]}$ and $\l \J_+$
%
%
%To solve the two equations \eqref{plhjtb} and \eqref{tvuconstr}, we make the Ansatz 
%% 
%\be
%\H_R =\frac{1}{\l^2} F \left(\l^2 \H_R^{(0)},  \l \J_+\right) \equiv \frac{1}{\l^2} F (x,\a)\;, \;\;\;\;\;\;\;\;\J_+ = \frac{\pi_1+ \phi'_1}{2} \label{HRansatz}
%\ee
- an Ansatz  motivated by dimensional analysis and the universality of the deformation - then the right-moving Hamiltonian density that solves the $J\bar T$ flow equation is given by 

\be
\H_R = \frac{2}{\l^2 \hat k} \left( 1 - \l \J_+ - \sqrt{(1-\l \J_+)^2 - \l^2 \hat k \H_R^{[0]}} \right) \label{solhr}
\ee
where we have reinstated an arbitrary level by rescaling $\l \r \l \sqrt{\hat k}$ and $\J_+ \r \J_+/\sqrt{\hat k}$ in the $\hat k=1$ result, while replacing $\H_R^{[0]}$ by the corresponding Hamiltonian density in the theory with level $\hat k$.

Note that, as in the case of $T\bar T$, this has the same functional form as the deformed energy formula \eqref{moder}, but applied at the level of the currents, rather than that of the global conserved charges. %\textcolor{blue}{Since   \eqref{eq:kutasov} only depends on $E_R^{(0)}$ and $Q^{(0)}_+$, we consequently assume that $\H_R$ only depends on $\H_R^{(0)}$ and $\J_+$. \emph{Was this actually used?}} 
%[Thus, the Hamiltonian of a rather general $J\bar T$ - deformed CFT is given in terms of the undeformed Hamiltonian and the $U(1)$ current via precisely \eqref{HRjtb}, where now $\H_R^{(0)}$ is the right-moving Hamiltonian  of the undeformed CFT and $\J_+ $ is given in \eqref{HRansatz}. ]
%
%\be
%\H_R = \frac{2}{\l^2} \left(1-\l J^{(0)}_+ -\sqrt{ \left(1-\l J^{(0)}_+\right)^2-\l^2 \H_R^{(0)}} \right)
%\ee
%It is not hard to check that the free-field deformed Hamiltonian takes precisely this form. 
%
%This expression also agrees with the general deformed energy formula \eqref{eq:kutasov} (obtained for constant current densities) if we set $k=2\pi$ in our conventions. 
 %{\color{ForestGreen}(is it true that here we assume $H_R^{[0]}$ does not also rescale i.e. no sugawara?)}
Since \eqref{solhr} is symmetric in $\pi_1\leftrightarrow\phi_1'$,  the $J\bar{T}$ deformation defined using $J^{sh}$ leads to the same (classical) deformed theory as the one defined using $J^{top} $. %\textcolor{red}{True this is the reason? $\H_R$ need not be symmetric...} The same holds if we define the classical Hamiltonian deformation using the chiral $U(1)$ current $K_\a$. \emph{\textbf{Check!!}} 
From here on, we will be exclusively using the latter definition, as the flow operator  and the charges  we construct starting with the next subsection do depend on the choice of current.  %{\color{ForestGreen}(so I removed the tildes in this section)} % In the next section,  we will be using $\tilde{J}\bar{T}$ in order to compute the flow operator for the energy eigenstates. 

%[Since it will appear a lot, it is useful to introduce a notation for:
%\begin{align}
%F:=\sqrt{(1-\lambda\mathcal{J}_+)^2-\lambda^2\mathcal{H}_R^{(0)}}=1-\lambda\mathcal{J}_+-\frac{\lambda^2\mathcal{H}_R}{2}]
%\end{align}
Given this solution, we can compute all components of the stress tensor and the remaining currents using \eqref{stresstcompham} and  \eqref{solhr} and re-express them in terms of $\J_\pm, \H_R, \mathcal{P}$.  For example, the components of the currents $K_\a, \bar K_\a$ introduced in \eqref{chircurjtb}, \eqref{kbarfb} are  %{\color{ForestGreen}(I think that for consistency here we should set $k=1$)} 
%
%\textcolor{red}{I think there should have been factors of $k$ in the Hamiltonian.}
%
\be \label{comprmcurrent}
K_t = K_\s =  \J_+ + \frac{\l  \hat{k}}{2} \H_R  \equiv \mathcal{K}_L \;, \;\;\;\;\;\;\; \bar K_t =  \J_- + \frac{\l  \hat{k}}{2} \H_R  \equiv \mathcal{K}_R\;, \;\;\;\;\; \bar K_\s =  K_\s - J^{top}_\s
\ee
where the new notation for the time components of the two currents has been introduced to harmonize with the notation for the Hamiltonians, and
\be \label{comptopcurrent}
J^{top}_{\sigma}=\mathcal{J}_+-\mathcal{J}_-+\frac{2\mathcal{J}_-+\lambda \hat{k}\mathcal{H}_R}{1-\lambda\mathcal{J}_+-\frac{\lambda^2  \hat{k}}{2}\mathcal{H}_R} \;, \;\;\;\;\;\; T_{\sigma V}=\frac{2\mathcal{H}_R}{1-\lambda\mathcal{J}_+-\frac{\lambda^2  \hat{k}}{2}\mathcal{H}_R}-\mathcal{H}_R
\ee
It is also interesting to  note  that the combination
\be
\H_L  - \frac{1}{\hat k} \, \mathcal{K}_L^2  =  \H_L^{[0]} - \frac{1}{\hat k} \,  \J_+^2
\ee
 is $\l$ - independent. Thus, we see that rather generally, the $J\bar T$ deformation induces a type  of  spectral flow in the left-moving sector.  In the right-moving one, one obtains instead \cite{Guica:2021pzy}
 
 \be
\H_R (1-\l \phi') - \frac{1}{\hat k} \, \mathcal{K}_R^2  =  \H_R^{[0]} - \frac{1}{\hat k} \,  \J_-^2
\ee
which simply follows from the previous relation, using $\mathcal{K}_L-\mathcal{K}_R = \J_+-\J_-=\phi'$.

Since $\H_R$ is determined by the undeformed currents via \eqref{solhr}, all Poisson 
 brackets of the various deformed currents are entirely determined by the undeformed ones and are universal, even if somewhat cumbersome. See e.g. \cite{Guica:2020uhm} for a comprehensive list of commutators. 
Using these universal Poisson brackets, 
one may easily check that  %\textcolor{red}{\emph{Is this a good place to have this discussion?}}

\be
 \{H, \H_L\} = - \p_\s \H_L\;, \;\;\;\;\; \{H,\mathcal{K}_L\}=-\partial_{\sigma} \mathcal{K}_L
  \ee 
which represents the conservation of   these holomorphic currents in Hamiltonian language.  Here, $H = \int_0^R d\s \H$ is the total Hamiltonian. Using this, one can also show that
%%This continues to hold  also if we multiply them by functions of $U$. Thus
\begin{align}\label{leftchrg}
Q_f \equiv \int_0^R d\sigma f(U) \, \mathcal{H}_L \;, \hspace{1cm}P_{\eta}\equiv \int_0^R d\sigma \, \eta(U)\, \mathcal{K}_L  %\left(\J_+ + \frac{\l \hat{k} \H_R}{2}\right)
\end{align}
are conserved  charges for any periodic $f, \eta$: $Q_f$ represents left-moving conformal symmetries, while $P_{\eta}$ generates left-moving affine transformations. 

On the other hand, the Poisson brackets of the Hamiltonian with the right-moving currents  are 
%{\color{ForestGreen}(same comment as before about k)}
%
\be
\{H,\mathcal{H}_R\}=\partial_{\sigma}\left( \mathcal{H}_R \frac{1+\lambda\mathcal{J}_++\frac{\lambda^2  \hat{k}}{2}\mathcal{H}_R}{1-\lambda\mathcal{J}_+-\frac{\lambda^2  \hat{k}}{2}\mathcal{H}_R}\right)\;, \;\;\;\;\;\; \{H, \mathcal{K}_R\}=\partial_{\sigma}\left(\mathcal{K}_R \frac{1+\lambda\mathcal{J}_++\frac{\lambda^2  \hat{k}}{2}\mathcal{H}_R}{1-\lambda\mathcal{J}_+-\frac{\lambda^2  \hat{k}}{2}\mathcal{H}_R}\right)
\ee
It is interesting to note that the combination that appears in parenthesis can be written as

\be
\frac{1+\lambda\mathcal{J}_++\frac{\lambda^2  \hat{k}}{2}\mathcal{H}_R}{1-\lambda\mathcal{J}_+-\frac{\lambda^2  \hat{k}}{2}\mathcal{H}_R} = \frac{1- \l \{ H, \phi\}}{1-\l\,  \p_\s\phi }= - \frac{\p_t v - \{ H, v\}}{\p_\s v}  \label{reqv}
\ee
where $v$ is the $J\bar T$ field-dependent coordinate \eqref{uvcoojtb}. This  is the Hamiltonian counterpart of the observation  \eqref{specialKb} made in the Lagrangian formalism, and is related to the construction of conserved  charges for field-dependent symmetries.  Interestingly, there are several possible choices for $v$ in the last term, related to different field-dependent choices of constant in \eqref{uvcoojtb}. Besides the na\"{i}ve possibility  $v_{naive} = V - \l \phi$, one can also add a multiple  of the spatio-temporal zero mode of $\phi$, $\varphi_0$, which satisfies  $\p_t \varphi_0 - \{H,\varphi_0\} = \p_\s \varphi_0=0$ and whose expression was worked out in \cite{Guica:2020eab}. As we will later discuss, the na\"{i}ve choice leads to a charge algebra that is inconsistent with charge and momentum quantization, while subtracting the zero mode of $v_{naive}$ leads to a consistent one. %\footnote{\textcolor{blue}{Note this notation is different from that of \cite{Guica:2020eab,Guica:2021pzy}, where $\bar Q_{\bar f}, \bar P_{\bar \eta}^{(KM)}$ denoted the (inconsistent) right-moving charges built from $v_{naive}$, while the  ones built from the correct field-dependent coordinate were denoted with calligraphic script: $\bar{\mathcal{Q}}, \bar{\mathcal{P}}$. %, with an additional tilde if these represented  flowed generators. Here, we have decided to drop this cumbersome notation, as only the charges whose action on phase space is consistent will appear, as well as only currents constructed from $\bar K_\a$, to which the `KM' superscript referred.}}.
%\begin{align}
%\bar{Q}_{\bar{f}}&=\int_0^R d\sigma \bar{f}(v)\mathcal{H}_R \hspace{1cm}\bar{P}_{\bar{\eta}}=\int_0^R d\sigma \, \bar{\eta}(v)\,  \mathcal{K}_R \label{RMcharges}
%\end{align}
%which are conserved for arbitrary periodic $\bar f(v),\bar{\eta}(v)$, provided the field-dependent coordinate $v$ satisfies 

%{\color{blue}As already discussed in the Lagrangian context, the obvious solution $v_{naive} = V - \l \phi$ does not lead to consistent commutators. However, the solution to \eqref{reqv} above is ambiguous up to the addition of a zero mode  satisfying  $\p_t \varphi_0 - \{H,\varphi_0\} = \p_\s \varphi_0=0$, for which  showed there is a non-trivial solution. Appropriately subtracting this zero mode from $v_{naive}$ leads to consistent commutators for the generators \eqref{RMcharges} that, in particular, respect charge and momentum quantization. }

\bigskip

\noindent \textbf{\emph{III. $JT^a$ - deformed QFTs}}

\medskip

\noindent  Let us now turn to the study of $JT^a$ - deformed QFTs,  which can be defined via an irrelevant SZ flow

\be
\frac{\p S_{JT_a}}{\p \l^a}=  \int d^2 \sigma \, e \, T^\a{}_a\,  \e_{\a\b} J^\b \label{eq:JTdef}
\ee
 where the coupling constant  $\mu^a$ is a vector of \emph{fixed} direction, but varying amplitude. This definition is taken to hold in both Euclidean and Lorentzian signatures, with the Euclidean and Lorentzian couplings  related via an appropriate analytic continuation.

We will concentrate  on  the finite-size spectrum and  the S-matrix, as well as the relation between them via TBA or via a  non-perturbative definition of the $JT^a$ deformation. As we will see, most relations are the direct counterpart of those discussed in the $T\bar T$ case, with the only major difference to adapt to being the lack of Lorentz invariance of these theories.  The general results on the  $J\bar T$ - deformed spectrum that we advertised earlier  can be easily derived in this context,  by simply setting the deformation parameter to be lightlike. 

%No particular gain by considering $J\bar T$, since initial theory is not conformal.  To explore connection between spectrum and S-matrix, as well as 

\subsubsection{General finite-size spectrum of $J T^a$ - deformed QFTs}

Consider an Euclidean $JT^a$ - deformed QFT on a circle of circumference $R$. The $JT^a$ coupling is a tangent-space vector, $\l^a$, which in general has both a spatial and a time component that could be varied independently. One thus obtains two sets of SZ flow equations \eqref{engflow} for the energy levels  %\textcolor{red}{\emph{Signs!}}

\be
\frac{\p E_n}{\p \l^{\tau_E}} = - R \bigl( \langle J_\tau \rangle \langle T_{\s \tau} \rangle - \langle J_\s \rangle \langle T_{\tau \tau} \rangle   \bigr)  \;, \;\;\;\;\;\;\;\;\frac{\p E_n}{\p \l^{\s}} =  R \bigl( \langle J_\tau \rangle \langle T_{\s \s} \rangle - \langle J_\s \rangle \langle T_{\tau \s} \rangle   \bigr)
\ee
which should determine the dependence of the finite-size deformed spectrum on both $\l^\tau, \l^\s$.  %\textcolor{red}{\emph{Do these collapse to a single equation in $J\bar T$?}} 
How to do this was first understood in  \cite{LeFloch:2019rut}, via a method based on coupling to background fields. Similar flow equations were obtained in \cite{Frolov:2019xzi} using uniform lightcone gauge.

To proceed, we first need to express all components of the stress tensor and the $U(1)$ current in terms of the conserved quantities associated to the state. 
The $\tau\tau, \tau\s$ and $\s\s$ components of the stress tensor are related to the finite-size energy and momentum as in \eqref{onepfep}, while the time component of the current is related to the $U(1)$ charge as in \eqref{expjt}. 
%\be
%E = - \int_0^R \!\! d\phi \, \langle T_{\tau\tau} \rangle\;, \;\;\;\;\;\; P = -i  \int_0^R \!\! d \phi \, \langle T_{\tau\phi} \rangle \;, \;\;\;\;\;\; Q = - i \int_0^R \!\! d\phi \, \langle J_\tau \rangle~,
%\ee
 %\footnote{We collect our conventions here. In any translationally invariant state $|n\rangle$: \begin{align}\langle n | T_{\tau\tau} | n \rangle = -\frac{E_n}{R} \;, \;\;\;\;\;\; \langle n | T_{\phi\phi} | n \rangle =- \frac{\p E_n}{\p R} \;,\;\;\;&\;\;\;\; \langle n | T_{\tau \phi} | n \rangle = \frac{i P_n}{R} \;, \;\;\;\;\;\; \langle n| T_{\phi\tau} |n \rangle =\frac{1}{R}\frac{\p E_n}{\p v}\;,\nonumber\\\langle n | J_{\tau} | n \rangle = \frac{ i Q_n}{ R}\;,\;\;\;\;\;&\;\;\;\;\; \langle n| J_{\phi} |n\rangle =\frac{1}{R}\frac{\p E_n}{\p a_\phi}~.\end{align}} 
The momentum is quantized in units of $2\pi/R$, where $R$ is the size of the spatial circle of the torus, whereas $Q$ is assumed to be integer quantized\footnote{Note that in the case of the $J\bar T$ deformation, the current whose charge is conserved  will not be chiral \cite{Chakraborty:2018vja,LeFloch:2019rut}. }. 
%P = \frac{2\pi p}{R} \;, \;\;\;\; Q,p \in \mathbb{Z}~.\label{eq:qpquant}
%\ee
 %
Since the $JT_a$ deformation explicitly breaks  Lorentz invariance, the conserved stress tensor is not symmetric along the flow.  Thus, we need a way  to also keep track of the $\s\tau$ component of the stress tensor, as well as of the spatial component of the current.

In  \cite{LeFloch:2019rut}, it was   proposed to introduce a coupling to an explicit  background  (Euclidean) vielbein $v_E$  and to a background gauge field $a^\s$  (coupling to the operator $J_\s$ in the Hamiltonian) - both constant - so that
\be
\langle T_{\s\tau} \rangle =\frac{1}{R}\frac{\p E}{\p v_E} \;, \;\;\;\;\;\;\; \langle J_{\s} \rangle =\frac{1}{R}\frac{\p E}{\p a^\s}
\ee 
%
 %that the energy levels will depend on, so that 
%
The energies will thus depend  on both deformation parameter(s) and the  background fields, namely

\be
E = E \left(\l^a_E, R,v_E,a^\s\right)~.
\ee
%, and subsequently derive the flow equations for the energy levels. %, which we then match with those previously derived in the literature, and more specifically in  \cite{Frolov:2019xzi}.
%
The background vielbein can be understood as being induced by a coordinate transformation acting on the tangent space indices %\textcolor{red}{\emph{Previous discussion?}} 
$x^a \r x^a + e^a{}_{\hat a} x^{\hat a}$, with 

\be
e^a{}_{\hat a} = \left(\begin{array}{cc} 1 & 0\\ v_E & 1 \end{array} \right)
\ee
where the metric in the hatted coordinates is the standard euclidean metric $\d_{\hat a \hat b}$. Since the standard results \eqref{onepfep} relating the stress tensor components to the energy and momentum of the state hold for the diagonal coordinates $x^{\hat a}$, we find an effective shift $T_{\tau \s} \r T_{\tau \s} +  v_E T_{\tau \tau}$ and $T_{\s \s} \r T_{\s \s} +  v_E T_{\s \tau}$ in the expectation values that enter the flow equations. Finally, one obtains\footnote{See also \cite{LeFloch:2019rut} and \cite{Anous:2019osb} for a discussion of the various conventions.} \cite{Frolov:2019xzi} 
\begin{align}
\frac{\p E_n}{\p\l^\tau_E}&=\frac{1}{R}\left[ -i Q_n\frac{\p E_n}{\p v_E}-E_n\frac{\p E_n}{\p a^\s}\right]~,\label{eq:mt}\\
 \frac{\p E_n}{\p\l^\s_E}&= -i Q_n\left(\frac{\p E_n}{\p R}-\frac{v_E}{R}\frac{\p E_n}{\p v_E}\right)+\frac{ i P_n +v_E E_n}{R}\frac{\p E_n}{\p a^\s}~,\label{eq:mp}
\end{align}
Notice that, despite appearances, the above flow equations are  real, as $\l^\s_E$ and $v_E$ are purely imaginary. %\textcolor{red}{\emph{Why?}} 
This can be made manifest by  letting $\l^\s_E =  i \l^\s$, $\l^\tau_E = \l^t$ and $v_E =- i v $, upon which we obtain

\be
\frac{\p E_n}{\p\l^t}=\frac{1}{R}\left[ Q_n\frac{\p E_n}{\p v}-E_n\frac{\p E_n}{\p a^\s}\right]\;,\;\; \;\;\;\;\;\frac{\p E_n}{\p\l^\s}=  Q_n\left(\frac{\p E_n}{\p R}-\frac{v}{R}\frac{\p E_n}{\p v}\right)+\frac{ - P_n +v E_n}{R}\frac{\p E_n}{\p a^\s} \label{flowlor}
\ee
Notice the above flow equations should hold in an arbitrary QFT, since their definition is insensitive to whether the seed theory is a CFT or a generic QFT, even a non-relativistic one.   A  more complicated set of flow equations for combined $T\bar T, JT^a, \tilde J T^a$ deformations is discussed in \cite{Frolov:2019xzi} (see also \cite{LeFloch:2019rut}).

 %Since we only consider deformations involving the current $J$, but not also the dual current $\tilde J_\a = \e_{\a\b} J^\b$, we set to zero the tilded conserved charge and its associated spacelike background field ($\tilde P_1 =0, u_1=0$ in Frolov's notation). Then, the flow equations  found in \cite{Frolov:2019xzi} simplify to \emph{Explain change of variables $\tilde{\mathbb{P}} \r a_\phi$. } 
%

%where we have not set to zero any anomaly terms, which would be obtained by setting the coefficients proportional to $a_\phi\rightarrow 0$.
%Translating  equations (\ref{eq:mt}-\ref{eq:mp}) into an equation on the partition function \eqref{ztrh}, we find precisely the flow equations \eqref{floweqsphys}. Thus, our flow equation agrees perfectly with the flow equations previously found in the literature for this particular case, which is a consistency check of the general definition of $JT_a$ deformed QFTs  we guessed (\emph{Couldn't we have derived it a la Cardy?}). 

To solve the above flow equations and obtain a solution for the deformed energies as a function of the undeformed ones and the other conserved charges, one proceeds in two steps.
The first step is to notice that, as was the case for $T\bar T$, the general solution to equations \eqref{eq:mt} - \eqref{eq:mp} is given simply by shifting the parameters according to   
\be\label{eq:edefiningeq}
E_n (\l^a,R, v,a^\s)=E_n^{[0]}\left(R+\l^\s Q_n,\frac{v R+\l^t Q_n}{R+\l^\s Q_n},a^\s+\frac{{\l}^\s(P_n+v E_n)-\l^t E_n}{R+\l^\s Q_n}\right)
\ee
%where from now on we work in terms of the Lorentzian parameters, which  are manifestly real, and we switched the sign of $P$ in \eqref{flowlor}, so as to match the conventions in the literature. \textcolor{red}{\emph{Keep?}}
Notice that if we write the solution not in terms of $R,v, a^\s$, but in terms of $R, Rv, a^\s R$ which are the combinations that enter more naturally the partition function of the theory \cite{Anous:2019osb}, the above shifts simply correspond to 

\be
R \r R + \l^\s Q \;, \;\;\;\;\; v R \r v R + \l^t Q \;, \;\;\;\;\; R a^\s \r R a^\s - \l^t E + \l^\s P + \l^\s v E \label{niceshifts}
\ee
%\textcolor{blue}{In section \ref{sec:smatrixtba}, we will see how these simple shifts are reproduced from the TBA equations.  }
%
Thus, if we know $E_n^{[0]}$ as a function of general background parameters, \eqref{eq:edefiningeq} provides us with an algebraic equation for the deformed energy levels, $E_n$. The second step consists in finding these expressions for $E_n^{[0]}$, which have been worked out explicitly in e.g. \cite{LeFloch:2019rut} for the special case of a seed CFT %
%A simple way to understand it is to notice that in absence of a vielbein but in presence of an external gauge potential $a_{\hat \s}$, the expression for the energy levels in a CFT on a circle of circumference $R$ is given by \textcolor{red}{\emph{Explain $a_{\hat \s}$ or remove}}
%
%\be
%E_n^{[0]} = \frac{2\pi\Delta_n}{R} + \frac{k R}{16 \pi} a_{\hat \s}^2
%\ee
%where $\Delta_n \in \mathbb{R}$ is the associated conformal dimension in the CFT on the plane and $k$ is the coefficient of the chiral anomaly. The second term represents the shift in the energy levels due to the external gauge potential. Once we turn on a background vielbein $v$, the expression becomes \cite{LeFloch:2019rut}: 

 \be
 E_n^{[0]} (R,v,a^\s)=\frac{1}{1-v^2} \left(\frac{2\pi \Delta_n}{R}+\frac{2\pi p_n}{R}v+ Q_n v \, a^\s  + \frac{k R}{16 \pi} (a^\s)^2 \right) \label{zeropeng}
 \ee
 where $\D_n $ is the associated conformal dimension in the CFT on the plane, $p_n = P_n R/2\pi$ is the number of momentum quanta and $k$ is the coefficient of the chiral anomaly. In absence of the background vielbein  $v$, the last term simply represents the shift in the energy levels due to the external gauge potential.

%where we have plugged in the expression \eqref{eq:qpquant} for $P_n$ and the third term can be understood as a shift in $a^\tau$ proportional to $v a_{\hat \phi} = v a^\phi $. 

Since we are interested in the spectrum of the deformed QFT with all background fields switched off, we should set $v = a^\s =0$ in \eqref{eq:edefiningeq}, which determines the potentials in the undeformed theory.
%
% $\mu^t = - \mu_t, \mu^\phi = \mu_\phi$. 
 Plugging \eqref{zeropeng} into \eqref{eq:edefiningeq} generically yields a quadratic equation %\footnote{Notice also that in the case $\mu_t =0$, the quadratic equation reduces to a linear one. \textcolor{red}{\emph{Keep?}}}
  for  the energy levels, %\footnote{More generally, ~~$k \gamma^2E_n^2+\left[\gamma(Q_n v-2kd)-\bar{R}\left(1-v^2\right)\right]E_n+E^{(0)}_n R+b\left(v+\frac{\gamma Q_n}{\bar{R}}\right)+kd^2=0$, where \be\gamma\equiv \mu^t -v\, \mu^\phi~,\quad\quad \bar{R}\equiv R+\mu^\phi\,Q_n~,\quad\quad d\equiv a^\phi\,\bar{R}+\mu^\phi P_n~,\quad\quad b\equiv R\,P_n + d Q_n~.\ee}: 
%
%\be
%\frac{k \mu_t^2  }{8 \pi} E_n^2 - 2   \left(R + \mu_\phi Q_n   -\mu_t \mu_\phi\, \frac{ k P}{8\pi}  \right) E_n + 2 E^{(0)}_n R + \frac{k \mu_\phi^2}{8\pi}  P_n^2 - 2 \mu_t P_n Q_n   =0
%\ee
with solution %\emph{same as Tarek with $P \r - P$ and $k \r k/(16\pi)$}
%
%We quote here the solution to \eqref{eq:quaden} in the limit $v\rightarrow 0$ and $a_\phi\rightarrow 0$ as it manifestly matches results found readily in the literature in more specific cases \cite{LeFloch:2019rut,Frolov:2019xzi,Guica:2019vnb}
\be
E_n = -\frac{\l_\s}{\l_t}P_n+\frac{8\pi}{k \l_t^2} \left[R + \l_\s Q_n - \sqrt{(R+\l_\s Q_n)^2-\frac{ k \l_t}{4\pi}  \left( R (\l_t E^{(0)}_n +\l_\s P_n)+P_n Q_n (\l_\s^2-\l_t^2)\right)}\,\right]
\ee
where $\l_t = - \l^t$ and the sign in front of the square root is fixed by the requirement that the spectrum reduces to the undeformed one as $\l_{t,\s} \r 0$. The spectrum  of the $J\bar{T}$ or $\bar{J}T$ deformations is then obtained by taking $\l_t=\mp \l_\s = \l$.  %\textcolor{red}{\emph{Why is $Q_n$ the chiral initial charge? Because of coupling in $E^{[0]}$? This should also agree with the CS analysis, but it doesn't.}}

It is interesting to note that the spectrum's dependence on the anomaly coefficient $k$ descendes entirely from the $k$ dependence of the undeformed CFT energy \eqref{zeropeng} in presence of the background fields, as the flow  equations \eqref{flowlor} do not contain any explicit factor of $k$. The same is true for the flow of the partition function we study in the following subsection\footnote{ This is quite different from the flow equation for $J\bar T$-deformed CFTs obtained by \cite{Aharony:2018ics}, which depends on $k$ in a rather complicated fashion. One possible reason for the different  flow equation we find is that while in \cite{Aharony:2018ics} the partition sum was over 
 the charge associated with the strictly chiral current, which satisfies a non-trivial flow equation itself, our partition function sums instead over an integer-quantized charge, which is associated with a possibly non-chiral current. One way to ensure that the current is exactly  chiral is to consider,  as in \cite{Frolov:2019xzi},  a joint $JT_a$ and $\tilde J T_a$ deformation with equal coefficients, where $\tilde J = \star J$ is the dual current to $J$. In this case,  the flow equations for the energy
  start depending explicitly on the anomaly coefficient \cite{Frolov:2019xzi,LeFloch:2019rut}, and likely the same is true of the flow equations for the partition function. }.  Note also that for large enough values of $|\l^a|$, the energies tend to become imaginary, raising questions about the existence of these theories all the way up into the UV\footnote{Our point of view on this issue is that the imaginary energies may be just be an artifact of (illegally) putting the theory at finite volume, see \cite{Cooper:2013ffa} for relevant comments.   }.

\subsubsection{Non-perturbative definition of $JT^a$ - deformed QFTs}

So far, we have been defining the $JT^a$ deformations via an irrelevant SZ flow. In view of the above comment
one may wonder whether, as was the case of $T\bar T$, there may exist an alternate,  non-perturbative definition of $JT_a$-deformed QFTs that simultaneously makes manifest their UV completeness and correctly reproduces the deformed spectrum. This is indeed the case, as shown in \cite{Anous:2019osb}, 
%
% As we already reviewed, in the  $T\bar T$ case this  coupling  was mainly designed to implement the dynamical change of coordinates between the analogues of conformal and static gauge. It turns out that  in the case of $J\bar T$, the effects of the deformation are well captured by a combination of a gauge transformation and  a coordinate change that only affects half of the components of the  vielbein \cite{Bzowski:2018pcy}.  It is thus natural to
who proposed a path integral definition of the $JT^a$ deformation in which  the original QFT is coupled to a flat external gauge field and `half' a flat dynamical vielbein.%\footnote{Recently, \cite{Aguilera-Damia:2019tpe} studied a path integral realisation of joint $J\bar T$, $T \bar J$ and $T\bar T$ deformations, where $J$ and $\bar J$ are  independent currents, in terms of coupling to a \emph{full} topological vielbein and \emph{two} independent gauge fields. This kernel is different from ours - roughly, it is a doubling of what we find - and it reflects the fact that in the proposal of \cite{Aguilera-Damia:2019tpe} it is impossible  to turn off one of the $J\bar T$ or $T\bar J$ couplings in the intermediate steps of the calculations. We comment more on the relation to our proposal in footnote \ref{ftnt:otherargentine}. } 

To motivate this, one starts from \eqref{eq:JTdef} and follows the Hubbard-Stratonovich procedure by coupling the $U(1)$ current to an external gauge field $\mathrm{a}_\a$ and  the stress tensor to  an external vielbein, see also section \ref{holodictjtbdtr}. Only the vielbein components parallel to the deformation  end up playing a role, so we can restrict the path integral to this  `half'  of the vielbein. It is clear from \eqref{generalmapdefundef} and \eqref{topcur} that, if one starts with a flat metric and zero gauge field, the saddle point values of these external fields will be pure gauge. % the only modes that survive in the path integral are the pure gauge modes of $\mathrm{a}_\a$  and those of the parallel vielbein. However, 
The strategy adopted in  \cite{Anous:2019osb} was  to simply guess that this would be the case and then show that the QFT so defined satisfies all the consistency requirements expected of it.

 Introducing a pair of auxiliary fields to enforce the flatness conditions, one is led to the following proposal:  % \textcolor{red}{\emph{Perhaps already intro $||$ notation}}
\be\label{eq:jtdefguess}
S_{JT_a}= \int d^2 \sigma \, e \left[ \mathcal{L}_{\rm QFT} (\varphi, \mathrm{a}_\a, e^{~~a}_\a)+\frac{1}{\l}\,\e^{\a\b} (\p_\a X^{||} - e^{~||}_\a)(\p_\b \Phi - \mathrm{a}_\b)  + \varpi^\a (e_\a{}^\perp - \d^{~\perp}_\a) \right]
\ee 
where we have let $\l^a = \l\,  n^a$ with $n_a n^a = \pm 1 \equiv \e$, $ X^{||}  \equiv \e\, n_a X^a$ - and similarly for $e^{||}$ - and $\varpi^\a $ are Lagrange multipliers.
%
%the vectors $\Lambda_a$ and  $\Lambda_a^\perp$  satisfy
%\begin{equation}
% 	\Lambda_a \mu^a =1~,\quad\quad\quad \Lambda_a^\perp \mu^a = 0
% \end{equation} 
%.  It will be useful to introduce the unit vectors $n^a$, so that $\mu^a = \mu\,  n^a$, $\Lambda_a = \s \mu^{-1} n_a$, where $\s = n_a n^a = \pm 1$ depends on whether the deformation is timelike or spacelike.
%We also define
%
%\be
% X^{||}  \equiv \s\, n_a X^a = \mu \, \Lambda_a X^a\;, \;\;\;\;\;\;\; e_\a{}^{||} \equiv \s \, e_\a{}^a n_a= \mu\,  \Lambda_a e_\a{}^a \label{Xpar}
% \ee
% where $\mu$ is the modulus of the vector. The null case can be treated separately. 
%
In analogy with the $T\bar{T}$ deformation, the $\Phi$ and  $X^{||}$ equations of motion set $\mathrm{a}_\a$ and $e_\a{}^{||}$ to have vanishing field strength, $\p_{[\a} \mathrm{a}_{\b]} = \p_{[\a} e^{~~||}_{\b]} =0 $, whereas the $\varpi^\a$ Lagrange multiplier sets the components of the vielbein that are perpendicular to $\l^a$ to be  trivial or, more generally, fixed to some pre-specified value. The classical equations of motion for the fields $\Phi$ and $X^a$ arise from varying the action with respect to the background fields $\mathrm{a}_\alpha$ and   $e_\alpha{}^{||}$%\footnote{ Our conventions for the variations in Lorentzian signature are: $\d S_{\rm QFT}=\int d^2x\, e\left[ T^\mu_{~~a}\d e_\mu^{~~a}-J^\mu\d a_\mu\right]$. }
\begin{equation}
\label{eq:eoms}
 	\p_\a \Phi=\mathrm{a}_\a+\l^a\e_{\a\b}T^\b_{~~a}~,\quad\quad\quad\quad \p_\a X^{||}=e_\a{}^{||}+\l \, \e_{\a\b} J^\b~
 	% \left(J^\b~  - \frac{k}{4\pi} \,\e^{\b\gamma} \mathrm{a}_\gamma~ \right)~,
 \end{equation}
%where $k$  is the coefficient of a  potential  anomaly for the $U(1)$ symmetry, and the last term arises from the variation of the path integral measure with respect to $\mathrm{a}_\a$. 
Note  that,  thanks to the conservation equations for the stress tensor and the current, these equations are entirely consistent with the  flatness of the gauge potentials $\mathrm{a}_\a$ and $e_\a^{~~a}$.

The interpretation of $X^{||}$ and $\Phi$ is again that of dynamical coordinates on physical space and on the $U(1)$ gauge orbit. This allows  one to interpolate between the original QFT, where $\mathrm{a}_\a =0$ and $e^{~~a}_\a = \d^{~~a}_\a$, and the $JT_a$-deformed one,  where $\Phi =0$ and the worldsheet coordinates are  aligned with the target space ones, $\s^\a=X^a$. This coordinate transformation is also responsible for the   S-matrix dressing,  as we discuss below.

As a first check of this  proposal, one can compute the partition function of the deformed QFT which, just like in $T\bar T$, is one-loop exact and turns out to satisfy an intrinsically  geometric flow equation 
\be
\frac{\p \Psi}{\p \l^a} = {\e}_{\a\b}  \frac{\p^2 \Psi}{\p L_\a{}^{a} \p \nu_\b} \label{eq:jtflow1}
\ee
where $\Psi = Z_{JT_a}/\mathcal{A}$, where $\mathcal{A}$ is the area of the torus on which the euclidean theory lives.   The $L_\a{}^a$ parametrise the lengths of the torus,  while $\nu_\a$ is the chemical potential. This flow equation gives rise to precisely the flow equations  \eqref{flowlor} for the energy levels, 
upon a proper identification of the parameters, thus confirming that the proposal is correct.

%\bigskip
%
%
%The above flow equations for the energy levels can be captured by a very simple and intuitive flow \cite{Anous:2019osb}  of the partition function of the $JT^a$ - deformed QFT, defined as usual via the Hilbert space trace 
%
%\be
%Z_{JT^a} = Tr \left[ e^{-\b E + i \theta P + i \eta Q} \right]
%\ee
%where $\b,\theta, \eta$ are the  chemical potentials that couple to energy, momentum and charge. Defining $\Psi = Z_{J T^a}/\mathcal{A}$, where $\mathcal{A}$ is the area of the torus on which the euclidean theory lives, $\Psi$ can be shown to satisfy 
%
%\be
%\frac{\p\Psi}{\p \mu^a}  = \e_{\a\b} \frac{\p^2 \Psi}{\p L_\a{}^a \nu_\b}
%\ee
%The $L_\a{}^a$ parametrise the lengths of the torus and are determined by $R, \b,\theta$ above, while $\nu_\a$ is the chemical potential, related to $\eta$ and the remanining parameters, as explained in \cite{Anous:2019osb}.
%
%\bigskip

The above general definition also allows us to compute the S-matrix dressing of $JT_a$-deformed QFTs, by an analogous calculation to that performed in \cite{Dubovsky:2017cnj} for $T\bar T$, which involves showing that the amplitudes defined with respect to $X^a$ are dressed versions of those defined with respect to the original QFT coordinates.  As expected, the deformed S-matrix only differs from the original one by a phase, schematically given by %\textcolor{red}{\emph{Notation!}}

\be
\mathcal{S}^{[\l^a]} (p_i^a,q_i)=e^{-i{\l_a} \sum_{i<j}\left(q_i p_j^a-q_j p_i^a\right)} \mathcal{S}_0(p_i^a,q_i) \label{drfac}
\ee
where $\mathcal{S}_0(p_i^a,q_i) $ is the S-matrix of the undeformed QFT, $p_i^a$ is the momentum vector of the $i^{\rm th}$ particle, $q_i$ is the its $U(1)$ charge, and the incoming (outgoing) particles are (anti-)ordered according to their rapidities. 
Assuming $\mathcal{S}_0$ describes a QFT with a CFT fixed point in the UV, the  exact expression above indicates that the deformed theory is also UV-complete, but no longer possesses a conventional UV fixed point.

\subsubsection{Change in the S-matrix via (mirror) TBA}

As already discussed in the $T\bar T$ case, if the QFT is integrable it should be possible to relate the scattering phase derived above to the change in the finite-size spectrum of the theory  via the TBA equations \cite{Zamolodchikov:1989cf}. In the case of a Lorentz-invariant theory, these equations are derived by interpreting the torus path integral of the QFT in two different ways in terms of the partition function of the Lorentzian theory. % on a torus whose sides have lengths  $R,L$ as either   the partition function of the   QFT on a circle of circumference $R$ and at a temperature $T=L^{-1}$, or as the partition function on a circle of size $L$ and at temperature $R^{-1}$. 

The main difference we need to take into account in considering $JT^a$ - deformed QFTs, which are not Lorentz-invariant, is that  the torus path integral  
now relates the partition functions of two in principle  different  Lorentzian  theories: that  of the original Lorentzian QFT on a circle of circumference $R$ and at a temperature $T=L^{-1}$, and that 
%
%%It is well known that in integrable theories, where scattering is elastic and the 
%%$2\r 2\,$ S-matrix elements   take the form of a phase shift $e^{i \d(\beta_i,\beta_j)}$, where $\b_i$ are the particle rapidities, the scattering phase can be related to the finite-size energy spectrum of the theory via the so-called Thermodynamic Bethe Ansatz (TBA) equations \cite{Zamolodchikov:1989cf}. 
%
%The derivation of these equations proceeds in two steps. First, one considers the partition function of the original Lorentzian QFT on a circle of circumference $R$ and at a temperature $T=L^{-1}$. This partition function can also
%be evaluated via an euclidean path integral on a torus whose sides have lengths $L$, $R$. The same path integral
%can be alternatively interpreted as the partition function 
%
%of an - in principle different - Lorentzian theory, called
 the so-called `mirror' theory on a circle of size $L$ and at temperature $R^{-1}$, where $L,R$ are the lengths of the two cycles of the torus.  Since they share the same euclidean path integral representation,
 %,  the original  and the mirror theory are related. 
%If the original QFT is relativistic, as considered in \cite{Zamolodchikov:1989cf}, then the mirror theory is identical to the original
%one. If, however, the original QFT is not Lorentz invariant, as will be the case for the $JT_a$ - deformed QFTs, then
the mirror theory is obtained via a double Wick rotation of the original one. We will label quantities such as the free energy or Hamiltonian of the mirror theory with tildes, to distinguish them from quantities in the original picture.
 As nicely explained in e.g. \cite{Arutyunov:2007tc}, this
double Wick rotation will in general affect the dispersion relation $E(p)$ of the asymptotic one-particle states; the mirror one can be obtained by the simple replacement $H \r i\tilde p$, $p \r i \tilde H$ in the dispersion relation of the original QFT.

 In the limit $L \r \infty$ , the partition function in the original picture is dominated by the finite-size ground state energy, $E_0(R)$, while in the mirror theory one considers the thermodynamic limit. The only difference with our previous analysis is that now one considers the statistics over mirror particles, with mirror anlogues of the quantization conditions \eqref{betheansatz} and  the pseudoenergy  \eqref{pseudoeng}. In presence of a chemical potential $\tilde \nu$ for the $U(1)$ charge in the mirror theory and one ($\tilde \theta$) for the mirror momentum, the latter is modified to (see \cite{Anous:2019osb} for further details)

\be
\varepsilon(\b)=   R m \cosh \b-i e\tilde \nu  - i \tilde \theta m \sinh \beta + \frac{1}{2\pi} \int d\b' \frac{\p \d(\b',\b)}{\p \b} \log\left(1-e^{-\varepsilon(\b')}\right)  \label{tba}
\ee
%which is defined via the ratio 
%%
%\be
%\frac{\rho_p}{\rho_l} \equiv \frac{1}{e^{\varepsilon}-1}~,
%\ee
%The free energy at the minimum turns out to only depend on this ratio, and reads  
%\be
%R \tilde f(R) =  \frac{m}{2\pi}  \int d \b\, \cosh \b \log \left(1- e^{-\varepsilon (\b)}\right)
%\ee 
%Using \eqref{eq:mirrorest}, this equals the vacuum energy $E_0(R)$ of the original theory.\footnote{While we have omitted it in the expression, it should be clear from \eqref{tba} that $\varepsilon$ depends on $R$.}  
%The charge in the original model is, similarly, simply given by \cite{Conti:2019dxg} 
%%
%{\be
%Q =  \frac{e}{2\pi} \int d\b  \log (1- e^{-\varepsilon (\b)})
%\ee}
%where we chose an appropriate normalization. 
%
In the case of $2 \r 2$ scattering, the additional shift \eqref{drfac}  in the scattering phase due to the  $JT_a$ deformation is %\textcolor{red}{\emph{Aren't these mirror variables?}}
%, where  the phase shift is given by \eqref{eq:jtdressing} 
%
\be 
\d_{JT_a}(\b,\b') =  m e \l_t (\cosh \b-\cosh \b') + m e \l^\s (\sinh \b - \sinh \b')  \label{phshift}
\ee
%Notice $\d_{JT_a}(\b,\b')$ is antisymmetric. Assuming the original S-matrix of the integrable QFT we deform is given by $e^{i \d (\b,\b')}$, the $JT_a$ deformation induces an additional phase shift $\Delta \d (\b,\b') = \d_{JT_a} (\b,\b')$ into the TBA equation \eqref{tba}. 
Plugging into \eqref{tba}, we find that the deformed pseudoenergy obeys
\bea
\varepsilon(\b) & = & R m \cosh \b- i \tilde \theta m \sinh \beta - ie\tilde \nu  - \frac{e m}{2\pi}  (\l_\s \cosh \b + \l_t \sinh \b )  \int d\b' \ln (1-e^{-\varepsilon(\b')}) +\nonumber \\
&& \hspace{15mm} +\,  \frac{1}{2\pi} \int d\b' \frac{\p \d^{[0]}(\b',\b)}{\p \b} \log\left(1-e^{-\varepsilon(\b')}\right)
\eea
where $\d^{[0]} (\b, \beta')$ is the phase shift in the undeformed (mirror) QFT. We immediately notice that the effect of the couplings $\l_{\s}, \l_t$ is to shift
{\be
R \r R - \l_\s Q \;, \;\;\;\;\;\; \tilde \theta \r \tilde \theta - i \l_t Q
\ee}
in the equation for the ground state energy. 
Noting that $\tilde \theta$ in the mirror picture simply corresponds to {$-v_E R = - i v R$} in the original one, this nicely reproduces the first two shifts in \eqref{niceshifts}. We used the fact that the mirror theory lives on a Minkowski background, so $\l^\s = \l_\s$ and $\l^t=-\l_t$.

The remaining shift in $ R a^\phi$ can be understood directly in the original theory, as being due to a constant shift in the quantization condition \eqref{betheansatz} for the particle momenta. This shift equals  the sum over all $\b'$  in the $JT^a$ phase shift 
\be
\d_{ct} = - e \l_t E -  e \l_\s P \label{ctshift}
\ee
where $E, P$ are the total energy and momentum in the original theory. The reason this shift can be interpreted as a shift in $a^\s$ is that a constant contribution to the momentum can be interpreted as a constant spatial Wilson line.

Notice that in the analysis above we have only exhibited the relation between the TBA and the shifts \eqref{niceshifts} for the ground state energy. However, as explained e.g. in \cite{Cavaglia:2016oda}, the excited levels can be easily captured using the same method.  %\textcolor{red}{\emph{Any additional literature on this?}}

%
%
%
%
%\subsubsection*{Generalised $T\bar T$}
%
%\bi
%\item relation b/w spectrum and S-matrix in the integrable theories (formal proof)?
%\ei
%\textcolor{blue}{In the original SZ paper, there is an argument that all CDD form factor deformations of an integrable S-matrix are related to the generalised $T\bar T$ deformations. The argument is roughly that, on the one hand, the change in the S-matrix is given by the variation of certain coefficients that multiply odd powers of cosh the rapidity. These coshes are in 1-1 correps with the spins of the integrals of motion. The variation in the S-matrix is also given by lowering the integrated deforming operator, sandwiched between insertions. The latter must give zero for inelastic processed. Looking also at integrable forms factor bootstrap, the argument was that the operator must be one of the integrability-preserving ones. \emph{Was this made precise in later papers, e.g. Tateo 19.04?}}

\subsection{Single-trace analogues of $T\bar T$ and $J\bar T$\label{strttbjtbsec}}

%\textcolor{blue}{They can also be defined via the flow equations, which can be shown to respect the SPO structure. In particular, degenerate states stay degenerate. \emph{Careful!!} If one knows the finite-size spectrum of the seed theory, then the one of the SPO simply follows from Bantay. Implies entropy.}

As detailed in the introduction, we are also interested in the single-trace versions of the $T\bar T$ and $J\bar T$ deformations. These simply correspond to the symmetric product orbifold (henceforth SPO) of  $T\bar T$ - or $J\bar T$ - deformed theories.
 
 Quite generally, the symmetric product orbifold  of some  QFT, $\mathcal{M}$, to which we will refer  as the `seed', corresponds to the $N$-fold tensor product $\mathcal{M}^N$, quotiented by the permutation group, $S_N$. The action - or Hamiltonian - of the theory simply corresponds to the sum of the actions/Hamiltonians in each copy. Consequently, a SPO of $T\bar T$ - deformed CFTs will satisfy the flow equation

\be
\p_\mu S = \sum_{i=1}^N \p_\mu S_i = \sum_{i=1}^N  T_i \bar T_i \label{defstttb}
\ee 
and similarly for $J\bar T$. Since the above is a single sum with respect to the  indices upon which the discrete gauge group $S_N$ acts, the appelative `single-trace' follows\footnote{ The first discussions of single-trace $T\bar T$ and $J\bar T$  deformations were in the string theory context \cite{Giveon:2017nie,Chakraborty:2018vja,Apolo:2018qpq}, where the relation to exact SPOs is somewhat delicate and discussed in sec \ref{holostr}. In these notes we stick to the  QFT definition above.  }. Conversely, one can start from a SPO of QFTs (or CFTs) and deform  it by the single-trace $T\bar T$ operator \eqref{defstttb}. Then, it is not hard  to show that the SPO structure of the seed theory is preserved,   so either definition can be used. 

The SPO procedure allows one to obtain new examples of theories that are UV complete, but  with a non-trivial and non-local UV behaviour. However, as in the case of standard (i.e., double-trace) SZ deformations, the structure of the deformed theory is highly rigid, being determined entirely by that of the undeformed theory \cite{Bantay:1998fy,Bantay:1999us,Bantay:2000eq,Chakraborty:2023wel}.
For example, the partition function of a symmetric product orbifold  is determined entirely by  that of the seed  theory,  
which in the case of SZ deformations is itself  fully determined by that of the undeformed QFT. Note one does not need conformal invariance for this to work - modular invariance is sufficient.  Knowledge of the partition function in turn  dictates the thermodynamics of the system, which only differs in minor ways from that of standard (double-trace) $T\bar T$ or $J\bar T$ - deformed CFTs. There are additional results on the symmetries and correlation functions. %, but these will be discussed in the dedicated subsections. 

%
%\bi
%\item comments on correlation functions? how only tori are ok?
%\item perhaps more credit given to worldsheet studies
%\ei

\subsubsection{The spectrum}

The Hilbert space of a symmetric product orbifold theory $\mathcal{M}^N/S_N$ is organized into twisted sectors \cite{Dijkgraaf:1996xw}, labeled by the conjugacy
classes of $S_N$. Each conjugacy class 
%
%\be
%\H (\mathcal{M}^N/S_N) = \oplus_{[g]} \H^{[g]}
%\ee
is entirely specified by the lengths $n$ and multiplicities $N_n$ of the cycles
of the permutation, with $\sum
 n N_n = N$.  The untwisted sector of
this Hilbert space, which corresponds to the conjugacy class of the identity, is simply $(\H_{seed})^N/S_N$ .
The twisted sectors are characterized by the basic fields having twisted boundary conditions around
the spatial cycle of the cylinder. %, and can be understood by mapping to corresponding covering spaces.

  Rather than considering
fields with twisted boundary conditions on the original cylinder, one can equivalently
work with fields with standard boundary conditions on the corresponding covering space. This is
particularly simple to implement for the torus partition function, as the relevant covering spaces
are again tori,  whose size and  modular parameters can be fully characterised using group-theoretical methods \cite{Bantay:1998fy,Bantay:1999us,Bantay:2000eq}. This allows one to express the partition function of the orbifold QFT solely in terms
of the seed partition function $Z^{seed}$.  The simplest way to  express this relation  is at the level of the grand canonical partition function of $S_N$ orbifolds

\be
\sum_{N=0}^\infty p^N Z^{S_N} (\tau, \bar \tau , R) = \exp  \left(\sum_{n=1}^\infty p^n \mathcal{Z}^{(n)} \right)
\ee 
where

\be
\mathcal{Z}^{(n)}  = \frac{1}{n} \sum_{\ell|n} \sum_{0\leq r <\ell} Z^{seed} \left( \frac{n \tau}{\ell^2} + \frac{r}{\ell},  \frac{n \bar \tau}{\ell^2} + \frac{r}{\ell} , R \ell \right) \label{defzn}
\ee
The expression for $Z^{S_N}$ in terms of $\mathcal{Z}^{(n)} $ is obtained by simply collecting  the coefficient of the $p^N$ term. The above formula is fully general and, as emphasized in \cite{Chakraborty:2023wel}, does not require conformal invariance, as can be seen from the fact that the partition function can in principle dependend on the radius of the $a$-cycle of the base torus\footnote{The same expression has been obtained by studying the long string partition function in the non-AdS setups that have been linked to single-trace $T\bar T$ and $J\bar T$ - deformed CFTs \cite{Hashimoto:2019wct,Hashimoto:2019hqo}. }.

To obtain the partition function of a single-trace $T\bar T$ or $J\bar T$ - deformed CFT, one simply needs to plug in the standard $T\bar T$ or $J\bar T$ - deformed partition function for $Z^{seed}$. Then, one can read off the spectrum of the single-trace  $T\bar T$ or $J\bar T$ - deformed CFT from the usual Hilbert space interpretation of the torus partition function 

\be
Z^{S_N} (\tau,\bar \tau, R) = \sum_n d_n e^{-\b E_n + i P_n \theta} \;, \;\;\;\;\;\;\;\;\; \b = R \tau_2 \;, \;\;\;\;\; \theta = R \tau_1 \label{zsnhilb}
\ee
It is then easy to show (from the radius dependence of \eqref{defzn}) that the energy contributions from the $w$ - twisted sectors are of the form 

\be
E_n^{(w)}  (R) = E_n^{seed} (R w) \label{twsecteng}
\ee
which one should then plug into the appropriate $T\bar T$ or $J\bar T$ - deformed energy formula. See  \cite{Chakraborty:2023wel} for a more detailed discussion. For single-trace $J\bar T$ - deformed CFTs, one can show that the twisted-sector conformal dimensions follow the well-known formula resulting from \eqref{twsecteng}  

\be
h^{(w)}_{J\bar T} (\bar p) = \frac{1}{w} h^{(seed)}_{J\bar T} (\bar p) + \frac{c}{24} \left( w - \frac{1}{w}\right)
\ee
where $h^{seed}_{J\bar T} (\bar p)$ is given by \eqref{jtbdims}.

\subsubsection{The entropy}

Given the partition functions of single-trace $T\bar T$ and $J\bar T$ - deformed CFTs, one can extract the degeneracy of states from it. Concentrating on a single twisted sector, it is not hard to see from  \eqref{zsnhilb} and \eqref{twsecteng} that the degeneracy will be the same as in the seed QFT at the same energy, but on a cylinder $w$ times larger.  One can then e.g. show  that the entropy of the maximally twisted sector in a single-trace $T\bar T$ - deformed CFT is 
\be
S^{(N)}_{s.tr\, T\bar T}  (E,R) = S_{T\bar T} (E, R N)  =  2 \pi \sqrt{ \frac{c}{6\pi} (E R N + \mu E^2)} \label{entstrttb}
\ee
where $c$ is the central charge of the seed CFT; the central charge of the undeformed CFT SPO is $c N$. One obtains the same leading density of states also in the untwisted sector 
\be
S^{(untw)}_{s.tr\, T\bar T}  (E,R) = N S_{T\bar T} (E/N, R )  =  2 \pi \sqrt{ \frac{c}{6\pi} (E R N + \mu E^2)}
\ee
where the contribution comes from N subsystems with equally partitioned energy $E/N$ \cite{Giveon:2017nie}.  As in the CFT case, the leading entropy at high energies is the same, as it is for $N/w$ twisted sectors of twist $w$. 

At large $N$, it is possible to estimate the degeneracy of the $T\bar T$ SPO also for smaller energies \cite{Apolo:2023aho}. The answer presents two regimes: a Hagedorn one below a certain critical energy, followed by the Cardy $\r$ Hagedorn behaviour characteristic of $T\bar T$, with a sharp transition between them.  The slope of the UV Hagedorn regime is always smaller than that of the IR one. See  \cite{Apolo:2023aho,Chakraborty:2023wel} for further details and discussion. 
\medskip

\begin{figure}[h]
\centering 
\includegraphics[height=3.5cm]{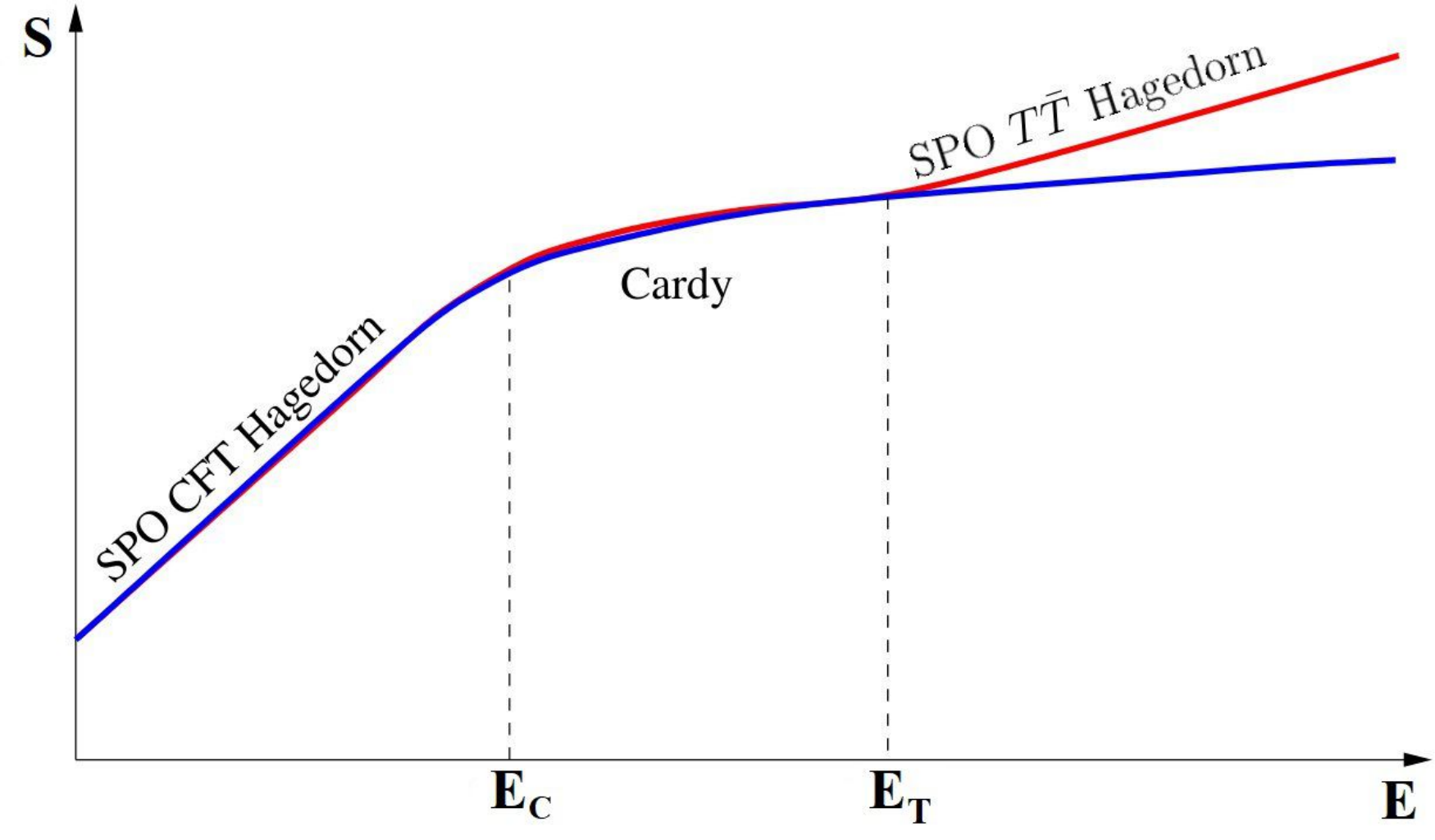}
\caption{ Plot of the entropy as a function of energy for a large N symmetric orbifold of $T\bar T $- deformed CFTs (red),
as compared to the entropy of a symmetric orbifold of CFTs with the same central charge (blue). The deformation parameter $\mu$ is small enough
so that there is a well-separated Cardy regime.}
\end{figure}

%As worked out in ASY, two regimes for the entropy of SPO $T\bar T$. One can show in the non-universal regime, this entropy is larger than $T\bar T$ of the HKS bound. Same leading entropy across various twisted sectors. 

\section{Holographic interpretation of $T\bar T$ and $J\bar T$ - deformed CFTs \label{holointdtr}}

So far, we have been focussing on basic quantum-field-theoretical properties of the $T\bar T$ and $J\bar T$ deformations, such as their definition, the deformed spectrum,the S-matrix. Quite remarkably, these observables are exactly computable at finite $\mu$, despite the deformation being irrelevant. 

In this section, we will discuss the $T\bar T$ and $J\bar T$ deformations in the holographic context. Consequently, we will restrict our attention to seed CFTs with a large central charge  $c \gg 1$ and a large gap in the spectrum of conformal dimensions, known as a holographic CFTs (see e.g. \cite{El-Showk:2011yvt} for a nice introduction). From the perspecive of a low-energy observer,
the holographic dual of such a CFT consists of Einstein gravity in AdS$_3$ coupled to  light matter fields\footnote{This  is appropriate for studying the universal subsector of the deformed theory; it would be interesting to  also study the effect of $T\bar T$ and $J\bar T$ double-trace deformations at the level of the string theory in the bulk; see e.g. comments in \cite{Giveon:2017nie}.}; for the case of $J\bar T$, these matter fields will of course need to include the holographic dual of the current used to define the deformation. The question that we would like to answer in this section  is:

\medskip

\begin{center} \emph{What is the holographic dual of the $T\bar T/J\bar T$ deformation of  a holographic CFT?}
\end{center}
\medskip

\noindent As we will see, the answer is extremely simple and predictable.  $T\bar T/J\bar T$ is a double-trace deformation in holographic parlance, and double-trace deformations have long been known to correspond to turning on mixed boundary conditions for the dual bulk fields. As we will show, the holographic dictionary for the $T\bar T$ and $J\bar T$ deformations can be derived at \emph{precision} level\footnote{The restriction to large $c$, large gap is mostly in order to have a manageable holographic dual; the effect of $T\bar T$ can likely be followed through exactly. }, and follows from a  straightforward application of the rules of holography in presence of double-trace deformations.

Of course, the subject of holography for  AdS$_3$ gravity with 
non-standard boundary conditions  has been intensely explored, see e.g. \cite{Sheikh-Jabbari:2025kjd} and references therein. There is, however, a very important conceptual distinction between the set-ups considered in this section and the remaining literature: while here, one starts from a boundary theory that has its own \emph{independent definition}, \eqref{ttbdefintro}, and \emph{derives} from it the mixed boundary conditions for the dual bulk theory, in the usual analyses one starts on the bulk side and simply \emph{guesses} a set of boundary conditions on the metric that satisfy certain consistency conditions, and then  \emph{assumes} that the gravity theory so defined is the low-energy limit of a consistent theory, with a meaningful holographic dual. Of course, the mere  classical consistency of certain boundary conditions in non-dynamical gravity, especially when the coupling to matter fields is ignored, is only a necessary, but far from sufficient condition for the existence of a holographic dual, 
%
% We, however, see no compelling reason that this assumption hold in general,
  as we discuss in a concrete example towards the end of this section.  Another interpretational distinction between the two ways of proceeding that follows from the above is that in the first case, bulk calculations can be used to perform precision \emph{tests} of the duality, whereas in the second, they are used to \emph{predict} the form of observables in the putative dual theory, which does not have an independent definition.

%The plan of this lecture is as  follows. We start by reviewing double-trace deformations in AdS/CFT and how the holographic dictionary for them is derived. In \ref{holodictttb}, we apply this procedure to the $T\bar T$ deformation and use it to \emph{derive} the holographic  dictionary. We then exemplify how this holographic dictionary  reproduces the finite-size energy spectrum \eqref{defengcft} in the $T\bar T$ - deformed CFT. In  subsection \ref{bcutoff}, we explain the relation between the holographic dictionary and an  earlier proposal that links the $T\bar T$ deformation to AdS$_3$ gravity in presence of a sharp radial cutoff. %, and show a better interpretation is to relate instead the $T\bar T$ observables (in typical high energy states) to the experience of an accelerated observer at a fixed radial position in the bulk.   

In this section, after a brief review of double-trace deformations in AdS/CFT, we apply the corresponding prescription to derive the dictionary for $T\bar T$ and $J\bar T$ - deformed holographic CFTs. As a check, we show that the dictionary correctly reproduces the thermodynamics and the deformed energy spectra. We then move on to study the asymptotic symmetries of these bulk theories with mixed boundary conditions, finding our first indication of the presence of an infinite set of symmetries in these theories. Along the way, we also comment on another holographic proposal for AdS$_3$ gravity with similar boundary conditions. 
%\bi
%\item general bla-bla about how this is top-down and can be used to test holographic dictionary
%%\item how about the holographic dictionary at the level of the worldsheet string theory?
%\ei

 \etocsetnexttocdepth{5}
    \etocsettocstyle{\subsubsection*{Contents of this section: }}{}
    \cftsubsubsecindent 34pt
    \localtableofcontents

\subsection{Double-trace deformations in holography}

\subsubsection{Brief review of usual AdS/CFT}

The AdS/CFT holographic dictionary states that the partition function of a $d$-dimensional  CFT in presence of sources, $J$, for its single-trace operators, $\O$, equals the partition function of the dual $d+1$-dimensional gravitational theory with prescribed (usually Dirichlet) boundary conditions on the dual bulk fields $\Phi$. Taking $\Phi$ to be a free scalar (which is usually a good approximation near the boundary), its asymptotic equation of motion fixes the radial dependence to be of the Fefferman-Graham form
\be
\Phi (z,x^\mu) = \phi^{(0)} (x^\mu) \, z^{d-\Delta} + \ldots + \phi^{(\D)} (x^\mu) z^\D+ \ldots 
\ee
where $z$ is the Poincar\'e radial coordinate, with the conformal boundary of AdS lying at $z=0$. The mode proportional to $\phi^{(0)}$ is non-normalizable, and is identified with the CFT source $J(x^\mu)$ via the AdS/CFT dictionary
\be
Z_{CFT}[J] = \int \mathcal{D} \psi\, e^{-S_{CFT}[\psi] - \int J \O[\psi]} = Z_{grav}[\phi^{(0)}=J]
\ee
Because it is non-normalizable, the coefficient $\phi^{(0)}$ is to be kept fixed and provides the boundary condition that the gravitational path integral obeys.

 The one-point function of $\O$ in the CFT is given by
\be
 \langle \O (x^\mu) \rangle = \frac{\d W[J]}{\d J (x^\mu)}
\ee
where $W[J] = - \ln Z_{CFT}[J]$ is the generating functional of connected CFT correlators. In the classical approximation, the gravitational path integral with fixed boundary conditions $\phi^{(0)}$ is approximated by the exponential of the holographically renormalized on-shell action $S^{ren}_{on-shell}[\phi^{(0)}]$, which is therefore  equated with  $W[J]$.   Taking the functional derivative, one finds that the  mode $\phi^{(\D)}(x^\mu)$, which  is normalizable and thus allowed to fluctuate, is identified\footnote{This is true for free scalars; more generally, $\langle\O \rangle$ will be a function of  the coefficients of both the normalisable and non-normalisable mode.} with the expectation value of the dual operator in the particular state that the CFT finds itself in 
\be
\frac{\d S_{on-shell}^{ren}[\phi^{(0)}]}{\d\phi^{(0)}(x^\mu)} = \phi^{(\D)}(x^\mu) = \langle \O (x^\mu) \rangle
\ee
Note that since our discussion is rather schematic, we have neglected all normalization factors. 

\subsubsection{Review of double-trace deformations in AdS/CFT}

According to the above discussion, if we would like to add a source $J$ for a single-trace CFT operator, all we need to do in the holographic dual  is to  impose the boundary condition $\phi^{(0)} = J$ when computing the gravitational path integral. In this subsubsection, we discuss what happens if we simultaneously add  a source for the double-trace operator $\O^2$ to the CFT action \cite{Witten:2001ua}.

The generating functional of connected correlators in the deformed CFT now reads, schematically

\be
e^{-W^{[\mu]} [J^{[\mu]}] }= \int \mathcal{D} \psi\, e^{-S_{CFT}[\psi] - \frac{\mu}{2} \int \O^2 - \int J^{[\mu]} \O} 
\ee
where $J^{[\mu]}$ is the (tunable) source for the single-trace operator $\O =\O[\psi]$ in the CFT deformed by the double-trace coupling $\mu \, \O^2/2$, where $\mu$ is kept fixed. As before,  a ${}^{[\mu]}$ or ${}^{[0]}$ superscript  indicates whether the corresponding quantity belongs to the deformed or undeformed CFT. 

The standard way to treat these deformations is to use the  Hubbard-Stratonovich method \cite{Gubser:2002vv}, which consists of first inserting a (formal) resolution of the identity $
1 =[ \det (- \textbf{1}/\mu)]^{1/2} \int \mathcal{D} \tilde \s \, e^{ \frac{1}{2\mu}\int  \tilde \s^2}
$ into the path integral, with an appropriately chosen integration contour, and then shifting the integration variable as $\tilde \s = \s' - \mu \, \O$. The result is
\bea
e^{-W^{[\mu]}\left[ J^{[\mu]}\right]} &= &[\det (- \textbf{1}/\mu)]^{1/2}   \int \mathcal{D} \psi \,\mathcal{D}\s'\, e^{-S_{CFT}[\psi]  - \int (J^{[\mu]}+\s')\, \O[\psi] + \frac{1}{2\mu} \int \s'^2} \nonumber \\
&   & \hspace{-1.5 cm}= \; (\det - \textbf{1}/\mu)^{-1/2} \int\mathcal{D}\s\, e^{ \frac{1}{2\mu}\int  ( \s-J^{[\mu]})^2}\, e^{-W^{[0]}[\s]} \approx e^{-W_0 [J^{[\mu]} + \mu \langle \O \rangle] + \frac{\mu}{2} \int \langle \O \rangle^2}
\eea
where the last step involves a saddle point approximation at large $N$, which yields the saddle-point value $\s_*=J^{[\mu]} + \mu \langle \O \rangle$. 
 Thus, we find that at large $N$, the source in the deformed theory is shifted with respect to the one in the undeformed CFT by the expectation value of the operator

\be
J^{[\mu]}= J^{[0]}-\mu \langle \O \rangle\;, \;\;\;\;\;\;\; W^{[\mu]}=W^{[0]}-\frac{\mu}{2} \int \langle \O \rangle^2 \label{newsvscalar}
\ee
while the generating function is shifted by \emph{minus} the double trace. The expectation value $\langle \O \rangle$ stays the same.  Notice that
\eqref{newsvscalar} follows from a purely field-theoretical argument. The large $N$ approximation is only used for the evaluation via saddle point, but exact results are in principle possible. 

The implications of the above shifts for the holographic dictionary in presence of the double-trace deformation are straightforward. Since all we did was to add the boundary term \eqref{newsvscalar}, the bulk theory is the same, and all that can change are the boundary conditions. Before the deformation, $J^{[0]}$ was identified with the non-normalizable mode $\phi^{(0)}$ of the bulk field, while $\langle \O \rangle$ was identified with the normalizable mode $\phi^{(\D)}$. %The gravitational path integral was to be performed with $\phi^{(0)}$ fixed and appropriate boundary conditions in the interior, which would determine  $\phi^\D$ through \eqref{}. \emph{Move at beginning?}
 Since the new source $J^{[\mu]}$ is a linear combination of the old source $J^{[0]}$ and the expectation value, we find that the deformed CFT corresponds to the same gravitational theory as before, but with \emph{mixed} boundary conditions on the bulk fields, namely the  linear combination $\phi^{[\mu]} \equiv \phi^{(0)} - \mu \, \phi^{(\D)} $ of the normalizable and the non-normalizable modes
is now held fixed. The holographic dictionary reads 

\be
Z^{[\mu]}_{QFT} \, [\, J^{[\mu]}\, ]  = Z_{grav} \, [\, \phi^{(0)}- \mu \, \phi^{(\D)}=J^{[\mu]}\,]
\ee
The functional derivative of $Z_{grav}$ with respect to the new source $\phi^{[\mu]}  $ will yield the  gravity dual of the expectation value of $\O$ in the deformed CFT which, in our simple scalar field example,  equals   $\phi^{(\D)}$.

\noindent

 To summarize,  finding the holographic dictionary in presence of double-trace deformations proceeds in two steps:
\begin{enumerate}
\item a \emph{purely field-theoretical} step, in which one uses the Hubbard-Stratonovich method and large $N$ to derive the relation between sources and expectation values before and after the deformation
\item the interpretation of the new sources and expectation values  in terms of the coefficients in the asymptotic expansion  of the dual bulk field, which  consists of simply plugging in the  \emph{undeformed} holographic dictionary into the results of step $1$  
\end{enumerate}

\subsubsection{Variational principle approach}

There exists an alternate - and, in practice, much simpler - approach for finding the relation between the deformed and undeformed CFT data, which we will denote as the `variational principle' approach. It is nicely explained in \cite{Papadimitriou:2007sj} and it  works  at the level of the \emph{classical} AdS/CFT dictionary, involving only (super)gravity fields. 

By definition, the variation of the undeformed/deformed generating functional  with respect  to a variation of  the corresponding source is 
\be
\d W^{[0]} = \int \langle \O \rangle_{[0]} \,  \d J^{[0]} \;, \;\;\;\;\;\;\; 
\d W^{[\mu]} = \int \langle \O \rangle_{[\mu]}\, \d J^{[\mu]} \label{delwmu}
\ee
where we indicated that the vevs in the deformed and undeformed theory could in principle be different, even though  here it is not the case. 
From \eqref{newsvscalar} we have that $W^{[\mu]} = W^{[0]} - S_{d.tr.}$.  %\emph{\textbf{How is this derived, generally?}} 
Taking the variation, we have
\be
\d W^{[\mu]} = \d W^{[0]} - \frac{\mu}{2}\,  \d \int \langle \O\rangle^2 = \int \langle \O \rangle \, \d \left(J^{[0]} - \mu \langle \O \rangle \right) 
\ee
Equating this with $\d W^{[\mu]}$ in \eqref{delwmu} and separately matching the terms inside the variation and their coefficients, we can effectively read off the same relations between the deformed and undeformed sources and vevs 
 as above.  Clearly, this is a much simpler way to derive the holographic dictionary. It %immediately detects cases when the vev also shifts (e.g. when studying the chance in the vev of the stress tensor due to the double-trace), 
is particularly useful  as we consider more complicated operators, e.g. those carrying spin, and expectation values that shift under the deformation.

%The Hubbard-Stratonovich procedure reviewed above involves a number of manipulations at the level of the path integral, which will only become more cumbersome to deal with  as we consider more complicated operators, e.g. those carrying spin, and expectation values that shift under the deformation. It
%
%In the very simple example reviewed above, the source for the operator $\O$ shifts as $\mu$ is turned on, but the expectation value stays the same. This will not be the case for more complicated operators or deformations, e.g. those with spin. Performing  the steps of the Hubbard-Stratonovich procedure can become rather cumbersome, and it
% would thus be useful to have a simpler method that yields directly the final answer.  

%Since for many purposes we are only interested on the effect of the double-trace deformation on the \emph{classical} AdS/CFT dictionary, i.e. at the level of the (super)gravity fields,  it is sufficient to study the effects of double-trace deformations on the \emph{classical} on-shell gravitational action  $S_{on-shell}[\phi^{(0)}]$ which, as we saw, equals the CFT generating functional $W[J]$ .

This exercise is very naturally phrased in bulk language. From the point of view of the canonical formulation of gravity in AdS in the radial Hamiltonian formalism, where $z$ plays the role of time,  $\phi^{(0)}$ should be viewed  as a generalized coordinate on the initial surface $z=0$, while $\phi^{(\D)}$ represents the (holographically renormalized) canonically conjugate momentum \cite{Papadimitriou:2004ap,Papadimitriou:2010as}, up to the appropriate rescalings. As before, $W[J]$ is identified with the \emph{classical} on-shell renormalized gravitational action  $S_{on-shell}^{ren}[\phi^{(0)}]$. The addition of the double trace boundary term induces a canonical transformation of the above phase space variables, which can be read off from the variational principle as above, under the identifications $W\leftrightarrow S, J\leftrightarrow \phi^{(0)}$ and $\O \leftrightarrow \phi^{(\D)}$.

%By definition we have 
%%
%\be
%\d S^{ren}_{on-shell}[\phi^{(0)}] = \int \phi^{(\D)} \d \phi^{(0)}
%\ee
%Using \eqref{newsvscalar}, it is easy to see that adding the double trace deformation to the boundary action simply corresponds to a canonical transformation%, which can be deduced from the variational principle obeyed by the deformed on-shell action \emph{Notation!}
%%
%\be
%\d S_{on-shell}^{[\mu]} = \d S_{on-shell}^{[0]} - \frac{\mu}{2} \d \int \O^2 = \int \O \, \d(J^{(0)} - \mu \O) \equiv \int \O^{(\mu)} \, \d J^{(\mu)}
%\ee
%Since in the deformed theory, $\d W_\mu = \int \O \d J^{[\mu]} = \d S^{[\mu]}$, 
%by requiring that the end result take the form: new vev $\times$ $\d$(new source), we can effectively read off the same holographic dictionary as above.  Clearly, this is a much simpler way to derive the dictionary, and it immediately detects cases when the vev also shifts (e.g. when studying the chance in the vev of the stress tensor due to the double-trace). 

The main message is that at large $N$, when $W[J]$ is well approximated by the classical on-shell action $S_{on-shell}^{ren}[\phi^{(0)}]$, rather than performing the steps of the Hubbard-Stratonovich procedure, followed by the saddle point approximation, to find the relation between the new sources and expectation values and the undeformed ones,  we can derive the same data from the (much simpler to handle) variational principle approach: these two procedures are equivalent at large $N$. %\emph{Proof equivalence?} 

\subsection{The holographic dictionary for $T\bar T$-deformed CFTs \label{holodictttb}}

We are now ready to derive the classical holographic dictionary for $T\bar T$-deformed holographic CFTs. This simply amounts to applying the recipe described in the previous section to the double-trace $T\bar T$ deformation\footnote{The standard story is slightly more complicated than this.  In the case of scalar deformations, one normally requires that $\O^2$ be a relevant or marginal operator, so that its effect is tractable and the perturbation does not affect the original AdS boundary. This implies that the dimension of $\O$ is $\leq d/2$. Such an operator corresponds to a bulk field quantized with Neumann boundary conditions, a.k.a. alternate quantization. The case of $T\bar T$ is in a certain sense simpler, in that one deforms the theory in the usual quantization. The reason that the effect of this \emph{irrelevant} double-trace operator is still tractable from a holographic point of view is that the dual bulk field is the $3d$ metric, which is not dynamical. In particular, its non-normalizable part is pure gauge and thus does not backreact on the local geometry, even at full non-linear level. }. 

As  explained in the previous subsection, the first step of the recipe is purely field-theoretical and uses the Hubbard-Stratonovich method to find the deformed sources and expectation values in terms of the original ones%\footnote{This was partly done in Cardy, though it is slightly different from us, because we are letting the background metric flow, whereas he was always off-setting it to be flat.}
.  Since for the purpose of deriving the classical holographic dictionary, the Hubbard-Stratonovich method is equivalent with the much simpler variational principle approach,  we will use the latter to find the desired field-theoretical relation,   following \cite{Guica:2019nzm}.

\subsubsection{Step 1: the relation between the deformed and undeformed sources and vevs}

The definition of the $T\bar T$ deformation  consists of incrementally adding to the QFT action the $T\bar T$ operator of the deformed theory %\emph{\textbf{Why equivalent with flat space $T\bar T$ definition?}}
%\textcolor{red}{\emph{Cross-check signs!}}
\be
\p_\mu S^{[\mu]}_{QFT} = - \frac{1}{2} \int d^2 x \sqrt{\g} (T^{\a\b} T_{\a\b} - T^2)_\mu
\ee
Here the source coupling to the stress tensor is the background metric $\g^{\a\b}$ and, by definition,  we have $\d S = \frac{1}{2} \int d^2 x \sqrt{\g} \, T_{\a\b} \d \g^{\a\b}$. These definitions make sense in the classical, large $N$ limit; indeed, outside this limit it is not clear if the operator can be defined on a non-flat background metric.  We are working in Euclidean signature and follow the conventions of \cite{Guica:2019nzm}. The variations of the generating functionals  in two nearby $T\bar T$ - deformed CFTs related  by infinitesimally changing $\mu \r \mu + \D \mu$ satisfy 
\bea
\d W^{[\mu+ \Delta \mu]} &=& \d W^{[\mu]} - \Delta \mu \, \d S_{(T\bar T)_\mu} = \frac{1}{2} \int d^2 x \sqrt{\g_{[\mu]}} \, T_{\a\b}^{[\mu]} \d \g^{\a\b}_{[\mu]} - \Delta \mu\,  \d S_{(T\bar T)_\mu} \nonumber \\
&\equiv &\frac{1}{2} \int d^2 x  \left( \sqrt{\g}\, T_{\a\b} \d \g^{\a\b}\right)_{[\mu+ \Delta \mu]}
\eea
In the limit $\Delta \mu \r 0$, this equation can be rewritten as 
\be
\p_\mu \left(\frac{1}{2} \int d^2 x \sqrt{\g_{[\mu]}}\, T_{\a\b}^{[\mu]} \d \g^{\a\b}_{[\mu]} \right) = \d\left( \frac{1}{2} \int d^2 x \sqrt{\g} (T^{\a\b} T_{\a\b} - T^2)  \right)_{[\mu]}
\ee
We need to solve this for an arbitrary variation of the sources $\g^{[\mu]}$. Separately equating the terms under the variations and their coefficients, one obtains 
 the following flow equations for the source and the expectation value of the stress tensor  
\be
\p_\mu \g_{\a\b} = -2 (T_{\a\b} - \g_{\a\b} T)\equiv  -2 \, \hat T_{\a\b}\;, \;\;\;\;\;\; \p_\mu \hat T_{\a\b} = - \hat T_{\a\g} \hat T_\b{}^\g \;, \;\;\;\;\;\;\; 
% \p_\mu (\hT_{\a\rho} \hat  T_{\b\s} \, \g^{\rho\s} ) =0
\p_\mu (\hat T_{\a\g} \hat T_\b{}^\g) = 0 \label{fleqttb}
\ee
This set of equations is trivial to integrate, and the solution is

\be
  \g_{\a\b}^{[\mu]} = \g_{\a\b}^{[0]} - 2 \mu \, \hat T_{\a\b}^{[0]} + \mu^2  \hat T_{\a\rho}^{[0]} \, \hat  T_{\sigma\b}^{[0]} \, \gamma^{[0] \rho \sigma}  \;, \;\;\;\;\;\;\;\;
   \hat T_{\a\b}^{[\mu]} = \hat T_{\a\b}^{[0]} - \mu \, \hat T_{\a\rho}^{[0]} \, \hat  T_{\sigma\b}^{[0]} \, \gamma^{[0] \rho \sigma}
\label{defmet}
\ee
%
%\begin{empheq}[box=\fbox]{align}
%&  \g_{\a\b}^{[\mu]} = \g_{\a\b}^{[0]} - 2 \mu \, \hat T_{\a\b}^{[0]} + \mu^2  \hat T_{\a\rho}^{[0]} \, \hat  T_{\sigma\b}^{[0]} \, \gamma^{[0] \rho \sigma}  \nonumber  \\[5pt] 
%& \hspace{8mm} \hat T_{\a\b}^{[\mu]} = \hat T_{\a\b}^{[0]} - \mu \, \hat T_{\a\rho}^{[0]} \, \hat  T_{\sigma\b}^{[0]} \, \gamma^{[0] \rho \sigma}
%\label{defmet}
%\end{empheq}
which represent the relations between the sources and expectation values in the $T\bar T$ - deformed CFT and the undeformed one. The expectation value of the deformed stress tensor is determined from that of $\hat T$ via 
\be
T_{\a\b}^{[\mu]} = \hat T_{\a\b}^{[\mu]} - \g_{\a\b}^{[\mu]}\,  \hat T^{[\mu]} \label{reltth}
\ee
Notice these relations are rather non-linear and the change in the expectation value of the stress tensor is non-trivial. We emphasize these relations follow from the definition of the $T\bar T$ deformation via a purely (large $N$) field theoretical derivation. The flow equations \eqref{fleqttb} can be additionally used to show that $\p_\mu (R\sqrt{\g})=0$, that the deforming operator does not flow, $\p_\mu (\sqrt{\g} \, \O_{T\bar T})=0$, and that the trace relation reads $T^{[\mu]} = c/24 \pi R^{[\mu]} - \mu\, \O_{T\bar T}^{[\mu]}$, which agrees with \eqref{ttbtrrel}, found via the Hamiltonian analysis, when the background metric is flat. %See \cite{Guica:2019nzm} for more details. 

One can in principle add sources for  `matter' fields, here taken for simplicity to be scalars. Since the double-trace deformation only involves the stress tensor, the variation with respect to the matter operator sources simply goes along for the ride -  at infinitsimal level at least - as  can be seen from the variational principle %in their presence
\be
\int \left( \sqrt{\g} \, \O \d J \right)_{[0]} = \int \left( \sqrt{\g} \, \O \d J \right)_{[\mu]} \label{varpmatt}
\ee
This implies that the sources for matter operators are unaffected by the deformation, while the expectation values are related via $\O^{[\mu]} = \O^{[0]} \sqrt{\g^{[0]}/ \g^{[\mu]}}$. % \emph{Careful non-infinitesimal!}
To extend this dictionary beyond leading order in the matter sources - as required, for example, for computing correlation functions of the matter operators -  one should proceed with care, as the presence of non-trivial operator sources  affects the  %(definition of the) 
$T\bar T$ flow.

\subsubsection{Step 2: the holographic dictionary}

Having found the relation between the deformed and undeformed stress tensor data at large $N$, the second step  of the holographic dictionary in presence of double-trace deformations consists in interpreting these data  in terms of the asymptotic values of the bulk fields in the dual asymptotically AdS$_3$ spacetime.  For this, we need to briefly review the standard holographic dictionary for the stress tensor in the context of AdS$_3$/CFT$_2$. 

As explained at the beginning of this section, the holographic dual to a large $c$, large gap  CFT$_2$ is $3d$ Einstein  gravity coupled to various light matter fields. % (whose mass is $\O(1)$ in units of the AdS length) of spin less than two. 
We will concentrate on the gravitational sector, which is the one that captures the dynamics of the stress tensor in the dual CFT. The asymptotic solution for the three-dimensional  metric is
simplest in the so-called Fefferman-Graham gauge ($g_{\rho\rho} = \ell^2/4 \rho^2$, $g_{\rho\a} =0$, where $\rho$ is the radial coordinate and $\ell$ is the AdS$_3$ length) and is given by the following expansion
\be
ds^2 =   \ell^2 \frac{d\rho^2}{4\rho^2}+ \left(  \frac{g^{(0)}_{\a\b}(x^\a)}{\rho} + g^{(2)}_{\a\b}(x^\a) + \ldots \right) \, dx^\a dx^\b \label{fg}
\ee
This expansion holds at \emph{non-linear level} provided the matter fields satisfy reasonable boundary conditions (i.e., the non-normalizable mode $\phi^{(0)}$ is set to zero beyond linearized level; however, arbitrary normalizable modes of the matter fields  are allowed at full non-linear level). The two leading terms written above are universal and correspond to the source and expectation value of the dual CFT$_2$ stress tensor. More precisely, 
\be
g^{(0)}_{\a\b} = \g^{[0]}_{\a\b} \;, \;\;\;\;\;\; g^{(2)}_{\a\b}= 8 \pi G \ell\, \hat T_{\a\b}^{[0]} \label{undefholdic}
\ee
The asymptotic Einstein equations fix the trace and divergence of $g^{(2)}$ in terms of $g^{(0)}$; these correspond to the holographic Ward identities of the CFT stress tensor - see e.g.  \cite{Skenderis:2002wp} for details.  The $\ldots$ correspond to terms that are subleading in the $\rho$ expansion.  They are non-universal and depend on the particular matter field expectation values that have been turned on.

We now have all the ingredients to describe the holographic dictionary for $T\bar T$ - deformed CFTs. The coefficients $g^{(0,2)}$ above encode the source and the expectation value of the stress tensor in the \emph{undeformed} CFT, so they should be identified with $\g^{[0]}$ and respectively $\hat T^{[0]}$ in \eqref{defmet}. The source for the deformed stress tensor (i.e., the background metric) in the $T\bar T$ - deformed CFT is $\g^{[\mu]}$, given in \eqref{defmet}. Using the undeformed holographic dictionary \eqref{undefholdic}, we find 
the following expression for $\g^{[\mu]}$ in terms of the coefficients appearing in the asymptotic  metric expansion
\be
\g_{\a\b}^{[\mu]} = g^{(0)}_{\a\b} - \hat \mu  g^{(2)}_{\a\b} + \frac{\hat \mu^2 }{4} \, g^{(2)}_{\a\g} g^{(0)\, \g\d} g^{(2)}_{\d\b} \;, \;\;\;\;\; \;\;\;\hat \mu \equiv \frac{\mu }{4 \pi G \ell}  \label{genbc} 
\ee 
Since we are supposed to keep  the source $\g^{[\mu]}$ in the deformed theory fixed, we find that the fluctuations of the dual bulk metric satisfy a mixed and rather non-linear boundary condition, given by the right-hand side of the above equation. Note that both the normalizable and non-normalizable (boundary metric) mode of the metric are allowed to fluctuate, as long as the above combination is held fixed. %The reason that it is ok for the non-normalizable mode to fluctuate is that the $3d$ metric is non-dynamical, so there is no backreaction of these fluctuations on the local geometry. 

 To summarize, the holographic dictionary for $T\bar T$-deformed CFTs is
\be
Z_{T\bar T \mbox{-} def.\, CFT} \, [\, J^{[\mu]}, \g^{[\mu]}\, ] = Z_{grav} \, [\, \phi^{(0)} = J^{[\mu]} , g^{(0)} - \hat \mu\, g^{(2)} + \frac{\hat \mu^2}{4} g^{(2)}g_{(0)}^{-1} g^{(2)}=\g^{[\mu]}\,] 
\ee
%where, to ease the notation, we have defined $\mu$ in units of $4\pi G \ell$. 
The first argument indicates that all matter fields have the same boundary conditions as before the deformation.
 Using \eqref{defmet} and \eqref{undefholdic}, we find that the expectation value of the  stress tensor in the deformed theory is given by \eqref{reltth}, with $\g^{[\mu]}$ given in  \eqref{genbc} and 

\be
\hat T_{\a\b}^{[\mu]} =  \frac{1}{8\pi G  \ell} \, g_{\a\b}^{(2)} -\frac{\mu }{(8 \pi G  \ell)^2} \, g^{(2)}_{\a\g} g^{(0)\, \g\d} g^{(2)}_{\d\b} \label{holoThat}
\ee
A few comments are in place:
\bi
\item[i)] just like in the undeformed case, the expectation value of the stress tensor \emph{only} involves the \emph{universal} asymptotic metric coefficients $g^{(0)}$ and $g^{(2)}$  that encode the energy and momentum density of the initial state in the undeformed CFT. %, and in particular does not depend at all on the non-universal, matter-field-profile dependent terms that follow in the FG expansion. It is important in that   the vev only depends on the asymptotic data, which 
This  will ultimately be responsible for the universality of the deformed energy formula, as we will show. 
  \item[ii)] unlike in the undeformed CFT case, $\hat T_{\a\b}$
bears a rather non-linear relation to the asymptotic data, and is computed by a different formula than in  AdS$_3$ with Dirichlet boundary conditions
\item[iii)] the boundary conditions for matter fields are unaffected, as follows from \eqref{varpmatt} 
\item[iv)] the non-locality of the boundary theory is realised only through the non-standard boundary conditions that bulk fields obey, the bulk theory and local asymptotic geometry being the same as in standard AdS$_3$/CFT$_2$
\ei

\subsubsection{Building the gravitational phase space}

Having established  what the sources and expectation values in the $T\bar T$ - deformed CFT correspond to in terms of the asymptotic behaviour of the metric components, the next natural question is to understand the phase space of the bulk theory, i.e.  what are the most general allowed  metric fluctuations for a given, fixed metric $\g^{[\mu]}$ in the $T\bar T$ - deformed  CFT.  Knowing the answer for arbitrary $\g^{[\mu]}$ 
is useful, e.g. for computing correlation functions of the stress tensor by repeated functional differentiation of the holographic one-point function. 

For $\mu =0$, the answer to this question is well-known: one allows for all tensors $g^{(2)}$ that are compatible with the holographic Ward identities, which fix their trace and divergence in terms of the boundary metric $g^{(0)}$, which is considered as given.
 For $\mu \neq 0$, the problem is significantly more complicated, since one needs to find the most general solution for $g^{(0,2)}$, subject to the holographic Ward identities  and the  non-linear boundary condition \eqref{genbc}. Solving this set of non-linear algebraic and differential equations for general $\g^{[\mu]}$ appears rather cumbersome. We will therefore concentrate on the much simpler problem of finding the most general $ g^{(0,2)}$ for which $\g^{[\mu]}_{\a\b} = \eta_{\a\b}$.  The price to pay is that, since we restrict the $T\bar T$ metric to be flat, we will not have access to arbitrary correlation functions of the stress tensor, but only to the one-point functions \eqref{holoThat}.

In the particular case  $\g^{[\mu]}_{\a\b} = \eta_{\a\b}$, i.e. when the $T\bar T$-deformed CFT lives on flat space, there is a trick to solve the general equations, which we will now explain.  Using the fact that the combination $R\sqrt{\g}$ is constant along the flow and that $R[\g^{[\mu]}] =0$, we conclude that $R[g^{(0)}]=0$, so $g^{(0)}$ and $\g^{[\mu]}$ are diffeomorphic to each other (this is not generally the case). Since the Ricci scalar of the boundary metric $g^{(0)}$ vanishes, this implies that there exists a coordinate system in which the metric $\g^{(0)}$ equals the two-dimensional Minkowski metric $\eta$.

 Let $U,V$ be the coordinates of the $T\bar T$-deformed CFT, in terms of which $\g^{[\mu]} = \eta$, and $u,v$ be the auxiliary set of coordinates, in  terms of which  $g^{(0)}=\eta$. The asymptotic solution for the bulk metric in the $u,v$ coordinate system is extremely simple, as the holographic Ward identities can be solved  explicitly in terms of two arbitrary functions $\L(u)$ and $\bar \L(v)$

\be
ds^2 = \frac{\ell^2}{4} \frac{d\rho^2}{\rho^2} + \frac{du dv}{\rho} + \L(u)\, du^2 + \bar \L(v)\, dv^2 + \ldots \label{asyeta}
\ee
where the $\ldots$ are $\O(\rho)$ and are non-universal. The functions $\L(u)$ and $\bar \L(v)$ are proportional to the expectation values of the holomorphic and, respectively, antiholomorphic stress tensor components $T_{uu}$ and $T_{vv}$. Plugging in the coefficients  $g^{(0,2)}$ read off from above into  \eqref{genbc}, we obtain the following expression for $\g^{[\mu]}$  in the $u,v$ coordinate system
\be
\g^{[\mu]}_{\a\b} dx^\a dx^\b = du dv - \hat \mu (\L(u) du^2 + \bar \L(v) dv^2) + \hat \mu^2 \L(u) \bar \L(v) du dv = \left(du - \hat \mu \bar \L(v) dv\right)\,\left(dv - \hat \mu \L(u) du\right)% = dU dV 
\label{gmuuv}
\ee
%where we have introduced the shortcut notation 
%
%\be
%\rho_c \equiv - \frac{\mu}{4\pi G \ell} \label{rhoc}
%\ee
The line element \eqref{gmuuv}   must equal the Minkowski line element $dU dV$ in terms of the $T\bar T$ coordinates $U,V$. % The solution can be easily read off from \eqref{gmuuv}
This yields the following relation between the two sets of coordinates %\textcolor{red}{\emph{Factor 2?}}

\be
U = u -\hat \mu \int^v \bar \L(v) dv \;, \;\;\;\;\; V = v - \hat \mu \int^u \L(u) du \label{defuv}
\ee
which is precisely the field-dependent coordinate transformation \eqref{fdepcootrfb}, now rederived from holography.  The constants of integration that enter the precise definition of the primitives 
 will be very important in the  calculation of the asymptotic symmetry algebra in subsection \ref{ttbasg}, but are not necessary for now.

Thus, the most general $g^{(0,2)}$ satisfying the holographic Ward identities and the boundary condition $\g^{[\mu]}=\eta$ are given by the asymptotic metric coefficients encoded in \eqref{asyeta}, translated back to the $U,V$ coordinate system. Explicitly, we have

\be
  g^{(0)}_{\a\b}\, dx^\a dx^\b = du dv = \frac{(dU+\hat \mu \bar \L(v) dV)(dV+\hat \mu \L(u) dU)}{(1-\hat \mu^2 \L(u)\bar \L(v))^2} \label{g0UV}
\ee
\be
g^{(2)}_{\a\b} \, dx^\a dx^\b = \L(u)du^2 + \bar \L (v) dv^2 = \frac{ \left(1+\hat \mu^2 \L(u) \bar \L(v)\right) ( \L(u)\, dU^2 +\bar \L(v) \, dV^2) +4 \hat \mu \L(u) \bar \L(v)\, dU dV }{\left(1-\hat \mu^2 \L(u) \bar \L(v)\right)^2} \nonumber
\ee
The expectation value of the stress tensor can be read off from \eqref{holoThat}, and is given by

\be
 T_{\a\b}^{[\mu]} dx^\a dx^\b=  (\hat T_{\a\b}^{[\mu]} - \eta_{\a\b} \, \eta^{\g\d} \hat T^{[\mu]}_{\g\d} )\, dx^\a dx^\b=\frac{\L(u) dU^2 + \bar \L(v) dV^2- 2 \hat \mu \L(u) \bar \L(v) dU dV }{8\pi G \ell(1-\hat \mu^2 \L(u)\bar \L(v))}  \label{stresst}
\ee
in the $U,V$ coordinate system. The energy and angular momentum of the configuration are simply given by the $\s$ integral of the appropriate component of the stress tensor on a constant time slice. 

Thus, we find that, just like in the case of AdS$_3$ with Dirichlet boundary conditions, the  space of bulk solutions satisfying the mixed boundary conditions with $\g^{[\mu]}=\eta$ is 
 parametrized by two arbitrary functions, $\L(u)$ and $\bar \L(v)$, where now the coordinates $u,v$ are field-dependent and are determined via \eqref{defuv}. In addition, one can have arbitrary matter expectation values turned on, which appear at subleading order. The solutions are still asymptotically locally AdS$_3$, since the non-normalizable mode of the metric \eqref{g0UV} is pure gauge.

\subsubsection{Checks of the correspondence}

In the first section, we derived the exact formula \eqref{defengcft} for the $T\bar T$-deformed energy spectrum on a cylinder, which is  a function of $\mu,R$ and only the initial energy and momentum of the state. Using the fact that the degeneracy of states does not change along the $T\bar T$ flow, this allowed us to compute the entropy in the deformed theory as a function of the deformed energy, leading to the characteristic Cardy $\r$ Hagedorn behaviour \eqref{defent}.  As a basic check of the proposed holographic dictionary, we would like to  reproduce these formulae from a holographic calculation. 

Since we need to compute the energies in finite volume, we take the AdS$_3$ boundary to be a cylinder, with $U,V = \s \pm t$ where $\s \sim \s + R$, and consider an energy-momentum eigenstate, characterized by $\L(u) = \L$ and $\bar \L (v) = \bar \L$, both constant. The deformed energy and angular momentum are given by 

\be
E_\mu = \int_0^R d\s \, T_{tt}^{[\mu]} = \frac{R}{8\pi G\ell} \frac{\L+\bar{\L}+2\hat\mu \L \bar{\L}}{1-\hat\mu^2 \L \bar{\L}} \;,\;\;\; \;\;\;\;\; P_\mu = \int_0^R d\s \, T_{t\s}^{[\mu]} =  \frac{R}{8\pi G\ell} \frac{\L-\bar{\L}}{1-\hat\mu^2 \L \bar{\L}} \label{emujmu}
\ee
where we simply plugged in the expression \eqref{stresst} for the stress tensor components. 

Checking the relation \eqref{defengcft} is not as simple as it may seem, since  one is comparing the energies of states in different theories, i.e. belonging to different gravitational phase spaces, where the asymptotic metric satisfies different boundary conditions.  One exception is the relation between the energies of the ground state in the two theories, which is an isolated state and can be easily identified holographically via the requirement that the geometry not exhibit a conical singularity.  Since the vacuum is a 
 solution of pure $3d$ gravity, for which the Fefferman-Graham expansion happens to truncate at $\O(\rho)$ \cite{Skenderis:1999nb}
\be
ds^2 = \frac{\ell^2}{4} \frac{d\rho^2}{\rho^2} + \frac{1}{\rho}\left(g^{(0)}_{\a\b} + \rho \, g^{(2)}_{\a\b} + \frac{\rho^2}{4} g^{(2)}_{\a\g}g^{(0)\, \g\d}g^{(2)}_{\d\b} \right) dx^\a dx^\b \label{fgexact}
\ee
the full bulk solution is entirely specified,  via \eqref{g0UV}, by the coefficients $\L_{vac},  \bar{\L}_{vac}$. These are, of course, equal and can be determined by requiring smoothness of the metric  near the place where the $\s$ circle shrinks to zero size, which yields 
\be
\frac{\L_{vac}}{(1-\hat \mu\, \L_{vac})^2}= \frac{4\pi G \ell}{R} E^{[\mu]}_{vac} \left(1 + \frac{\mu E_{vac}^{[\mu]}}{R}\right) = -  \left( \frac{\pi \ell}{R}\right)^2
\ee
whose solution is the correct vacuum energy of a $T\bar T$ - deformed CFT of central charge $c= 3 \ell/2G$.

Another observable that can be easily checked is the entropy-to-energy relation \eqref{entropyttb} in the deformed CFT. This relation is valid in the high-energy regime, where typical configurations are expected to correspond to black holes in the bulk. Again, these are solutions to pure $3d$ gravity characterised by constant parameters $\L, \bar \L$, this time taken to be positive.  In this case, there is a horizon located at  $\rho_h= (\L \bar\L)^{-1/2}$ and its area is given by
\be
\mathcal{A}_\mu =R \frac{\sqrt{\L}+\sqrt{\bar{\L}}}{1-\hat \mu \sqrt{\L \bar{\L}}}
\ee
Rewriting $\L, \bar \L$ in terms of the  energy and momentum \eqref{emujmu} in the deformed theory, one finds precisely the $T\bar T$ relation \eqref{entropyttb}. 

Finally, let us discuss how one may reproduce the formula \eqref{defengcft}, which  expresses the energy of an energy-momentum eigenstate in the deformed CFT in terms of the energy of the `same' eigenstate in the undeformed one, where by `same' we mean that the two states are connected by adiabatic evolution as $\mu$ is increased from zero to its final value.  The proposal of \cite{Guica:2019nzm} is to deduce the left/right-moving energies %($\L_0, \bar \L_0$)
 in the undeformed CFT by focussing on 
%
%
%
%Let the corresponding state in the undeformed CFT be  characterized by the left/right-moving energies $\L_0, \bar \L_0$. In order to reproduce 
%\eqref{defengcft},   we  need to find a way to relate $\L_0, \bar \L_0$ to $\L_\mu,\bar \L_\mu$.
% to identify which state in the undeformed CFT corresponds to the deformed state characterized by $\L_\mu,\bar \L_\mu$.
 %
% A good strategy is to look for
  quantities that are invariant along the $T\bar T$ flow, such as the momentum, which is quantized, %which gives one relation: $J_\mu = J_0 = R(\L_0-\bar \L_0)/(8\pi G\ell)$.
%
%Another quantity that is invariant along the flow is 
%
and the degeneracy of states  around the state of interest. For high energies, this degeneracy is computed by the horizon area of the corresponding black hole. The resulting equations completely fix the relation between the deformed and undeformed energies to the $T\bar T$ relation \eqref{defengcft}. 
%
%labeled by the CFT conformal dimensions $h,\bar h$; this was discussed in the first lecture. 

The latter calculation is of course not independent of the previous one, as one is effectively \emph{defining} the undeformed energy as the square of the black hole entropy, as follows from Cardy's formula. For typical states dual to black holes, this argument boils down to an adiabaticity one, which allows one to track sets of states. However, if one would like the argument to apply also  to atypical states, then one would  need to additionally assume a one-to-one relation between the deformed and undeformed energies; only then will the deformed energy formula be universal. In the $T\bar T$ case, this information is provided by SZ's argument that  the energy of eigenstates flows according to a universal equation.  A practical way to achieve this from a bulk perspective would be to consider exactly the same  configuration   as before the deformation (written in terms of $u,v$), and then perform a change of coordinates to the dynamical coordinates $U,V$ that render the $T\bar T$ metric \eqref{genbc} flat.   This is a rather close analogue to the field-theoretical interpretation of  the effect of the $T\bar T$ deformation \cite{dubovskytalk}. While this proposal works at the classical level, one should  understand better how to deal with quantum states in the bulk.

These caveats in mind, we find that the proposed holographic dictionary perfectly reproduces the QFT answer for the deformed energies. This match works for both signs of $\mu$. Remember from our discussion in section \ref{ttbsubsec}  that for $\mu>0$, the vacuum and the states close to it can acquire complex energy, whereas for $\mu<0$ all states above a certain energy \eqref{emax} do so. From the point of view of our calculation, what accounts for this behaviour is that in these parameter ranges, there is no real, CTC-free bulk metric that satisfies the mixed boundary conditions. This is simply a bulk manifestation of the fact that for $\mu>0$, the deformed theory cannot be   put on a circle of radius smaller than $R_c$, whereas for $\mu<0$, it cannot be put in finite volume at all.

 %Second, as explained in section \ref{}, the vacuum and the steas close to it for $\mu >0$ and most of the high energy states for $\mu<0$, the energy becomes imaginary.  In our calculation, this translates into the fact that there is no real/CTC-free metric solution \emph{Check!} that satisfies the boundary conditions. As we explained, the deformed theory does not make sense in finite size if $R<R_c$ for $\mu>0$, and for any $R$ if $\mu <0$. Thus, the boundary theory does not make sense, and the complex metrics are simply a bulk manifestation of this sickness. \emph{Careful?}

It is important to emphasize that the deformed energy only depends on the \emph{asymptotic value} of the metric, encoded by $\L$, $\bar{\L}$, and so this derivation  works also  if  matter fields are turned on, under the above assumptions for identifying the energies. This makes perfect sense from the point of view of the deformed energy, as a \emph{universal} energy formula in the QFT can only depend on the \emph{universal} asymptotic data in the bulk.  

\bigskip

Another check one may perform is the following. As is well known, the temperature of a black hole is encoded in the identifications of the euclidean solution. Written in terms of the $u,v$ coordinates, the black hole metric is nothing but BTZ, and the identifications of the rescaled coordinates $\hat u = u/R_u$ and $\hat v = v/R_v$, where $R_{u,v}$ are the circumferences of the field-dependent coordinates \eqref{defuv}, that ensure smoothness of the euclidean solution read  %\textcolor{red}{\emph{Define $R_{u,v}$ ?}}
\be
\hat u \sim \hat u +m + i n \frac{\pi \ell}{R_u \sqrt{\L}} \;, \;\;\;\;\;\; \hat v \sim \hat v +m - i n \frac{\pi \ell}{R_v \sqrt{\bar{\L}}} \label{idhatuvttb}
\ee
To the extent that states in the undeformed theory are to be identified with states in the deformed one by performing just the field-dependent coordinate transformation \eqref{defuv}, the above euclidean time identifications should simply correspond to the temperatures of the state in the undeformed CFT; more precisely, the coefficients of $i\, n$ should correspond to $\pm  1/R T_{L,R}^{[0]}$. The temperatures in the deformed theory are encoded in the identifications of the dynamical coordinates $U,V$ which, using  \eqref{idhatuvttb} and \eqref{defuv}, read
\be
U \sim U + (R_u -\hat\mu \bar \L R_v) m  + i n \pi\, \ell \left(\frac{1}{\sqrt{\L}} +\hat \mu \sqrt{\bar \L}\right)  \;, \;\;\;\;\;\; R_{u,v}  =  R + \mu (E\mp P) 
\ee
and the identification of $V$ is obtained by simply replacing $u \leftrightarrow v$ and $\L \leftrightarrow \bar \L$. It is not hard to check that the spatial identification of $U$ is by an integer multiple of the circumference since $ R_u -\hat\mu  \bar \L R_v=R$,  as expected from consistency. The temporal identification of $U$ yields 
\be
\frac{1}{T_L} = \pi \ell \left(\frac{1}{\sqrt{\L}} +\hat \mu \sqrt{\bar \L}\right) = \frac{R_u}{R \, T_L^{[0]}} + \frac{2\mu E_R}{R T^{[0]}_R}
\ee
and a corresponding expression for $T_R$, with left and right inverted.  It is easy to check that differentiating the entropy \eqref{entropyttb} with respect to $E_L$ and then replacing the left/right entropies that appear in the denominators by the undeformed entropies, which are simply proportional to $T_{L,R}^{[0]}$, yields precisely the expression above. 

It is an interesting question whether states in the quantum theory  in the bulk (e.g., a scalar quantum localized in AdS$_3$) are again related to the corresponding state in the $T\bar T$ - deformed CFT via a field-dependent coordinate transformation, as above, and whether one can test the operatorial nature of the transformation in such a setup.

%\textcolor{red}{\emph{Stress tensor operator vs its expectation value? Should be op.}}

\subsubsection{Demystifying the finite bulk cutoff proposal \label{bcutoff}}

In the above, we   presented a first principles \emph{derivation} of the large $N$ holographic  dictionary for $T\bar T$-deformed CFTs, for both signs of $\mu$, by simply applying the well-known rules of AdS/CFT to this particular case. We would now like to comment on the relation between this dictionary and an earlier proposal by \cite{McGough:2016lol}, according to which  $T\bar T$ - deformed CFTs with $\mu<0$ (in our conventions) are dual to AdS$_3$ gravity with a sharp radial cutoff.  

The proposal of \cite{McGough:2016lol}, who worked exclusively with pure $3d$ gravity,  was motivated by the perfect match between  various $T\bar T$ observables - the deformed energy spectrum, the speed of sound, the thermodynamic relations - and  the measurements of a bulk observer  sitting at a fixed radial position $r_c = 1/\sqrt{|\hat \mu|}$ %\textcolor{red}{\emph{Numerical prop. constant?}} 
in the background of a BTZ black hole. They also found a perfect match between the Hamilton-Jacobi equation obeyed by the bulk classical on-shell action with  a finite radial cutoff and the $T\bar T$ trace relation \eqref{ttbtrrel}, including an expected contribution from  the conformal anomaly.

Note that none of the above matches implies that the portion of the bulk lying outside the $r_c$ surface should be removed, which is the geometric cutoff proposal proper. The latter  is instead  motivated by finding a cure to the presence of complex energy states  in $T\bar T$ - deformed CFTs  with  $\mu<0$ in finite size. 
More concretely, remember  that for $\mu<0$ there is a maximum energy, $E_{max}$ \eqref{emax}, above which all states acquire imaginary energies. The authors of \cite{McGough:2016lol} made the interesting observation that for a black hole whose mass is at the threshold $E_{max}$, its Schwarzschild radius reaches the $r_c$ surface. Thus, by excluding black holes supported on  radii larger than $r_c$,  one can exclude the complex energy states from the spectrum, which in the case of pure gravity only consists of black hole states and the vacuum. The energy $E_{max}$ was interpreted as UV cutoff in the deformed CFT, and  the finite number of states below this cutoff would  correspond to the finite number of states of bulk quantum gravity in a finite region.

Thus, the  proposal and results of \cite{McGough:2016lol} consist of  two logically distinct themes: 
\bi
\item[i)] the (highly non-trivial) observation that various $T\bar T$ observables match the measurements of a bulk observer  at a fixed radial position $r_c = 1/\sqrt{|\mu|}$% in the background of a BTZ black hole% (energy, etc). Note  $r_c = 1/\mu$ in Schwarzschild coordinates. Were able to precisely reproduce energy spectrum, , WdW eqn, providing a highly non-trivial check of the proposal. 

\item[ii)] the proposal that the bulk degrees of freedom outside the $r_c$ surface should be removed%:  this is the geometric cutoff proposal proper, which is  motivated by the presence of complex energies for $\mu<0$. 
\ei
In the following, we would like to  comment on both these aspects from the perspective of the  holographic  dictionary we have derived. 

\vskip2mm

\noindent \textbf{i).} One can show that when $\mu<0$ and one concentrates on \emph{on-shell} solutions of \emph{pure gravity}, for which the full bulk solution takes the special form \eqref{fgexact},  the mixed boundary conditions \eqref{genbc} effectively reduce to  Dirichet boundary conditions at the bulk radius $\rho_c = - \hat \mu$ , which is independent of the energy. Likewise, the deformed stress tensor \eqref{stresst} precisely coincides with the Brown-York stress tensor on the $\rho=\rho_c$ surface, supplemented by the same counterterm as at infinity.
Consequently, all $T\bar T$ observables coincide with the measurements of an observer at this fixed radius. 

Thus, all the checks performed in step i) of \cite{McGough:2016lol}'s proposal are simply checks of this effective Dirichlet boundary condition, which follows from the mixed ones under the particular circumstances listed above.  From the point of view of the  mixed boundary conditions, the match to the observations at finite bulk radius is a pure coincidence - albeit a fascinating one - having to do with the particular way the asymptotic solution is extended into the bulk.    This coincidence no longer happens once matter field profiles are turned on, as was explicitly shown in \cite{Guica:2019nzm}.  

Note, however, that typical high-energy states in a holographic two-dimensional CFT are expected to be described by just a BTZ black hole, so the coincidence of $T\bar T$ observables with the measurements of an observer located at fixed radius in the bulk is expected to hold for the vast majority of CFT high-energy states, as depicted in the figure below.
\vskip3mm

\begin{figure}[h]
\begin{minipage}{0.43 \textwidth}
\centering
\includegraphics[width=3.8 cm]{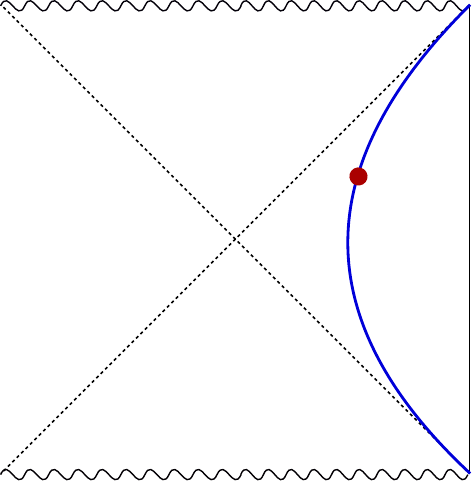} %\label{accobs}
\end{minipage}
\hspace{1.5 cm}
\begin{minipage}{0.38 \textwidth}
In \emph{typical} high energy states, (but not for general states), the $T\bar T$-deformed CFT with $\mu<0$ can be interpreted as describing the experience of an accelerated observer's laboratory, located at fixed radial coordinate $r_c= 1/\sqrt{-\hat \mu}$ in the bulk. 
\end{minipage}
\end{figure}

\vskip3mm

\noindent \textbf{ii).}  The evidence for a sharp geometric  cutoff in the bulk dual to a $T\bar T$ - deformed CFT is uncompelling:
\bi
\item %from a field-theoretical perspective,
 the geometric cutoff proposal - which, according to the holographic RG \cite{
 deBoer:1999tgo,Heemskerk:2010hk,Faulkner:2010jy} intuition, should correspond to integrating out UV degrees of freedom - is in tension with the integrability and expected UV completeness of $T\bar T$ - deformed CFTs
\item as discussed in section \ref{ttbsubsec}, $\mu<0$ $T\bar T$ is inconsistent in finite size, due to the appearance of CTCs. It is absolutely unclear why removing the associated imaginary energy states by hand should produce a consistent quantum theory\footnote{Note that the UV-completeness referred to above is expected of the theory on the plane, which has a chance of being consistent \cite{Cooper:2013ffa}, whereas the surgery of the imaginary-energy states is for the theory on the cylinder. }

\item  it can be  shown that, once matter fields are turned on, a holographic dictionary of the form suggested by \cite{McGough:2016lol}
\be
Z_{T\bar T}[\, \mu\, ] = Z_{grav}[\, r \leq r_c\,]
\ee
cannot hold: the energies do \emph{not} match and, moreover, the onset of the complex energy states can be shown to have nothing to do with a bulk distance or the presence of a  horizon \cite{Guica:2019nzm}. Thus, a precise relation between $\mu <0$ $T\bar T$ - deformed CFTs and gravity with a sharp radial cutoff can only hold when the latter theory only contains pure gauge degrees of freedom. 
\ei
Of course, for a  bulk theory with only boundary degrees of freedom, the issue of the radial cutoff becomes rather immaterial, as there is no  immediate way to probe whether the region  outside the $r_c$ surface has been removed or not. 
  If one insists upon the presence of propagating degrees of freedom in the bulk (i.e., low-lying operators in the dual CFT), then one  must choose between i) having an exact $T\bar T$ description, case in which the correct description of the bulk is only via the mixed boundary conditions, and ii) defining  the theory via the bulk with a sharp radial cutoff, case in which additional double-trace operators corresponding to matter fields must be added to the boundary action \cite{Kraus:2018xrn}.  Unlike the SZ definition of $T\bar T$, this procedure is only defined at leading order in $1/N$, by effectively transposing the gravitational picture in field-theoretical language, and there is no reason to expect that the theory resulting from  these combined irrelevant deformations is UV - complete.

%to the extent that the central charge of the ASG we discussed is a measure of the number of degrees of freedom in the $T\bar T$ - deformed CFT, the fact that it is a $\mu$ - independent constant that equals the original CFT central charge is  nicely consistent with  integrability, but not clearly consistent with a UV cutoff interpretation

Despite of the above issues, the relation between a sharp radial cutoff and $T\bar T$  has found many applications and  generalisations, such as to higher dimensions  \cite{Hartman:2018tkw,Taylor:2018xcy} (where the higher-dimensional analogue to $T\bar T$ is defined by requiring Dirichlet boundary conditions for  gravity in AdS$_{d>3}$ at a fixed finite radius), to de Sitter spacetimes \cite{Gorbenko:2018oov} (where, in addition to cutting off the energy spectrum, one  further modifies the energies via a $\mu$ - dependent cosmological constant %\textcolor{red}{\emph{Check!}}
), to tensor networks  \cite{Caputa:2020fbc} and attempts to build holographic examples with emergent time \cite{Araujo-Regado:2022gvw}.  In all of these cases, the `$T\bar T$ - deformed CFT' is defined via the bulk at large $N$ and does not possess an independent boundary definition; 
in a certain sense, it is just a new name given to the gravitational path integral with a finite radial cutoff. It would of course be valuable if one could put the boundary theory  on a firmer footing, in view of  all these potentially interesting applications. 
%
%\noindent The relation between a radial cutoff in the bulk and a UV cutoff in the boundary has a long history in terms of the holographic renormalization group \cite{deBoer:1999tgo,Heemskerk:2010hk,Faulkner:2010jy} : integrating out degrees of freedom above some UV scale in the boundary would correspond to integrating out the fluctuations of the bulk fields outside a given radius. However, the relation between the field-theory cutoff and the bulk radius  was never made precise.
 %
%Notice that for the case at hand,  the reduction in the number of degrees of freedom implicit in holographic  RG  would be in conflict with the integrability and likely UV completeness of  $T\bar T$ - deformed \emph{two-dimensional} CFTs we have been arguing for. Therefore, from this point of view,  the proposal that the dual bulk theory should have bulk  degrees of freedom removed  is at least counterintuitive.
%
For further discussion of these points, see  \cite{Guica:2019nzm,cernlect}. % and \emph{Keep?}}the lecture notes \cite{}. \textcolor{red}{ \emph{Comment entangl?}}

\subsection{The holographic dictionary for $J\bar T$-deformed CFTs\label{holodictjtbdtr} }

In this subsection, we apply the  exactly the same series of steps to derive the holographic dictionary for $J\bar T$ - deformed CFTs. The first version of this holographic dictionary, which holds when $J$ is a chiral current, was  proposed in \cite{Bzowski:2018pcy}. We will instead follow the more recent version of this dictionary presented in \cite{Georgescu:2024ppd}, in which the $U(1)$ current is non-chiral, and the analysis of anomalies is clarified.

\subsubsection{Step 1: the relation between the deformed and undeformed sources and vevs}

The minimal set of phase space variables that we need to consider are the stress tensor $T^\a{}_a$ and current $J^\a$, which are canonically conjugate to the boundary vielbein $e_\a{}^a$ and gauge field, $\mathrm{a}_\a$.  Here, latin indices denote the tangent space and greek ones are spacetime indices. The reason  we prefer the vielbein formulation is that the deformed theory is not Lorentz invariant;  consequently, the conserved stress tensor is not  symmetric and it  naturally couples to the vielbein, and not the metric. The variation of the on-shell action (identified with the classical limit of the deformed generating functional)  reads 

\be
\d S = \int d^2 x\, e \, \left( T^\a{}_{a}\, \d e_{\a}^a + J^\a \d \mathrm{a}_\a \right) \label{varactjtbs}
\ee
where $e = \det e^a_\a$ and all quantities are in the deformed theories at finite $\l$ and should, in principle, carry this label that we momentarily omit.  

The  $J\bar T$ coupling  is naturally a (null) vector, $\l^a$,  in tangent space\footnote{The manipulations to follow apply equally well to $JT^a$ - deformed QFTs, for which $\l^a$ need not be null. }.  We will write $\l^a = \l \hat \l^a$, where  $\hat \l^a$ is a unit
tangent space vector or its null counterpart. 
The variation of the Lorentzian on-shell action when $\l$ is infinitesimally changed is
given by minus the double-trace deformation of the boundary  action

\be
\p_{\l^a} S^{[\l^a]} = - \int d^2 x \, e \, \O_{J\bar T}^{[\l^a]} = - \hat \l^a \int d^2 x \, e \, \e_{\a\b} J^\a T^\b{}_a
\ee
where the $JT^a$ operator has been  appropriately covariantized.  %Since the coupling parameter is a dimensionful null vector, in  an arbitrary background it makes the most sense to keep the coupling with tangent space indices fixed, so $\mu^a= \mu \, \d^a_-$, where $\mu$ is a constant with dimensions of length and $x^-$ is a null direction along the boundary.  The covariantized multitrace operator is then
%
%\be
%\Delta S_{J\bar T} = \pm \Delta \l^a \int d^2 x \, e \, \e_{\a\b} J^\a  T^\b{}_a 
%\ee

In order to ensure the deformation is well-defined also in presence of external sources, we
assume the current $J^\a$ is exactly conserved, i.e. it is not anomalous. The requirement that
$S^{[\l^a+\Delta \l^a]}$ have a good variational principle translates into being able to write

\be
\d S^{\l^a+\Delta \l^a} = \d S^{\l^a} - \Delta \l^a \int d^2 x \,\d \left( e \, \e_{\a\b} J^\a T^\b{}_a\right)
\ee
again in the form \eqref{varactjtbs}, for some slightly modified canonical variables (the variation $\Delta \l^a$ is infinitesimal). Plugging in, we have
\bea \label{newsourcevev}
\d S^{[\l^a+\Delta \l^a]}\!\!  &= & \!\! \int d^2 x \left[
%\left [e \, T^\a{}_a \d e_\a{}^a + e \, J^\a \d \mathrm{a}_\a - \Delta \l^a \d (e  \,\e_{\a\b} T^\a{}_a J^\b) \right] \\&= & 
e \, T^\a{}_a (\d e_\a{}^a + \Delta \l^a \e_{\a\b} \d J^\b)  +    e J^\a (\d \mathrm{a}_\a  -   \Delta \l^a  \e_{\a\b} \d T^\b{}_a)  + \Delta \l^a T^\a{}_a \d(e \e_{\a\b}) J^\b \right] \nonumber \\
& & \hspace{-2.7cm}= \int d^2 x \left[ e \, T^\a{}_a \d (e_\a{}^a +\Delta \l^a \e_{\a\b}  J^\b)  +   e J^\a \d ( \mathrm{a}_\a  -   \Delta \l^a  \e_{\a\b} T^\b{}_a) +   \Delta \l^a T^\a{}_a  \left( \d (e\e_{\a\b}) - 2e \d\e_{\a\b} \right) J^\b \right]
\eea
Using $\e_{\a\b} = e \hat \e_{\a\b}$, where $\hat \e_{\a\b}$ is the Levi-Civita symbol, one can show the term in the last paranthesis vanishes, and thus the variation of the action $\d S^{[\l^a + \Delta \l^a]}$ again takes the form \eqref{varactjtbs}, with slightly modified sources and vevs. From \eqref{newsourcevev}, we can easily read off the  change in these data with the flow parameter $\l$, defined as above\footnote{These solutions are rather similar to those obtained in \cite{Bzowski:2018pcy}, if one particularises $\l^a$ to be a null vector, additionally assumes that $J^\a = \e^{\a\b} J_\b$, %(which implies, in particular, that $e^{[\l]} = e^{[0]}$) %,  a condition that is easily checked to be preserved along the restricted flow, 
and considers  a coupling of the current to $\e^{\a\b}\mathrm{a}_\b$ instead. See \cite{Bzowski:2018pcy,Georgescu:2024ppd} for more details. } %\emph{Ambiguities?} %which  are extremely simple
%
%\emph{\textcolor{red}{Recheck signs!!!}}{\color{ForestGreen}(checked)}
\be \label{defundefrel}
\p_\l e_\a{}^a = \hat \l^a \e_{\a\b} J^\b \;, \;\;\;\;\; \p_{\l} \mathrm{a}_\a = -   \hat \l^a \e_{\a\b} T^\b{}_a \;, \;\;\;\;\; \p_\l (e J^\a) = \p_\l (e T^\a{}_a) =0
\ee
%where $\hat{\l}^a=\l^a/\l$. \emph{How do you define $\l$?}(if the vector is not null it's its norm; for the null vector we  can bring it in the form $(\lambda,\lambda)$ up to signs, right?)
These flow equations are trivially solved by

\be\label{generalmapdefundef}
e^{[\l]\, a}_\a = e_\a^{[0] \, a} + \l^a \e_{\a\b} J^\b \;, \;\;\;\;\; \mathrm{a}_\a^{[\l]} =  \mathrm{a}_\a^{[0]}  -   \l^a \e_{\a\b} T^\b{}_a 
\ee

\be
(e\, J^\a)^{[\l]} = (e \, J^\a)^{[0]} \;, \;\;\;\;\; (e \, T^\a{}_a)^{[\l]} = (e\,  T^\a{}_a)^{[0]} \label{relvevs}
\ee
The expectation values  that shift the sources in \eqref{generalmapdefundef} can be considered in either the undeformed or deformed theory, as implied by \eqref{relvevs}. 
% The only expression that looks rather different from \cite{Bzowski:2018pcy} is that  for $T^{\a}{}_a$, which in our current derivation is significantly simpler. It should though be noted that the positioning of the tangent space and spacetime indices in the non-linear expression obtained in \cite{Bzowski:2018pcy} is opposite from ours, which partly explains the difference.  \emph{Old below:}

\subsubsection{Step 2: the holographic dictionary}

In order to write down the holographic dictionary, one first needs to correctly model the $U(1)$ currents, and in particular their anomalies, in the undeformed CFT. We assume that the current $J$ that enters the definition of the $J\bar T$ operator is not anomalous, and thus we can globally write 

\be
J^{\a} = \e^{\a\b} \p_\b \phi \label{topcur}
\ee
for some scalar field $\phi$, defined via the above bosonisation of the current $J$.  While this definition holds in principle in the deformed theory, the map \eqref{relvevs} implies it it will be true also in the undeformed CFT to which we map, where the vielbein is $e^{[0]}$. Note that for almost all this subsection we will be working in the undeformed CFT, though we will mostly omit the corresponding labels ${}^{[0]}$. 

The Hodge dual $J'= \star J$, which is analogous to the usual shift current,  is then anomalous
\be \label{anomalyshift}
\nabla_\a J'^{[0] \a} = -\frac{k}{2\pi} \e^{\a\b}_{[0]} \p_\a \mathrm{a}_\b^{[0]} 
\ee
These properties of the two  currents  can be easily modeled in holography  using two three-dimensional Chern-Simons gauge fields \cite{Banados:2006fe}, whose (bulk) anomaly reproduces the field theory one %\textcolor{red}{\emph{Conventions $\e$?}}

\be \label{proposalactioncs}
S_{CS} = - \frac{k}{8\pi} \int d^3 x \sqrt{g} \, \e^{\mu\nu\rho} (A_\mu+B_\mu) \p_\nu (A_\rho-B_\rho)
\ee
This action is by construction invariant under equal and opposite gauge transformations of $A,B$, but picks up a boundary term under equal shifts of the gauge potential; thus, only half of the bulk gauge symmetry is preserved at the boundary. %, and one expects  the associated current to be non-anomalous. 
The perhaps unusual minus sign in front of the action is due to our unusual conventions  for the $\e$ tensor.  The Chern-Simons equations of motion imply as usual that the gauge connections are flat; choosing the radial gauge $A_\rho=B_\rho =0$ moreover implies that the remaining components of the bulk fields, $A_\a, B_\a$ are independent of $\rho$,    and thus effectively two-dimensional. 

As is well-known, the action \eqref{proposalactioncs} does not have a well-defined variational principle. To make it so, we need to add a boundary term \cite{Kraus:2006wn}. A choice that yields - in Minkowski space -  a chiral current coupling to $A_-$ and an anti-chiral one coupling to $B_+$ is  %\textcolor{red}{Notation bnd?}
\begin{align}
S_{bnd}&=  \frac{k}{16\pi}\int_{\partial} d^2x \sqrt{-\gamma}\gamma^{\alpha\beta}(A+B)_{\alpha}(A+B)_{\beta}
\end{align}
Only this term will contribute to the stress tensor, since \eqref{proposalactioncs} is topological. The variation of the action is: %\emph{\textcolor{red}{You can just put it on-shell, directly. In addition, the first line is not defined, likely wrong}}
%{\color{ForestGreen}(checked)}
\begin{align}\label{varactcs}
\delta S&=  \delta S_{CS}+\delta S_{bnd}= -\frac{k}{8\pi}\int d^3x\sqrt{-g}\epsilon^{\mu\nu\rho}(F^A_{\mu\nu}\delta A_{\rho}-F^B_{\mu\nu} \delta B_{\rho})+ \nonumber\\
&+ \frac{k}{8\pi}\int_{\partial} d^2x \sqrt{-\gamma}\bigg[(A_{\alpha}+B_{\alpha})(\gamma^{\alpha\beta} -\epsilon^{\alpha\beta})\delta A_{\beta}+(A_{\alpha}+B_{\alpha})(\gamma^{\alpha\beta}+\epsilon^{\alpha\beta})\delta B_{\beta}\bigg]+ \nonumber\\
&+\frac{k}{16\pi}\int_{\partial}d^2x \sqrt{-\gamma}\bigg[(A+B)_{\alpha}(A+B)_{\beta}-\frac{\gamma_{\alpha\beta}}{2}\gamma^{\mu\nu}(A+B)_{\mu}(A+B)_{\nu}\bigg]\delta\gamma^{\alpha\beta}
\end{align}
where we have taken into account the fact that the outward-pointing unit normal to the AdS boundary is $- (2\rho/ \ell)\, \p_\rho$.  %{\color{ForestGreen}(I agree, the boundary is at $z=0$ and the normal is positive inward)} \emph{Correct? Factors $\ell$ ok?}{\color{ForestGreen} (yes, the unit vector is $-\frac{z}{\ell}\partial_z$ but we also get an $\ell$ when we go from $\sqrt{-g}$ to $\sqrt{-\g})$} 
 From the second line above, we immediately read off the components of the two chiral currents, which are  linear combinations %{\color{red}$\frac{1}{2}(J \pm \star J)$  \emph{Check!}  of the deforming current } 
  of the form $\frac{1}{2}(\star J\pm J)$. 
%
%{\color{ForestGreen} On-shell, we can write:
%\begin{align}
%\hspace{-1.5cm}\delta S=\frac{k}{8\pi}\int d^2x \sqrt{-\g}\epsilon^{\alpha\beta}(A+B)_{\beta}\bigg[\epsilon_{\alpha\rho}(\delta A+\delta B)^{\rho}-(\delta A-\delta B)_{\alpha}\bigg]-\frac{k}{16\pi}\int d^2x\sqrt{-\g}\bigg[(A+B)_{\alpha}(A+B)_{\beta}-\frac{\gamma_{\alpha\beta}}{2}(A+B)^2\bigg]\delta\gamma^{\alpha\beta}
%\end{align}
%(quick check: in flat space we get from the first part $\frac{k}{2\pi}(A_+\delta A_- - B_-\delta B_+)$)}
It is then natural to identify  the topological current \eqref{topcur} as %\emph{Superscript $[0]$ on $\e$?}
\be
J^{[0] \a} =  \frac{k}{4\pi} \e^{\a\b} (A_\b + B_\b)  \label{holocur}
\ee  
since the right-hand side is conserved, using the Chern-Simons equations of motion, for \emph{any} flat connections $A, B$, just like the left-hand side is conserved with no restriction on $\phi$. This identification implies  of course that 
\be
J'^{[0]}_\a =   \frac{k}{4\pi} (A_\a+B_\a)% =  \p_\a \phi 
\label{holocurprime}
\ee
To identify the field theory source that couples to the topological current \eqref{holocur}, we simply interpret the second  line in the action variation \eqref{varactcs} as $\int d^2 x \sqrt{-\g} J^{\a} \d \mathrm{a}_\a$, with $J^\a$ given in \eqref{holocur}, plus some extra contribution proportional to the variation of the projectors.  Concretely, on-shell the action variation \eqref{varactcs} becomes
\bea
\delta S^{on-shell} & = &\frac{1}{2}\int_{\partial} d^2 x \sqrt{-\gamma}\, J^{[0]\alpha}\delta[A_{\alpha}-B_{\alpha}-\epsilon_{\alpha\beta}(A^{\beta}+B^{\beta})] - \\
&&\hspace{1cm} -\frac{k}{16\pi}\int_{\partial} d^2 x \sqrt{-\gamma}\bigg[(A+B)_{\alpha}(A+B)_{\beta}-\frac{\gamma_{\alpha\beta}}{2}(A+B)^2\bigg]\delta\gamma^{\alpha\beta} \nonumber 
\eea
This immediately leads to the identification
{\color{red} 
}
\be
\mathrm{a}_\a = \frac{1}{2} [ A_\a - B_\a - \e_{\a\b} (A^\b+B^\b)] \label{holosource}
\ee
The variation of the action also allows us  to read off  the Chern-Simons contribution to the energy-momentum tensor\footnote{ Note that a na\"{i}ve computation using just the last term in \eqref{varactcs} would yield \emph{minus} the above answer, 
 as $T_{\alpha\beta}=-\frac{2}{\sqrt{-\gamma}}\frac{\delta S }{\delta \gamma^{\alpha\beta}}$ in Lorentzian signature. The contribution of the $\e$ tensor variation is \emph{essential} for obtaining the correct relative sign between the stress tensor and the $U(1)$ charge, to which the $J\bar T$-deformed spectrum is highly sensitive. %\emph{Correct?}} {\color{ForestGreen}(yes, keep the comment. we got the correct signs only after taking into account this contribution from the projectors)}\emph{Comment literature!}
 }
\begin{align}\label{energymomtensor}
T^{CS}_{\alpha\beta}&=\frac{k}{8\pi}\bigg[ (A_\a+B_\a)(A_\b+B_\b) - \frac{1}{2} \g_{\a\b} (A+B)^2\bigg]
\end{align}
Note that, since it only depends on the sum $A+B$, the Chern-Simons contribution to the stress tensor is also fully determined by $\phi$, i.e. it only involves the expectation values of the currents. The full stress tensor is given by the  sum of the gravitational and the Chern-Simons contributions

\be
T_{\a\b} = T_{\a\b}^{g} + T_{\a\b}^{CS} \;, \;\;\;\;\; \;\; T_{\a\b}^{g}  = \frac{1}{8\pi G \ell} g^{(2)}_{\a\b}
\ee
where $g^{(2)}$ is the first subleading Fefferman-Graham coefficient in \eqref{fg}, and we have assumed that the boundary metric $g^{(0)} \leftrightarrow \g^{[0]}$ is flat, in order  to drop an additional potential contribution to $T^g$ proportional to its Ricci scalar.

To summarize, in terms of bulk fields, the undeformed sources $\g^{[0]}$ and $\mathrm{a}_\a^{[0]}$ correspond to $g^{(0)}$ in the Fefferman-Graham expansion and, respectively,  the combination \eqref{holosource} of the bulk gauge fields in radial gauge. The expectation values of the dual currents  are encoded in \eqref{holocur} and the above expression for the holographic stress tensor.  The bulk field combinations that correspond to the deformed sources and vevs are obtained by simply plugging this undeformed dictionary into \eqref{generalmapdefundef} - \eqref{relvevs}. 

\subsubsection{The gravitational phase space}

To build the gravitational phase space corresponding to the deformed theory, we set the sources in the deformed theory to zero and look for the most general solution to the bulk equations of motion satisfying the corresponding boundary conditions. Using \eqref{generalmapdefundef}, the latter read

\be
e_\a^{[0]\, a} = \d_\a^a - \l^a \e_{\a\b} J^\b \;, \;\;\;\;\; \mathrm{a}_\a^{[0]} =   \l^a \e_{\a\b} T^\b{}_a \label{sourcesflatsp}
\ee
%to which we need to find the most general solution. The conservation equations of the various currents/stress tensor are encoded in the bulk Einstein/CS equations. 
%
Plugging in \eqref{topcur}, the CFT vielbein \eqref{sourcesflatsp} then takes the form 

\be
e_\a^{[0]\, a} = \d_\a^a - \l^a \p_\a \phi = \begin{pmatrix}
1 & -\l \partial_U\phi\\
0 & 1-\l \partial_V\phi
\end{pmatrix} \label{explbackviel}
\ee
The most general solution to three-dimensional Einstein gravity with this boundary condition is simply obtained by acting with the diffeomorphism 
\be
u \r U \;, \;\;\;\;\;\; v \r V -\l \phi + const \label{jtbgendiffeo}
\ee
on the most general Ba\~{n}ados metric, which is parametrised by two arbitrary functions $\L(u), \bar \L(v)$ that are assumed to be given.  

Let us now determine the bulk gauge fields, in  radial gauge. Using the relation \eqref{relvevs} between the expectation value of  $J$  in the deformed and the undeformed theory, we find  that the current in the undeformed CFT on the background  \eqref{explbackviel} is %\textcolor{red}{\emph{Check!}}

\be
J^{\a}_{[0]} = \frac{e^{[\l]}}{e^{[0]}} J^\a_{[\l]} = \left( - \frac{2\partial_V\phi}{1-\l\partial_V\phi},  \frac{2\partial_U\phi}{1-\l\partial_V\phi} \right) \; \;\;\; \Rightarrow \;\;\;\;%J^{[0]}_\a = \left( \partial_U\phi\frac{1+\l\partial_V\phi}{1-\l\partial_V\phi} , -\p_V \phi\right)
J'^{[0]}_{\a} = \e_{\a\b}^{[0]} J^{[0]\, \b} =  (\p_U \phi, \p_V \phi)
\ee
Using  the Chern-Simons holographic dictionary \eqref{holocurprime}, this determines $A_\a+B_\a= \frac{4\pi}{k}\p_\a \phi$. % (still unkown), sinc
%
%\be
%J'^{\a}_{[0]}  = \e_{\a\b}^{[0]} J^{[0]\, \b} =\left(\frac{2\p_V\phi}{1-\l\partial_V\phi},\frac{2\p_U\phi(1+\l \p_V\phi)}{(1-\l\p_V\phi)^2}\right)  \;, \;\;\;\;\;\; J'^{[0]}_{\a} = (\p_U \phi, \p_V \phi) \label{exprjp}
%\ee
To determine the individual $A$ and $B$ we use holographic expression \eqref{holosource} for $\mathrm{a}_\a$ and the relation \eqref{sourcesflatsp}, whose right-hand-side is known because the stress tensor only depends on  $A+B$. The expressions simplify if we work in terms of the tangent space components $a = \pm$.  For the stress tensor, we find
%
%The expression for the stress tensor simplifies if we write it in tangent space where, in  components, we have 
%
\be \label{energymomtg}
T_{++} = \frac{\mathcal{L} }{8\pi G \ell}+\frac{2\pi}{k}\frac{(\partial_U\phi)^2}{(1-\lambda\partial_V\phi)^2} \;, \;\;\;\;\;\; T_{--} = \frac{\bar{\mathcal{L}}}{8\pi G \ell\,}+\frac{2\pi}{k}\frac{(\partial_V\phi)^2}{(1-\lambda\partial_V\phi)^2}
\ee 
and $T_{+-}=T_{-+} =0$. For the gauge fields

\be \label{sourcestgspace}
\mathrm{a}_+ = - B_+= 0 \;, \;\;\;\;\;\; \mathrm{a}_- = A_- =  -\l T_{--}% = {\color{red} -} \l \left( \frac{\ell \, \bar \L}{8\pi G}  + \frac{2\pi}{k}\bigg(\frac{\partial_V\phi}{1-\lambda\partial_V\phi}\bigg)^2 \right)
\ee
The remaining components, $A_+$, $B_-$, are determined from knowledge of $A+B$ in terms of the presumably known functions $\bar \L, \phi$. The spacetime components of the holographic stress tensor and gauge fields can be obtained by simply multiplying with $e^{[0]\, a}_\a$, given in \eqref{defmet}.  See \cite{Georgescu:2024ppd} for the complete expressions. 

%According to   \eqref{sourcesflatsp}, the gauge field source \eqref{holosource} should be identified with the stress tensor. In tangent space (where the calculation is simpler), we have

%with $T_{--}$ given in \eqref{energymomtg}. Together with \eqref{sumofcurrents}, these  determine $A_{\pm}, B_{\pm}$ individually . The spacetime components are obtained by multiplying with the vielbein \eqref{defmet}, and we obtain   %{\color{red} Redo/check signs!} 
%
%{\color{ForestGreen}(checked)}
%
%\be \label{gaugefieldA}
%A_U = \frac{4\pi}{k}\frac{\p_U\phi}{1-\l \p_V\phi} + \l^2 \p_U\phi \,T_{--} \;, \;\;\;\;\; A_V = -\l (1-\l \p_V\phi) T_{--} 
%\ee
%%
%\be  \label{gaugefieldB}
%B_U = \l \p_U\phi \left(-\frac{4\pi}{k}\frac{\p_V\phi}{1-\l \p_V\phi} -\l T_{--}\right)  \;, \;\;\;\;\;\; B_V = \frac{4\pi}{k}\p_V \phi +\l (1-\l \p_V\phi) T_{--} 
%\ee
Note that so far in our holographic analysis, the field $\phi$ could be arbitrary.  Nonetheless, 
the flatness conditions for the  gauge fields result in 
an  equation of motion for it.  
 Even though there are two gauge fields, one obtains a single condition on $\phi$ since, as we already discussed, the flatness condition for $A+B$ does not yield a constraint. The flatness condition for $A-B$ can be written, using 
\eqref{holosource}, as
%
%\be
% \e^{\a\b} \p_\a (A_\b -B_\b) = 2 \nabla_\a (\e^{\a\b} \mathrm{a}_\b) + \frac{4\pi}{k} \nabla_\a (A^\a+B^\a)=  -2 \l^a \nabla_\a T^\a{}_a + \frac{4\pi}{k} \Box \phi =0
%\ee
\be
 \e^{\a\b} \p_\a (A_\b -B_\b) = 2 \nabla_\a (\e^{\a\b} \mathrm{a}_\b) +  \nabla_\a (A^\a+B^\a)=  2 \l^a \nabla_\a T^\a{}_a + \frac{4\pi}{k} \Box \phi =0
\ee
Since the stress tensor is conserved, the first term may be dropped, and we simply find that $\phi$ satisfies the free wave equation in the non-trivial metric \eqref{defmet}. The latter takes precisely the form of the  equation of motion for a $J\bar{T}$-deformed free boson 
%{\color{ForestGreen}(rechecked)}
%
\begin{align}\label{eomdfboson}
\partial_V\left(\frac{\partial_U\phi}{1-\l\partial_V\phi}\right)=0
\end{align}
%{\color{blue}[Thus,  in the absence of the source, we obtain that $\phi$ satisfies the eom of a free scalar in a $J\bar{T}$-deformed CFT. When the sources are turned on, it is not obvious that this is still true.] } 
whose general solution takes the form 
\be  \label{scalarfieldonshell}
\phi(U,V) = f(U) + g (v)  \;, \;\;\;\;\;\; v = V-\l \phi + const
\ee
It immediately follows that the quantities

\be\label{tangtspparam}
 \frac{\p_U\phi}{1-\l \p_V\phi}= f'\equiv  \J(U)     \;, \;\; \;\;\; \frac{\p_V\phi}{1-\l \p_V\phi} =g'  \equiv   \bar \J(v)
\ee
are only functions of $U$ and, respectively, the field-dependent coordinate $v$. Note these quantities correspond precisely to the tangent space components of the current combinations $\frac{J'\pm J}{2}$, which should be identified\footnote{To obtain the components of $K_\a, \bar K_\a$ in the deformed theory, one should first compute the components of  $ \frac{1}{2}(J'^{[0]} \pm J^{[0]})^\a$ by multiplying with $e^{[0] \a}_a$, and then translate the currents to the deformed theory by multiplying by $e^{[0]}/e^{[\l]}$. \label{currcompcomp} } with  $K_\a, \bar K_\a$  \eqref{eq:JTbarKUKV} - \eqref{kbarfb} in   the $J\bar T$ - deformed free boson case, and the more general currents \eqref{comprmcurrent} discussed in   Hamiltonian language. %\emph{Mention gauge field gen gauge transf.}

As the most general $J\bar T$ - deformed background metric can be obtained by applying the diffeomorphism \eqref{jtbgendiffeo} to a general Ba\~{n}ados geometry, the general solution for the gauge fields can also be obtained from that for  Chern-Simons  fields  with standard boundary conditions

\be
A_u = \frac{4\pi}{k} \p_u \phi \;, \;\;\;\;\; A_v = 0 \;, \;\;\;\;\; B_u =0\;, \;\;\;\;\;\; B_v = \frac{4\pi}{k} \p_v \phi 
\ee
via the same diffeomorphism, together with a field dependent gauge transformation 

\be
A_\a \r A_\a - \p_\a \Lambda\;, \;\;\;\;\;\; B_\a \r B_\a + \p_\a \Lambda \;, \;\;\;\;\; \Lambda = \l \int^v dv' T_{--}(v') \label{fielddepgaugetr2}
\ee
where $T_{--}$ is given in \eqref{energymomtg} and is purely a function of $v$.

To summarize, the most general solution to the gravity - Chern-Simons equations of
motion that satisfies the $J\bar T$- deformed boundary conditions \eqref{sourcesflatsp} is parametrized by
four arbitrary functions $\L (U), \bar \L (v), \J (U), \bar \J (v)$ of the left-moving coordinate $U$ and, respectively, of the
field-dependent `right-moving' coordinate $v$.  The latter is itself  determined by  $\J, \bar \J$ via the defining equation \eqref{defnv}, with solution \eqref{fielddepcoordinitial}.
%and \eqref{tangtspparam}. 
The locally AdS$_3$ metric and flat gauge fields are explicitly given by %\textcolor{red}{\emph{Conventions?}}

\begin{align}\label{deformedmetricsec6}
ds^2&=\ell^2 \frac{ d\rho^2}{4 \rho^2}+\bigg(\mathcal{L}+\frac{\l^2\bar{\mathcal{L}}\mathcal{J}^2}{(1+\l\bar{\mathcal{J}})^2}-\frac{1+ \rho^2\mathcal{L}\bar{\mathcal{L}}}{\rho}\frac{\l\mathcal{J}}{1+\l\bar{\mathcal{J}}}\bigg)dU^2
+\bigg(\frac{1+ \rho^2\mathcal{L}\bar{\mathcal{L}}}{\rho (1+\l\bar{\mathcal{J}})}-\frac{2\l\bar{\mathcal{L}}\mathcal{J}}{(1+\l\bar{\mathcal{J}})^2}\bigg)dUdV+\frac{\bar{\mathcal{L}}}{(1+\l\bar{\mathcal{J}})^2}\, dV^2 \nonumber
\end{align}
%and of the gauge fields \eqref{gaugefieldA}, \eqref{gaugefieldB} 
%\textcolor{red}{Factors $2\pi$ and $k$, maybe also $\ell$?}
\be
A_U = \mathcal{J} + \frac{\l^2\mathcal{J}(\kappa\bar{\mathcal{L}}+\bar{\mathcal{J}}^2)}{2(1+\l\bar{\mathcal{J}})}\;, \;\;\;\;\;\;\; A_V = -\frac{\l(\kappa\bar{\mathcal{L}}+\bar{\mathcal{J}}^2)}{2(1+\l\bar{\mathcal{J}})}\;, \;\;\;\;\;\;\;
B_U = -\frac{\l\mathcal{J}}{1+\l\bar{\mathcal{J}}}\bigg[\bar{\mathcal{J}}+\frac{\l(\kappa\bar{\mathcal{L}}+\bar{\mathcal{J}}^2)}{2}\bigg] \nonumber \ee

\be B_V = \frac{\bar{\mathcal{J}}}{1+\l\bar{\mathcal{J}}}+ \frac{\l(\kappa\bar{\mathcal{L}}+\bar{\mathcal{J}}^2)}{2(1+\l\bar{\mathcal{J}})}\;, \;\;\;\;\;\;\;\; \kappa\equiv \frac{k}{16\pi^2 G\ell} 
\label{paramgaugefields}
\ee
The latter have been rescaled  by an overall factor of $\frac{k}{4\pi}$ with  respect to the previous subsection, to simplify the notation. The relations \eqref{relvevs} map the expectation values in the deformed theory without sources to the holographic
expectation values encoded in this bulk solution as
\be
T^{[\lambda]}_{UU}%=T_{++}
 =  \frac{\mathcal{L} }{8\pi G \ell}+\frac{2\pi}{k}\J^2 \;, \;\;\;\;\;\;\; T^{[\lambda]}_{VU}=0 \;, \;\;\;\;\;
T^{[\lambda]}_{UV}=%\lambda\partial_U\phi T_{--} =
\l \J T_{VV}^{[\l]}  \;, \;\;\;\;\;\; T^{[\lambda]}_{VV}=%(1-\lambda\partial_V\phi)T_{--}=
\frac{1}{1+\l \bar{\J}}\bigg(\frac{\bar{\mathcal{L}}}{8\pi G \ell\,}+\frac{2\pi}{k}\bar{\mathcal{J}}^2\bigg)  \nonumber
\ee
\be
K_\a \equiv \frac{1}{2} (J'^{[\l]} + J^{[\l]})_\a = \left(\J,\, 0\, \right) \;, \;\;\;\;\;\bar K_\a \equiv \frac{1}{2} (J'^{[\l]} - J^{[\l]})_\a = \left(\frac{\l \J \bar \J}{1+\l \bar \J},\, \frac{\bar \J}{1+\l \bar \J}\, \right)\label{energymomdefth}
\ee
where the computation of the current components is explained in the footnote \ref{currcompcomp}, and also \cite{Georgescu:2024ppd}. While this  calculation was performed in pure $3d$ gravity coupled to Chern-Simons gauge fields (a nondynamical theory), the asymptotic expansions of the metric and gauge fields obeying the boundary conditions \eqref{sourcesflatsp} is unaffected by the inclusion of matter field expectation values, which change the form of the expansion at subleading orders only.

% \textcolor{red}{Anything about the fact this is just asymptotic, and how to include matter?}

\subsubsection{Checks of the dictionary}

As in the $T\bar T$ case,  the simplest check is to compute the entropy-to-energy relation for black hole backgrounds, parametrised by constant $\L, \bar \L, \J, \bar \J$. Integrating the expressions \eqref{energymomdefth} over $\s$, we obtain\footnote{The same expression for the energies and conserved charges can also be obtained using the covariant phase space formalism, yielding  a non-trivial check of the holographic formulae \eqref{energymomdefth}.} 
%{\color{red}\emph{Work out Factors!!}}
%
\be \label{spectrumbulk}
E_L = \int d\s T_{t U}^{[\l]} = \left(\frac{\L}{8\pi G\ell} +\frac{2\pi}{k}\J^2 \right)R \;, \;\;\;\;\;\;E_R = - \int d\s T_{t V}^{[\l]} =  \left(\frac{\bar \L}{8\pi G\ell} +\frac{2\pi}{k} \bar \J^2\right) R_v
\ee
\be
 Q_L \equiv \int d\s K_t =  \J R \;, \;\;\;\;\;\;  Q_R \equiv \int d\s \bar K_t=  - \bar \J R_v
\ee
where we introduced the  convenient notation $R_v$, standing for  `radius of the field-dependent coordinate'

\be
R_v = R-\l w \;, \;\;\;\;\; w \equiv \int d\s \, \p_\s \phi = \frac{R (\J + \bar \J)}{1+\l \bar \J} = Q_L - Q_R
\ee
As its definition indicates, $w$ is the winding of the field $\phi$, an integer. We note that the parametrisation in terms of $\J, \bar \J$ is not particularly adept at capturing this simple fact.
%, $w=Q_L-Q_R$. Here, we have simply defined $Q_{L,R}$ as the conserved charges associated with the chiral combinations of the currents.  

For  a black hole, $\L, \bar \L $ are positive. The horizon is located at $\rho = (\ell^2 \L \bar \L)^{-1/2}$  %\textcolor{red}{\emph{Factors $\ell$!}} %\textcolor{red}{Correct?}{\color{ForestGreen}(yes)} 
and the Bekenstein-Hawking entropy reads %{\color{red}%\emph{Factors!!}}
\be \label{entropyform}
S = \frac{1}{4 G}(\sqrt{\L}R + \sqrt{\bar \L} R_v ) = \sqrt{\frac{\pi\ell}{2 G}}\bigg( \sqrt{R E_L-\frac{2\pi}{k}Q_L^2} + \sqrt{E_R R_v -\frac{2\pi}{k} Q_R^2} \bigg)
\ee
where in the second step we have rewritten the answer in terms of the conserved charges \eqref{spectrumbulk}. Trading $\ell/G$ in the prefactor for the   Brown-Henneaux central charge $c=\frac{3\ell}{2G}$, we find  precisely the universal entropy formula \eqref{entropyJTbarFT} for a $J\bar T$-deformed CFT as a function of the (deformed) conserved charges.  Note this formula makes no use of, nor reference to,  which charges are quantized, so our  identification of the currents in \eqref{energymomdefth} with the ones whose associated charges are $Q_{L,R}$ was important.

\medskip

While the above match is satisfactory, one would also like to  obtain the relation between the deformed and undeformed observables, such as the energies and the charges. For this, one may use the fact that
the entropy, the angular momentum and the winding charge must not vary with $\l$: the first, because the deformation is adiabatic and does not change the number of states, while the latter two, because the corresponding charges are quantized. To fully determine the deformed spectrum, which depends on four parameters,  one needs  one more $\l$ - independent quantity.  The proposal of \cite{Georgescu:2024ppd} is  that this is the charge associated to the combination $J'^{[0]\a}+ \frac{k}{2\pi} \e_{[0]}^{\a\b} \mathrm{a}_\b$ which, using \eqref{anomalyshift},  is exactly conserved. One thus  expects its charge  to stay constant along the flow, and equal the quantized shift charge in the undeformed CFT. %\textcolor{red}{\emph{Argument more?}}{\color{ForestGreen}(I think we have comments about this in section 5.1, I wrote something in green)}

Denoting the conserved charges in the deformed theory as $Q_L, Q_R$ and in the undeformed one as $J_0, \bar J_0$, the relations below follow %\emph{Factors!}
\be
Q_L-Q_R = w = J_0 -\bar J_0 \;, \;\;\;\; Q_L+Q_R - \frac{\l k}{2\pi} E_R = J_0 +\bar J_0
\ee
which allow us to derive  the known expressions for $Q_{L,R}$ in the deformed theory in terms of the quantized charges $J_0,\bar J_0$. It turns out that, in the bulk, the above conditions are equivalent to requiring that the holonomies of the gauge fields $A, B$ around the $\s$ direction 
\begin{align}\label{holonomies}
\frac{k}{4\pi}\oint A_{\sigma}&=Q_L-\frac{\lambda k}{4\pi} E_R=J_0\hspace{1cm}\frac{k}{4\pi}\oint B_{\sigma}=-Q_R + \frac{\lambda k}{4\pi}=-\bar{J}_0
\end{align}
do not change with $\l$.  See \cite{Georgescu:2024ppd} for more details. The formula for the deformed energies in terms of the undeformed ones follows from the four constraints we described.

\medskip

Since  the bulk spectrum perfectly matches that of a $J\bar T$ - deformed CFT, obviously the thermodynamics should also match. The relationship between the thermodynamic quantities of the deformed theory - which are to be read off from the smoothness conditions for the analytically continued deformed  spacetime - and those of the undeformed one - read off from the smoothness of the euclideanised charged BTZ solution - can be  obtained by noting the two solutions are  simply related by the coordinate transformation \eqref{jtbgendiffeo}, accompanied by the gauge transformation \eqref{fielddepgaugetr2}.

In the bulk, the fact that the deformed solution written in terms of the field-dependent coordinates is identical to the undeformed one with some parameters $\L, \bar \L$ implies that the identifications of the field-dependent rescaled coordinates  $\hat{u}=\frac{U}{R},\hat{v}=\frac{v}{R_v}$ are identical to the identifications of the corresponding rescaled coordinates in the undeformed geometry 
 %\textcolor{red}{Careful factors of $2\pi$, probably best to divide by them below}
\begin{align}
\hat{u}\sim \hat{u}+ m +\frac{i n}{R T_L^{[0]}}\;, \hspace{1cm}\hat{v}\sim \hat{v}+ m -\frac{i n}{R T_R^{[0]}} \;, \;\;\;\;\;\; m, n \in \mathbb{Z}
\end{align}
where  $T^{[0]}_{L,R}$ are related to the parameters of the background in the standard way, fixed by smoothness 
\be
\frac{1}{T_L^{[0]}}=\frac{\pi\ell}{\sqrt{\mathcal{L}}}\;, \;\;\;\;\; \frac{1}{T_R^{[0]}}=\frac{\pi\ell R }{R_v\sqrt{\bar{\mathcal{L}}}}
\ee
Remember $v$ is related to $V$ via the shift $v = V -\l \phi + const$, where for constant solutions 
\begin{align}\label{constantsolparam}
\phi(U,V)&=\frac{\frac{Q_L R_v}{R}U -Q_R V}{R-\lambda Q_L}
\end{align}
which simply follows from requiring that the charges of the conserved currents associated with $\phi$ be given by $Q_{L,R}$. % \emph{Correct?}} {\color{ForestGreen}(yes, we get $\frac{Q_L}{R}$ for $\frac{\partial_U\phi}{1-\lambda\partial_V\phi}$ and $-\frac{Q_R}{R_v}$ for $\frac{\partial_V\phi}{1-\lambda\partial_V\phi}$)}
%
%{\color{blue}
%In the bulk, we can extract the temperatures from the identifications of the euclidean geometry required for smoothness. Expressed in the coordinates $\hat{u}=\frac{U}{R},\hat{v}=\frac{v}{R_v}$, \textcolor{red}{Notation!!!!} the metric is just BTZ, for which the smoothness of the euclidean geometry requires:
%Using these notations (better place to move it?):
%For the stationary solution we can simply write:  \textcolor{red}{Notation!!!}
%\begin{align}
%\hat{u}=\hat{U}=\frac{U}{R}\hspace{1cm}\hat{v}=\frac{V-\frac{\lambda Q_L}{R} U}{R-\lambda Q_L}
%\end{align}}
Then, using the relation between the two sets of coordinates: $\{\hat{u},\hat{v}\}$ and $\{U,V\}$, which  can be conveniently written as 
\begin{align}
V&=(R-\lambda Q_L) \hat{v}+\lambda Q_L \hat{u}\hspace{1cm}U=\hat{u} R
\end{align}
one can  immediately read off the deformed temperatures, associated to the periodicities of the  latter. More precisely,  the periodicity of the $U$ coordinate yields
\begin{align}
T_L=T_L^{[0]}=\frac{\sqrt{\mathcal{L}}}{\pi\ell}
\end{align}
From the periodicity of $V$ we obtain: 
\begin{align}
V+  m(R-\lambda Q_L +\lambda Q_L)-in\bigg( \frac{R-\lambda Q_L}{R T_R^{[0]}}-\frac{\lambda Q_L}{R T_L^{[0]}}\bigg)=V+ m R-\frac{i n }{T_R}
\end{align}
The middle terms trivially agree, while the identification of the last terms gives: 
\begin{align}
\frac{1}{T_R}&=\frac{R-\lambda Q_L}{R T_R^{[0]}}-\frac{\lambda Q_L}{R T_L^{[0]}}
\end{align}
which is precisely the relation \eqref{fieldtheorytempapp} between deformed and undeformed temperatures in the boundary theory. One can also match the
 chemical potentials to their field-theoretical values \eqref{fieldtheorypotentialapp}.

\subsection{Asymptotic symmetries\label{asysymmdtr}}

In this section, we would like to determine the asymptotic symmetries of AdS$_3$ gravity with the  mixed boundary conditions  that correspond to $T\bar T$ and, respectively, $J\bar T$ - deformed boundary CFTs. This will give us a first glimpse into the  extended symmetries of these very special non-local field theories. 

Since the boundary conditions on the bulk fields are known - being fully determined by the dual field theory - determining the asymptotic symmetry group should  just  be a simple exercise in three steps:

\begin{enumerate}
\item determine the most general diffeomorphisms that obey the boundary conditions
\item compute the conserved charges, using one's favourite formalism
\item compute the charge algebra by varying the charge with respect to another diffeomorphism
\end{enumerate}
In the case of $J\bar T$, where gauge fields must be present in the bulk, one must also take into account the corresponding gauge transformations in the first step, in addition to diffeomorphisms.

Despite the apparent simplicity of the above recipe, there are several technical - as well as conceptual - complications.   For example, since 
the holographic analyses we performed  gave us  access to the expectation values of the field theory currents (namely, \eqref{stresst} for $T\bar T$ and \eqref{energymomdefth} for $J\bar T$), one may think the simplest way to compute the conserved charges is to dot these currents into the components of the allowed diffeomorphisms or gauge transformations, as we did for computing the global conserved charges.  However, for a general allowed diffeomorphism the resulting current is not conserved,  and so \cite{Guica:2019nzm}%,
, where this method was used, had to propose a  related, but different vector in order to obtain a conserved charge. The same method was used in  \cite{Bzowski:2018pcy} for $J\bar T$.   While their answer for the charges was ultimately correct, it seems important to be able to derive the expression for the charges from first principles and  fully trace the link between bulk diffeomorphisms/gauge transformations and the functions that parametrise the conserved charges.

For this reason, we turn to the covariant phase space formalism, since it provides an entirely algorithmic  procedure, with minimal ambiguities, for constructing the conserved charges,  once the action and the   boundary conditions on the fields are given.  As we will see, the results - first worked out in the appendix of \cite{Georgescu:2022iyx} for $T\bar T$, and in \cite{Georgescu:2024ppd} for $J\bar T$ - will turn out to perfectly match an entirely independent field-theoretical analysis that we present in section \ref{infsymmsec}.

%
%More precisely, given an action, one may construct the allowed diffeos/gauge transformations. From the Lagrangian, one may construct the conserved charges as 
%
%\be
%\backslash \!\!\! \d Q_\xi = \int k_\xi (\Phi, \d \Phi)
%\ee
%and trivial ones don't contribute. Thus, the cahrges are constructed directly from the diffeos, which in turn are directly determined by the boundary conditions.

\subsubsection{$T\bar T$ - deformed CFTs \label{ttbasg}}

%We can equally well perform the ASG analysis for the mixed boundary conditions associated to $T\bar T$-deformed CFTs. As we saw, the most general bulk solution is parametrized by two arbitrary functions $\L(u), \bar \L(v)$ of the auxiliary coordinates $u,v$ defined in \eqref{defuv}. The diffeomorphisms that are allowed yet non-trivial act on  the coordinates $U,V$ of the $T\bar T$ - deformed CFT by shifts depending on arbitrary functions of these state-dependent coordinates, i.e.  

Let us now apply the above steps to the holographic dual to $T\bar T$ - deformed CFTs, namely AdS$_3$ with the mixed boundary conditions \eqref{genbc}. We work, for simplicity, with pure $3d$ gravity, though our analysis - which is performed at infinity - should be equally valid in presence of non-trivial expectation values for the matter fields.

\bigskip

\noindent \emph{1. The allowed diffeomorphisms}

\medskip

\noindent The gravitational phase space is given in \eqref{g0UV} which, under our assumption above, determine the full metric via \eqref{fgexact}. Thus, the most general bulk solution is parametrized by two arbitrary (periodic) functions $\L(u), \bar \L(v)$ of the auxiliary coordinates $u,v$ defined in \eqref{defuv}. %It will be useful to rescale the functions $\L, \bar \L$ by a factor of  $(4\pi G\ell)$ \textcolor{red}{\emph{Should I rescale from the beginning?}} so that the field-dependent coordinates are simply given by 
%
%\be
%U = u -\mu \int^v \bar \L(v')\,  dv' \;, \;\;\;\;\;\; V = v - \mu \int^u \L(u') \, du'  \label{deffdcoord}
%\ee
%\textcolor{blue}{where $U,V = \s \pm t$ are the fixed $T\bar T$ coordinates}. 
%
The most general diffeomorphisms that preserve the radial gauge of the background and the boundary conditions \eqref{genbc} read \cite{Guica:2019nzm}
%
%, which are similar to diffeos preserving flat metric on finite cutoff. These were worked out in \cite{mirage} and read
%
\be
\xi^U = f(u)+ \hat \mu \bar \L_{\bar f}(v)+ \frac{\ell(\rho+\hat \mu)(\rho \bar \L f''(u) - \bar f''(v))}{8\pi G(1-\rho^2 \L \bar \L)}\nonumber %\label{xiU}
\ee

\be
\xi^V = \bar f(v)+\hat \mu \L_f(u) + \frac{\ell(\rho+\hat \mu)(\rho \L \bar f''(v) - f''(u))}{8\pi G(1-\rho^2 \L \bar \L)} \label{xiV} \;, \;\;\;\;\;\; \xi^\rho = \rho (f'(u)+\bar f'(v))
\ee
where $f, \bar f$ are arbitrary functions of the field-dependent coordinates and $ \L_f$, $\bar \L_{\bar f}$ are defined as

\be
\L_f (u) = \int^{u}  \!\!\!\!du' \L (u')  f'(u')\;,\;\;\; \;\;\;\;\;\; \bar \L_{  \bar f}(v) =  \int^{v}  \!\!\!\!dv'\bar \L (v') \bar f'(v') \label{defLf}
\ee
%As before, these primitives are only defined up to an overall Fourier zero mode, which will be denoted as $c_{\L_f}, c_{\bar \L_{\bar f}}$.  
%
Intuitively, these are the diffeomorphisms that leave invariant the metric on the $\rho = - \hat \mu$ surface, and have precisely this bulk interpretation if $\mu <0$.

In principle,  $\L_f, \bar \L_{\bar f}$ are only defined up to an additive constant, as is the relation \eqref{defuv} between $U,V$ and $u,v$. We denote these constants as $c_{\L_f}$, $c_\L$ and their barred counterparts, whereas the primitives are defined to not have a zero mode. These constants, which are in general field-dependent, turn out to play an important role in the integrability of the conserved charges and in the charge algebra.  

 %Then, the precise meaning of $\int^u \L$ is  that, as a function of $u$, it has Note we have assumed $\L$ to be periodic. }

% The function $\L(u)$ is assumed to be  periodic, and thus its primitive consists of  a  term linear in $u$ that is proportional to the zero mode of $\L$, a  set of non-zero Fourier modes, as well as a possible  constant term that will be denoted $c_{\L}$. The same applies to the primitive of $\bar \L(v)$, where the constant term is denoted $c_{\bar \L}$. 

 An important observation made in \cite{Georgescu:2022iyx} is that, since the  dual $T\bar T$ - deformed CFT can be pictured as living on the $\rho=-\hat \mu$ surface, 
%
%The analysis of \cite{lindasg} made two important observations, which were essential and not observed in \cite{mirage}. First,
%
 the diffeomorphisms \eqref{xiV} act on the $T\bar T$ coordinates $U,V$  as 

\be
U \r U + f(u) + \hat \mu \bar{\L}_{\bar f} (v) \;, \;\;\;\;\;\;\; V \r V + \bar f(v) + \hat \mu \L_f(u)
\ee
obtained by setting $\rho=-\hat \mu$ in \eqref{xiV}. Since the identifications of $U,V$ are fixed, this implies that $f, \bar f$ have non-trivial  winding,  proportional to that of $\bar{\L}_{\bar f}$ and $\L_f$. The latter is determined by the zero modes of the integrands in \eqref{defLf}, which turn out to be proportional to the conserved charges associated to the diffeomorphisms parametrised by $f'$ and, respectively, $\bar f'$.

\bigskip

\noindent \emph{2. The conserved charges}

\medskip

\noindent  The next step is to plug in the above expressions for the diffeomorphisms into the covariant phase space formula for the (difference of) conserved charges, which takes the schematic form 

\be
\not{\!\d} Q_\xi  = \oint \boldsymbol k_\xi (g, \d g) \label{varchg}
\ee 
where $k_\xi$ is a one-form whose expression for Einstein gravity is well-known (see e.g.  \cite{Compere:2018aar}).  Importantly, this expression only  yields  the difference in the corresponding charge between two nearby backgrounds, which is not automatically integrable (hence the $\not{\!\d}$), but integrability needs to be separately checked. 

To evaluate this expression, one needs to carefully keep track of the periodicity and winding of the various contributions. As mentioned above, the functions $f,\bar f$ consist of a periodic and a winding part

\be
f(u) = f_p(u) + w_f \, u \;, \;\;\;\;\;\; \bar f(v) = \bar f_p(v) + w_{\bar f} \, v
\ee
with the windings determined as discussed. Second, while the functions $\L (u), \bar \L(v)$ must be periodic,  their variation - which enters   \eqref{varchg} -
%
%The second obsrevation was that integrability and periodicity were extremely important. For example, because $\L(u)$,the variation 
 can be split into an intrinsic part and one due to the variation of the field-dependent coordinate
\be
\d [\L (u)] = [\d \L_{int}] (u) + \L'(u) \d u   \;, \;\;\;\;\; \d \L_{int} (u) = \d \L^p_{int} (u) - u \frac{\d R_u}{R_u} \L'(u) \;, \;\;\;\; \d R_u = 2 \mu \d H_R \label{defndLint}
\ee
which includes a winding piece. The intrinsic part, which  depends in fact on $\hat u = u/R_u$, can itself be split into a periodic and a winding part; the former corresponds to simply varying the Fourier coefficients in the Fourier expansion of $\L(u)$. Plugging these decompositions  into the covariant phase space  formula for the charge difference, carefully taking care of  winding terms, zero modes,  and many  integrations by parts, one finally finds  that \eqref{varchg} reduces to %\textcolor{red}{\emph{Notation! $\xi_{f,\bar f}$?}}

\be
\not{\!\d} Q_{\xi_{f,\bar f}} = \d Q_{f_p} + \d \bar Q_{\bar f_p} + \pi \left(c_{\bar\L_{\bar f}} \d R_v + \d c_{\bar \L} w_{\bar f}  R_v- c_{\L_f} \d R_u - \d c_\L w_f  R_u\right) \label{finaldelQ}
\ee
where 
\be
\d Q_{f_p} = \d \left[ \frac{1}{2}\oint d\s \p_\s u \, f_p (\hat u)  \L (u) \right] \;, \;\;\;\;\;\;\; \d \bar Q_{\bar f_p} =  \d \left[- \frac{1}{2}\oint d\s \p_\s v \, \bar f_p (\hat v) \bar \L (v) \right]% = - \frac{1}{2} \oint d \s \bar f_p(\hat v) \p_\s v \d \bar \L^p
\label{quasilocgenasg}
 \ee
and we explicitly assumed that $f_p, \bar f_p$ depended only on $\hat u$ and, respectively, $\hat v$, but not on any additional field-dependent constant.

There are several important observations to make. First, even though the functions $f,\bar f$ that parametrise the diffeomorphism have winding, the conserved charges  only depend on their periodic parts, $f_p, \bar f_p$. Their expression coincides with that obtained via an educated guess in \cite{Guica:2019nzm}. Having established the expression for the charges, the winding pieces of $f, \bar f$ can be written in terms of them as

\be
w_f = \frac{\mu R_v}{\pi R R_H} \, \bar Q_{\bar f_p'}\,  +  \frac{\mu^2 H_R}{\pi^2 R R_H} Q_{f_p'}\;, \;\;\;\;\; w_{\bar f} = -  \frac{\mu R_u}{\pi R R_H} \,  Q_{ f_p'}\,  -  \frac{\mu^2 H_L}{\pi^2 R R_H} \bar Q_{\bar f_p'} \label{fdepwind}
\ee
as advertised, where $R_H=R+2\mu H$. 

Second, the charges can be made integrable by appropriately choosing the constants $c_\L$, etc. Clearly, if $f_p, \bar f_p$ are just functions of $\hat u, \hat v$, with no additional dependence on field-dependent constants, then an integrable choice is to set all these constants to zero. We will discuss another interesting choice shortly.

%Clearly, a solution is setting all the integration constants to zero. Another possibility would be to take $f = R_u \hat f$, case in which the $c's$ are different. 

\bigskip

\noindent \emph{3. The charge algebra}

\medskip

\noindent In the covariant phase space formalism, the charge algebra \eqref{chalgcovphs} is given by $\d_\chi Q_\xi$: the change in the charge associated to an allowed diffeomorphism, $\xi$, between
two backgrounds that differ by another diffeomorphism, $\chi$.  In our case, both diffeomorphisms are
of the form \eqref{xiV}. We will denote the two functions that label $\chi$, which acts on the background, as $g(u), \bar g (v)$. 

The technical subtlety one encounters in this computation is the following: given an initial background parametrised by two functions $\L(u)$, $\bar \L(v)$, the change in it induced by acting with a diffeomorphism $\chi$ is fully determined by the Lie derivative; thus $\d_\chi \L$ and $\d_\chi \bar \L$ are known, and their explicit expression can already be found in \cite{Guica:2019nzm}.  However, what enters the charge variation is the periodic part of $\d \L_{int}$ defined in \eqref{defndLint}, which requires knowledge of how to split the total known $\d_\chi \L$ into an intrinsic contribution, and one due to the change in the field-dependent coordinate.  While the functional dependence of the two terms leads to a natural choice, there is still an ambiguity in shifting $\d u$ by a constant, $\d c_u$, which leads to a shift by  $- \L' \d c_u $ in $\d \L_{int}$. A similar ambiguity exists for the right-movers.

The point is now that these constant ambiguities in $\d_\chi u, \d_\chi v$ enter the variation of the integration constants $c_\L$, etc. we have introduced above; the precise relations read
\be
\d_{\chi} c_\L = c_{\L_g} + \frac{ Q_{g_p} }{\pi R_u}+ \frac{\d_\chi c_v}{\mu} - \frac{H_L}{\pi R_u} \d_\chi c_u \;, \;\;\;\;\;\;\d_{\chi} c_{\bar \L} = c_{\bar \L_{\bar g}} - \frac{\bar Q_{\bar g_p}}{\pi R_v} + \frac{\d_\chi c_u}{\mu} - \frac{H_R}{\pi R_v} \d_\chi c_v \label{eqdcuv}
\ee
We immediately note that it is not possible to set all these constant ambiguities to zero.
For our previous choice of integration constants,  $\d c_\L = c_{\L_f} =0$ (which worked when $f_p$ was only a function of $\hat u$), we can  immediately determine 
$\d_\chi c_u , \d_\chi c_v$ in terms of the conserved charges $Q_{g_p}, \bar Q_{\bar g_p}$.   These, in turn, will contribute to the charge algebra via their contribution to $\d_\chi \L_{int}$, and additional contributions will come from the winding terms of $g, \bar g$, which involve $Q_{g'_p}, \bar Q_{\bar g'_p}$, as per \eqref{fdepwind}.  The end result, worked out in the appendix of \cite{Georgescu:2022iyx}, is the following \emph{non-linear} algebra\footnote{To present it, it is useful to split the commutation relation between left-movers and right-movers, by associating $Q_f$ to the case where $f_p \neq 0$ and  $\bar f_p =0$, and the reverse for $\bar Q_{\bar f}$; note that in both cases, the winding parts of $f$ and $\bar f$ are non-zero, being determined by the periodic parts via \eqref{fdepwind}.  }
%
%Using  this definition and labeling the charges by their periodic part ??? only\footnote{More precisely, $Q_f$ corresponds to $f_p \neq 0$ and $\bar f_p =0$, with the winding parts of $f,\bar f$ determined as usual via \eqref{windingffb}; the reverse holds for $\bar Q_{\bar f}$.}, 
%The charge algebra that we find is  %\textbf{\emph{Careful periodic!}}

\be
\{Q_f, Q_g\}  =   Q_{f_p g_p'-g_p f_p'}   + \frac{\mu^2 H_R}{\pi^2 R R_H } (Q_f Q_{g'} - Q_g Q_{f'})  - \frac{\ell}{16\pi G } \oint d u \, f_p( u)g_p'''( u) \nonumber
\ee

\be
\{ \bar{Q}_{\bar f}, \bar{Q}_{\bar g}\}  =   \bar{Q}_{\bar f_p \bar g_p'-\bar g_p \bar  f_p'}   - \frac{\mu^2 H_L}{\pi^2 R R_H } (\bar{Q}_{\bar f} \bar{Q}_{\bar g'} - \bar{Q}_{\bar g} \bar{Q}_{\bar f'})  - \frac{\ell}{16\pi G } \oint d v \, \bar f_p( v)\bar g_p'''( v) \label{nonlineartt}
\ee
%where $\mathcal{K}_{f,g}$ is the central extension in \eqref{delgQf}, and $\bar{ \mathcal{K}}_{\bar f,\bar g}$ is its right-moving counterpart. For the left-right commutator we find %\textbf{\emph{Re-check relative sign!}}

\be\label{non-linearttmixed}
\{Q_f,\bar Q_{\bar g}\} = \frac{2\mu}{2\pi R R_H} (R_u Q_{f'} \bar Q_{\bar g} + R_v Q_f \bar Q_{\bar g'} )
\ee
It is useful to rewrite these charges and their commutators in terms of a Fourier basis, where $f_n = e^{i n u/R_u}$ and $\bar f_n = -e^{-i n v/R_v}$. We find 

\be
i \{ Q_m, Q_n \} = \frac{1}{R_u} (m-n) Q_{m+n} + \frac{\mu^2 H_R}{\pi^2 R R_H R_u} (m-n) Q_m Q_n + \frac{c\, m^3}{12 R_u^2} \d_{m+n}\nonumber
\ee

\be
i \{ \bar Q_m, \bar Q_n \} = \frac{1}{R_v} (m-n) \bar Q_{m+n} + \frac{\mu^2 H_L}{\pi^2 R R_H R_v} (m-n)\bar  Q_m \bar  Q_n + \frac{c \, m^3}{12 R_v^2} \d_{m+n}\nonumber
\ee

\be
i \{  Q_m, \bar Q_n \} = - \frac{\mu (m-n)}{\pi R R_H} Q_m \bar Q_n \;, \;\;\;\;\;\;\;\; c=\frac{3\ell}{2G} \label{clsttbnlalg}
\ee
The semi-classical commutation relations are obtained by the replacement $i   \{\;, \} \r \hbar^{-1} [\; , ]$.  The result is a Poisson algebra that is a non-linear modification of two copies of the centrally-extended Virasoro algebra, to which it reduces when $\mu=0$. To see the non-linearities, it is important to be working in finite size.  Note that, since we obtained this result from a semiclassical calculation, we expect the above to yield only the first term in the expansion of the full quantum commutator with respect to $\hbar$, as in principle a Poisson algebra can contain arbitrary powers of the generators; in particular, in the full quantum theory the ordering of the generators that appear in the non-linear correction  would need to be specified. %This issue is further discussed in \cite{clsqsymmttb}. 
%\textcolor{red}{\emph{More/better comments?}}

%
% This, together with the equation above, determines the values of $\d_\chi c_{u,v}$ in \eqref{dLintdiff}.  Given the expression \eqref{dLintdiff} for $\d \L_{int}$, we can obtain - using \eqref{defLp} - the  periodic quantity $\d \L^p(u)$ that enters the formula for charge difference
%%
%\be
%\d \L^p (u) =  2 \L g_p' + \L' g_p +w_g \L - \frac{\ell}{8\pi G}\, g_p''' -\L' \d c_u
%\ee
%where $g_p$ is the periodic part of $g$. A similar expression holds for the right-movers. \emph{Explain how this works!}

Another interesting choice of integration constants is  $\d c_{u,v} = \d c_{\L, \bar \L} =0$, case in which \eqref{eqdcuv} implies

\be
c_{\L_{f}} = - \frac{Q_f}{\pi R_u} \;, \;\;\;\;\; c_{\bar \L_{\bar f} } = \frac{\bar Q_{\bar f}}{\pi R_v} \label{altsolcLf}
\ee
The expression \eqref{finaldelQ} now appears to no longer be integrable; however, it is easy to see that, by simply multiplying $f_p$ by $R_u$ and $\bar f_p$ by $R_v$, the charge associated to these rescaled diffeomorphisms will be integrable. The algebra of these charges associated to the rescaled diffeomorphisms will be simpler to compute, since the $\d c_{u,v}$ contribution is absent; as it turns out, the winding contributions also cancel, and the end result is simply two commuting copies of the Virasoro algebra % \textcolor{red}{\emph{Can I put the $R_u^2$ in the diffeo?}}

%It is interesting to work out the algebra of the rescaled charges $Q_{R_u f}$. Remember that, due to the integrability constraint, the variation of this charge was not simply equal to $R_u \d Q_f$, which implies that the algebra of these charges will be different. The main difference is that now the value \eqref{altsolcLf} for the integration constants $c_{\L_f}, c_{\bar \L_{\bar f}}$ when plugged into \eqref{eqdcuv} with $\d_\chi c_{\L, \bar \L} =0$ sets $\d_\chi c_{u,v} =0$. This new solution affects the charge variation,  which now reads

\be
\d_{R_u g} Q_{R_u f} = R_u Q_{R_u(f_p  g_p'-  g_p f_p')} - \frac{R_u^2 \ell}{16 \pi G} \oint du f_p g_p''' %+ w_g R_u Q_f + Q_f \d_g R_u
\ee
%Since $\d_g R_u = - w_g R_u$, the last two terms cancel and we are left with  a Virasoro algebra 
with the same central extension as the undeformed CFT. %,  precisely reproducing the prediction of \cite{Guica:2021pzy}.
 It is to be expected that this algebra will not receive further quantum corrections. In the semiclassical limit, the Virasoro generators are simply rescaled versions, by a factor of $R_u$, of the first ones we discussed, \eqref{quasilocgenasg}; however, their relation could  become more complicated in the full quantum theory.   

Note that from a field-theoretical perspective, it is obvious
that multiplying the generators by a factor of the field-dependent radius, which depends on the Hamiltonian, will change their algebra. However, in the covariant phase space formalism, this multiplicative factor is simply seen as a $c$-number, and it is the integrability of the conserved charges that ensures that the correct algebra of the rescaled generators is reproduced, via the subtle contributions of integration constants that it constrains. %integrability. %, as we explained. \textcolor{red}{Some proper discussion and advertisement!!}

Thus, the ASG analysis of AdS$_3$ gravity with the mixed boundary conditions that correspond to the $T\bar T$ deformation of the dual CFT  shows that in a certain basis, the asymptotic symmetries organise themselves into two commuting copies of the Virasoro algebra, with the same central extension as in the undeformed CFT. Since in the holographic context, the ASG of the bulk gravitational theory is identified with the symmetries of its boundary QFT dual, this calculation strongly suggests that at least in the classical large $c$ limit,   $T\bar T$-deformed CFTs possess full Virasoro $\times$ Virasoro symmetry. These symmetries are however unusual, in that they depend on the field configuration, and also that in the natural Fourier basis for the generators, the symmetry algebra corresponds to a non-linear modification of Virasoro$^2$.  If the existence of these symmetries is confirmed at non-perturbative level, it could lead to an  interpretation of $T\bar T$ - deformed CFTs as non-local generalizations of  usual CFTs. %] %, rather than theories of quantum gravity.}

\subsubsection{$J\bar T$ - deformed CFTs}

The analysis of asymptotic symmetries  in the  spacetimes dual to $J\bar T$-deformed CFTs is entirely analogous to the one above, though slightly more complicated due to the larger number of fields involved. 

We are interested in the asymptotic symmetries of Einstein gravity in AdS$_3$ coupled to the two Chern-Simons gauge fields \eqref{proposalactioncs}, subject to the mixed boundary conditions \eqref{sourcesflatsp}. As discussed, the most general  background satisfying these boundary conditions is  conveniently  parametrized by four arbitrary \emph{periodic} functions $\mathcal{L}(U),\bar{\mathcal{L}} (v),\mathcal{J}(U),\bar{\mathcal{J}}(v)$, where $v$ is a field-dependent coordinate  that satisfies

\be \label{defnv}
\p_U v  = - \l \p_U \phi = - \frac{\l \mathcal{J}(U)}{1+\l\bar{\mathcal{J}}(v)}
 \;,  \;\;\;\;\;\p_V v = 1-\l \partial_V\phi=\frac{1}{1+\l\bar{\mathcal{J}}(v)}
\ee
as follows from \eqref{scalarfieldonshell} and \eqref{tangtspparam}. 
The scalar field $\phi$ is the bosonisation \eqref{topcur} of the  $U(1)$ current entering the $J\bar T$ deformation. 
%
%, which on-shell consists of a left-moving piece parametrised by $\J(U)$ and a right-moving one, encoded in $\bar \J (v)$, both introduced in \eqref{tangtspparam}.
%The derivatives of the scalar are related to the currents as 
%\be
%\p_U \phi = \frac{\mathcal{J}(U)}{1+\l\bar{\mathcal{J}}(v)} \; , \;\;\;\;\;  \partial_V\phi=\frac{\mathcal{J}(v)}{1+\l\bar{\mathcal{J}}(v)}
%\ee
One may formally integrate the above equations and 
 express $v$ in terms of the primitives of $\mathcal{J}(U),\bar{\mathcal{J}}(v)$  as
 \begin{align} \label{fielddepcoordinitial}
v&=V-\l\left(\int^U dU'\mathcal{J}(U') + \int^v dv'\bar{\mathcal{J}}(v') + c_v\right)
\end{align}
where  the integrals are defined to not contain any constant mode,  placing instead this ambiguity in the integration constant $c_v$ which, \emph{a priori}, can be field-dependent% (with a non-zero variation in field space)
. %This field-dependence is in principle fixed by the definition of $v$, even though we will need to wait until section \ref{section63:asysymalgebra} to access this information. 
%\textcolor{blue}{ Note that the Fourier zero modes of  $\mathcal{J},\bar{\mathcal{J}}$ will lead to terms  linear in $U,v$, while the remaining contributions are periodic. }

\bigskip
\noindent \emph{1. Allowed diffeomorphisms and gauge transformations}
\medskip

\noindent We work in radial gauge for both the metric and the gauge fields. The most general such diffeomorphism is parametrised by  three functions of $U,V$ only

\be
\xi =  z F_z(U,V) \p_z + \left(F_U(U,V) + \ldots \right) \p_U + (F_V(U,V) + \ldots ) \p_V
\ee
where the $\ldots$ stand for terms proportional to derivatives of $F_z$ and we changed the radial coordinate to $z = \sqrt{\rho}$. To keep the gauge fields in radial gauge, one needs to perform a compensating gauge transformation that depends on the above functions (and the background); in addition, one can perform $z$ - independent gauge transformations, which will be parametrised by two additional functions, $G_{A,B}(U,V)$. 

The requirement that the background generated by this diffeomorphism and gauge transformations satisfy the mixed boundary conditions associated to $J\bar T$ fixes the form of these functions to
\be
F_U(U,V)=f-\frac{\ell^2 f''}{2\mathcal{L}}\;, \;\;\;\;\;
 F_V(U,V)=\bar{f}-\l \eta-\l \bar{\eta}-\frac{\l \ell^2\mathcal{J}f''}{2\mathcal{L}}-\frac{(1+\l\bar{\mathcal{J}})\ell^2\bar{f}''}{2\bar{\mathcal{L}}} \;, \;\;\;\;\;
F_z(U,V)=\frac{f'+\bar{f}'}{2} \nonumber% \label{solutiondiffeo}
\ee 
\be
G_A(U,V)=\frac{\mathcal{J}\ell^2 f''}{2\mathcal{L}}-\frac{\l \bar{\mathcal{J}}^2 \ell^2\bar{f}''}{4\bar{\mathcal{L}}}+\eta-\l\int^{v} \bar{\mathcal{J}}\bar{\eta}'-\frac{\l}{2}\int^{v} (\kappa\bar{\mathcal{L}}+\bar{\mathcal{J}}^2)\bar{f}'
\nonumber
\ee

\be
G_B(U,V)=\frac{\bar{\mathcal{J}}(2+\l\bar{\mathcal{J}})\ell^2\bar{f}''}{4\bar{\mathcal{L}}}+\bar{\eta}+\l\int^{v} \bar{\mathcal{J}}\bar{\eta}'+\frac{\l}{2}\int^{v} (\kappa\bar{\mathcal{L}}+\bar{\mathcal{J}}^2)\bar{f}' \label{solutiongauge}
\ee
where $f(U),  \eta(U), \bar f (v), \bar \eta(v)$ are arbitrary functions of the indicated coordinates. 

As before, it is important to determine the periodicity conditions on these functions. The main condition stems from the requirement  that the holonomies of the gauge fields $A,B$ around the $\s$ circle be fixed; as we saw in the previous subsection, these holonomies are related to the undeformed charges $J_0, \bar J_0$, which are quantized and  should thus not be affected by continuous symmetry transformations.  This in turn implies that the functions $G_{A,B}$ have no winding.  Taking the sum and difference of the equations \eqref{solutiongauge}, we find that $\eta$ and $\bar \eta$ have equal, but opposite winding 
% 
%
%Any winding terms in the above would contribute to the holonomy of the fields $A,B$ and in turn imply that $J_0,\bar J_0$ are not fixed. 
%
%In the following, we  denote by $w_{\eta}$ the winding of $\eta$ which, by \eqref{nowindetacomb},  is minus that of $\bar{\eta}$

\be
\eta=\eta_p(\hat{U}) + w_{\eta}\, \hat{U}\hspace{1cm}\bar{\eta}=\bar{\eta}_p(\hat{v})-w_{\eta}\, \hat{v}
\ee 
with  %\textcolor{red}{\emph{Check factor $R_v$ in front.  Silvia says not.}}
\be\label{windingeta}
w_{\eta}=\frac{1}{[1+\l\bar{\mathcal{J}}]_{zm}}\left(\frac{\lambda}{2}[(\kappa\bar{\mathcal{L}}+\bar{\mathcal{J}}^2)\bar{f}']_{zm}+\l[\bar{\mathcal{J}}\bar{\eta}'_p]_{zm}\right)=-\frac{1}{R_Q}\left(\frac{\lambda \hat{k}}{2}\bar{Q}_{\bar{f}'}+\l\bar{P}_{\bar{\eta}'}\right)
\ee
where $\eta_p, \bar \eta_p$  are periodic, $\hat{U}=U/R,\hat{v}=v/R_v$ and $[~~]_{zm}$ denotes the zero mode of the corresponding quantity, which we will soon show is proportional to the conserved  charges associated to these symmetries.  % (here we need to write in terms of hatted variables in order for the overall combination to have the correct periodicity). Finally, from $G_A-G_B$ we find that: \textcolor{red}{Where does the denominator come from?}
%%
%Plugging this decomposition into \eqref{nowindcond2}, the no-winding condition becomes
%
%\textcolor{blue}{where in the second equality we used the expressions for the conserved  charges that  will be derived in the next subsection, which can be viewed just as a notation, for now. }
 Thus, the requirement that the charges $J_0, \bar J_0$ be fixed translates into a charge-dependent winding for these functions. One can easily see that $f, \bar f$ are both periodic.

\bigskip

\noindent \emph{2. The conserved charges}
\medskip

\noindent The conserved charges are obtained by simply plugging in the above  general allowed diffeomorphism and gauge transformation in the appropriate covariant phase space formula for the conserved charges, which now receives contributions from both the Einstein and the Chern-Simons terms. See \cite{Georgescu:2024ppd} for the precise formulae. It is useful to split the conserved charges in four categories, labeled by the four functions we have introduced: %\textcolor{red}{\emph{Notation winding!}}

\bi
\item \emph{Left affine charges: }~$\xi_{\eta}=-\frac{\l \hat{k}}{2}\eta_p(U) \partial_V\,, \; \Lambda_A=\frac{\hat{k}}{2}\eta_p(U)\,, \;\Lambda_B=0\,; \;P_{\eta}=\int d\sigma \; \eta_p(U)\mathcal{J}(U)$ 

\vskip -11.2mm

\be \label{leftaffinejtb}\ee

\vskip1mm

\item \emph{Left conformal charges: }~ only $f(U)$ turned on in \eqref{solutiongauge}; ~~ $  Q_f=\int d\sigma f(U)\left(\frac{\mathcal{L}}{8\pi G \ell}+\frac{2\pi}{k}\mathcal{J}^2\right)$

\vskip1mm

\item \emph{Right affine charges: } ~ $\eta = w_\eta \hat U, \; \bar \eta = \bar \eta_p  - w_\eta \hat v, \; f=\bar f=0$; ~~ $\bar{P}_{\bar{\eta}}=-\int d\sigma \bar{\eta}_p \bar{\mathcal{J}}\partial_{\sigma}v$

\vskip1mm

\item \emph{Right pseudoconformal charges: }~ $\eta = w_{\eta} \hat U, \; \bar \eta =   - w_{\eta} \hat v, \; f=0$; ~ $\bar{Q}_{\bar{f}}=-\int d\sigma \partial_{\sigma} v \bar{f}\left(\frac{\bar{\mathcal{L}}}{8\pi G \ell}+\frac{2\pi}{k}\bar{\mathcal{J}}^2\right)$
\ei
where $w_\eta$ is determined by the right-moving function that is turned on via the general formula \eqref{windingeta}.
As before, only the periodic part of the functions enters the conserved charges. The way the latter were computed  is of course via the charge differences $\slash{\!\!\!\delta}  Q$, but these can be shown to be integrable on the phase space%\footnote{In the case of the right-moving charges, there are some issues with a non-periodic contribution to the charges. This turns out to be identical to a similar contribution in the field theory calculation, which can be discarded by choosing a different current representative. \textcolor{red}{\emph{Understand!}}}
.  Quite naturally, each type of function leads to a conserved charge, except that for the right-moving asymptotic symmetries we are forced to turn on equal and opposite windings for both $\eta, \bar \eta$ due to the charge constraint.  

 Note also that the left-moving affine transformations, where the transformation was short enough for us to write it down explicitly, consists of both the expected gauge transformation and a right-moving translation. This is quite natural from the Hamiltonian perspective, since the left affine currents \eqref{comprmcurrent} contain a factor of the right-moving Hamiltonian; it will also be important shortly, when we compare this analysis with that of the CSS boundary conditions.   

\bigskip

\noindent \emph{3. The asymptotic symmetry group}
\medskip

\noindent The charge algebra is computed via the same steps as for the holographic dual to $T\bar T$. One again encounters the subtlety of how to split the variations of $\d \bar \J$ and $\d \bar \L$, if applicable, into an `intrinsic' part and one due to the variation of the field-dependent coordinate, resulting in an ambiguity parametrised by a field-dependent constant. This ambiguity can be fixed by requiring that the bracket be antisymmetric. Taking into account also the winding contributions and decomposing the functions into the appropriate Fourier modes, $e^{2\pi i m U/R}$ and $e^{-2\pi i n v/R_v}$, one finds % I multiplied $\bar Q_m$ with a minus sign!
\be
\{P_m,P_n\}=-\frac{ i m  \hat k}{2}  \delta_{n+m,0} \;, \;\;\;\;\; \{Q_m,Q_n\}= - \frac{2\pi i(m-n)}{R} Q_{m+n} \;, \;\;\;\; \{Q_m,P_n\}= \frac{2\pi i n}{R} P_{m+n}  \nonumber
\ee
\be
\{ \bar P_m, \bar P_n\} = -\frac{i m \hat k}{2} \delta_{m+n,0}-\frac{i m \hat k\lambda }{2 R_Q} \bar{P}_m \delta_{n,0}+\frac{i n \hat k\lambda}{2R_Q}\bar{P}_n \delta_{m,0} \;, \;\;\;\;\; \{\bar{Q}_m,\bar{P}_n\}=\frac{2\pi i n }{R_v}\bar{P}_{m+n}+\frac{2\pi i n\lambda  }{R_Q R_v}\bar{P}_n\bar{P}_m-\frac{i m \hat k \lambda}{R_Q}\bar{Q}_m\delta_{n,0} \nonumber
\ee

\be
\{\bar{Q}_m,\bar{Q}_n\}=-\frac{2\pi i(m-n)\bar{Q}_{m+n}}{R_v} - \frac{2\pi i m \lambda}{R_Q R_v}\bar{Q}_m \bar{P}_n+\frac{2\pi i n \lambda}{R_Q R_v}\bar{Q}_n \bar{P}_m + \frac{c}{12} \frac{in^3}{(R_v/2\pi)^2}\delta_{m+n,0} \nonumber
\ee
where $R_Q \equiv R-\l Q_L$ and $R_v = R - \l w$. 
The mixed left-right commutators are 
\be
\{ Q_m, \bar Q_n\}= \frac{2\pi i n \lambda}{R R_Q}\bar{Q}_n P_m \;, \;\;\;\;\; \{P_m,\bar{P}_n\}=\frac{i n \hat k\lambda}{2 R_Q}\bar{P}_n\delta_{m,0}\nonumber
\ee

\be
\{P_m,\bar{Q}_{n}\}=\frac{ 2\pi i n \l}{R_Q}\bar{Q}_{n}\delta_{m,0}\;, \;\;\;\;\; \{Q_m,\bar{P}_{n}\}=\frac{2\pi i n \lambda}{R_Q R} \bar{P}_{n}P_m \label{jtbquasilocalg}
\ee
The algebra of the left-moving generators is simply a Virasoro-Kac-Moody algebra, with the same central extensions as in the undeformed CFT. The algebra of the right-movers is a non-linear modification of Virasoro-Kac-Moody (to which it reduces when $\l=0$). This algebra has the special property that,  if ones defines the new generators  ($H_R=\bar Q_0$)

\be
\tilde Q_m = Q_m - \frac{\l H_R}{R} P_m + \frac{\l^2 \hat k H_R^2}{4 R} \d_{m,0} \;, \;\;\;\; \;\;\;\; \tilde P_m = P_m - \frac{\l \hat k H_R}{2} \d_{m,0} \nonumber
\ee

\be
\tilde{\bar{Q}}_m = \frac{R_v}{R} \bar Q_m - \frac{\l H_R}{R} \bar  P_m + \frac{\l^2 \hat k H_R^2}{4 R} \d_{m,0} \;, \;\;\;\; \;\;\;\;\tilde{\bar{P}}_m =  \bar P_m - \frac{\l \hat k H_R}{2} \d_{m,0} \label{relflunfljtb}
\ee 
their algebra %\footnote{Computed with just commutators, not changing integrability constants. \textcolor{red}{Make comment in $T\bar T$! Poisson vs choice.}}
 consists  precisely of two commuting copies of the Virasoro-Kac-Moody algebra, with the same central extensions as in the undeformed spacetime. Hence, as in the $T\bar T$ case, the asymptotic symmetry algebra of the spacetimes dual to holographic $J\bar T$ - deformed CFTs is a  non-linear algebra, but one that can be brought to Virasoro form via a simple non-linear redefinition. In the present case, this non-linear map formally resembles  a spectral flow by $\l H_R$. %\textcolor{red}{\emph{More bla-bla?}}
 As we will show, this result matches perfectly that of the field theory  analysis in the next section.

\subsubsection{Comments on AdS$_3$ gravity with CSS boundary conditions}

Throughout this section, we have been emphasizing the fact that, unlike most holographic studies of AdS$_3$ with non-standard boundary conditions that exist in the literature, the two holographic dictonaries we have presented are top-down, in the sense of being \emph{derived} from boundary QFTs that have their own, \emph{independent definition}. In particular,  the boundary conditions that the bulk fields obey are dictated by the dual field theory and automatically lead to conserved charges that are finite, integrable and conserved, %\emph{Careful $J\bar T$!} 
 all while allowing for full dynamics for the matter fields that may be present in the bulk.  This should be contrasted with the bottom-up approaches, where the boundary conditions on the bulk fields are the result of an educated guess, one may or  not be able to include dynamical matter, and the
  existence of a boundary theory that posseses the properties one  infers from the bulk is always an \emph{assumption}, whose validity depends on the consistency of the setup and of the boundary conditions. 

To illustrate this point, we would like to end this section with a comparison between the $J\bar T$ holographic dictionary we have derived and a \emph{bottom-up} holographic proposal for AdS$_3$ with very similar non-standard boundary conditions, known as Comp\`ere-Song-Strominger (CSS)  ones \cite{Compere:2013bya}. More precisely, the backgrounds allowed by the CSS boundary conditions  correspond to a truncation of the allowed backgrounds \eqref{paramgaugefields}  to the subsector where $\bar \J=0$ and $\bar \L$ is required to be constant. The map between the parametrisation  of \cite{Compere:2013bya} and ours is 
\begin{align}
\Delta \leftrightarrow \bar{\mathcal{L}}\;, \hspace{1.5cm}\bar{L}(t^+)\leftrightarrow\mathcal{L}(U)\;, \hspace{1.5cm}\partial_+\bar{P}(t^+)\leftrightarrow -\lambda\mathcal{J}(U)
\end{align}
% \textcolor{blue}{dropping an irrelevant factor of   $ \ell/4G$}. %The factor of $\l$ could be absorbed into the definition of $\J$. 
%
% {\color{ForestGreen}(we keep $k$ for our level)} \textcolor{red}{What is $\l$ for them?} {\color{ForestGreen}($\lambda$ does not appear in their analysis, we could also absorb it in a redefinition of $\J$ and $\l \bar{\mathcal{J}}$, since all we do is a coordinate tr $V-\l\phi$ so we can absorb $\l$)}
Note, however, that the Chern-Simons gauge fields \eqref{paramgaugefields} do not vanish when $\bar \J=0$, whereas in the  analysis of \cite{Compere:2013bya}, they are simply absent. 
  
One of the main reasons to study   \cite{Compere:2013bya}'s setup is their proposal that Einstein gravity with    CSS boundary conditions   is  holographically dual to a  warped CFT \cite{Hofman:2011zj, Detournay:2012pc}. Remember from section \ref{wads3toym} that warped CFTs are \emph{putative} %\footnote{Simple examples of such theories have been constructed in \cite{}. \textcolor{red}{Are any interacting examples known?}} 
two-dimensional QFT with $SL(2,\mathbb{R})_L \times U(1)_R$ global symmetry that is  enhanced to a left-moving Virasoro-Kac-Moody algebra, with the peculiarity that the Kac-Moody factor enhances right translations \ref{wcftenh}. %\textcolor{red}{Say somewhere that Hofman-Strominger is only for local theories, whereas all holographic examples are non-local.} 
Or,    \cite{Compere:2013bya} found precisely this symmetry enhancement pattern in the asymptotic symmetries of pure $3d$ gravity with CSS boundary conditions.  However, several peculiarities were present, such as the fact that  the $U(1)$ level was found to be background dependent (proportional to $\D$), and generally negative.  %\textcolor{blue}{ The unitarity issues related to the negative $U(1)$ level  need  then to be addressed, see e.g. \cite{Apolo:2018eky, Aggarwal:2022xfd} for discussions. } 
Second, the requirement that $\bar \L = \D$ be fixed (or, at most a varying constant) is a very severe restriction on the dynamics of the theory:  in a CFT, this would correspond to requiring that all the non-zero Fourier modes of the right-moving stress tensor  vanish\footnote{In particular, 
 states of the form $\O (U, V) | 0\rangle $, which would yield  a non-trivial $V$-dependent profile for the expectation value of the right-moving stress tensor, would not be allowed for any operator $\O$ that has a non-trivial right-moving scaling.}.  How to include dynamical matter  in the CSS phase space is thus not entirely clear.  %For these reasons, the construction of the phase space put forth in \cite{Compere:2013bya} appears somewhat unnatural. 
 %\textcolor{red}{More on profile of RM stress tensor disallowed, only chiral matter allowed, unnatural.}

%that gets enhanced to a left-moving Virasoro-Kac-Moody one, where the zero mode of the $U(1)$ symmetry corresponds to right-moving translations. We discussed these in  

%  comment on the relation between our analysis and the one of, who considered pure gravity in $AdS_3$ with non-standard, chiral boundary conditions  and obtained an asymptotic symmetry algebra consisting of a \emph{single} $U(1)$ Virasoro-Kac-Moody factor, with a background-dependent $U(1)$ level that is negative for black holes. 
  
Given that the backgrounds considered in \cite{Compere:2013bya} correspond to a %(inconsistent) 
truncation of the backgrounds \eqref{paramgaugefields}, let us try to understand the CSS ASG calculation from the  perspective of $J\bar T$ holography, which is on a significantly firmer footing. 
%
%The relation between our present work and \cite{Compere:2013bya} is that the backgrounds allowed by their boundary conditions  correspond to a truncation of the phase space we consider.  We can thus interpret their analysis   from the perspective of ours, which  has, by now, a very  solid foundation in $J\bar T$ holography. While the above-mentioned  truncation is consistent at the level of the metric (in the sense that it is obtained by fixing some of the independent functions that parametrize our solution), it is not so at the level of the gauge fields, which are simply absent in \cite{Compere:2013bya}, but do not become zero on our restricted solution. In particular, we can link the appearance of the background-dependent $U(1)$ level in \cite{Compere:2013bya} to missing gauge field contributions to the conserved charges. The reason that, in \cite{Compere:2013bya}, these missing contributions do   not lead to inconsistent results is   the very restricted nature of the phase space variations considered.  Nonetheless, the price to pay for these restricted variations is an important loss of dynamics of the theory.  
%\textcolor{blue}{
The allowed transformations on the restricted phase space with $\bar \J=0$ and $\bar \L = const.$ consist of  left affine and  left conformal transformations, as well as a constant right-moving  translation. %} %\textcolor{red}{Are our restricted diffeos the same as theirs?} {\color{ForestGreen}(if we just set $\bar{\J}=0$, we can do left affine, left conformal and some combination of right affine and right pseudoconformal such that $\delta\bar{\J}=0$. They additionally fix $\bar{\L}$, which kills right pseudoconformal except for constant ones and automatically kills right affine except for constant ones).}
It is rather clear that the Virasoro-Kac-Moody symmetries \cite{Compere:2013bya} find correspond to the $J\bar T$ left-moving Virasoro-Kac-Moody symmetries, though with several differences

\bi
\item the \emph{interpretation} of the Kac-Moody symmetries is different, as already remarked in \cite{Bzowski:2018pcy}:  while in $J\bar T$, the Kac-Moody is an affine $U(1)$  symmetry associated to the left $U(1)$ current, whose bulk implementation consists of the standard gauge transformations \emph{and} an accompanying right translation dictated by the mixed boundary conditions, in  \cite{Compere:2013bya},  since the Chern-Simons gauge fields are absent, only  the shift of the right-moving coordinate is visible, which is then interpreted as the    enhancement of right translations to left Kac-Moody that is characteristic of warped CFTs
\item the calculations leading to the Virasoro - Kac-Moody symmetries are different; in particular, the $J\bar T$ calculation leads  to a KM level that is constant and positive (the same as in the undeformed CFT), whereas the level found by CSS is state-dependent and usually negative.
\ei  
To understand how the calculation of \cite{Compere:2013bya}  works from the $J\bar T$ perspective, one can  separate the calculation of the Kac-Moody conserved charges \eqref{leftaffinejtb} into the contribution coming from the metric sector, which is the same as the one in \cite{Compere:2013bya}
%
% write  below the contributions from the metric and from the gauge fields to our final expressions for the conserved charges $Q_f,P_{\eta}$. For the left-moving affine transformations we obtain, for $\bar{\J}=0$
\be
[\slash{\!\!\!\delta} P_{\eta}]_{metric} =\int d\sigma\frac{ \eta(U)}{2}\bigg(- 2\lambda^2 \bar{\mathcal{L}}\delta\mathcal{J}+\l (1-\lambda\mathcal{J})\delta\bar{\mathcal{L}} \bigg) \label{kmchmetonly}
\ee
and the contribution of the Chern-Simons terms in the action %to $k_{\xi, \Lambda}$
\be
[\slash{\!\!\!\delta} P_{\eta}]_{CS}=\int_0^R d\sigma\frac{\eta(U)}{2}\bigg(-\lambda(1-\lambda\mathcal{J})\delta\bar{\mathcal{L}} +2(1+\lambda^2\bar{\mathcal{L}})\delta\mathcal{J}\bigg)
\ee
%
%\begin{align}
%[\slash{\!\!\!\delta} P_{\eta}]_{metric} &=-\int_0^R d\sigma\frac{\lambda \eta(U)}{2}\bigg( 2\lambda \bar{\mathcal{L}}\delta\mathcal{J}-(1-\lambda\mathcal{J})\delta\bar{\mathcal{L}} \bigg)\\
%[\slash{\!\!\!\delta} P_{\eta}]_{CSdiff}&=\int_0^R d\sigma\frac{\lambda^2\eta(U)}{2}\bar{\mathcal{L}}\delta\mathcal{J}\\
%[\slash{\!\!\!\delta} P_{\eta}]_{CSgauge}&=\int_0^R d\sigma\frac{\eta(U)}{2}\bigg(-\lambda(1-\lambda\mathcal{J})\delta\bar{\mathcal{L}} +(2+\lambda^2\bar{\mathcal{L}})\delta\mathcal{J}\bigg)
%\end{align}
%where we split the contributions from the Chern-Simons gauge fields into the part associated to the diffeomorphisms and the part associated to the gauge transformations. %Note the sum of the three is the simple expression \eqref{resultleftaffineph}, which leads to the same $U(1)$ level as in the undeformed CFT. {\color{ForestGreen}(here the same comment applies as in the main text, we probably need to rescale, but it will be for sure the same $k$)} 
If  one only considers the metric contribution, then the charge variations are not integrable and one is forced to fix $\bar{\mathcal{L}}$ (or $\mathcal{J}$, but we do not consider this possibility), %\textcolor{red}{So, no issue with finiteness?} {\color{ForestGreen}(no, they are separately finite)}
as \cite{Compere:2013bya} did. This seemed overly restrictive, though, and the authors contrived a way to make the charges integrable for $\Delta$ a variable constant. The resulting asymptotic symmetry algebra is a $U(1)$ Kac-Moody algebra with a $\Delta$ ($=\bar \L$) - dependent level, which is negative for positive $\bar{\L}$, as can be seen from \eqref{kmchmetonly}.

If, on the other hand, one adds in  the contribution from the gauge fields, the terms proportional to $\delta\bar{\mathcal{L}}$ and $\bar{\mathcal{L}} \d \J$  automatically cancel, so that there is no issue regarding integrability and one can freely vary $\bar{\mathcal{L}}$ in the phase space. 
%Moreover, the terms proportional to $\bar{\mathcal{L}} \d \J$ cancel and 
The final variation is given by the simple expression \eqref{leftaffinejtb}, which leads to a Kac-Moody algebra with the same $U(1)$ level as in the undeformed CFT. A similar story holds for the left-moving conformal symmetries, see \cite{Georgescu:2024ppd} for details.

Thus, we see that the inclusion of the Chern-Simons gauge fields in the low-energy effective bulk  action, with appropriate boundary conditions   dictated by the dual $J\bar T$ deformation, solves both the non-dynamics and the negative $U(1)$ level problems associated to the CSS boundary conditions.  
 Their
%
 %\emph{independent} QFT definition. \textcolor{red}{Understandable?} Using this field-theoretical definition, we are able to the dual gravitational low-energy action and boundary conditions for the dual bulk fields.  In particular, the action must contain 
%
%Our analysis (see also \cite{Bzowski:2018pcy}) suggests a very simple solution to both problems: that the low-energy description of a gravitational theory allowing for such  non-standard boundary conditions must contain other contributions to the conserved charges besides the metric, for example from a
% Chern-Simons gauge fields whose
  inclusion%, from the point of view of the analysis of \cite{Compere:2013bya}
  : i) permits  relaxing the boundary conditions, so that arbitrary matter field excitations (with reasonable near-boundary behaviour)  are allowed, thus solving the dynamics problem, and ii)  gives additional contributions to the charges, so that the negative background-dependent $U(1)$ level never shows up.  This points towards the conclusion that  the framework  of pure $3d$ Einstein gravity  is simply not sufficient
  to properly model a consistent phase space 
 for  the  non-standard boundary conditions  put forth in \cite{Compere:2013bya}.  
  % 
%  It is not unreasonable to conclude that, in presence of such non-standard boundary conditions,  the inclusion of other fields in the low-energy action besides the metric is necessary in order to have a 
  %
The holographic setup associated with the $J\bar T$ deformation does provide such a consistent framework, which preserves the full dynamics of the theory; however, this new framework requires the inclusion of a Chern-Simons gauge field dual to the $U(1)$ current used to define the $J\bar T$ deformation. If the original CFT does not contain a $U(1)$ current, then we see no obvious way to construct a well-defined double-trace operator that would drive the deformation of the boundary conditions from Dirichlet to CSS  by  just using the stress tensor\footnote{Taking the square root of the left stress tensor to produce a dimension one operator is not a legit quantum operation.}.     Another consequence of having to use the $J\bar T$ framework in order to implement boundary conditions of the type considered in \cite{Compere:2013bya} is that the correct interpretation of the left Kac-Moody asympotic symmetries is as the affine $U(1)$ algebra that enhances the left $U(1)$ current, rather than the Kac-Moody enhancement of right translations that characterises a warped CFT. 
  The new  dual interpretation of CSS-like boundary conditions is within a class of field theories that are significantly better defined and understood than warped CFTs, for which the only known concrete examples are free  \cite{Hofman:2014loa,Castro:2015uaa}. As we just saw, the problem with  holographic definitions of interacting (strongly-coupled) warped CFTs is that one first needs to ascertain the consistency of the holographic setup, which in general is hard to do. The ultimate test whether a given set of boundary conditions in a low-energy effective field theory - assumed, as usual, to admit a consistent UV completion - are correct or not is whether they match the predictions from an independently-defined and consistent  dual field theory. The existence of this dual would likely impose many more constraints on the gravitational phase space than are currently known.

\section{Infinite pseudoconformal symmetries of $T\bar T$ and $J\bar T$ - deformed CFTs\label{infsymmsec}}

As reviewed in section \ref{microads3}, in holography, the asymptotic symmetries of a given spacetime correspond to the global symmetries of the dual field theory. The ASG analyses of the previous section  thus make  a rather remarkable prediction: that  both $T\bar T$ and $J\bar T$ - deformed CFTs should posses an infinite number of symmetries, which obey a certain non-linear modification of the Virasoro $\times$ Virasoro or (Virasoro-Kac-Moody)$^2$ algebra.
Moreover, a simple non-linear change of the basis of symmetry generators turns the algebra into two commuting copies of the Virasoro or Virasoro-Kac-Moody algebra, the same as the symmetry algebra of the undeformed CFT. 
 %Nonetheless, in a natural basis, the algebra is a non-linear modification thereof. 
 
 The goal of this section is to uncover these symmetries and their peculiar features also from the purely field-theoretical point of view.  We start with a Hamiltonian analysis, which is conceptually the simplest, and then translate the results to the Lagrangian formalism, which is closer to holography.

%In section \ref{}, we noted the existence of dynamical/field-dependent corodinates in both $T\bar T$ and $J\bar T$ - deformed CFTs and remarked that there exist conserved currents associated to arbitrary functions of the field-dependent coordinates, and also the holographic analysis implied the existence of an infinite set of symmetries in these theories, which would be a generalization of the conformal symmetries of $2d$ CFTs. We call these symmetries pseudoconformal. As we will see, they are somewhat more involved than what the above observation would suggest, and it turns out that a Hamiltonian perspective is particularly useful in understanding them.  To effectively deal with a QM system, we place the theory on the cylinder and  study the flow of the symmetry generators under the $T\bar T$ and $J\bar T$ deformations.

\vskip-2mm

 \etocsetnexttocdepth{5}
    \etocsettocstyle{\subsubsection*{Contents of this section: }}{}
    \cftsubsubsecindent 34pt
    \localtableofcontents

\subsection{General  argument for Virasoro symmetry\label{genargsymm}}

 The simplest argument that shows the existence of extended symmetries in $T\bar T$ and $J\bar T$ - deformed CFTs is also very general and valid at the full quantum level: consider a family of non-local two-dimensional QFTs, assumed to be UV complete, which are obtained via an integrable irrelevant deformation of a  CFT$_2$ (of central charge $c$), labeled by a continuous parameter $\l$. One then places these QFTs on a cylinder of circumference $R$, and assumes they can be simply treated as  quantum-mechanical systems.   The deformation is assumed to not change the Hilbert space, but only the Hamiltonian  and its associated eigenstates, in an adiabatic fashion. The latter should satisfy a smooth flow equation  

\be
\p_\l |n_\l\rangle = \hat{\mathcal{X}} |n_\l\rangle \label{flowstates}
\ee 
where $ \hat{\mathcal{X}}$ is entirely determined by the (known) deformation of the Hamiltonian, $\p_\l H$, via%, which equals the integral of the deforming operator %which is in turn % well-defined \emph{Ambiguities?} as it is  fixed by the $T\bar T$/ $J\bar T$ operator%. More precisely, its definition is

\be
\p_\l H = %\int d\s \O_{J^A \wedge J^B} = 
D - [H, \hat{\mathcal{X}}] \label{defchi}
\ee
where $D$ represents the diagonal matrix elements of $\p_\l H$ in the instantaneous energy eigenbasis. One may then use $\hat{\mathcal{X}}$ to  formally define a set of generators  $\widetilde L_m^\l$ as \cite{LeFloch:2019rut,Guica:2020eab,Guica:2021pzy}

\be
\mathcal{D}_\l \widetilde L_m^\l \equiv \p_\l \widetilde L_m^\l - [\hat{\mathcal{X}}, \widetilde L_m^\l] =0 \;,\;\; \;\;\;\;\; \widetilde L_m^{\l=0} = L_m^{CFT} \label{virfl}
\ee
Of course, there will be one such generator for every generator of extended symmetries in the original CFT; the properties of $\widetilde L_m^\l$  below will thus hold, with the appropriate modifications, also for the right-moving Virasoro generators, as well as for possible Kac-Moody symmetries that may be present. By construction, the $\widetilde L_m^\l$ : 
\begin{enumerate}
\item[i)] are well-defined, since $\hat{\mathcal{X}}$ is well-defined
\item[ ii)]  obey a  Virasoro \emph{algebra}, with the same central extension, $c$, as the seed CFT 
\item[ iii)] are conserved, and therefore correspond to \emph{symmetries} of the theory
\end{enumerate}

\noindent To check conservation, one should first define $\widetilde L_n^\l (t)$ by using the flow operator for the eigenstates on a $t \neq 0$ time slice. Since the time dependence of energy eigenstates is simply $ e^{- i E_n t}$, it immediately follows that $\hat{\mathcal{X}} (t) = \hat{\mathcal{X}} (0) -i D t $. Even though $\widetilde L_n^\l (t)$ is a Schr\"{o}dinger picture operator, it has explicit time dependence because $\mathcal{X} (t)$ used in its definition does. It can be easily checked  then that  the $\l$ derivative of the conservation equation  is %\textcolor{red}{\emph{Check i and $\hbar$!}}
 \be
\frac{d}{d\l} \left( \p_t \widetilde{L}_{m}^{\l} + \frac{i}{\hbar} [H,\widetilde L_{m}^\l ] \right) = \bigg[\p_t \hat{\X} (t) + \frac{i}{\hbar} \p_\l H - \frac{i}{\hbar} [\hat{\X}, H], \widetilde L_{m}^{\l}  \bigg] + \biggl[\hat{\X}(t), \p_t \widetilde L_m +\frac{i}{\hbar} [H, \widetilde L_m^\l]\biggr] \label{consflow}
 \ee
 Since the 
  first term on the right-hand-side vanishes via the definition \eqref{defchi}, the above implies that the charge will stay conserved, if it was initially so. %, since t
 %the superscripts $S,H$ refer to the Schr\"{o}dinger and, respectively, Heisenberg picture. This formula generalizes the standard requirement that the symmetry generators commute with the Hamiltonian to the case of time-dependent symmetries. This is zero, and thus the $L_m$ stay conserved. 

This remarkably simple argument shows that all extended symmetries of the undeformed CFT can be transported along the irrelevant flow to the deformed theory, leaving the symmetry algebra  unchanged. One may nonetheless
feel uncomfortable with the generality of the argument, which only uses adiabaticity, and none of the special properties of the $T\bar T$ or $J\bar T$ deformation. In particular, this argument seems to apply to any irrelevant or relevant deformation of a CFT, as long as it is adiabatic,
% object that this is simply a formal definition, without necessarily a physical meaning, view supported by the fact that nowhere did we use the special properties of our theories, except adiabaticity, to arrive at the conclusion that the symmetries are preserved, and that the argument seems to work for any deformation, be it irrelevant or massive, 
predicting in each case the preservation of the Virasoro symmetries, which is puzzling.

While we currently do not have a precise understanding of when and why this type of argument should hold or fail, let us make a few potentially relevant remarks. First, the construction of the conserved charges via the flow we have described does not guarantee that the result is the integral of a local current. As we will show in the next subsection, $T\bar T$ and $J\bar T$ are special in that one can solve for the (classical) flowed currents explicitly and keep track of these non-localities, at least at the classical level. As we will show, the result is more non-local than the expected non-localities of $T\bar T$/$J\bar T$ deformed CFTs.  It is thus possible that in more general theories, the symmetries so defined mix IR and UV modes in a way that becomes intractable from the IR  point of view.
%
% and we have additional evidence that the symmetries make sense via the holographic analysis that we have performed. In any case, it would be instructive to understand precisely  
Second, we have been using a purely quantum-mechanical setup to make the argument; however, it is not clear whether the 
%
%Note that the symmetries we constructed rely on the flow picture, which in turn rests on the adiabaticity of the deformation. However, it is not clear whether the 
UV divergences that are characteristic of QFT may  affect the argument \cite{Balasubramanian:2014bfa}. 
%
%(as opposed to QM) do not affect the argument. Also, not clear whether UV completeness of the deformation is necessary for the results, and hard to predict a priori whether results are trustworthy. 
For the special case of the $T\bar T$ and $J\bar T$ deformations, the consequences derived from the argument above will yield charges that are in perfect agreement with the holographic computations of the previous section. Nonetheless, it would be valuable to have a direct QFT appraisal of its validity.

\bigskip

\noindent In the remainder of this section, we will explicitly solve for the flowed currents in \emph{classical} $T\bar T$ and $J\bar T$ - deformed CFTs and derive the classical symmetries they generate. But first, let us mention a few more quantum results in $T\bar T$ and $J\bar T$ - deformed CFTs that one can derive from the flow above:

\bi
\item the fact that $\widetilde{L}^\l_0$  and $\widetilde{\bar{L}}^\l_0$  are entirely determined by the Hamiltonian, momentum and charge operators of the deformed theory. More precisely, in $T\bar T$ - deformed CFTs, one finds \cite{LeFloch:2019rut} %\textcolor{red}{\emph{Tildes?}}
\be
\widetilde{L}_0^\mu = H_L (1+ 2 \mu H_R) \;, \;\;\;\;\;\;\widetilde{\bar{L}}_0^\mu = H_R (1+ 2 \mu H_R) %\;, \;\;\;\;\; H_{L,R} \equiv \frac{H\pm P}{2} 
\label{rell0httb}
\ee
whereas in $J\bar T$, where the %\textcolor{red}{\emph{Check! Currents too?}}
 flowed generators are $\widetilde L_m^\l, \widetilde{\bar L}_m^\l, \widetilde K_m^\l, \widetilde{\bar K}_m^\l$, the zero modes read 
\be
\widetilde L_0^\l = H_L - \l J_0 H_R - \frac{k \l^2}{4} H_R^2 \;, \;\;\;\; \widetilde K_0 = J_0 \;, \;\;\;\;\; \widetilde{\bar L}_0^\l = H_R- \l J_0 H_R - \frac{k \l^2}{4} H_R^2\;, \;\;\;\; \widetilde{\bar{K}}_0= \bar J_0  \label{rell0hjtb}
\ee
\item the  exact time dependence of the Schr\"{o}dinger picture generators $\widetilde{L}^\l_n$ entering \eqref{consflow} %, which follows from that of $\hat{\X}(t)$ 
\ei
The expressions \eqref{rell0httb}-\eqref{rell0hjtb}  can be derived by either noting that the states $\widetilde L^\l_0 | n_\l \rangle$ and  $| n_\l \rangle$ flow in the same way with $\l$, implying that the constant of proportionality between them is the same as in the undeformed CFT, i.e. the undeformed left energy. Expressing it in terms of the action of the deformed $H, P$ etc. and using genericity of the undeformed CFT spectrum, one arrives at these relations \cite{LeFloch:2019rut}. 
%
% Then,  the genericity of the spectrum of the CFT upon which the deformation acts and the fact that the undeformed eigenvalues are a function of the deformed ones allows one to write $\widetilde{L}_0^\l$ in terms of $H, P$. \textcolor{blue}{The same relation can be inferred from the $T\bar T$ finite-size spectrum, using the argument of \cite{LeFloch:2019rut} to extract the relation between $H, P$ and $L_0^\mu, \bar L_0^\mu$.} %That   $\widetilde{L}_0^\l \neq (H+P)/2$ is obvious from the definition of $\mathcal{X}$, which equates $\mathcal{D}_\l H = D$, not zero. \emph{Can one also show this directly?}
%
An alternate derivation  \cite{Guica:2022gts,Guica:2021pzy} uses  the flow equation for the Hamiltonian that follows from \eqref{defchi} %\textcolor{red}{\emph{Define $\mathcal{D}_\mu$!}} %, which read  in $T\bar T$ and, respectively, $J\bar T$

\be
\mathcal{D}_\mu H_{T\bar T} =  D_{T\bar T} = - \frac{4 H_L H_R}{R+2\mu H} \;, \;\;\;\;\;\;\;\; \mathcal{D}_\l H_{J\bar T}= D_{J\bar T}=  2 \frac{H_R Q_L}{R-\l Q_L} \label{Dopttbjtb}
\ee 
while $\mathcal{D}_\mu P=0$, where $\mathcal{D}_\l$ is defined in \eqref{virfl}.  The expressions for the diagonal elements in $T\bar T$ and $J\bar T$ simply follow by re-expressing $\p_\l E$ in the corresponding theory in terms of the deformed energy, momentum and charge of the state. %, because $P$ does not flow and $\X$ is $\s$ - independent. 
One can then easily check that the combinations \eqref{rell0httb} in $T\bar T$ and \eqref{rell0httb} in $J\bar T$ are covariantly constant along the flow. %, qed. % and equal the left/right hamiltonian in the undeformed CFT, implying the LHS must be $\widetilde L_0$. 

%\textcolor{blue}{Because of the simplicity of the deformation, we can work out exact time dependence of the conserved charge. Agrees in classical limit with analysis next. }

As for the exact time dependence, it follows from the commutator $[H,\widetilde L_m^\l ]$, as is clear from  \eqref{consflow}. 
%
%Another computation we can perform exactly to all orders in $\hbar$ is that of the time dependence of the symmetry generators.  
Using the fact the $\widetilde L_m$ satisfy a Virasoro alegebra  and the explicit expression for $\widetilde L_0$ in terms of the Hamiltonian, it immediately follows that \cite{Guica:2021pzy} %\textcolor{red}{\emph{Notation!}}
\be
[ \widetilde{L}_m^\l,H] = \a_m(H,P,Q) \widetilde{L}_m^\l \;\;\;\;\; \Rightarrow \;\;\;\;\; \widetilde{L}_{m,S}^\mu (t) = e^{i \a_m(H,P,Q) t} \widetilde{L}_{m,S}^\mu(0) \label{defamcomm}
\ee  
The operators  $\a_m$ can be computed to all orders in $\hbar$.  
%
%\be
%L_{m,S}^\mu (t) \equiv e^{i \a_m(H,P) t} L_{m,S}^\mu(0) \;, \;\;\;\;\;\; \bar L_{m,S}^\mu (t) \equiv e^{i \bar \a_m(H,P) t} \bar L_{m,S}^\mu(0) 
%\ee
%then have conserved expectation value in any state, where for 
In $T\bar T$ - deformed CFTs %\textcolor{red}{\emph{Factors R!}}
\be
\a_m(H,P) = \frac{ \sqrt{(R+2\mu H)^2 + 4 \mu m \hbar (R+2\mu P)/R + 4 \mu^2 m^2 \hbar^2/R^2} - (R+2\mu H)}{2\mu} \approx \frac{m \hbar}{R} \cdot \frac{R+2\mu P}{R+2\mu H} +\O(\hbar^2) \label{amttb}
\ee
and $\bar \a_m$  (appearing in the commutator of the flowed right-moving generators with $H$) is  given by the same expression, but with $P \r - P$.  In $J\bar T$ - deformed CFTs, we have % there are different commutation relations for the left and right-movers. 
\be
\a_m = \frac{ m \hbar}{R} \;, \;\;\;\;\;\;\;\;\; \bar{\a}_{m} (H_R) =  2\,\frac{R - \l Q_L  - 
   \sqrt{ (R -\l Q_L )^2 -\hbar k m \l^2}}{k \l^2} \approx \frac{m \hbar}{R -\l Q_L} + \O(\hbar^2) %\pm \frac{k m^2 \l^2 \hbar^2}{4(R-\l Q_K)^3}+ \O(\hbar^3) 
   \label{amlr}
\ee
%
%\be
%[H, \widetilde L_m^\l] = - m \hbar \widetilde L_m^\l
%\ee
%where we used the fact that both $H_R$ and $J_0$ commute with the left generators. Note this is precisely the commutator one obtains in a CFT. The conserved Schr\"{o}dinger-picture generators are then 
%
%\be
%\widetilde L_{m, S}^\l(t) = e^{i m t} \widetilde L_{m, S}^\l(0) 
%\ee
%which satisfy \eqref{} by construction. Thus, we find that $\a_m = m \hbar$ when symmetry is local. One finds a similar commutator for th left flowed  Kac-Moody generator.  The commutators of the right-moving generators with $H_R$ take the form
%
%\be
%[\widetilde{\bar L}_m^\l, H_R] = \widetilde{\bar L}_m^\l \a_m^r = \a_m^l \widetilde{\bar L}_m^\l
%\ee
The value of $\a_m$ reflects the locality of   $J\bar T$ - deformed CFTs on  the left. The  leading term in $\bar \a_m$ will be reproduced by the classical analysis in the next subsection. See \cite{Guica:2021pzy} for more details, including how the expression for $\a_m$ changes depending on whether it appears to the left or the right of $\widetilde L_m^\l$ in \eqref{defamcomm}. 

The relations \eqref{rell0httb}-\eqref{rell0hjtb} provide a way to phrase the difference between these theories and a standard, local $2d$ CFT:   while they all possess a Virasoro symmetry algebra, in CFT the zero modes  of the  Virasoro generators coincide with the left/right Hamiltonian, whereas in $T\bar T$ and $J\bar T$ - deformed CFTs, the relation between these zero modes and the Hamiltonian is non-linear.

Finally, let us comment that  one can also use the covariant derivative to flow the KdV charges in a $T\bar T$/$J\bar T$ - deformed CFT, as well as other observables of interest. Similarly, one can apply the flow argument to single-trace $T\bar T$ and $J\bar T$ - deformed CFT, where the flow in each twisted sector can be mapped to a standard $T\bar T$/$J\bar T$ flow equation  on the covering space \cite{Chakraborty:2023wel}, with very similar conclusions as in the double-trace case.

\subsection{Classical limit of the generators and emergence of the field-dependent coordinate\label{clslimflow}}

The arguments above  did not use the explicit form of the flow operator $\hat{\mathcal{X}}$, but only its existence. Nonetheless, to have a better handle over the symmetries, one would need  explicit expressions for the flow operator and the conserved charges. As it turns out,  these can be obtained without much effort in  the classical limit of $T\bar T$ and $J\bar T$ - deformed CFTs, in which operator ordering issues can be neglected and  the commutators are replaced by Poisson brackets. As we already pointed out,   all Poisson brackets of the currents are  known exactly in these theories, for finite deformation parameter, as follows from the fact that the deformed Hamiltonian density is a known function of the undeformed Hamiltonian, momentum and current densities, whose Poisson brackets are universal.

Concretely, in this subsection we use the explicit form of the $T\bar T$ and $J\bar T$ operators, which determine the Hamiltonian deformation, 
to  obtain an explicit expression for the corresponding flow operators  in the classical limit. Then, we solve the classical counterpart of \eqref{virfl} for the flowed conserved currents and show how the  field-dependent coordinates \eqref{fdepcootrfb} and \eqref{uvcoojtb}
that have previously appeared in our discussions of these deformations 
\emph{emerge} from the flow.

\subsubsection{Classical limit of the flow operator}

The flow  operator, $\hat{\X}$ is defined in \eqref{defchi}. In the classical limit, we let $\hat{\X} \r_{cls} i \X_{cls}$. For a general Smirnov-Zamolodchikov deformation,  $\p_\l H$ is minus the spatial integral of the SZ operator. To find $\X_{cls}$, it is useful to write $\p_\l H$ as \cite{Kruthoff:2020hsi}

\be
\p_\l H = - \int d\s  \e^{\a \b} J^A_\a (\s)  J^B_{\b} (\s) = - \int d\s d\tilde \s  \e^{\a \b} J^A_\a (\s) \d(\s-\tilde \s) J^B_{\b} (\tilde \s) \label{szdefsplit}
\ee
Introducing the Green's function on the cylinder via

\be
\p_\s G(\s - \tilde \s) = \d(\s-\tilde \s) - \frac{1}{R} \label{greensfcyl}
\ee
and integrating by parts, we find %($\e^{\s t} =1$) \textcolor{red}{\emph{Check signs!}}
\bea
\p_\l H & =  & - \frac{1}{R} \e^{\a\b} \int d\s J^A_\a (\s) \int d\tilde \s J^B_\b (\tilde \s) + \int d \s d\tilde \s G(\s-\tilde \s) (\p_\s J^A_\s J^B_t + J^A_t \p_{\tilde \s} J^B_\s) \nonumber \\
& = & -  \frac{1}{R} \e^{\a\b} \int d\s J^A_\a (\s) \int d\tilde \s J^B_\b (\tilde \s) - \left\{ H, \int d \s d\tilde \s G(\s-\tilde \s) J^A_t(\s) J^B_t(\tilde \s)\right\}
\eea
where in the second line we have used current conservation.  Comparing with \eqref{defchi}, the last term in the Poisson bracket yields part of the contributions to $\X$, while the rest will come from the off-diagonal components of the first term\footnote{In the quantum theory, since the SZ operator is only defined up to total derivatives, there may be   additional contributions to $\X$ coming from the  total time derivative ambiguities (the spatial ones drop upon integration). The definition \eqref{szdefsplit} is definitely fine for classical purposes. We ignore operator ordering issues for the same reason. }. To understand them, we particularize to the specific deformations.

 In the  $T\bar T$ case, we have  
\be
\p_\mu H_{T\bar T} = - \int d\s \O_{T\bar T} = - \frac{1}{R} \left( H \int T_{\s\s} - P^2\right) + \left\{ H, \int d\s d\tilde \s G(\s-\tilde \s) \H(\s) \P (\tilde \s) \right\}
\ee
Remember from section \ref{ttbsubsec} that, at least classically, $T\bar T$ - deformed CFTs satisfy the trace relation  %\textcolor{red}{\emph{Factors 2!}}

\be
T_{\s\s} -\H = - 2 \mu \O_{T\bar T}  \label{eqntssttb}
\ee 
Plugging this into the integral above, we find \cite{Guica:2022gts}

\be
\p_\mu H = - \frac{H^2-P^2}{R+2\mu H} +  \left\{ H,  \frac{R}{R+2\mu H} \int d\s d\tilde \s G(\s-\tilde \s) \H(\s) \P (\tilde \s) \right\} = D_{T\bar T} + \left\{ H, \X^{cls}_{T\bar T}\right\} \label{chittb}
\ee
which is precisely the classical counterpart of \eqref{defchi}, where $D$ takes the form \eqref{Dopttbjtb} and the term inside the Poisson bracket with $H$ can be identified with $\X^{cls}_{T\bar T}$. %, yielding

% In the quantum case, we of course have to be careful about operator ordering. Promoting the Poisson bracket to a commutator, we immediately note the above expression precisely splits into a diagonal part (in the energy eigenbasis) and a commutator with $H$, and thus we find that, in the classical limit  
% 
% \be
% \X^{cls}_{T\bar T} =  \frac{R}{R+2\mu H}\int d\s d\tilde \s\, G(\s-\tilde \s) \H(\s) \P(\tilde \s)
% \ee 
%Of course, there is some ambiguity of functions that commute with $H$, but the above minimal choice would lead to zero expectation value in the energy eigenstates. Note that total time derivative ambiguities in the $T\bar T$ operator would contribute to $\X_{T\bar T}$, but are not present classically. 
%
The same derivation can be used for the $J\bar T$ flow operator. Taking the $U(1)$ current to correspond to a topological current \eqref{topcur}, 
the flow of the Hamiltonian is
\be
\p_\l H=  - \int d\s \O_{J\bar T} = - \frac{1}{R} \e^{\a\b} \int d\s \tilde J_\a \int d\s T_{\b V} - \left\{H, \int d\s d\tilde \s\, G(\s-\tilde \s) \phi'(\s) \H_R (\tilde \s)\right\}
\ee
One can  use the identity $T_{\s V} = \H_R - \l \O_{J\bar T}^{cls}$, together with some additional tricks \cite{Georgescu:2024ppd} to write $\p_\l H_{J\bar T}$ in the form \eqref{defchi}, where the diagonal elements are given by \eqref{Dopttbjtb} and  the classical limit of the $J\bar T$ flow operator is 
\be
\X_{J\bar T}^{cls} = \int d\s d\tilde \s\, G(\s-\tilde \s) \phi'(\s) \H_R (\tilde \s) - \frac{w \l}{R_v} \int d\s \H_R \hat \phi_{nzm} - \frac{\l H_R R}{R_v (R-\l Q_L)} \int d\s (\J_- + \frac{\l}{2}\H_R) \hat \phi_{nzm} \label{chijtb}
\ee
where $\hat \phi_{nzm}$ denotes the scalar field with its winding and zero mode removed. Note both \eqref{chittb} and \eqref{chijtb} represent $\X^{cls} (0)$; to obtain $\X^{cls} (t)$ one should add the appropriate time dependence, $- D t$, with $D$ given in \eqref{Dopttbjtb}.

\subsubsection{Flow of the currents}

We are now ready to study the flow of the symmetry generators in the classical limit. We consider the flow of the chiral CFT currents rather than that of the charges, which are simply the Fourier modes of the former. Note that chirality is trivially preserved along the flow. 

\bigskip

\noindent \emph{$T\bar T$ - deformed CFTs}

\medskip

\noindent Let us start with $T\bar T$ - deformed CFTs. Our task is to find two functions on phase space that satisfy

\be
\Dmu \mathscr{H}_{L,R} =0 \;, \;\;\;\;\; \left. \mathscr{H}_{L,R} \right|_{\mu=0} = \H^{[0]}_{L,R}
\ee
where $\H^{[0]}_{L,R}$ are the Hamiltonian densities in the undeformed CFT. All the above functions depend on $\s$ and are defined on a $t=$constant slice. 

As it turns out, it is simpler to guess the solution that to iteratively integrate the flow equations.  A natural  starting point  are the Hamiltonian densities $\H_{L,R}$ at finite $\mu$. Their flow equations read \cite{Guica:2022gts} %\textcolor{red}{\emph{Factors R!}}
\be
\mathcal{D}_\mu \H_L = \frac{R}{R_H} \p_\s \left[2 \H_L \left(\hat \chi_R + \frac{ \mu \X_{T\bar T}}{R}\right) -\mu \p_\mu \H \left(\hat \chi_L - \hat \chi_R - \frac{2 \mu \X_{T\bar T}}{R}\right)\right] - \frac{2 H_R}{R_H} \H_L \label{flowHLttb}
\ee
with a similar expression for $\mathcal{D}_\mu\H_R$, 
where  
\be
\p_\s \hat \chi_{L,R} \equiv \H_{L,R} - \frac{H_{L,R}}{R}
\ee
and are defined to have no constant mode. 
It is not hard to see that the last term in \eqref{flowHLttb} can be absorbed into a rescaling of $\H_L$ by $R_u \equiv R+2 \mu H_R$, so now the RHS is a total $\s$ derivative. To understand what we should add to $R_u \H_L$ in order to make the RHS zero, let us try to understand first the $\mu \r 0$ limit, in which the latter equals $\p_\s(2 \H_L \hat \chi_R)$. If we subtract from this $\Dmu \p_\s[ 2 \H_L (\mu \hat \chi_R)]$, then to zeroth order in $\mu$ only the $\p_\mu(\mu \hat \chi_R)$ term contributes. It is then possible to extend the analysis to all orders. Including also the time dependence inside the  operator with which we flow, the solution is given in terms of   the following quantities %\textcolor{red}{\emph{If we use flow by time-dependent op }}

\be
\D \hat u \equiv \frac{2\mu R}{R_u} \left(\hat \chi_R + \frac{\mu \X_{T\bar T}(0)}{R}\right) - \frac{4\mu H_R}{R_H} t\;, \;\;\;\;\;\; 
\D \hat v \equiv \frac{2\mu R}{R_v} \left(\hat \chi_L - \frac{\mu \X_{T\bar T}(0)}{R}\right) + \frac{4\mu H_L}{R_H} t \label{ttbfdepshifts}
\ee
where $R_{u,v} \equiv R + 2 \mu H_{R,L}$, and formally reads %\footnote{The way this is proven is by noting that, if $\Dmu (R_u \H_L) = \p_\s \mathcal{A}$, then $\mathcal{D}_\mu (\H_L R_u (\Delta \hat u)^n)=\p_\s [\mathcal{A} (\D \hat u)^n] + n (\D \hat u)^{n-1} \mathcal{A}$. } 
\be
\mathscr{H}_L = \sum_{k=0}^\infty \frac{(-1)^k}{k!} \p_\s^k [R_u \H_L (\D \hat u)^k] \;, \;\;\;\;\;\;\;\mathscr{H}_R = \sum_{k=0}^\infty \frac{(-1)^k}{k!} \p_\s^k [R_v \H_R (\D \hat v)^k]
\ee
Let us now consider the conserved charges, which are simply the Fourier modes of the above currents
\be
\hspace{-2mm} \tilde Q_m \equiv \int d \s e^{\frac{2\pi i m (\s+t)}{R}} \mathscr{H}_L = R_u \!\int \! d \s e^{\frac{2\pi im (\s+t)}{R}} \sum_{k=0}^\infty \frac{(-1)^k}{k!} \p_\s^k [\H_L (\Delta \hat u)^k] = R_u \!\int d\s \!\, e^{\frac{2\pi i m (\s +t + \D \hat u)}{R}} \H_L
\ee
upon integrating by parts and resumming the result into an exponential. Since $\H_L$ is the quasilocal Hamiltonian current in the deformed theory, this charge can be interpreted as generating a field-dependent coordinate transformation of the form   $U \r U + R_u f(\hat u)$, where 

\be
\hat u = \frac{\s +t + \Delta \hat u}{R}  \equiv \frac{u}{R_u} \label{defuhatttb}
\ee
The reason for introducing the rescaled field-dependent coordinate $u$, instead of the $\hat u$ that emerges from the flow,  is that $\p_\s u = 1 + 2\mu \H_R$, in analogy with \eqref{fdepcootrfb}, which naturally leads to a field-dependent radius $R_u$. A similar definition holds for the right-movers, with $v \equiv R_v (\s -t +\Delta \hat v)/R$ and a completely analogous expression for the charges.  Note also that the time dependence of both coordinates perfectly matches that  \eqref{amttb} of the flowed quantum generators to $\O(\hbar)$.

To summarize, the conserved charges corresponding to the Fourier modes of the flowed currents can be naturally rewritten as an integral over the left-moving  Hamiltonian times a function of a certain \emph{field-dependent} coordinate that can be seen as \emph{emerging} from the $T\bar T$ flow. This field-dependent coordinate is completely unambiguous; in particular, its field-dependent zero modes are fully fixed by our derivation; this would not be the case if we  simply tried to guess a field-dependent coordinate that would lead to conserved right-moving charges. %, along the lines of \eqref{}.   %\footnote{They wouldn't be fixed if we simply tried to guess a conserved charge associated to right-moving transformations, with important consequences for the charge algebra.}. 
%
%. We thus see that up to an overall coefficient, the charges can be interpreted as generating a \emph{field-dependent} coordinate transformation,. However, the correct interpretation is subtle and will be discussed in the next section. 
%
The algebra of the above conserved charges consists, by construction, of two commuting copies of the Witt algebra, as can also be checked explicitly. As expected, the central extension is invisible in the classical limit. % We may want to work in terms of the rescaled generators, which satisfy instead a non-linear algebra. 

\bigskip

\noindent \emph{$J\bar T$ - deformed CFTs}

\medskip

\noindent One can perform the same exercise in $J\bar T$ - deformed CFTs \cite{Georgescu:2024ppd}.  There are now four chiral currents of interest in the undeformed CFT: the left and right-moving affine $U(1)$ currents $\J_\pm$, %\textcolor{red}{\emph{Notation!}}, 
and the left and right Hamiltonian densities $\H_{L,R}^{[0]}$.  One would then like to find the expressions  for their flowed counterparts, generated using the $J\bar T$ flow operator  \eqref{chijtb} in the classical limit, in terms of the quasilocal currents \eqref{comprmcurrent} and $\H_{L,R}$ in the deformed theory. 

In practice, it is again simpler to guess the answer for the combinations that flow covariantly.  For the left-movers, it  is trivial to see, by starting with a trial $\mathcal{K}_L, \H_L$ and then correcting the current until the covariant $\mathcal{D}_\l$ derivative is zero,   that the covariantly constant combinations are %\textcolor{red}{\emph{Notation!}} %, using the fact that the chiral current $K_U$ satisfies

\be
\widetilde{\mathcal{K}}_L = \mathcal{K}_L %\J_+ + \frac{\l \hat k}{2} \H_R 
- \frac{\l \hat k H_R}{2 R} \;, \;\;\;\;\;\; \widetilde \H_L = \H_L - \frac{\l H_R}{R} \mathcal{K}_L + \frac{\l^2 \hat k H_R^2}{4 R^2} \label{flowlmcjtb}
\ee
We immediately note that the flowed left-moving currents are \emph{not} the quasilocal currents $\H_L, \mathcal{K}_L$ that generate the left-moving affine and conformal symmetries, but rather a set of non-local currents (in the sense that $H_R$ is an integral over all space) that are related to the quasilocal ones via what resembles a `spectral flow by $\l H_R$'.

%\be
%\mathcal{D}_\l (K_U - \l E_R/2) =0 \;, \;\;\;\;\;\; K_U = \J_+ + \frac{\l}{2} \H_R
%\ee 
The situation is of course more complicated for the right-moving currents, which reside on the non-local side of the theory. As in the $T\bar T$ case, one can find the solution to the flow equation in two steps: first, identify a quantity whose $\mathcal{D}_\l$ derivative is a total derivative with respect to $\s$  and, second, identify another quantity, whose $\mathcal{D}_\l$ derivative times the first quantity has the properties \eqref{flrmcurrjtb}. The result of the first step is 
\be
\widetilde{\mathcal{K}}_R \equiv \mathcal{K}_R %\J_- + \frac{\l \hat k}{2} \H_R 
- \frac{\l \hat k H_R}{2 R_v } (1- \l  \phi')   \;, \;\;\;\;\; \mathcal{D}_\l \widetilde{\mathcal{K}}_R = - \p_\s \left(\widetilde{\mathcal{K}}_R \frac{\Lambda}{\Sigma}\right)
\ee
where  $R_v \equiv R - \l w$, with $w$ the winding of the scalar field around the $\s$ direction, and the expressions for $\Lambda, \Sigma$ can be found in \cite{Georgescu:2024ppd}. In the second step, one finds another quantity, $\l \Phi$, so that\footnote{A possibly useful intermediate step is the identity $\mathcal{D}_\l [(\l \Phi)^n \widetilde{\mathcal{K}}_R] = n (\l \Phi)^{n-1}  \widetilde{\mathcal{K}}_R \frac{\Lambda}{\Sigma} - \p_\s [(\l\Phi)^n \widetilde{\mathcal{K}}_R \frac{\Lambda}{\Sigma} ]$, which follows from the first equation and implies the second. } 
\be
\mathcal{D}_\l (\l \Phi) =  \frac{\Lambda}{\Sigma } (1- \l  \p_\s \Phi) \; \;\;\;\; \Rightarrow \;\;\;\;\; \widetilde{\mathscr{K}}_R \equiv \sum_{n=0}^\infty \frac{1}{n!} \p_\s^n (\l^n \Phi^n \widetilde{\mathcal{K}}_R)  \;\; \; \mbox{satisfies} \;\;\; \mathcal{D}_\l  \widetilde{\mathscr{K}}_R =0 \label{flrmcurrjtb}
\ee
where the explicit expression for $\Phi$  can be found in \cite{Georgescu:2024ppd}.  Plugging  the flowed  $\widetilde{\mathscr{K}}_R $ current into the right affine conserved charge and integrating by parts, one again finds that the derivative contributions to the current can be resummed into a shift of the exponent
%
%Thus, we have been able to find a right-moving current that flows with the $J\bar T$ flow operator. Its Fourier modes are of course nothing but the flowed charges, which take the form
%
\be
\tilde{\bar{P}}_m =  \int d\s \,  e^{2\pi i m \frac{t- \s}{R}}  \widetilde{\mathscr{K}}_R = \int d\s \,  e^{2\pi i m \frac{t- \s}{R}}  \sum_{n=0}^\infty \frac{\left( 2\pi i m \l \Phi\right)^n}{ R^n \, n!}   \widetilde{\mathcal{K}}_R  %\int d\s \,  e^{- i m \s} \sum_{n=0}^\infty \frac{(i m)^n \l^n}{n!} \left(  \frac{\hat \phi}{R_v} - \frac{\tilde \phi_0}{R-\l Q_K}\right)^n\mathcal{K}_-  \nonumber\\
=\int d\s \, e^{- 2\pi i m \frac{\s-t - \l \Phi}{R}} \widetilde{\mathcal{K}}_R  %\left[\J_- + \frac{\l}{2} \H_R-\frac{\l E_R}{2} \left(1-\frac{\l \hat \phi'}{R_v}\right) \right]
\ee
which behaves effectively as an \emph{emergent} field-dependent right-moving coordinate 

\be
\hat v = \frac{\s -t - \l \Phi}{R} = \frac{\s - t - \l (\phi-\varphi_0)}{R_v} \equiv \frac{v}{R_v} \label{truedefv}
\ee
In the second step, we used the explicit expression for $\Phi$ derived in \cite{Georgescu:2024ppd}. The phase space function $\varphi_0$ turns out to correspond to the spatio-temporal   zero mode of $\phi$, i.e. it is $\s$ - independent and satisfies

\be
\p_t \varphi_0 -\{ H,\varphi_0\} =0
\ee
Thus, we see that not only is the field-dependent coordinate emergent from the $J\bar T$ flow, but also the constant ambiguity we have been finding in trying to define it is entirely fixed: what appears in $v$ is the scalar field with its spatio-temporal zero mode removed -  an entirely non-local procedure. This is important for the charge algebra, as otherwise the action of the right-moving  generators would be changing the global charge $J_0+\bar J_0$   \cite{Guica:2020eab}, which we are assuming is quantized.  %\textcolor{red}{\emph{Corresponding algebra vs z.m. remarks $T\bar T$!}}

For completeness, let us transcribe the expression for $\varphi_0$ in the Hamiltonian formalism  \cite{Guica:2020eab}
\be
\varphi_0= \frac{R_v}{R -\l Q_L} \left[ \phi_0 - \frac{\l R}{R_v} \int d\s \mathcal{K}_R \hat \phi - \left(\frac{Q_L}{R} +\frac{Q_R}{R_v}\right) t\right]  \label{varphi0}
\ee
where $\phi_0$ is simply the spatial zero mode of $\phi$. From this, one can easily check that the time-dependence of the generators agress with the $\O(\hbar)$ prediction from \eqref{amlr}.

One can perform an entirely analogous analysis  for the flow of the right-moving Hamiltonian. Again, it is useful to define a quasilocal flowed current 

\be
\widetilde{\H}_R = \frac{R_v \H_R}{R} - \frac{\l H_R}{R} \mathcal{K}_R + \frac{\l^2 \hat k H_R^2}{4 R R_v} (1-\l \phi') 
\ee
whose covariant derivative along the flow is a total $\s$ derivative. 
The flowed chiral right-moving Hamiltonian again takes the form 

\be
\widetilde{\mathscr{H}}_R \equiv \sum_{n=0}^\infty \frac{1}{n!} \p_\s^n (\l^n \Phi^n \widetilde{\mathcal{H}}_R)  \;\; \;\; \mbox{with}\; \;\;\; \mathcal{D}_\l  \widetilde{\mathscr{H}}_R =0
\ee
leading to the same emergent right-moving coordinate.  By construction, the algebra of the  generators consists of two commuting copies of the Witt-Kac-Moody algebra.

\subsection{Flowed versus quasi-local generators}

Let us recapitulate our findings above. In $T\bar T$ - deformed CFTs, we have found two commuting sets of Witt algebra generators, whose form is 
\be
\widetilde{Q}_f = R_u \int d\s f(\hat u) \H_L \;, \;\;\;\;\; \;\widetilde{\bar{Q}}_{\bar f} = R_v \int d\s \bar f(\hat v) \H_R 
\ee
with periodic $f, \bar f$. We will refer to these as the flowed generators, since they satisfy $\mathcal{D}_\mu  \widetilde{Q}_f = \mathcal{D}_\mu  \widetilde{\bar Q}_{\bar f} =0  $.  

From the point of view of the generated symmetries,  it seems more natural to consider instead  the `quasi-local' generators 
 \be
Q_f =  \int d\s f(\hat u) \H_L \;, \;\;\;\;\; \;\bar{Q}_{\bar f} = \int d\s \bar f(\hat v) \H_R  \label{quasilocgenttb}
\ee
whose algebra (in a Fourier basis for $f, \bar f$) is the non-linear modification \eqref{clsttbnlalg} of the Witt algebra, upon discarding the central terms, which are a quantum effect.  Let us explain our terminology. The currents $\H_{L,R}$ are quasilocal, in the sense that, to any finite order in $\mu$, they only contain a finite number of derivatives of the basic fields. The field-dependent coordinates $\hat u, \hat v$ are themselves non-local, due to the zero mode that has been subtracted from the field $\chi_{R,L}$ used in their construction. In this sense, both $Q_f$ and $\widetilde{Q}_{f}$ are non-local generators for general $f$. However,  $\widetilde{Q}_f$ are arguably more non-local than $Q$ due to the factor of $R_u = R+2\mu H_R$, which involves an integral over all space. When comparing the action of the two sets of generators on fields, the former has additional non-localities, as we will see explicitly in the next subsection.

\bigskip

\noindent The above two types of generators also appear in the case of $J\bar T$ - deformed CFTs.  The flow argument leads to the following expressions for the \emph{flowed} charges  %\textcolor{red}{\emph{Notation!}}

\vskip7mm

%\begin{table}[h]
%{\tabulinesep=1.35mm
\begin{tabular}{|c|c|}
%\caption{List of all the flowed generators}
\hline
\emph{\small{affine LM}}    &  $\widetilde{P}_{\eta}=\int d\sigma\;\eta(\hat{U})\left(\mathcal{J}_+ + \frac{\lambda\hat{k}}{2}\mathcal{H}_R-\frac{\lambda \hat{k} H_R}{2R}\right)$  \\[2pt] \hline
\emph{\small{affine RM}}    &  $\widetilde{\bar{P}}_{\bar{\eta}}=\int  d\sigma\;\bar{\eta}(\hat{v})\left[\mathcal{J}_- + \frac{\lambda \hat{k}}{2}\mathcal{H}_R-\frac{\lambda \hat{k}H_R}{2}\left(\frac{1}{R}-\frac{\lambda\hat{\phi}'}{R_v}\right)\right]$              \\[2pt] \hline
\emph{\small{conformal LM}} &   $\widetilde{Q}_{f}=\int d\sigma f(\hat{U})\left[\mathcal{H}_L -\frac{\lambda H_R}{R}\left(\mathcal{J}_+ + \frac{\lambda \hat{k}}{2}\mathcal{H}_R\right)+\frac{\lambda^2 \hat{k}H_R^2}{4R^2}\right]$    \\[2pt] \hline
\emph{\small{(pseudo)conformal RM}} & $\widetilde{\bar{Q}}_{\bar{f}}=\int d\sigma \bar{f}(\hat{v})\left[\frac{R_v\mathcal{H}_R}{R} -\frac{\lambda H_R}{R}\left(\mathcal{J}_- + \frac{\lambda \hat{k}}{2}\mathcal{H}_R\right)+\frac{\lambda^2\hat{k} H_R^2}{4R}\left(\frac{1}{R}-\frac{\lambda\hat{\phi}'}{R_v}\right)\right]$     \\        \hline
%\label{table:generators}
\end{tabular}
%}
\vskip-2mm
\be\label{tablejtbflgen}
\ee
%\vskip3mm

\noindent whose covariant derivative $\mathcal{D}_\l$ vanishes. Looking at these expressions, it is natural to define the $J\bar T$ quasi-local generators as %\textcolor{blue}{\emph{Mention periodic}}
\be
P_\eta \equiv \int d\sigma\;\eta(\hat U)\bigg(\mathcal{J}_+ + 
\frac{\lambda\hat{k}}{2}\mathcal{H}_R \bigg) \;, \;\;\;\;\;\;\;\; Q_f \equiv \int d\s f(\hat U) \H_L
\ee
\be
\bar{P}_{\bar{\eta}}\equiv \int  d\sigma\;\bar{\eta}(\hat{v})\left(\mathcal{J}_- + \frac{\lambda \hat{k}}{2}\mathcal{H}_R\right)\;, \;\;\;\;\;\;\;\; \bar Q_{\bar f} \equiv \int d\s \bar f(\hat v) \H_R \label{rmquasilocjtb}
\ee
where $\eta, \bar \eta, f, \bar f$ are all periodic functions of their respective arguments.

The distinction between the action of the flowed versus the quasi-local generators is easiest to draw for the left-moving ones. The quasi-local generators $P_\eta, Q_f$ act locally on the fields in the theory, as can be noted from \eqref{leftmovingtransf} below.  Meanwhile, the action \eqref{actionflgn}, \eqref{actionflgn2} of the flowed generators $\widetilde{P}_\eta, \widetilde{Q}_f$ is not local, because removing the zero mode of the symmetry parameter involves an integral over all space.

As for the right-moving generators, the distinction is less clear, since the associated transformations are both non-local. If we consider the ones \eqref{rightafftransf}, \eqref{rightpseudocftr} generated by the quasi-local charges, there are two sources of non-locality: the field-dependent coordinate $v$ (which is non-local, due to the fact that it involves \eqref{truedefv} the scalar field with its zero mode removed) and the coefficient of the so-called `compensating transformation', defined in the next subsection, which is an integral over all space. However, the flowed right-moving generators, whose action can be found in \cite{Georgescu:2024ppd},   act even more non-locally on the fields, effectively by subtracting the zero modes of the symmetry parameters. This hopefully justifies our choice of terminology, which effectively ignores the first two types of non-locality.

The change of basis \eqref{relflunfljtb} between the flowed and the quasi-local generators in $J\bar T$ - deformed CFTs takes precisely the form of a spectral flow, but with an operator-valued parameter $\l H_R$, proportional to the right Hamiltonian. %In a Fourier basis, we have \cite{Guica:2020eab,Guica:2021pzy}
%
%\be
%\widetilde P_n = P_n - \frac{\l \hat k}{2} H_R \, \d_{n,0} \;, \;\;\;\;\;\; R\, \widetilde Q_n = R \, Q_n - \l H_R \, P_n + \frac{\l^2 \hat k}{4}  H_R^2\, \d_{n,0}
%\ee
%
%\be
%\widetilde{\bar P}_n = \bar P_n - \frac{\l \hat k}{2} H_R \, \d_{n,0} \;, \;\;\;\;\;\; R\, \widetilde{\bar Q}_n = R_v \, \bar Q_n - \l H_R \,\bar P_n + \frac{\l^2 \hat k}{4}  H_R^2\, \d_{n,0} \label{rmquasilocjtb}
%\ee
Since  we are now starting from the flowed generators, the algebra of the quasi-local ones can be computed using \eqref{relflunfljtb} and the fact that the algebra of the flowed generators consists of two commuting copies of the Witt-Kac-Moody algebra, by construction.   The resulting commutation relations are precisely  \eqref{jtbquasilocalg}, which we found from a completely independent holographic analysis.  

An interesting question is whether we can go beyond the classical algebra and propose a \emph{quantum} set of quasilocal generators. As emphasized in the first subsection of this section, the definition of the flowed generators holds at full quantum level, and their algebra will always be identical to that in the undeformed CFT. The question is whether the relation between the flowed and quasi-local generators can be found at quantum level.

In the case of $J\bar T$ - deformed CFTs, it is rather natural to suppose that  the relations between the quantum generators will still correspond to the  spectral flow \eqref{relflunfljtb} 

\be
K_n^\l = \widetilde K_n^\l + \frac{\l \hat k}{2} H_R\, \d_{n,0} \;, \;\;\;\;\;\; L_n^\l = \widetilde{L}_n^\l + \l \widetilde K_n^\l H_R + \frac{\hat k \l^2}{4} H_R^2 \, \d_{n,0} \nonumber
\ee

\be
\bar K_n^\l = \widetilde{\bar K}_n^\l + \frac{\l \hat k}{2} H_R\, \d_{n,0} \;, \;\;\;\;\;\; \bar L_n^\l = \widetilde{\bar L}_n^\l + \l : \widetilde{\bar K}_n^\l H_R : + \frac{\hat k \l^2}{4} H_R^2 \, \d_{n,0} \label{flvsqlquantum}
\ee
where the factor of $R_v$ has been absorbed into the definition of $\bar L_n$%\textcolor{red}{\emph{This is not terribly natural.}}
, (so $\bar L_0 = R_v H_R$) and the normal ordering puts $H_R$ to the left if $n>0$, and to the right if $n<0$ \cite{Guica:2021pzy}. Using this definition, the fact that the flowed generators satisfy an exact (Virasoro-Kac-Moody)$^2$ algebra, as well as the commutators  \eqref{defamcomm}, \eqref{amlr} of the generators with $H_R$, one can compute the algebra of the so-defined quasi-local generators to all orders in $\hbar$. Of course, the left generators will still satisfy a Virasoro-Kac-Moody algebra, since they commute with $H_R$, and  spectral flow does not change the algebra. The non-linear results for the right-right and left-right commutators can be found in \cite{Guica:2021pzy}. 

Unfortunately, such a natural guess does not seem to exist in $T\bar T$ - deformed CFTs, though we clearly expect that corrections to \eqref{clsttbnlalg} to all orders in $\hbar$ exist. Most likely, there exists a better way to think about the $T\bar T$ symmetries where the definition of the true, quasi-local symmetries is natural. We hope future work will uncover the necessary better way to think about them.
%
%\bi
%\item compute algebra, emphasize finite $R$ and discuss quantum corrections.
%\ei
%\textcolor{blue}{Why did we have to go through the flowed ones to find symmetries? Seems answer is that it's hard to guess directly the field-dependent coordinate, because of the zm issue. Clearly, $H$ has a defn via flow. }

\subsection{Lagrangian realisation of the symmetries}

We would now like to understand the field-dependent symmetries in the Lagrangian language. The idea is very simple. We first compute the variation of a basic field, $\phi$, under the action of the symmetry generators of the previous subsection 

\be
\d_\e \phi = - \{ Q_\e, \phi\}
\ee
where the right-hand side depends on the fields and their conjugate momenta. We
then translate the result to Lagrangian language using $\dot \phi = - \{ H, \phi\} $  to replace momenta by appropriate functions of the velocities,   and  then interpret it.  For simplicity, we concentrate on a $T\bar T$ or $J\bar T$ - deformed free boson, but our analysis can easily be generalised. We  will start our discussion with the $J\bar T$ case, which is simpler, being half-local, and then move on to $T\bar T$. %\textcolor{blue}{ Note Lagr and Ham transf only need agree on-shell.}

\subsubsection{$J\bar T$ - deformed CFTs}

In $J\bar T$ - deformed CFTs, the relation between time derivatives and momenta is
\be
\dot \phi = -\{ H, \phi\} = 2 \frac{\J_- + \frac{\l \hat{k}}{2}\H_R}{1-\l \J_+ - \frac{\l^2 \hat k}{2} \H_R} + \phi' \;, \;\;\;\;\; \p_{U/V} \phi = \frac{1}{2} (\phi' \pm \dot \phi)
\ee
 In the following,   we will treat each type of symmetry in part,  concentrating on the quasi-local symmetry generators which, as explained, act more naturally on the fields of the theory than the flowed ones. %, as we exemplify below for the case of left-moving symmetries. 

%More precisely, {\color{ForestGreen}
%We find it useful to summarize here the locality properties of the various transformations derived in this section. There are two sources of non-locality in the symmetry transformations: the different effect of constant transformations compared to the others and the explicit presence of conserved charges (integrated quantities). Nevertheless, the latter can be tracked back to the extraction of the zero mode in the field-dependent coordinate. The LM quasilocal affine and conformal generators act locally on $\phi$. Their flowed counterparts act differently depending on whether they are constant or not, thus being non-local. The action of all RM transformations is non-local. In the case of the quasilocal ones, if treating $v$ as a usual coordinate, the only source of non-locality is the presence of conserved charges. In the case of the flowed ones, additional non-locality comes from the different effect of constant ones with respect to the rest. 

\bigskip

\noindent \emph{Left affine and conformal symmetries}

\medskip

\noindent The generators of these symmetries are 
\be \label{leftmovingch}
P_\eta = \int_0^R d\s  \, \eta_p(U)  \left(\J_+ + \frac{\l \hat{k}}{2} \H_R \right) \;, \;\;\;\;\;\; Q_f = \int_0^R d\s \, f(U) \H_L
\ee
where we have added a subscript, `$p$', to the function labeling the affine charges, to emphasize the fact that it is periodic - a specification that will  be useful later. %\textcolor{blue}{ However, we do not write this label on the $\eta$ appearing in $P_\eta$,  since for it there is no possibility of confusion: this function is always   periodic.}
  We compute  
\be \label{leftmovingtransf}
\d_\eta \phi =-\{P_\eta, \phi\} =  \frac{\hat{k}}{2}  (1-\l \p_V \phi)  \eta_p(U) \;, \;\;\;\;\;\;\;\d_f \phi=- \{ Q_f,\phi\} =   \p_U \phi f(U)
\ee
The action of $Q_f$ is noting but the standard action of a left-moving diffeomorphism $\xi = - f(U) \p_U$. The action of $P_\eta$
can be understood as the  the standard affine shift by an arbitrary (periodic) function of the left-moving coordinate, accompanied by a right diffeomorphism  

\be
\phi \r \phi +\frac{\hat{k}}{2}\eta_p(U) \;, \;\;\;\;\;\; V \r V   +\frac{\l \hat k}{2} \eta_p (U) \label{actlmsymm}
\ee 
Note this leaves unchanged the combination   $ V - \l \phi$.  
%\textcolor{red}{I feel we should be able to change our definitions so that these ugly signs disappear.}
%{\color{ForestGreen}Since the zero mode of $\phi$ shifts with the zero mode of the function $\eta$, let's denote it $[\eta]_{zm}$, we obtain that the field-dependent coordinate shifts $v\rightarrow v+ \lambda [\eta]_{zm}$.(because $V$ shifts like this, while $(\phi-\varphi_0)$ does not shift) If we want to have the field dependent coordinate without zero mode after the transformation, we should do an extra isometry $V\rightarrow V-\lambda [\eta]_{zm}$, but this would bring an extra contribution to the charges proportional to $E_R$ which would amount for the spectral flow. (is there a problem to induce a zero mode to $v$ as long as it's not the zero mode of $\phi$ which is by definition ``field-dependent"? this should not affect the algebra in any way, right?)}  {\color{red}???} {\color{ForestGreen}(here I just wanted to say that not any zero mode in $v$ is problematic, but only the zero mode of $\phi$ in $v$. For ex, the transformation above clearly induces a zero mode in $v$, which is $[\eta]_{zm}$)}

%{\color{red} I think we should add a line with the explicit action of $\widetilde{P}_{\eta}$ and $\widetilde{Q}_{f}$. In fact, I thought we already had something like this - did it get erased?} {\color{ForestGreen}(I think we only had it for RM)}

Clearly, these transformations act locally on the scalar field. By comparison, the action of the flowed left-moving affine and conformal charges given in table \eqref{tablejtbflgen} is non-local, as we can see from the explicit expressions for the transformations they generate
\be\label{actionflgn}
\widetilde{\delta}_{\eta}\phi=-\{\widetilde{P}_{\eta},\phi\}=\frac{\hat{k}}{2}\big[\eta_p(U)-\lambda\partial_V\phi(\eta_p(U)-[\eta_p]_{zm})\big]
\ee
\be\label{actionflgn2}
\widetilde{\delta}_{f}\phi=-\{\widetilde{Q}_{f},\phi\}=\partial_U\phi f(U)-\frac{\lambda\hat{k} H_R}{2R}\big[f(U)-\lambda\partial_V\phi(f(U)-[f]_{zm})\big]+\frac{\lambda \partial_U\phi\partial_V\phi}{1-\lambda\partial_V\phi}[f]_{zm}
\ee
where $[\;]_{zm}$ denotes the  spatial zero mode of the corresponding function. 
Their subtraction shows explicitly the action of the left flowed generators 
is non-local, as does the  explicit presence of the integrated Hamiltonian $H_R$ in $\widetilde{\d}_f \phi$. Similar additional non-localities can be seen in the action of the flowed versus quasi-local right-moving generators \cite{Georgescu:2024ppd}.  The flowed left-moving affine charges leave invariant 
%, the shift in $V$ is absent from \eqref{actlmsymm} when $\eta(U) = const.$, which leads to the fact that 
the field-dependent coordinate
%
%Thus, the action on the scalar field is not local,  {\color{red}Can we make this fully sharp?} but the field-dependent coordinate {\color{ForestGreen}(since in $v$ we remove the zero mode of $\phi$, does it make sense to say that $v$ itself is non-local? and then automatically the flowed generators which leave $v$ invariant are non-local)}
%
$
v = V-\l (\phi - \varphi_0)
$
including when $\eta_p(U) = const.$, which %is left invariant by all the \textcolor{red}{flowed} left affine transformations, including the constant ones. 
%This
 is important for having consistent  commutators of  the shift charge with the right-moving generators. % in section \ref{section6:asg}.
 %
 %is required by having zero commutation relations between $Q_0$ and the field-dependent charges {\color{ForestGreen}(RM charges)}; clearly, the currents are invariant under constant shifts of $\phi$, so all we need is that $v$ be also invariant.}  Conversely, we could have inferred the above expression for the field-dependent coordinate from the requirement that it be invariant under constant shifts of $\phi$. 
%
%Note that for the flowed left-moving  conformal transformations, the non-locality is also manifest  in the explicit presence of $E_R$ \textcolor{red}{\emph{Notation!}} in the symmetry transformation.

\bigskip

\noindent \emph{Right affine symmetries}

\medskip

\noindent The expression for  the right-moving affine generators is 
%{\color{ForestGreen}(same comment as before if we want to keep $k$ arbitrary; we can introduce $\hat{k}$)} 

\be \label{rmaffch}
\bar P_{\bar \eta} = \int_0^R d\s \, \bar \eta_p \left(\hat v\right) \left( \J_- + \frac{\l \hat{k}}{2} \H_R \right) \;, \;\;\;\;\; \hat v \equiv \frac{v}{R_v} = \frac{V - \l \phi + \l \varphi_0}{R_v}
\ee
where the expression for $\varphi_0$ is given in \eqref{varphi0} and we have again added a subscript `$p$' to the function parametrizing the transformation, to emphasize it is a periodic function of its argument (lest $\bar P_{\bar \eta}$ is not conserved). %{\color{ForestGreen}(only in the case of RM affine we will write this p index, since all the other functions are periodic; hmm actually not true since $\eta$ can be nonperiodic but only when compensating.)}
The change of the scalar under this transformation is  

\be
\d_{\bar \eta} \phi = -\{\bar{P}_{\bar \eta} , \phi\} = \frac{ \hat{k}}{2}  (1-\l \p_V \phi) \bar \eta_p - \l \bar{P}_{\bar \eta'} \left\{ \frac{\varphi_0}{R_v} ,\phi\right\}
\ee
where the prime in $\bar{P}_{\bar{\eta}'}$ denotes a derivative with respect to $\hat{v}$. The first term is just the standard transformation of the scalar under a right affine transformation, similar to the one above. The second term corresponds specifically to the contribution of the zero mode of $\phi$ that had to be   subtracted from the field-dependent coordinate. 
%{\color{ForestGreen}We notice that this term is charge-dependent. Having charge-dependent coefficients in the symmetry transformations leads to non-linear symmetry algebras.} \textcolor{red}{Is this a general statement?}  {\color{ForestGreen} (I think so because the Poisson bracket is defined as a variation of a charge. The charge depends on the field, so if the field variation has some charge-dependent coefficient it will go out of the integral, while the rest should integrate to another charge in the theory, assuming that the algebra closes. In this way, we get some QQ terms.)} {\color{red} I'm not sure we need to bring up this non-linear algebra stuff here.} {\color{ForestGreen}(I agree, but maybe we can say something about non-locality here since we have charge-dependent transformations. I would say we have two notions of non-locality, one is when the transformation acts differently on zero modes than on the rest and another one is when we have charge-dependent transformations)}

The corresponding commutator evaluates to 
\be \label{largeaffinetr}
% \left\{ \frac{\widetilde \phi_0 - \a t}{R-\l Q_K } ,\phi\right\} = 
 \left\{ \frac{\varphi_0}{R_v} , \phi \right\} 
 =  \frac{\hat{k}}{R_Q} (1-\l \p_V\phi) \, \hat{\mathcal{T}} \;, \;\;\;\;\;\; \hat{\mathcal{T}} \equiv \frac{1}{2} \left(\frac{U}{R}-\frac{v}{R_v}\right)
\ee
which may be interpreted as a large affine transformation that is being performed simultaneously on the left and on the right%\footnote{Note that, due to the non-linearity, the coefficient of the $\eta, \bar \eta$ winding term that enters in the symmetry generator (subscript of $\d$ below) is different from the coefficient of the linear terms which shift $\phi$ %\textcolor{ForestGreen}{(checked)}
%%
%\be
% \d_{\bar \eta = \hat{v}} \phi - \d_{\eta = \hat{U}} \phi  =  \hat{k}  (1-\l \p_V \phi)\hat{\mathcal{T}} + \frac{\l Q_{\bar K}}{R_v} \{ \varphi_0, \phi\} = \hat{k}  (1-\l \p_V \phi) \hat{\mathcal{T}} \left(1 + \frac{\l Q_{\bar K}}{R_Q}\right)=\frac{\hat{k}R_v}{R_Q}(1-\l \p_V \phi) \hat{\mathcal{T}}
%\ee
%{\color{blue}Note the coefficient of this transformation is $\l \bar P_{\bar \eta'}/R_v$, due to the non-linearity. } \textcolor{red}{Please mark if  you move  things around. } }
, in such a way that no additional  spatial winding is introduced in the scalar field. The field-dependent time coordinate $ \hat{\mathcal{T}} $ introduced above reduces to the standard CFT time, $t$, in the $\l \r 0$ limit.

The full expression for the right-moving affine transformation is then given by  
%{\color{red} Signs!} {\color{ForestGreen}(? i think it's correct)}
\be \label{rightafftransf}
\d_{\bar \eta} \phi =  \frac{\hat{k}}{2}  (1-\l \p_V \phi) \left(\bar \eta_p  - \frac{2\l \bar P_{\bar \eta'}}{R_Q} \hat{\mathcal{T}} \right)%,\hspace{1.5cm}\hat{\textbf{t}}=\frac{1}{2}\left(\frac{U}{R}-\frac{v}{R_v}\right)
\ee
Note that, at the level of the action on $\phi$ (but not at the level of the functions that enter the generator, which do need to be periodic) this behaves as if the periodic function that parametrises the affine transformation, $\bar \eta_p$, has picked up a  winding term proportional to the affine charge associated with $\bar \eta_p'$, and similarly for the previously absent left-moving affine transformation, leading to the `total' affine transformations 
%
%{\color{red} Check signs!}{\color{ForestGreen}(plus in $\bar{\eta}$)}
%
\be  \label{comptransformations}
\bar \eta (\hat v) = \bar \eta_p  (\hat v) + \frac{\l \bar P_{\bar \eta'}}{R_Q}\,  \hat v \;\;\;\;\;\;\;\; \eta (U) =  -  \frac{\l \bar P_{\bar \eta'}}{R_Q} \, \hat U \;, \;\;\;\;\;\; R_Q \equiv R-\l Q_L
\ee
Note these are \emph{precisely} the affine gauge transformations that we found in the holographic analysis of section \ref{asysymmdtr}, with a winding determined by the right affine charge via \eqref{windingeta}. 
% The fact that the field-dependent transformation needs to be accompanied by a  large compensating transformation with a charge-dependent coefficient is entirely analogous    to the way the field-dependent symmetries are implemented in $T\bar T$ - deformed CFTs  \cite{Guica:2022gts}.
% {\color{ForestGreen}A difference with respect to the case of $T\bar{T}$ is that here by construction the winding, thus also the field-dependent radius, will commute with everything.} \textcolor{red}{Later}. 

%Consequently, 
%
%\be
%\d_{\bar \eta} \phi =  - \frac{k}{2}  (1-\l \p_V \phi) \bar \eta + \frac{\l \bar{\mathcal{P}}_{\bar \eta'}}{R_v}   (\d_{\bar \eta = \frac{v_{imp}}{R_v}} \phi - \d_{\eta = \frac{U}{R}} \phi )
%\ee

 Thus, the transformation induced by $\bar{P}_{\bar \eta}$ consists of a field-dependent affine transformation, accompanied by a large shift in the value of the scalar field. %The latter is necessary in order to keep the $U(1)$ charge unchanged, as required by charge quantization. 
The origin of the large affine compensating  transformation %\eqref{largeaffinetr}
 can be understood from the fact that 
 
 \be \label{relationundefdef}
P_0= Q_L = J_0 + \frac{\l \hat{k} H_R}{2} \;, \;\;\;\;\;\;\;\;\bar P_0= \bar Q_{R} = \bar J_0    + \frac{\l  \hat{k} H_R}{2}
 \ee 
 are the coefficients of the terms linear in $U$ and, respectively, $v$, in e.g. the $J\bar T$ - deformed free boson solution. Under a right-moving affine transformation, $\d H_R \propto \bar P_{\bar \eta'}$, while we are requiring that $\d J_0 = \d \bar J_0 =0$ to respect shift charge quantization, resulting in  a change in the linearly growing modes. %The qualificative `compensating' refers to the fact that, hadn't this linearly growing transformation been included, one would have effectively changed the quantized charges $J_0, \bar J_0$. 
 See the appendix of \cite{Georgescu:2024ppd} for more details.

\bigskip

\noindent \emph{Right pseudoconformal transformations}

\medskip 

\noindent Finally, the transformation of $\phi$ under the right-moving pseudoconformal symmetries $\bar Q_{\bar f}$ in \eqref{rmquasilocjtb} takes the form, 
%%
%\be
%\bar{Q}_{\bar f} = \int_0^R d \s \bar f (\hat{v}) \H_R 
%%\;, \;\;\;\;\;\; v_{imp} = \s -t - \l (\phi - \varphi_0)
%\ee
after an almost identical computation to that for the right-moving affine symmetries 
%\be
%\d_{\bar f} \phi \equiv -\{ \bar{Q}_{\bar f} ,\phi\} = -\bar f (\hat{v}) \p_V \phi- \l \bar Q_{\bar f'} \left\{ \frac{\varphi_0 }{R_v} ,\phi\right\}
%\ee
%Using \eqref{largeaffinetr}, we obtain  
\begin{align}\label{rightpseudocftr}
\delta_{\bar{f}}\phi&=-\bar{f}(\hat v)\partial_V\phi -\frac{\lambda \hat{k} \bar{Q}_{\bar{f}'}}{R_Q}(1-\lambda\partial_V \phi)\hat{\mathcal{T}}
\end{align}
The first term is simply the transformation of the scalar under a field-dependent  change of the $V$ coordinate. The second term corresponds, as before, to an  accompanying large affine transformation whose coefficient is proportional to the change, $\bar Q_{\bar f'}$, in the right-moving energys and ensures, as before, that the first transformation does not affect the quantized $U(1)$ charge. 

%because the corresponding expansion coefficients depend on $H_R$, whose variation is precisely $\bar Q_{\bar f'}$, and not taking this into account would yield to a violation of $U(1) $ charge quantisation. 

%This corresponds to a RM field-dependent transformation, accompanied by a winding term and an affine transformations with  $\bar \eta$ and $\eta$ not  periodic, but have a winding term proportional to the charge {\color{ForestGreen}(mention that in 4.9 we already considered this?)}. This is more natural from the holographic perspective. In the appendix, we work out the action of the above generators on a free $J\bar T$ - deformed scalar, and try to understand there the physical interpretation of the compensating transformation. 
%Let us also compute the corresponding flowed transformations 
%\bea \label{rmpseudoflowed}
%\widetilde{\delta}_{\bar{f}}\phi&=&-\frac{R_v}{R}\bigg(\bar{f}\partial_V\phi+\frac{\lambda \hat{k} \bar{Q}_{\bar{f}'}}{R_Q}(1-\lambda\partial_V \phi)\hat{\mathcal{T}}\bigg)+\frac{\lambda\partial_V\phi}{R}\bar{P}_{\bar{f}}- \nonumber\\&& \hspace{1.2cm}-\frac{\lambda E_R}{2R}\bigg[\bar{f}-\lambda \partial_V\phi(\bar{f}-[\bar{f}]_{zm})+(1-\lambda \partial_V\phi)\frac{\lambda\bar{P}_{\bar{f}'}}{R_Q}\hat{\mathcal{T}}\bigg]
%\eea
%which suffers from the same types of non-locality as  the right-moving affine case  discussed.

More generally, the large affine transformations are an integral part of the symmetry transformations of the fields; without them, the charge commutators that we previously found in the Hamiltonian formalism would not be reproduced. See \cite{Georgescu:2024ppd} for a thorough discussion of their properties from the Lagrangian viewpoint, some of which are slightly non-standard. %Another peculiarity they present is that, if we would like their associated Noether current to coincide with the current in the Hamiltonian formalism, then  the divergence of this current does not equal $E \d_\varepsilon \phi$

 Equations \eqref{leftmovingtransf}, \eqref{rightafftransf} and \eqref{rightpseudocftr} tell us, in principle, 
how the field in the Lagrangian formalism should change under each type of transformation.  One should note, however, that Hamitonian and Lagrangian transformations only need to agree on-shell, whereas we are of course looking for the off-shell transformation of the fields. One concrete question that arises is whether the charge-dependent coefficients of the compensating large affine transformations, which are time-independent on-shell, should be mapped to the off-shell charge at some particular time $t$, which is in principle time-dependent, or to the time average of the charge, which is by construction a constant. %\textcolor{blue}{ As in \cite{Guica:2022gts} for $T\bar T$, we will be 
The choice of \cite{Georgescu:2024ppd} was the latter option, which greatly simplifies the variation of the action. It would nonetheless be worthwhile to   understand the  reasons   behind this choice from first principles.

%
% 
%\bi
%%\item discuss how compensating transformation is necessary for the correct charge algebra
%\item discuss problem with current and fix identical to holographic side
%\ei

\subsubsection{ $T\bar T$ - deformed CFTs}

The procedure for determining the action of the extended symmetries in the  $T\bar T$ - deformed CFTs is identical. Concentrating again on the $T\bar T$ - deformed free boson,  the action of the left-moving $T\bar T$ quasilocal generators \eqref{quasilocgenttb} on it is given by 

\be 
\{ Q_f, \phi\} = -  f(\hat u)\, \frac{ \phi'+\p_\pi \H }{2} + \int d \tilde \s \hat f'( \hat{\tilde u} )\tilde \H_L \left \{ \frac{\tilde u}{R_u},\phi \right\} 
\ee
where the first term comes from the commutator of $\phi$ with $\H_L$. The equal-time commutator of the $T\bar T$ field-dependent  coordinate $u$ is evaluated using its explicit expression \eqref{ttbfdepshifts} and the known Poisson brackets with $\phi$. The end result for the symmetry variation generated by $Q_f$ \cite{Guica:2022gts}

\be
\d_f \phi = - \left(  f(\hat u) - \frac{2\mu \D \hat u}{R_H} Q_{\hat f'}\right)\, \frac{\phi'+\p_\pi \H}{2} %+ \frac{2\mu Q_{f'}}{R_u}\left( \frac{R_v \Delta \hat v}{R_H} -2t\right)\frac{\phi'-\p_\pi \H}{2}
 + \frac{2\mu}{R_u} \left[ \mathcal{F}_f + Q_{\hat f'} \left( \frac{R_v \Delta \hat v}{R_H} -2t\right)  \right]\, \frac{\phi'-\p_\pi \H}{2} \label{varphi}
 \ee
where we have defined

\be
\mathcal{F}_f (\s) \equiv \int d\tilde \s f'(\hat {\tilde u}) \tilde \H_L  [G(\tilde \s -\s) - \Delta \hat{\tilde u}] \label{Ff}
\ee
We would now like to translate these symmetry transformations to the Lagrangian formalism. Upon identifying $\p_\pi \H = \dot \phi$, the transformation simply corresponds to a   diffeomorphism, with 
 components 
%Let us now consider the diffeomorphisms \emph{Switch sign!}
%
\be
\xi^U =  f(\hat u) - \frac{2\mu Q_{\hat f'}}{R_H} (\hat u - U) \;, \;\;\;\;\; \xi^V = -\frac{2\mu}{R_u} (\mathcal{F}_f- Q_{\hat f'} U)- \frac{2\mu Q_{\hat f'}}{R_u R_H} (R_v \hat v +2\mu H_R V ) \label{newdiff}
\ee
where we used the definition \eqref{defuhatttb} of the field-dependent coordinate and its right-moving analogue.  It is instructive to compute the derivatives of the function $\mathcal{F}_f$ which read, using \eqref{greensfcyl} 
\be
\p_\s \mathcal{F}_f = - \hat f'(\hat u) \H_L + Q_{\hat f'} \;, \;\;\;\;\; \frac{d \mathcal{F}_f}{dt} = \p_t \mathcal{F}_f - \{ H, \mathcal{F}_f\} =  - \hat f' (\hat u) \H_L \frac{1+2\mu \P}{1+2\mu \H} + Q_{\hat f'} 
\ee
Note that we have used the on-shell expression for the time derivative, as only then need the Hamiltonian and Lagrangian symmetry transformations   agree. %\footnote{\textcolor{blue}{This fact can already be seen in the case of a two-dimensional free scalar transforming under e.g. left conformal transformations, under which $\d \phi = - f(U) (\pi+\phi')$, $\d \pi = - \p_{\s} [f(U) (\pi+\phi')]$. Even though $\pi = \dot \phi$ only using the definition of the velocity, we note that $\d_f \pi = \d_f \dot \phi$ only upon using \emph{both} Hamiltonian equations of motion.}}. 
It is perhaps more instructive to write the derivatives with respect to $U,V$
%Replacing $\H_{L,R}$ by their Lagrangian expressions \eqref{HLRlagr} and using the expressions for the stress tensor components given e.g. in \cite{Guica:2020uhm}, we deduce that in the Lagrangian formalism
%
% Writing thus $\H_L = \L \p_\s u \p_u U $, which is only valid on-shell, the other term is $\L \p_t u $ in the Lagrangian sense. We may therefore guess that in the Lagrangian formalism  \emph{Write this better}
%
\be
\p_U \mathcal{F}_f = %- \frac{1}{2} \hat f' \L \p_U u \p_u U  + Q_{f'} = 
- \hat f' T_{UU}+ Q_{\hat f'}\;, \;\;\;\;\;\;\;\;  \p_V \mathcal{F}_f = %- f' \L \p_V u \p_u U = 
 \hat f' T_{UV}
\ee
where we used \eqref{eqntssttb} to %identify the trace of the stress tensor with the particular combination of
determine all components of the stress tensor in terms of 
 $\H, \P$. Further rewriting the stress tensor components in terms of the $\L, \bar{\L}$ parametrisation \eqref{stresst} we used when discussing holography, we find these diffeomorphisms agree perfectly, on-shell,  with the $U,V$ components of \eqref{xiV}, restricted to the $\rho=-\hat \mu$ surface, including the winding terms, upon taking into account the fact that $Q_{f'} = \frac{1}{R_u} Q_{\hat f'}$.  %\textcolor{red}{\emph{Check!}}

To also reproduce the radial component of the bulk diffeomorphism, note 
that the variation of the $T\bar T$ - deformed action under the above two-dimensional diffeomorphism is given by
%\textcolor{blue}{} %\eqref{varactdiff}. , we compute
%
%\be
%T_{UU} \p_V \xi^U + T_{UV} \p_U \xi^U =  \frac{2\mu Q_{\hat f'}}{R_H} T_{UV}
%\ee
%%
%\be
%T_{UV} \p_V \xi^V + T_{VV} \p_U \xi^V = - \frac{2\mu \hat f'}{R_u} (T_{UV}^2 - T_{UU} T_{VV}) - \frac{4 \mu^2 H_R Q_{\hat f'}}{R_u R_H} T_{UV} =- \frac{\hat f'}{R_u} T_{UV} -  \frac{4 \mu^2 H_R Q_{\hat f'}}{R_u R_H} T_{UV}
%\ee
%for a total of
%
\be
T_{\a\b}\, \p^\a \xi^\b =-2\left(  f'(u)- \frac{2\mu Q_{ f'}}{R_H}\right)  T_{UV} 
\ee
where we assumed, 
 for simplicity, that $Q_{f'}$ is constant. 
However, since the overall non-invariance of the action is proportional to the trace of the stress tensor, one immediately notes that this non-zero variation  it can be absorbed by a Weyl transformation $g_{\mu\nu} \r \Omega^2 g_{\mu\nu}$, with
\be
\Omega^2 = 1+ \left( f' - \frac{2\mu Q_{f'}}{ R_H}\right) \label{weyl}
\ee
where $f$ is assumed to be infinitesimal. 
The last term precisely agrees with the radial component of the  diffeomorphism  \eqref{xiV} in the dual space-time, which is known to implement  Weyl transformations in the boundary theory.  The right-moving symmetries are treated in a directly analogous manner.

%In fact, it can just be written as $2 f' + w_f + w_{\bar f}$, whose first term disagrees by a factor of two with the expectation. 

Thus, we find that  in the Lagrangian formalism,  the symmetries  generated by $Q_f, \bar Q_{\bar f}$ are given by the field-dependent coordinate transformation \eqref{newdiff}, accompanied by a Weyl rescaling \eqref{weyl} of the metric. 
 These symmetry transformations are in perfect agreement with the results of the holographic analyses. They also turn out to lead to well-defined charge variations, directly in the Lagrangian formalism. Note that from the point of view of the Lagrangian variations, the charges $Q_{f'}$ that enter the diffeomorphism much be taken to be space-time constants, e.g. by projecting onto the temporal zero mode. This is related to the fact that the Hamiltonian and Lagrangian symmetries only need to agree on-shell. 

 In conclusion,  we were able to find the explicit form of the extended symmetries of $T\bar T$ and $J\bar T$ - deformed CFTs on a cylinder in the Lagrangian formalism. The form of the symmetries associated to field-dependent transformations is somewhat peculiar, in that it involes non-local contributions meant to restore charge and momentum quantization, which would be good to better understand. Nonetheless, this form of the symmetries  perfectly matches the result of the holographic analyses in section \ref{asysymmdtr}.

%\newpage

\section{On correlation functions  in $T\bar T$ and $J\bar T$ - deformed CFTs \label{corrfsec} }

One of the most important observables in QFT are correlation functions of an appropriately chosen set of operators, such as local operators in the case of local QFTs: they are the objects used to define the QFT in axiomatic approaches, they encode the S-matrix when the latter exists, they can be used to characterise a conformal field theory, a.s.o. 

Given the non-standard UV behaviour of $T\bar T$ and $J\bar T$ - deformed QFTs, the study of correlation functions in these theories is a very interesting and important question, both for enlarging the set of  observables one can compute  in these particular set-ups, but also for understanding what modifications with respect to the standard QFT  behaviour these non-local - yet seemingly consistent - theories entail. This question is particularly interesting in view of possible extensions of $T\bar T$ and $J\bar T$ - deformed QFTs to more general theories with similar UV behaviour.   Since CFTs are the simplest (interacting) QFTs as far as the behaviour of the correlation functions is concerned, one would like to  concentrate on $T\bar T$ and $J\bar T$ - deformed CFTs as the simplest examples of theories displaying the UV behaviour of interest.

Despite all these captivating questions, in this section we will be brief. Part of the reason is that,  many interesting developments notwithstanding, a fully satisfactory understanding of correlation functions in $T\bar T$ and $J\bar T$ - deformed CFTs has yet to emerge. In particular, the current approaches for studying correlators in  $T\bar T$ and $J\bar T$ - deformed CFTs  rely on the existence of the field-dependent coordinates, in terms of which the dynamics - operators included - can be reduced to 
that of the undeformed CFT. Therefore, these methods do not appear easy to generalise
%. This way of appr the correlators obscures - in our opinion - the  ways of defining and calculating them that may be universal and applicable also
% 
to other non-local QFTs with similar UV asymptotics. 
 Third, the recent review \cite{He:2025ppz} does cover the known facts about correlation functions in $T\bar T$ - deformed CFTs, so the reader can simply consult this reference for details. 

 We will therefore  concentrate mostly on correlation functions in $J\bar T$ - deformed CFTs, which have not been covered by a review yet, and where the definition of the operators   may have a higher chance of being generalisable to other dipole CFTs.  In addition, the form of the $J\bar T$ correlators is needed to come full circle on the Kerr/`CFT' story that threads this review. To introduce the problem, we will make a few general comments about computing correlation functions in these theories,  then flash the known results for $T\bar T$ - deformed correlators, whose form we find very suggestive.

%Correlation functions is a very interesting topic, and essential for precision holography in these backgrounds. However, a bit outside our main interest, because they are very specific to $T\bar T/J\bar T$, and may not generalise to other theories in the universality class of interest. 

 \etocsetnexttocdepth{5}
    \etocsettocstyle{\subsubsection*{Contents of this section: }}{}
    \cftsubsubsecindent 34pt
    \localtableofcontents

\subsection{General considerations}

A basic  issue when discussing  correlation functions is choosing an appropriate basis of  operators. In a local QFT, it is natural to choose a basis of local operators - though interesting non-local operators such as Wilson loops may also exist. In a $2d$ CFT, it is natural to work with Virasoro primary operators, whose correlation functions determine those of any local descendant operator, etc.  In non-local theories such as $T\bar T$ or $J\bar T$ - deformed CFTs, it is \emph{a priori} not clear which basis of operators to use to compute correlation functions\footnote{In early studies of the $T\bar T$ deformation, in which it was proposed as a theory of $2d$ quantum gravity, the question of existence of correlation functions was also discussed \cite{Dubovsky:2012wk}. }. 

First, some kinematic considerations. 
 Since both  theories are translationally invariant, though non-local,  it is natural to work in a basis of momentum-space operators, $\O(p,\bar p)$. Given that $T\bar T$ - deformed CFTs are Lorentz-invariant, correlation functions can depend on the dimensionless scalar quantity $\mu p \bar p$, whereas in $J\bar T$, they will be general functions of $\l \bar p$. Since the latter theory is local and conformal on the left, it is also natural to consider operators in a mixed  basis $\O(z, \bar p)$. The chosen operators may, in addition, be required to be primary with respect to $SL(2,\mathbb{R})_L$. Note, however, that their left conformal dimension will, in principle, be a function of the right-moving momentum, $h(\l \bar p)$, as we already saw in \eqref{jtbdims}. In $T\bar T$ or on  the right-moving side of $J\bar T$, there is no \emph{a priori}  clear notion of dimension. %\textcolor{red}{\emph{Talk about renormalisation and lack of knowledge of allowed counterterms!}}

Another issue is how to relate operators in two different theories related by a deformation. If a fundamental description of the theory is available, one may find it natural  to keep the same expression for the operators (in terms of the fundamental fields) as one modifies the action. However, sometimes a more natural basis of operators is obtained by changing the corresponding expression, and one needs some criteria that the operators should obey in order to decide what is their best definition. % in terms of the fundamental fields. %, see e.g. \cite{} for an example. 

%sometimes one may want to change the expression of the preferred operators in terms of fundamental fields; an example is given in non-commutative $\N=4$ SYM, where Tr$[\Phi^k]$ is no longer a gauge invariant operator, so its expression needs to be modified \cite{}. 

There are two main approaches for computing correlation functions in the $T\bar T$/$J\bar T$ literature: using path integrals methods, and studying the flow of operators in the Hilbert space. 

\bigskip

\noindent \emph{Conformal perturbation theory}

\medskip

\noindent In the path integral approach, the change in correlation functions is given by integrating the correlator with an additional insertion of the deforming operator \cite{Cardy:2019qao}
 \be
 \d_\mu \langle \O(x_1) \ldots  \O(x_n) \rangle = \int d^2 x \langle \O(x_1) \ldots  \O_{J^A\wedge J^B}(x) \ldots \O(x_n) \rangle_\mu  \label{basiccpt}
 \ee
where  we  used the definition of the path integral with fixed expressions for the operators.  We consider a generic SZ operator $J^A \wedge J^B$ built from two currents. The latter may even be two $U(1)$ currents, which preserve the 
CFT symmetries and locality, but may be used to test our various proposals. 

 To fully define the right-hand-side, one needs a prescription for how to deal with the singularities that arise in integration. %\textcolor{red}{\emph{In $J\bar J$, the expression for the vertex operator changes - is that also seen in CPT, with mixing due to the divergences?}} 
 For CFTs, this has been discussed in  e.g. \cite{Ranganathan:1992nb,Ranganathan:1993vj}. %, and this is also the method used in the $T\bar T/J\bar T$ case.
Since SZ operators are built from conserved currents, whose OPEs with general operators are fixed, at the CFT point, by the corresponding Ward identities, one obtains certain universal shifts in the correlators. These take the form of log corrections  that resemble an anomalous dimension proportional to $\mu p^2$ in $T\bar T$ \cite{Cardy:2019qao}, and  of shifts in the $SL(2,\mathbb{R})_L$ conformal dimensions in $J\bar T$ that precisely reproduce \eqref{jtbdims} - \eqref{chshift} to leading order in $\l$ \cite{Guica:2019vnb}.

 Of course, the more interesting question is whether one can define and compute the correlation functions in the deformed theory at finite $\mu$. Starting from the path integral definition \eqref{basiccpt}, Cardy \cite{Cardy:2019qao} used the special fact that a Smirnov-Zamolodchikov operator can be written as a total derivative to reduce its action to  one that acts  on each operator insertion
 \be
\d_\mu \langle \O(x_1) \ldots  \O(x_n) \rangle = \sum_i \langle \O(x_1) \ldots  \d_\mu \O(x_i) \ldots \O(x_n) \rangle
 \ee
allowing one to relate the correlation function of the above operators in the deformed theory to the correlator of operators in the undeformed CFT\footnote{These usually non-local operators are built from members of the same conformal family as the operator under study. }. Note that, even though the operators are labeled by a point, their correlation function is not local. To concretely evaluate the correlator, one needs to regulate and control the UV divergences that appear in the computation. If the underlying theories are non-local, such as $T\bar T$, the structure of divergences is, too, and one generally needs mode-dependent renormalisation. Since the form of the allowed  counterterms is then not known in  general, 
 this procedure tends to introduce ambiguities in the final result. %\textcolor{red}{\emph{Correct?}} \textcolor{blue}{\emph{Also talk about non-locality despite $x_i$ labels.}}

 % It is important to note that whenever one deals with $T\bar T$/ $J\bar T$ at finite deformation, one is led to work in momentum space, because the theory is non-local (but translation invariant). 

\medskip

\noindent \emph{Using the path-integral definition of the deformation}

\medskip

\noindent Another path integral-based approach that has been developed specifically for $T\bar T$ - deformed CFTs uses their path integral description in terms of coupling to topological gravity \cite{Aharony:2023dod}. In this formulation of the theory, the authors propose a natural set of operators, which are just local operators on the flat target space, treated as deformations of the original  CFT ones.  Correlation functions of these operators can then be computed by performing the path integral over the fields appearing in \eqref{ttbpathintdef}, with the corresponding additional insertions.  An advantage of this method is that  the calculations are performed directly at finite $\mu$. %, and there is a natural definition of operators. \textcolor{red}{\emph{Redundant!}}
 However, there are again UV divergences that need to be dealt with, and the classical definition of the operators does not fix the renormalisation.  It is interesting to ask whether this definition of the operators is equivalent to the previous one, given that  SZ and  path integral definitions of the $T\bar T$ deformation should be equivalent.

% Now path integral - which is over a certain, different set of fields than above -  performed at finite $\mu$. Same issues with renormalisation. 

\bigskip

\noindent \emph{Operator flow in the Hilbert space}

\medskip

\noindent In the Hilbert space approach, the flow of the energy eigenstates of the theory on the cylinder is captured  by the $\hat{\X}$ operator  in \eqref{flowstates}. In principle, one can use it to
 define operators that are covariantly constant along the flow - termed `dressed' operators in  \cite{Kruthoff:2020hsi} - as

%One can then formally define a set of  as solutions to the flow equation

\be
\mathcal{D}_\l \widetilde{\O}(w,\bar w) =0 \;, \;\;\;\;\;\left. \widetilde{\O}(w,\bar w) \right|_{\l=0} = \O_{CFT} (w,\bar w) \label{dressop}
\ee
The main advantage of this approach is that, at least na\"{i}vely, the issue of UV divergences may be avoided by viewing the above as the flow of some big matrices that act on the Hilbert space. Since the deformation simply acts by conjugation on the initial CFT operators by the same unitary that mixes the energy eigenstates, we find  %\textcolor{red}{\emph{Notation!}}

\be
\langle \widetilde{\O}(w_1, \bar w_1) \ldots  \widetilde{\O}(w_n, \bar w_n) \rangle_\l = \langle \O(w_1, \bar w_1) \ldots  \O(w_n, \bar w_n) \rangle_0 
\ee
where the former correlator is computed in the deformed CFT vacuum\footnote{For well-definiteness, we consider the correlators on the cylinder. We could also consider them on the plane, viewed as the infinite-radius limit of the cylinder. It is not clear whether considering the flow of the states directly on the plane  without this limiting procedure is well-defined, as this is not a discrete quantum-mechanical system. %\textcolor{red}{\emph{Correct?}}
}. Note that, even in the case of exactly marginal (SZ) deformations where the theory stays a CFT, the operator $\widetilde{\O}(w,\bar w)$ is not a local operator, despite the fact that its correlation functions are identical with those in the undeformed CFT\footnote{For an exemplification of this fact, see the appendix of \cite{Guica:2021fkv}.}. In fact, we have already seen this phenomenon in $J\bar T$ - deformed CFTs, where the parallel transport of the left-moving CFT currents along the flow are not the quasi-local left-moving currents in the deformed theory, but certain non-local shifts thereof  \eqref{flowlmcjtb}.  Therefore, the challenge in this approach is to find how to  define a meaningful basis of operators. As we show in \ref{defprimjtb}, in $J\bar T$ - deformed CFTs it is possible to use the $\widetilde{\O}(w,\bar w)$ as auxiliary constructs in building the operators of interest, which may be fixed by requiring they satisfy appropriate Ward identities. The main advantage of this method is, again, that one can avoid dealing with  UV divergences, at least na\"{i}vely. %, if implementing it with care.  

%) wrt the flow of the energy eigenstates. 
%Hidden problem of UV divergences. 
% In practice, one requires that the resulting object obey the exptected Ward identities wrt the symmetry generators. It is used for $J\bar T$ in xxx and xxx. 
% \bi
% \item can one show equivalence of the two methods in CFTs? How about $J_1 \wedge J_2$?
% \item in CPT, what's the difference in the connnection for operators vs energy eigenstates?
% \ei 

\bigskip

\noindent \emph{Using worldsheet string theory}

\medskip

\noindent Last but not least, there is an indirect method for computing correlation functions in $T\bar T$ and $J\bar T$ - deformed CFTs, which  corresponds to evaluating  correlation functions of vertex operator in a certain string-theoretical background. The relation between string theory in this background and  the deformations of interest is somewhat subtle, and will be discussed in  section \ref{holostr}. Using the fact  that worldsheet string theory in this background  is related via a non-local coordinate redefinition to string theory in AdS$_3$, one can compute \cite{Azeyanagi:2012zd,Cui:2023jrb} the deformed  correlators while avoiding  the issue of UV divergences. String theory also provides a natural normalisation of the operators.

\subsection{Expression for correlation functions in $T\bar T$ - deformed CFTs\label{ttbcorrsec}}

As discussed, in a non-local theory such as $T\bar T$, it is natural to work in a basis of momentum-space operators. One would like to know how the correlation functions in $T\bar T$ - deformed CFTs at finite $\mu$ depend on the momentum, in particular in the UV limit, in which the theory is expected to behave fully non-locally. The most general analysis to date is that performed in \cite{Aharony:2023dod}, who used the path integral definition of the $T\bar T$ deformation and the choice of operators mentioned above to compute arbitrary correlation functions in the deformed CFT. The answer they find for the momentum-space operators is remarkably simple 
\be
\langle \O_1(p_1) \ldots \O_n(p_n) \rangle \propto \int \prod_{i=1}^n d^2 x_i \, C\, e^{i \sum_i p_i \cdot x^i}  (\Lambda |x_{ij}|)^{2 \mu p_i \cdot p_j}\langle \O_1(x_1) \ldots \O_n(x_n) \rangle_{CFT} \label{aharonybarel}
\ee
being entirely determined by the original CFT correlators via a universal transform. %  resembles a Fourier transform. %, but with a  kernel that softens the UV divergences as operators approach each other. \textcolor{red}{\emph{Correct? Check Cardy?}} 
Above,  $\Lambda$ is a UV cutoff, and we have changed the notation to fit our conventions. $C$ represents some factors that may depend on the momentum, but which are subleading when the latter is large. %\textcolor{red}{\emph{What do they depend on?}} 
 Note that translation invariance of the integrand ensures momentum conservation, as usual. 

In order to not have to deal with the function $C$, \cite{Aharony:2023dod} only estimate the leading UV behaviour of their correlators, via a saddle point approximation at large $p_i$, to which $C$ does not contribute. Before discussing their result, it is useful to remind the reader  the form of the Euclidean momentum-space two-point function for a scalar operator of dimension $\D$ in a $2d$ CFT\footnote{$\mathcal{G}_E(p)$ is related to the two-point function via $\langle \O(p) \O(q)\rangle_E = (2\pi)^2 \d^{(2)} (p+q) \mathcal{G}_E(p)$.}

\be
  \mathcal{G}_E(p)  =  \frac{\pi \Gamma(1-\D)}{2^{2\D-1} \Gamma(\D)} (p^2)^{\D-1} =  \frac{\pi^2}{2^{2\D-1} \Gamma(\D)^2 \sin (\pi \D)} (p^2)^{\D-1} \label{euclcft2pf}
 \ee
Interestingly, a computation \cite{Cui:2023jrb} of the two-point function of  long string vertex operators with unit winding in a particular string background, discussed in the next section, which is conjectured to capture observables in single-trace $T\bar T$ - deformed CFTs, yields a result that is precisely the above momentum-space CFT two-point function, but with $\D$ replaced by $\D +\mu p^2$.  Or, this is precisely what  \eqref{aharonybarel} yields if $C=1$, since the prefactor $|x_{12}|^{-2\mu p^2}$ simply shifts the exponent of the CFT two-point function, $|x_{12}|^{-2\D}$. %\textcolor{blue}{Or, this is very similar to the behaviour we will be seeing for $J\bar T$ in the next subsection.} 
 It would be  nice if, for example, a symmetry principle could be invoked to arrive at this value of $C$.

 One of the main differences between the two methods  for computing  the correlators  is that in   \cite{Aharony:2023dod}, the operators need to be renormalised. Given that the theory is non-local, this must be done in a mode-dependent fashion,  with associated momentum-dependent ambiguities. By contrast, in the  computation of  \cite{Cui:2023jrb}, string theory provides a natural normalisation of the  operators and the computation can be performed so that  no  UV divergences are encountered, %, as one is effectively performing a spectral flow. \textcolor{red}{\emph{Check!}} On the other hand, in
resulting in an expression that can be trusted at all momenta. Thus, a very interesting and important question in $T\bar T$ - deformed CFTs is to understand the structure of divergences and the allowed counterterms. 
Given the extremely special properties these theories have, one may hope a set of rigid rules exists, possibly based on a yet-to-be discovered symmetry principle, that would lead to a universal, unambiguous expression for the correlators. 

Returning to our previous discussion, due to the  momentum-dependent ambiguities in the renormalisation of the operators mentioned above, \cite{Aharony:2023dod} only estimate the high-momentum behaviour of the two-point function, finding agreement to the large $p$ behaviour of  \eqref{euclcft2pf} with $\D \r \D +\mu p^2$. The correlator is exponentially damped at high momentum, $\mathcal{G}_E \sim (p^2)^{-\mu p^2}$, a behaviour for which it is essential to take into account the behaviour of the $\Gamma$ functions in the denominator. A similar behaviour was  obtained previously in \cite{Cardy:2019qao} by resumming the leading logs in the correlation function, computed using conformal perturbation theory. However,  this resummation  ommitted the $\Gamma$ function contributions, thus obtaining  an exponential growth at high momentum, rather than exponential damping. See \cite{Aharony:2023dod} for further discussion.

Regarding the special nature of the $T\bar T$ - deformed operators, several recent works \cite{Hirano:2024eab,Chen:2025jzb} arrive at the conclusion  that they  can be understood as operators in the original one, evaluated at field-dependent values of the coordinates.
This interpretation follows naturally from the work of \cite{Cardy:2019qao} and   is compatible, at least in spirit,  with the proposal of  \cite{Aharony:2023dod} for defining the operators.  %It is also naturally implemented via BRST methods \cite{grassi}. 
It appears, though, that a full quantum-field-theoretic understanding of the operators is still required, in addition to resolving the 
%
% \textcolor{red}{\emph{Check! Too classical issues?}} Unfortunately, these works are still subject to 
 ambiguities related to their renormalisation. %, but one hopes the issue will be soon under control. 
Assuming these issues will soon be under control, we note that correlation functions in $T\bar T$ - deformed CFTs    follow the pattern obeyed by  all $T\bar T$ observables as simple, universal, yet non-local deformations of the original CFT ones.  %\textcolor{blue}{Mention also Gandalf!}%Particularising to the two-point function of scalar operators of dimension $\D$, the result is  \textcolor{red}{\emph{Discussion, looks weird...}}

\subsection{Definition and correlators of primary analogues in $J\bar T$ - deformed CFTs \label{defprimjtb}}

As emphasized in the previous section, one of the difficulties in computing correlation functions in non-local theories such as $T\bar T$ and $J\bar T$ - deformed CFTs is the lack of precise knowledge of how to renormalise the operators. In the case of $J\bar T$ - deformed CFTs, \cite{Guica:2021fkv} succeeded in avoiding this issue by using the fact the deformation can be phrased in terms of a certain `spectral flow by the right-moving Hamiltonian', which allows one to define operators in the deformed theory and compute their correlation functions via purely Hilbert space flow methods.

As a warm-up, it is instructive to explain how this method works for the case of exactly marginal $J_1 \wedge J_2$ SZ deformations, for which the deformed theory is still a CFT and where the deformed vertex operators can be explicitly constructed. 

\subsubsection{$J_1 \wedge J_2$ warmup}

Since  the $J_1 \wedge J_2$  deformation is a SZ deformation, one can readily sole for the deformed finite-size energies and deformed charges. Let us introduce the (integer-quantized) shift and winding charges for the two currents
\be
n^a = \int d\s J_t^a \;, \;\;\;\;\;\;\; w^a = \int d\s \tilde J_t^a \;,\;\;\;\;\; a=1,2
\ee
where $\tilde J^a$ is simply the topological current  \eqref{topcur} associated with the respective boson.  Since  the theory stays a CFT, the energies of states on the cylinder can be mapped to operator dimensions on the plane

\be
h_{L,R} = h^{[0]}_{L,R} \pm \l \e_{ab} q^{a\,[0]}_{L,R} n^b + \frac{\l^2 k}{4} n_a n^a \;, \;\;\;\;\; q^a_{L,R} = q^{a\, [0]}_{L,R} \pm \frac{\l k}{2} \e^{a b} n_b \label{jjbdefdimsch}
\ee
where $ q^{a\,[0]}_{L,R} = (n^a\pm w^a)/2$ are the undeformed chiral charges, and $q^a$ the deformed ones. The above formulae take precisely the form of a spectral flow by $\pm \l \e_{ab} n^b$. 

It is also straightforward to work out the spectral flow operator $\hat{\X}$ and use it to build the flowed currents, with the result %\textcolor{red}{\emph{Rather write $L_n$?}}
\be
\widetilde \H_L = \H_L - \l \e_{ab} \J_L^a \Pi^b + \frac{\l^2 \hat k}{4} \Pi^b\Pi_b \;, \;\;\;\;\;\;\;\widetilde{\J}_L^a = \J_L^a- \frac{\l \hat k}{2} \e^{ab} \Pi_b\ee
and similarly for the right-movers, where $\Pi_a$ is the corresponding $U(1)$ charge operator (whose eigenvalue is $n_a$).  Note that, again, these current operators are non-local, due to the explicit presence of the charge operator, which is an integral over all space. 

One would now like to compute correlation functions of operators in the theory, whose dimensions and charges are given by \eqref{jjbdefdimsch}.  This can be done using conformal perturbation theory \cite{Cardy:2019qao}, but the procedure introduces UV divergences. Alternatively \cite{Guica:2021fkv}, one can start  from the dressed operators \eqref{dressop} and then `correct' them until one reaches the local operators of interest. Concretely, this can be achieved by requiring that the corrected operators obey the expected primary Ward identities with respect to the true conformal symmetry generators in the deformed theory; note that 
the flowed operators obey, by construction, the corresponding Ward identities with respect to the flowed Virasoro generators, which differ from the former by a spectral flow by $\l \e_{ab} \Pi^b$. It is not hard to show that the resulting relation between the two sets of operators in the deformed theory  on the cylinder is
\be
\O(w,\bar w) =  \e^{ A_{\mbox{\tiny{$\O$}}} w + B_{\mbox{\tiny{$\O$}}} \bar w} e^{\eta_\O^a \sum_{n=1}^\infty \frac{1}{n} e^{n w} \tilde J^a_{-n} + \bar \eta^a_\O \sum_{n=1}^\infty \frac{1}{n}  e^{n \bar w} \tilde{\bar J}^a_{-n}} \tilde \O(w,\bar w) \, e^{-\eta_\O^a \sum_{n=1}^\infty \frac{1}{n} e^{- n w} \tilde J^a_n - \bar \eta_\O^a \sum_{n=1}^\infty \frac{1}{n} e^{-n \bar w} \tilde{\bar J}^a_n} \label{reloot}
\ee
where $\eta^a_\O = \l \e^{ab} n_b = - \bar \eta^a_\O$, where $n_a$ is the charge of the operator, and the expressions for $A_\O, B_\O$ can be found in \cite{Guica:2021fkv}.  This relation can also be easily derived for explicit vertex operators, where it simply implements the shift of the charge in the exponent of the  vertex operator.  Correlation functions can be   computed by sandwiching several such operators between the vacuum, and then commuting through the positive $J_n$ modes to the right, and the negative ones to the left, where they annihilate the respective in/out vacuum, with the result
\be
 \langle \O_1(w_1) \ldots \O_n(w_n) \rangle =  \prod_{i<j} \left(e^{\frac{\pi w_{ij}}{R}}-e^{-\frac{\pi w_{ij}}{R}}\right)^{\frac{2}{k} (q_i q_j -  q_i^{[0]}  q_j^{[0]})} \!\! \left(e^{\frac{\pi\bar w_{ij}}{R}}-e^{-\frac{\pi \bar w_{ij}}{R}}\right)^{\frac{2}{k} (\bar q_i \bar q_j -  \bar q_i^{[0]} \bar q_j^{[0]})}  \!\! \langle \tilde \O_1(w_1)\ldots \tilde \O_n(w_n) \rangle \label{4pfjjb}
\ee
The correlation function of dressed operators equals, by construction, the undeformed CFT correlator. 
Thus, the correlation functions of primary operators in the deformed theory can be rather simply expressed in terms of the undeformed correlators; the reason for this simplicity is the fact that the deformed and undeformed theories are related via a kind of spectral flow.  Using the fact that  Virasoro-Kac-Moody four-point blocks can be written in terms of a spectral-flow invariant block and an affine $U(1)$ one \cite{Song:2017czq,bombini}, the change in the correlation function is entirely encoded into a change in the $U(1)$ block, which only depends on the charges of the external operators. Note this implies the CFT OPE coefficients are unchanged by the deformation; using the above-mentioned property of Virasoro-Kac-Moody blocks, all bootstrap relations can be shown to be trivially satisfied in the deformed CFT, if they hold in the original one. 

\subsubsection{$J\bar T$ - deformed CFTs}

One would now like to apply the same idea to the case of $J\bar T$ - deformed CFTs. As discussed, the relation between flowed and quasi-local symmetry generators is given by \eqref{flvsqlquantum}. One again defines dressed/flowed operators using \eqref{dressop}, which by construction  satisfy CFT-like Ward identities with respect to the flowed generators $\widetilde{L}_n$, etc. Note that these operators are not local, despite the appearance of $w, \bar w$, which can be considered simply labels.  One then tries to modify them in such a way that the new operators are primary, to the largest extent possible, with respect to the quasi-local generators $L_n, \bar L_n, K_n, \bar K_n$. 
 On the left-moving side, where the relation between the symmetry generators is almost  a standard spectral flow, one  requires that the new operators satisfy the standard Ward identities for an operator  with left-moving dimension \eqref{jtbdims} exactly, which reduces to the previous calculation for $J_1 \wedge J_2$ deformations. On the right-moving side, it is not clear what commutation relation with the  quasilocal generators $\bar L_n, \bar K_n$ one should  impose. One thus resorts to simply guessing an appropriate right-moving factor, which corresponds to the same spectral flow transformation as for the left-movers %work out its properties, and then justify our choice \emph{a posteriori} via its rather reasonable predictions in the $R \r \infty$ limit.  

\be 
\O(p,\bar p) = \int d w d \bar w \, e^{-p w- \bar p \bar w } \e^{A_\O w + B_\O \bar w} e^{ \eta_\O \sum_{n=1}^\infty \frac{1}{n} (e^{n w} \tilde J_{-n} +e^{n \bar w} \tilde {\bar J}_{-n})} \tilde \O(w,\bar w) e^{-\eta_\O \sum_{n=1}^\infty \frac{1}{n} ( e^{-n w} \tilde J_n  + e^{- n \bar w} \tilde{\bar J}_n )}\label{otry}
\ee
with the spectral flow parameter $\eta_\O = \l \bar p$. 
The commutators of $\O(p, \bar p)$ with the quasilocal  generators
$\bar L_n$, 
  resemble a modified conformal Ward identity for an operator of right dimension  $\bar h_i$

\be
 h_i (\bar p_i) =  h_i^{[0]} + \l  q_i^{[0]} \bar p_i +\frac{\l^2 k}{4} \bar p_i^2 \;, \;\;\;\;\;\;\;  \bar h_i (\bar p_i) = \bar h_i^{[0]} + \l  {\bar q}_i^{[0]} \bar p_i + \frac{k \l^2}{4} \bar p_i^2 \label{momdephhb}
\ee
where we also recalled the left conformal dimension, for completeness. The commutators with the affine $U(1)$ generators $K_n, \bar K_n$ resemble those of operators with charges 

\be
q_i (\bar p_i) = q_i^{[0]} + \frac{\l \bar p_i k}{2} \;, \;\;\;\;\;\;\; \bar q_i (\bar p_i) =\bar  q_i^{[0]} + \frac{\l \bar p_i k}{2} 
\ee
Correlation functions are computed, as in the $J_1 \wedge J_2$ case, by sandwiching the operators in the deformed vacuum and commuting through the current modes. This computation is performed on the cylinder; however the end result   simplifies in the $R \r \infty$ limit. %\textcolor{red}{\emph{What were the subtle issues here?}} 
One again obtains a series of universal prefactors times the original CFT correlation function, now evaluated on the plane %. \textcolor{red}{\emph{Check formula!}}

\be
 \langle \O_1(p_1) \ldots \O_n(p_n) \rangle = \int \prod_i d^2w_i \, e^{i p^i w_i} \prod_{i<j} w_{ij}^{\frac{2}{k} (q_i q_j -  q_i^{[0]}  q_j^{[0]})} \bar w_{ij}^{\frac{2}{k} (\bar q_i \bar q_j -  \bar q_i^{[0]} \bar  q_j^{[0]})}  \langle \tilde \O_1(w_1)\ldots \tilde \O_n(w_n) \rangle 
\ee
which directly follows from  the $R\r \infty$ limit of an expression very similar to \eqref{4pfjjb}.  One can easily check that the momentum-space 
  two and three-point functions of the operators \eqref{otry} are precisely given %(up to some trivial shifts in the arguments)
   by  a momentum-space CFT two/three-point function, but with the conformal dimensions  replaced by their momentum-dependent counterparts\footnote{
Note that when writing this expression, one should replace the dimensions by their momentum-dependent counterparts not only in the functional part of the correlator, but also in the prefactors, which  contain factors of e.g. $\Gamma(h)$.  This resonates with the behaviour of the   correlation functions discussed in the single-trace version of $T\bar T$-deformed CFTs. %, which are computed    using worldsheet string theory \cite{Asrat:2017tzd}.
} \eqref{momdephhb}.
  Remarkably, this is exactly the same behaviour that we observed  in  \eqref{bh2pf} for scattering amplitudes  off near-extremal black holes. 

%The same is true for three-point functions; note that the appropriately normalised OPE coefficients stay fixed.  \textcolor{red}{Spell out disagreement with three-point functions. \emph{Or, perhaps they agree?}}

It is of course interesting to understand the operators directly in position space, as well as how the symmetry generators act on them; \cite{Chen:2025umo} presents work in this direction, though most of their analysis is classical and the precise relation to \cite{Guica:2021fkv}, whose framework the authors use, is not explained. In particular, their proposed preferred operator has a slightly different right-moving conformal dimension than \eqref{momdephhb}. %\textcolor{blue}{One obtains a position space interpretation of our operator, in terms of the original CFT operators evaluated at a field-dependent value of the coordinate (the momentum-space expression we have makes this precise), plus a little extra that corresponds exactly to the $H_R$ - dependent large affine transformation, except that one should have different coefficients $q$ and $\bar q$. Note this is an \emph{alternative} to the spectral flow interpretation (the parameter of the spectral flow tells us whether there is a charge shift or a coordinate shift).  \textcolor{red}{\emph{Correct?}} We thus find some complication for charged operators with $\tilde q \neq \tilde{\bar q}$, but everything should be straightforward in the case of the currents. We concentrate on the free boson for simplicity. }

The above analysis of correlation functions can be extended to single-trace $J\bar T$ - deformed CFTs, where one uses the spectral flow, now on a twisted sector-by-sector basis \cite{Chakraborty:2023wel}. The resulting momentum-space correlators take the form of those in a symmetric product orbifold CFT, but with dimensions shifted as in the single-trace analogue of \eqref{momdephhb}.

%\bi
%\item mention flow single-trace $J\bar T$
%\item comment on discrepancies with Wei's results in field theory 2511.xxxx
%\item the Ward identities are only partly understood - does Wei et al study them?
%\item how does the action of conformal transformations on $\O_{CFT}$ translate on $\O$?
%\item can we use this to derive the momentum-dependent shift in the conformal dimensions?
%\item can on use this to propose right-moving Ward identities?
%\item include discussion of the mode expansion and action of the symmetries on the modes?
%\item is there a simple proposal for the operators e.g. in $J\bar T$ - deformed free boson?
%\item is there a simple form/solution of the position space Ward identity (where $h$ $\r$ operator)? 
%\ei
%

\section{Non-AdS holography and single-trace $T\bar T$/$J\bar T$  \label{holostr}}

In  section \ref{holointdtr}, we discussed the holographic dictionary for the  usual, \emph{double-trace} $T\bar T$ and $J\bar T$ deformations, whose main effect was to change the asymptotic boundary conditions for the non-dynamical metric (and gauge fields). The new boundary data are simply a reinterpretation of the usual AdS/CFT ones, which explains the universality of the deformed spectrum from a holographic perspective.  The price to pay for this universality is that  the holographic dictionary is slightly unexciting,  at least within the context  of the low-energy gravity approximation, since the geometry remains locally AdS$_3$.

If we would like the geometry to be genuinely non-AdS, we need the irrelevant deformation to be `single-trace', so as to correspond to a new non-normalizable perturbation in the bulk. A natural candidate is the single-trace $T\bar T/J\bar T$ deformation, whose properties were discussed in section \ref{strttbjtbsec}. As we already explained, to define this deformation one needs that the seed CFT take the form of a  symmetric product orbifold.

In this section, we discuss certain string-theoretical backgrounds, whose holographic dual has been proposed to be related to single-trace $T\bar T$ or $J\bar T$ - deformed CFTs. After a very brief review of the D1-D5 CFT and its holographically dual spacetime, we discuss the relation between the decoupled NS5-F1 background and single-trace $T\bar T$, and between a warped AdS$_3$ background supported by purely NS flux and single-trace $J\bar T$. As we will try to explain, this relation is somewhat subtle, and only indicates the dual is in the same `universality class'  as single-trace $T\bar T$ or $J\bar T$. Finally, we discuss the relation between more general backgrounds with the same asymptotics, and the above putative universality classes.

 \etocsetnexttocdepth{5}
    \etocsettocstyle{\subsubsection*{Contents of this section: }}{}
    \cftsubsubsecindent 34pt
    \localtableofcontents

\subsection{The D1-D5 setup}

\begin{figure}[h]
\hspace{8mm}
\begin{minipage}{0.44\linewidth}
%\flushright
\vspace{3mm}
\noindent \emph{Schematic picture of the D1-D5 moduli space.} \vskip2mm
The full  CFT becomes a SPO $\mathcal{M}^N/S_N$ at the symmetric orbifold point, conjecturally dual to the $k=1$ string background.   There are also singular points for $k>1$ where one has an effective CFT description of the perturbative string sector by an SPO of the form $\mathcal{M}'^p/S_p$, deformed by a twist two operator. 
\end{minipage}
\hspace{5mm}
\begin{minipage}{0.45\linewidth}
\flushright
\includegraphics[height=5.5cm]{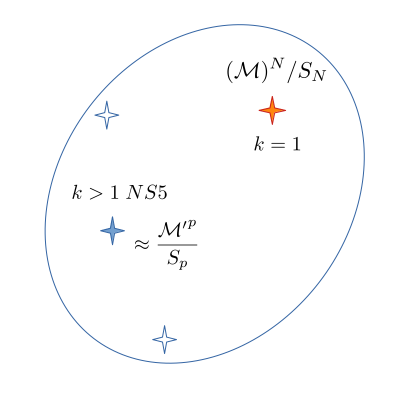}
\caption{The D1-D5 moduli space}
\label{potato}
\end{minipage}
\end{figure}

We will henceforth concentrate on the D1-D5 system - or its NS5-F1 presentation -  which provides the primary string-theoretical realisation of the AdS$_3$/CFT$_2$ correspondence \cite{Maldacena:1997re,Aharony:1999ti,David:2002wn}. The D1 - D5 CFT is obtained in the infrared limit of the theory living on a stack of $p$ D1 and $k$ D5 branes compactified on  K3 or $T^4$ (or their S-dual F1 and NS5 counterparts). It enjoys two-dimensional  $(4,4)$ superconformal symmetry and possesses  a large moduli space (84 - and, respectively, 20 - dimensional). %It is widely believed that \emph{Status?} that there exists 
Brane configurations with the same $N \equiv k p$ are believed to be part of the same moduli space, as they are  related to the $k=1, p=N$ background via U-dualities \cite{Larsen:1999uk}. %\textcolor{red}{\emph{Any issue with non-mutually prime?}}
 Within this  moduli space, one expects there is a point where the CFT takes the form of a symmetric product orbifold  $\mathcal{M}^N/S_N$, where $\mathcal{M}$ is the sigma model on K3 or $T^4$  and $N = kp$,  see figure \ref{potato}. The bulk dual of the CFT at this point is  highly stringy, and has been argued to correspond to the $AdS_3 \times S^3\times M_4$ background supported by a single unit ($k=1$) of magnetic NS flux\footnote{More precisely, this string background is conjectured to be holographically dual   to a grand canonical ensemble of SPO CFTs with fixed chemical potential for the F1 charge \cite{Kim:2015gak,Aharony:2024fid}.}, as supported by the  match of the spectrum and     correlation functions at this point  to those in the free symmetric orbifold CFT \cite{Eberhardt:2018ouy,Eberhardt:2019ywk}.  Irrelevant deformations of the CFT at this point in moduli space are also  holographically dual to highly stringy spacetimes, but are nonetheless also accessible via worldsheet techniques \cite{Dei:2024sct}. 

\medskip

\noindent We will, however, be primarily interested in supergravity solutions, where the curvature radius of the bulk spacetime  is large compared to the string length. For purely NS backgrounds, this can be ensured by considering a large number ($k\gg 1$) of  NS5 branes. By considering an even larger number of F1 strings, the string coupling can be made  everywhere small (and thus string perturbation theory is valid). 

This system can be analysed using worldsheet techniques \cite{
Giveon:1998ns,Kutasov:1999xu,deBoer:1998gyt,Maldacena:2000hw}.  %\textcolor{red}{\emph{More concrete details?}} 
It is found to contain two types of excitations: short strings, localised near the centre of AdS  and long ones, which have a continuous spectrum and live mostly near the AdS boundary. The spectrum of long strings is approximately described by that of a symmetric product orbifold (SPO), of the form  \cite{Argurio:2000tb}
\be
\frac{(\mathcal{M'}_{6k})^p}{S_p} \label{spod1d5}
\ee
where $\mathcal{M'}_{6k}$ is a (non-compact) CFT of central charge $6k$. The SPO structure is however broken by a marginal deformation involving a twist $2$ operator \cite{Eberhardt:2021vsx}. This deformation is  non-normalisable;  despite this, one can  still formally  match correlation functions in this theory to those derived from the worldsheet CFT, see e.g. \cite{Knighton:2024pqh} for a discussion. %\textcolor{red}{\emph{Also mention short strings.}}
 The breaking of the symmetric orbifold structure can also be understood from the fact that short strings, whose spectrum is clearly not governed by a SPO,  appear as poles in the scattering of long ones \cite{Aharony:2004xn}.

 It is important to note that the deformed SPO is not expected to be the full holographic dual to quantum gravity in AdS$_3$ supported by  $k>1$ units of NS flux; rather, it is only expected to capture the perturbative 
 (long) string subsector of the full dual \cite{Chakraborty:2025nlb}. %\emph{Is it supposed to somehow also describe short strings in some non-perturbative fashion?} 
This description of the bulk is only expected to exist at singular   loci in moduli space, where the underlying branes can separate \cite{Seiberg:1999xz}. These 
 %
% It is sometimes referred to as a `singular CFT'. It posseses a continuous spectrum - the explicit realisation of the $\mathcal{M}_{6k}$ factor being non-compact - and there are missing chiral primary states from its spectrum. The associated singular 
 are clearly different from the SPO point we discussed previously. 
 
 %; the resulting CFT itself is not an SPO\footnote{There are proposals in the literature that  the CFT could insted be given by a deformation of an SPO of a non-compact theory of the form linear dilaton $\times X$ by a twist two operator; while correlation functions on `both' sides formally match, the deformation is non-normalisable for $k>1$ and it's not clear whether  this describes the full theory, or just the perturbative string subsector. \emph{Understand!}}. 
 
 This is the main message of this subsection:  for pure NS AdS$_3$ backgrounds with $k>1$, the holographically dual CFT is a somewhat singular theory that is \emph{not} a symmetric product orbifold. Rather, the relation between this CFT and a SPO is that the former  contains a subsector - capturing perturbative long string states in the dual bulk, but not black hole microstates \cite{Giveon:2005mi} -  that is only  approximately a SPO. The lack of an SPO structure is evident at the level of the short string perturbative bulk spectrum, but can also be seen in higher correlators of long strings.  If one would like the dual CFT to be an exact SPO, then one should consider the AdS$_3$ background with a single unit of NS5 brane flux ($k=1$). 
 
 % not an SPO (only approximately so) and the proposals in the literature are only supposed to . \emph{Recheck!}
 
 The above string-theoretical setting, usually  with $k>1$, is the  starting point 
for what is called a \emph{single-trace} variant of the $T\bar T$ \cite{Giveon:2017nie}  and $J\bar T$ deformations  \cite{Apolo:2018qpq,Chakraborty:2018vja} in holography. These  correspond to non-normalizable deformations of this  AdS$_3$ background supported by pure NS flux that lead to different asymptotics. 
It should be clear from what we said so far that the holographic duals cannot be \emph{exactly} the single-trace $T\bar T$ or $J\bar T$ - deformed CFTs that we  analysed  from a purely field-theoretical viewpoint in section \ref{strttbjtbsec}, due to the broken symmetric product orbifold structure; nonetheless, the dual theories still share many features of these deformations, as we now  explain.  %\textcolor{red}{\emph{Write better!}}

%At first, the construction of \cite{Giveon:2017nie}  appeared to provide a \emph{tractable} example of { non-AdS} holography. While the situation is a bit more nuanced, it is still very interesting because: \emph{Why???} i)  it provides a rare example of \emph{tractable non-AdS} holography; ii) it shows that there exist deformations with very similar properties, particularly in what concerns the  UV behaviour, to $T\bar T$, which are less universal  and thus  more interesting for gaining insight into \emph{general} asymptotically fragile theories; iii) it provides a concrete description of a two-dimensional compactification of little string theory.  [While not tractable, clearly closely related to $T\bar T$.]

 %Then, we briefly sketch the construction of the single-trace analogue of the $T\bar T$ operator and list a number of checks and predictions. 

\subsection{The asymptotically linear dilaton background and single-trace $T\bar T$\label{aldttbsec}}

In this section, we start by reviewing the relevant string theory setup, which is the 
NS5-F1 system in the NS5 decoupling limit. After describing this limit and the effect of adding the F1  strings, we discuss the various (kinds of) relations that have been found between the corresponding decoupled  supergravity  background and single-trace $T\bar T$. 

\subsubsection{The  decoupling limit of  NS5 branes}

NS5 branes are solitonic objects in string theory that are magnetically charged under the $B$-field. In type IIB string theory, they are related to D5-branes via S-duality, while in type IIA they are related to M5-branes via uplift. It is interesting to ask whether there exists a limit in which the modes on the NS5 branes decouple from gravitational physics. Unlike for D-branes, where this limit takes the string length to zero ($\a'\r 0$), for NS5 branes the limit is, rather\footnote{This is related via S-duality, which sends $g_s \r \frac{1}{g_s}, \a' \r \a' g_s$ to the  standard D5 decoupling limit, $\a' \r 0$,  $\a' g_s$ fixed. } \cite{Seiberg:1997zk}
\be
g_s \r 0 \;,\;\;\;\;\;\; \a' \;\; fixed
\ee 
The worldbrane theory obtained in this limit is known as \emph{little string theory} (LST). It is a strongly-coupled theory of strings with tension $(\a' k)^{-1}$, which for $k\gg 1$ are  much lighter   than the fundamental string. It  is also not gravitational -  no masslesss spin two excitation is present in its spectrum - non-local and displays an intriguing form of UV/IR mixing \cite{Aharony:2004xn}. In particular, LST exhibits T-duality, since NS5-branes  turn into NS5 under  T-duality along their worldvolume, and $g_s=0$ is a fixed point of this transformation. Some of the properties of LST depend strongly on whether one is working in type IIA or type IIB string theory%\footnote{For example, in type IIB it it possible to see, using the S-duality between D5 and NS5 branes, that at low energies the $6d$ gauge coupling on NS5 branes is $g_6^2 = \a' = finite$.}
, see \cite{Kutasov:2001uf} for a review.  Also, given its strongly-coupled nature, 
many of its properties, such as the fact that it exhibits a Hagedorn growth of states at high energies, are  inferred from  the holographic dual, to which we now turn. %, as are most known properties of LSTs. % are in fact inferred from the holographic dual. 

The background that is holographically dual to LST \cite{Aharony:1998ub} is obtained  by taking the decoupling limit of    the backreacted solution for $k$ NS5-branes.  In string frame, the latter reads %\emph{Check!}

\be
ds^2 = dx^\mu dx_\mu + \left(1+ \frac{k \a'}{r^2}  \right) (dr^2 + r^2 d\Omega_3^2) \;, \;\;\;\;\; e^{2\Phi} = g_s^2\left(1 + \frac{k \a'}{r^2} \right) \;, \;\;\;\;\; H = \star_4 d\Phi
\ee
where $x^\mu \in \mathbb{R}^{1,5}$ span the directions along the brane and $r$ is the radial coordinate in the transverse $\mathbb{R}^4$.  In the decoupling limit,  the value of the dilaton at infinity, $g_s$, is sent to zero, with $\tilde r \equiv r/g_s$ fixed. %\emph{How to see low-energy limit?}
 The spacetime becomes 
\be
\mathbb{R}^{5,1} \times \mathbb{R}_\phi \times S^3 \;, \;\;\;\;\;\;\;\;\; \phi \equiv \sqrt{k \a'} \,  \ln \tilde r
\ee
with a linear dilaton, $\Phi = - \phi/\sqrt{k \a'}$. 
In the asymptotic region, the string coupling is small %(the theory is free as $r \r \infty$)
, so  string perturbation theory is valid. The worldsheet CFT is exactly solvable in this region. However, as we approach $r \r 0$, the dilaton grows without bound and the string description breaks down. % Note that an understanding of what happens to e.g. correlation functions require that string modes be allowed to interact in the stronger coupling region, where higher-genus worldsheets would also contribute. The description of the physics in the strongly coupled region depends sensitively on whether we are in type IIA or IIB.   
As shown in  \cite{Maldacena:1997cg}, one way to go around this problem is to consider the system at finite temperature.

%As mentioned above, it is easy to exhibit the Hagedorn growth of states using the holographic dual.
 If the NS5 branes are made near-extremal, the metric becomes%\footnote{From now on, $r$ is rescaled by a factor of $g_s$ with respect to its value at asymptotically flat infinity. }

\be
ds^2 = - f_{\mathcal{E}} dt^2 + ds^2_{\mathbb{R}^5} + k \a' \left( \frac{ dr^2}{r^2 f_{\mathcal{E}} } +  d\Omega_3^2 \right) \;, \;\;\;\;\;\; f_{\mathcal{E}}  = 1- \frac{r_0^2}{r^2} \label{haglst}
\ee
where we have dropped the tilde from the rescaled radial coordinate, and $r_0$ is the location of the horizon. It is assumed to be large enough so that the string coupling, $k \a'/r_0^2$, is still small there and the supergravity solution can be trusted. 
By analytically continuing to euclidean time,  the temperature associated to this black brane can be shown to be independent of $r_0$ and is given by $T^{-1}_H = 2 \pi \sqrt{k \a'}$, a factor of  $\sqrt{k}/2$ smaller than in usual string theory.  A temperature that is independent of the energy implies that $S \propto E$, i.e. Hagedorn behaviour. %[Notice that the Hagedorn temperature, so the strings of little string theory are  indeed diffrerent.]

\subsubsection{The interpolating AdS$_3 \r $ asymptotically linear dilaton (ALD) solution }

So far for the little strings. Let us now turn to the setup of \cite{Giveon:2017nie}. The first step is to do away with the strong coupling region at small $r$. One way to achieve this also in the vacuum is to add $p$ F1 strings to the system, which extend along one of the NS5 brane directions, denoted $\s \sim \s +R$, and compactify the remaining directions of the NS5 branes on $M_4 = T^4$ or K3. %\textcolor{red}{\emph{Notation!}} 
 The metric and dilaton then become %\emph{Check!}

\be
ds^2 = \frac{-dt^2 + d\s^2}{f_1} + k \a' \left(\frac{dr^2}{r^2} +  d \Omega_3^2\right) + ds^2_{T^4} \;, \;\;\;\;\;\; e^{2\Phi} = \frac{k \a'}{r^2 f_1} \;, \;\;\;\;\; f_1= 1+ \frac{r_1^2}{r^2}\;, \;\;\;\;\; r_1^2 = \frac{p \a'}{v} \label{f1ns5b}
\ee
where $v$ is the volume of the $T^4$ in units of $(2\pi)^4 \a'^2$. The solution  is also supported by $k$ units of $H_3$ - form flux.  Since the string coupling at $r=0$ is inversely proportional to  $p$ , for a very large number of F1 strings the coupling is small everywhere and we can trust string perturbation theory.  

The background \eqref{f1ns5b} interpolates between AdS$_3\times S^3 \times T^4$ in the IR $(r \r 0)$ and $\mathbb{R}_\phi \times \mathbb{R}_t \times S^1_{\s} \times S^3 \times T^4$ in the UV ($r \r \infty$). The radius of the AdS$_3$ in string units is $\sqrt{k}$, and so the supergravity approximation is valid when $k\gg 1$, which we will generally assume from now on. 

The holographic dual to the IR region is the D1-D5 CFT at a singular point in its moduli space, as discussed in the introduction, while 
the deep UV corresponds to a two-dimensional compactification of LST on K3 or  $T^4$, which is a non-gravitational UV-complete theory that is intrinsically non-local.  By expanding e.g. the dilaton in \eqref{f1ns5b} around the AdS$_3$  background in the IR, one can easily show that   infinitesimally away from AdS$_3$, the deformation corresponds to  turning on a non-normalizable mode, which corresponds to a source for an operator of dimension $(2,2)$. This deformation moreover preserves maximal (Poincar\'e) supersymmetry.  If the irrelevant deformation could somehow be solved, one would obtain a tractable description of compactified LST, which would be highly valuable. 

\vskip2mm

\begin{figure}[h] 
\centering
\includegraphics[height=5.5cm]{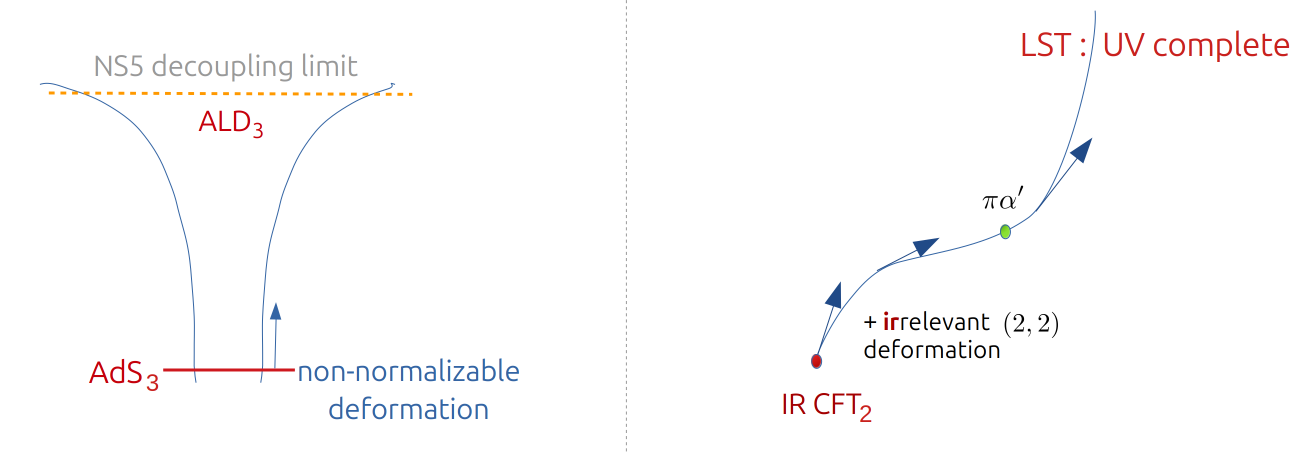}
\caption{\small{The decoupling limit that yields the AdS$_3 \r $ ALD$_3$ interpolating geometry, and its dual interpretation.}}
\end{figure}

\noindent The background \eqref{f1ns5b} has many nice features. One is that worldsheet string theory is known  for the full interpolating geometry \cite{Forste:1994wp}. There are many methods that can be used to study it, such as by using a coset description \cite{Giveon:2017myj} or using the fact the background is related via a so-called TsT transformation to AdS$_3$ supported by NS flux \cite{Apolo:2019zai}, which implies its worldsheet description is the same as that for AdS$_3$, but with twisted boundary conditions  \cite{Alday:2005ww}.

% There are several ways to think about this background. The first is as the NS5 decoupling limit $(g_s\r 0)$ of the NS5-F1 system, as we have just described.
%
%

Let us briefly comment on this TsT transformation. Starting with the extremal near-horizon F1-NS5 solution \eqref{btzinnh}, namely AdS$_3\times S^3 \times T^4$, one performs the following  set of dualities and coordinate shifts:

\be
T: ~ \mbox{T-duality along} ~ \s \;, \;\;\;\;\; s: ~ \mbox{a shift} ~ t \r t-\l \tilde \s \;, \;\;\;\;\; T: ~ \mbox{T-duality back along} ~ \tilde \s
\ee
where $\tilde \s$ is the T-dual coordinate to $\s$. The result precisely agrees with  \eqref{f1ns5b} for $\l=1/2$ \cite{Apolo:2019zai}. % \textcolor{red}{\emph{I'm also getting that $t$ is rescaled by $\l$, is this correct? No }}
%
%
% However, it is also interesting to note that the full background  \eqref{f1ns5b} can be obtained via a TsT transformation of the . % \emph{Check $\#$ is actually one!}
%Here, TsT stands for:
% T-duality along $x_1$,  a shift s: $t+\l \tilde x_1$, where $\tilde x_1$ is the T-dual coordinate to $x_1$, followed by a T-duality back. \emph{Determine the value of  $\l$}. % for the background \eqref{f1ns5b}.  
% 

One may also consider black holes in this spacetime. They are simply obtained by taking the  NS5 decoupling limit on finite-energy-momentum  solutions. The metric and dilaton read % \textcolor{red}{\emph{Notation!}}

\be
ds^2 = \frac{1}{f_1} \left(d\s^2-dt^2 + \frac{r_0^2}{r^2} (\cosh \d_n dt + \sinh \d_n d\s)^2 \right)  + \a' k \left( \frac{d r^2}{ r^2 - r_0^2 } + d \Omega_3^2\right) + ds^2_{M_4}  \label{aldbh}
\ee

\be
e^{2\Phi} = \frac{k \a'}{r^2 f_1} \;, \;\;\;\;\;\; f_1 = 1 + \frac{r_0^2 \sinh^2 \d_1}{r^2} \;, \;\;\;\; r_0^2 \sinh 2 \d_1 = \frac{ 2  \a' p}{v} \;, \;\;\;\; r_0^2 \sinh 2 \d_n = \frac{ 2  \a'^2 n}{R^2 v} 
\ee
%and the relation between $r_0, \d_i$ and the quantized charges is given by \eqref{bhcharges} \textcolor{red}{\emph{Careful these are in the D1-D5 frame!}}. 
One may also try to generate this solution by starting with BTZ $\times S^3\times T^4$, for which $f_1 \propto r^{-2}$, and performing the above-mentioned TsT transformation. While the metric  is correctly generated,  %\textcolor{red}{\emph{Check! up to rescaling of space and time, I find}},
 the dilaton  differs  by a temperature-dependent factor \cite{Apolo:2019zai}. %\textcolor{red}{\emph{Work it out!}}
  Thus,  the equivalence between the decoupled background and TsT of AdS$_3 \times S^3 \times T^4$ is not valid at finite temperature, but can be reinstated by manually correcting the dilaton. 
%
%{\color{blue}
%More precisely, performing TsT we find  ($B_{t\s} = \frac{p/v}{f_1 r^2} = \left( 1 -\frac{1}{f_1} \right) \frac{p}{v r_1^2} $)
%
%\be
%\frac{d\s^2-f dt^2}{f_1} \;\r_{T,s} \;  \left( 2 \l + \frac{\l^2 v r_0^2}{p} +  \frac{p}{vr^2}\right) \left[d\tilde \s^2 + \frac{(\varepsilon + \frac{\l v r_0^2}{p}) dt}{f_1'} \right]^2 - \frac{f dt^2}{f_1'}
%\ee
%where $f_1'$ is the first coefficient function. Rescaling $r = \tilde r/\tilde r_1 \sqrt{p/v}$ reproduces finite-temperature metric (note we don't rescale $t,\s$). \emph{Check both metric and dilaton!} }

Given the almost precise match between the decoupled NS5-F1 background and the TsT-transformed  AdS$_3$/BTZ one, the relation between string theory in this background and single-trace $T\bar T$ is sometimes referred to as the TsT/$T\bar T$ correspondence \cite{Apolo:2019zai}, especially when keeping the TsT parameter (which is proportional to the irrelevant `$T\bar T$ coupling') variable. While this way of viewing the correspondence is useful for matching long string observables with those in single-trace $T\bar T$, the fact that TsT does not yield the same temperature-dependence of the background as the decoupling limit means that one cannot trust the TsT point of view for the purpose of matching the thermodynamics. 
%
%\textcolor{red}{\emph{Careful,  extra continuous parameter. Also triality.}} 
%Note, however, that the  background that is holographically dual to compactified LST at  finite temperature is the one given by the decoupling limit, while the
%
The TsT picture is thus useful, but simply coincidental. 

The background with $r_0=0$ corresponds to the Ramond ground state of the system, a non-normalisable deformation of massless BTZ.  By taking $r_0^2 <0$  and demanding the absence of conical singularities, it is possible to also obtain the ALD completion \eqref{glald} of global AdS$_3$. %\emph{Does the $\l=1/2$ background at which this breaks down coincide with decoupled global AdS $\r$ ALD?}
Interestingly, if one computes the energy of this spacetime, one precisely recovers the ground state energy in a single-trace $T\bar T$ - deformed CFT %, for $\mu = xxx $ \textcolor{red}{\emph{Compute this!}}
 \cite{Apolo:2019zai}. % \textcolor{blue}{Smoothness requires $R |r_0|/(r_1 \sqrt{\a' k}) = 2\pi$, but need eom to fix $r_1$ in terms of $r_0$}. We have $r_1^4-r_0^2 r_1^2 = 1/4 b^2 k^3 \a'^3 = p^2 \a'^2 $
%
%A nice property of the background \eqref{} is that the worldsheet sigma model for a string propagating in it is known exactly, as we will review shortly.
%
%\bi
%\item is there a reliable worldsheet description at finite temperature?
%\item discuss the various methods to underst wsheet CFT: 
%\ei

%\noindent This result has been used in \cite{} for a simple derivation of the spectrum. 
%\be
%ds^2_{nh} = - \frac{dt^2 + dx_1^2}{r^2} + N_5 \a' \frac{dr^2}{r^2}+ \ldots \;, \;\;\;\;\; B= \# r^{-2} dt \wedge dx_1
%\ee 
%More precisely, after T-duality along $x_1$ the various fields are
%
%\be
%d\tilde s^2 = -\frac{dt^2}{r^2} + r^2 (d\tilde x_1 + dt/r^2)^2 =  r^2 d\tilde x_1^2 + 2 d\tilde x_1 dt  \r r^2 d\tilde x_1^2 + 2 d\tilde x_1 (dt + \l d\tilde x_1)   \;, \;\;\;\; \tilde B =0
%\ee
%where in the second step we performed a shift, $t \r t + \l \tilde x_1$. Performing a T-duality back and choosing $\l$ judiciously, we obtain precisely the background \eqref{f1ns5b}. This $\l$ equals the $\l$ of the $J^-\bar J^-$ deformation. 
%

\bigskip

\noindent \textbf{\emph{Relation(s) between the interpolating background and single-trace $T\bar T$}}

\medskip

\noindent Shortly after the $T\bar T$ deformation was defined, \cite{Giveon:2017nie} put forth a very interesting proposal, which related the  interpolating AdS$_3 \r $ ALD background described above to  the single-trace $T\bar T$ deformation of the D1-D5 CFT at the singular locus in moduli space that corresponds to a pure NS background. Since a single-trace $T\bar T$ - deformed CFT is essentially a solvable theory, this proposal could provide not only  one of the first examples of tractable - and even precision - non-AdS holography, but also the first tractable approach to LST, in this compactified setting. 

As usual, the physics is much more than meets the eye. While there does, indeed, appear to be a deep link between the interpolating AdS$_3 \r$ ALD background \eqref{f1ns5b} and a single-trace $T\bar T$ - deformed CFT, it is not true that there is an exact holographic duality  between them (for $k>1$). To explain their actual relation, it is useful to note that the various pieces of evidence given in \cite{Giveon:2017nie} in favour of a correspondence are of two different natures: 

\bi
\item[i)] a worldsheet argument, which refers to the perturbative long string subsector of the theory
\item[ii)] a supergravity argument, which refers to the high-energy states, dominated by black holes 
\ei
%
% This subject may be slightly confusing, in that there are two types of relations, of a somewhat different nature, between the AdS$_3 \r$ ALD background and single-trace $T\bar T$.
 %
 In this subsection, we will  separately review the above two types of arguments, before drawing a conclusion from them. 
 
 % made in favour of these relations: worldsheet and, respectively, supergravity ones. 

 %[The relation proposed in \cite{} between the interpolating  (ALD$_3$) background and single trace $T\bar T$ is argumented as follows: The IR AdS$_3$ region is described by a dual CFT$_2$, which is argued to have an SPO structure $\mathcal{M}^p/S_p$. Deforming it by a spt operator that is argued to correspond to single-trace $T\bar T$, one lands precisely on the exactly marginal worldsheet deformation that produces the ALD background].

\subsubsection{Relation to single-trace $T\bar T$ i) : worldsheet argument}

%\medskip

%\noindent

 The worldsheet argument of \cite{Giveon:2017nie} is based on the fact that the worldsheet CFT for a string propagating in the full interpolating geometry can be written down explicitly. %, as we now review.
Let us first concentrate on the  $r\r 0$ IR region, where the geometry becomes AdS$_3 \times S^3 \times T^4$, supported by purely NS-NS three-form flux.  
%Since the background \eqref{f1ns5b} is weakly coupled, it can be studied using peturbative worldsheet  techniques.
 The worldsheet CFT  describing this background is an $SL(2,\mathbb{R}) \times SU(2) \times U(1)^4$ WZW model in  the bosonic sector, plus fermions. % As discussed in the introduction, for $k>1$ the perturbative dynamics of strings in this background can be captured by a boundary CFT that is  a deformation of an SPO of the form 
%
%The spectrum one obtains depends on the type of $SL(2,\mathbb{R})$ representation: continuous or discrete, leading to long or short string states. As it turns out, the spectrum of the long strings is compatible with that of a SPO \emph{Under which approximations?}/the CFT corresponds to an SPO + deformation. \emph{Relation to the spectrum?} [For $p$ very large, the dual CFT is conjectured to be a free symmetric product orbifold CFT, of the form
% $(\mathcal{M}_{6k})^{p}/S_{p}$.  \emph{Any difference between massless BTZ and global AdS?}
 %, where $\mathcal{M}_{6N_5}$ is a CFT with central charge $6 N_5$.  \emph{What is this large $p$ argument?}]   
%This statement will need much qualification, but for now we will simply proceed.  %In the present case, $\mathcal{M}_{6k} = ???$ 
%[This conjecture was motivated by the fact that the worldsheet spectrum has a free orbifold structure  (so, it only works in a certain subsector?).] 
%
%As we already mentioned, the spectrum of excitations falls into two classes: short strings and long strings, distinguished by the type of worldsheet $SL(2,\mathbb{R})$ representation involved: continuous or discrete \cite{Maldacena:2000hw}.  The spectrum of the long strings - but not that of short ones - displays a  SPO structure. 
%
We will henceforth denote the compact part of the spacetime manifold as $\mathcal{N}$,  which can in principle be more general than $S^3 \times T^4$. 
The holographic dictionary \cite{deBoer:1998gyt,Kutasov:1999xu}  takes the schematic form 
\be
 \O_{bnd}(x) = \int d^2 z  \,  \Phi_{wsh}(x, z)  \mathscr{O}(z)\label{obndwsh}
\ee
where $ \O_{bnd}(x)$ is a gauge-invariant operator in the boundary CFT  (with coordinates $x$),  $ \Phi_{wsh}(x, z) $ is a natural vertex operator in the worldsheet $SL(2,\mathbb{R})$  WZW model (with coordinates $z$)  that describes AdS$_3$, and
 $ \mathscr{O}(z)$ is a primary operator in the  worldsheet theory on $\mathcal{N}$. 

The worldsheet theory describing the entire interpolating background \eqref{f1ns5b} corresponds to an exactly marginal deformation of this WZW model  by the $J^- \bar J^-$ operator, built from a certain null component of the $SL(2,\mathbb{R})$  current.  
%where $x$ represents the coordinates in the boundary CFT, while $z$ denotes the worldsheet ones.   
\cite{Giveon:2017nie} were able to rewrite the $J^- \bar J^-$ worldsheet deformation as an integrated deformation of the boundary CFT, which they then interpreted as a single-trace $T\bar T$ deformation.

% The  holographic dictionary \cite{deBoer:1998gyt,Kutasov:1999xu}  provides  a map between gauge-invariant operators in the boundary CFT %(where `single-trace' is defined with respect to this SPO structure)
%

%where $\Phi_h$ is an eigenfunction of the Laplacian on AdS$_3$, with Casimir $h(h-1)$ (thus, the dual operator has dimension $(h,h)$) which in the quantum t
More concretely, the stress tensor of the boundary CFT is given by the following expression 

\be
T(x) =\frac{1}{2 k} \int d^2 z \bigl(\p_x J (x,z) \p_x \Phi_1 (x,z) + 2 \p_x^2 J(x,z) \Phi_1 (x,z)\bigr) \bar J (\bar x,\bar {z})
\ee
where $J(x,z) \equiv e^{-x J_0^-} J^+(z) e^{x J_0^-} = x^2 J^- - 2 x J^3 + J^+$ is a convenient packaging of the worldsheet $SL(2,\mathbb{R})$ WZW currents, with spacetime scaling dimension $(-1,0)$, and $\Phi_h(x,z)$ is an operator with spacetime scaling dimension $(h,h)$, which is a primary both on the worldsheet and in the  boundary CFT, see \cite{Giveon:2017nie} for the explicit expressions.  It is then easy to check  that the right-hand-side has the correct dimension $(2,0)$ from the boundary point of view. %Also, since a single worldsheet integral is involved,  it corresponds to a `single-trace' boundary operator. 
Assuming  the dual CFT has a symmetric product orbifold structure of the form \eqref{spod1d5}, \cite{Giveon:2017nie}  argue the boundary representation of this bulk operator takes the form
%
%. In that case, the dual stress tensor  takes the form
%
\be
T(x)\;\;  \longleftrightarrow \;\; \sum_{i=1}^{p} T_i (x)
\ee
where $T_i$ is the stress tensor in a single copy of the $\mathcal{M}_{6k}$ CFT. A  similar expression holds for the antiholomorphic stress tensor.  Of course, as explained, this representation of the stress tensor is approximate at best, given that the boundary CFT is not exactly a SPO.

In the dictionary \eqref{obndwsh}, single worldsheet integrals correspond to single trace operators, while double integrals correspond to double-trace ones. Accordingly,  the standard double-trace $T\bar T$ operator 
$T\bar T = \sum_i T_i \sum_j \bar T_j $ would correspond to a double worldsheet integral. 
To find the worldsheet dual of the would-be single-trace $T\bar T$ operator $\sum_i T_i \bar T_i$ needed to drive the geometry away from AdS,  \cite{Giveon:2017nie}  proposed  the following deforming operator is, written as  a worldsheet integral

\be
D(x) =   \int d^2 z (\p_x J \p_x \Phi_1 + 2 \p_x^2 J \Phi_1) (\p_{\bar x} \bar J \p_{\bar x} \Phi_1 + 2 \p_{\bar x}^2 \bar J \Phi_1) 
\ee
This operator has the correct spacetime scaling dimension, $(2,2)$, and also its OPE with other operators. %\emph{With all operators? Is this an exact statement?Do the OPEs have the same form at finite $\l$?}
 behaves as that of 
 $T$ on the left and $\bar T$ on the right. %; however,  it is single-trace because it involves a single worldsheet integral. 
  Given these properties, \cite{Giveon:2017nie} 
%Its integral $\int d^2 x D(x) = \int d^2 z J^- \bar J^-$, as required. Notice it \emph{does not} correspond to the double-trace $T\bar T$ deformation, since that would involve two worldsheet integrals.  \cite{kutasov} went on further to
naturally conjectured that in the dual CFT, this operator corresponds to 
\be
D(x) \longleftrightarrow  \sum_{i=1}^{p} T_i \bar T_i
\ee
%which has the correct OPEs \emph{which were checked?} and is single-trace. 
%
Next, one would like to add this operator, integrated, to the action of the boundary CFT. 
It is useful to let the parameter of the deformation be a tunable parameter, $\mu$, even though in \eqref{f1ns5b} it has a fixed value. The identification of the leading deforming operator performed above holds for $\mu$ infinitesimal. Remarkably, the results of this analysis can be meaningfully extended to finite $\mu$. On the worldsheet side, the reason for this is that the integrated operator  satisfies

\be
 \int d^2 x D(x) = \int d^2 z\,  J^- \bar J^-
\ee
which corresponds %where $J^-$ is a  null component of the $SL(2,\mathbb{R})$ current.This  $J^- \bar J^-$ deformation is 
an exactly marginal deformation that can be turned on a finite amount. On the boundary side, if the deformed theory were truly single-trace $T\bar T$ - deformed CFT, then the deformation at finite $\mu$ would also be clearly well-defined and one would also have a well-defined framework for computing various observables.  %This leads to the conjecture (supported a posteriori by checks) that the holographic dual to the background \eqref{f1ns5b} is the symmetric product orbifold of $T\bar T$ - deformed CFTs
%
%\be
%Z_{string}[\mbox{F1-NS5}] \leftrightarrow Z\left[(T\bar T\mbox{-}def. \;\mathcal{M}_{6 N_5})^{N_1}/S_{N_1}\right] 
%\ee
The value of the irrelevant deformation parameter $\mu$ that is needed to match the background \eqref{f1ns5b} is 
\be
\mu =  \pi \a'
\ee
As explained, there is no reason that the dual theory be exactly single-trace $T\bar T$, as the undeformed AdS$_3$ background does not have the required SPO structure. In the following, we will ignore these cautioning remarks and simply test to what extent  observables computed from the bulk agree with those computed using single-trace $T\bar T$.

%[The same background can be described via a coset CFT, starting from $\mathbb{R}^{1,1} \times AdS_3 \times S^3 \times T^4$ and gauging the null current $\p x^- + \l J^-$, as well as its RM counterpart.]

%\bigskip

\subsubsection{The deformed string spectrum}

%\medskip

\noindent As a first check of the proposed duality, one can compare the string  spectrum in the background \eqref{f1ns5b}, as  computed using worldsheet techniques,  with that of the boundary symmetric product orbifold QFT.  The string spectrum can be computed either using a null coset construction of the worldsheet CFT \cite{Giveon:2017myj}, or by using the Wakimoto representation and flowing \cite{Chakraborty:2018vja},  %that corresponds to the background \eqref{f1ns5b}, 
or  by applying  the TsT transformation to the AdS result \cite{Apolo:2019zai}. %\textcolor{red}{\emph{Or, some simple flow on the worldsheet, }} % , which is known to only change the boundary conditions on the worldsheet fields. 
It is important to differentiate between long string and short string states, which correspond to continuous and, respectively, discrete representations of the worldsheet $SL(2,\mathbb{R})$.

Let us start by considering worldsheet vertex operators of type II superstring theory in global $AdS_3\times\mathcal{N}$ in the presence of pure NS-NS flux. Here $\mathcal{N}$ is a 7-dimensional compact manifold. % with a left moving $U(1)$ at level $k'$. \emph{Why primed?}
 Long strings correspond to worldsheet vertex operators  that belong to the continuous series representation of $SL(2,\mathbb{R})$. %Such vertex operators are dual via AdS/CFT to operators of the boundary field theory with left and right moving dimensions $(h,\bar{h})$. 
Their left worldsheet dimension
%s of such physical (on-shell) worldsheet vertex operators are 
is given by 
\begin{equation}
\begin{aligned}\label{vir1}
&\Delta=-\frac{j(j+1)}{k}-w\left(h+\frac{k w}{4}\right)+\Delta_{\N} +N \\
%&\bar{\Delta}=-\frac{j(j+1)}{N_5}-w\left(\bar{h}+\frac{N_5 w}{4}\right)+ \bar{\Delta}_{\N} +\bar{N}
\end{aligned}
\end{equation}
with a similar expression on the right-moving side. 
Here,
% the rightmost equality is the physical on-shell condition for the vertex operator,  
$j\in -1/2+i\mathbb{R}$ labels the Casimir of the global $SL(2,\mathbb{R})$ algebra, $w\geq 1$ denotes the integer spectral flow in $AdS_3$ - identified with the winding of the long string around the AdS$_3$ boundary, $\Delta_{\N}%\bar{\Delta}_{\N})
$ is the left  vertex operator dimension of the worldsheet CFT in $\N$, $N$ %$(N,\bar{N})$ 
is the left oscillator number in $AdS_3$ and, finally, $h$  represents the eigenvalue of the $J^3_0$  mode of the worldsheet  $SL(2,\mathbb{R})$, which for continuous series representations is unrelated to the eigenvalue of the Casimir. In global AdS$_3$, $h$ and its right-moving counterpart are identified with the left/right energies of the state on the cylinder, and thus to the dual operator dimensions  via the standard state-operator map. The physical on-shell condition for these superstring vertex operators is $\Delta=\bar \Delta = 1/2$.

The relation \eqref{vir1} also holds  for short strings, which correspond to worldsheet vertex operators  belonging to the discrete series of representations of $SL(2,\mathbb{R})$. In this case, $j$ is real and positive and $w$ can be zero. Unlike for the long strings, $j$ and $h,\bar{h}$ are not independent variables, but rather $h=j+m,\bar{h}=j+\bar{m}$ where $m,\bar{m}$ are positive integers. %In what follows, we will restrict ourselves to long strings, for comparison with our symmetric product orbifold results.
The on-shell condition then reduces to a quadratic equation for $h$ \cite{Maldacena:2000hw}.

%Let us consider worldsheet vertex operators of string theory in $AdS_3\times\mathcal{N}$ in the presence of pure NS-NS flux. Here $\mathcal{N}$ is a 7-dimensional compact manifold with a left moving $U(1)$ at level $k'$. In the discussion that follows, we will restrict ourselves to long strings. 

%These form the continuous series states of the  boundary theory. Note that for the case of continuous states (long strings), the $j$ and $h,\bar{h}$ are independent variables. This is not the case for the discrete states (short strings). 

The  effect of the $J^-\bar J^-$ deformation that turns on  the
asymptotically linear dilaton background is to slightly change % can both be obtained from AdS$_3$ $\times \; \mathcal{\N}$ via a transformation known as TsT: T-duality, shift, T-duality. In both cases, the effect of the TsT transformation can be encoded in a non-local coordinate transformation, which is equivalent to twisting the  boundary conditions of the fields on the  AdS$_3$ worldsheet theory in a charge-dependent way \cite{Alday:2005ww}. This mildly affects 
the relationship between $\Delta$ and $h$, by adding %{\color{ForestGreen}(we need to check again $J\bar{T}$)}

%\textbf{\emph{Factors $\pi$! $J\bar T$ looks wrong!!!!}}
%\cite{Apolo:2019zai,Cui:2023jrb}

\be\label{deltah}
\d_{ALD} \Delta =   \d_{ALD}\bar  \Delta  = - \frac{\mu}{\pi}\, h \bar{h} %\;, \;\;\;\;\;\; \d_{J\bar T} \Delta =   \d_{J\bar T}\bar  \Delta = \l\bar{p}\left(q^{[0]}+\frac{\lambda k \bar{p}}{4}\right)
\ee
on the right-hand side of \eqref{vir1}. Since $(h, \bar h)$ are interpreted as the deformed left/right global energy, and $j,w,N, \D_{\N}$ are held fixed in the process, then these shifts  precisely reproduce the relation (given by \eqref{defengeqttb} with the replacement $R\r R w$) between deformed and undeformed energies in a winding $w$ sector of a single-trace $T\bar T$-deformed CFT, for $R=2\pi$.  %\textcolor{red}{\emph{Write it!}}

The same argument, applied to the short string  sector,  \emph{does not} yield a match to single-trace $T\bar T$, because $j$ is no longer independent of $h$.  

%\bigskip

\subsubsection{The deformed correlation functions}

%\medskip

\noindent A slight variation on the above can be used to compute correlation functions of operators in the deformed boundary theory, which are related to worldsheet vertex operators via \eqref{obndwsh},  assumed to still hold. Since the boundary theory becomes non-local, it is natural to consider the correlation functions of momentum-space operators, obtained by Fourier-transforming \eqref{obndwsh} with respect to $x$, and also place the theory on the plane. The zero modes of the $SL(2,\mathbb{R})$ current $J_0^3$  are now identified with the momenta along the  boundary directions.  

The change in the correlator of $\O_{bnd}(p)$ will come from the change in the correlator of $\Phi(p,z)$. Concentrating on two-point functions, let us write the total worldsheet dimension \eqref{vir1} as 

\be
\D = \D_h +\D_{\mathscr{O}} \;,\;\;\; \;\;\;\;\; \D_h \equiv -\frac{j(j+1)}{k}-w\left(h+\frac{k w}{4}\right) \label{defdeltah}
\ee
and similarly on the right. 
%
%One can use the holographic map between boundary and worldsheet operators to study correlation functions in the deformed boundary CFT using the well-controlled worldsheet techniques. For a large class of boundary operators, this map takes the form  %\emph{\textbf{How does one know it's not deformed?}}
%%
%%\be
%%\O(x) = \int d^2 z \, \Phi_h (x,z) \mathcal{V}(z) \label{holomap}
%%\ee
%%where $\mathcal{V}$ is a vertex operator associated with the part of the worldsheet CFT that describes the internal space. The above operator satisfies the mass-shell condition
%%
%%\be
%%- \frac{h(h-1)}{N_5} +\D_{\mathcal{V}} = 1/2 \label{masssh}
%%\ee
Before the deformation, and concentrating on scalar operators characterised by $\Phi_h(x,z)$, the position-space correlator reads
\be\label{vv2pt}
\langle \Phi_h (z_1, x_1) \Phi_h(z_2, x_2) \rangle = \frac{1}{z_{12}^{2\Delta_h} \bar z_{12}^{2\bar \Delta_h} x_{12}^{2 h} \bar x_{12}^{2 \bar h}}
\ee
with Fourier transform \eqref{euclcft2pf}
\be\label{VVws2pt}
\langle \Phi_h (z_1,p) \Phi_h(z_2,-p) \rangle =\frac{2\pi^2 (p \bar p)^{2h -1} }{2^{4h} \sin (2\pi h) \Gamma(2h)^2}   \frac{1}{z_{12}^{2\Delta_h} \bar z_{12}^{2\Delta_h}} 
\ee
where we stripped off the momentum-conserving $\d$ function. The change in the above correlator will come from the change in $h$ triggered by adding to $\D_h$ of the extra contribution \eqref{deltah} associated with the $J^-\bar J^-$ deformation. Since $\D_{\mathscr{O}}$ is not expected to change,  nor does the mass-shell condition $\D=1/2$, one must have
\be
\D_{h^{[\mu]}} + \frac{\mu}{\pi} p \bar p = \D_{h^{[0]}}
\ee 
taking into account the fact that the additional contribution  \eqref{deltah} is always proportional to the product of the left and right momenta, which in global coordinates translate into a shift proportional  to the product of energies, but in Poincar\'e ones to a shift constructed from  $p$. The different sign corresponds to working in Euclidean, rather than Lorentzian, signature. %\textcolor{red}{\emph{Check!}}

The actual change in the correlation function depends on whether one is computing correlation functions of long or short string vertex operators. For long strings, $j$ is independent of $h$ and considered fixed, so one simply obtains \cite{Cui:2023jrb}  
\be
h^{[\mu]}  = h^{[0]} + \frac{\mu}{\pi w} p\bar p
\ee
a replacement to be plugged into the $z$-independent part of %the correlator
 \eqref{VVws2pt}. For $w=1$, this is expected to coincide with the double-trace $T\bar T$ correlator, as discussed in section \ref{ttbcorrsec}, but has not been proven (yet).

The case of short strings has been studied in  \cite{Asrat:2017tzd,Giribet:2017imm}.  Taking $w=0$ in \eqref{defdeltah} and using the fact that, for discrete representations, $j=h-1$,  one finds  the following shift  %\textcolor{red}{\emph{Conventions and normalisation!}}

\be
h^{[\mu]}(h^{[\mu]}-1) = h^{[0]}(h^{[0]}-1) + \frac{\mu k}{\pi} p \bar p
\ee
%
%to study the two-point function of operators in the deformed theory. Since the deformation is irrelevant, the correlator becomes non-local. It is thus best described in momentum space, using the Fourier-transformed version  of the map \eqref{holomap}. From a worldsheet perspective, the deformation only modifies the mass shell condition \eqref{masssh}, and amounts to replacing
%
%where $p$ is the momentum of the Fourier-transformed boundary operator, $\O(p)$. 
The the two-point function of such short string operators  is then again given by the first factor in \eqref{VVws2pt}, with the replacement $h \r h^{[\mu]} (p)$.   %Due to the square root appearing in the solution for $h^{[\mu]}(p)$, the ana
%
%same formula as the Fourier transform of the CFT two-point function ($\propto (p^2)^{2h-1}$ times an $h$-dependent prefactor), but now with $h\r h(p)$ as read off from above. 
This two-point function is well-defined and smooth for real Euclidean momenta; however, in  Lorentzian signature it has a branch cut for timelike ones.  The same behaviour of the correlator can be found by studying supergravity modes in the ALD background, where it is associated to the presence of momentum-dependent radial falloffs for the fields.  %\textcolor{red}{Still 1to1, if different.}

%, whose interpretation remains to be understood. 
%
%\bi
%\item talk about analytic structure
%\item say that short ones don't match, but long ones do
%\ei

% They can be constructed explicitly \cite{Kutasov:1999xu}, and their correlation function takes the form\footnote{In this article, we normalize the worldsheet operators such that the two-point function of the dual CFT operators takes the form $x_{12}^{-2h}\bar{x}_{12}^{-2\bar{h}}$.  Note this normalization  is different from the standard convention in string theory in $AdS_3$.  }

%where $x, \bar x$ are auxiliary coordinates that become identified with the space where the dual CFT lives. Integrating over the worldsheet coordinates, one obtains the standard correlation function of CFT operators on the boundary. One may also perform a Fourier transform of the latter, to obtain the momentum-space boundary correlator.    %{\color{ForestGreen}{ For the computation of correlation using worldsheet techniques we would like to restrict ourselves to the case $h=\bar{h}$.
% For example, the momentum space two-point function takes the form %\emph{\textbf{Check!}}

%which will be useful in this section.  

The natural conclusion of these analyses is that, while the long string subsector, and in particular the spectrum of long strings in the interpolating AdS$_3 \r $ ALD geometry show a good match to single-trace $T\bar T$, the short string subsector does not.  Consequently, there cannot be an exact duality between type II string theory in the interpolating background and the single-trace $T\bar T$ deformed D1-D5 CFT. That such an exact duality could not be possible was already clear from the fact that the original CFT dual to the undeformed background is not itself a SPO.  Even in the long string subsector, the match between observables computed in string theory and single-trace $T\bar T$ ones is unlikely to hold beyond the spectrum, due to the interactions between long and short strings. Thus, even the validity of  the perfect match we found for the long string spectrum and correlators is somewhat restricted. The exception is for $k=1$, where the discrete states are absent, the continuous ones are restricted to zero radial momentum ($j=1/2$) \cite{Eberhardt:2018ouy} and the full theory is described by perturbative strings. % \cite{Giveon:2005mi}. \textcolor{red}{\emph{Check both!} } 
 Also note that, already in AdS$_3$, the long strings are expected to be lifted %\textcolor{red}{\emph{Mention before!}} 
 upon turning on a RR perturbation  and the spectrum is supposed to become discrete. It is unclear what will happen to the match to single-trace $T\bar T$ if one turns on such a perturbation in the ALD background, namely whether the continuous  set of states for which the match is present will completely disappear, or not. 
 
 The above discussion raises the question: is there any deep relation between the ALD background and $T\bar T$, that would apply to: i) the entire theory, not just a subsector and ii) at a generic point in the moduli space of the D1-D5 CFT? %\textcolor{red}{So, match may just be a coincidence. \emph{Could it be explained by TsT structure?}}

%\bigskip

\subsubsection{Relation to single-trace $T\bar T$ ii) : black hole entropy}

The second check that \cite{Giveon:2017nie} perform of their proposed relation between the interpolating AdS$_3 \r $ ALD background and single-trace $T\bar T$ is the  perfect match of the Bekenstein-Hawking entropy of black holes in this space-time and the universal asymptotic density of states \eqref{entstrttb} in a single-trace  $T\bar T$ - deformed CFT. Note that, unlike the previous check, this one refers to the \emph{full} theory. 

%A seemingly related check, but of actually a completely different nature, is the match of balck hole entropy.   More about this later. 

The (string frame) metric of black hole solutions in the ALD spacetime is given by \eqref{aldbh}, obtained by taking the NS5 decoupling limit of the near-extremal NS5-F1 black brane system. 
%
%%An important check of the proposed duality was to show that the symmetric product orbifold  of $T\bar T$ - deformed CFTs correctly reproduces the entropy of a black hole in the bulk. 
%
%On the gravity side, we consider the non-extremal NS5-F1 solution . The string frame metric reads
%
%\be
%ds^2 = - \frac{f_\mathcal{E}}{f_1} dt^2 + \frac{dx_1^2}{f_1} + \frac{N_5 \a'}{f_\mathcal{E}} \frac{dr^2}{r^2}\;, \;\;\;\;\;\;e^{2\Phi} = g_s^2 \frac{N_5 \a'}{r^2 f_1}%\;, \;\;\;\;f_1 = 1+\frac{N_1 \a' g_s^2}{v\, r^2} \;, \;\;\;\; f_5 = \frac{N_5 \a'}{r^2} \;, \;\;\; f_{\mathcal{E}} = 1 - \frac{r_0^2}{r^2}
%\ee
%where $f_1,f_\mathcal{E}$ are defined in \eqref{haglst} and \eqref{f1ns5b} and 
% $r_0^2$ is proportional to the mass of the black hole. 
%%
%\be
%\mathcal{E} = \frac{R v r_0^2}{4\pi g_s^2 \a'}
%\ee
The Bekenstein-Hawking entropy of these black holes, with horizon topology $S^1_\s \times  S^3 \times M_4$, is 

\be
S_{BH} = \frac{ e^{-2\Phi(r_0)}\mathcal{A}_H}{4G_{10}} =  \frac{ R v  r_0^2\sqrt{k}\cosh \d_1 \cosh \d_0}{\a'^{3/2}} %= \frac{R \sqrt{r_0^2r_1^2 +r_0^4}}{4 G_3 N_5 \a'} %\;, \;\;\;\;\; G_3 = \ldots
\label{bhent}
\ee
%where $v$ is the volume of the internal manifold in string units, $V_{M_4} = (2\pi)^4 v \a'^2$, $R$ is the circumference of the $\s$ circle, and the $10d$ Newton's constant is given by $16\pi G_{10} = (2\pi)^7 $.
 One would like to rewrite this formula in terms of the energy above extremality and momentum of these black holes,  given by  \cite{Chakraborty:2020swe}
\be
E = \frac{R v r_0^2}{2 \a'^2} (e^{-2\d_1}+\cosh 2 \d_n)\;, \;\;\;\;\; P=  \frac{R v r_0^2}{2 \a'^2} \sinh  2 \d_n
\ee
%
%In the dual theory, the maximal entropy is obtained for equal repartition of the energy between the $N_1$ copies of the $T\bar T$ - deformed $\mathcal{M}_{6N_5}$ CFT, each of which carries energy $E/N_1$. Using \eqref{defent} for each copy, we find 
%
%\be
%S= N_1 S_0 = 2\pi N_1 \sqrt{\frac{ N_5}{\pi} \left(\frac{E R}{N_1}  + \l \frac{E^2}{N_1^2} \right)} =2 \pi \sqrt{\frac{ N_5}{\pi} (E R N_1 + \l E^2) } \label{spoent}
%\ee
%which is in perfect agreement with the gravity calculation for $\l=??$.
It is then not hard to show that, in terms of the left/right-moving energies $E_{L,R} = (E\pm P)/2$, the entropy takes the form\footnote{To show this, it is useful to note that the combinations $E_{L,R} + \frac{\a'}{R \, p}  E_L E_R= \frac{R p \cosh^2(\d_1 \pm \d_n)}{\a' \sinh^2 2 \d_1}$.}

\be
S(E) = 2 \pi \sqrt{  p k E_L R/2\pi + \a' k E_L E_R }+ 2 \pi \sqrt{  p k E_R R/2\pi  + \a' k E_L E_R } \label{ldentropy}
\ee
which precisely agrees with the entropy of a single-trace $T\bar T$ - deformed CFT, where the seed CFT has central charge $c=6 p k$, $\mu = \pi \a'$ and the symmetric orbifold is with respect to $S_p$ permutations\footnote{The black holes \eqref{aldbh} are thermodynamically stable. Nonetheless, as we already saw in section \ref{ttbsubsec}, the asymptotic Hagedorn behaviour can easily lead to negative specific heat upon including corrections. %\textcolor{blue}{One should also worry about corrections to LST thermodynamics. There can be $\a'$ correction (same as 1-loop?) which may make the specific heat slightly negative. According to Barbon, a much bigger corrections comes from thermal graviton gas in the throat, which behaves as gas in flat space. This has infinite entropy at infinite vaolume, so one should regulate it. It may give a growing contribution to the entropy. \emph{Check!?} Then, the black hole may jump and eat the gas box, as in flat space, apparently.}
}. 

 Assuming adiabaticity of the irrelevant deformation, one may equate $S(E)$ above with the Cardy entropy in the undeformed CFT for some energy $E^{[0]}$ and momentum $P=E_L-E_R$. Of course, the relation between $E^{[0]}$ and $E_{L,R}$ that  results from this is precisely the single-trace $T\bar T$ relation \cite{Chakraborty:2023mzc}. Nonetheless, as in the case for the holographic dictionary for double-trace $T\bar T$- deformed CFTs, one cannot conclude from this exercise that the individual energy levels will flow according to the $T\bar T$ formula; for that, one needs an \emph{independent} field-theoretical argument. In the case of double-trace $T\bar T$ - deformed CFTs, such an argument was provided by Smirnov and Zamolodchikov.  In the LST case, no such argument is currently known, and moreover the low-energy states corresponding to short strings clearly do not follow the $T\bar T$ formula for their energies, indicating one should not expect it to  hold.

%Given that $S(E)$ perfectly matches the entropy in  a single-trace $T\bar T$ - deformed CFT, and that can be read off from this equality to be given precisely by the $T\bar T$ deformed energy formula. 

%
%\bigskip
%
%\noindent \begin{minipage}[t]{0.05\textwidth}  \emph{Exercise:} \end{minipage} \hspace{7mm} \begin{minipage}[t]{0.9\textwidth} Express $r_0^2$ in terms of the mass of the black hole and use it to show that the boundary entropy \eqref{spoent} agrees with the bulk Bekenstein-Hawing entropy \eqref{bhent}  for the appropriate value of $
%\l$.  Check that this value agrees with that used to match the Hagedorn temperature, computed in a previous exercise. 
%\end{minipage}

 %\textcolor{red}{\emph{Comment on the fact it's wrong to use this to infer a relation between the deformed and undeformed energies!}}

Thus, even though - as we saw - the ALD background and a single-trace $T\bar T$ - deformed CFT  cannot be exactly dual to each other, the asymptotic degeneracy of states in the theory dual to the ALD background takes precisely the form  of that in a  single-trace $T\bar T$ - deformed CFT. This suggests it belongs to a  universality class  of theories that also includes single-trace $T\bar T$.  In the following, we present more evidence in this direction.

%\bigskip

\subsubsection{Relation to single-trace $T\bar T$ ii): Asymptotic symmetries}

%\medskip

\noindent As discussed throughout this review, a very important observable in holography are the asymptotic symmetries of the bulk spacetime, which should map to standard symmetries of the dual theory. % These symmetries can be very powerful in constraining the dynamics of the theory, especially when infinite-dimensional.
If it is true that the interpolating AdS$_3 \r$ ALD spacetime is  related to  single-trace $T\bar T$, then one would expect  the single-trace version %\textcolor{red}{\emph{Talk about it in that section!}} 
of the intricate  extended $T\bar T$ symmetries \cite{Chakraborty:2023wel}, discussed at the end of \ref{genargsymm},  to appear as asymptotic symmetries of the ALD spacetime.

However, determining the asymptotic symmetry group generators is a delicate problem since, as reviewed, they often depend quite sensitively on the boundary conditions one imposes%\footnote{\textcolor{blue}{Remember that, for example, in stringy warped AdS, one could get one or two Virasoro, or warped conformal symmetries, depending on the boundary conditions imposed. }}
. When we were able to show, in section \ref{asysymmdtr}, that AdS$_3$ with mixed boundary conditions precisely reproduces the $T\bar T$ symmetries, it was largely thanks to the fact that the boundary conditons on AdS$_3$ were known, having been derived from the field theory; the derivation of the charges and the algebra  then followed algorithmically. For the case of the ALD background, we do not \emph{a priori} know what the boundary conditions should be, and  the main challenge in studying the asymptotic symmetries of this spacetime is to find them.  %\textcolor{red}{\emph{Rewrite!}}

Boundary conditions in ALD spacetimes were discussed in \cite{Marolf:2007ys}. As is standard in the ASG literature, their proposal corresponds to a reasonable guess based on the asymptotic fall-offs of known solutions.  Given that  the  metric of the ALD black holes \eqref{aldbh}   is asymptotically independent of the temperature, \cite{Marolf:2007ys} proposed that that one should fix the metric at spatial infinity, %in Einstein frame, \emph{Does this matter?} 
while letting the subleading components in the $1/r$ expansion fluctuate. 
While such boundary conditions (possibly supplemented by a more careful analysis of subleading fall-offs) do appear consistent, it is not clear whether they are \emph{the} boundary conditions  that the holographic dual to compactified LST obeys, or a more restrictive set. Ideally, one would like  to be maximally agnostic and let the system \emph{indicate} the boundary conditions  it  obeys. %Of course, these boundary conditions should be such that the black hole solutions \eqref{} correspond to states in the bulk theory. The same boundary conditions also determine the ASG generators, which are the most general diffeomorphisms that preserve the boundary conditions on the metric, while leading to generically finite conserved charges.

Of course, we know from the decoupling limit that, as far as solutions that are independent of the boundary coordinates $\s, t$  are concerned, the black hole solutions \eqref{aldbh} must be allowed. One would like to understand what should be the allowed coordinate-dependent fluctuations thereof.  
%
%The black hole solutions only tell us about (definitely) allowed boundary-coordinate-independent  fluctuations/solutions. How about more general fluctuations? 
%
To answer this question while making a minimal set of assumptions, let us remember the basic requirements for having a consistent phase space.

% Here, we will be much more general/maximally agnostic, and make the minimally intrusive assumption that the symplectic form obeys the conditions below, which ensure charge conservation and finiteness.

In general, in order  for a theory to have a well-defined phase space, one requires that the symplectic current  $D-1$ form  $\boldsymbol \om$ associated with any two allowed perturbations vanish asymptotically. More precisely, in three dimensions one imposes that

 \be
\boldsymbol \om_{ab} [\Phi, \d_1 \Phi, \d_2 \Phi] = o(r^0), \qquad\boldsymbol \om_{r a} [\Phi, \d_1 \Phi, \d_2 \Phi] = o(r^{-1})\label{phs}
\ee
where $a$ denotes the tangent indices to the boundary located at $r \rightarrow \infty$, $\Phi$ stands  generically for the fields in the theory, $\d \Phi$ are arbitrary variations thereof and the notation $o(r^{-c})$ signifies that the fall-off must be faster than the indicated power of $r$. The first condition above ensures that the symplectic flux is conserved, whereas the second leads  to normalizability of the symplectic form at infinity. The boundary conditions are chosen such that these requirements are satisfied. Since $\boldsymbol \om$ is only defined up to the addition of an exact form,  $\boldsymbol \om \r \boldsymbol \om + d \boldsymbol \om_Y $, where $\boldsymbol \om_Y$ is antisymmetric in the variations,  the above fall-off conditions only need to be satisfied for some choice of this boundary term.   Also, since they are local conditions, they could be violated by total spatial derivative terms that drop out of the integrated symplectic form. Note also that the analysis of the symplectic form will not necessarily indicate all the modes that need to be fixed, but rather which modes should not be allowed simultaneously. 

When one of the perturbations is generated by a diffeomorphism, $\d \Phi= \L_\xi \Phi$, the symplectic form can be related on-shell to a $D-2$ form $ \boldsymbol k_\xi$
\be
\boldsymbol \om [\Phi, \d_\xi \Phi, \d \Phi] = d \boldsymbol k_\xi [\Phi, \d\Phi]
\ee
whose spatial integral at infinity yields the conserved charge difference \eqref{covphspch} associated to the corresponding diffeomorphism. From the point of view of the  charges, \eqref{phs} ensure conservation and, respectively, finiteness. 

The observation of \cite{Georgescu:2022iyx} is that it is possible to constrain the diffeomorphisms $\xi^{ASG}$ that correspond to  asymptotic symmetries by requiring that the symplectic product of the change in the metric they generate with a perturbation that is known to belong to the phase space obey the fall-off conditions \eqref{phs}.  For the latter perturbation one can take, for example, one that slightly changes the mass and angular momentum  of the black holes \eqref{aldbh}.

More concretely, 
%Let us now apply the conditions \eqref{phs} to linearised perturbations of the ALD black hole backgrounds \eqref{aldbh}, following \cite{Georgescu:2022iyx}. 
it is useful to parametrize  the black hole solutions as  %\textcolor{red}{\emph{Mention consistent truncation}}
\be
ds^2 = \frac{r^2}{ r^4 + \b r^2/(\a' v ) + L_u L_v} \left(  r^2 dU dV + L_u dU^2 + L_v dV^2 + \frac{L_u L_v}{r^2} dU dV\right) + \a' k \left( \frac{d r^2}{ r^2 } + d \Omega_3^2\right) + ds^2_{M_4}  \nonumber
\ee
\be
e^{2\Phi} =  \frac{k r^2}{ \a' ( r^4 + \b r^2/(v\a') + L_u L_v)}   \;, \;\;\;\;\;\b = \sqrt{ p^2 + 4 \a'^2 v^2 L_u L_v } \label{aldbhluv}
\ee
where\footnote{ $L_{u,v}= \frac{r_0^2}{4 \a'^2} e^{\pm 2\d_n} $ in terms of the horizon radius $r_0$  in \eqref{aldbh}. The $r$ coordinate  above is different from that in \eqref{aldbh}.} $L_{u,v}$ encode the energy and momentum of the black hole solutions  (see \cite{Georgescu:2022iyx} for more details) and have a good $\a' \r 0$ limit.  To simplify the calculations, one can also work in a consistent truncation to three dimensions   \cite{Georgescu:2022iyx} that only keeps the three-dimensional part of the metric and the dilaton. To find the constraints on $\xi^{ASG}$, one first fixes a gauge, such as radial gauge in \emph{Einstein} frame\footnote{%If we work with a diffeomorphism in string frame, the symplectic form has  log divergences that constrain $F_r$ to be at most linear in $U,V$. We also find $r^2$ divergences that depend on the second derivatives of $F_r$, though they vanish if $F_r$ takes the form \eqref{formFr}, $C_{ij}$ satisfies \eqref{mixedbc}  and $C_r =0$. While fixing $F_r =0$ in string frame would resolve these issues, it would also impair our ability to find a central extension to the asymptotic symmetry algebra we discuss in the next section.//
As discussed in \cite{Georgescu:2022iyx}, the choice of frame is important, as
the structure of normalisable and non-normalisable modes is different in Einstein vs.  string frame radial gauge.
% the change of coordinates from radial gauge in string frame to the same gauge in Einstein frame seems to induce non-normalisable terms
In particular, the radial component of the diffeomorphism, as well as  the  change in the dilaton it induces, is  more subleading at large $r$ in Einstein frame  than for string frame radial gauge perturbations. This opens the possibility that it be an allowed mode, while its string frame counterpart is not.   }. %We find it is parametrised by 
The most general diffeomorphism that preserves this gauge is %e radial metric in Einstein frame reads
\bea
\xi_{rad}^{Einst.} &= & \left( F_U (U,V) + \frac{k  (r^2 \,  \p_V F_r - L_v\,  \p_U F_r)}{r^4-L_u L_v} \right) \p_U  +  \left( F_V (U,V) + \frac{k  (r^2 \, \p_U F_r - L_u \,\p_V F_r)}{r^4-L_u L_v} \right) \p_V  + \nonumber \\
&+& \frac{r^3  F_r (U,V)}{\a' r^4 + r^2 \b + \a' L_u L_v} \p_r \label{xiradeinst}
\eea
where the functions $F_U, F_V$ and $F_r$ are arbitrary at this stage. Imposing 

\be
\boldsymbol \om_{ab} [\Phi, \d_\xi \Phi, \d_{L_u} \Phi] =  \boldsymbol \om_{ab} [\Phi, \d_\xi \Phi, \d_{L_v} \Phi] =o(r^0) \label{condsymplf}
\ee
%as well as the normalizability condition \textcolor{red}{\emph{Check!}}
 fixes the form of the functions $F_U, F_V$ to be

\be
 F_U = f \left(u \right) + \frac{2 \a' L_v}{p+\b} \, \bar f \left(v\right) + c_U \;, \;\;\;\;\;  F_V =  \bar f \left(v \right) + \frac{2 \a' L_u}{p+\b}\, f \left(u\right) +c_V\label{solFUV}
\ee
where $c_{U,V}$ are integration constants and the `field-dependent coordinates' $u,v$ are defined as

\be
u \equiv  \frac{(p+\b) \, U+2 \a' L_v V}{2p} \;, \;\;\;\;\;\; v \equiv  \frac{(p+\b)\, V + 2 \a' L_u U}{2p} \label{fdepcoordst}
\ee
Quite remarkably, these components of the allowed diffeomorphisms \emph{precisely coincide} with the  asymptotic symmetry generators  \eqref{xiV}  of the  AdS$_3$ background with mixed boundary conditions that is holographically dual to double-trace $T\bar T$ - deformed CFTs,  particularized  to constant parameters and with the radial component removed. This is despite the fact that the two backgrounds, \eqref{aldbhluv} and \eqref{fg} with \eqref{g0UV}, have completely different asymptotics. As noted in \cite{Georgescu:2022iyx}, despite this fact, the details of the charge and ASG computations follow closely those of AdS$_3$ with mixed boundary conditions.

Note that in order to obtain sufficient constraints on $\xi$, it was important to consider the most  general black hole solutions and to consider variations with respect to all their parameters.  By imposing the zero flux condition on the symplectic product of two such diffeomorphisms also fixes $F_r$ to take the form

%where their overall normalization has been chosen for later convenience. Note these
% coordinates $u,v$ have identifications
% %
% \be
% u \sim u + 2 \pi R \, r_u  \;, \;\;\;\;\;\; v \sim v + 2 \pi R \, r_v
% \ee
%  where
%%
% \be
% r_u = \frac{\b+p+2\a' L_v}{2p} = 1 + \frac{\a' H_R}{p R} \;, \;\;\;\;\; r_v = \frac{\b+p+2\a' L_u}{2p} = 1 + \frac{\a' H_L}{p R} \label{defruv}
% \ee
%which follows from the fact that $\p_\s u = r_u$ and $\p_\s v = r_v$. 
%where we used \eqref{engmombhsols}. The reason for using the notation $H_{L,R}$ for the left/right-moving energies, rather than the previously used $E_{L,R}$, is to emphasize the operatorial origin  of the field-dependent radius of these coordinates; the two quantities are identical as far as the classical analysis of this article is concerned. Acting with these diffeomorphisms produces a change in the components of the asymptotic metric that precisely obeys \eqref{mixedbc}, as one can easily check, where now $C_{ij}$ is space-time dependent.

%
%
%
%%Finally,  writing $C_{uu}, C_{vv}$ in terms of $C_{uv}$ using the boundary conditions \eqref{mixedbc} and plugging in  the solution for $F_{U,V}$, we find the term proportional to $\hat C_{uv}$ is %emph{Keep?}
% 
% \be
% \frac{k \left(\alpha'  L_u F_r^{(0,2)}(U,V)+\alpha'  L_v F_r^{(2,0)}(U,V)-\beta  F_r^{(1,1)}(U,V)\right)}{\b } \hat  C_{uv} \label{constrradcomp}
% \ee
%If $\hat C_{uv} \neq 0$, then  the most general solution for $F_r$ is a linear combination of  two arbitrary functions of the field-dependent coordinates \label{fdepcoordst} 

\be
F_r = f_r (u) + \bar f_r (v) \label{formFr}
\ee
for two  new functions $f_r, \bar f_r$ of the same field-dependent coordinates. It is not \emph{a priori} clear whether these functions are completely independent of the previous two, or simply the symplectic-form-based
 procedure  is weaker than a set of boundary conditions. For example, if one tried to infer the asymptotic symmetry generators \eqref{asgads3} of asymptotically AdS$_3$ spacetimes by requiring the analogue of the conditions \eqref{condsymplf} hold on the set of BTZ metrics, we would again obtain four functions $f(U), \bar f(V), f_r(U), \bar f_r(V)$, but with no relation between $f$ and $f_r$, whereas Dirichlet boundary conditions imply that $f_r = - f'/2$ (for $\rho= r^{-2}$). % \textcolor{red}{\emph{Check factor!}} 
 Below, we discuss how the ASG analysis may suggest precisely this type of relation.

The action of the above allowed diffeomorphisms on the metric lead to the following relations between its asymptotic components  % \textcolor{red}{\emph{Compare Marolf}}
\be
\d g_{UU} = \frac{2 \a' L_u}{\b}\, \d g_{UV} \;, \;\;\;\;\;\;\; \d g_{VV} = \frac{2 \a' L_v}{\b}\, \d g_{UV} \label{aldbndcond}
\ee
which are \emph{not} asymptotically Dirichlet, since $g_{UV}$ fluctuates at leading order. Since they were obtained by making a minimal set of assumptions lends credence to the fact that they may be the correct boundary conditions for the ALD space-time.  %\textcolor{red}{\emph{Note $T\bar T$ metric is fixed, so map to bulk non-trivial.}}

%
%We can get a glimpse into the  boundary conditions that in principle give rise to these asymptotic symmetries by computing   \textcolor{red}{\emph{Need Einstein radial gauge!}}
%
%\be
%h_{ij} =_{r\r \infty}  \left[ \frac{1}{\a'} \left(\begin{array}{cc} \p_{{U}} F_V &  \frac{\p_U F_U + \p_V F_V}{2}    \\  \frac{\p_U F_U + \p_V F_V}{2}   & \p_V F_U  \end{array} \right) - 2 k \ln r  \left(\begin{array}{cc} \p_U^2 F_r  &  \p_U \p_V F_r \\  \p_U \p_V F_r  &  \p_V^2 F_r \end{array} \right)   \right] \label{effasymet}
%\ee
%%where the log term comes from solving the non-homogenous differential equation for $h$. Both parantheses receive corrections at $\O(r^{-2})$ and we have ommitted the radial components of the metric perturbation, which are zero in this gauge. 
%%
%Note the effective boundary condition is \textcolor{red}{What??}
%

It is important to note that the above are boundary conditions for linearised perturbations around the \emph{constant} black hole  backgrounds, which is what the above method allows one to compute. In principle, one would like to obtain boundary conditions for finite perturbations. While the condition  \eqref{condsymplf} can in principle be solved order by order in perturbation around the background \eqref{aldbhluv}  (and this is necessary to do explicitly to at least first order in order to compute the charge algebra), the very strong field dependence of the asymptotic diffeomorphisms makes it difficult to infer the full solution.

% (not even sufficient to extract charge algebra, for which one has to go to x order). How about general boundary conditions?  This question can be addressed perturbatively in the inhomogeneity. \emph{Does the metric falloff wrt $r$ change in this case?}

Asymptotic symmetries in hand, one can compute the conserved charges using the covariant phase space formalism. In order for the generic charges to be non-zero, one needs to compute them on backgrounds that are slightly perturbed away from the constant ones. One also needs to compute the first order corrections to the allowed diffeomorphisms \eqref{solFUV} about these backgrounds, which turn out to display  winding modes with exactly the same structure as \eqref{fdepwind}; these winding modes are required for consistency of the asymptotic
coordinate transformations. The charge algebra that one obtains at the first non-trivial order is non-linear in the conserved charges $Q_f, \bar Q_{\bar f}$; the source of non-linearity is the presence of charge-dependent parameters, namely the windings. This algebra precisely matches the algebra of the quasi-local generators   in single-trace $T\bar T$- deformed CFTs, which is given by \eqref{clsttbnlalg} with $R \r R p$. More precisely, the $U,V$ components of the asymptotic symmetry generators reproduce the algebra \eqref{clsttbnlalg}, but without the term proportional to the central charge. The  components of the asymptotic diffeomorphisms labeled by $f_{r}(u)$ produce charges  that commute with each other, but  have the following commutators with the charges labeled by $f(u)$

\be
\{ Q_f, Q_{f_r}\} = \frac{ k}{2\pi} \int d\s \p_\s u f_r''(u) f (u) \label{aldcentralterm}
\ee
and similarly for the right-movers. By choosing $f_r = -p f'/4$, one can also reproduce the central term in \eqref{clsttbnlalg}, with the correct coefficients. While this is a strong hint in favour of making this choice, the symplectic form analysis is not able to impose it. We note this algebra was also studied from a worldsheet perspective in \cite{Du:2024bqk}, who found a similar structure in the perturbative subsector of the bulk string theory, which is different from the typical high-energy states we discussed herein.

 The calculation described above establishes an additional important
link between asymptotically linear dilaton backgrounds and the single-trace $T\bar T$ deformation, by
showing that the full doubly-infinite set of extended symmetries of the two theories are identical
to the order  checked.  Thus, we see that  another  universal quantity - the symmetries - perfectly matches on the two sides, supporting the idea that, while  the two theories are clearly not the same, they do belong to the same universality class. 

%Thus, we see another extremely strong hint of a match between the ALD background and single-trace $T\bar T$ at the level of another , namely the extended symmetries. 

% \textcolor{red}{\emph{More?}}

%
%\bi
%\item is $\omega_r =0$ throughout, or just asymptotically?
%%\item what would be the algebra if we didn't make choice for $f_r$?
%%\item setup perturbative ASG and perfect match to $T\bar T$
%%\item ambiguity assigning functions
%%\item comment recent paper Wei ASG from worldsheet
%\ei

%We thus take a diffeo in Einstein frame radial gauge (important!) and the constant modes that simply vary the black hole parameters. There are some total derivative terms that violate the above falloff conditions which are thrown out, nevertheless. 

%The asymptotic diffeos are parametrised by four functions $f(u), \bar f(v)$ of the field-dependent coordinates. These precisely agree with the functions xxx in the constant $L_u, L_v$ limit. 

%Note, unlike in double-trace $T\bar T$, where the boundary conditions are known and the vevs derived, here we are building the allowed solution starting with the constant one. The diffeos we found are only good to leading order in the perturbation around constant. \emph{What are the boundary conditions for the ALD metric in Einstein frame?}

\subsubsection{Other observables: entanglement entropy}

Another interesting universal probe into the structure of a QFT is given by entanglement entropy, which has an extremely simple, geometric holographic implementation.  Systems with Hagedorn behaviour are particularly interesting to study from an information-theoretic perspective, given that the split property of local QFT is expected to fail. The entanglement entropy between an interval of length $L$ in the boundary QFT and its complement was computed holographically in \cite{Chakraborty:2018kpr},  using the    Ryu-Takayanagi formula. Their results nicely display non-local features, as expected of the non-local boundary theory. 

More concretely, just like in AdS, the area of the extremal surface is divergent, and needs to be regulated by a short-distance cutoff $\e$. The non-locality of the boundary theory manifests itself in that the Ryu-Takayanagi surface stops existing when the length of the boundary interval is smaller than 
a minimum length $L_{min}$ set by the little string length

\be
L_{min} = \frac{\pi}{2} \sqrt{k \a'} = \frac{1}{4} \b_H
\ee
where $\b_H$ is the inverse Hagedorn  temperature of LST.  The expression for the vacuum entanglement entropy when the length of the boundary interval, $L$, is large is given by
\be
S_{EE} = \frac{2\, c\, L_{min}^2}{3\pi^2 \e^2} +\frac{c}{3}\, \ln \frac{L}{\e} + \O (L_{min}^2/L^2)
\ee
%where $\b_H$ is the Hagedorn inverse temperature $2\pi \sqrt{k \a'}$. 
For $L \approx L_{min}$, \cite{Chakraborty:2018kpr} find
\be
S_{EE} = \frac{2\,c\, L_{min}^2}{3\pi^2 \e^2} +\frac{c}{6} \ln \frac{4 L_{min}(L-L_{min})}{\e^2} + \O(L-L_{min})
\ee
The second term in the large $L$ expansion is similar to the vacuum entanglement entropy in  a CFT$_2$. The first term - which does not depend on the length of the interval, and would thus drop out from the renormalized entanglement entropy - could be due to contact terms.  At finite temperature, the prefactor of the log term is found to be temperature-dependent. %\textcolor{red}{\emph{Expectations from field theory?}}

This computation has been extended in \cite{Asrat:2020uib} to two intervals on the  boundary of the ALD spacetime. In particular, the mutual information was computed, which is expected to be finite when the intervals are disjoint.  Introducing again a near-boundary cutoff $\e$ to regulate the area computation, the authors find indeed that the holographic mutual information is independent of $\e$; they nonetheless find a divergence of the form $(d-L_{min})^{-1}$ as the intervals approach each other, with $d$ the distance between them. This signals the expected failure of the split property in theories with a Hagedorn density of states.

%\bi
%\item comment Hagedorn behaviour from algebraic field theory
%\ei

\subsection{Warped AdS$_3$ and single-trace $J\bar T$ \label{wads3stjtbsec}  }

There is an entirely analogous story  that relates  warped AdS$_3$ backgrounds  in string theory supported by purely NS-NS flux and a single-trace $J\bar T$-deformed CFT.

\subsubsection{Worldsheet considerations}

One again starts from AdS$_3 \times S^3 \times T^4$ supported by $k$ units of purely NS-NS flux, which is holographically dual to the D1-D5 CFT at a singular point in its moduli space. Using  the holographic dictionary between (current) operators in the boundary CFT and worldsheet vertex operators, one again constructs a spacetime operator that has the same OPEs as some $U(1)$ current, $J$, on the left and the right-moving stress tensor, $\bar T$ on the right, and is given by a single worldsheet integral. The $U(1)$ current can either be an isometry of the $T^4$ \cite{Chakraborty:2018vja}, or be associated to the  $U(1)$ Hopf fibre when the $S^3$ is viewed as a fibration over $S^2$ \cite{Apolo:2018qpq}. Using the same logic and with the same caveats as in the previous subsection, one then argues that the resulting operator should be thought of as single-trace $J\bar T$, $\sum_{i=1}^p J_i \bar T_i $ in the approximate symmetric product orbifold CFT \eqref{spod1d5} that describes the D1-D5 CFT on the singular locus. 

Adding the integrated operator to the boundary CFT action can again be mapped to an exactly marginal deformation of the worldsheet WZW model, of the form $K \bar J^-$, where $K$ is a worldsheet current that either belongs to the $U(1)^4$ factor describing the $T^4$, or the third component of the left-moving $SU(2)$ current describing the sphere, while  $\bar J^-$ is, as before, the corresponding component of the $SL(2,\mathbb{R})$ current. Since the deformation is exactly marginal, it corresponds to a consistent new string background and can be turned on a finite amount, $\l$. The corresponding  deformation of the massless BTZ background reads 
\be \label{wadsbckgnds}
ds^2 = \sum_{i=1}^4 dy_i^2 + + \frac{1}{4} \left(d\theta^2 +\sin^2\theta d\phi^2 + (d\psi +\cos \theta d\phi)^2\right)  +  k \frac{dr^2}{r^2} + r^2 dx^+  \left\{ \begin{array}{c}  dx^- + \l (d\psi + \cos \theta d\phi)
%\,, & U(1) \in T^4
\\[3pt] dx^- + \l dy_1  % \, , & U(1) \in S^3
 \end{array} \right. 
\ee
for $U(1) \in S^3$ and, respectively, $U(1) \in T^4$, and the geometry is supported by purely NS-NS flux.  If one completes the square in the compact part of the geometry, $T^4 \times S^3$, one obtains precisely a null warped AdS$_3$ factor in the remaining non-compact directions. Thus, these setups can be used as  toy models for the Kerr/`CFT' correspondence.  While the geometry for $U(1) \in T^4$ is simpler, being given, at least in the vacuum case, by a simple shift $x^- \r x^- +\l y_1$, the $U(1) \in S^3$ setup is more closely related  to  Kerr/`CFT' proper, which generally involves a squashing of the sphere. In fact, the upper background \eqref{wadsbckgnds} has been previously studied in the Kerr/`CFT' context in  \cite{Azeyanagi:2012zd}.  %\textcolor{red}{\emph{Or, was this background related by a shift?}}

The  deformed worldsheet theory is again solvable, using the fact it is a current-current deformation. % \cite{Chakraborty:2018vja} shows the deformation precisely reproduces the single-trace $J\bar T$ deformation of the spectrum in the Ramond sector. 
%
%. Its spectrum has been solved for using various methods. \emph{Describe!}
 The conclusions  of its study are exactly the same as in the  case of the ALD background 

\bi
\item the  long string spectrum \emph{perfectly matches} that of a  single-trace $J\bar T$-deformed CFT \cite{Apolo:2018qpq,Chakraborty:2018vja} %spectrum in the long string sector
\item the deformed  short string spectrum  does not match
\item   long string correlation functions  almost match \cite{Chakraborty:2023wel}  $J\bar T$ ones, while those of short strings do not  %, as the mome-dep  $SL(2,\mathbb{R})_L$ dimensions are easily computed 
\ei
The latter were computed in  \cite{Azeyanagi:2012zd}, using the fact that the deformed background is related via a TsT transformation to the original AdS$_3$ one. The vertex operators in the deformed theory may, as in the ALD case, be expressed in terms of the vertex operators in the undeformed one, with some appropriate dressing. %This has been worked out in  \cite{Azeyanagi:2012zd} for the  warped $AdS_3$ background  and in  \cite{Cui:2023jrb} for the asymptotically linear dilaton one. % From the OPE of the deformed energy-momentum tensor with the deformed vertex operators one can deduce that the dressing factors induce a shift in conformal dimensions, which is referred to as spectral flow \footnote{This spectral flow should not be confused with the usual spectral flow automorphism on the worldsheet.}.  Very recently, \cite{Cui:2023jrb} has been able to compute correlation function of the associated vertex operators in ALD. 
Their correlation functions  still take the form \eqref{VVws2pt}, %{\color{red}{(Comment: I think the worldsheet vertex operator correlation functions in the deformed theory can only be written in momentum space. I don't know of a neat way of writing worldsheet correlation function in x-space.) }} 
but the relationship between $h$ and $\Delta_h$ in \eqref{defdeltah} is modified by the shifts \cite{Azeyanagi:2012zd}% \textcolor{red}{\emph{Notation k!}}
%\textbf{\emph{Check factors!}}  
%{\color{ForestGreen}(we need to recheck factors)}
%
\be
\d_{wAdS} \Delta_h =   \d_{wAdS}\bar  \Delta_h = \l\bar{p}\left(q^{[0]}+\frac{\lambda \hat k \bar{p}}{4}\right)
\ee
This relation holds also
for long string operators, for which $h$ is not related to the $SL(2,\mathbb{R})$ Casimir parametrised by $j$. In this case, it simply translates into a shift of $h, \bar h$ that takes the form 

\be
%&T\bar T &: h \r h + \frac{\mu}{w\pi} p \bar p \;, \;\;\;\; \hspace{1.8cm}\bar h \r \bar h + \frac{\mu}{w\pi} p \bar p \nonumber \\[2pt]
%J\bar T & : 
 h \r h+\frac{\lambda q^{[0]}\bar{p}}{w}+\frac{\lambda^2 k \bar{p}^2 }{4 w}\;, \;\;\;\;\;\; \;\; \bar h \r  \bar h+\frac{\lambda q^{[0]}\bar{p}}{w}+\frac{\lambda^2 k \bar{p}^2 }{4 w}
\ee
%and can be obtained by combining \eqref{vir1} with the appropriate shift in \eqref{deltah}.
Note this shift is different from the one \eqref{momdephhb} obtained using the primary analogues proposal of \cite{Guica:2021fkv,Chakraborty:2023wel}.  It would be interesting to understand the origin of this mismatch. Of course, it could simply be that the string theory vertex
operators  are a different basis of operators from the of the primary analogues of \cite{Guica:2021fkv}. Since the latter are fixed by their  Ward 
identities with the right-moving charges, which were simply postulated in \cite{Guica:2021fkv},  it would be interesting to better understand the Ward identities  in the string theory setting and how they originate. 

%\textcolor{blue}{ Remember, in par-ticular, that in our field-theory analysis we did encounter shifts of the right-movers that involved J0 instead of $\bar J0$; however, the discrepancies produced by these terms disappeared in the $R \r \infty$limit27. It would be interesting to perform a more careful worldsheet analysis, possibly on the cylinder, in order to track this discrepancy. Of course, it could also be that the string theory vertex operators simply are a different set of operators from those for which we computed correlation functions in field theory. Our criterion for fixing the right-moving piece of the operators in field theory was based on symmetries, i.e. by requiring that they satisfy CFT-like Ward identities with respect to the right-moving generators (3.9), which were related to the flowed right-moving Virasoro generators by an operator-dependent spectral flow. Since these Virasoro symmetries are not yet understood from the worldsheet perspective, the analogous way to fix the operator basis is not yet available on the string theory side. Conversely, it could be that consistency of the worldsheet vertex operators (e.g., mutual locality) would single out a different set of constraints on the operators that would be natural to impose. In any case, it would be interesting to further explore the properties of the two sets of operators and check whether one may be preferred to the other.}

For completeness, let us mention that exactly the same method may be used to compute correlation functions of the short string vertex operators; one simply needs to reinterpret the relation between $\D_h$ and $h$ for the discrete $SL(2,\mathbb{R})$ representations on the worldsheet. For  $w=0$ short strings, %.  Since now the representation is  lowest-weight, the spacetime dimension is related to the worldsheet Casimir as $h=j+1$. Taking into account the shift in the relation between $\D$ and the worldsheet dimension, 
one finds 
\be
%\hspace{-0.01cm}h_{ALD} = \frac{1}{2} + \sqrt{\left(h-\frac{1}{2}\right)^2+\frac{\mu N_5}{2\pi}\,p\bar{p}}\;, \;\;\;\;\;
 h_{wAdS} =\frac{1}{2}  + \sqrt{\left(h-\frac{1}{2}\right)^2 + \lambda k  q^{[0]}\, \bar{p}+\frac{\lambda^2 \hat k k}{4}\, \bar{p}^2}  \label{hwads}
\ee
As before, the perfect match between the spectrum of long strings in the backgrounds \eqref{wadsbckgnds} and that of  a single-trace $J\bar T$ - deformed CFT for the corresponding choice of $U(1)$ current  does not extend to the full theory. %\textcolor{red}{\emph{Coincidence?}} 
Consequently, there cannot be an exact holographic duality between these backgrounds and single-trace $J\bar T$. One may nonetheless ask whether a relation between universal quantities such as the entropy and the extended symmetries of these particular warped AdS$_3$ spacetimes and single-trace $J\bar T$  can be found, as was the case for the AdS$_3 \r $ALD background and $T\bar T$.

% does not lead to any conclusion about the holographic dual to the full theory, which clearly cannot be a SPO (except for $k=1$). 

%
%\bi
%\item worldsheet analysis for finite $T$? Subtleties interpretation? 
%\ei

We end by noting that  the worldsheet analysis of the long string spectrum  can be generalised to mixed $T\bar T, J\bar T$ etc combinations, and again a perfect match is found to  field theory results \cite{Chakraborty:2019mdf}. % (for $w=1$). 

\subsubsection{Thermodynamic considerations}

In order to check the entropy-to-energy relation at high energies, one needs to first write down the corresponding warped AdS$_3$ black hole solutions.   Unlike the  ALD background, which could be obtained from a  decoupling limit, including for finite-temperature states, the backgrounds \eqref{wadsbckgnds} are only obtained from AdS$_3$ by solution generating techniques, such as combinations of coordinate shifts and T - dualities, and no decoupling limit to $w$AdS$_3 \times S^3$ supported by purely NS-NS flux is known at this stage. This should be contrasted with the case where the $w$AdS geometry is supported by RR three-form flux, where the decoupling limit is the dipole one \cite{Bergman:2000cw}. 

The problem with this state of affairs is that it is unclear how to select the finite-temperature background to study. For example, for the case where the deforming $U(1)$ current is an isometry $\in T^4$, the relation between the zero-temperature background and AdS can be seen as either a shift, $x^- \r x^- + \l y^1$, or a TsT transformation, where the shift is accompanied by a T-duality along $y_1$, before and after. While the effect of these transformations is the same when applied to  Poincar\'e AdS$_3$, they yield different results on BTZ.  Moreover, multiplying the dilaton by an arbitrary function of the temperatures (independent of the coordinates) still solves the equations of motion, and represents an important ambiguity in identifying the  background; indeed, the entropy-energy relation can be entirely modified by such a multiplication. Similar problems exist in identifying the $U(1)$ charges.   Note that, for understanding the thermodynamics of the system, it is important to turn on \emph{completey general} charges.

Nonetheless, \cite{Apolo:2021wcn} were able to devise a criterion that was likely to correspond to a meaningful geometry. 
 Since in $J\bar T$ - deformed CFTs there are two $U(1)$ charges for $J$ non-chiral, they first generated a warped AdS$_3$ black hole by applying a TsT transformation to the charged BTZ black hole $\times S^3 \times T^4$, supported by purely NS-NS flux. This solution is characterised by four parameters: two temperatures, $T_{L,R}$ and two $U(1)$ charges, $Q_{L,R}$ of the original charged BTZ, in addition to the shift parameter, $\l$. One problem it presents is that the number of F1 strings in the background, counted by $\int e^{-2\Phi} \star H$, is not integer-quantized. The fix proposed by  \cite{Apolo:2021wcn} was to simply shift the value of the dilaton by a (parameter-dependent) constant, so that the F1 charge coincides with that of the undeformed background. Since the background is purely NS, this constant shift does not affect the equations of motion. Quite remarkably, upon making this simple modification to the solution, the entropy-energy relation \emph{precisely becomes} the single-trace $J\bar T$ one, \eqref{entropyJTbarFT}.   The identification of the $U(1)$ charges is 

\be
Q_L \;\;\leftrightarrow \;\;\; \frac{1}{2} (Q_{\p_y} - Q_{dy})\;, \;\;\;\;\;\;\;\; Q_R  \;\;\leftrightarrow \;\;\; - \frac{1}{2} (Q_{\p_y} + Q_{dy}) + \frac{\l}{2} Q_{-\p_{V}}
\ee
 following \cite{Georgescu:2025jlx}, where the solution for just the $J\bar T$ case is isolated from the more general analysis of \cite{Apolo:2021wcn}. $Q_{\p_y}$ and $Q_{-\p_{V}}$ are the global conserved charges associated with translations along $y$ and, respectively, $V$,  while $Q_{dy}$ is the charge associated with the gauge transformation of the $B$ - field given by $\Lambda = dy$. 

%,  one can easily generate a warped AdS$_3$ background from AdS via a TsT along one of the null AdS$_3$ directions and a compact $U(1) \in S^3 \times T^4$, as already discussed in section \ref{wads3toym}.

Of course, since the background of \cite{Apolo:2021wcn} is still obtained via a solution-generating technique, rather than a decoupling limit, it is not clear there exists a consistent   theory dual to it. However, the fact that i) the F1 charge is quantized and ii) the thermodynamics is identical to that of single-trace $J\bar T$ - deformed CFTs, which are well-defined  QFTs with an explicit construction, gives credence that this particular procedure of generating the background yields a meaningful answer.

%\noindent As discussed in the ALD case, a much more far-reaching piece of evidence for a holographic link between the ALD spacetime and $T\bar T$ was the perfect match of black hole entropy. However, this comparison is hard to make precise for the wAdS$_3$ backgrounds, because it is not known how (and if) to obtain them from a decoupling argument.  The importance of the decoupling argument is that it indicates which quantities (time, deformation parameters) are expected to be fixed at finite temperature. While the TsT does produce a solution with the desired asymptotics, its parametrisation may be off, as we already saw in the case of ALD, where both methods are available and can be compared. 
%
%The proposal of \cite{} on how to fix the parametrisation is to modify the temperature/charge dependence of the dilaton in such a way that the number of F1 strings remains quantized - a rather sensible requirement. The solution is 
%
%\be
%ds^2 = xxx \;, \;\;\;\; e^{2\Phi} = x \cdot x
%\ee
%Remarkably, with this choice of parametrisation, the entropy-to energy and charge relation becomes precisely the $J\bar T$ one, \eqref{}. \emph{How about the charge quantization conditions? What would be $S(E,Q)$ for TsT only? Comment on previous Wei-Luis paper?}

\subsubsection{Asymptotic symmetries}

The asymptotic symmetries of the warped AdS$_3$ background put forth in \cite{Apolo:2021wcn} were studied in \cite{Georgescu:2025jlx} using the symplectic form approach we discussed in the ALD case. More precisely, the ASG generators were fixed by the requirement that 

\be
\om_{ab...} (\d_{\xi, \Lambda} \Phi, \d_{T_{L,R}} \Phi) =\om_{ab...} (\d_{\xi,\Lambda} \Phi, \d_{Q_{L,R}} \Phi) = 0 
\ee
namely, the symplectic product between an allowed diffeomorphism or gauge transformation and a constant variation in the parameters of the background should vanish, as they  both correspond to allowed perturbations inside the phase space. 
To simplify the problem, \cite{Georgescu:2025jlx} set to zero a potential radial component in the diffeomorphisms considered, which is the likely reason that the charge algebra resulting from this analysis is not centrally extended.
 Quite remarkably,  the asymptotic symmetries obtained via this method precisely agree with the the allowed transformations for the locally AdS$_3$ bulk dual of a $J\bar T$ -deformed CFT; in particular, the winding modes of the affine transformations are visible, with the expected \eqref{windingeta} charge-dependent coefficients. These  in turn result in the asymptotic symmetry algebra being non-linear.
The result is precisely consistent with the single-trace $J\bar T$ algebra of the the quasi-local generators (namely, \eqref{jtbquasilocalg} with $R \r R p$) to the order in linearized perturbations about a constant background to which it was computed. %\textcolor{red}{\emph{Any issues?}}

%\noindent The asymptotic symmetries of this pure NS wAdS background have been studied many times. Obtained: Virasoro, VKM, worldsheet inconclusive. \emph{New worldsheet?}

%In this section, we would like to discuss the approach of \cite{}, who did the symplectic form. Perfect match to $J\bar T$. \emph{True? Issues?}

This perfect match to $J\bar T$ extended symmetries would not have been obtained if the dilaton were not shifted so as to reproduce $J\bar T$ thermodynamics. %\textcolor{red}{\emph{True?}}.
 There does thus appear to be a link between extended symmetries and the thermodynamics of the system, that would be interesting to explore further. %\textcolor{red}{\emph{Later?}}

%\newpage

\subsection{General comments on the interpretation of these correspondences\label{gencommentsint}}

While the initial suggestion of \cite{Giveon:2017nie,Apolo:2018qpq,Chakraborty:2018vja} was that string theory in the ALD and warped AdS$_3$ backgrounds \eqref{wadsbckgnds} is holographically dual to a single-trace $T\bar T$ and, respectively, $J\bar T$ - deformed CFT, the relations between the two sides have turned out to be more subtle, as  reviewed in this section.

First, if $k>1$, an exact duality relation between string theory in these backgrounds and a symmetric product orbifold of $T\bar T/ J\bar T$ - deformed CFTs is not possible, for the very simple reason that, even before the irrelevant deformation, the CFT dual to AdS$_3$ does not take a SPO form. Rather, the SPO \eqref{spod1d5} is only meant to describe certain properties of long strings in AdS$_3$; a more complete description thereof that takes into account interactions  breaks the SPO structure by a twist two operator \cite{Eberhardt:2021vsx}, and even this corrected theory is only able to capture perturbative string states in AdS$_3$, but not black hole microstates\footnote{Only in the tensionless limit, $k=1$, is it in principle possible to have an exact duality \cite{Dei:2024sct}. We concentrate on $k>1$. } \cite{Chakraborty:2025nlb}. 

Since the undeformed CFT does not have a SPO structure, it is not clear what is meant by the single-trace $T\bar T$ or $J\bar T$ operator at the level of the full theory; what is clear, though, is that there exist gauge-invariant operators in this CFT of dimension $(2,2)$ and, respectively, $(1,2)$ that are dual to the non-normalizable AdS$_3$ deformations that seed the ALD and warped AdS$_3$ spacetimes\footnote{Sometimes, these irrelevant deformations are called `single-trace $T\bar T/ J\bar T$' in the sense that modes of the bulk fields are dual to single-trace operators. We find this terminology confusing, since single-trace $T\bar T/J\bar T$ have a very clear definition, \eqref{defstttb}, which requires a SPO structure and implies the theory is solvable; neither of these properties seem applicable here. %We will therefore not use this terminology. % It would be interesting to undestand whether a more general notion of `single-trace' holds throughout the D1-D5 moduli space.
 }.  These operators, nonetheless, only appear to act as single-trace $T\bar T/J\bar T$ on the long string subsector of the CFT. Accordingly, once the irrelevant deformation is turned on,  the deformed long string spectrum matches a single-trace $T\bar T/J\bar T$ - deformed one, and one can obtain valuable suggestions for the form of correlation functions in $T\bar T/J\bar T$ - deformed CFTs. These perfect matches of certain observables in the long string subsector sometimes go under the name of TsT/$T\bar T$  correspondence \cite{Du:2024bqk}.  At the same time, if one considers
%
%If one continues to concentrate on perturbative string states, but includes
 the short strings, then the breaking of the SPO structure in the seed CFT is explicit, and one does not find any match between  observables (spectrum, correlators) computed via string theory in the deformed background and single-trace $T\bar T/J\bar T$.

Thus, one does not have an exact duality between the proposed two sides for $k>1$. Given this state of affairs, it is interesting to inquire into the reason that the long string spectrum  matches. It appears likely that this is simply due to a coincidence, related to the fact that both the vacuum ALD and warped AdS spacetimes are related via a TsT transformation to AdS$_3$, and the associated non-local change of coordinates that relates the worldsheet theory to the one describing strings in AdS$_3$ has a similar effect on long strings as the $T\bar T$ deformation, as hinted in \cite{Sfondrini:2019smd}. It is not clear whether any of the $T\bar T/J\bar T$ special structure associated with long string survives once  moving off the singular locus in the moduli space since, at least from the  point of view of the IR AdS$_3$, these states are expected to lift.

 A likely more fundamental link between string theory on the ALD/warped AdS$_3$ backgrounds and single-trace $T\bar T/J\bar T$ is given by the match of universal quantities on the two sides, namely the entropy and the extended symmetries. As we mentioned,  the current status of the entropy match for the ALD versus warped AdS$_3$ pure NS backgrounds is slightly different, in that the ALD background is obtained via a well-defined decoupling limit, whereas warped AdS$_3$ is only produced via solution-generating techniques. Consequently, one exactly knows the parametrisation of the AdS$_3 \r$ ALD background and can unambiguously compute its entropy-to-energy relation, finding a highly non-trivial match to single-trace $T\bar T$. The decoupling limit also indicates the background is dual to a UV complete theory, which starts of as a $(2,2)$ irrelevant deformation of a CFT.  In the warped AdS$_3$ case, the finite-temperature solution is simply generated from charged BTZ via TsT, and it is \emph{a priori} unclear whether there exists a consistent dual theory  to this background\footnote{As we already mentioned, applying TsT  along the boundary directions to BTZ $\times S^3 \times T^4$ with NS flux almost produces the decoupled finite-temperature  ALD background, but with the wrong temperature-dependence for the dilaton \cite{Apolo:2019zai}, and thus the wrong thermodynamics.}. However, \cite{Apolo:2021wcn} found that a simple charge-dependent rescaling of the dilaton, which is fixed by requiring that the F1 charge be quantized, produces a system that has precisely single-trace $J\bar T$ thermodynamics. Since $J\bar T$ currently provides the only concrete example of a $2d$ dipole CFT, this match is highly suggestive. It is absolutely remarkable that, once the backgrounds are fixed and their thermodynamics reproduces that of single-trace $T\bar T/J\bar T$, the asymptotic symmetries, fixed via the minimalistic assumption \eqref{phs}, precisely reproduce the intricate extended symmetry structure of $T\bar T/J\bar T$ - deformed CFTs. These  matches are all the more interesting since they involve quantities that capture the \emph{full} theory, not just the partially decoupled long string subsector.

To summarize, one finds a perfect match, at the level of the full theory, between \emph{universal} quantities in string theory on the ALD/warped AdS$_3$ backgrounds and  single-trace $T\bar T/J\bar T$-deformed CFTs,  such as the entropy and symmetries; non-universal observables such as the low-lying spectrum generally do not match, except for a particular subsector where the agreement may well be accidental. 
This observation suggests that

\bi
\item $\exists$ non-trivial generalisations of  single-trace $T\bar T/J\bar T$-deformed CFTs  with the same UV properties %(corresponding to $2d$ compactifications of LST and generalisations thereof)
\item only \emph{universal} observables such as the asymptotic density of states  and symmetries  are expected to be the same; non-universal ones, such as the  spectrum, need not take a $T\bar T/J\bar T$ form
%\item one can also include RR backgrounds with ALD asymptotics as long as their thermodynamics and symmetries are $T\bar T$ -like (only sugra is required to establish this)
\ei
In other words, these suggest the existence of  \textbf{universality classes} of $T\bar T/J\bar T$ - deformed CFTs, for which the theories studied in section \ref{strttbjtbsec} are the absolutely simplest (and somewhat boring) representatives. The fact that the above irrelevant deformations are built upon a point in the D1-D5 moduli space that is not described by a SPO suggests the latter structure is not essential in defining such theories; it would though be interesting to understand which structures are.

%full holographic dual to string theory in these backgrounds, which is what the two above computations refer to, there is a link to single-trace  $T\bar T/J\bar T$  at the level of  only. 
 
From this perspective, one would expect that any consistent  ALD or warped AdS background, whether or not it is supported by pure NS or also RR flux, could belong to the above universality classes.  To test this conjecture, one should consider more general backgrounds, supported by both types of fluxes, with these asymptotics, and compute the entropy-to-energy relation and asymptotic symmetries. This is the subject of the next subsection.

\subsubsection{Generalised ALD spacetimes and Little String CFTs }

In this subsection, we would like to discuss additional examples \cite{Georgescu:2024iam} of dualities between a space-time with linear dilaton asymptotics and a generalisations of single-trace $T\bar T$. These will be generalisations of the setup of \cite{Giveon:2017nie} that also include RR fluxes. 

We start by considering $\frac{1}{2}$ BPS black string solutions in type IIB string theory compactified on K3. The resulting six-dimensional theory has a large moduli space, parametrized by $105$ scalars. These scalars are subject to an attractor mechanism \cite{Ferrara:1995ih}, whereby $21$ of them must reach prespecified values at the horizon,  determined by the charges of the black string. The remaining $84$ scalars are not fixed at the horizon and parametrise the moduli space of the D1-D5 CFT. 

From the viewpoint of the near-horizon AdS$_3$, the $21$ fixed scalars correspond to massive fields that are holographically dual to operators of dimension $(2,2)$ that preserve maximal supersymmetry. One additional $(2,2)$ operator\footnote{All these operators are descendants of the 22 single-trace $(1,1)$ chiral primaries in the D1-D5 CFT.} is associated with changes in the size of the $S^3$.  Given the standard identification between the radial direction in the bulk and the energy
scale in the boundary theory, as well as between the supergravity moduli and the coupling
constants on the brane worldvolume, the attractor flow may be seen, from the boundary
perspective, as a flow towards low energies that ends at an IR fixed point \cite{deBoer:2008ss}, where $22$ of the
couplings of the original theory become irrelevant in the IR. The approach of \cite{Georgescu:2024iam} was to look at this attractor flow in reverse, and ask which of the irrelevant deformations seen from the IR near-AdS$_3$ analysis could be followed up into the UV. 

Focussing on the NS5-F1 string, the $22$ maximally supersymmetric single-trace irrelevant deformations of its near-horizon geometry  (supported by purely self-dual NS flux) correspond to either turning on more self-dual three-form NS flux, $H_3^+$, or one of the $21$ anti-self-dual field strengths, $F_{\bar{\imath}}$, which can originate from either the NS three-form field, the RR one, or the reduction of the type IIB self-dual five-form   on the $19$ anti-self-dual two-cycles of K3. From the point of view of the supergravity solution describing the black string, whose metric is entirely determined by a set of harmonic functions, $H_\Lambda$

\be
ds^2= \frac{1}{\sqrt{H_\Lambda H^\Lambda}} (-dt^2+ d\s^2) + \sqrt{H_\Lambda H^\Lambda } (dr^2+r^2 d\Omega_3^2) \;, \;\;\;\;\;\; H_\Lambda = \frac{\a' q_\Lambda}{r^2} + c_\Lambda \label{bsmet}
\ee
they correspond to simply turning on the constants $c_\Lambda$ in the harmonic functions, which takes the geometry away from AdS$_3 \times S^3$. The irrelevant couplings $\l_+$ and $\l_{\bar \imath}$ of the operators map to 

\be
\l_+ \propto  c_\Lambda q^\Lambda \;, \;\;\;\;\;\;\;\; \l_+^2 - \l_{\bar \imath}\l_{\bar \imath} \propto  c_\Lambda c^\Lambda
\ee
and are all proportional to $\a'$. 

It is interesting to ask what the purely NS background of \cite{Giveon:2017nie} corresponds to in this language. It is easy to see that the answer is $\l_+ = \l_{\bar 1}$ and all other $\l_{\bar \imath}$ zero, where  $\l_{\bar 1}$ is the irrelevant coupling corresponding to $H_3^-$ flux. The ALD asymptotia of the solution are related to the fact that $c_\Lambda c^\Lambda = \l_+^2 - \l_{\bar 1}^2=0$. It is then quite natural to define the most general form of `ALD' asymptotics by the requirement that the asymptotic metric \eqref{bsmet} degenerate, namely $ c_\Lambda c^\Lambda =0$. There are $21$ irrelevant deformations that span the   spacetimes with such asymptotics, for which the self-dual irrelevant coupling $\l_+$ is determined by the remaining $\l_{\bar \imath}$ via 

\be
\l_+^2 = \l_{\bar \imath} \l_{\bar \imath} \label{constrirr}
\ee
Of course, the analysis of the irrelevant couplings only holds near the IR fixed point. To argue that the irrelevant deformations satisfying \eqref{constrirr} lead to  UV-complete theories, \cite{Georgescu:2024iam} identified various decoupling limits of NS5-branes, also in presence of RR fluxes, that are dual to decoupled spacetimes with the expected degenerate asymptotics. The corresponding boundary theories corresponds to various generalisations of LST, known as the $OD1$ (open D1-brane) and $OD3$ theories \cite{Gopakumar:2000ep,Harmark:2000ff}, compactified to $2d$. The $OD1$ theory is S-dual to the D5 NCOS (non-commutative open string) theory, a space-time non-commutative theory of open strings only, obtained by tuning the electric field on the D5 worldvolume to a critical value \cite{Seiberg:2000ms,Gopakumar:2000na}. This gives a glimpse into the physical properties expected of the holographic duals to ALD spacetimes.

%
%\begin{figure}[h]
%    \centering
%     \captionsetup{width=.8\linewidth}
%        \subfloat{{\includegraphics[width=6.7cm]{aldv3} }}%
%    \qquad\qquad\qquad
%    \subfloat{{\includegraphics[width=4.9cm]{mlstv2} }}%
%    \caption{\footnotesize{The asymptotics of the black string solutions degenerate on a codimension one subspace of the moduli space at infinity - $\mathcal{M}_{LST}$ - which is spanned by known decoupling limits of string theory that yield little string theory and  deformations thereof. It corresponds to turning on $21$ of the possible $22$ maximally supersymmetric irrelevant couplings. }}%
%    \label{fig:3}%
%\end{figure}

%\medskip

\begin{figure}[h]
\begin{minipage}{0.45 \linewidth}
\centering
\includegraphics[height=4.5cm]{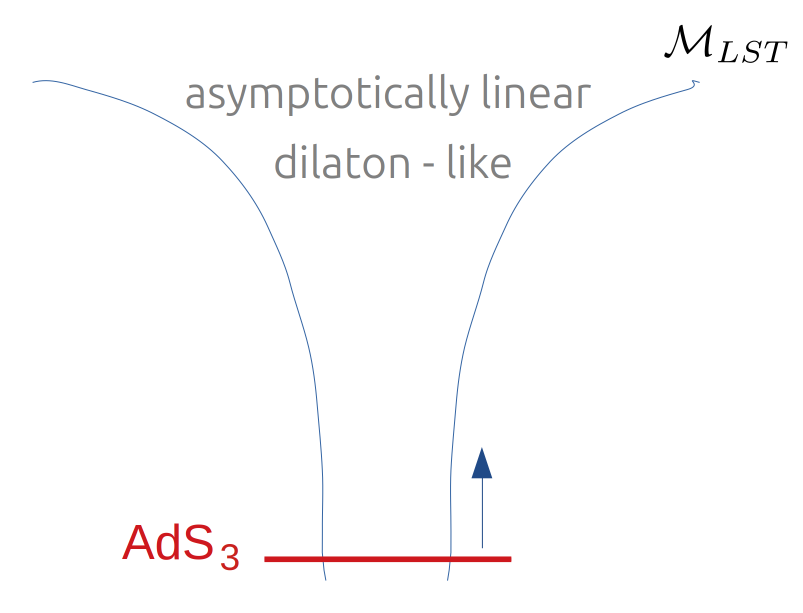}
%\caption{The dimensionless energy $E R$ as a function of the dimensionless coupling $\mu/R^2$ for $\mu>0$. }
\end{minipage}
\hspace{1cm}
\begin{minipage}{0.45 \linewidth}
\centering
\includegraphics[height=5cm]{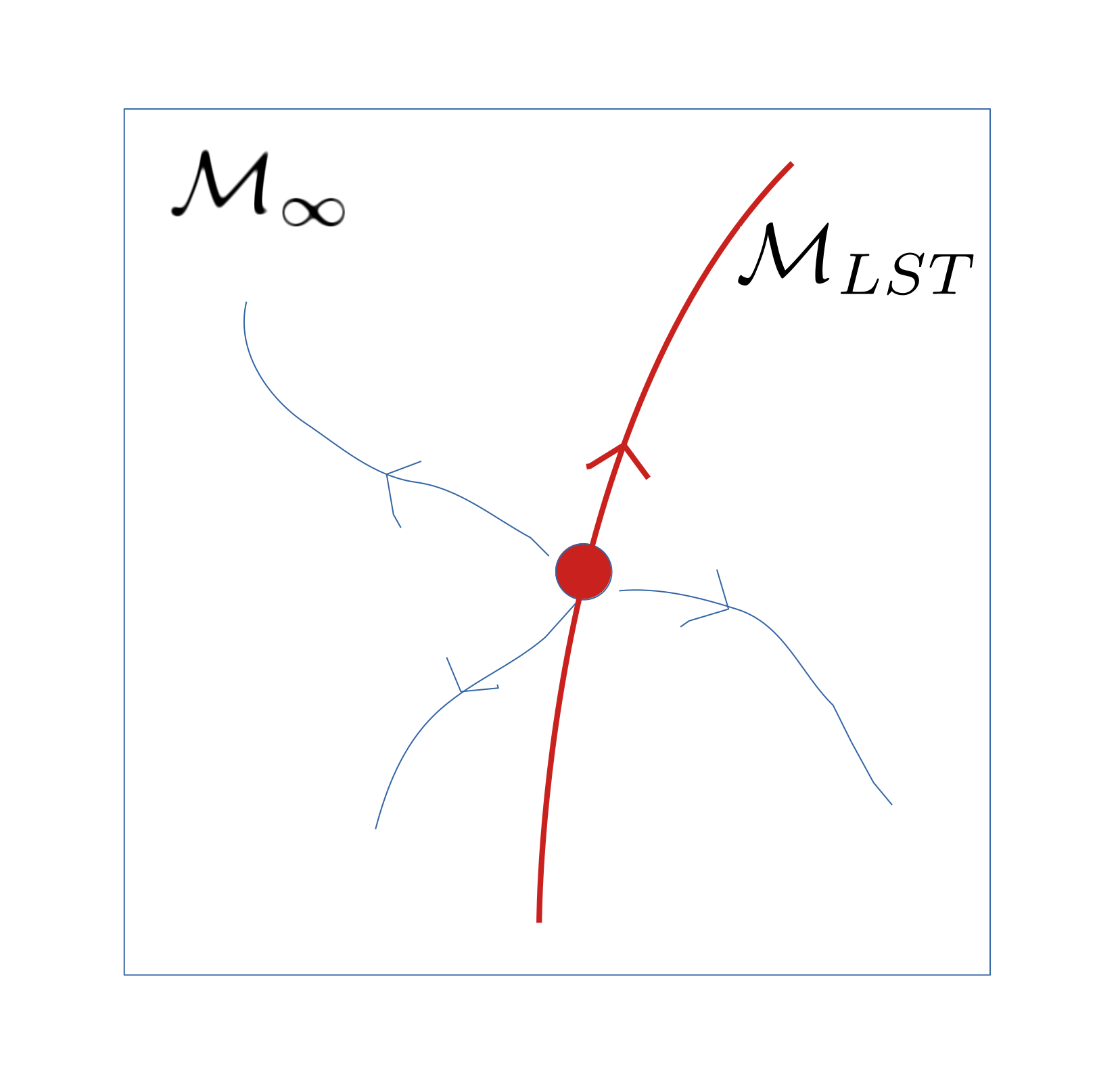}
%\caption{The dimensionless energy $E R$ as a function of the dimensionless coupling $\mu/R^2$ for $\mu<0$. }
\label{plotemuneg}
\end{minipage}
\caption{\footnotesize{The asymptotics of the black string solutions degenerate on a codimension one subspace of the moduli space at infinity - $\mathcal{M}_{LST}$ - which is spanned by known decoupling limits of string theory that yield little string theory and  deformations thereof. It corresponds to turning on $21$ of the possible $22$ maximally supersymmetric irrelevant couplings. }}
\end{figure}

\noindent Consequently, the theories labelled by $\l_{\bar \imath}$ and satisfying \eqref{constrirr} are all expected to correspond  to  UV-complete $2d$ non-local theories. Since they are all strongly coupled, their properties are only accessible via holography. One may consider  the finite-temperature case, obtained by taking the decoupling limit on the corresponding nearly-extremal (rather than exactly extremal) brane system.  Quite remarkably, after carefully keeping track of the decoupling limit (which is somewhat subtle in the NCOS case, especially if one uses TsT to generate the background), \cite{Georgescu:2024iam}  showed that they all follow the Cardy $\r$ Hagedorn behaviour that is characteristic of single-trace $T\bar T$, with the same irrelevant coupling $\a' k$.

 Thus, all these theories - which are best defined as compactifications  of various generalisations of LST to $2d$ -  appear to fall into the same universality class as single-trace $T\bar T$.  We propose to call them \emph{little string CFTs} - LS-CFTs in short - as we further elaborate in the next section. 

 %{\color{blue}%One of the main goals of this article is to argue that this picture extends to the full moduli space of the D1-D5 system. Namely, we show that all instances in which the asymptotics of the general half-BPS black string solution degenerate to an ALD-like spacetime - which happens on a codimension one subspace of the moduli space at infinity - correspond to a known decoupling limit of string theory. We thus expect that the irrelevant deformations whose coefficients satisfy the corresponding field-theoretical constraint  yield a  \emph{UV-complete} theory (see figure \ref{fig:3}). 
%These deformations span a $21$-dimensional subspace of the $22$ possible maximally supersymmetric irrelevant deformations of the system.
 The remaining irrelevant direction - parametrised by $\l_+$ only -  turns on the deformation to six-dimensional asymptotically flat space, for which a rigorous decoupling limit is not known.  It would be very interesting to be able to differentiate this deformation from the rest from an IR perspective.

\subsubsection{Warped AdS$_3$ backgrounds and dipole CFTs\label{wadsvsdipcft}}

In this subsection, we would like to address the question: to what extent do generalisations of pure NS the warped AdS$_3$ backgrounds \eqref{wadsbckgnds} fit into a would-be single-trace $J\bar T$ universality class?

A classification of warped AdS$_3$ solutions of type IIB string theory on K3 that is  similar to the one above was performed in \cite{ElShowk:2011cm}, concentrating on solutions that preserve an $SL(2,\mathbb{R})_L \times SU(2)_R \times U(1)^2$ subgroup of the original $AdS_3 \times S^3$ isometries that are more interesting from the point of view of the Kerr/`CFT' correspondence. These deformations can be classified into whether they correspond to adding more self-dual flux, or one of the $21$ anti-self-dual fluxes, for a total of $22$ deformations (they are built from the same $(1,1)$ chiral primaries that are involved in the ALD case). An important difference with the ALD solutions we discussed is that, for all but one of these backgrounds, it is not known how to obtain them from a decoupling limit. The known case corresponds to the (untractable) dipole deformation  \cite{Bergman:2000cw} of the D1-D5 CFT, which is simply obtained by applying a TsT transformation to the purely RR D1-D5 background.

Finite-temperature generalisations of these solutions can be obtained using a combination of  TsT transformations, S-duality, and $6d$ electromagnetic duality \cite{Bena:2012wc}. The thermodynamics of a two-parameter set of these solutions was studied in \cite{Detournay:2012dz}. From our renewed perspective, this analysis has two drawbacks: i) it is unclear whether the temperature-dependence of the backgrounds is correct, given that it is not known how to obtain them from a decoupling limit and ii) the possible $U(1)$ charges are turned off. Or, we recently emphasized the importance of turning on arbitrary $U(1)$ charges in order to properly test the entropy formula.   It would be interesting to revisit this work from the new perspective of \cite{Apolo:2021wcn}. %\textcolor{red}{\emph{Note that if we simply strip off a factor from the dilaton, the entropy looks like Cardy. How come for $\l_2 =0$?}}

As mentioned, the only known decoupled geometries in the set are the  dipole-deformed D1-D5 backgrounds, whose thermodynamic  analyses  are therefore  trustworthy. Moreover, \cite{Georgescu:2025jlx} has recently studied the entropy and asymptotic symmetries of the dipole background  with the deforming $U(1) \in T^4$
%\textcolor{red}{\emph{Clarify definitions!}}
 also in presence of $U(1)$ charges for the current involved in the TsT. It was found that the entropy of this black hole was given precisely by the charged Cardy formula \eqref{schcardy}, and not the $J\bar T$  one. Moreover, the analysis of the asymptotic symmetries selected by the radial symplectic form simply yielded a completely standard (Virasoro-Kac-Moody)$^2$ algebra, implemented by diffeomorphisms that depend on the standard left or right-moving coordinates on the boundary.

Therefore, it is not clear whether the dipole-deformed D1-D5 CFT is captured by the single-trace $J\bar T$ universality class, or not.  The form of the entropy and asymptotic symmetries still allow for the possibility of $J\bar T$ universal behaviour, but with respect to some other $U(1)$ current, whose charges were not turned on in \cite{Georgescu:2025jlx}, leading to no change with respect to the AdS case.   Another possibility is that the dipole-deformed D1-D5 CFT does not belong to the $J\bar T$ universality class, but to one where the entropy in this non-local theory takes an exactly charged Cardy form (for generic charges)  and the symmetry generators correspond to the standard, local Virasoro ones. It is not clear to us whether such a scenario is possible from a field-theoretical perspective. 
%\bi
%\item example dipole, pure Virasoro
%\item for $J\bar T$ classify backgrounds via soln gen, but no decoupling arg $\Rightarrow$ not established if leads to UV complete
%\ei

 %for dipoles, where the entropy is exactly Cardy, this method simply yields non field-dependent ASG \emph{True?} yileding two commuting Virasoros. \emph{True? (even without tweaking the symplectic form?)}

%\newpage

\section{From solvable irrelevant deformations back to black holes\label{backtobh}}

Having reviewed the progress, in the past almost ten years, in the $T\bar T$ and $J\bar T$ deformations both on the field theory and the gravity side, in this section we would like to take stock of what we have learned about the problems posed at the beginning of this review, namely the Kerr/CFT correspondence and
the microscopics of non-extremal black holes. Along the way, we will also comment on other non-AdS holographic dualities, as well as  possibly interesting  future directions.

 \etocsetnexttocdepth{5}
    \etocsettocstyle{\subsubsection*{Contents of this section: }}{}
    \cftsubsubsecindent 34pt
    \localtableofcontents

\subsection{New perspectives on old conjectures \label{newperspold}}

In this subsection, we discuss  what have we learned about the Kerr/CFT correspondence and the microscopic description of more general black holes from the study of $T\bar T, J\bar T$ - deformed CFTs, as compared to the status - reviewed in section \ref{bhstoirrelsec} - before these theories were  put forth.

\subsubsection{A $J\bar T$ perspective on Kerr/`CFT'}

Let us start by  reminding the reader the conclusion of the discussion in section \ref{bhstoirrelsec}. We argued that, among the various toy models that have been proposed for the Kerr/`CFT' correspondence, the most
  trustworthy ones  are the  ones derived from string theory, for which the holographic dual corresponds to a \emph{dipole CFT}:  a UV-complete $2d$ QFT that is local and conformal on the left, non-local on the right, and can be viewed as an irrelevant deformation of an IR CFT$_2$  by a leading operator of dimension  $(1,2)$. The properties of these theories were exclusively inferred from the dual spacetime, leading to  a Cardy form of the entropy, correlation functions that took a CFT form - but with momentum-dependent conformal dimensions - and an inconclusive result for the extended symmetries, 
   due to the lack of knowledge of the correct  boundary conditions to impose on metric fluctuations.  In the most common proposals, though,  the symmetries contain a Virasoro$_R$ factor that enhances the right-moving translations and  appears to be  in tension with the obvious non-locality of the dipole CFT on the right-moving side. 

 $J\bar T$ - deformed CFTs are concrete examples of dipole CFTs that can be solved exactly. In their context, several of the puzzles posed by the Kerr/`CFT' correspondence are  dispelled. In particular, they provide

\bi
\item an explicit construction showing that the Virasoro symmetry \emph{can} coexist with non-locality
\item a natural basis of operators,  whose correlators take precisely the form  of CFT momentum space correlators, but  with momentum-dependent conformal dimensions 
\ei
Thus,  the apparently incongrous properties   of dipole CFTs suggested by the bottom-up holographic analyses can be realised in  QFT and can be consistent with each other, at least in principle. This compatibility is implemented in a 
non-trivial fashion, and is heavily reliant on the special properties of the $J\bar T$ extended symmetry generators. %, whose structure is still far from being well-understood. 

In the figure below, we compare the properties of holographic dipole CFTs, as inferred from the bottom-up analyses of the dual spacetimes mentioned in section \ref{bhstoirrelsec},  with the corresponding observables in $J\bar T$ - deformed CFTs.   

%\vskip2.3cm

\begin{figure}[t]
\begin{minipage}{0.45 \linewidth}
\flushleft
\includegraphics[height=5.5cm]{kcft_triangle1}
\caption{Properties of Kerr/`CFT' }
\end{minipage}
\hspace{1cm}
\begin{minipage}{0.45 \linewidth}
\flushright
\includegraphics[height=5.5cm]{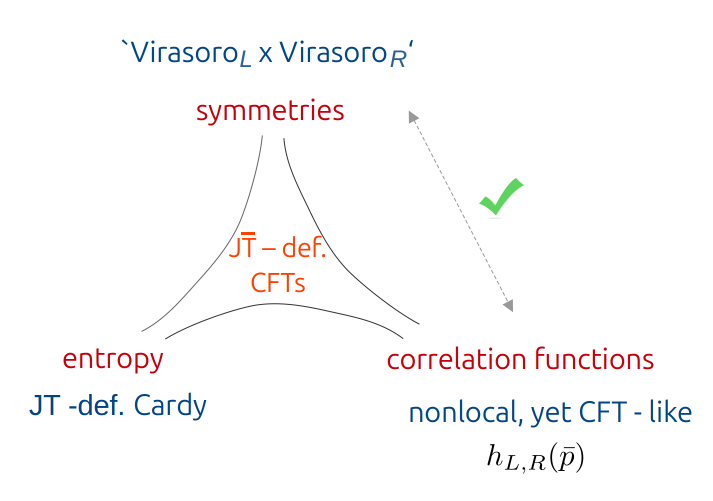}
\caption{Properties of $J\bar T$ - deformed CFTs.  }
\label{compkcftjtb}
\end{minipage}
%\caption{}
\end{figure}

It is  important to note that, at a more fine-grained level,  
$J\bar T$ - deformed CFTs  possess features that have not been previously encountered in studies of the Kerr/`CFT' correspondence: 
\bi
\item the right-moving extended symmetries act in a non-standard fashion; in the most natural basis for them, the algebra is not (Virasoro-KM)$^2$, but rather a non-linear modification thereof \eqref{jtbquasilocalg}
\item the entropy is \emph{not} given by the charged Cardy formula, but by its $J\bar T$-deformed analogue \eqref{entropyJTbarFT}
\item the functional form of $h(\bar p)$ is not the same, in $J\bar T$ it is given by \eqref{jtbdims}, while in Kerr/`CFT' it takes a form closer to \eqref{hwads}
\ei 
To see these features of $J\bar T$ - deformed CFTs, it is  important to  turn on arbitrary charges: indeed, only then  is one  able to distinguish the $J\bar T$ - deformed Cardy formula from the standard Cardy one and  see the field-dependence of the asymptotic symmetry generators, which is only present if the relevant $U(1)$ current has a non-zero expectation value. It is also important to know which charges are quantised, as this  determines the non-linear terms in the asymptotic symmetry algebra, as we explicitly saw in section \ref{asysymmdtr}.

%\textcolor{blue}{The technical details through which these conclusions/results are interesting, and a full understanding is still missing. Concretely: $\exists$ two bases of symmetry generators: flowed (Virasoro by construction) and quasilocal (integrals of the stress tensor) as the more appropriate (true?) basis of symmetry generators. The two sets are related by a non-linear redefinition, which in particular implies that the algebra of the quasilocal generators is non-linear.  In the Lagrangian language, the symmetries have the characteristic (in finite size) that they necessitate the addition of a compensating large diffeo or affine transformation in order for charge quantization to be obeyed; this in turn can be seen as leading to the non-linearity of the algebra.  The operator basis is singled out by the quasilocal generators. \emph{Is this the same notion that is given by path integral analysis?} All these technical details, necessary for the consistency of the full picture, are carefully worked out and checked for the $J\bar T$ deformation. Another lesson is the huge importance of knowing the boundary conditions and (relatedly) the parametrisation.}

%\subsubsection{A roadmap for the Kerr/`CFT' correspondence}

Since $J\bar T$ - deformed CFTs are the only example of dipole CFTs where these properties could be derived from a QFT perspective, it is natural to conjecture that the dipole CFTs that are dual to warped AdS$_3$ spacetimes - and, by extension, to near-extremal black holes - belong to the same `universality class' as 
%
%Given these results, a natural hypothesis one should test is whether the Kerr/'CFT' correspondence\footnote{In this subsection, by Kerr/`CFT'  we mean string-theoretical toy models that can be carefully controlled.} fits within a universality class of 
%
single-trace $J\bar T$ - deformed CFTs, which are characterised by the properties summarised in figure \ref{compkcftjtb}.
%
% The reason this conjecture is natural is the explicit existence of QFTs with this set of properties.  
%
%from the viewpoint that there exists at least one class of $2d$ QFTs with these properties. 
%
%What do these results imply about warped AdS$_3$ holography, at least those stringy toy models that should be described by dipole CFTs? A natural assumption  is that they have the same ``QFT structure" as $J\bar T$. \emph{Picture!}
%
Of course,  since $J\bar T$ - deformed CFTs are extremely special, one should first specify which conditions \emph{generic} representatives of this putative universality class should satisfy. Based on our experience with ALD spacetimes, it seems reasonable to require that universal observables such as the entropy and asymptotic symmetries take  a single-trace $J\bar T$ form, but not necessarily non-universal ones such as the spectrum of momentum-dependent conformal dimensions, $h(\bar p)$. It would of course be very interesting to understand if there exist general constraints on the possible momentum-dependence of these dimensions.

In order to test this hypothesis, it is important to study solutions with generic $U(1)$ charges, as  explained above. %otherwise it may not be possible to tell apart a $J\bar T$ - deformed Cardy formula from a simple Cardy one, and a field-dependent coordinate from a standard one.
 Studies in presence of arbitrary $U(1)$ charges are relatively few, the ones mentioned in section \ref{wadsvsdipcft}, which  have been mostly motivated by the connection to $J\bar T$, being an example. It is also important to work with geometries that are obtained via decoupling limits, rather than just simple solutions to supergravity obtained via solution-generating techniques; indeed, without full control over the parametrisation, one cannot infer the correct entropy formula. %, nor the asymptotic symmetry group, especially if deduced using the radial symplectic form. 

%A major problem one encounters in these studies is the
%
%At the precision level at which we are currently working, the scattering amplitudes do have the expected property. As far as the entropy is concerned,note that in $J\bar T$ - deformed CFTs the entropy, though universal, is not given by Cardy's formula; instead, corrections away from it can be seen, provided the backgrounds have the appropriate $U(1)$ charges turned on.  
%The problem we encounter is that of a 
%lack of knowledge of the correct parametrisation of the backgrounds. This problem stems from the fact that most  examples of warped AdS$_3$ backgrounds in string theory  are obtained from solution generating techniques, and not  from decoupling limits.  These often have the feature of not singling out the correct, fixed coordinates and coupling constants in the dual theory, which does have consequences for the estimation of the entropy\footnote{One such example is that of the NS-NS background generated via TsT, where shifting the dilaton by a constant by temperature/charge-dependent factor completely changes the thermodynamics to be that of $J\bar T$.}.  In absence of a decoupling limit, it is hard to check this should be the correct parametrisation.  

Knowing the correct parametrisation is  a subcase of knowing  the correct boundary conditions; more precisely, it is related to the boundary conditions for fluctuations that are independent  of the boundary coordinates. As we saw in our asymptotic symmetry group analyses of the ALD and  the warped AdS$_3$ backgrounds supported by pure NS-NS flux, the boundary conditions for  modes that depend on the  boundary coordinates can be more lax than those for constant ones. It is of course of central importance in this bottom-up holographic program to understand how the boundary conditions for generic fluctuations can be determined, e.g. whether they can be derived from the decoupling limit when the latter is known. It would also be interesting to establish whether the method based on the symplectic form  presented in section \ref{holostr}%(also previously used in \cite{} \textcolor{red}{\emph{Find ref!}})
, which has the advantage of being based on a procedure, rather than a simple guess,  is \emph{the} way to at least partially determine the boundary conditions in a given spacetime.% \textcolor{blue}{\emph{Mention previous work by Geo.}} 

%  The latter are obviously related to the ASG, which in turn is related back to the parametrisation via the particular method we have chosen to build it. \emph{Can one perhaps derive the bnd cond if the decoupling limit is known?}

The above-suggested exercise in bottom-up holography, if properly executed, should yield a definite answer: yes or no,  to the question whether the toy\footnote{By this we mean string-theoretical toy models of Kerr/`CFT' that can be carefully controlled.} Kerr/`CFT' correspondence  fits within the $J\bar T$ universality class.  The currently existing evidence does not point in any preferred direction, yet. For example,  the asymptotic analysis of \cite{Georgescu:2025jlx} of the charged dipole-deformed D1-D5 backgrounds yields a purely charged Cardy entropy and asymptotic symmetries parametrised by field-independent  functions. While there is a valid possibility that there exists some other $U(1)$ current, not turned on in the analysis of \cite{Georgescu:2025jlx}, that plays the role of the current entering the effective $J\bar T$ description, it can also be that this example does not fall under the $J\bar T$ universality class. In this interesting scenario, which suggests at least two  different universality classes for dipole CFTs, 
one should understand again, from a field theory perspective, how the symmetries and the entropy can be reconciled with the  non-locality of the theory.

%This being said, for the decoupled dipole background, the entropy is Cardy and the ASG Virasoro, \emph{True?} even when some RR $U(1)$ charge is turned on. However, this does not exclude the possibility that the entropy may become $J\bar T$ - deformed Cardy if some other, appropriate, $U(1)$ charge is turned on. [Current trials gave infinite charges].  So, this is the first lesson we learn: it is important to turn on all possible parameters in the theory in order to check form of the entropy. 

\subsubsection{Lessons for non-extremal black holes}

 In section \ref{nonextrsec}, we presented various bottom-up holographic studies of non-extremal black holes. These  suggested that their microscopic duals   correspond to  special irrelevant deformations of a two-dimensional CFT that preserve the entropy of thermal states. The latter property was  suggested, in particular,  by the  `subtraction procedure' for studying non-extremal black hole microscopics, which consists of changing a particular conformal factor in the black hole metric. This procedure   is interpreted as  `placing the black hole in a conformal box', which changes the surrounding environment of the black hole, while preserving its intrinsic properties, such as the entropy  and the temperature.  
 %
%  reviewed the subtraction procedure, and the picture it suggests for the holographic dual to non-extremal black holes: irrelevant deformations that change the theory, but not the entropy\footnote{From subtracted geometries section: This may seem a bit puzzling, since if two systems have the same $S$ and $T$, one may be lead to conclude that $S(T)$ is the same (and thus $C$), which is clearly not the case. We further discuss this point in section \ref{}.}..
  It holds for general classes of black holes embedded in string theory, and roughly corresponds to dropping the constants in the harmonic functions that support the solution.% \textcolor{red}{\emph{Write better!}}\textcolor{red}{Also mention problem charge}.
  
In this subsection, we revisit this picture in the simple case of the NS5-F1 system, which has been linked to the $T\bar T$ deformation. 
The asymptotically flat NS5-F1 geometry can indeed be thought of as turning on various irrelevant deformations in the CFT dual to the near-extremal, near-horizon AdS$_3$ geometry, which roughly correspond to putting back the ones in the $f_{1,5}$ harmonic functions. Thinking sequentially, putting back the one in the $f_1$ harmonic function changes the asymptotics to ALD and turns on the irrelevant deformation to compactified LST; additionally turning on the constant in $f_5$ changes the asymptotics to flat space and corresponds to an additional irrelevant deformation; it is not known whether the associated theory can be obtained via a  decoupling limit. 

The first of these irrelevant deformations is under relatively good control, especially via the link to $T\bar T$. As discussed throughout this review, the entropy changes from Cardy in the IR to the form \eqref{ttbenttemp} specific to\footnote{Let us try to understand this from the point of view of the subtraction procedure, applied directly to the ALD background \eqref{aldbh}, rather than the asymptotically flat NS5-F1 one. We concentrate on zero momentum, for simplicity, and use  the more formalised procedure of \cite{Baggio:2012db}, in which the charges are explicitly fixed  and only the constants in the harmonic functions can vary. The decoupled background takes the form  %\textcolor{red}{\emph{Check!}}
\be
ds^2 = \frac{1}{f_1} \left(-\frac{f}{f_n} dt^2 + f_n d\s^2 \right) + f_5 \left(\frac{dr^2}{f} + r^2 d\Omega_3^2\right) + ds^2_{M_4}\;, \;\;\;\; f_1 = c_1 + \frac{r_1^2}{r^2} \;, \;\;\;\; f_5 = \frac{k\a'}{r^2} \;, \;\;\;\;\; f_n = c_n\;, \;\;\;\; f= 1-\frac{r_0^2}{r^2}
\ee
Charge quantization requires that 
\be
r_1^2 = \frac{c_1 r_0^2}{2} \left( \sqrt{1+ \left(\frac{2 p  \a'}{v r_0^2 c_1}\right)^2} -1 \right)
\ee The entropy is given by $S \sim r_0^3 \sqrt{f_1 f_5 f_n} $, while $T^{-1} = 2\pi r_0 \sqrt{f_1 f_5 f_n} $. The subtraction procedure consists in this case of matching the system with $c_1=c_n=1$ with one where $c_1 \r 0$ and $c_n$   chosen so that the entropy and temperature match at fixed $r_0$. It it not hard to see this leads to the condition that $f_1 (r_0) f_n$ is the same, implying that
\be
 c_n^{(subtr)} \, \frac{2p  \a'}{v r_0^2} = \sqrt{1+  \left(\frac{2 p  \a'}{v r_0^2 }\right)^2} +1
\ee
which can only be solved if either $c_n^{(subtr)} $ (which parametrizes the radius of the $\s$ circle) or the parameters $g_s, \a', v$ depend on $r_0$. Therefore, one can indeed write the entropy of the ALD black holes in Cardy form throughout, but at the expense of making some of the parameters of the system temperature-dependent. } $T\bar T$.  Since the theory is defined via the LST decoupling limit in the UV, one can no longer invoke the freedom to redefine the irrelevant operator sources from an IR point of view to argue for constancy of the entropy. %\textcolor{red}{\emph{Correct?}}
Also, the nature of the excitations whose entropy these formulae  count seems to change, from massless modes of the 1-5 strings in the IR to excitations of the little strings in the UV. Thus, the view according to which the change in the conformal factor associated to the irrelevant deformation is to only  change the environment in which the black hole is immersed, without affecting its `intrinsic' degrees of freedom, does not appear supported in this example. 
%
%
%To understand what happens, one could try applying the subtraction procedure directly to the ALD background \eqref{}, rather than the asymptotically flat NS5-F1 one  (we set the momentum to zero, for simplicity). Using  
%\textcolor{red}{\emph{only for $\d_0=0$, it seems}} and 
%\be
%H  = xx\;, \;\;\;\; e^{2\Phi} = g_s^2 \frac{f_5}{f_1}
%\ee
%where we now allow for arbitrary constants in the harmonic functions: $f_1 = c_1 + \frac{r_1^2}{r^2}, f_n = c_n$ etc. and $c_5 \r 0$. \textcolor{red}{\emph{Correct?}} 
%Charge quantization requires that 
%
%\be
%r_1^2 = \frac{c_1 r_0^2}{2} \left( \sqrt{1+ \left(\frac{2 p g_s^2 \a'}{v r_0^2 c_1}\right)^2} -1 \right)
%\ee
%etc. The entropy is given by $S \sim r_0^3 \sqrt{f_1 f_5 f_n} $, while $T^{-1} \sim r_0 \sqrt{f_1 f_5 f_n} $, with $f_5 = k\a'/r_0^2$.  
%
On the other hand, given the field-dependent coordinates \eqref{fdepcoordst} that seemigly emerge in the study of the asymptotic symmetries of the ALD background, it is not excluded that some effective CFT dynamics that could be seen through the prism of these dynamical coordinates is at play, and its Cardy entropy does correspond to the entropy of the system, in a possibly non-linear ensemble. In other words, since the entropy of compactified LST does match that of $T\bar T$, and degeneracy in the latter is in one-to-one correspondence to that in a CFT, but measured with respect to a different notion of energy, there may ultimately be a link to \cite{Cvetic:2011hp}'s picture of changing the definition of energy via the change of asymptotics. It would be interesting to understand if such a picture can be made precise.

\medskip

The above discussion is in the context of a sequential application of the two irrelevant deformations at play. But how should one interpret asymptotically flat NS5-F1 black strings where the constants in the $f_{1,5}$ harmonic functions are both non-zero? An interesting proposal, put forth in \cite{Chakraborty:2020nme} for the case of non-extremal D3-branes and  adapted to  the non-extremal  NS5-F1 brane system in \cite{Georgescu:2024iam} suggests that the effective $T\bar T$ or LST description may still be useful, even in absence of a decoupling limit, at least for a certain range of energies.

\begin{figure}[h]
\begin{minipage}{0.45 \linewidth}
\centering 
\includegraphics[height=3.4cm]{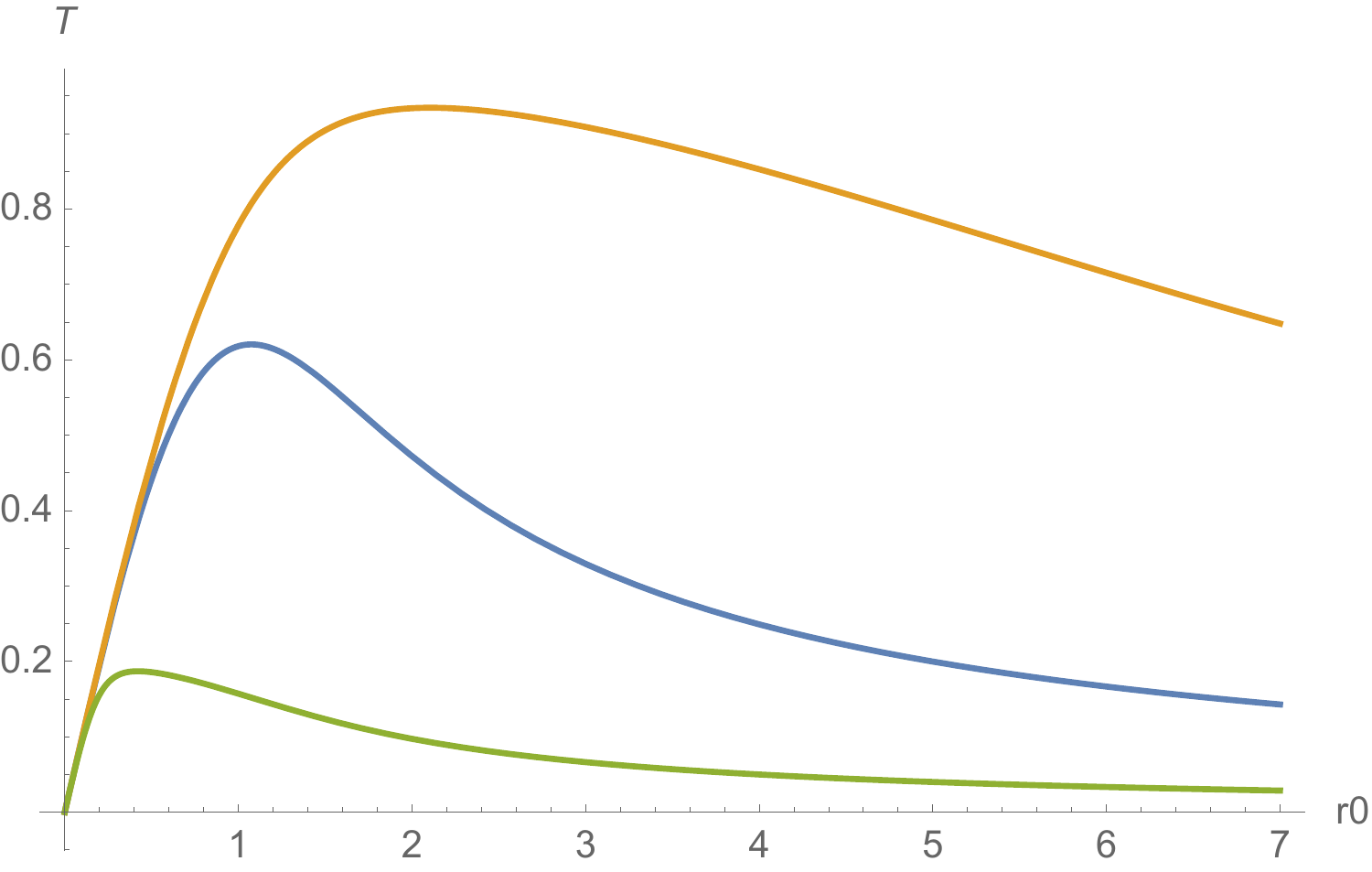} 
\caption{$T$ as a function of  $r_0/g_6$ for $\varkappa = 5$ (green), $\varkappa = 1$ (blue) and $\varkappa = 0.2$ (orange).
}
\label{tvsens5f1}
\end{minipage}
\hspace{1cm}
\begin{minipage}{0.45 \linewidth}
$T_{max} = 
\left\{\begin{array}{ccc} \frac{1}{2\pi \sqrt{\a' k}} & \mbox{for} & \varkappa \ll 1 \\[9pt]  \frac{1}{2\pi \varkappa \sqrt{\a' k}} & \mbox{for} & \varkappa \gg 1 
 \end{array} \right.
 $
\end{minipage}
\end{figure}

As is the case for many charged non-extremal black holes, the temperature-to-energy relation in these systems has the shape in figure \ref{tempvsengnonextr},  discussed in section \ref{nonextrsec}: there is a maximum temperature and below it, there are two possible black hole solutions - one small and stable, and one large and unstable.  
%
%Despite this, the (special) irrelevant deformation picture for the microscopic dual remains valid, even if $S(T)$ may not be unchanged (still universal and admitting a physical explanation). So, the question still remains, what type of theory explains the entropy beyond the very low energy regime? The $T\bar T$ work reviewed in this article gives a (potential) partial answer. 
%
%In this review, we have amply discussed UV-complete theories starting off as irrelevant deformations of a CFT (possibly not the best possible description). Precisely in this contect, would like to comment on a picture suggested/proposed in \cite{} for non-extremal D3-branes, which is readily adapted to  the non-extremal D1-D5/ NS5-F1 brane system. The $T(r_+)$ takes the same shape, \ref{}, as for any non-extremal charged black hole or black brane. \emph{Careful check planar vs global D3.}  
%
In the NS5-F1 system, the position of the maximum depends on the moduli at infinity, characterised by $\varkappa \equiv g_6/g_6^\star$ in \cite{Georgescu:2024iam}, where $g_6^\star=\sqrt{k/p}$ is the attractor value of the six-dimensional string coupling constant, which is very small for $p \gg k$. A plot of the maximum temperature as a function of the horizon radius divided by $g_6 \sqrt{\a' p}$ is reproduced in figure  \ref{tvsens5f1}, for various values  of the relative coupling $\varkappa$.  As $g_s \r 0$ ($\varkappa \r 0$), the  position of the maximum in terms of this rescaled coordinate is pushed  to infinity, % \emph{Review map between constants and couplings}
 the behaviour becomes purely Hagedon with the expected LST Hagedorn temperature, so %(since the $C<0$ region disappears) % \emph{Discuss with Rabinovici?}
   it is very clearly captured by little strings. As $g_s $ - or $\varkappa$ - is increased,  the descending part of the curve (corresponding to negative specific heat) appears and the maximum temperature displays a clear $\varkappa$ dependence. 

\noindent 
  
The suggestion of  \cite{Chakraborty:2020nme} is to keep on identifying $T_{max}$ with the Hagedorn temperature (for little strings, in this case) even without taking the decoupling limit. On the stable, small black hole branch, these strings are expected to capture the black hole entropy. Thus, at least for energies smaller than $E_{max}$ -  the energy corresponding to the maximum temperature in figure \ref{tvsens5f1}, these should be able to explain the entropy of non-extremal black holes. %\textcolor{red}{\emph{How large is the departure from extremality for them?}} 
  % 
%   \cite{} basically proposes to identify $T_{max}$ in the non-extremal system with some sort of Hagedorn temperature for little strings (open strings in the D3 picture \emph{Can this be argued at strong coupling?}). These (little) strings capture the entropy up to $E_{max}$, and the dual gravitational picture are the small black holes with $C>0$. 
Note that the maximum temperature does receive corrections as $g_s$ is increased, as indicated by the gravity solution,  and can become significantly smaller. 

For higher energies, one reaches the unstable branch, and the discussion in section \ref{nonextrsec} regarding ensembles applies. It is likely that new degrees of freedom are then 
%
%   How about higher energy black holes? The suggestion of \cite{} appears to be that new degrees of freedom (beyond the open/little strings) are 
 %  
   necessary to capture the entropy in this regime. Since, as we discussed, there is no known decoupling limit of the NS5-F1 system to flat space (a similar discussion holds in the D3-brane case, see \cite{Chakraborty:2020nme}), it is plausible that these new degrees of freedom may  be gravitational. 
One may nonetheless hope there exists some \emph{universal} mechanism that may explain $C<0$ regimes in generic black holes. If this is the  case, then one can look for inspiration in the existing work on the microscopic description of small black holes in AdS. For example, one  (very qualitative) microscopic model for negative specific heat \cite{Berenstein:2018hpl} shows that, if one starts with a system displaying Hagedorn behaviour and perturbs it by general interactions of a certain kind, one  may produce $C<0$.   A specific example considered therein to yield $C<0$ simply redefines the energy by adding a small quadratic term in it, which is vaguely reminiscent of $\mu<0$ $T\bar T$. %\textcolor{red}{\emph{Mention speculative.}}

%\textcolor{blue}{  One may inquire more  generally  about the microscopic description of black holes  with a negative heat capacity, hoping that a universal mechanism may be at play - in this case, the  work of \cite{} on small black holes in AdS is relevant. }

More evidence of a putative connection to $T\bar T$ can be obtained from an asymptotic symmetry group analysis of the NS5-F1 black string spacetime, also performed in \cite{Georgescu:2024iam} (for $\varkappa=1$, for simplicity). Using   the method based on the vanishing of the radial symplectic flux, one again finds a set of strongly background-dependent asymptotic symmetries. At a very heuristic level, the  ASG generators indicate the presence of two $T\bar T$ - like irrelevant deformations, with qualitatively different behaviour:   one that behaves as $\mu>0$  $T\bar T$ and is presumably related to the LST description, and another that behaves in a qualitatively different fashion, and is presumably associated with the departure of the self-dual irrelevant coupling $\l_+$ from the extremal value \eqref{constrirr}. Interestingly, the expressions for the background-dependent asymptotic symmetry generators break down precisely at $E_{max}$, 
   %
   %However, it is absolutely not clear what type of properties these new dof ( which may be non-decoupled gravitational degrees of freedom) should have in order to produce $C<0$ with the expected energy dependence. This  $E_{max}$ picture appears to be corroborated by an ASG analysis in the non-extremal NS5-F1 background, nameky one finds a $T\bar T$ - like ASG below this energy, and the ASG stops making sense beyond $E_{max}$,
     beyond which the effective $T\bar T$ description is no longer expected to be  valid. %It would though be interesting to investigate whether the description in terms of effective $\mu<0$ $T\bar T$  

\medskip

To conclude, the recent work on  $T\bar T$ and $ J\bar T$ - deformed CFTs and their holographic interpretation has improved our understanding of extremal and non-extremal black holes in several different ways:  it
 i) gave reassurance that the seemingly incompatible properties of their dual theories  inferred from  gravitational analyses (e.g., the extended symmetries and the correlation functions) could in fact be compatible; ii) indicated that 
the entropy of warped AdS black holes may be   captured by  a universal formula that is similar to, but different from Cardy's formula  and  iii) suggested a picture based on partial decoupling for describing the small  black hole
branch of non-extremal black holes, which may be modelled by generalisations of $T\bar T$ and $J\bar T$-deformed CFTs and whose range of validity could extend all the way up to $E_{max}$ in figure \ref{tempvsengnonextr}; it would be very interesting to test whether such a scenario holds. Beyond this energy, it is not clear whether one can have a decoupled dual (except perhaps at infinite $N$) that captures the black hole degrees of freedom, and whether it'd admit a simple, universal description.

% and suggests UV complete non-gravitational non-local theories to describe .   \emph{In fact, can prob compare LST with self-dual coupling - how does $\d$ depend on it?}

%[ When the full deformation to AF is turned on, at least in this system there doesn't appear to be any decoupling limit. But, perhaps, this can be understood as an irrelevant deformation that yields to a non-decoupled theory that can be modeled as ``$\mu <0\;\; T\bar T$. The specific heat is again affected. 

%For more general near-extremal black holes: what if a decoupling limit to warped AdS (to a theory, not a state) is not possible? Does it make sense to take about a non-decoupled description valid in just a certain energy range (which)? Is this useful for explaining the entropy in the given range (even if modular invariance may not be applicable?)? ]

\subsection{Towards a general definition of dipole and `little string' CFTs}

The $T\bar T$ and $J\bar T$ deformations are universal irrelevant deformations of $2d$ QFTs that lead to non-local but UV-complete, fully controllable theories. As a consequence of this amount of  universality and control, the spectrum of the deformed  theory, the deformed S-matrix and likely also the correlation functions are rigidly determined in terms of their undeformed counterparts. In fact, most effects of these deformations can be rephrased in terms of studying the original QFT via the prism of certain universal, state-dependent coordinates.

On the other hand, studies of the QCD string \cite{Dubovsky:2015zey,Donahue:2019adv}, as well as analyses of the non-AdS backgrounds discused in section \ref{holostr}, strongly suggest the existence of generalisations of the   $T\bar T$ and $J\bar T$ deformations that may play an important role in describing rather diverse physical phenomena, from certain aspects of confining strings to the near-horizon dynamics of general black holes. The holographic studies we reviewed show that, while universal observables such as the entropy and the symmetries resemble those of  $T\bar T$ and $J\bar T$-deformed CFTs, less universal ones such as the deformed spectrum do not, bringing  hope of  more interesting, less rigid dynamics. 

In this subsection, we would like to discuss how one may approach the problem of such generalisations. Our main guiding examples will be restricted to those that appear in the holographic context of ALD and warped AdS$_3$ spacetimes, though there may exist  other hints in different set-ups. 

In our attempt to characterise the theories we are after, let us start by giving them a name. Those connected to warped AdS$_3$ holography have already been named in section \ref{bhstoirrelsec} : \emph{dipole CFTs}. The ones hinted at from ALD holography will be called \emph{little string CFTs} (LS-CFTs in short), with the following tentative characterisation
  
  \vskip6mm
  
  \noindent \emph{Little String CFTs:} Poincar\'e invariant $2d$ QFTs that are non-local, yet UV-complete and can be written 
  
  \hspace{2.7cm} as a finely-tuned irrelevant deformation of a CFT with coupling $[\mu] = (length)^2$

\vskip3mm

\noindent Given a coupling of this dimension, the leading irrelevant operator in the IR will have dimension $(2,2)$, and the ones to follow, $(n,n)$. Note  the above is far from being a definition: the UV properties of these theories  are still to be understood, while attempting to define them via an irrelevant flow that starts in the IR  looks difficult and extremely unnatural. Clearly, developing a proper QFT framework for understanding LS-CFTs, which may  exhibit UV/IR mixing \cite{Aharony:2004xn},  is of great interest.

Note that, for the purposes of our limited discussion, these theories do have a definition \cite{Georgescu:2024iam}: they correspond to various decoupling limits \cite{Gopakumar:2000ep} of NS5-branes compactified on K3 or $T^4$, sometimes with critical RR fluxes on their worldvolume, and in presence of F1 strings. These decoupling limits yield the entire flow between compactified LST in the UV and the D1-D5 CFT in the IR, or vice-versa. Note that the UV and the IR perspectives on this flow are not symmetric: while from the IR perspective, there are $21$ independent $(2,2)$ irrelevant deformations one can turn on, the UV growth of states  is characterised by a single length scale ($\sqrt{k \a'}$)%\textcolor{red}{\emph{Check!}}
, with the ratios of the various irrelevant couplings behaving as moduli. Thus, in a certain sense, all the examples studied in  \cite{Georgescu:2024iam} correspond to a single theory, LST/K3. Our approach in this section will be to draw inspiration from this limited example in order to explore how more general theories in the LS-CFT `universality class' may be defined.

%By taking a very general approach, one may nonetheless hope to make some more generally applicable statements.  \textcolor{red}{\emph{Complain about very limited discussion because so many things not understood; will work with existing hints, and hopefully better discussion in the future.}}

Our chosen terminology reflects the expectation that the above theories will involve stringy degrees of freedom. In the concrete holographic examples dual to ALD spacetimes, this is strongly suggested by the Hagedorn behaviour of the entropy. However, in general it is not clear whether UV-completeness and the presence of a single length scale of the given dimension are sufficient to ensure this growth;  one may then want to specify the Hagedorn behaviour as part of the definition of these theories. 

The use of `CFT' in the name is less obvious, as all these theories have an explicit length scale, at which the physics becomes non-local. However, the ASG analyses of the ALD backgrounds indicate the presence of a  modified Virasoro $\times$ Virasoro symmetry,  of the type characteristic to $T\bar T$. These symmetries are indicative of a rather constrained structure - possibly as much as 
that of CFT$_2$ - that is yet to be understood. %) that could be almost as constrained as .
Finally,  the  qualifier  `little' was used in reference to our guiding LST example, and also because the SCFT abbreviation is already taken\footnote{We do not find a terminology based on $T\bar T$ useful, as current-current deformations appear to be too specific examples of the theories we are looking for: in the LST example, most deforming operators are built from fields from a non-trivial twisted sector, so the common denominator does not appear to be the bilinear current structure. Also, the universal $T\bar T$ spectrum appears too special, as does the the relation to the original CFT via a coordinate redefinition.}.

%
%The LST setup we use as guidance is also reflected in 
%\textcolor{red}{\emph{Shorten and check previous:}}\bi 
%\item \emph{little string:} in the stringy examples, the holographic decoupled backgrounds indicate Hagedorn entropy for all the candidate deformations, which strongly suggests the presence of . The ``little'' is because the specific example of LST  .
%\item \emph{CFT:} of course, none of these theories are CFTs, having an explicit length scale, at which the theory becomes non-local.  \ei.

%The holographic/decoupled backgrounds studied in this work suggest the existence of two classes of two-dimensional theories, which we denote as ``dipole'' and, respectively, ``little string'' CFTs. The definition that we would like to use (despite it being totally unnatural) is:

%\emph{Definition:} Theories obtained via finely-tuned irrelevant deformations of a $2d$ CFT that start either as $(1,2)$ or $(2,2)$ and lead to UV-complete non-gravitational theories (hard to establish in $2d$). UV completeness is inferred from decoupling limit in the stringy examples. 

How can general LS-CFTs be defined?  Currently, it appears that the most concrete approach  is to start in the IR and attempt to characterise the irrelevant flow to the UV%\footnote{This is not to say that the existence of an irrelevant flow from a CFT should be a requirement on the existence of the UV theory, just that this is the only way we can currently think of defining the deformation which, as we will see, is very constrained and non-generic.  }
. A first requirement is that the undeformed theory  contain an operator of dimension exactly $(2,2)$. Barring the universal current-current $T\bar T$ deformation, generic CFTs will not usually contain such an operator. In the D1-D5 CFT, their existence  is linked to supersymmetry, as all the $(2,2)$ leading irrelevant operators are supersymmetric descendants of $(1,1)$ chiral primaries \cite{deBoer:2008ss}, whose dimension is protected by supersymmetry.  
Thus, one relevant question in defining these deformations is that of genericity: does one need supersymmetry to define LS-CFTs? We do not know the answer to this question: while in our concrete LST example  the answer is clearly yes, the situation is less clear for generic black hole set-ups, where one may be able to identify  effective $(2,2)$ irrelevant deformations of an underlying CFT also in absence of supersymmetry (the same issue is present in trying to model non-supersymmetric near-extremal black holes with dipole CFTs). % \textcolor{red}{\emph{Any chance that generic large $N$, large gap CFTs contain a $(2,2)$ operator? Does one require an on-shell realisation of the subtracted geometry to see it/ is a low-energy limit sufficient?}}
One possibility is of course that general black holes are dual to irrelevant deformations of only special CFTs, which contain $(2,2)$ and $(1,2)$ operators. 

For the maximally supersymmetric LS-CFTs that constitute our concrete example, one may try to test an interesting proposal that Intriligator  has put forth  \cite{Intriligator:1999ai} for the case of the maximally supersymmetric irrelevant deformation of $\N=4$ SYM that drives the theory to flat space, namely that it may be sufficient to add  just  the leading irrelevant operator (built from the  appropriately-corrected supersymmetry generators)  to define the theory. While the status of this proposal %for $\N=4$ SYM
 is  currently unclear \cite{Caetano:2020ofu}, let us note that the above deformation of $\N=4$ SYM is analogous to the self-dual deformation of \cite{Georgescu:2024iam}, %which drives the theory to flat space and 
for which no decoupling limit is known; by contrast, for the deformations  \eqref{constrirr} that lead to LS-CFTs, this UV limit is at least expected to exist.  It would be of course very interesting to understand, from the irrelevant flow perspective, what differentiates the $\l_+$ and the constrained deformations, so they lead to such different UV behaviour.

%In the LS-CFT case, these are irrelevant deformations by some specific combo of $(2,2)$ operators. The presence of multitrace (related also to issue of boundary conditions) or of descendants of long operators not clear. The UV completeness is also not clear, and in what are distinguished the decoupled $(2,2)$ deformations with respect to the undecoupled one (the entropy of non-susic states distinguishes them, but other,  more tractable, obsrevables?).  All the other known stringy examples appear to be (non-commutative) deformations of LST, possibly belonging to the moduli space of the latter theory. 

%Alternative possible definition: given the susies preserved by the deformations, maximal in the $(2,2)$ case, it is interesting to ask whether they could fix the theory, as has been previously proposed by e.g. intriligator (only $G's$ change, not the chiral primary). Not clear this is true (should check for $T\bar T$). Supersymmetric observables may perhaps be computed only knowing the descendants of the BPS $(2,2)$ operators. 

Another approach that one may imagine, also in the irrelevant deformation expansion around the IR CFT, is a bootstrap one. Starting from an IR CFT that satisfies crossing symmetry, one may try to set up a momentum-space bootstrap \cite{Gillioz:2025yfb} problem that may  constrain the possible irrelevant deformations of the CFT data, perturbatively in $\mu p^2$; % Of course, in order to do this, one would need to face the  hurdles of momentum-space bootstrap \cite{} \textcolor{red}{\emph{Ref?}}, but some hope exists .  \textcolor{red}{\emph{Check!}} 
%Of course, the correlation functions of $T\bar T$ - deformed CFTs are expected to correspond to a simple solution \textcolor{red}{\emph{Can one explicitly check?}} %(possibly via some sort of spectral flow interpretation, as in $J\bar T$ \cite{Guica:2021fkv}),
%; the interesting question is whether additional solutions  exist, as well as 
it would be interesting to understand
 how constraining the bootstrap approach can be. In any case, if such an approach can be developed, it may be able to  provide an 
 axiomatic definition to LS-CFTs. 

So far for the IR-based considerations. A more sensible approach would be  to define LS-CFTs via their symmetries, as is standard in local QFT. As we extensively discussed,  (single-trace) $T\bar T$ - deformed CFTs display an intricate pattern of extended symmetries, and there are holographic indications, in the form of the bottom-up  ASG analyses of  section \ref{aldttbsec},  that  LS-CFTs may display a similar pattern. A natural approach would then be to understand how to define these symmetries in general terms (e.g., via adiabatic flow of the Virasoro generators% for the flowed ones, but the quasilocal ones are harder to pinpoint
) and how constraining they are in practice; in particular, whether they fix the momentum-dependence of the correlation functions.   A particularly interesting question is whether emergent field-dependent coordinates  also exist in this more general setting; the ASG analysis of the ALD  backgrounds suggests that the answer is yes. If this is case,  there appears to be a tension between the expected universality of all  observables, associated with the existence of dynamical coordinates, and the genericity properties of LS-CFTs\footnote{It is also unclear why a set of quasi-local field-dependent coordinates should emerge from the irrelevant flow triggered by more general deformations than $T\bar T$:  in our analysis of section \ref{infsymmsec}, the properties of the emergent field-dependent coordinates seemed intimately tied to the very special properties of $T\bar T$.  The current evidence  for them is consistent with this description only existing at high energies, or in some special decoupled subsector. }. This could perhaps be better understood by studying observables that are not expected to be universal. %, so as to test  the generiticity properties  

%
%\emph{Candidate theories}: stringy examples based on decoupling limits, barring single-trace $T\bar T / J\bar T$ that we know how to construct. 
%
%Properties: could be that Virasoro can again be flowed, but quasilocal generators may be less universal than before. Not clear then if there exist emergent coordinates. Axiomatic definition? 
%
 %More seriously, if adiabaticity is taken at face value, a set of Virasoro generators can always be constructed. \textcolor{red}{\emph{Careful UV divergences!} } The real question is what is the relation between these "spectrum generating"/flowed generators and the `actual' symmetry generators of the theory. In the case of $T\bar T/ J\bar T$, because of the simple and universal expression for $D$, a simple relation could be found. It is less clear what happens in the general case, i.e. whether the Virasoro can be found starting from a geometric symmetry. It is not clear what the answer should be in general, though
   %The way the  ASG computation was performed suggests a link  between the  entropy of the system and its extended  symmetries, at least as far as their action on high-energy states is concerned. 
 
\medskip

%\noindent \textcolor{red}{------ to do ------ }

\noindent Most of these approaches can be adapted to the dipole case.  Recall these theories are defined as

  \vskip6mm
  
  \noindent \emph{Dipole CFT:} {\normalsize{$SL(2,\mathbb{R}) \times U(1)$}} invariant $2d$ QFT that is non-local, yet UV-complete and can be written
  
  \hspace{1.7cm}   as a finely-tuned irrelevant deformation of a CFT with a null vector coupling  $[\l] = length$

\vskip3mm

\noindent Dimensional analysis and the broken Lorentz invariance of the problem  imply that the leading deforming operator around the IR CFT will have dimension $(1,2)$, while subsequent ones - $(1,n)$.  Besides the universal $J\bar T$ deformation and its single-trace counterpart,  there exists only one other example of a dipole CFT in string theory, namely the holographic dual of the decoupled dipole-deformed D1-D5 background, where the UV-completeness is suggested by the existence of the  decoupling limit. 

Perhaps surprisingly, currently  dipole CFTs are less understood than  the LS-CFTs we have just reviewed.  First, while there exist additional warped AdS$_3$ backgrounds in string theory \cite{ElShowk:2011cm} that correspond to turning on   other $(1,2)$ deformations of the D1-D5 CFT  (which can be written as supersymmetric descendants of the same $(1,1)$ chiral primaries as in the LS-CFT/AF case), it is presently not known how to obtain them from decoupling limits. Thus, it is not clear whether these irrelevant deformations lead to UV - complete theories.

Second, it is not clear whether there exists a single, universal asymptotic entropy in dipole CFTs, or there can be different possibilities.  That is, while in $J\bar T$ - deformed CFTs the entropy takes on a universal $J\bar T$ - deformed Cardy form, holographic studies of the decoupled D1-D5 dipole background currently only indicate a standard charged Cardy formula. 

The various approaches that we mentioned in the LS-CFT case can also be attempted in that of dipole CFTs. In particular, one can investigate whether the
 addition of just the $(1,2)$ (half-supersymmetric) irrelevant operator may be sufficient to define the theory. One can also try to define dipole CFTs via their extended  symmetries. Since the latter appear connected to the form of the entropy via the symplectic form argument we have been advocating for, it appears that understanding the entropy in these theories should be resolved first.

 %As discussed, so far only one non-trivial warped AdS$_3$ background is known how to be obtained from a decoupling limit of string theory, lending credence to the UV-completeness claim.   It would be nice to have this information for all deformations of the D1-D5 CFT that start as a $(1,2)$  one, and . One additional hurdle with respect to the LS-CFTs is that the form of the entropy in these theories has not yet been established, as it is not clear, given this one decoupled example, whether it should be Cardy or $J\bar T$ - deformed Cardy.  

%A similar logic also holds in the dipole case, except that here only one background is known to be obtained from a decoupling limit.
%
%Let us explain the chosen terminology. \textcolor{red}{\emph{Here or footnote?}} 
%\bi
%\item \emph{dipole}: the deformation is a null vector (dimension length). In stringy example the excitatons are truly extended dipole-like degrees of freedom with charges at the ends. It need not be null, and can also be higher dim'l. E.g. $JT^a$. By this definition, $J\bar T$ is part of this class.  
%\ei
% % To note also that it's not clear the entropy in standard $T\bar T$ is related to counting stringy dof, though in LST it appears so. 

\subsection{Comments on asymptotically flat spacetimes with and without a linear dilaton\label{afvsald}}

%\subsection{An ALD perspective on flat holography}

One of the main motivations  for studying the asymptotically linear dilaton (ALD) backgrounds is as an intermediate set-up between the well-understood AdS holography and the more physically relevant asymptotically flat (AF) case. %, all while maintaining a string-theoretical handle over the duality.  
While  with the latter  it shares the asymptotic flatness of the metric (in a specific sense), an important relative advantage of ALD holography is that it can be derived from a decoupling limit of string theory, whereas in AF holography, all dynamical constructions are bottom-up.  

In this review, we have discussed at length the holographic description of ALD spacetimes, reviewing top-down evidence that they are related to non-local, non-gravitational theories of strings, and bottom-up evidence they share the universal properties of  $T\bar T$ -deformed CFTs. Given the similarities between ALD and AF spacetimes, it is worthwhile to ask: to what extent  could the ALD perspective be useful for AF holography, and vice-versa? 
%
%Given the progress in ALD$_3$ holography over the past few years, most notably its connection to $T\bar T$, it is worthwhile to ask to ask what lessons one may draw from it concerning AF holography. 
For example, while the standard approach to AF holography is centered around null infinity, alternate approaches \cite{Marolf:2006bk} that exploit the similarity between AF and ALD spacetimes  are more naturally focussed on spatial infinity. It is therefore interesting to understand the asymptotic symmetries and other bottom-up observables in AF spacetimes from this perspective. %In the following, we make a number of assorted comments on this topic that the reader may (or may not) find interesting.

% It has been remarked many times that ALD spacetimes are quite similar to AF and have sometimes been proposed as toy models/inspiration for the latter. In this section, we'd like to discuss/comment on the similarities and differences between the two, and explore  the question: to what extent  could the ALD perspective be useful for AF holography \& vice-versa? \textcolor{red}{\emph{Say smth above about lack decoupling AF \& necessity bottom-up.}}

\subsubsection{Asymptotically flat spacetimes and the celestial holography programme}

The Penrose diagram of a spacetime -  in particular, its boundary -  often plays an important role in guiding the choices made in bottom-up holography. In the well-known Penrose diagram of flat space, the smoothest part of the boundary is represented by the null infinities, $\mathcal{I}^\pm$. Their structure, together with that of $i^0$, is expected to be preserved   when  non-vacuum configurations are also considered. 

An asymptotic analysis of the symmetries of AF spacetimes shows they do not simply consist of the  expected Poincar\'e  generators, but instead comprise an infinite-dimensional extension known as the BMS group \cite{Bondi:1962px,Sachs:1962wk}, generated by translations along null infinity that are parametrised by an arbitrary function on the celestial sphere. Quite remarkably, in \cite{Strominger:2013jfa} it was shown that  BMS transformations at $\mathcal{I}^+$ and $\mathcal{I}^-$ that are related via a suitable `antipodal' matching condition in the vicinity\footnote{More precisely, between the infinite past of $\mathcal{I}^+$ and the infinite future of $\mathcal{I}^-$.} of $i^0$ are symmetries of the S-matrix, and their action can be directly related to Weinberg's soft graviton theorem \cite{He:2014laa}. One may additionally consider an infinite-dimensional extension of  Lorentz transformations known as superrotations \cite{Barnich:2010eb}, whose algebra in four dimensions consists of two copies of the Witt algebra These symmetries can be related to subleading soft theorems \cite{Cachazo:2014fwa}.  

The leading holographic proposal for   AF spacetimes is a bottom-up approach known as the `celestial holography' programme. Concentrating on four dimensions - where the richest structure unfolds - this programme  interprets $4d$ scattering amplitudes as correlation functions in a two-dimensional `celestial' CFT that lives on the celestial sphere, see \cite{Raclariu:2021zjz,Pasterski:2023ikd} for recent reviews.  This  is enabled by the identification of the $4d$ Lorentz group with the $2d$ conformal group and the fact that the soft charge associated to superrotations - which enhance these symmetries -  can be manipulated into a candidate two-dimensional stress tensor,  with the standard action on boost eigenstates. 
%Despite much progress in this subject,  many properties of celestial CFTs are still unclear, in particular whether they can carry independent information from the S-matrix. \textcolor{red}{\emph{Be nice!}}
It is important to note that celestial CFTs ($\mathcal{C}$CFTs) are physically different from standard $2d$ CFTs, as many of their  properties are determined not by intrinsic $2d$ QFT considerations, but by those of the higher-dimensional S-matrix they encode. 

Even though the celestial CFT  lives on a codimension two surface with respect to the AF spacetime whose gravitational dynamics it is conjectured to encode,  its structure is intimately tied  to that of null infinity, where the massless states that scatter are prepared and 
the symmetry generators  that are central to  this proposal are constructed. 
There also exists a `Carollian' version of this programme (see \cite{Nguyen:2025zhg} for a recent review), where null infinity makes an  explicit appearance.

\subsubsection{The $i^0$ point of view}

However, as already emphasized in \cite{Witten:1998qj},  the conformal boundary of a spacetime need not play a special role in holography. In AdS/CFT, its usefulness is associated to the fact that it provides a convenient parametrisation of the boundary conditions for bulk fields, which are then related to dual operators. However, in other spacetimes, e.g. the Schr\"{o}dinger and warped AdS spacetimes discussed in section \ref{wads3toym}, the na\"{i}ve conformal boundary is not useful for this.  From the point of view of defining the gravitational theory, one is rather looking for a natural place where boundary conditions for the various fields should be imposed. In flat space, this role is played by (the resolved) spatial infinity, $i^0$,  where the variational principle is well-defined and one can  define charges that are integrable and conserved - unlike those defined at null infinity, which are affected by the passage of flux. The null infinities $\mathcal{I}^\pm$ are, rather, hypersurfaces where incoming and outgoing states are defined. A quick parallel with the case of timelike boundaries (AdS) indicates that possibly the most natural way 
to think about the  asymptotic symmetries at null infinity is by transporting them from near $i^0$.

The structure flat space near $i^0$ is best seen in a foliation of $d+1$ - dimensional Minkowski spacetime with $d$ - dimensional de Sitter slices  %hyperboloid of spacelike directions)

\be
ds^2 = d\rho^2 + \rho^2 ds^2_{dS_d} \; , \;\;\;\;\;\; \;\;ds^2_{dS_d} = - d\tau^2 + \cosh^2 \tau\,  d\Omega_{d-1}^2 \label{flatdsslice}
\ee
These coordinates do not cover the full Minkowski space; rather,  in the vicinity of $i^\pm$, Euclidean AdS$_d$ slices can be used, see figure \ref{penrosemink}. One can then generalise this to AF spacetimes by employing  a so-called Beig-Schmidt decomposition near $i^0$. In $4d$, this takes the form \cite{Beig:1982ifu,Compere:2011ve} %\emph{On which of these boundary fields do the BMS charges depend?}
%\textcolor{red}{\emph{Fix refs!}}

\be
ds^2 = \left(1+ \frac{2\s}{\rho} + \ldots \right) d\rho^2 + \rho^2 \left(h_{ab} + \frac{f_{ab}}{\rho} + \ldots \right)(dx^a + N^a d\rho) (dx^b + N^b d\rho)
\ee
where the various coefficient functions only depend on $x^a$. The asymptotic Einstein equations and boundary conditions place  various constraints constraints on them: for example,  $h_{ab}$ is the metric on a positive-curvature Einstein space, $\s$ satisfies a certain wave equation on this space, etc. 
%
%\be
%R_{ab}[h_{ab}] = 2 h_{ab} \;, \;\;\;\;\; (D^2+3)\s=0
%\ee
%etc. \emph{Which are actually important asymptotically?}  
Usually, in $4d$, one considers a fixed de Sitter asymptotic metric - choice linked to  allowing BMS transformations, but not superrotations. %\textcolor{red}{\emph{Check!}}
%The boundary conditions considered in \cite{} fix $h_{ab}$. This seems to disallow superrotations.
% \textcolor{red}{\emph{Check this and also Lorentz!}}

An important point  is that solutions allowed by Bondi gauge at null infinity %,  which are required to fulfill specific falloffs as $|u| \r \infty$ \textcolor{blue}{(no radiation in this limit, as a condition on matter stress tensor)},  
map to just a \emph{subset} of the solutions in Beig-Schmidt gauge, which %\textcolor{blue}{are homogenous at leading order} \textcolor{red}{\emph{Check!}} and
 have definite parity under $\tau \r -\tau$ and an antipodal flip of the celestial sphere \cite{Compere:2011ve,Troessaert:2017jcm,Capone:2022gme}.  This can be established by  comparing the expansions around null and spacelike infinity in an appropriate regime of overlap\footnote{The relation between Bondi and $i^0$ coordinates is
$
u = - \rho e^{-\tau} , \, r= \rho \cosh \tau
$
and solutions can be expanded in the double limit
$
r\gg |u|\gg 1, \, \rho \gg e^\tau \gg 1
$.}. These parity properties are ultimately  responsible for the  antipodal matching to Bondi data on $\mathcal{I}^-$ . %, and the diagonal BMS symmetry of scattering. 
However, if one restricted the analysis to $i^0$ only, it appears there is no reason  not to keep both parities, resulting in an enlargement of the ASG at spatial infinity  \cite{Compere:2011ve,zwikel}. 
%
%The result of of \cite{} is that the Bondi data at $\mathcal{I}^+$, which are required to fulfill specific falloffs as $|u| \r \infty$ (no radiation in this limit, as a condition on matter stress tensor), map to homogenous solutions to the BS equations. \emph{True? Or, the leading solutions are homogenous?} These solutions have specific parity properties under $\tau \r -\tau$, which makes them map to Bondi data on $\mathcal{I}^-$. This is how a single, diagonal BMS is found. \emph{Clear that two of them exist at $i^0$ in $4d$?} To
%\textcolor{red}{ Check statement that these more restrictive falloffs follow from evolution of (compactly supported?) or, in any case, reasonable initial data on $\mathcal{I}^-$.  }

% [It was found in \cite{nguyen} that matching to $\mathcal{I}^+$ data selects a certain subset of solutions of the phase space at $i^0$, which happen to satisfy a certain parity condition, ultimately responsible for antipodal matching. Note this analysis fixes the boundary metric in Beig-Smidt gauge, and superrotations are not allowed. \emph{Re-check!} The punchline would be that, if we stayed at $i^0$, the ASG would be bigger (double), but the requirement that it be extendible to null infinity (where it can be used to predict the very nice symmetries of the S-matrix) cuts it in half. The question is then whether the symmetries of the ``theory'' should be doubled BMS, or just the extendible piece?   Is it a necessary requirement that the ASG be extendible to where the states are defined?]

The same phenomenon happens in the much simpler setting of $3d$ AF spacetimes, studied in \cite{Compere:2017knf} and reviewed below. There, superrotations (whose action in $3d$ is perfectly well-defined, unlike in $4d$)  are easily allowed by letting the trace of the boundary Einstein metric fluctuate. The ASG analysis at spatial infinity singles out \emph{two} supertranslation and \emph{two} superrotation extended symmetries; the requirement of a smooth extension to null infinity  then  projects them to  the well-known definite-parity combinations.

%We will shortly see a similar phenomenon in $3d$, where a doubled BMS$_3$ can be easily constructed. 

\subsubsection{AF vs. ALD spacetimes}

\begin{figure}[t]
\begin{minipage}{0.45 \linewidth}
\centering
\includegraphics[height=4cm]{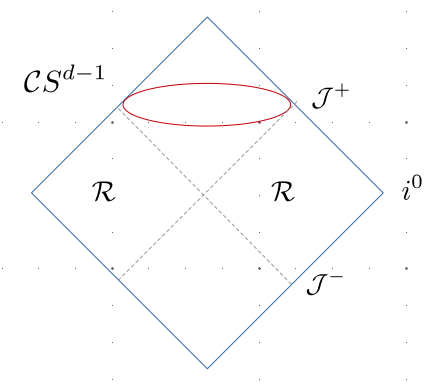}
\caption{\small{Minkowski$_{d+1}$ Penrose diagram}}
\label{penrosemink}
\end{minipage}
\hskip 0.05 \linewidth
\begin{minipage}{0.45 \linewidth}
\centering
\includegraphics[height=4cm]{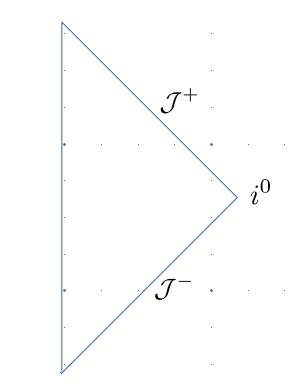}
\caption{\small{Penrose diagram for global ALD$_3$}}
\label{penroseALD}
\end{minipage}
%\hskip 0.05 \linewidth
%\begin{minipage}{0.3 \linewidth}
%\centering
%\includegraphics[height=4cm]{PenrosediagramMinkowski}
%\caption{  $5d$ flat c)}
%\label{penrose5dflat}
%\end{minipage}
\end{figure}

As reviewed in section \ref{holostr}, the string-frame metric of ALD spacetimes, which for  a stack of decoupled  NS5-branes reads 
\be
ds^2 = dx_6^2 + k\a' \left( \frac{dr^2}{r^2} + d\Omega_3^2 \right) \;, \;\;\;\;\;\;\; e^{2\Phi} = \frac{k\a'}{r^2} \label{decns5sol}
\ee
becomes flat asymptotically, upon defining $\phi = - \sqrt{k\a'} \ln r$. Since this spacetime is obtained via a string-theoretical  decoupling limit, one  concludes that the dual LST lives at  spatial infinity $\mathbb{R}^{1,5}$, and not the conformal boundary of the spacetime, illustrated in figure \ref{penroseALD}, for te NS5-F1 system, for which the interior of the spacetime can also be studied in the supergravity approximation.

%Even though the Einstein frame metric is not asymptotically flat according to the standard definition, 
%Since ALD spacetimes have an asymptotically flat metric, in addition to the linear dilaton, the two setups appear close enough to each other to deserve a comparison. % In particular, it seems the causal structure of the black hole backgrounds is very similar to that of flat space. 
%
 As remarked by Marolf  \cite{Marolf:2006bk}, the similarity between the asymptotic structures of ALD and AF spacetimes is most visible  from the point of view of the asymptotic expansion in the vicinity of $i^0$. %, the structure of  ALD spacetimes is rather similar to that of   AF spacetimes in BS gauge. 
 More precisely, if one %starts with the $10 d$ decoupled NS5 background \eqref{} \textcolor{red}{\emph{Emphasize LST doesn't live on conformal boundary.}}
translates the metric \eqref{decns5sol}  to Einstein frame, $ds_E^2 = e^{-\Phi} ds^2_{str}$ and defines $\rho^2 =4  r \sqrt{k\a'}$, then the latter takes the form 

\be
ds_E^2 = d \rho^2 + \rho^2 \left(dx'^2_6 + \frac{1}{4}d\Omega_3^2\right)  \;, \;\;\;\;\; x'^\mu = \frac{x^\mu}{2\sqrt{k\a'}}
\ee 
which is similar to  the Minkowski metric  in dS slicing \eqref{flatdsslice}, the only difference being that the slices are flat, rather than de Sitter. %\textcolor{blue}{Note this is different from the naive taking the Minkowski construction for $dx_5^2 + d \ln r^2$, as the rotation symmetry is broken in the interior.  } 
% Given this similarity of  the metric structure near $i^0$, as well as the associated Penrose diagrams, the idea put forth in 
 %
%  perhaps the holographic descriptions should also be similar. 
   Note that one should not identify the metric that multiplies $\rho^2$ above with that of the spacetime where the dual theory lives: only the  $dx_6^2$ part represents the directions of the dual LST, but not the  $S^3$, despite the fact that it appears on the same footing from the point of view of the asymptotic expansion\footnote{
One is naturally led to wonder  \cite{Marolf:2006bk} % \textcolor{red}{\emph{Check!}} 
whether the $S^{d-1}$ appearing in the flat space expansion \eqref{flatdsslice} may  have a similar fate. Perhaps the question can be answered in  the BFSS model \cite{Banks:1996vh}, which captures the symmetries of scattering in M-theory \cite{Tropper:2023fjr,Herderschee:2023bnc}, but is not extended along the sphere directions. }. 
%In any case, the comparative holographic analyses performed in \cite{Marolf:2006bk}  did, indeed, show similarities, though not to the desi

Given this similarity of  the asymptotic expansions near $i^0$, 
 \cite{Marolf:2006bk}  suggested that the holographic analyses of the two spacetimes could proceed in analogy to each other, finding, indeed, many  echoing results. In the following we, too,  perform a comparison of  various observables in certain AF and ALD spacetimes related to the NS5-F1 system, hoping the  perspectives it offers will help shed light on both.
 
 %detailed comparison in $3d$, where we can be entirely explicit.   The hope is that the better understanding of ALD would shed light on AF. In this subsection, we would like to discuss this comparison, in particular in $3d$. We focus on several different aspects. 

\bigskip

\noindent\emph{Penrose diagrams}

\medskip

\noindent From a simplified $2d$ perspective, the structure of the Penrose diagrams for AF and ALD spacetimes is extremely similar. This is less so if we also consider how the compact directions are being fibered over this $2d$ section. For concreteness, we consider the following three geometries: 
\bi
\item $3d$ Minkowski space, with $ds^2 = - dt^2 + dr^2 + r^2 d\s^2$ and $\s \in (0,2\pi)$. In the Rindler patch

\be
ds^2  = d\rho^2 + \rho^2 (-d\tau^2+\cosh^2 \tau d\s^2) = d\rho^2 + \frac{\rho^2}{\cos^2 T} (-dT^2 + d\s^2)
\ee
where $
t= \rho \sinh \tau = \rho \tan T, \; r= \rho \cosh \tau = \frac{\rho}{\cos T}
$ with   $T \in \left(-\frac{\pi}{2}, \frac{\pi}{2}\right)$. 
\item the global vacuum of the interpolating AdS$_3 \r $ ALD$_3$ spacetime\footnote{Taking $r_0^2 <0 $ in \eqref{aldbh} with $\d_n=0$ and $r_1=r_0 \sinh \d_1$, smoothness near $r =0$ fixed $\frac{r_1}{|r_0|} =\frac{R}{2\pi \sqrt{\a' k}}$, while the equations of motion  fix $r_1 \sqrt{r_1^2-|r_0^2|} = \frac{p \a'}{v}$. We have then rescaled $r$ by a factor of $|r_0|$.
} %\textcolor{red}{\emph{geodesic completeness?}}
\be
\label{glald}
ds^2 = \frac{r^2}{r^2 + \upsilon^2} \left[d\s^2 - \left(1+\frac{1}{r^2}\right) dt^2)\right] + k \a' \, \frac{dr^2}{r^2+1} %+d\Omega_3^2\right) 
\;, \;\;\;\;\;\;\upsilon = \frac{R}{2\pi \sqrt{k\a'}}
\ee
\item the asymptotically   $\mathbb{R}^{4,1} \times S^1$  $6d$ black string (we only write the extremal solution, for simplicity) %\textcolor{red}{ASG?}
\be
ds^2 = \frac{r^2}{r^2 + p\a'/v} (d\s^2 -  dt^2) + \left(1 + \frac{k \a'}{r^2}\right) \left( dr^2+ r^2 d\Omega_3^2\right)  \label{afblackstr}
\ee
\ei
The last example is included in order to explore whether AF holography  can be adapted to describing the NS5-F1 system studied herein, namely a slightly modified version of AF holography where the flux through the celestial $S^3$ is held fixed. That is, one is looking for a holographic description not of any object with asymptotically flat asymptotics, but only of those having the same charges as the NS5-F1 system. Since there is no decoupling limit, there is no energetic reason to separate system into `background' and `excitations'. Taking the analogue of the $N \r \infty$ limit may help. %\textcolor{red}{\emph{Is this $k \r \infty$?}}

 Let us start by comparing three-dimensional Minkowski spacetime with the global ALD$_3$ spacetime \eqref{glald}, whose Penrose diagrams are depicted in figures \ref{penrosemink} and, respectively, \ref{penroseALD}. 
\noindent Near $r \approx 0$, the latter looks like flat space, with $\s \sim \s +R$ the angular coordinate. However, asymptotically, the $\s$ circle has fixed size and factors out, while in the Minkowski diagram in figure \ref{penrosemink}, its size grows and it remains part of the boundary. As $r\r \infty$,  the rest of the metric becomes conformal to Rindler space, whose boundary corresponds to  $r \r \infty, t\r \pm \infty$, resulting in the Penrose diagram of figure \ref{penroseALD}. %, \textcolor{blue}{which corresponds to roughly one quarter of the Minkowski Penrose diagram}.  
 One may also think of the Rindler time coordinate as being replaced by $\tau=\sqrt{t^2-\s^2}$ on the boundary, since one recovers $2d$ Lorentz invariance as $r\r \infty$; in any case, the effect of $\s$ is  very small, as far as the Penrose diagram is concerned, since $|t| \r \infty$. 
%. \textcolor{blue}{Alternatively, this looks a bit like Rindler space, where the null line is parametrised by $\l = \sqrt{t^2-\s^2}$.}

 Another difference between the two spacetimes is that in \eqref{glald}, the natural time coordinate  near $i^0$, namely $t$, 
 is a good coordinate throughout, whereas in the Minkowski case, the  time coordinate that is good near $i^0$ is $\tau$ or $T$, which corresponds to  Rindler time from the point of view of the global spacetime.  Heuristically, we can  think of the Penrose diagram \eqref{penroseALD} of the ALD spacetime as corresponding to the Rindler wedge of the Minkowski Penrose diagram \eqref{penrosemink}. Thus, while in the ALD case,  boundary time  provides a good notion of global time, in Minkowski the preferred time coordinate is, in a certain sense, emergent from the point of view of the time coordinate that is natural near $i^0$, suggesting the natural Minkowski vacuum  is an entangled state from the point of view of the theory defined at $i^0$ \cite{Melton:2023dee}.  %  \textcolor{red}{\emph{Refs?}}

It is also interesting to contrast the ALD spacetime $\times S^3$ with the asymptotically $\mathbb{R}^{5,1} \times S^1$  spacetime describing the black string \eqref{afblackstr}.  Since we are only interested in the asymptotics near $i^0$, the corresponding piece of the Penrose diagram for \eqref{afblackstr} will look like that in figure \ref{penrosemink}, except that the celestial sphere is an $S^3$, and there is an overall multiplication by an $S^1$ of fixed asymptotic size. Thus, asymptotically, the size of the $S^3$ will grow much faster than that of the $S^1$.  Since the celestial sphere is threaded by three-form flux, the asymptotic analysis is likely different from standard analyses of AF spacetimes, as in this particular system it makes sense to work in a fixed charge sector. In particular, the ASG generators  selected by the vanishing of the symplectic flux, discussed in section \ref{newperspold} and that roughly take the form \eqref{strasgald}, correspond to a set of field-dependent diffeomorphisms that mostly live along and depend on the $U,V= \s \pm t$ directions associated with the NS5-F1 string. While, from a top-down perspective, these should be the coordinates involved in the holographic dual, if it exists, none of them makes a natural appearance from the point of view of either $i^0$ or $\mathcal{I}^\pm$ (though $t$ seems more closely related to Minkowski time).  Neither is the obvious Lorentz invariance of the system along the $\s, t$ directions evidenced in the Penrose diagram. 
 
 On the other hand, the asymptotic structure of the ALD background \eqref{glald} $\times S^3$ - say, in Einstein frame -  yields a `boundary' metric of the form $d\s^2-dt^2 + d\Omega_{S^3}^2$, where the $S^3$ and $S^1$ have fixed size\footnote{ If we  turn on the $1$ in the $f_5$ harmonic function, the size of the $S^3$  will start to grow much faster than that of the $S^1$, yielding the previous case.}. Remember,  though, that the $S^3$ is  not part of the LST description, while the $S^1$ is. The natural time coordinate at $i^0$ in the ALD spacetime does now coincide with the boundary time (of the decoupled NS5-F1 theory), and the Lorentz invariance of the  $\mathbb{R}^{1,1}$ parametrised by $\s, t$ is now manifest at the level of the boundary metric, though it is still obscured at the level of the Penrose diagram. The coordinates along  $\mathcal{I}^\pm$ of the ALD spacetime do not appear natural from the point of view of the boundary theory. 
 
If a conclusion can be drawn from the above discussion, it is that the coordinates that naturally parametrise the boundaries of the Penrose diagram, either $i^0$ or $\mathcal{I}^\pm$, are not necessarily simply related to those in the dual field theory. While in the ALD spacetime, it makes sense to specify the boundary  data at $i^0$, since we know this is where the dual theory lives, in AF spacetimes neither $i^0$, nor  $\mathcal{I}^\pm$, seem preferred from the point of view of the coordinates along the branes.  Nonetheless, from the point of view of the AF gravitational theory, boundary conditions do need to be specified (at $i^0$); it would be interesting to understand this issue better. %\textcolor{red}{\emph{Mention brane-antibrane?}}

 %. The asymptotic structure of the Minkowski Penrose diagram  one obtains obscures the Lorentz invariance along the string, which one expects to be preserved by the associated irrelevant deformation.  The natural time coordinate on the branes is $t$, which is related to the Minkowski/Schwarzschild time, but not to any of the natural time coordinates  along $i^0$ or $\mathcal{I}^\pm$. \textcolor{red}{\emph{Recheck!}} Thus, if there is any holographic dual to this spacetime - not clear, because of the absence of decoupling - it doesn't appear associated to the boundaries of the Penrose diagram. The fate of the $S^3$ is, again, unclear, since it is clearly not part of the boundary theory in the intermediate ALD decoupling limit.  \textcolor{red}{\emph{Re-think and rewrite!}}

% First, we should multiply the middle Penrose diagram by $S^3$, and work in terms of the rescaled (Einstein frame) metric. The asymptotic factor will contain   as boundary metric. However, we are quite sure the $S^3$ is not part of the boundary theory. In the $5d$ analysis, however, the $S^3$ is the celestial sphere. This is slightly different from standard case, as it is threaded by a fixed amount of three-form flux, which is possibly what one should consider. In any case, now the $S^3$ grows faster than the $\mathbb{R}^{1,1}$, which motivates rescaling the time in order for it to participate to the conformal boundary.  

\bigskip

\noindent\emph{Asymptotic symmetries}

\medskip

\noindent Let us now contrast the  asymptotic boundary conditions and symmetries for the three spacetimes we are considering. Since, in the latter two cases, the ASG analyses were performed for the black hole spacetimes only, we will need to consider the finite-temperature versions  of \eqref{glald} and \eqref{afblackstr}.

%It is more interesting now to consider black hole spacetimes for b) and c). 

%  In this subsection, we would like to perform a detailed comparison of the interpolating ALD$_3$ spacetime  we have been studying with $3d$ flat spacetime in pure $3d$ Einstein gravity. 

Let us start by recalling the results of \cite{Compere:2017knf} for the asymptotic symmetries of $3d$ Minkowski from the point of view of $i^0$. 
The most general metric near $i^0$ in a Beig-Schmidt-type expansion is

\be
ds^2 = d\rho^2 + (\rho^2 h_{ab}^{(0)} + \rho h_{ab}^{(1)} + h_{ab}^{(2)}) dx^a dx^b
\ee
where $h^{(0)}$ is  a  locally dS$_2$ metric, $h^{(1)}$ encodes the conserved stress tensor, while $h^{(2)}$ is entirely determined by the previous two.  As for pure gravity in AdS$_3$, this expansion is exact. 
To obtain interesting  superrotations, \cite{Compere:2017knf} only fix the `boundary' metric $h^{(0)}$ up to a conformal factor, while letting its conformal mode fluctuate. Working in terms of null coordinates $x^\pm = T \pm \s$, the most general allowed metric is
%
%boundary condition is $h^{(0)}_{++}=h^{(0)}_{--} =0$, while $h^{(0)}_{+-}$ is 
given by an arbitrary (anti)holomorphic coordinate transformation $x^\pm \r X^\pm(x^\pm)$ on dS$_2$

\be
%ds^2= d\rho^2 - 2 h_{+-}^{(0)} \left( \rho dx^+ + \frac{\Xi_-(x^-)}{h_{+-}^{(0)}} dx^- \right)  \left( \rho dx^- + \frac{\Xi_+(x^+)}{h_{+-}^{(0)}} dx^+ \right)
h^{(0)}_{++} = h^{(0)}_{--} =0  \;, \;\;\;\;\;\;\;
 h_{+-}^{(0)} = \frac{\p_+ X^+ \p_- X^-}{1+\cos (X^++X^-)} \label{afbndcond}
\ee
In principle, one can also consider conical defect solutions, which slightly generalise the form that $h^{(0)}_{+-}$ can take, but we will not do so.   \cite{Compere:2017knf} additionally fix the trace of $h^{(1)}$, as required by the vanishing of the radial symplectic flux when tr$\, h^{(0)}$ is allowed to fluctuate. The most general solution to the equations of motion is then parametrised by the other two components of $h^{(1)}$, which correspond to two (anti)holomorphic functions, $\Xi^\pm (x^\pm)$.

 % What is fixed instead (something must be, to have a well-defined variational principle) is the trace of the stress tensor/$h^{(1)}$. They then find the most general solution in their phase space to be parametrised by two arbitrary holomorphic coordinate transformations in $dS_2$, $X^+(x^+)$ and $X^-(x^-)$, as well as two holomorphic stress tensor components $\Xi^\pm (x^\pm)$. 
% \emph{Or, they just affect the boundary conditions/ can be absorbed by rescalings?} Let us 

Note that the AF boundary conditions \eqref{afbndcond} near $i^0$ are rather similar to the ones \eqref{aldbndcond} we found via the symplectic form analysis of the ALD backgrounds \eqref{aldbh}, if we ignore the field-dependence, e.g. by setting\footnote{The asymptotic symmetries of  \cite{Georgescu:2022iyx} were nonetheless derived under the specific assumption that $L_{u,v} \neq 0$.} $L_{u,v} =0$.  In that case, we expect $h^{(0)}_{\pm\pm}$ to be fixed, while only the trace, $h^{(0)}_{+-}$, fluctuates. %Of course, its functional dependence is different from that in AF spacetimes.
 It would be interesting to directly study the asymptotic symmetries of the  vacuum ALD spacetime, for a more precise comparison, though it is clear that the field-dependence is only present in this case, but not in the AF one.
%
%The boundary conditions imposed by \cite{Compere:2017knf} consist of fixing the metric, up to a conformal factor, and the trace of the stress tensor.  \emph{Check!} In a certain sense, they are similar to ours . 
%Namely, they fix two components of the boundary metric, while letting the remaining one (the determinant of $h^{(0)}$) fluctuate. For the ALD background, the boundary conditions were
%
%\be
%C_{uu} = \frac{2\a' L_u}{\b} C_{uv} \;, \;\;\;\;\;\; C_{vv} = \frac{2\a' L_v}{\b} C_{uv} \;, \;\;\;\;\;\; \b = \sqrt{p^2 + 4 \a'^2 L_u L_v}
%\ee
%where $C_{uv}$ is allowed to fluctuate; these sound extremely similar around the vacuum, had we studied it. 

%  Concretely, they study pure $3d$ gravity with no cosmological constant. To fix the notation, the metric on $3d$ Minkowski space is The change of coordinates between the two is 

%Concentrating on the spacetimes with $X^\pm=x^\pm$. \emph{Is this necessary?} 
The asymptotic symmetries associated with the above boundary conditions at $i^0$ are given by superrotations 
\be
\xi_{R} = R^+(x^+) \p_+ + R^- (x^-) \p_- \;, \;\;\;\;\;\;\;\; Q_{R^\pm} = \int d\s R_\pm \Xi_\pm \label{3dsrrot}
\ee
%with associated conserved charges %[ \emph{Are there other ALD analyses with what conclusions?} . We compare with the analysis of \cite{} of the ASG of $3d$ flat space at spatial infinity, where conformal (\emph{Name?}) boundary conditions were imposed, which did allow for both supertranslations and superrotations, which in $3d$ are not singular. \emph{Correct?} One can check this ASG is compatible with the rotating conical defects. \emph{Re-check!} The phase space was parametrised by two functions of $x^+$ and two of $x^-$, and the aymptotic charges were given by ]
%
%\be
%Q_R = \int d\s (R_+ \Xi_++R_- \Xi_-) \;, \;\;\;\;\; 
%\ee
and supertranslations,  parametrised by  two other  functions, $T^\pm (x^\pm)$, %with  \textcolor{red}{\emph{Write diffeo instead!}}
which for $X^\pm = x^\pm$ simplify to 
%\be
% \xi^\rho =\om = - \frac{1}{2} (T^+ \p_+ X^+ + T^-\p_- X^-) \tan \frac{X^++X^-}{2} - \frac{1}{2} \frac{\p_+ (T^+ \p_+ X^+ ) }{\p_+ X^+} - \frac{1}{2} \frac{\p_- (T^- \p_- X^- ) }{\p_- X^-}
%\ee
%
\be
\xi_{T} = \left( - \frac{1}{2} (T^+ + T^-) \tan T - \frac{1}{2} (T'^+ + T'^-) \right) \p_\rho + \O(\rho^{-1}) \;, \;\;\;\;\;\;
%\ee
%
%The associated conserved charges are given by 
%\be
Q_{T^\pm} = \int d\s T_\pm \Theta_\pm  %+ T_- \Theta_-) \;, \;\;\;\;\; \Theta_+ = f (\p_+ X^+)
\ee
where $\Theta_\pm$ are specific functions of $\p_\pm X^\pm$, whose form can be found in \cite{Compere:2017knf}. The algebra of the associated conserved charges consists of two commuting copies 
 of the BMS$_3$ algebra

\be
[R_m^\pm, R^\pm_n] = (m-n) R^\pm_{m+n} \;, \;\;\;\;\; [R^\pm_m, T^\pm_n] = (m-n) T^\pm_{m+n} \;, \;\;\;\;\;[T^\pm_m,T^\pm_n] =0 \label{bms3alg}
\ee
Thus,  the ASG  of three-dimensional flat space at $i^0$ consists of \emph{two} copies of the BMS$_3$ algebra, which is double the standard ASG for this  spacetime. 

Let us now compare this to the asymptotic symmetry generators of the ALD$_3$ spacetime at the corresponding spatial infinity. Due to the  method used to find the generators, the analysis was performed at finite temperature only, but we will be ignoring the associated field-dependence for the purposes of our comparison. %\textcolor{red}{\emph{Can't we find  bnd condition in vac and simply settle this?}}
 The  ASG generators \eqref{solFUV} and \eqref{formFr} were parametrised by four functions, two of $u$ and two of $v$, and took the schematic form 

\be
\xi_{ASG} = [\, f(u) +\ldots] \, \p_U +[\, \bar f (v) +\ldots]\, \p_V  + \frac{1}{r} \left[f_r(u) + \bar f_r (v)\right] \p_r + \O(r^{-2})
\label{strasgald}
\ee
where the $\ldots$ stand for additional field-dependent pieces we will not track down.
Thus, the diffeomorphisms parametrized by $f$ and, respectively $\bar f$, appear analogous to the superrotations at flat spatial infinity and also form two copies the Witt algebra, while the radial functions $f_r, \bar f_r$, appear analogous to the two supertranslations\footnote{At this very rough level of comparison, we are completely neglecting the field dependence and the associated non-linearities in the charge algebra.}.  Their associated conserved charges commute with each other, the only non-zero commutator found at leading order in the perturbative analysis about the constant backgrounds is the zero mode \eqref{aldcentralterm}   in the $\{ Q_f, Q_{f_r}\}$ commutator and its right-moving counterpart. This is of course qualitatively different from the corresponding $[R, T]$ commutator \eqref{bms3alg} in AF space. 
%
% One pair shifting the boundary coordinates $U,V$, and the other corresponding to radial shifts. The first pair satisfied a non-linear realisation of two copies of Witt, and thus appear analogous to the two superrotations found at spatial infinity of flat space. The other transformations commute with each other, and seem to give a central extension of the form 
%\be
%[R'_m, T'_n] \sim  k  n^2 \d_{m+n} \;, \;\;\;\;\; [T'_m, T'_n] = 0
%\ee
%ignoring field-dependent factors and at  leading order. Left-right generators in principle commute, up to the field-dependence. 
Thus, we see that up to this and issues related to the field-dependence of the asymptotic symmetry generators, which is present in ALD, but completely absent in Minkowski spacetimes with the given set of boundary conditions,  the asymptotic symmetries generators at $i^0$ in the two spacetimes are  similar to each other, and in one-to-one correspondence.

 It is also interesting to compare the asymptotic symmetries of the ALD spacetime with those of the asymptotically flat black string solution, obtained  in \cite{Georgescu:2024iam} via the condition of vanishing symplectic flux. They are found to have the same structure as \eqref{strasgald}, though with somewhat different field-dependence.  From the point of view of the asymptotic $\mathbb{R}^{1,4} \times S^1$, they only act along the 
%
% As for the $5d$ AF spacetime, we see that the asymptotic symmetries that are connected to the entropy (for the spherically symmetric case) only live in the 
 $\s, t,r$ directions, and not at all on the celestial $S^3$. The dynamics along the $S^1_\s$ direction is essential for seeing the associated  asymptotic symmetries (which act rather naturally along the common NS5-F1 directions); if one performed instead only a $5d$ asymptotic analysis, one would need to consider  symmetries that act non-trivially on the various KK modes. Symmetries acting on the $S^3$ in the ALD background  would  correspond to extended affine symmetries related to $SU(2)_L \times SU(2)_R$ currents that likely exist in this compactification of LST; it would be interesting to work them out and compare them to  the asymptotic symmetries of $\mathbb{R}^{4,1}$ \cite{Fuentealba:2022yqt}.

\medskip 

Back to the comparison between $3d$  flat space and the ALD$_3$ background, we have
so far remarked that, without any further assumptions, their asymptotic symmetries at $i^0$  are similar, and in one-to-one correspondence.   However, in both cases there are reasons to cut this initial ASG in two.

 In the AF case, the projection is motivated by mapping to the data a null infinity \cite{Compere:2017knf}. More precisely, if one starts with the general solution allowed by the boundary conditions at $i^0$ and maps it to $\mathcal{I}^{\pm}$, one finds that the boundary conditions of \cite{Barnich:2006av} on the allowed metric fluctuations at null infinity are violated%\footnote{Actually, for $X^\pm = x^\pm$, only the $g_{\s\s}$ component has a term proportional to $\Xi_++\Xi_-$ and $\sqrt{r}$, instead of being $\O(r)$, but this should be fine. \textcolor{red}{\emph{Does the violation come from non-trivial $X^\pm$?}} }
 .  This led \cite{Compere:2017knf} to impose a parity condition on  $X^\pm$, which implies an antipodal identification of the superrotation generators  (which excludes e.g. global de Sitter translations). As a consequence of this constraint,  $\Theta_+=\Theta_-$, which implies that supertranslations  with $T^++T^- =0$ are trivial, resulting in the antipodal identification of also the supertranslation generators. Supertranslations that satisfy this constraint can only generate  backgrounds with  $\Xi^+(x^+) + \Xi^-(x^-)=0$, though one could, in principle, also  consider  backgrounds with general $\Xi^\pm$, even if they differ by a trivial diffeomorphism. 
 %
% half of the superrotation charges \eqref{3dsrrot} become trivial.
% 
 % and $\Xi^\pm$. \textcolor{red}{\emph{Or, does the condition on $\Xi^\pm$ follow from that on $X^\pm$?}} In the latter case, 
  Note the condition $\Xi^+(x^+) + \Xi^-(x^-)=0$ translates, as $T \r \frac{\pi}{2}$, into a condition that relates their Fourier modes as $\Xi_{-n}^-= (-1)^{n+1} \Xi_n^+$, which amounts to the well-known antipodal identification. %The same occurs for the supertranslation ones, after $\Theta^\pm$ become identified due to the projection on $X^\pm$. 
% 
% , is being able to map to $\mathcal{I}^\pm$ that imposes a projection (\emph{What would be the falloffs at null infinity if we didn't make the projection?}). 
%  In AF spacetimes, the above asymptotic symmetries get reduced to just one set upon imposing parity conditions that appear linked to being able to define null infinity; a better understanding of these issues seems necessary. We actually need transformation to Bondi coordinates, $u=t-r$, for $u <0$ (as we will be matching to the $u \r -\infty$ limit). This is
%
%\be
%%u=-\rho \, e^{-\tau} = \rho \left(\tan T - \frac{1}{\cos T}\right) \;, \;\;\;\;\; r = \rho \cosh \tau = \frac{\rho}{\cos T}
%%\ee
%\textcolor{blue}{
% However, requiring smooth extension to $\mathcal{I}^\pm$ restricts the functions to $X^-=-X^+$, which in turn implies $R_-=-R_+$ and $\Theta_+=\Theta_-$, which in turn implies $\Xi_+-\Xi_-$ and $T^+-T^-$ are trivial. }  \textcolor{red}{\emph{Check!}}
  Nonetheless, even if we relax the parity conditions, it seems the asymptotic charges at null infinity are still conserved. %\textcolor{red}{\emph{Check!}} 
  In this case, it would be interesting to better understand the physics of these more general spacetimes.

 This should be contrasted to the  choice we made in the ALD space-time, namely to relate $f_r$ to $f$ in a specific way, $f_r = - \frac{p}{4} f'$. This choice  was motivated by the connection to $T\bar T$  and the fact that in AdS$_3$, a very similar relation between these functions follows from 
 the boundary conditions, even though the symplectic-form-based analysis is not able to capture this fact.  Recombining the ASG generators as $Q_f +  Q_{f_r= - pf'/4}$,  allows one to recover the central extension of the asymptotic symmetry algebra, which precisely matches the expected central extension of the $T\bar T$ - deformed algebra - a tantalising hint. It would be good to ascertain whether this reduction of the ASG is in place, e.g. by establishing the boundary conditions the various perturbations obey.

Neither of the above reasons to cut the number of ASG generators found at $i^0$ in two appears natural from the other's point of view. The  ALD   ASG generators (barring global translations), even when ignoring the field-dependence, to not appear to admit a smooth extension\footnote{ Ignoring the field dependence, the ALD generators take the form $e^{i n (t\pm \s)}$, while the limit to $\mathcal{I}^+$   fixes $r e^{- \sqrt{t^2-\s^2}} \approx r e^{-t}$ as $r \r \infty$.  In particular,  they all seem to have different falloffs at $\mathcal{I^+}$.} to $\mathcal{I}^\pm$,  so there is no particular reason to be imposing a parity condition. % Conversely, it does not appear sensible to impose parity conditions on the ALD  ASG generators:  first, because their field-dependence may impede it and second, because the requirement of a smooth action   on $\mathcal{I}^\pm$ does not appear to be satisfied for any of the non-constant generators.  
On the other hand, a recombination of the generators  of the form $R_m + \#  m T_m$ would not appear natural from the point of view of BMS$_3$: first, due to the different commutation relations of the ASG generators the algebra would still consist of two copies of the Witt one; second, one would also be discarding the constant mode of supertranslations ($T_+=T_-= const$) which is nothing but global time translations in Minkowski spacetime.

In conclusion, while the ASG generators at $i^0$ resemble each other for $3d$ AF and ALD spacetimes, the end proposals for the asymptotic symmetries are very different.    The fact the ALD generators  do not have a smooth extension\footnote{ The structure of null infinity in ALD spacetimes needs to be further studied, as it is not clear whether any of the nice properties of $\mathcal{I}^\pm$ in the AF case will extend to it. } to $\mathcal{I}^\pm$ need not preclude their symmetry interpretation at $i^0$.  
    If this is the case, then it is not clear why in the AF case, one should be imposing the requirement of a smooth extension to $\mathcal{I}^\pm$. 
 %
% , where the radial functions are related to the main ones if the central extension is to work out.
%
%
%Diffeomorphisms respecing this structure are 
%
%\be
%\xi^a = R^a (x^a) + \ldots \;, \;\;\;\;\; \xi^\rho = \om (x^a)
%\ee
%the first of which are related to superrotations, and the second to Spi supertranslations. \emph{Check!} 
%
% On the other hand, in the ALD background one has two sets of field-dependent transformations, two associated with the components of the diffeos that are parallel to the boundary, and two that are related to the radial component. In an AdS-like setting, the radial components would be fixed in terms of the other two; in the BMS-like setting, it seems parity relates the two memebers of each pair to each other. In ALD, it is not clear whether one will get more constraints from imposing proper boundary conditions, as opposed to working with the symplectic form. 
%
% The analogous procedure in AF would be to consider
% 
% \be
%% T_m = m R_m \;, \;\;\;\;
% R'_m  = R_m + \a m  T_m
% \ee
% which still satisfies a centerless Virasoro algebra, and to ignore the $T_m$ as 
% independent generators. Note, however, this throws out the $T_0 $ generator, which is nothing but global time translations in Minkowski 
% 
%\be
%\p_t = - \sinh \tau \p_\rho + \frac{\cosh \tau}{\rho} \p_\tau = - \tan T \, \p_\rho + \frac{1}{\rho} \p_T
%\ee
Apart from this general comment, it is not clear how far the analogy between the two spacetimes can be pushed. % very far. % or  whether it is useful. % o useful, and they are fundamentally different, despite some similarities.  %\textcolor or{red}{\emph{Rewrite!}}  \textcolor{blue}{Note, however, that analogue of Bondi coordinates along scri not very natural. Conformal symmetries are also broken in the interior of ALD. Not clear any of the nice properties of scri in the AF case will still hold in ALD, b/c of the presence of matter asymptotically. Map from spatial to null infinity hard.}

\bigskip

\noindent\emph{Scattering, correlation functions and boundary non-locality}

\medskip

\noindent As discussed, the main focus of the celestial holography programme in dimensions higher than three  is  the S-matrix. However, as pointed out in \cite{Marolf:2006bk}, by working with $i^0$ one may access different observables, such as correlation functions. The latter display strongly non-local features with respect to the `boundary' at $i^0$, in ALD as well as in AF spacetimes.  It would thus be interesting to study them further - at least for the purpose of enlarging the set of observables one can compute in AF spacetimes - and to understand their exact connection to the S-matrix. 

%
%\bi
%\item do they carry independent information from the S-matrix, or is it recoverable?
%\ei

An interesting question  that arises concerns the interplay of the asymptotic symmetries at $i^0$ with these observables. For example, for three-dimensional flat spacetimes, one may wonder whether correlation functions transform under the doubled BMS group we have advocated for%(the analogous question in $4d$ would be whether they transform under logarithmic supertranslations)
. While  one does not generally consider scattering in $3d$ flat space, as gauge fields and the metric have no  propagating degrees of freedom, one could in principle study massless scalar scattering and investigate  what its most general symmetries are. One may try to test, for example,  whether only a subset of the symmetries at $i^0$ act on the S-matrix, suggesting the restricted symmetries are related to restricted observables. Another advantage of the $3d$ setup would be that superrotations are not singular in this case. %,  though it is of certainly lesser physical interest. 

%However, one may equally well consider boundary to boundary propagators anchored at $i^0$, as Marolf points out.  Unfortunately, his analysis is inconclusive (the Feynman propagator is a purely divergent term, which gets thrown out upon "renormalisation". However, seems to reach same conclusion in ALD, which seems wrong. He may get decent results from summing modes). A more in-depth analysis by Suvrat. \emph{Conclusions? Explain non-locality sources!} The conclusion is that from the point of view of the theory living on $i^0$, correlation functions, as computed using the boundary-to-boundary propagator, are extremely non-local.  In particular, bulk-to-boundary propagator blows up outside local support of the source. 

% This could be interesting, at least given that superrotations are perfectly well-defined in this case: much cleaner,  though of obviously less physical interest. 

Another interesting avenue is to study the interplay of scattering and asymptotic symmetries in the ALD spacetime. Given the similarities between the ASG at spatial infinity and the symmetries of $T\bar T$, one may expect to obtain holographic Ward identities for correlation functions that resemble those of $T\bar T$. 
%
%One can also study scattering in ALD. In fact, supergravity calculations readily give answers for correlation functions (computed at $i^0$), which have the expected non-localities. It would be interesting to carefully contrast them with $T\bar T$. 
%In ALD, momentum-dependent correlators computed at $i^0$. \emph{True that not same results as in flat space because we're solving wave equation in a different metric?}
 One may also in principle compute S-matrix elements in this theory, by studying the scattering of long strings. %, which are presumably are prepared at null infinity. \textcolor{red}{\emph{Correct?}}
   Given that the extended asymptotic symmetries do not appear to survive at $\mathcal{I}^\pm$, it would be interesting to understand  how to recover the expected symmetries  of scattering amplitudes in ALD. 
%Observables: scattering, or also correlation functions? In ALD long strings - what happens to them if we turn on RR flux?} %$3d$ scattering: massless scalars only?

%Marolf studied this and found nonsense. 

%%\subsubsection*{Correlation functions and boundary non-locality}
%
% 
%\bi
%%\item can the S-matrix characterise a theory?
%%\item review Marolf argument about boundary non-local response 
%%\item review ALD momentum-dependent correlators
%\item difference from S-matrix in ALD? Are these different observables? LST dual of S-matrix?
%\ei
%One point is that, if ALD is $T\bar T$ with the expected symmetries, then expect them to act in some way on correlation functions.  
%Then, same question in AF at spacelike infinity. For simplification purposes, it would be good to do $3d$. Could it be that correlators transform under two copies of BMS, while S-matrix only under one?

\bigskip

\noindent\emph{Conclusions}

\medskip

\noindent  The asymptotic structure of AF and ALD spacetimes appears very similar, as can be noted from the Penrose diagrams \ref{penrosemink}-\ref{penroseALD}. If the latter were to dictate how the holographic dictionary is organised, then one would expect very similar outcomes in the two cases. 

This, however, does not appear to be the case. For one, the NS5 decoupling limit identifies $i^0$ of the ALD spacetime (and not $\mathcal{I}^\pm$) as the spacetime where the boundary theory lives. In AF spacetime, the analogous time coordinate near $i^0$ would be that of the de Sitter Rindler slices; however, this is not obviously related to the time coordinate on the branes' worldvolume in the string theory realisation. Of course, given that there is no known decoupling limit to flat space, the relation between the bulk and the boundary description is not  well understood.

Another superficial resemblance between the two spacetimes are the asymptotic symmetries defined at spatial infinity, which in both cases consist of two copies of Witt generators, plus some additional affine $U(1)$ ones. Of course, the asymptotic symmetry algebras are not the same, as one set of generators is field-dependent, and the other not, and the commutation relations between the two types of generators are also somewhat different. In the AF case, the BMS generators at spacelike infinity are projected  to a particular parity, which matches smoothly to the generators on $\mathcal{I}^\pm $, which are thus antipodally identified. However, it is not clear why the existence of a smooth extension to $\mathcal{I}^\pm $ should be a requirement from the point of view of the symmetries of the theory; in particular, there is no such smooth extension for any non-trivial generator in the ALD case, yet the symmetries are highly appealing through their perfect match to $T\bar T$.

%Similarities: causal structure (Penrose diagram), asymptotic metric. More precisely, ALD in standard coordinate very similar to Minkowski in dS slicing. Analogy  $i^0$, Penrose diagram. \emph{Global ALD vacuum?}  symmetries

 Finally, correlation functions at $i^0$ should also be part of the boundary data. In ALD, they should naturally reveal the constrained non-local structure of the boundary theory; in AF, it is less clear what this structure is, but it also appears to be non-local \cite{Cotler:2025npu}. 
%
%Differences: symmetries, presence stable black holes, decoupling vs non-decoupling. 
%
%Another difference that is interesting is the fact that ALD is obtained from a decoupling limit, which strongly suggests a duality to the non-gravitational LST living on flat space, at the $i^0$ of ALD. Note this is not the conformal boundary of the ALD spacetime, and puts under a question mark the standard assumption that the holographic dual to flat space lives on the null boundary/celestial sphere. It is thus not clear to which extent   the asymptotic structure of a spacetime is supposed to guide holography. This question is important, because it affects the expected structure of the theory (non-local, vs Carrollian, or whatever.)
%
%In any case, the ALD framework is under much better control. An interesting question is what are the observables: correlation functions and/or scattering? A comparison between correlation functions in ALD and flat has been made by Marolf, who highlighted inherent non-locality. However, note ALD wave equation was not the correct one. \emph{True?} 
%
%Another obvious observable are the black holes, which are stable in the ALD case, at least in a first approximation.   It would be interesting if the effective irrelevant deformation picture could apply to both ALD and AF black holes, as suggested by their similar symplectic-form-based analyses. 
It would be very interesting if a set of universal rules for (non-AdS) holography could be extracted through the study of such asymptotically similar spacetimes.  So far, it is not clear what form such putative universal rules would take. 

%We already discussed potential interpretation of AF ones in terms of $\mu <0$ $T\bar T$.
%
%\bi
%\item compare ASG between ALD and flat Compere-Fiorucci
%\item also compare ASG above AF D1-D5 background vs  standard AF black holes
%\ei
%
%If this correspondence is better understood (in particular, one has an independent definition for dipole/LS-CFTs), then may be able to infer rules for holography in non-AdS spacetimes. % In this section (also the intro) we mostly wanted to emphasize the choices that more-or-less secretly underlie each step.

\subsubsection*{Acknowledgements}

The author is grateful to Alejandra Castro, Jos\'e Barb\'on, Alex Belin, Sergei Dubovsky, Silvia Georgescu, Victor Gorbenko,  Sean Hartnoll, Bob  Knighton,  Per Kraus,  Mukund Rangamani,  Shiraz Minwalla, Ruben Monten,  Anthony  Speranza, Toby Wiseman and Sasha Zhiboedov for stimulating discussions, and especially to Alejandra Castro, Silvia Georgescu and Ruben Monten for useful comments on the draft. She would also like to thank the `Centro de Ciencias de Benasque Pedro Pascual' and the `Physics Sessions Initiative' at the Pollica Physics Centre,   where part of this work was completed, for hospitality and a
wonderful discussion environment. %She also acknowledges support Exact methods in low-dimensional quantum systems

\end{document}